\documentclass[10pt]{iopart}
\newcommand{\Review}[1]{#1}

\usepackage{amsfonts}
\usepackage{amssymb}
\newcommand{\hp}{\hphantom}
\newcommand{\text}{\mbox}
\usepackage{verbatim}
\usepackage{ifthen}
\usepackage{lipsum}

\Review{\newcommand{\showreview}{***REVIEW***}} 
	{\ifthenelse{\isundefined{\showlecture}}%
	{\expandafter\comment}%
	{}%
}%
{%
	\ifthenelse{\isundefined{\showlecture}}%
	{\expandafter\endcomment}%
	{}%
}
\newenvironment{reviewKay}%
    {\ifthenelse{\isundefined{\showreview}}%
    	{\expandafter\comment}%
        {}%
    }%
    {\ifthenelse{\isundefined{\showlecture}}%
        {\expandafter\endcomment}%
        {}%
}      
\newenvironment{widefigure}
{
\Review{\begin{figure*}}
}
{
\Review{\end{figure*}}
}

\Review{\newcommand{\komma}{,}
\newcommand{\punkt}{.}
}

\usepackage{graphicx}
\usepackage{times}
\usepackage{color}

\usepackage{tikz}
\usetikzlibrary{arrows,shapes,positioning,decorations.pathmorphing}
\usetikzlibrary{decorations.markings}
\usetikzlibrary{snakes}
\tikzset{snake it/.style={decorate, decoration=snake,segment length=3mm}}
\tikzstyle arrowstyle=[scale=1]
\tikzstyle directed=[postaction={decorate,decoration={markings,
    mark=at position .65 with {\arrow[arrowstyle]{stealth}}}}]
\tikzstyle endreversedirected=[postaction={decorate,decoration={markings,
    mark=at position 1.0 with {\arrow[arrowstyle]{stealth}}}}]
\tikzstyle enddirected=[postaction={decorate,decoration={markings,
    mark=at position 1.0 with {\arrow[arrowstyle]{stealth}}}}]
\tikzstyle reverse directed=[postaction={decorate,decoration={markings,
    mark=at position .65 with {\arrowreversed[arrowstyle]{stealth};}}}]
\usetikzlibrary{decorations.markings}
\tikzset{->-/.style={decoration={
  markings,
  mark=at position #1 with {\arrow{>}}},postaction={decorate}}}

\newcommand{\figinfo}[1]{}

\definecolor{grey}{cmyk}{.4,.2,0,0}

\usepackage[colorlinks=true,linkcolor=blue,urlcolor=blue,citecolor=blue]{hyperref}

\definecolor{labelkey}{cmyk}{.4,.2,0,0}
\newcommand{\blue}{\color{blue}}

\newcommand{\new}{\color{black}}
\newcommand{\rot}{\color{red}}

\graphicspath{{./figures/},{}}

\Review{
\evensidemargin 25pt
\oddsidemargin  25pt
}

\definecolor{exercisecolor}{cmyk}{1,.8,0.2,0}

\Review{\newcommand{\highlight}[1]{#1}}
\newlength{\highlightlength}
\newlength{\highlightlengthbis}

\Review{\newcommand{\hl}[1]{#1}}

\newcommand{\percent}{\protect\%}
\newcommand{\Mathematica}[1]{}
\newcommand {\ds}{\displaystyle}

\Review{\newcommand{\dotsb}{\ldots}}
\usepackage{pifont}

\newcommand{\half}{\frac{1}{2}}
\newcommand{\ts}{\hskip0.1ex\raisebox{-1ex}[0ex][0.8ex]{\rule{0.1ex}{2.75ex}\hskip0.2ex}}
\newcommand{\lts}{
   {\raisebox{0ex}{\parbox{0.5ex}{
      \setlength{\unitlength}{0.5ex}
      \begin{picture}(1,12)
         \thinlines
         \put(0.5,-0.25){\line(0,1){12}}
      \end{picture}
   }}}
}

\newcommand{\f}{{\rm f}}

\newcommand{\fig}[2]{\includegraphics[width=#1]{#2}}
\newcommand{\Fig}[1]{\includegraphics[width=0.485\textwidth]{#1}}

\newlength{\bilderlength}
\newcommand{\bilderscale}{0.35}

\newcommand{\bilderskip}{\hspace*{0.8ex}}

\newcommand{\diagram}[1]{%
\settowidth{\bilderlength}{\bilderskip%
\includegraphics[scale=\bilderscale]{#1}\bilderskip}%
\parbox{\bilderlength}{\bilderskip%
\includegraphics[scale=\bilderscale]{#1}\bilderskip}}

\newcommand{\nn}{\nonumber}\newcommand {\eq}[1]{(\ref{#1})}

\newcommand {\Eq}[1]{Eq.\hspace{0.55ex}(\ref{#1})}
\newcommand {\REF}[1]{Ref.~\cite{#1}}
\newcommand {\Eqs}[1]{Eqs.\hspace{0.55ex}(\ref{#1})}

\newcommand{\ket}[1]{\left|#1\right>}

\newcommand{\E}{\varepsilon }
\newcommand{\R}{\mathbb{R}}

\newcommand{\beq}{\begin{equation}}
\newcommand{\eeq}{\end{equation}}
\newcommand{\be}{\begin{equation}}
\newcommand{\ee}{\end{equation}}
\newcommand{\bea}{\begin{eqnarray}}
\newcommand{\eea}{\end{eqnarray}}

\newcommand{\tf}{t_{\rm f}}
\newcommand{\ti}{t_{\rm i}}
\newcommand{\tm}{t_{\rm m}}
\newcommand{\xfi}{x_{\rm f}}
\newcommand{\xin}{x_{\rm i}}

\newcommand{\ca}[1]{{\cal #1}}

\newcommand{\KPZ}{{\rm \scriptscriptstyle KPZ}}

\newbox{\expbox}
\newlength{\explength}

\newbox{\atbox}
\newlength{\atlengtha}
\newlength{\atlengthb}

\newcommand{\bn}{:\hspace*{-0.5ex}}
\newcommand{\en}{\hspace*{-0.5ex}:}
\newcommand{\ah}{{\hat a}}
\newcommand{\ha}{{\hat a}}
\newcommand{\ad}{{\hat a}^\dagger}

\newcommand{\LH}{{\parbox{1.15cm}{{\begin{tikzpicture}
\coordinate (x1) at  (0.35,0) ; 
\coordinate (x2) at  (1.35,0) ; 
\coordinate (x3) at  (0.85,0.866025) ; 
\fill (x1) circle (2pt);
\fill (x2) circle (2pt);
\fill (x3) circle (2pt);
\draw  (x2) arc(60:120:1);
\draw  (x2) arc(-60:-120:1);
\draw  (x1)--(x3)--(x2);
\end{tikzpicture}}}}}

\newcommand{\Ione}{{\parbox{1.15cm}{{\begin{tikzpicture}
\coordinate (x1) at  (0.35,0) ; 
\coordinate (x2) at  (1.35,0) ; 
\fill (x1) circle (2pt);
\fill (x2) circle (2pt);
\draw   (x2) arc(60:120:1);
\draw   (x2) arc(-60:-120:1);
\end{tikzpicture}}}}}

\newcommand{\ITP}{{\parbox{0.82cm}{{\begin{tikzpicture}
\coordinate (x1) at  (0,0) ; 
\fill (x1) circle (2pt);
\draw  (x1) arc(-90:270:0.4);
\end{tikzpicture}}}}}

\newcounter{exercisecounter}

\begin{document}

\makeatletter
\def\@fnsymbol#1{^{\arabic{footnote}}\relax}
\makeatother
\setcounter{footnote}{0}

\Review{

\vspace*{-2cm}

\title{\sffamily    Theory and Experiments for Disordered Elastic Manifolds, Depinning, Avalanches, and Sandpiles}

\author{\sffamily\bfseries\normalsize Kay J\"org Wiese}
\address{Laboratoire de physique, D\'epartement de physique de l'ENS, \'Ecole normale sup\'erieure, UPMC Univ. Paris 06, CNRS, PSL Research University, 75005 Paris, France}

\enlargethispage{5.5cm}

\begin{abstract}
 Domain walls in magnets, vortex lattices in superconductors, contact lines at depinning, and many other systems  can   be modeled as an
elastic system subject to quenched disorder.  The ensuing field theory possesses a well-controlled perturbative 
expansion around its upper critical dimension. Contrary to standard field theory,  the renormalization group flow involves a function,  the   disorder correlator $\Delta(w)$, and is   therefore termed the functional renormalization group (FRG). $\Delta(w)$ is a physical observable,   the auto-correlation function of the center of mass of the elastic manifold.
In this review, we give a pedagogical introduction into its phenomenology and  techniques. This allows us to treat both equilibrium (statics), and depinning (dynamics). Building on these techniques,  avalanche observables are accessible: distributions of size, duration, and velocity,  as well as the  spatial and  temporal shape.
Various equivalences between disordered elastic manifolds, and sandpile models exist: an elastic string driven at a point and the Oslo model; disordered elastic manifolds   and  Manna sandpiles; charge density waves and  Abelian sandpiles or loop-erased random walks. 
Each of the mappings between these systems requires specific techniques, which we develop, including modeling of discrete stochastic systems via coarse-grained stochastic equations of motion, super-symmetry techniques, and cellular automata. Stronger than quadratic     nearest-neighbor interactions  lead  to directed percolation, and non-linear surface growth with  additional KPZ terms. On the other hand, KPZ without disorder can be mapped back to disordered elastic manifolds, either on the directed polymer for its steady state, or a single particle for its decay.
Other topics covered are the relation between functional RG and replica symmetry breaking, and random field magnets. Emphasis is given to numerical and experimental tests of the theory. 

\end{abstract}

\bigskip

\noindent
\hspace*{-.7ex}{{\fig{17.5cm}{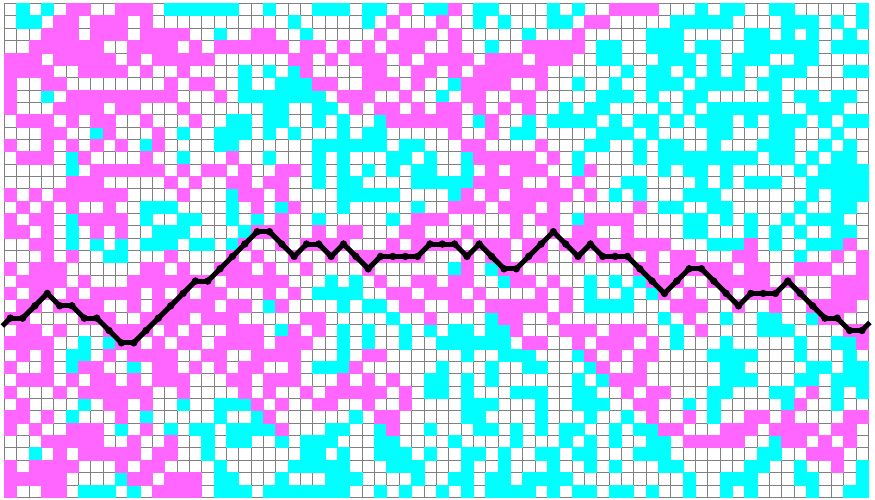}}}
Anisotropic depinning with its relation to directed percolation, explained in section \ref{s:qKPZ}.

\bigskip
\bigskip

\centerline{\blue{\huge~~~~~~~~~~~~~~~~~~~~~~~~} VERSION 3.0 --- 12 August 2022, close to the published version in 
Rep. Prog. Phys. 85 (2022) 086502}

\def\ioptwocol
{\setlength\hoffset{-0.5in}\setlength\voffset{-0.5in}\setlength\textwidth{6.75in}
\setlength\columnsep{0.2in}\setlength\textheight{9.25in}\mathindent=0in\twocolumn}

\ioptwocol \tableofcontents

\title{\sffamily    Theory and Experiments for Disordered Elastic Manifolds, Depinning, Avalanches, and Sandpiles}

}

\section*{Foreword}
This review grew out of lectures the author gave in the ICTP master program at ENS Paris. While the beginning of each section is  elementary, later parts are   more specialized and can be skipped at first reading. Beginners wishing to enter the subject are encouraged to start reading  sections \ref{intro} (introduction), \ref{s:General remarks about renormalization}-\ref{beyond1loop} (equilibrium/statics), and \ref{s:Phenomenology}-\ref{s:dep-loops} (depinning/dynamics).
An introduction to avalanches is given in sections \ref{s:Observables and scaling relations}-\ref{s:ABBM}, \ref{s:BFM}-\ref{s:Avalanche-size distribution in the BFM}.  The remaining sections are more specialized: Sandpile models and anisotropic depinning are treated in sections \ref{s:sandpiles} and \ref{s:CSPI}. An introduction to the KPZ equation and its relation to disordered elastic systems is given in section \ref{s:KPZ, Burgers, and the directed polymer}. Section \ref{s:links} discusses links between a class of theories encompassing loop-erased random walks, charge density waves, Abelian sandpiles, and $n$-component $\phi^4$ theory with $n=-2$, linked by supermathematics. 
Further developments and ideas are collected in section \ref{s:Further developments and ideas}. The appendix \ref{s:Appendix: Basic Tools} contains useful basic tools.

\section{Disordered Elastic Manifolds: Phenomenology}
\label{intro}
\subsection{Introduction}

Statistical mechanics is by now a rather mature branch of physics.
For pure systems like a ferromagnet, it allows one to calculate with precision
details as the behavior of the specific heat on approaching the
Curie point. We know that it diverges as a function of the distance in 
temperature to the Curie temperature, we know that this divergence has
the form of a power law, we can calculate the exponent, and we can do
this with at least 3 digits of accuracy using the perturbative renormalization group   \cite{Amit,Zinn,CardyBook,KardarBook,BrezinBook,Vasilev2004,ParisiBook,PelissettoVicari2002}, and even more precisely with the newly developed conformal bootstrap \cite{El-ShowkPaulosPolandRychkovSimmons-DuffinVichi2014,El-ShowkPaulosPolandRychkovSimmons-DuffinVichi2012,ChesterLandryLiuPolandSimmons-DuffinNing-Su2019}. Best of all, these findings
are in excellent agreement with the most precise simulations \cite{FerrenbergXuLandau2018,ClisbyDunweg2016,Clisby2017}, and experiments \cite{LipaSwansonNissenGengWilliamsonStrickerChuiIsraelssonLarson2000}. This is
a true success story of statistical physics.  On the other hand, in
nature no system is really pure, i.e.\ without at least some disorder
(``dirt'').  As experiments (and theory)   suggest, a little bit
of disorder does not change much. Otherwise experiments
on the specific heat of Helium\footnote{Even though there is some tension between values obtained in a space-shuttle experiment \cite{LipaSwansonNissenGengWilliamsonStrickerChuiIsraelssonLarson2000} on one side, and simulations \cite{Hasenbusch2019} and the conformal bootstrap \cite{ChesterLandryLiuPolandSimmons-DuffinNing-Su2019} on the other hand.}   would not so extraordinarily well
confirm theoretical predictions. But what happens for strong disorder?
By this we mean that disorder   dominates over entropy, so effectively the system is at zero temperature. Then
already the question: ``What is the ground state?'' is no longer
simple. This goes hand in hand with the appearance of {\em 
metastable states}. States, which in energy are   close to the
ground state, but which in configuration space may be far apart. Any
relaxational dynamics will take an enormous time to find the correct
ground state, and may fail altogether, as can be seen in
computer simulations as well as in experiments, particularly in glasses \cite{FranzJacquinParisiUrbaniZamponi2012}. This means that our
way of thinking, taught in the treatment of pure systems, has to be
adapted to account for disorder. We will see that in contrast to pure
systems, whose universal large-scale properties can be ``modeled by
  few parameters'', disordered systems demand to model the
whole disorder-correlation function (in contrast to its first few
moments). We show how universality nevertheless emerges.

Experimental realizations of strongly disordered systems are glasses,
or more specifically spin glasses, vortex glasses, electron glasses
and structural glasses  \cite{MuellerWyart2014,NattermannScheidl2000,KierfeldNattermannHwa1997,CarpentierLedoussalGiamarchi1996,CuleShapir1995,HwaFisher1994a,HwaFisher1994b,Balents1993,FranzJacquinParisiUrbaniZamponi2012}.  Furthermore random-field
magnets \cite{Feldman2002,MiddletonFisher2002,DahmenSethnaKuntzPerkovic2001,DahmenSethnaPerkovic2000,BricmontKupiainen1987,Imbrie1984,ParisiSourlas1979,LeDoussalWiese2005b,TissierTarjus2011,TarjusTissier2008,TissierTarjus2008b,TarjusTissier2005,TarjusTissier2006,TarjusTissier2004}, and last not least elastic systems subject to disorder, sometimes termed {\em disordered elastic systems} or {\em disordered elastic manifolds} \cite{HusemannWiese2017,WieseHusemannLeDoussal2018,WieseLeDoussal2006,HuiTang2006,FedorenkoLeDoussalWiese2006b,Wiese2004,Wiese2005,RepainBauerJametFerreMouginChappertBernas2004,BolechRosso2004,LeDoussalWiese2003b,LeDoussalWiese2003a,Wiese2003a,Wiese2002,RossoKrauth2001b,CuleHwa1998}, on which we   focus below. 

What is our current understanding of disordered   systems?
There are a few exact solutions, mostly for idealized or toy systems \cite{Derrida1980}, there are phenomenological approaches 
(like the droplet model \cite{FisherHuse1986}, section \ref{s:droplet}), and there is a mean-field approximation,
involving a method called replica-symmetry breaking (RSB) \cite{MezardParisiVirasoroBook}. This method
  predicts the properties of infinitely connected systems, as e.g.\ the  
Sherrington-Kirkpatrick (SK) model \cite{KirkpatrickSherrington1978,SherringtonKirkpatrick1975}. The  solution   proposed in 1979 by G.~Parisi \cite{Parisi1979} is parameterized by a function $q(x)$, where $x$ ``lives between replica indices $0$ and $1$''.  Today we have a much better understanding of this solution \cite{MezardParisiSourlasToulouseVirasoro1984,MezardParisiVirasoro1985,CugliandoloKurchan1993}, and many features can  be proven rigorously  \cite{Guerra2003,Talagrand2011a,Talagrand2011b,Panchenko2013}. The most notable feature is the presence of an extensive number of ground states arranged in a hierarchic way (ultrametricity). 
On the other hand, this solution is inappropriate for systems in which each degree of freedom is coupled only to its neighbors, as is e.g.\ the case in short-ranged magnetic systems.

While the RSB method mentioned above is intellectually challenging and  rewarding, its complexity  makes intuition difficult, and performing a field theoretic expansion around this mean-field solution has proven too challenging a task. Random-field models, which can be recast in a $\phi^4$-type theory are seemingly more tractable, but still the non-linearity of the  $\phi^4$-interaction makes progress difficult. What one would like to have is a field theory which in absence of disorder is as simple as possible. 
The simplest such system certainly is a non-interacting, Gaussian, i.e.\ free  theory, to which one could then add disorder. Actually, experimental systems of this type are abundant:
Magnetic domain walls in presence of disorder a.k.a.\ Barkhausen noise  \cite{Barkhausen1919,CizeauZapperiDurinStanley1997,DurinBohnCorreaSommerDoussalWiese2016}, a contact line wetting a disordered substrate  \cite{LeDoussalWieseMoulinetRolley2009},   fracture in brittle heterogeneous systems \cite{PonsonBonamyBouchaud2007,BonamyPonsonPradesBouchaudGuillot2006,PonsonBonamyBouchaud2006}, or earthquakes \cite{GutenbergRichter1956} are good examples for  elastic systems subject to quenched disorder.  
They have a quite different phenomenology from mean-field models, with notably a single ground state. Asking questions about this ground state, or more generally the probability measure at a given temperature, is termed {\em equilibrium}. It supposes that if external parameters change, they change so slowly that the system has   enough time to explore the full phase space (ergodicity), and  find the ground state. 

In the opposite limit, notably if there are no thermal fluctuations at all, is {\em depinning}: 
Increasing an external applied field yields   jumps in the center-of-mass of the system (the total magnetization in a magnet). These jumps are  termed shocks or avalanches. While one can show that the sequence of 
avalanches is deterministic given a specific disorder (see below), we are more interested in typical behavior, i.e.\ an average over disorder. The latter average can often be obtained by watching the system for an extended time; one says that  the system is {\em self-averaging}\footnote{\new In contrast to disordered elastic manifolds, some disordered systems such as   long-range spin glasses are not self-averaging, which leads to  replica-symmetry breaking and a hierarchic organization of states, see \cite{BinderYoung1986} for a review, and  section \ref{s:RSB} for a discussion in our context. The presence of a finite correlation length as given in \Eqs{xi-stat}, \eq{def:nu} and \eq{xi-m}   insures self-averaging.}.

In these lectures combined with a review, I aim at explaining the field theory behind these phenomena. 
All key ingredients   are in addition derived analytically in well-chosen toy models.  Theoretically most exciting are the connections between seemingly unrelated models. Finally, all main theoretical concepts are checked in  experiments.
While the field theory has   been developed for more than thirty years,   no comprehensive and pedagogical introduction is yet available. It is my aim to close this gap.
Despite the more than 700 references included in this review, I am aware of   omissions. 
My apologies to all  colleagues whose work is not covered in depth. 
Luckily, some of them have written reviews or lectures themselves, and we refer the reader to   
\cite{NATOASISeries1995,Kardar1997,GiamarchiLeDoussalBookYoung,DSFisher1998,NattermannScheidl2000,BrazovskiiNattermann2004,LeDoussal2008,PruessnerBook}  for complementary presentations.

\subsection{Physical realizations, model and observables}\label{model}
\begin{widefigure}
\centerline{\fig{0.25\textwidth}{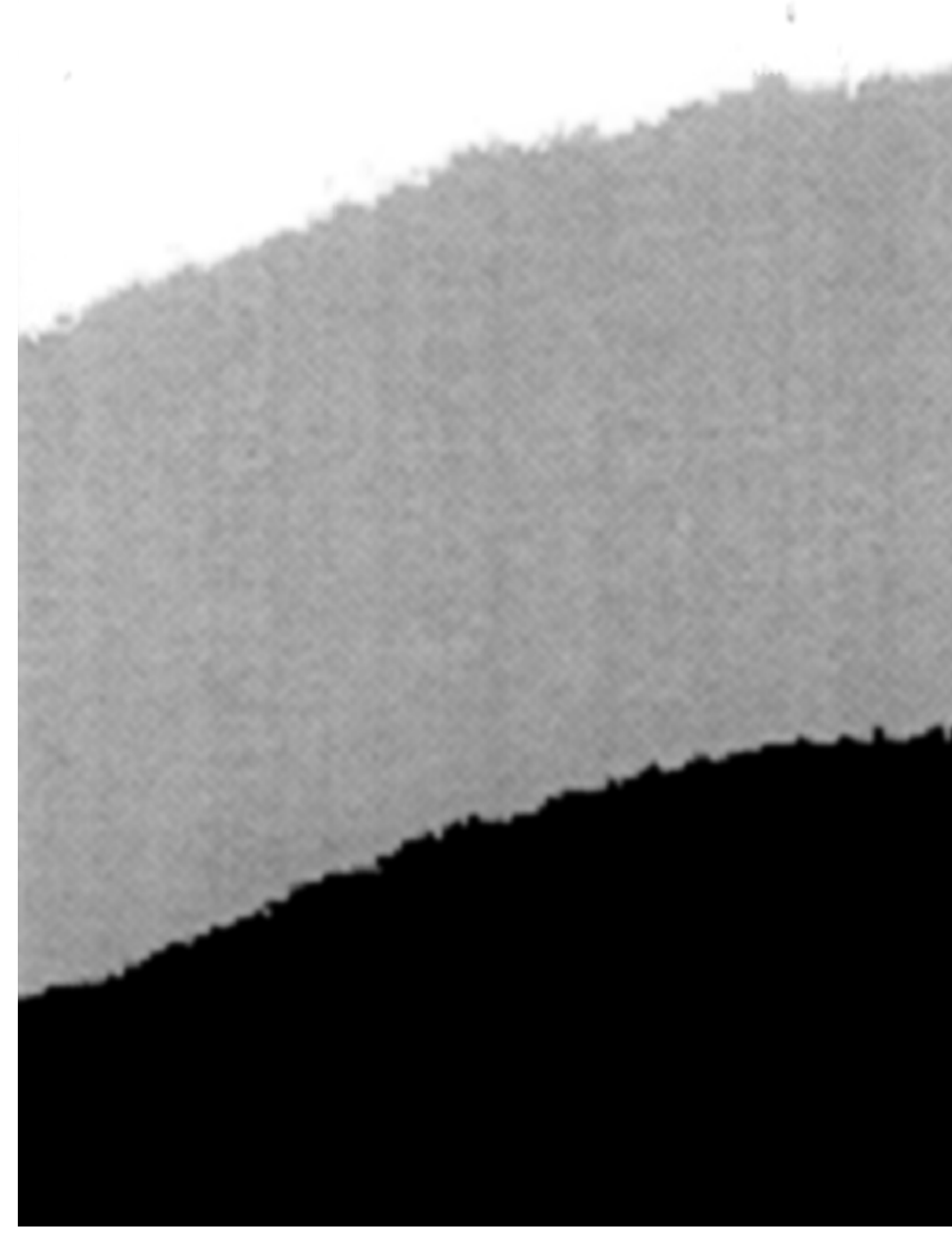}~~~~~~\fig{0.7\textwidth}{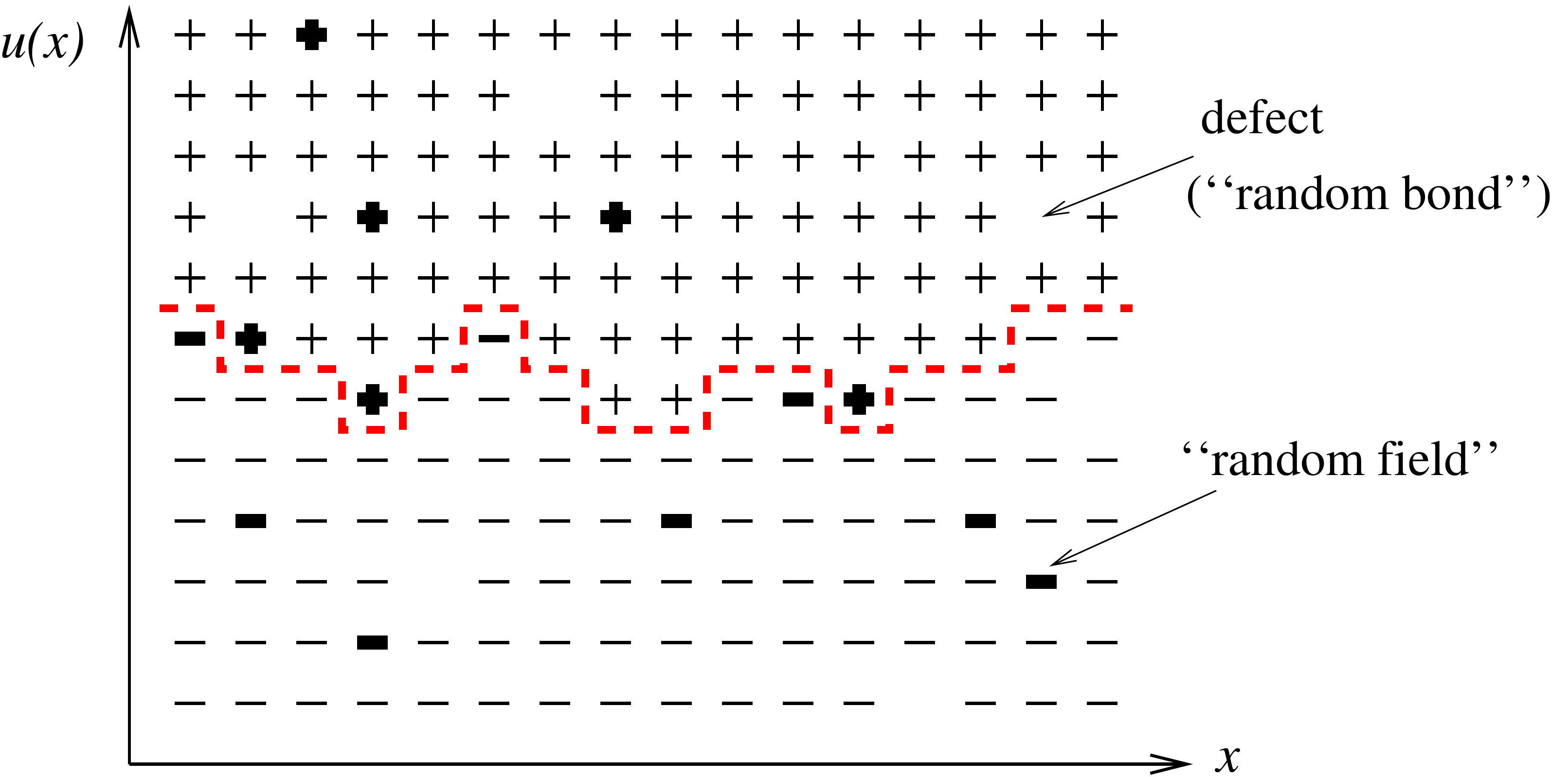}}
\caption{\new An Ising magnet with up (``$+$'') and down (``$-$'') spins at low temperatures forms a domain wall
described by a function $u (x)$ (right) (Fig.~from \cite{WieseLeDoussal2006}).   Two types of disorder are observed: missing spins, weakening the effective nearest-neighbor interactions (``random-bond disorder''), and frozen in magnetic moments aligning its immediate neighbor (``random field disorder''), indicated by thick $\pm$ signs. An experiment on a thin
Cobalt film (left)
\protect\cite{LemerleFerreChappertMathetGiamarchiLeDoussal1998}; with
kind permission of the authors.}
\label{exp:Magnet}
\centerline{{\parbox{0.52\textwidth}{\fig{0.52\textwidth}{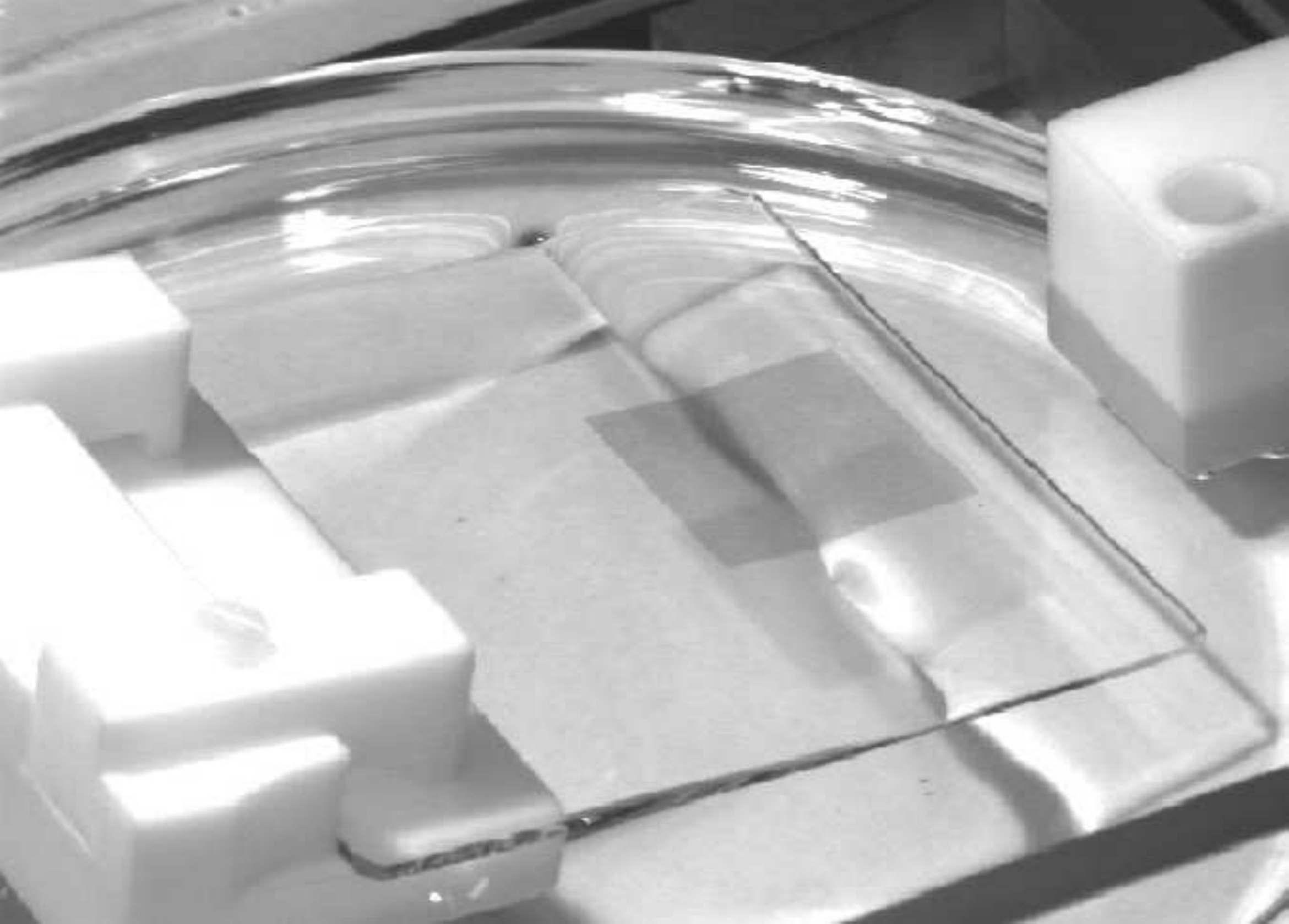}}}~~~~
{\parbox{0.44\textwidth}{\begin{minipage}{\textwidth}
\Fig{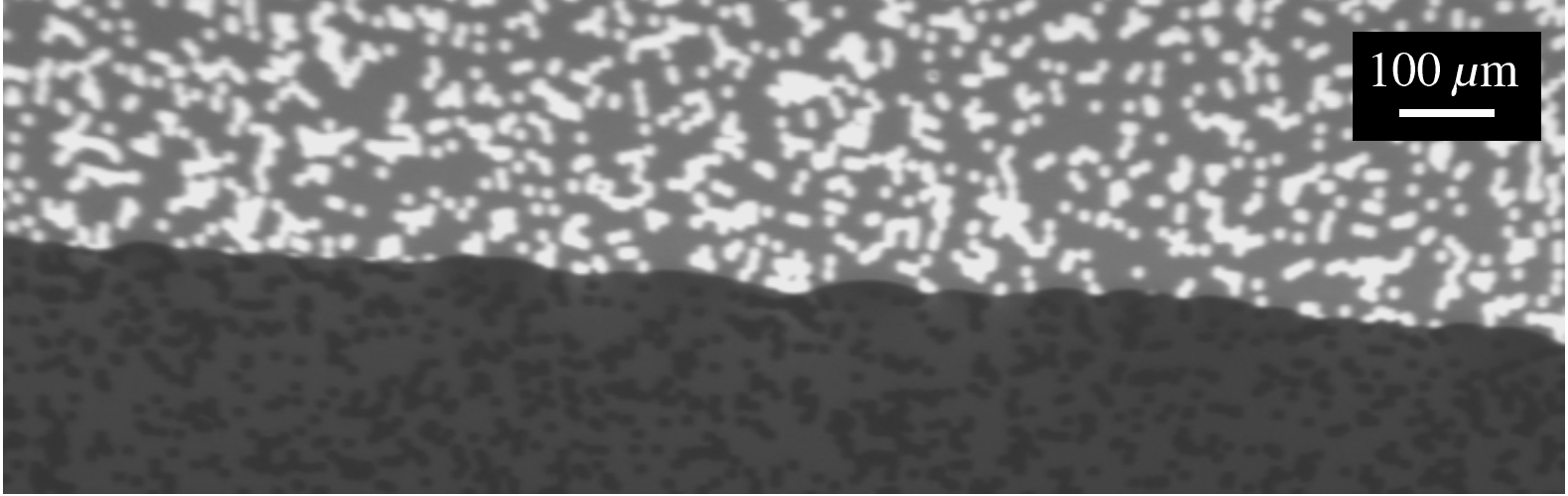}\\
\Fig{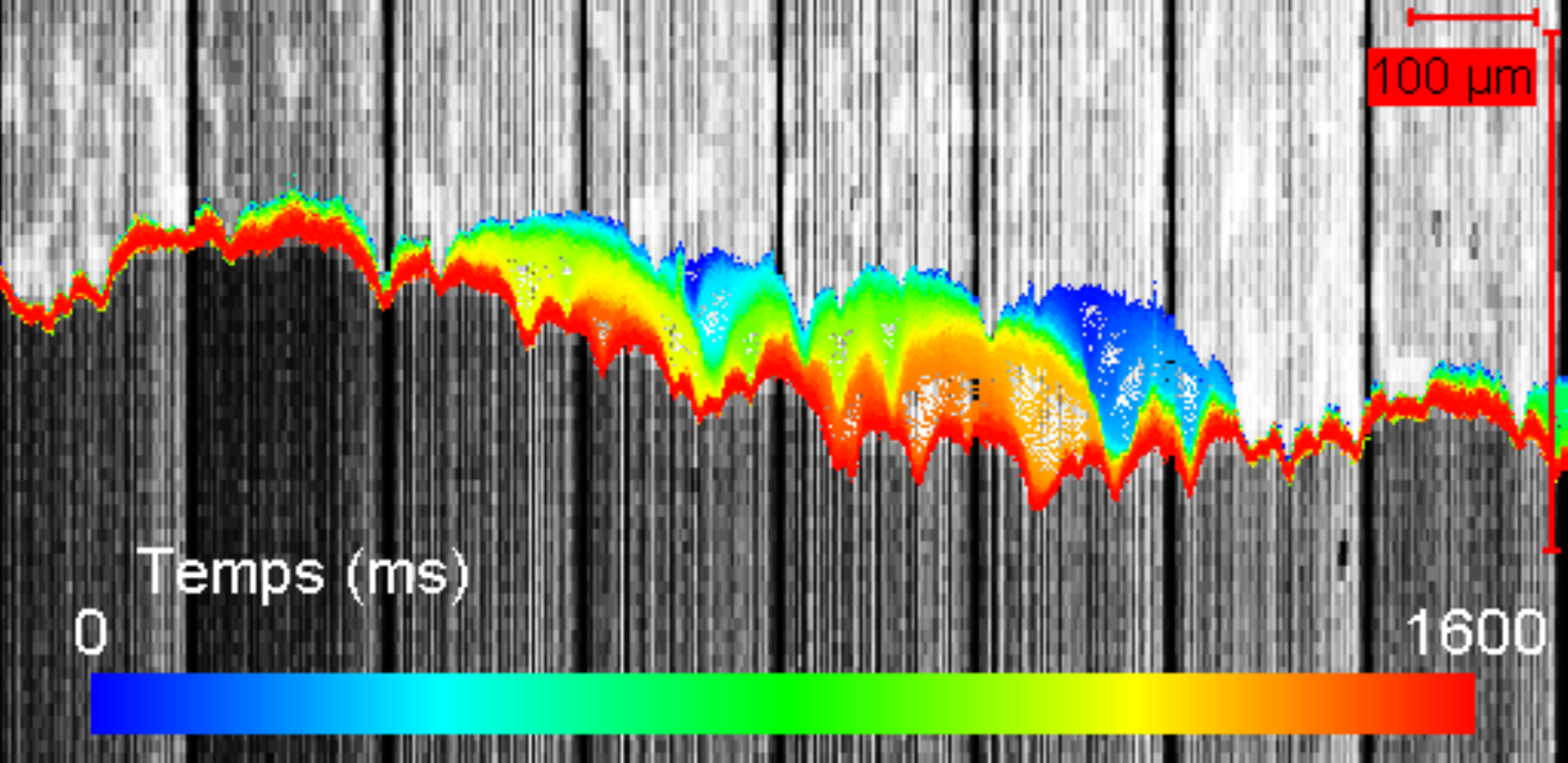}
\end{minipage}}}}
\caption{A contact line for the wetting of a disordered substrate by
Glycerine \protect\cite{MoulinetGuthmannRolley2002}. Experimental setup
(left). The disorder consists of randomly deposited islands of
Chromium, appearing as bright spots (top right), \new with a correlation length of about $10\mu m$. Temporal evolution of
the retreating contact line (bottom right). Note the different scales
parallel and perpendicular to the contact line. Pictures courtesy of
S.~Moulinet, with kind permission.}  \label{exp:contact-line}
\end{widefigure}
\begin{figure}\label{f:vortex-lattic}
\centerline{\mbox{\setlength{\unitlength}{0.75cm}
\begin{picture}(10.8,5.3)
\put(9.3,3.4){$\vec u(x,y,z)$}
\put(8.8,5){$  (x,y,z)$}
\put(0,0){{\fig{10\unitlength}{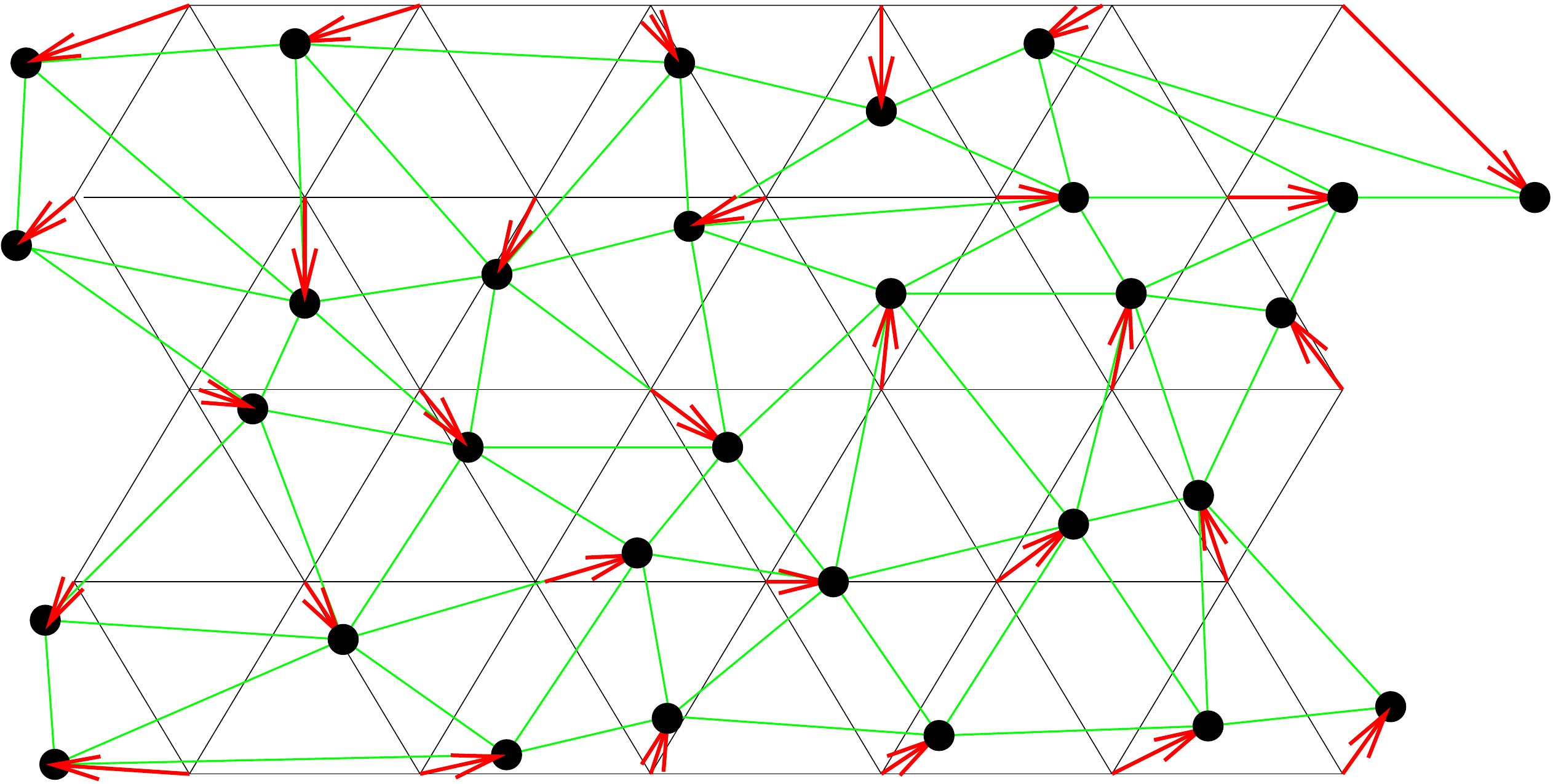}}}
\end{picture}
}}
\caption{A vortex lattice is described by a deformation of a lattice point $(x,y,z)$ to $(x,y,z)+\vec u(x,y,z)$.
Shown is a cartoon of a single layer, i.e.\ fixed $z$. The vortex lines continue perpendicular to the drawing (Fig. from \cite{WieseLeDoussal2006}).}
\label{sketch-Bragg-glass}
\end{figure}

Before developing the theory to treat elastic systems in a disordered
environment, let us give some physical realizations. The simplest one
is an Ising magnet. Imposing boundary conditions with all spins up at
the upper and all spins down at the lower boundary (see figure \ref{exp:Magnet}), at
low temperatures, a domain wall separates a region with spins  up from a
region with spins down. In a pure system at temperature $T=0$, this
domain wall is   flat.  Disorder can deform the domain wall,
making it eventually rough. 
Figure 1 shows, how the domain wall is described by a
displacement field $u (\vec x)$.
Two types of disorder are common:
\begin{enumerate}
\item[(i)] {\em random-bond} disorder (RB), 
where the bonds between neighboring sites are random. 
On a course grained level this also represents missing
spins. The correlations of the random potential are short-ranged. 
\item[(ii)] 
{\em random-field} disorder (RF), i.e.\ coupling of the spins to an external random
magnetic field. This disorder is ``long-ranged'', as the random potential is the sum over the random fields below the domain wall, i.e.\ effectively has the statistics of a random walk. 
 Taking a derivative of the potential,  one obtains  short-ranged correlated random forces. 
\end{enumerate} 
Another example is the contact line of
a liquid (water, isobutanol, or liquid helium), wetting a rough {\new (ideally scale-free)} substrate, see figure
\ref{exp:contact-line}. Here,   elasticity becomes   {\em long-ranged}, see   \Eq{Hel-contact-line} below. 

A realization
with a 2-parameter displacement field $\vec{u} (x,y,z) $ is the
deformation of a vortex lattice, see figure \ref{sketch-Bragg-glass}: the position of each vortex is
deformed from the 3-dimensional vector $\vec x = ( x,y,z)$ to $\vec x + \vec u (\vec x)$, with $\vec u \in \mathbb R^2$ (its $z$-component is $0$).  
Irradiating the sample produces line defects. They allow experimentalists to realize \cite{FedorenkoLeDoussalWiese2006b}
\begin{enumerate}
\item[(iii)]  generic {\em long-range} (LR) correlated disorder. The most extreme example are 
\item[(iv)] {\em random forces with the statistics of a random walk}. This model, the Brownian force model (BFM) of section \ref{s:BFM},  plays an important role as its center-of-mass motion advances as a single degree of freedom, known as the {\em ABBM}  or  {\em mean-field} model (section \ref{s:ABBM}), often used to describe avalanches. 
\end{enumerate}
Another   
example are charge-density waves, first predicted by Peierls \cite{Peierls1955}: They can spontaneously form in certain semiconductor  devices, where a uniform charge density is unstable towards a super lattice in which   the underlying lattice is periodically  deformed, and the charge density of the globally neutral device becomes \cite{FukuyamaLee1978,LeeRice1979,Gruner1988,Monceau2012}.
\be
\rho(\vec x) =\rho_0 \cos(\vec k \vec x)\punkt
\ee 
Adding disorder, the latter locally deforms the phase, modifying the charge density to  
\be
\rho\big(\vec x, u(\vec x)\big) =\rho_0 \cos\big(\vec k \vec x+2 \pi u(\vec x) \big)\punkt
\ee
As the charge density is invariant under $u(\vec x) \to u(\vec x)+1$, we find another disorder class, 
\begin{enumerate}
\item[(v)] {\em random periodic} disorder (RP).
\end{enumerate}
\medskip
All these models have in common  that they can be described
by a displacement field
\begin{equation}\label{u}
\vec x\in \R^d \ \longrightarrow\  \vec u (\vec x) \in \R^N
\punkt
\end{equation}
For simplicity, we suppress the vector notation wherever possible, and mostly consider $N=1$.  After some initial
coarse graining, the energy ${\cal H}={\cal H}_{\mathrm{el}}+{\cal H}_{\rm conf} + {\cal
H}_{{\rm dis}}$ consists   of three parts: the elastic energy
\begin{equation}
\label{Hel}
\highlight{ {\cal H}_{\mathrm{el}}[u] = \int \rmd ^d x \, \half\! \left[ \nabla u
(x)\right]^2 }\komma 
\end{equation}
the confining potential
\be\label{Hconf}
\highlight{ {\cal H}_{\rm conf}[u]=    \int \rmd ^d x \, \frac{m^2}2\left[ u(x)-w\right]^2} \komma 
\ee
and the disorder
\begin{equation}\label{HDO}
\highlight{  {\cal H}_{{\rm dis}}[u] = \int \rmd ^{d} x \, V \big(x,u (x)\big)}\punkt
\end{equation}
In order to proceed, we need to specify the  correlations of
disorder. Suppose that fluctuations of $u$   scale  as
\begin{equation}\label{roughness}
\highlight{  \overline{ \left< \left[u (x)-u (y) \right]^{2} \right>}  \sim  |x-y|^{2\zeta }}\punkt
\end{equation}
Notations are such that $\left< ... \right>$ denotes thermal averages, i.e.\ averages of an observable using the weight $\rme^{-\beta H}$, properly normalized by the partition function $Z = \left < \rme^{-\beta \ca H}\right>$. 
{\new At zero temperature, this reduces to the contribution of a single state, the ground state.}
Overbars denote the average over disorder. 
This defines  a {\em roughness-exponent} $\zeta$. 
Starting from a disorder
correlator
\begin{equation}
{  \overline{V (x,u)V (x',u')} = R (u-u') f (x-x') }
\end{equation}
with both $R(u)$ and $f(x)$ vanishing at large distances, 
 for each   rescaling  in the RG procedure by $\lambda$ in the  $x$-direction one   rescales by $\lambda^\zeta$ in the  $u$-direction. As long as $\zeta<1$, this    eventually
reduces $f (x)$ to a $\delta $-distribution, whereas the structure
of $R (u)$ may remain visible.  We therefore choose as our
starting correlations for the disorder
\begin{equation}\label{DOcorrelR}
\highlight{\overline{V (x,u)V (x',u')} :=  R (u-u') \delta ^{d } (x-x')}
\punkt
\end{equation}
As we do not consider higher cumulants of the disorder, this implicitly assumes that the distribution of the disorder  is Gaussian\footnote{\label{foot:KardarLH1994}For the concept of cumulants see e.g.~Ref.~\cite{KardarLH1994}.}.

There are a couple of useful observables. We already mentioned the
roughness exponent $\zeta $. The second is the renormalized
(effective) disorder  $R (u)$.  

Noting by $F(x,u):= -\partial_u V(x,u)$ the disorder forces, the  corresponding force-force correlator can be written as
\be \label{Delta-def}
\highlight{\overline{\left<F(x,u) F(x',u') \right>} = \Delta (u-u')\delta ^{d } (x-x') \punkt}
\ee
Since 
$
\overline{\left<F(x,u) F(x',u') \right>}= \partial_u\partial_{u'}\overline{V (x,u)V (x',u')}  =-  R'' (u-u')\delta ^{d } (x-x') $, we 
identify 
\be\label{Delta-R-rel}
 \highlight{ \Delta (u) = -R''(u)\punkt}
\ee

\subsection{Long-range elasticity (contact line of a fluid, fracture, earthquakes, magnets with dipolar interactions)}
\label{s:LR-elasticity}
\begin{figure}
\centerline{\fig{8.5cm}{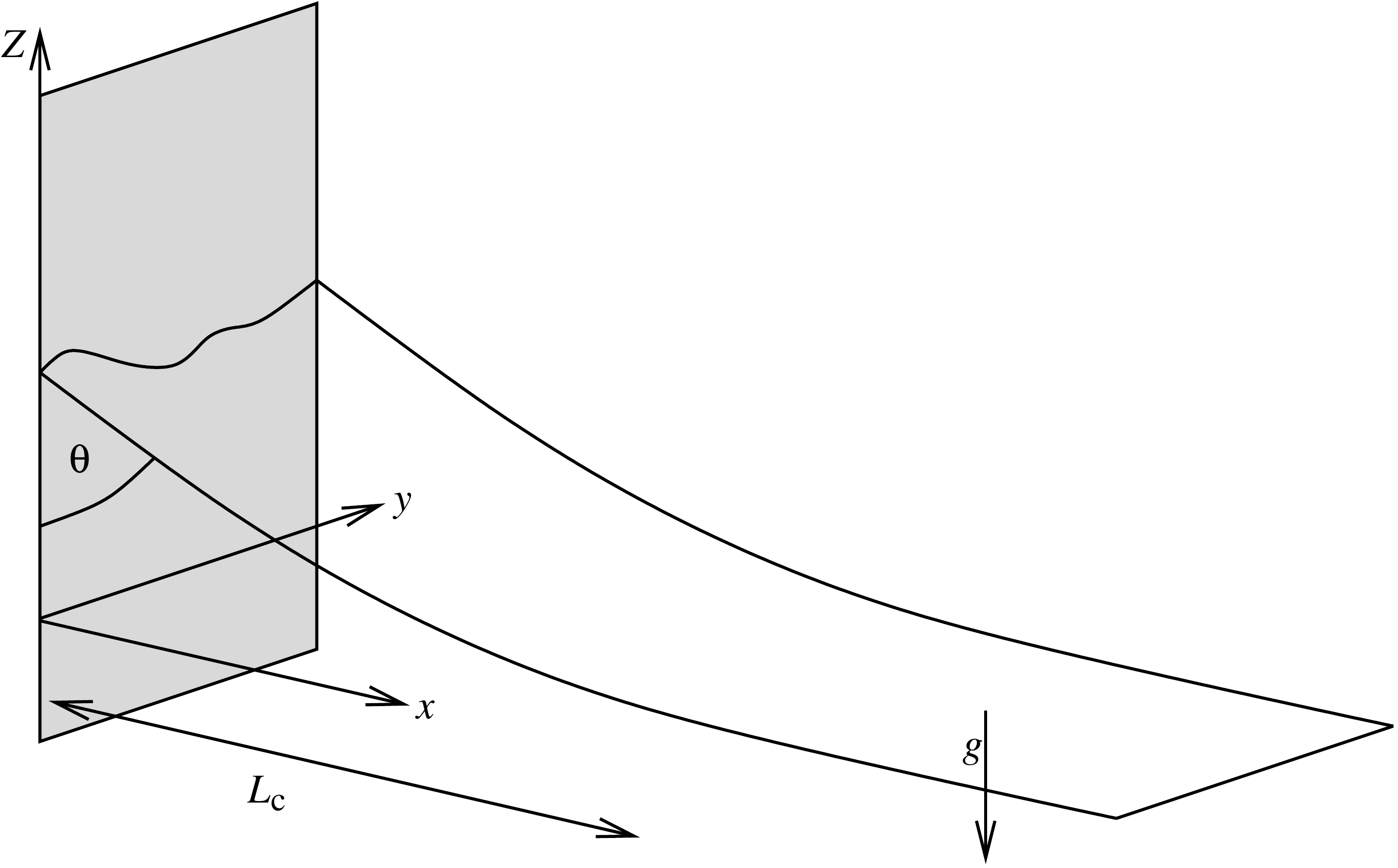}}
\caption{The coordinate system for a vertical wall.The air/liquid interface becomes flat for large $x$. The height $h(x,y)$ is along the $z$-direction.
\protect\figinfo{figure similar to EPL 87 (2009) 56001 and Phys. Rev. E 75 (2007) 031601.}}
\label{f:contact-line-energy}       \end{figure}
There are several relevant experimental systems for which the elasticity is different from \Eq{Hel}. This mostly happens when the elasticity of a  lower-dimensional subsystem  is mediated by the surrounding bulk.
The simplest such example is a contact line \cite{BrochardGennes1991} in a coffee mug or water bottle, i.e.\ the line where coffee, cup and air meet. A laboratory example is shown in Fig.~\ref{exp:contact-line}. For fracture this was introduced in \cite{Rice1985}.

Consider  a liquid with height $h(x,y)$,  defined in the half space $x\ge 0$ (see Fig.~\ref{f:contact-line-energy}). Its elastic energy is surface tension times   surface area, i.e.
\bea\label{Hel-liquid}
{\cal H}_{\rm el}^{\rm liquid}[h] &=& \gamma \int_y \int_{x>0}  \sqrt{ 1+ [\nabla h(x,y)]^2} \nn\\
&\simeq& \mbox{const.}+ \int_y \int_{x>0} \frac\gamma2 [\nabla h(x,y)]^2
\eea
We wish to express this as  a function of the height $u(y):=h(0,y)$ on the boundary at $x=0$. A minimum energy configuration satisfies 
\be\label{4.13}
0   = \frac{\delta {\cal H}_{\rm el}^{\rm liquid}[h]}{\delta h(x,y)} = -\gamma \nabla^2 h(x,y)\punkt
\ee
This is achieved by   the ansatz
\be
h(x,y) = \int \frac{\rmd k}{2\pi}\tilde u(k) \,\rme^{i k y - |k| x},  
\ee
which decays to zero at large $x$.
On the boundary at $x=0$ this is the standard Fourier transform of the height $u(y)$. 
Integrating by parts, the elastic energy as a function of $\tilde u(k)$ becomes with the help of \Eq{4.13}
\Review{\bea\label{Hel-contact-line}
&& {\cal H}_{\rm el}^{\rm liquid}[u] = \int_y \int_{x\ge 0} \frac\gamma2 [\nabla h(x,y)]^2  \nn\\
&& = 
 \frac{\gamma} 2\left[   \int_y \int_{x\ge 0} \nabla\Big( h(x,y) \nabla h(x,y)\Big) -h(x,y) \nabla ^2 h(x,y) \right] \nn\\
&&= - \frac \gamma 2 \int_y h(x,y) \partial_x h(x,y)\Big|_{x=0} \nn\\
&&=   \frac \gamma 2 \int \frac{\rmd k}{2\pi }\, |k|\, \tilde u(k) \tilde u(-k).
\eea}
In generalization of \Eq{Hel-contact-line} one can write
\be\label{Hel-alpha-k}
{\cal H}_{\rm el}^{ \alpha} [u] = \frac12 \int \frac{\rmd^d k}{(2\pi)^d }\, |k|^\alpha\, \tilde u(k) \tilde u(-k)\punkt
\ee
For $\alpha=2$, this is equivalent to the local interaction of \Eq{Hel}. For $\alpha<2$, the interaction is non-local in position space, 
\numparts
\bea\label{Hel-alpha-pos}
{\cal H}^{ \alpha} _{\rm el} [u] = \frac{\ca A_d^{\alpha}}2 \int  {\rmd^d \vec x}\int  {\rmd^d \vec y}\, \frac{ \left[  u(\vec x)-u(\vec y) \right]^2}{|\vec x- \vec y|^{d+\alpha} },\\
\label{A-d-alpha}
\ca A_d^{\alpha} = -\frac{2^{\alpha -1} 
   \Gamma  (\frac{d+\alpha
   }{2} )}{ \pi ^{\frac d 2} \Gamma
    (-\frac{\alpha
   }{2} )}.
\eea
\endnumparts
For $d=\alpha=1$ this yields
\be\label{Hel-alpha=1-pos}
{\cal H}^{\alpha= 1} _{\rm el} [h] = \frac{1}{4\pi}\int  {\rmd  x}\int  {\rmd  y} \,\frac{ \left[  u(x)-u(y) \right]^2}{|x-y|^{2} }.
\ee
Note that for $\alpha\to 2$, $\ca A_d \sim (2-\alpha)$, reducing the long-range kernel to the  short-range one.

\Eq{Hel-liquid} is an approximation, as higher-order terms are neglected. The latter can be generated efficiently \cite{BachasLeDoussalWiese2006}, and may change the physics of the system \cite{LeDoussalWieseRaphaelGolestanian2004}.  
When the contact angle  is different from the inclination of the wall, the elastic energy is further modified  \cite{LeDoussalWiese2009a}.

The theory we   develop below works for arbitrary (positive) $\alpha$, with $\alpha=2$ for standard {\em short-ranged} elasticity, and $\alpha=1$ for (standard) {\em long-ranged} elasticity. Apart from contact lines, long-ranged elasticity with $\alpha=1$ appears for a $d$-dimensional elastic object (a surface), where the elastic interactions are mediated by a   bulk material of higher dimension $D>d$.
Important   examples are the displacement of tectonic plates relevant to describe   earthquakes ($d=2$, $D=3$) \cite{GutenbergRichter1944,GutenbergRichter1956,DSFisher1998} and fracture ($d=1$, $D=2$ or $D=3$) \cite{BonamyPonsonPradesBouchaudGuillot2006,PonsonBonamyBouchaud2006}.

For magnetic domain walls ($d=2$) with dipolar interactions, the interactions are also long-ranged. The elastic kernel is given by \cite{ZapperiCizeauDurinStanley1998} (page 6357)
\bea
\ca H_{\rm el}[u] = \gamma \int \rmd^2 \vec r_1\int \rmd^2 {\vec r_2}\, \frac{\partial_{x_1} u(\vec r_1) \partial_{x_2} u(\vec r_2)}{|\vec r_1-\vec r_2|} , \nn \\
\vec r_1=(x_1,y_1), \quad \vec r_2=(x_2,y_2). 
\eea
In Fourier space, this reads
\be
\ca H_{\rm el}[u]  = \frac{\gamma}{2\pi}\int \rmd^2 \vec k\, \tilde u(\vec k) \tilde u(-\vec k) \frac{k_x^2}{|\vec k|}.
\ee

\subsection{Flory estimates and bounds}\label{a1}
Above, we distinguished four types of disorder, resulting in four different
universality classes:
\begin{itemize}
\item [ (i)] Random-Bond disorder (RB): short-range
correlated potential-potential correlations, i.e.\ a short-range
correlated $R (u)$.
\item [ (ii)] Random-Field disorder (RF): a short-range
correlated force-force correlator $\Delta (u):= -R'' (u)$. As the name
says, this disorder is relevant for random-field systems where the
disorder potential is the sum over all magnetic fields below a domain wall.
\item [(iii)] Generic long-range correlated disorder (LR): $R (u)\sim |u|^{-\gamma
}$.
\item [(iv)] Random-Periodic disorder (RP): Relevant when the disorder couples
to a phase, as e.g.\ in charge-density waves. $R (u)=R (u+1)$,
supposing that $u$ is periodic with period 1.
\end{itemize}
To get an idea how large the roughness $\zeta$ becomes in these
situations, one compares the contributions of elastic energy and
disorder, and demands that they scale in the same way. This estimate
has first been used by Flory \cite{Flory1953} for self-avoiding polymers, and is therefore
called the Flory estimate
\footnote{For disordered systems this type of argument was   employed by Harris \cite{Harris1974} and Imry and Ma \cite{ImryMa1975}, and the reader will find reference to them as well.}. Despite the fact that Flory estimates are
conceptually crude, they often give a decent
approximation. For RB disorder, this gives for an $N$-component field $u$:
$\int_{x} u |\nabla|^\alpha u \sim \int_{x} \sqrt{\overline{VV}}$, or $
L^{d-\alpha} u^2 \sim L^{d} \sqrt{L^{-d}u^{-N}} $, i.e.\ $u \sim L ^{\zeta
}$ with
\begin{equation}\label{a2}
\zeta_{\mathrm{Flory}}^{\mathrm{RB}} = \frac{2\alpha-d}{4+N} \stackrel{\alpha\to 2}{=} \frac{4-d}{4+N}\punkt
\end{equation}
For RF disorder   $\Delta(u) = -R''(u)$   is short-ranged, and  
\begin{equation}\label{a3}
\zeta_{\mathrm{Flory}}^{\mathrm{RF}} = \frac{2\alpha-d}{2+N} \stackrel{\alpha\to 2}{=} \frac{4-d}{2+N}\punkt
\end{equation}
For generic LR correlated disorder 
\begin{equation}\label{a4}
\zeta_{\mathrm{Flory}}^{\mathrm{LR}} = \frac{2\alpha-d}{4+\gamma} \stackrel{\alpha\to 2}{=}  \frac{4-d}{4+\gamma }\punkt
\end{equation}
For RP disorder the field $u$ cannot be rescaled or one would break   periodicity,  and thus 
\be
\zeta^{\mathrm{RP}}=0
\ee
exactly. We will see below in section \ref{s:FRG-fixed-points} that these estimates are a decent approximation, and even exact for RF at $N=1$, or for LR disorder.

\subsection{Replica trick and basic perturbation theory}
\label{s:replicas}
In disordered systems, a particular configuration  strongly depends on the disorder, and therefore  statements about a specific configuration are in general meaningless. What one needs to calculate are averages, of the form (``gs'' denotes the ground state)
\bea
\lefteqn{\overline{ {\cal O}[u]} := \overline{ \frac{\left<{ \cal O}[u]\,\rme^{-  {\cal H}[u]/T}\right>}{\left<\rme^{-{\cal H}[u]/T}\right>}}} \nn\\
&& \stackrel{T\to 0}{-\!\!\!-\!\!\!\longrightarrow}
\overline{ \frac{ { \cal O}[u_{\rm gs}]\,\rme^{- {\cal H}[u_{\rm gs}]/T}}{ \rme^{-  {\cal H}[u_{\rm gs}]/T} }} \equiv
\overline{ { \cal O}[u_{\rm gs}]}\punkt
\eea
Note that division by the partition function ${  Z}={\left<\rme^{-  {\cal H}[u]/T}\right>}$   is crucial. This is particularly pronounced in the limit of $T\to 0$, where ${  Z} = \rme^{-  {\cal H}[u_{\rm gs}]/T} $ diverges or vanishes when $T\to 0$, except if by chance ${\cal H}[u_{\rm gs}]=0$. 
Thus the denominator can not be replaced by its mean. This is a difficult situation: while integer powers ${  Z}^n$, with $n\in \mathbb N$ can be obtained by using $n$ {\em copies} or {\em replicas} of the system, $1/  Z$ cannot. 
On the other hand, we  observe that, {\em } independent of $n$, 
\be
\overline{ {\cal O}[u] }= \overline{ \frac{\left<{ \cal O}[u]\,\rme^{- {\cal H}[u]/T}\right> {  Z}^{n-1}}{{  Z}^n}} \punkt
\ee
The replica-trick \cite{Brout1959,EdwardsAnderson1975}\footnote{\label{f:Brout}It is not quite clear who ``invented'' the replica trick. In Ref.~\cite{Brout1959} R.~Brout stresses that $\ln Z$ has to be averaged over disorder,  not $ Z$. Brout considers a cluster expansion for a quenched disordered system, organizing his expansion in powers of $n$, equivalent to  a cumulant expansion, or  {\em sums over independent replicas},   concepts we use below.} consists in doing the calculations for arbitrary $n$. This is  possible    in perturbation theory, as  results there    are polynomials in $n$. It may become troublesome for exact solutions (notably leading to replica-symmetry breaking \cite{MezardParisiVirasoroBook}). Knowing the dependence on $n$, the idea is to set  $n\to 0$ at the end of the calculation,  thus eliminating the denominator,
\be
\overline{ {\cal O}[u] }= \lim _{n\to 0}\overline{  {\left<{ \cal O}[u]\,\rme^{- {\cal H}[u]/T}\right> {  Z}^{n-1}}} \punkt
\ee
Since   thermal averages over distinct replicas factorize, we   write their joint measure as 
\Review{
\bea
\overline{  {\left<{ \cal O}[u] \,\rme^{- {\cal H}[u]/T}\right> {Z}^{n-1}}} = \overline {  \left<  { \cal O}[u_1]\prod_{a=1}^n \rme^{-  {\cal H}[u_a]/T} \right>  } \nn\\
=  \left<  \overline {  { \cal O}[u_1] \rme^{-\sum_{a=1}^n   {\cal H}[u_a]/T}} \right>  \nn \\
=  \left<  { \cal O}[u_1] \,  \overline { \rme^{ - {\frac1T} \sum_{a=1}^n \ca H_{\rm el}[u_a]+ \ca H_{\rm conf}[u_a] + \ca H_{\rm dis}[u_a] }}\, \right> \punkt   
\eea}
Note that in the second equality we have exchanged thermal and disorder averages. We   also allowed for different positions $w$ of the parabola for the different replicas, denoted $w_a$\footnote{The partition function for each of these replicas may be different. The formalism takes this into account.}.
Finally,  we assume for simplicity of presentation  that ${\cal O}[u]$ does not explicitly depend on the disorder. 
Since only the last term in the exponential depends on $V(x,u)$, and since $V(x,u)$   is Gauss distributed,  
\bea
 \overline { \rme^{ - {\frac1T} \sum_{a=1}^n  \ca H_{\rm dis}[u_a] }} = 
\overline { \exp \!\left ( {-\frac1T  \int_x \sum_a V(x,u_a(x))} \right) } \nn\\
= \exp \!\left ( {\frac{1}{2T^2} \int_x \int_{y} \sum_{a,b=1}^n \overline{V(x,u_a(x)) V(y,u_b(y))}  } \right) \nn\\
 = \exp \!\left ( {\frac{1}{2T^2}  \int_x \sum_{a,b=1}^n  R\big(u_a(x) - u_b(x)\big)  } \right)\punkt
\eea
In the second step we used that $V$ is Gaussian; in the last step we used the correlator \eq{DOcorrelR}.

To summarize: to evaluate the expectation of  an observable, we take averages with measure $\rme^{- {\cal S}_{\rm rep}[u]}$ and {\em replica Hamiltonian} or {\em action} 
\bea \label{H}
\highlight { {\cal S}_{\rm rep}[u] &:=&      \frac1T\sum_{a=1}^n \int_x \left\{ \half [\nabla u_a(x)]^2 + \frac{m^2}2 [u_a(x)-w_a]^2 \right\}\nn\\
 && - {\frac{1}{2T^2} \int_x  \sum_{a,b=1}^n  R\big(u_a(x) - u_b(x)\big)  }    \punkt   }
\eea
Note that each replica sum comes with a factor of $1/T$. If the disorder had a third cumulant, this would appear  as a triple replica sum, and a factor${}^{\ref{foot:KardarLH1994},\ref{f:Brout}}$ of $1/T^3$.

Let us now turn to perturbation theory. 
The free propagator, constructed from the first line of \Eq{H}, and indicated by the index ``0'', is (first in Fourier, than in real space)
\bea\label{CorTk}
&&\highlight {  \left< \tilde u_a (-k) \tilde u_b(k) \right>_0  = T \delta_{ab} \tilde C(k) } \komma  \\
\label{CorT}
&&\highlight {  \left< u_a (x) u_b(y) \right>_0  = T \delta_{ab} C(x{-}y)\punkt} \eea
\Review{\numparts
Noting $C(x-y)$   the Fourier transform of $\tilde C(k)$, and $S_d=2 \pi^{d/2}/\Gamma(d/2)$   the area of the $d$-sphere, we have
\bea\label{m:cor1bis}
  \tilde C(k) &= \frac1{k^2+m^{2}} ,\\
 C(x) &= \int \frac{\rmd^d k}{(2\pi)^d} \frac{\rme^{i k x}}{k^2 + m^2}\nn\\
&\simeq\frac1{(d-2)S_d} |x|^{2-d} ~\mbox{ for }~ x\to 0\punkt
\eea\endnumparts}{\new On the other hand, for large $x$, the correlation function decays exponentially $\sim \rme^{-m|x|}$, which we associate with a correlation length}
\be\label{xi-stat}
\xi = \frac1 m.
\ee
\Eq{m:cor1bis} allows us to  calculate   expectation values in the full theory. As an example consider 
\bea\label{uu-correl-0}
\overline{
\big< [u (x)-w_1] \big>_{w_1} \big< [u (z)-w_2] \big>_{w_2} }^{\,\rm c} \nn\\
\equiv 
\big< [u_1 (x)-w_1]  [u_2 (z)-w_2] \big>  _{\cal S_{\rm rep}} \nn\\
= -\int_y C(x-y) C(z-y) R''(w_1-w_2)+ ...
\eea
Let us clarify the notations: Firstly, $\big< [u (x)-w_1] \big>_{w_1}$ is the thermal average of $u (x)-w_1$, obtained by evaluating the path integral   for a fixed disorder configuration $V$,   at a position of the parabola given by $w_1$. 
This procedure is   repeated for $\big< [u (z)-w_2] \big>_{w_2}$, with the same $V$, and parabola position $w_2$. Finally the average over the disorder potential $V$ is taken. According to the calculations   above, this can be evaluated with the help of the replica action ${\cal S}_{\rm rep}[u]$, represented by $\big< [u_1 (x)-w_1]  [u_2 (z)-w_2] \big>  _{\cal S_{\rm rep}}$. The latter is already averaged over disorder. 
The  last line shows the leading order in perturbation theory, dropping terms of order $T$ and higher. 

Finally, let us integrate this expression over $x$ and $z$, and multiply by $m^4/L^d$. This leads to 
\bea\label{uu-correl}
\frac{m^4}{L^d} \int_{x,z}
\overline{
\big< [u (x)-w_1] \big>_{w_1} \big< [u (z)-w_2] \big>_{w_2} }\nn\\
 = - {R''(w_1-w_2)} + ...
\eea
Let us understand the prefactor on the l.h.s.:
The combination $m^2[u (x)-w_1] $ is the force acting  on point $x$ (a density), its integral over $x$ the total force acting on the interface. Force correlations are short ranged in $x$, leading to the    factor of $1/L^d$. 
Note that the thermal 2-point function \eq{CorT} is absent, as we consider two distinct copies of the system.

\subsection{Dimensional reduction}\label{dimred}

\Review{It is an interesting exercise to show that  for $w_1=w_2 $  no perturbative corrections to \Eq{uu-correl} exist in the limit of $T\to0$, as long as one supposes that $R(w)$ is an analytic function. Similarly, one  shows that in the same limit $\left< u u u u\right>^{\rm c} =0$, and the same holds true for higher connected expectations. }
Thus $u$ is a Gaussian field with correlations
\bea
\overline{\left< \tilde u(k)\right>\left< \tilde u(-k)\right> }  
= \overline{\left< \tilde u(k) \tilde u(-k) \right>}  \nn\\
= \overline{ \tilde u(k) \tilde u(-k) } = -\frac{R''(0)}{(k^2+m^2)^2}\punkt
\eea
In the third expression we suppressed the thermal expectation values since at $T=0$ only a single ground state survives\footnote{For disordered elastic manifolds with continuous disorder, the ground state is almost surely unique. This is in strong contrast to mean-field spin glasses, where it is highly degenerate, see e.g.\ \cite{MezardParisiVirasoroBook}.}.
Fourier-transforming back to position space yields (with some    amplitude $\cal A$, and in the limit of $m x\to 0$)
\be\label{332}
\half \overline{  \big[ u(x) - u(y)\big]^2 } = - {R''(0)} \, {\cal A}|x-y|^{4-d}\punkt
\ee
This looks very much like the thermal expectation \eq{CorT}, except that the dimension of space has been shifted by $2$.
Further, both theories are seemingly Gaussian, i.e.\ higher cumulants vanish. 

We have just given a simple version of a  beautiful and rather
mind-boggling theorem relating disordered systems to pure ones
(i.e.\ without disorder). The theorem applies to a large class of systems,
even when non-linearities are present in the absence of disorder. It is
called dimensional reduction \cite{AharonyImryMa1976,EfetovLarkin1977,Young1977}.  We  formulate it  as
follows:

\noindent {\underline{``Theorem'':}} {\em A   $d$-dimensional disordered
system at zero temperature is equivalent to all orders in perturbation
theory to a pure   system in $d-2$ dimensions at finite temperature. 
}

We give in section \ref{a:susy} a proof of this theorem using a supersymmetric   field theory introduced in Ref.~\cite{ParisiSourlas1979}. The proof {\em implicitly assumes that $R(u)$ is analytic}, thus all derivatives can be taken.
The equivalence is  rather powerful, since the supersymmetric theory knows   about different replicas, and   allows   one to calculate even away from  the critical point.

However, evidence from experiments, simulations, and analytic solutions show that the above ``theorem'' is actually {\em wrong}. 
 A prominent counter-example is the 3-dimensional
random-field Ising model at zero temperature \cite{BricmontKupiainen1987}; according to the theorem
it should be equivalent to the pure 1-dimensional Ising-model at
finite temperature. While it was shown rigorously \cite{BricmontKupiainen1987} that the former
has an ordered phase,  the latter is disordered  at finite temperature \cite{Ising1925}. So what went
wrong? Let us stress that there are no missing diagrams or any such
thing, but that the problem is more fundamental: As we will see later,
the proof makes the assumption  that $R(u)$ is analytic. While this assumption is correct in the microscopic model, it is not valid at large scales. 

 Nevertheless,
the above ``theorem'' remains important since it has a devastating
consequence for all perturbative calculations in the disorder: However
clever a procedure we invent, as long as we perform a perturbative
expansion, expanding the disorder in its moments, all our efforts are
futile: dimensional reduction tells us that we get a trivial and
unphysical result. Before we try to understand why this is so and how
to overcome it, let us give one more counter-example. Dimensional reduction
allowed us in \Eq{332} to calculate the roughness-exponent $\zeta $ defined in
equation (\ref{roughness}), as 
\be\label{zetaDR}
 \zeta_{\rm DR} =\frac{4-d}{2}\punkt
\ee
On the other hand, the directed polymer in dimension $d=1$ does not have  a roughness exponent of $\zeta_{\rm DR}=3/2$, but 
\cite{Kardar1987}
\be
\zeta_{d=1}^{\rm RB} =\frac23\punkt
\ee
Experiments and simulations for disordered elastic manifolds   discussed below in sections \ref{s:Simulations in equilibrium},
\ref{s:Experiments at equilibrium}, \ref{s:Simulation strategies}, 
\ref{s:Characterisation of the  1-dimensional string},
\ref{s:Experiments on contact-line depinning}, \ref{s:fracture},
\ref{s:Experiment on RNA-DNA unzipping}, and 
\ref{s:Experiments on thin magnetic   films} all violate dimensional reduction.

\subsection{Larkin-length, and the role of temperature}\label{Larkin}
To understand the failure of dimensional reduction, let us turn to crucial arguments given by Larkin \cite{Larkin1970}. He considers a
piece of an elastic manifold of size $L$. If the disorder has
correlation length $r$, and characteristic potential energy $\bar {\cal E}$, there are $(L/r)^d$ independent degrees of freedom, and according to the central-limit theorem
this piece of size $L$ will typically see a potential energy of amplitude
\begin{equation}\label{335}
\ca E_{{\rm dis}} = \bar {\cal E} \left(\frac{L}{r} \right)^{\!\frac{d}{2}}\punkt
\end{equation}
On the other hand, the  elastic energy scales as
\begin{equation}
\ca E_{\mathrm{el}} = c\, L^{d-2}\punkt
\end{equation}
These energies are balanced at the  {\em Larkin-length} $L=L_{\rm c}$
with
\begin{equation}
L_{\rm c} = \left(\frac{c^{2}}{\bar {\cal E}^{2}}r^{d} \right)^{\frac{1}{4-d}}
\punkt
\end{equation}
More important than this value is the observation that in all
physically interesting dimensions $d<d_{\rm c}=4$, and at scales $L>L_{\rm c}$, the disorder energy \eq{335} wins; as a consequence the 
manifold is pinned by disorder, whereas on small scales the elastic energy
dominates. For long-ranged elasticity, the same argument implies
\be
d_{\rm c} = 2 \alpha,   \mbox{ and disorder relevant for }d<d_{\rm c}.
\ee
Since the disorder has many minima which are far apart
in configurational space but close in energy (metastability), the
manifold can be in either of these minima, and local minimum does not imply global minimum. However, the existence of exactly one minimum is assumed in e.g.\ the proof of
dimensional reduction, {\new even though formally, the field theory sums over all saddle points.}

Another important question is the role of temperature. In
Eq.~(\ref{roughness}) we had supposed that $u$ scales with the system
size as $u\sim L^{\zeta}$. Demanding that the action \eq{H} be dimensionless, the first
term in \Eq{H} scales as $L^{d-2+2 \zeta}/T$. This implies that 
\begin{equation}\label{a8}
  T\sim a^{\theta}\komma \qquad  \theta =d-2+2 \zeta\komma 
\end{equation}
where $a$ is a microscopic cutoff with the dimension of $L$, to compensate the factor of $L^{d-2+2 \zeta}$. 
For completeness, we also give the result for generic LR-elasticity, 
\be
\theta_{\alpha} = d-\alpha +2 \zeta\punkt
\ee
The thermodynamic limit is obtained by taking $L\to \infty$. 
Temperature is thus irrelevant when $\theta >0$, which is the
case for $d>2$, and when $\zeta >0$ even below. As a consequence, the RG fixed point we
are  looking for   is at zero temperature \cite{Nattermann1985}.
The same argument applies to the  free energy 
\be\label{free-energy-scaling}
\ca F [u] = -\frac1T \ln ( Z[u]) \sim \left(\frac La\right) ^{\!\theta}\punkt
\ee
We   added $u$ as an argument to $\ca F[u]$, as e.g.\ in the directed polymer the partition function is the weight of all trajectories arriving at $u$. This is  important   in section \ref{s:KPZ} when considering the KPZ equation.

    From the second term in Eq.~(\ref{H}) we conclude that the (microscopic) disorder scales
as
\begin{equation}\label{R-scaling}
R\sim a^{2 \theta -d} = a^{d-4+4\zeta} \punkt
\end{equation}
For $\zeta=0$, this again implies that $d=4$ is the upper critical dimension. More thorough arguments are presented in the next section, where we will construct an $\epsilon=4-d$ expansion for the RG flow of $R(u)$.

\section{Equilibrium (statics)}
\label{s:The field-theoretic treatment}

{\new
\subsection{General remarks about renormalization}\label{s:General remarks about renormalization}
In the next section \ref{s:deriveRG} we   derive the central renormalization group equations for disordered elastic manifolds. These equations are obtained in a controlled $\epsilon=4-d$ expansion \cite{WilsonFisher1972} around the upper critical dimension. Retaining in this expansion only the leading divergences which show up as poles in $1/\epsilon$, by  using {\em minimal subtraction}, this expansion is unique. This is a deep result, ensured by  the {\em  renormalizability} of the theory (see e.g.\ \cite{BogoliubovParasiuk1957,Hepp1966,Zimmermann1969,BergereLam1975,RivasseauBook}). We consider it a {\em gift}: However we set up our RG scheme, we  always get the same result. 
This allows us to choose one scheme, and switch to a different one whenever its particular features help us in our reasoning. 
The  schemes in question are 
\begin{itemize}
\item[(i)] Wilson's momentum-shell scheme. This scheme goes back to K.~Wilson, who suggested to integrate over the {\em fast} modes, i.e.\ modes $k$ contained in a momentum shell between $\Lambda(1-\delta)$ and $\Lambda$, with $\delta\ll 1$. Doing this incrementally is interpreted as a flow equation for the effective parameters of the theory.  The process stops when one reaches the scale one is interested in, which is zero for correlations of the center of mass. While intuitive, this technique is cumbersome to implement, especially at subleading order. 
We refer to the classical text \cite{WilsonKogut1974} for an introduction. 
\item[(ii)] Field theory as used in  high-energy physics. This is the standard technique to treat critical phenomena, and is explained in many classical texts \cite{Amit,Zinn,KardarBook,BrezinBook,Vasilev2004,ParisiBook}. A well-oiled machinery, especially for higher-order calculations. 
\item[(iii)] The  operator product expansion  as explained in \cite{CardyBook}, or section 3.4 of \cite{WieseHabil}. Realizing that the dominant contributions in schemes (i) and (ii) come from large momenta implies that they must come from short distances in position space. It is not only very efficient at leading order \footnote{In $\phi^4$-theory it gives the 2-loop correction to $\eta$ from a single integral, see section 3.4 of \cite{WieseHabil}.}, it also explains why counter-terms are local (see below).  
\item[(iv)] Non-perturbative functional RG: A rather heavy machinery, which we believe should be restricted to cases where   other schemes fail (see section \ref{s:Random-field magnets} on   Random-Field magnets). 
\item[(iv)] The experimentalist's point of view:   If all RG procedures are equivalent, then we can {\em choose} to study the flow equations by reducing an experimentally relevant parameter, here the strength $m^2$ of the  confining potential. As we show below in sections \ref{s:shocks}-\ref{measurecusp}, the theory can be defined at any $m^2$, e.g.\ by doing an experiment or simulation at this scale. This definition does not make reference to any perturbative calculation. The latter can then be viewed as an efficient analytical {\em tool} to predict in an experiment or a simulation the consequences of a change of the parameter $m^2$.  
\end{itemize}
If we think about standard perturbative RG for   $\phi^4$ theory, we remark that the parameter $\epsilon$ controls the order of perturbation theory necessary\footnote{\new As a rule of thumb:  order $n$ in $\epsilon$ necessitate   order $n$ in the interaction  $\int_x\phi^4(x)$.}, and that at leading order $\ca O(\epsilon)$ the differences boil down to a choice of how to evaluate the elementary integral \eq{I1}. For disordered systems, there is an additional quirk: The {\em interaction} termed $R(u)$ in  \Eq{H} is a function of the field differences, and we have no a-priory knowledge of its form. It will turn out in the next section \ref{s:deriveRG} that we can write down a flow equation for the function $R(u)$ itself. 
We would already like to stress that similar to $\phi^4$-theory, the fixed point for $R(u)$ is of order $\epsilon$, thus the calculation remains perturbatively controlled.  
}

 \subsection{Derivation of the functional RG
equations}\label{s:deriveRG} In section \ref{Larkin}, we had seen
that 4 is the upper critical dimension for SR elasticity, which we treat now. As for standard critical
phenomena \cite{Amit,Zinn,CardyBook,KardarBook,BrezinBook,Vasilev2004,ParisiBook}, we   construct 
an $\epsilon= (4-d)$-expansion.  Taking the dimensional-reduction result
(\ref{zetaDR}) in $d=4$ dimensions tells us that the field $u$ is
dimensionless there. Thus, the width $\sigma = -R''(0)$ of the disorder is
not the only relevant coupling at small $\epsilon$, but any function
of $u$ has the same scaling dimension in the limit of $\epsilon=0$,
and might   equivalently contribute. The natural conclusion is to
follow the full function $R(u)$ under renormalization, instead of just
its second derivative $R''(0)$. 

Such an RG-treatment is most easily
implemented in the replica approach: The $n$ times replicated
partition function 
led after averaging over disorder to a path integral with weight $\rme^{-{\cal S}_{\rm rep}[u]}$, with action \eq{H}.
Perturbation theory is constructed as follows:  The
bare correlation function for replicas $a$ and $b$, graphically depicted as a solid line, is
with momentum $k$ flowing through, see \Eqs{CorTk}--\eq{m:cor1bis}, 
\begin{equation}
\left< \tilde u_a(k) \tilde u_b(k) \right>_0 = 
T \times~{\parbox{2.2cm}{{\begin{tikzpicture}
\coordinate (x1t1) at  (0,0) ; 
\coordinate (x1t2) at  (1.5,0) ; 
\node (t1) at  (-.25,0)    {$\!\!\!\parbox{0mm}{$\raisebox{-1mm}[0mm][0mm]{$\scriptstyle a$}$}$};
\node (t2) at  (1.75,0)    {$\!\!\!\parbox{0mm}{$\raisebox{-1mm}[0mm][0mm]{$\scriptstyle b$}$}$};
\draw [thick] (x1t1) -- (x1t2);
\end{tikzpicture}}}} =   T\delta_{ab}\tilde C(k)\punkt
\end{equation}
Note that the  factor of $T$ is   explicit in our graphical notation, and not included in the line. 
The disorder vertex is (we added an index $R_0$ to $R$ to indicate that this is the microscopic (bare) disorder)
\begin{equation}
\frac1{T^2}\times ~~
{\parbox{0.7cm}{{\begin{tikzpicture}
\coordinate (x1t1) at  (0,0) ; 
\coordinate (x1t2) at  (0,.5) ; 
\node (x) at  (0,0)    {$\!\!\!\parbox{0mm}{$\raisebox{-3.5mm}[2.5mm][2.5mm]{$\scriptstyle x$}$}$};
\node (t1) at  (-.25,0)    {$\!\!\!\parbox{0mm}{$\raisebox{-1mm}[0mm][0mm]{$\scriptstyle b$}$}$};
\node (t2) at  (-.25,0.5)    {$\!\!\!\parbox{0mm}{$\raisebox{-1mm}[0mm][0mm]{$\scriptstyle a$}$}$};
\fill (x1t1) circle (2pt);
\fill (x1t2) circle (2pt);
\draw [dashed,thick] (x1t1) -- (x1t2);
\end{tikzpicture}}}}=
\frac1{T^2}\times  R_{0}\Big(u_{a}(x)-u_{b}(x)\Big)\punkt
\end{equation}
The rules of the game are to find all contributions which correct $R$,
and which survive in the limit of $T\to 0$. At leading order, i.e.\ order
$R_{0}^{2}$, counting of factors  $T$ shows that we can use at most two correlators, as each contributes a factor of $T$. On the other hand, $\sum_{a,b}
R_{0}(u_{a}-u_{b})$ has two independent sums over replicas\footnote{The concept of {\em sums over independent replicas} already appears in the   work by R.~Brout \cite{Brout1959}, see footnote \ref{f:Brout}.}. Thus at order
$R_{0}^{2}$ four independent sums over replicas appear, and in order to
reduce them to two, one needs at least two correlators (each
contributing a $\delta_{ab}$). Thus, at leading order, only diagrams
with two propagators survive. 

Before writing down these diagrams, we need to see what   Wick-contractions do on functions of the field. 
To see this, remind that a single Wick contraction (indicated by 
{\Review{\setlength{\unitlength}{0.85cm}}
\fboxsep0mm\mbox{\begin{picture}(1.2,0.25)
\put(0.0,0.25){\line(1,0){1.2}}
\put(0.0,0.25){\line(0,-1){0.1}}
\put(1.2,0.25){\line(0,-1){0.1}}
\end{picture}}}\; sitting on top of the fields to be contracted)
\bea
\Review{\setlength{\unitlength}{0.85cm}}
\fboxsep0mm
\mbox{\begin{picture}(2.4,0.5)
\put(0,0){$ u_a(x)^n  u_b(y)^m $}
\put(0.1,0.5){\line(1,0){1.2}}
\put(0.1,0.5){\line(0,-1){0.1}}
\put(1.3,0.5){\line(0,-1){0.1}}
\end{picture}} \nn\\
= n u_a(x)^{n-1} \times m u_b(x)^{m-1}\times T \delta _{ab}C(x-y)\punkt
\eea
Realizing that $n   u^{n-1} = \partial_u u^n$, we can write the Wick contraction for an arbitrary function $V(u)$ as 
\bea
\Review{\setlength{\unitlength}{0.85cm}}
\fboxsep0mm
\mbox{\begin{picture}(3.3,0.5)
\put(0,0){$ V\big(u_a(x)\big)   V\big( u_b(y)\big) $}
\put(0.65,0.5){\line(1,0){1.7}}
\put(0.65,0.5){\line(0,-1){0.1}}
\put(2.35,0.5){\line(0,-1){0.1}}
\end{picture}} \nn\\
= V'\big(u_a(x)\big) \times  V'\big(u_b(y)\big) \times T \delta _{ab}C(x-y)\punkt
\label{45}
\eea
Graphically we   have at second order for the correction of disorder
\bea\frac1{2T^2} \delta R  = \frac1{2!}
\left[ 
\frac{1}{2T^2}  ~
{\parbox{0.6cm}{{\begin{tikzpicture}
\coordinate (x1t1) at  (0,0) ; 
\coordinate (x1t2) at  (0,.5) ; 
\node (x) at  (0,0)    {$\!\!\!\parbox{0mm}{$\raisebox{-3.5mm}[2.5mm][2.5mm]{$\scriptstyle x$}$}$};
\node (t1) at  (-.25,0)    {$\!\!\!\parbox{0mm}{$\raisebox{-1mm}[0mm][0mm]{$\scriptstyle b$}$}$};
\node (t2) at  (-.25,0.5)    {$\!\!\!\parbox{0mm}{$\raisebox{-1mm}[0mm][0mm]{$\scriptstyle a$}$}$};
\fill (x1t1) circle (2pt);
\fill (x1t2) circle (2pt);
\draw [dashed,thick] (x1t1) -- (x1t2);
\end{tikzpicture}}}} \right]
~~
{\parbox{2.2cm}{{\begin{tikzpicture}
\coordinate (x1t1) at  (0,0) ; 
\coordinate (x1t2) at  (1.5,0) ; 
\coordinate (x1t1b) at  (0,0.6) ; 
\coordinate (x1t2b) at  (1.5,0.6) ; 
\node (T1) at  (0.75,0.7)    {$\!\!\!\parbox{0mm}{$\raisebox{-1mm}[0mm][0mm]{$T$}$}$};
\node (T2) at  (0.75,0.1)    {$\!\!\!\parbox{0mm}{$\raisebox{-1mm}[0mm][0mm]{$T$}$}$};
\node () at  (1.75,0)    {$\!\!\!\parbox{0mm}{$\raisebox{-1mm}[0mm][0mm]{$\scriptstyle f$}$}$};
\node () at  (1.75,0.6)    {$\!\!\!\parbox{0mm}{$\raisebox{-1mm}[0mm][0mm]{$\scriptstyle e$}$}$};
\node () at  (-.2,0)    {$\!\!\!\parbox{0mm}{$\raisebox{-1mm}[0mm][0mm]{$\scriptstyle f$}$}$};
\node () at  (-.2,0.6)    {$\!\!\!\parbox{0mm}{$\raisebox{-1mm}[0mm][0mm]{$\scriptstyle e$}$}$};
\draw [thick] (x1t1) -- (x1t2);
\draw [thick] (x1t1b) -- (x1t2b);
\end{tikzpicture}}}} 
\left[ 
\frac{1}{2T^2}  ~
{\parbox{0.6cm}{{\begin{tikzpicture}
\coordinate (x1t1) at  (0,0) ; 
\coordinate (x1t2) at  (0,.5) ; 
\node (x) at  (0,0)    {$\!\!\!\parbox{0mm}{$\raisebox{-3.5mm}[2.5mm][2.5mm]{$\scriptstyle y$}$}$};
\node (t1) at  (-.25,0)    {$\!\!\!\parbox{0mm}{$\raisebox{-1mm}[0mm][0mm]{$\scriptstyle d$}$}$};
\node (t2) at  (-.25,0.5)    {$\!\!\!\parbox{0mm}{$\raisebox{-1mm}[0mm][0mm]{$\scriptstyle c$}$}$};
\fill (x1t1) circle (2pt);
\fill (x1t2) circle (2pt);
\draw [dashed,thick] (x1t1) -- (x1t2);
\end{tikzpicture}}}}\right]\nn\\
\eea
We have explicitly written all factors: a $1/2!$ from the expansion of the exponential function $\exp({-\ca S_{\rm rep}[u]})$, a factor of $1/(2T^2)$ per disorder vertex, and a factor of $T$ per propagator.
Using these rules, we obtain  two distinct contributions
\begin{eqnarray}\label{80cis}
\delta R^{(1)} = \half ~~{\parbox{2.2cm}{{\begin{tikzpicture}
\coordinate (x1t1) at  (0,0) ; 
\coordinate (x1t2) at  (0,.5) ; 
\coordinate (x2t3) at  (1.5,0) ; 
\coordinate (x2t4) at  (1.5,0.5) ; 
\node (x) at  (0,0)    {$\!\!\!\parbox{0mm}{$\raisebox{-3.5mm}[2.5mm][2.5mm]{$\scriptstyle x$}$}$};
\node (y) at  (1.5,0)    {$\!\!\!\parbox{0mm}{$\raisebox{-3.5mm}[2.5mm][2.5mm]{$\scriptstyle y$}$}$};
\node (t1) at  (-.25,0)    {$\!\!\!\parbox{0mm}{$\raisebox{-1mm}[0mm][0mm]{$\scriptstyle b$}$}$};
\node (t2) at  (-.25,0.5)    {$\!\!\!\parbox{0mm}{$\raisebox{-1mm}[0mm][0mm]{$\scriptstyle a$}$}$};
\node (t3) at  (1.75,0)    {$\!\!\!\parbox{0mm}{$\raisebox{-1mm}[0mm][0mm]{$\scriptstyle b$}$}$};
\node (t4) at  (1.75,0.5)    {$\!\!\!\parbox{0mm}{$\raisebox{-1mm}[0mm][0mm]{$\scriptstyle a$}$}$};
\fill (x1t1) circle (2pt);
\fill (x1t2) circle (2pt);
\fill (x2t3) circle (2pt);
\fill (x2t4) circle (2pt);
\draw  (x1t1) -- (x2t3);
\draw  (x1t2) -- (x2t4);
\draw [dashed,thick] (x1t1) -- (x1t2);
\draw [dashed,thick] (x2t3) -- (x2t4);
\end{tikzpicture}}}}
\\
= \half \int_{x}
 R_{0}''\big(u_{a}(x) {-}u_{b}(x)\big)R_{0}''\big(u_{a}(y) {-}u_{b}(y)\big)  C(x{-}y)^{2}\komma \nn
\\
\delta R^{(2)} =~{\parbox{2.2cm}{{\begin{tikzpicture}
\coordinate (x1t1) at  (0,0) ; 
\coordinate (x1t2) at  (0,.5) ; 
\coordinate (x2t3) at  (1.5,0) ; 
\coordinate (x2t4) at  (1.5,0.5) ; 
\node (x) at  (0,0)    {$\!\!\!\parbox{0mm}{$\raisebox{-3.5mm}[2.5mm][2.5mm]{$\scriptstyle x$}$}$};
\node (y) at  (1.5,0)    {$\!\!\!\parbox{0mm}{$\raisebox{-3.5mm}[2.5mm][2.5mm]{$\scriptstyle y$}$}$};
\node (t1) at  (-.25,0)    {$\!\!\!\parbox{0mm}{$\raisebox{-1mm}[0mm][0mm]{$\scriptstyle a$}$}$};
\node (t2) at  (-.25,0.5)    {$\!\!\!\parbox{0mm}{$\raisebox{-1mm}[0mm][0mm]{$\scriptstyle a$}$}$};
\node (t3) at  (1.75,0)    {$\!\!\!\parbox{0mm}{$\raisebox{-1mm}[0mm][0mm]{$ \scriptstyle b$}$}$};
\node (t4) at  (1.75,0.5)    {$\!\!\!\parbox{0mm}{$\raisebox{-1mm}[0mm][0mm]{$\scriptstyle a$}$}$};
\fill (x1t1) circle (2pt);
\fill (x1t2) circle (2pt);
\fill (x2t3) circle (2pt);
\fill (x2t4) circle (2pt);
\draw  (x1t1) -- (x2t4);
\draw  (x1t2) -- (x2t4);
\draw [dashed,thick] (x1t1) -- (x1t2);
\draw [dashed,thick] (x2t3) -- (x2t4);
\end{tikzpicture}}}}
\\
= - \int_{x} R_{0}''\big(u_{a}(x) {-}u_{a}(x)\big)R_{0}''\big(u_{a}(y) {-}u_{b}(y)\big)
C(x{-}y)^{2} \punkt\nn
\label{81}
\end{eqnarray}
Note that all factors of $T$ have disappeared, and only two replica sums (not written explicitly) remain. Each $R_0(u_a-u_b)$ has been contracted twice, giving rise to two derivatives. In the first diagram, since once $u_a$ and once $u_b$ has been contracted, each $R_0''$ comes with an additional minus sign; these cancel. In the second diagram, there is a minus sign from the first $R''_0$, but not from the second; thus the overall sign is negative. 

Note that the following diagram also contains two correlators (correct
counting in powers of temperature), but is not a 2-replica but a
3-replica sum,
\begin{equation}\label{339}
{\parbox{2.2cm}{{\begin{tikzpicture}
\coordinate (x1t1) at  (0,0) ; 
\coordinate (x1t2) at  (0,.5) ; 
\coordinate (x2t3) at  (1.5,0) ; 
\coordinate (x2t4) at  (1.5,0.5) ; 
\node (x) at  (0,0)    {$\!\!\!\parbox{0mm}{$\raisebox{-3.5mm}[2.5mm][2.5mm]{$\scriptstyle x$}$}$};
\node (y) at  (1.5,0)    {$\!\!\!\parbox{0mm}{$\raisebox{-3.5mm}[2.5mm][2.5mm]{$\scriptstyle y$}$}$};
\node (t1) at  (-.25,0)    {$\!\!\!\parbox{0mm}{$\raisebox{-1mm}[0mm][0mm]{$\scriptstyle b$}$}$};
\node (t2) at  (-.25,0.5)    {$\!\!\!\parbox{0mm}{$\raisebox{-1mm}[0mm][0mm]{$\scriptstyle a$}$}$};
\node (t3) at  (1.75,0)    {$\!\!\!\parbox{0mm}{$\raisebox{-1mm}[0mm][0mm]{$\scriptstyle c$}$}$};
\node (t4) at  (1.75,0.5)    {$\!\!\!\parbox{0mm}{$\raisebox{-1mm}[0mm][0mm]{$\scriptstyle a$}$}$};
\fill (x1t1) circle (2pt);
\fill (x1t2) circle (2pt);
\fill (x2t3) circle (2pt);
\fill (x2t4) circle (2pt);
\draw  (x2t4) arc(60:120:1.5);
\draw  (x2t4) arc(-60:-120:1.5);
\draw [dashed,thick] (x1t1) -- (x1t2);
\draw [dashed,thick] (x2t3) -- (x2t4);
\end{tikzpicture}}}}.
\end{equation}
In a renormalization program, we are looking for   divergences of
these diagrams. These divergences are localized at $x=y$: indeed the integral over the difference $  z:=y-x$, is in radial coordinates with $r=|z|$,  $\epsilon = 4-d$, and for $m\to 0$ (up to a geometrical prefactor)
\bea\label{2.8}
\int _{ z} C( z) ^2 \sim \int_a^L \frac{\rmd r}{r} {r^d} r^{2 (2-d)} = \int_a^L \frac{\rmd r}{r}  r^{4-d}\nn\\
= \frac 1\epsilon \left( L^\epsilon - a^\epsilon\right)\punkt
\eea
Note that for $\epsilon \to 0$ each scale contributes the same: from $r=1/2$ to $r=1$ the same as from $r=1/4$ to $r=1/2$, and again the same for $r=1/8$ to $r=1/4$. Thus the divergence comes from small scales, 
which allows us 
to approximate $R_{0}''(u_{a}(y)-u_{b}(y)) \approx R_{0}''(u_{a}(x)-u_{b}(x))$. This is formally an analysis of the theory via an operator product expansion. For an introduction and applications see \cite{CardyBook,WieseHabil}.

\Eq{2.8} is regularized   with cutoffs $a$ and $L$. It is   convenient to use $\epsilon>0$ (what we need anyway), which allows us to take $a\to 0$ and $L \to \infty$ while keeping $m$ finite, as the latter appears   as 
 the harmonic well introduced in section \ref{measurecusp}. 
The integral in that limit becomes  
\bea\label{I1}
\highlight{ I_{1}:=\Ione=\int_{x-y} C(x-y)^{2} = \int_{k} \frac 1 {(
k^{2}+m^{2})^{2}} \nn\\
~~~~~= \frac {m^{-\epsilon}}\epsilon  \frac{2\Gamma(1+\frac\epsilon2)}{(4 \pi)^{{d/2}}}}
\Review{\punkt}
\eea  It is the standard 1-loop diagram of massive
 $\phi^{4}$-theory\footnote{The trick to calculate integrals of this type is to write 
\be\nn
\frac1{(k^2+m^2)^2}= \int_0^\infty \rmd s\, s\rme^{-s(k^2+m^2)}.
\ee The integral over $k$ is then the 1-dimensional integral to the power of $d$. Finally one integrates over $s$.}. 

Setting $u=u_a(x)-u_b(x)$,  we obtain for the effective disorder correlator $  R(u)$ at 1-loop order 
with all combinatorial factors as given above, 
\be\label{351}
 R(u) =R_{0}(u)+\left[ \half R_{0}''(u)^{2} - R_{0}''(u) R_{0}''(0) \right]  I_{1} + ...
\ee
We can now study its flow, by taking a derivative w.r.t.\ $m$, and replacing on the r.h.s.\ $R_{0}$ with $ R$, as given by the above equation. This leads to 
\begin{equation}\label{RG1loop-bare}
-m\frac{\partial}{\partial m}   R(u) = \left[ \half   R''(u)^{2} -  R''(u)   R''(0)\right] \epsilon I_{1}\punkt
\end{equation}
This equation still contains the factor of $\epsilon I_{1}$, which has both a scale $m^{{-\epsilon}}$, as a finite amplitude. There are two convenient ways out of this: 
We can   parameterize the flow by the integral $I_{1}$ itself,   defining 
\bea
\label{RG1loop-no-rescale}
\highlight{\partial_{\ell }   R(u) := -\frac{\partial}{\partial I_{1}}   R(u) =  \half   R''(u)^{2} -  R''(u)   R''(0) \punkt}
\eea
This is convenient to study the flow numerically. 

To arrive at a fixed point one needs to  rescale both $  R$ and $u$, in order to make them dimensionless. The field $u$ has  dimension $u \sim L^\zeta\sim m^{-\zeta}$, whereas the dimension of $ R$ can be read off from \Eq{80cis}, namely $ R(u)\sim  R''(u)^2 m^{-\epsilon}$, equivalent to $ R\sim m^{\epsilon-4 \zeta} $. The dimensionless effective disorder $\tilde R$, as function of the dimensionless field $\bf u$ is then defined as
\be\label{Rtillde-def}
\tilde R({\bf u}) := \epsilon I_{1} \,m^{4\zeta}  R({ u}= { \bf u}\, m^{-\zeta})\punkt
\ee
Inserting this into \Eq{RG1loop-no-rescale}, we arrive at\footnote{$\ell$ in \Eqs{RG1loop-no-rescale} and \eq{RG1loop} is different.}
\bea\label{RG1loop}
\highlight{\partial_\ell \tilde R({\bf u}):=-m\frac{\partial}{\partial m} \tilde R({\bf u}) \\
= (\epsilon -4 \zeta) \tilde R({\bf u}) + \zeta
{\bf u} \tilde R'({\bf u}) + \half\tilde R''({\bf u})^{2} -\tilde R''({\bf u})\tilde R''(0)\punkt}\nn
\eea
This is the functional RG flow equation for the renormalized dimensionless disorder $\tilde R(u)$, first derived in Ref.~\cite{DSFisher1986} within the Wilson scheme\footnote{The RG flow equation \eq{RG1loop} is at this order  {\em  independent} of the RG scheme. Universal quantities are scheme-independent to all orders \cite{Zinn,BogoliubovParasiuk1957,Hepp1966,Zimmermann1969,WieseHabil}.}.
We will in general set ${\bf u}\to u$ in the above equation, to simplify notations, and suppress the tilde as long as this does not lead to confusion.

{\new We would like to stress what we already said in section \ref{s:General remarks about renormalization}, namely that the flow equations we derived as functions of $m$ have a very intuitive interpretation: Since the strength $m^2$ of the confining potential \eq{Hconf} is a parameter of the experimental system which does not renormalise (see the next section \ref{s:tilt-symmetry}), the RG equation can be taken quite literally: what happens if in an experiment or a simulation the confining potential is weakened? In a peeling or unzipping experiment (sections \ref{s:Experiments at equilibrium} and \ref{s:Experiment on RNA-DNA unzipping}) this even happens during the experiment. The answer is that for $m\to \infty$, one sees the microscopic disorder, while for smaller $m$ an effective scale-dependent disorder is measured. This is explained in detail in section \ref{measurecusp}. Before doing this, let us first ensure that $m$ does not renormalize (next section \ref{s:tilt-symmetry}), and then study what happens if $m$ is lowered (section \ref{s:cusp}).}

\subsection{Statistical tilt symmetry}\label{s:tilt-symmetry}
We claim that there are no renormalizations of the quadratic parts of the action which are replicated copies of 
\be
\ca H_{\rm 0}[u] := {\cal H}_{\mathrm{el}}[u] + {\cal H}_{\rm conf}[u]
\ee given in \Eqs{Hel} and \eq{Hconf}. This is  due to the  
{\em statistical tilt symmetry} (STS)
\be\label{STS}
u_a (x)\to u_a (x)+ \alpha x\punkt
\ee 
As the interaction is proportional to $R(u_a(x)-u_b(x))$, the latter is invariant under the transformation \eq{STS}.
The change in $\ca H_0[u]$ becomes 
\bea\label{a11}
  \delta {\cal H}_0[u] &=& c\int\rmd^{d}x \left[ \nabla u (x) \alpha+\half
\alpha^{2} \right]  \nn\\
&+&  {m^{2}}\int\rmd^{d}x \left[u (x) \alpha x +
\frac{1}{2}\alpha^{2}x^{2} \right]\punkt
\eea
To render the presentation clearer, the elastic constant $c$ set to
$c=1$ in equation (\ref{H}) has been introduced. The important
observation is that all fields $u$ involved are {\em large-scale}
variables, which are also present in the renormalized action, where they  change 
according to ${\cal H}^{\mathrm{ren}}[u]\to {\cal
H}^{\mathrm{ren}}[u]+\delta \ca H^{\rm ren}[u]$. 
Since one can either first renormalize and then tilt, or first tilt and then renormalize, we obtain 
$
\delta \ca H_0^{\rm ren}[u]  = \delta \ca H_0^{\rm bare}[u]
$. This means that neither the elastic constant $c$, nor $m$ change under renormalization.

\subsection{Solution of the FRG equation, and cusp}\label{s:cusp}
\begin{widefigure}
\centerline{\fig{8.5cm}{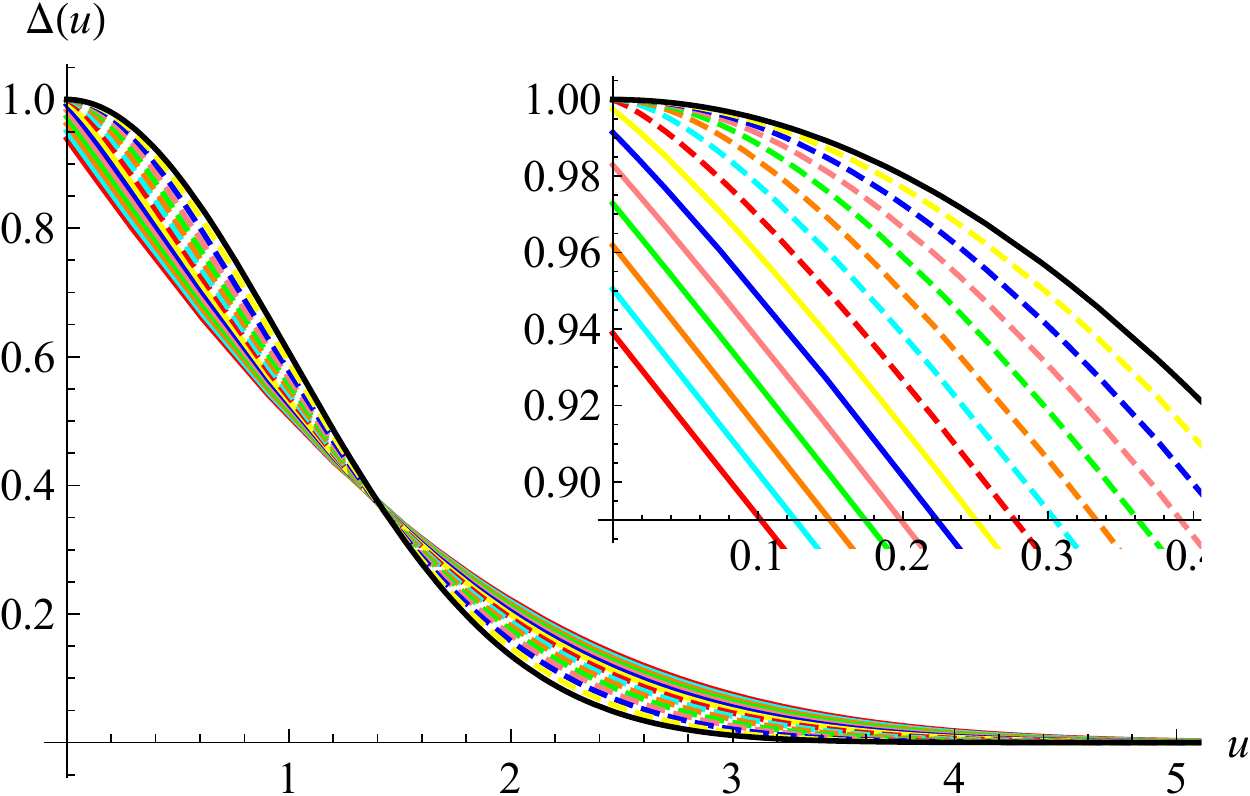}~~~~\fig{8.5cm}{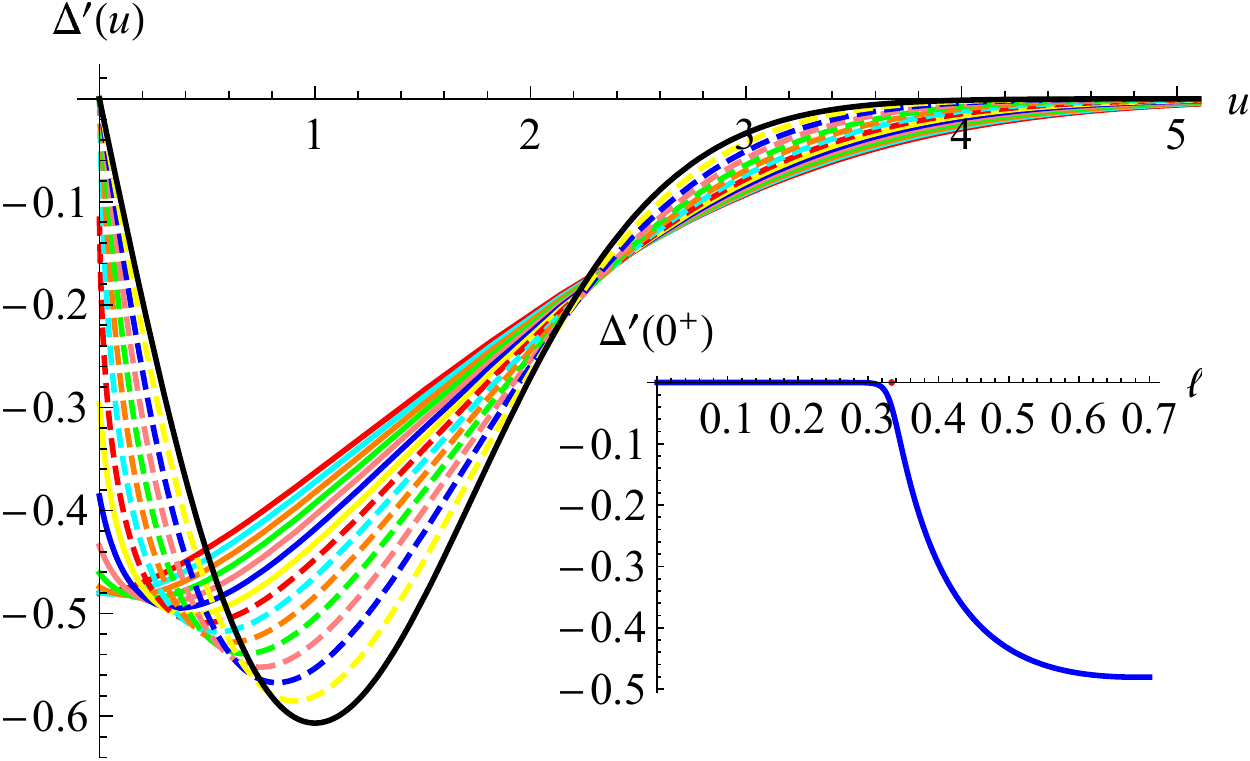}}\smallskip

\centerline{\fig{8.5cm}{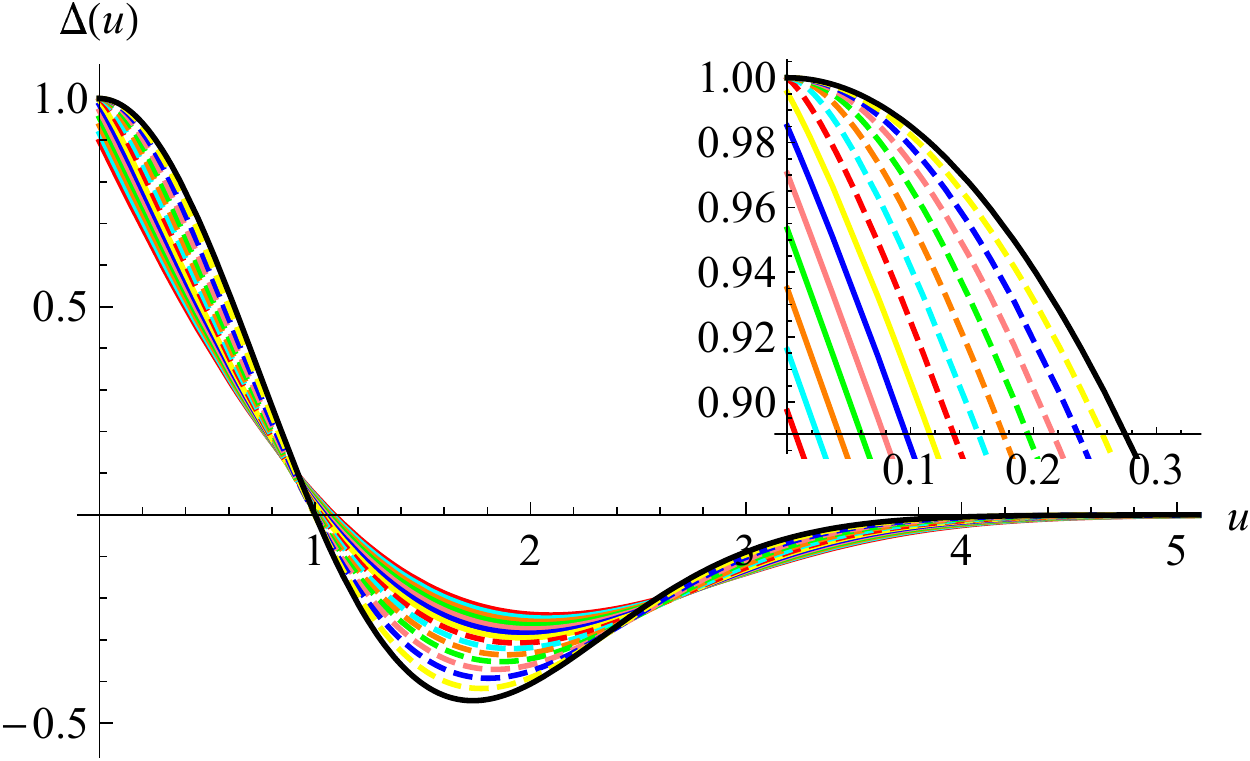}~~~~\fig{8.5cm}{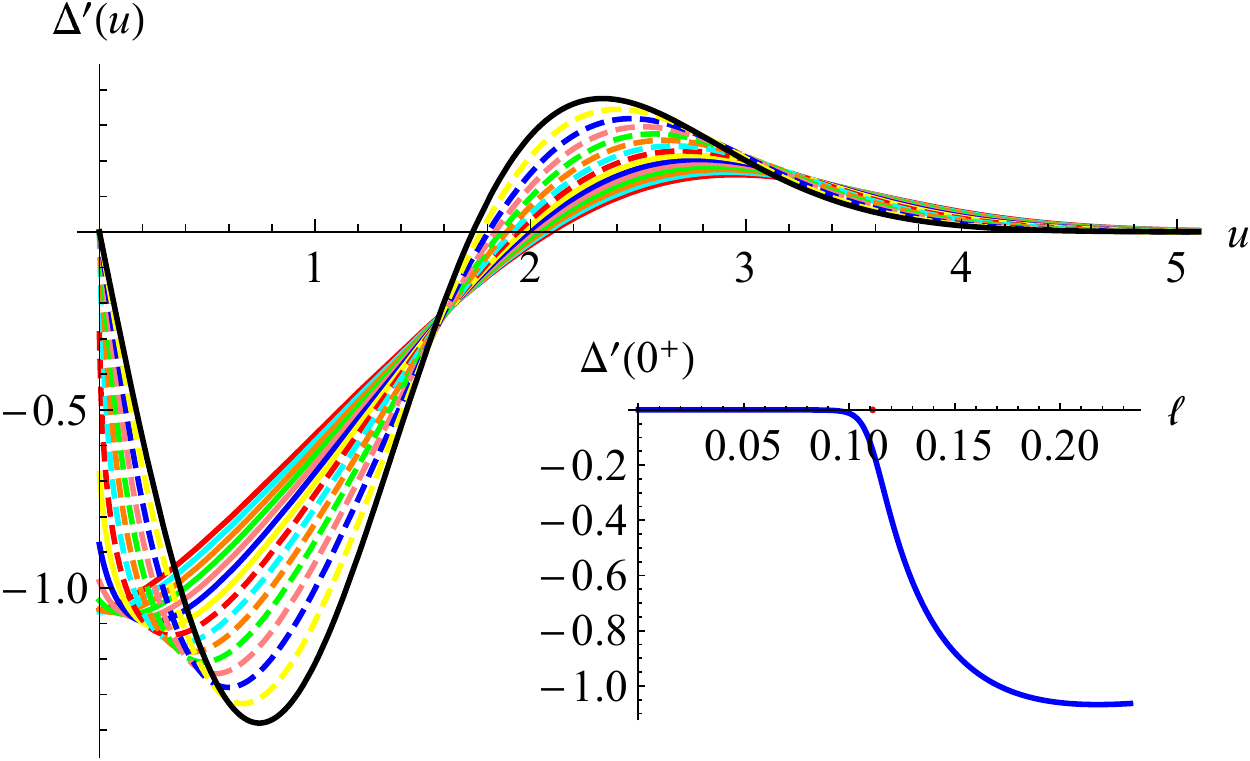}}
\caption{Top: Change of $\Delta(u):=-R'' (u)$ under renormalization and formation of
the cusp. Left: Explicit numerical integration of \Eq{RG1loop-no-rescale}, starting from $\Delta(u)= \rme^{-u^2/2}$ (in solid black, top curve for   $u\to 0$).  The function at scale $\ell $ is shown in   steps of $\delta \ell=1/20$. Inset: blow-up. Right: plots of $\Delta'(u)$. Inset: $\Delta'(0^+)$ as a function of $\ell$. The cusp appears for $\ell=1/3$ (red dot);     dashed lines are before appearance of the cusp, and   solid lines after. 
Bottom row: the same plots for RB disorder, starting from $R(u)= \rme^{-u^2/2}$; the cusp appears for $\ell=1/9$; $\delta \ell = 1/60$.} \label{fig:cusp}
\end{widefigure}
We now analyze the FRG flow equations \eq{RG1loop-no-rescale} and \eq{RG1loop}. To simplify our arguments, we first derive them twice w.r.t.\ $u$, to obtain flow equations for $\Delta(u)\equiv -R''(u)$. This yields
\bea\label{flow-Delta-no-rescale}
\mbox{no rescaling:~~~~}&&\highlight{\partial_\ell  \Delta(u)  =- \partial_u^2\, \half \big[ \Delta(u)  - \Delta (0) \big] ^2 \komma  }\\
\label{flow-Delta}
\mbox{with rescaling:~~~~}&&\highlight{\partial_\ell \tilde \Delta({ u}) =  (\epsilon -2 \zeta) \tilde\Delta({ u}) + \zeta
{  u} \tilde\Delta'({ u}) \Review{\nn\\&&~~~~~~~~~~~~~}
- \partial_u^2 \,\half \big[ \tilde\Delta ({ u}) - \tilde\Delta(0) \big]^2\punkt}
\eea
For concreteness, consider \Eq{flow-Delta-no-rescale}, and start with an analytic function, 
\be\label{Delta-initial}
  \Delta_{\ell = 0}(u)= \rme^{-u^2/2}\punkt
\ee 
According to our classification, this is microscopically RF disorder. 
Since $  \Delta(u)  -  \Delta(0)$ grows quadratically in $u$ at small $u$, the r.h.s.\ of \Eq{flow-Delta-no-rescale}  also grows $\sim u^2$    at $u=0$, and both $ \Delta (0)$ as well as $ \Delta'(0^+)$ do not flow in the beginning. This can be seen on the plots of figure \ref{fig:cusp}.

Integrating further, a cusp forms, i.e.\ $ \Delta''(0)\to \infty$, and as a consequence $ \Delta'(0^+)$ becomes non-zero. This is best   seen by taking two more derivatives of 
\Eq{flow-Delta-no-rescale}, and then taking the limit of  $u\to 0$,
\be\label{64}
\partial_\ell  \Delta''(0^+) = -3  \Delta''(0^+)^2 - 4  \Delta'(0^+)  \Delta'''(0^+) \punkt
\ee
Since in the beginning $ \Delta'(0^+)=0$, only the first term survives. Its behavior crucially depends on the sign of $  \Delta''(0)$. In the example \eq{Delta-initial}, $  \Delta_{\ell=0}''(0)<0$. This is true in general, as can be seen by 
rewriting \Eq{Delta-def} for the unrescaled microscopic disorder correlator at  $x=x'$, as  
\be
\Delta(0)-\Delta (u{-}u') = {  \half}\overline{ \left<\big[ F(x,u)-F(x,u') \big]^2\right>} \ge 0.
\ee
Developing the l.h.s. for small $u-u'$ with a vanishing first derivative implies that $\Delta''(0)<0$, valid also for the rescaled  $  \Delta''(0)$. 

Integrating \Eq{64} with this sign yields
\bea
 \Delta_\ell''(0) = \frac{ \Delta_0''(0) }{1+3  \Delta_0''(0) \ell} = -\frac13 \frac{1}{\ell_{\rm c}- \ell} \komma  
\qquad \nn\\
\ell_{\rm c} =-\frac1{3 \Delta_0''(0)} ~~\stackrel{ \Delta_0''(0)=-1}{-\!\!\!-\!\!\!-\!\!\!-\!\!\!-\!\!\!-\!\!\!\longrightarrow} ~~\frac{1 }3 \punkt \eea 
In the last equality we used the initial condition \eq{Delta-initial}. With this,  $ \Delta_\ell''(0)$ diverges at  $\ell=\frac13$, thus $ \Delta_\ell(u)$ acquires a cups, i.e.\ $ \Delta_\ell'(0^+)\neq 0$ for all $\ell>1/3$. 
Physically, this is the scale where multiple minima appear. In terms of the  Larkin-scale $L_{\rm c}$ defined in section \ref{Larkin} 
\be
\ell_{\rm c}= \ln (L_{\rm c}/a)\punkt
\ee
Our numerical solution shows the appearance of the cusp only approximately, see the inset in the top right plot of figure \ref{fig:cusp}. This discrepancy comes from discretization errors. It is indeed not simple to numerically integrate   equation \eq{flow-Delta-no-rescale} for large times, as $ \Delta_\ell''(0)$ diverges at $\ell=\ell_{\rm c}$, and all further  derivatives at $u=0^+$ were extracted from numerical {\em extrapolations} of the obtained functions, in the limit of $u\to 0$. 

Interpreting derivatives in this sense is   an {\em assumption}, to be justified,  without which one cannot continue to integrate the flow equations.  
In this spirit, let us again look at the flow equation for $ \Delta(0)$, now including the rescaling terms,  
\be
\partial_\ell \tilde  \Delta (0) = (\epsilon-2 \zeta) \tilde \Delta(0)- \tilde\Delta'(0^+)^2 \punkt 
\ee
This equation tells us that  as long as  $\Delta'(0^+)=0$, 
\be
\zeta_{\ell <\ell_{\rm c}} \simeq \zeta_{\rm DR}=\frac\epsilon 2  = \frac{4-d}2 \komma 
\ee
the dimensional-reduction result. Beyond that scale, we have  (as long as we are at least close to a fixed point)
\be\label{zeta-bound}
\zeta_{\ell >\ell_{\rm c}} = \frac\epsilon 2 -\frac{ \Delta'(0^+)^2}{\Delta(0)} < \frac\epsilon 2\komma
\ee
since both $ \Delta'(0^+)^2$ and  $ \Delta(0)$ are positive. 

Let us repeat our analysis for RB disorder, starting from the microscopic disorder 
\be
 R_0(u) = \rme^{-u^2/2} \quad \Longleftrightarrow \quad\Delta(u) =  \rme^{-u^2/2}(1-u^2)\punkt
\ee
This is shown on the bottom of figure \ref{fig:cusp}. Phenomenologically, the scenario is rather similar, with a critical scale $\ell_{\rm c}=1/9$ instead of $1/3$.

\subsection{Fixed points of the FRG equation}
\label{s:FRG-fixed-points}
We had seen in the last section that integrating the flow equation explicitly is rather cumbersome; moreover, an estimation of the critical exponent $\zeta$ will be rather imprecise. For this purpose, it is better to directly search for a solution of the fixed-point equation \eq{flow-Delta}, i.e.\ $\partial_{\ell}\tilde \Delta(u)=0$,
\be\label{RF-FP}
\highlight{ 0 = (\epsilon -2 \zeta) \tilde \Delta({ u}) + \zeta
{  u} \tilde \Delta'({ u}) - \partial_u^2 \half \left[ \tilde \Delta ({ u}) - \tilde \Delta(0) \right]^2\!.~}
\ee
We start our analysis with situations where $u$ is unbounded, as for the position of an interface. 
Then  the fixed point is not unique; indeed, if $\tilde \Delta(u)$ is solution of \Eq{RF-FP}, so is 
\be\label{marginal-mode}
\tilde \Delta_{\kappa}(u) := \kappa^{-2}\tilde \Delta(\kappa u)\punkt
\ee
\subsection{Random-field (RF) fixed point}
There is one solution we can find analytically: To this purpose integrate \Eq{RF-FP} from 0 to $\infty$, assuming that $\tilde \Delta(u)$ has a cusp at $u=0$, but no stronger singularity,
\bea \label{80}
0= 
\int_0^\infty &(\epsilon -2 \zeta)  \tilde \Delta({ u})   +  \zeta  
{  u} \tilde \Delta'({ u})   \nn\\
&- \partial_u^2 \half \left[ \tilde \Delta ({ u})  - \tilde \Delta(0) \right]^2\rmd u\punkt
\eea
Integrating the second term by part, and using that the last term is a total derivative which vanishes both at $0$ and at $\infty$ yields
\be\label{357}
0 = (\epsilon - 3 \zeta) \int_0^\infty \tilde \Delta(u) \,\rmd u\punkt
\ee
This equation has two solutions: either the integral vanishes, which is the case   for RB disorder\footnote{For RB disorder $$\int_0^\infty \rmd u\, \tilde \Delta (u) =-\int_0^\infty \rmd u \,\tilde R'' (u) = \tilde R'(0) - \tilde R'(\infty)=0. $$}, or 
\be\label{zeta-RF-1loop}
\highlight{ \zeta_{\rm RF} = \frac{\epsilon}{3}\punkt}
\ee 
This  is the  exponent \eq{a3} (at $N=1$) predicted by a Flory argument. 
Let us remark that \Eq{80} remains valid to all orders in $\epsilon$, as long as $\Delta(u)$ is the second derivative of $R(u)$, s.t.\ the additional terms at 2- and higher-loop order are all total derivatives, as is the last term in \Eq{RF-FP}. 

Let us pursue our analysis with the   solution \eq{zeta-RF-1loop}. Inserting \Eq{zeta-RF-1loop} into \Eq{RF-FP}, and setting  
\be\label{Delta=y...}
\tilde \Delta(u)=\frac\epsilon{3}y(u)
\ee yields 
\be\label{2.33}
\partial_u \left[ u y(u) -\half \partial_u \Big(y(u)-y(0)\Big)^2 \right]=0\punkt
\ee
This implies that the expression in the square bracket is a constant, fixed to 0 by considering either the limit of $u\to 0$ or $u\to \infty$. Simplifying yields
\be
u y(u) + \big[y(0)-y(u)\big] y'(u) =0\;.
\ee
Dividing by $y(u)$ and integrating once again gives
\be
\frac{u^2}{2}-y(u)+y(0) \ln (y(u)) = \mbox{const.}
\ee
Let us now use \Eq{marginal-mode} to set $y(0)\to 1$. This fixes the constant to   $-1$. Dropping the argument of $y$, we obtain
\be\label{RF-FP-y}
y - \ln(y) = 1+\frac{u^2}2\punkt
\ee
\begin{widefigure}
\fig{0.49\textwidth}{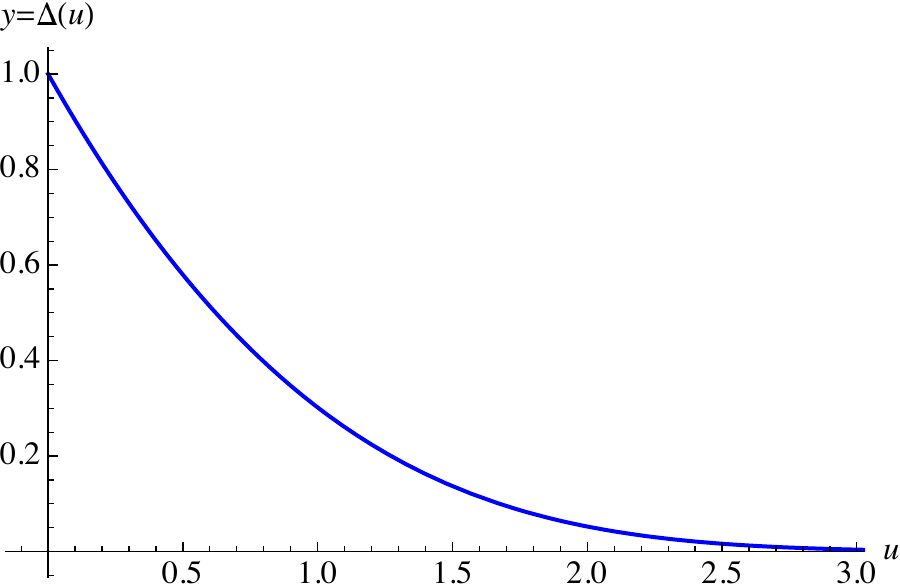}~~~~\fig{0.49\textwidth}{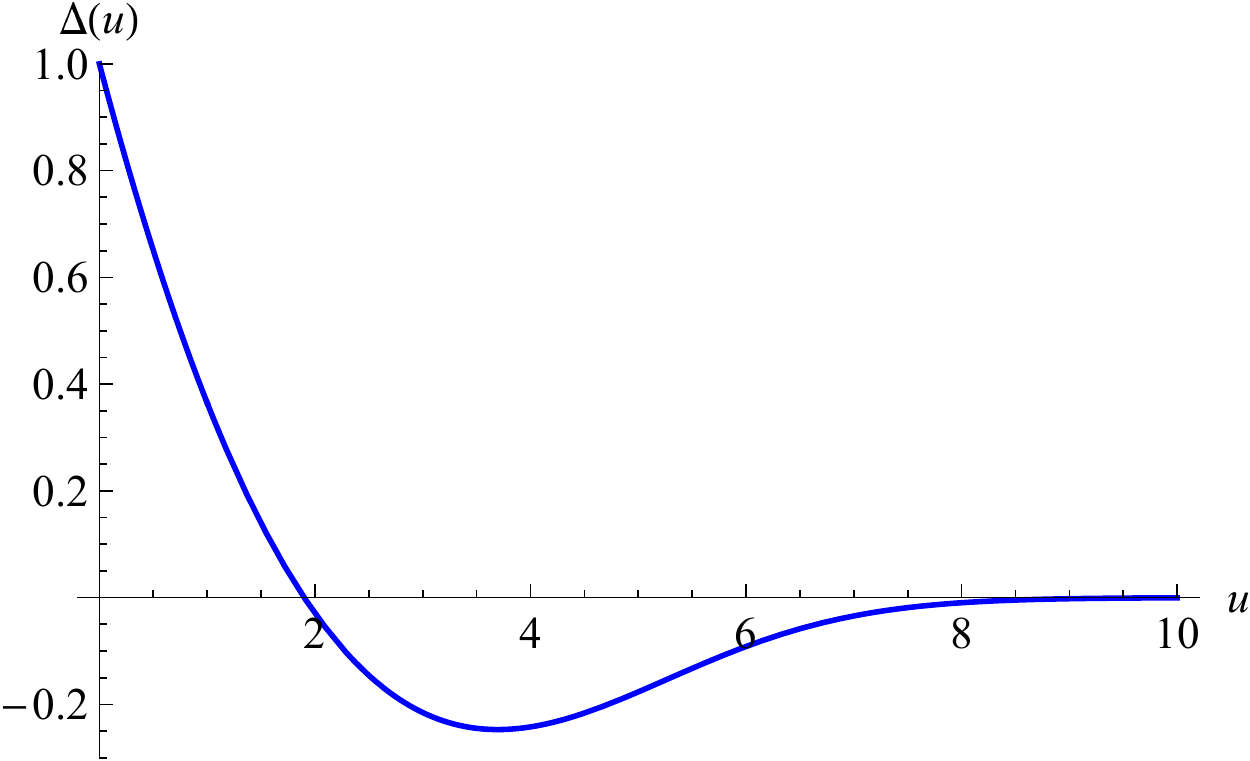}
\caption{Left: The RF fixed point \eq{RF-FP-y} with $\zeta_{\rm RF}=\frac \epsilon 3$. Right: The RB fixed point \eq{zeta-RB}, with $\zeta_{\rm RB}=  0.208298 \epsilon $.}
\label{f:RF-FP}
\end{widefigure}
This is plotted on figure \ref{f:RF-FP}.

\subsection{Random-bond (RB) and tricritical fixed points}
The other option for a fixed point is to have the integral in \Eq{357} vanish,
\be\label{rel-RB}
\int_0^\infty \tilde \Delta_{\rm RB}(u) =0.
\ee
 A numerical analysis of the fixed-point equation 
\eq{RF-FP} proceeds as follows: Choose $\tilde \Delta(0)=1$; choose $\zeta$; solve the differential equation \eq{RF-FP} for  $\tilde \Delta''(u)$. Integrate the latter from $u=0$ to $u=\infty$. In practice, to avoid numerical problems for $u\approx0$, one first solves the differential equation in a Taylor-expansion around 0;   as the latter does not converge for large $u$ one then solves, with the information from the Taylor series evaluated at $u=0.1$, the differential equation numerically up to $u_\infty\approx30$. 
One then reports, as a function of $\zeta$, the value of $\tilde \Delta(u_{\infty})$. As in quantum mechanics, one finds that there are several discrete values of $\zeta$ with $\tilde \Delta(u_{\infty})=0$. The largest value of $\zeta$ is the one given in \Eq{zeta-RF-1loop}, where $\tilde \Delta(u)$ has no zero crossing. The next smaller value of $\zeta$ is 
\be\label{zeta-RB}
\hl{ \zeta_{\rm RB}=  0.208298 \epsilon}\punkt
\ee 
The corresponding function is plotted on figure~\ref{f:RF-FP} (right).  It  has  one zero-crossing. 
Consistent with \Eq{357}, it integrates to zero. This is the random-bond fixed point. It is   surprisingly close, but distinct, from the Flory estimate (\ref{a2}), $\zeta=\epsilon/5$.

For $\epsilon=3$ we have the  directed polymer ($d=1$) in dimension $N=1$, which has roughness $
\zeta_{d=1}^{\rm RB} =\frac23$. Our result \eq{zeta-RB} yields $\zeta(d=1) =0.624894+{\cal O}(\epsilon^2)$. This is quite good, knowing that $\epsilon=3$ is rather large. This value   gets improved at 2-loop order (see section \ref{beyond1loop}), with $\zeta(d=1)=0.686616+{\cal O}(\epsilon^3)$. 
Despite the ``strange cusp'', it seems  the method works!

The next solution is  at \be
\hl{ \zeta_{3\rm crit}=0.14366 \epsilon}.
\ee 
It has two zero-crossings, and corresponds to a tricritical point. We do not know of any physical realization. 

\subsection{Generic long-ranged fixed point}
If $\zeta$ is not one of these special values, then the  solution of the fixed-point \Eq{RF-FP}  decays algebraically: Suppose that $\Delta(u)\sim u^{\alpha}$. Then the first two terms of \Eq{RF-FP} are dominant over the last one, as long as $\alpha<2$. Solving \Eq{RF-FP} in this limit  one finds 
\be\label{zeta-LR}
\hl{\Delta_{\zeta} (u)\sim u^{2-\frac{\epsilon}{\zeta}} \quad \mbox{for}\quad u \to \infty}\punkt
\ee
An important application are the ABBM and BFM models   discussed   in sections \ref{s:ABBM} and \ref{s:BFM}, for which 
\be\label{FB-ABBM}
\zeta_{\rm ABBM}=\epsilon\komma\quad 
\Delta_{\rm ABBM}(0)-\Delta_{\rm ABBM}(u)=\sigma |u|\komma
\ee 
such that the correlations of the random forces have the statistics of a random walk. One easily checks that the flow equation \eq{RG1loop-no-rescale}   vanishes for all $u>0$. In this case $\Delta(0)$ is formally infinite, s.t.\ the bound \eq{zeta-bound} does not apply. Generically, however,   \Eq{zeta-bound} applies,   implying that the exponent in \Eq{zeta-LR} is negative, and  $\Delta_{\zeta} (u)$ decays  algebraically. 
This is what we mostly see    in   numerical solutions of the fixed-point \Eq{RF-FP}.

\subsection{Charge-density wave (CDW) fixed point}
\label{s:Charge-density wave (CDW) fixed point}
In the above considerations, we had supposed that $u$ can take any real value. There are    important applications where the disorder is periodic, or $u$ is a phase between $0$ and $2\pi$. This is the case for the CDWs introduced above.  To be consistent with the standard conventions employed in the literature \cite{NarayanMiddleton1994,NarayanDSFisher1992b,NarayanDSFisher1992a,LeDoussalWieseChauve2003,LeDoussalWieseChauve2002,ChauveLeDoussalWiese2000a}, we  take the period of the disorder to be $1$.  
One checks that the following ansatz is  a fixed point of the FRG equation \eq{RF-FP}
\bea\label{RP-fixed-point}
\zeta_{\rm RP} = 0,  \qquad \Delta_{\rm RP}(u) = \frac g {12}- \frac g 2 u(1-u),\nn\\
0\le u\le 1\punkt 
\eea
This ansatz is unique, due to the following three constraints: (i) $\zeta=0$, as the period is fixed and cannot change under renormalization. (ii) $\Delta(u)=\Delta(-u)=\Delta(1-u)$ due to the symmetry $u \to -u$, and periodicity. Thus $\Delta(u)$ is a polynomial in $u(1-u)$. (iii) a polynomial of degree $2$ in $u$ closes under RG. (iv) the integral $\int_0^1\rmd u\,\Delta(u)=0$, since $\Delta(u) = -R''(u)$, and $R(u)$ itself is periodic. 
The fixed point has     \bea\label{RP-fixed-point-2}
 g&=&  \frac{\epsilon }{3}+...
\eea
Instead of a universal scaling exponent $\zeta$, the latter vanishes, $\zeta=0$. As a consequence,  the 2-point function is logarithmic in all dimensions, with a universal amplitude given in \Eqs{universal-amplitude-1}-\eq{universal-amplitude-2b}. Apart from geometric prefactors, this amplitude is simply the fixed-point value $g$.

\subsection{The cusp and shocks: A  toy model}\label{s:shocks}
\begin{widefigure}
\centerline{\fig{0.7\textwidth}{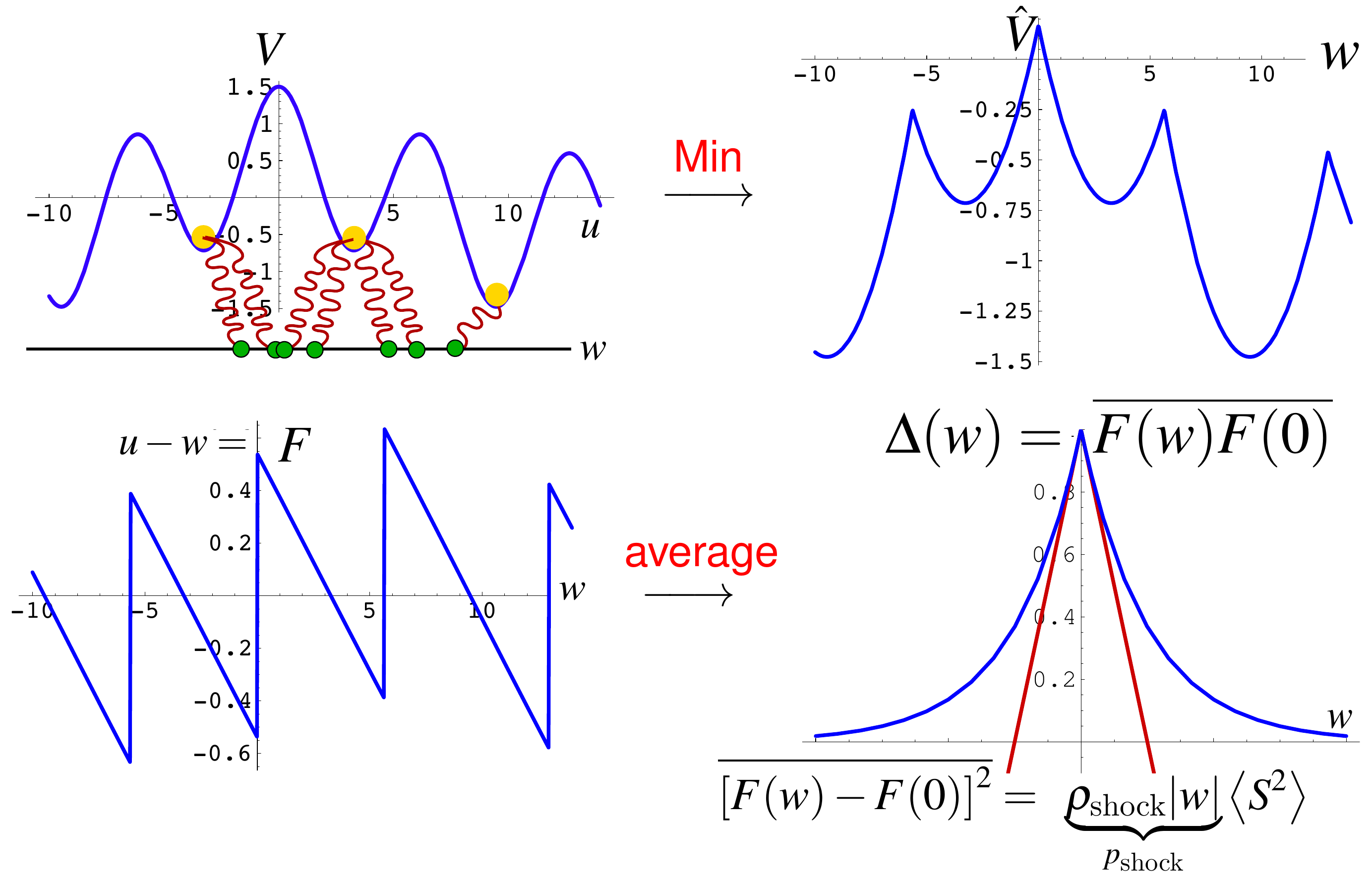}}
\caption{Generation of the cusp, as explained in the main text.} \label{minfig}
\end{widefigure}

Let us give a simple
argument  why a cusp is a physical necessity, and not an
artifact. The argument is quite old and appeared probably first in the
treatment of correlation-functions by shocks in Burgers
turbulence. It became popular in \cite{BalentsBouchaudMezard1996}. 
We want to solve the problem for a single degree of freedom which sees both disorder and a  parabolic trap centered at $w$, which we can view as a spring attached to the point $w$. 
This is graphically represented on figure
\ref{minfig} (upper left), with the quenched  disorder realization having roughly a sinusoidal shape.     For 
a given disorder realization $V(u)$, the minimum of the potential as a function
of $w$ is 
\be\label{5.37}
\hat V(w):= \min_u \left[ V(u) + \frac {m^2}{2}(u-w)^2\right] .
\ee
This is
reported on figure \ref{minfig} (upper right). Note that it has
non-analytic points, which mark the transition from one minimum to
another. The remaining parts are parabolic, and stem almost entirely from the spring, as long as the minima of the disorder are sharp,  i.e.\ have a high curvature as compared to the spring. This is   rather natural, knowing that the disorder varies on microscopic scales, while the confining potential   changes on macroscopic scales.

Taking the derivative of the potential leads to the force in
figure \ref{minfig} (lower left). It is characterized by almost linear pieces,
and shocks (i.e.\ jumps).  
Let us now calculate the  correlator  of forces $F(u):=- \nabla \hat V(u)$,
\be\label{Delta-F}
\Delta(w):= \overline {F(w') F(w'-w)}^{\,\rm c}\punkt
\ee
Here the average is over disorder realizations, or equivalently $w'$, on which it should not depend. 
Let us analyze its behavior at small distances,  
\bea
\label{Delta(0)-Delta(w)}
\Delta(0)-\Delta(w)   &=& \half\overline {\left[ F(w') -F(w'-w)\right]^2} \nn\\
&=& \half p_{\rm shock}(w) \left< \delta F^2\right>  + {\cal O}(w^2)
\eea
As written, the leading contribution is proportional to the probability to have a shock (jump) inside the window of size $w$, times the expectation of the  second moment of the force jump   $\delta F$. If shocks are not dense, then the probability to have a shock is given by the 
density $\rho_{\rm shock}$ of shocks times the size  $w$ of the window, i.e. 
\be\label{rho-shock}
p_{\rm shock}(w) \simeq  \rho_{\rm shock} |w| \punkt
\ee
Let us now relate $\delta F$ to the change in $u$; as the spring-constant is $m^{2}$, 
\be
\delta F = m^2 \delta u \equiv m^2 S\punkt
\ee
Here we have introduced the {\em avalanche size} $S:=\delta u$. Putting everything together yields
\be\label{369}
\Delta(0)-\Delta(w) = \frac{m^4}2 \left <S^2 \right> \rho_{\rm shock} |w| + {\cal O}(w^2)\punkt
\ee
We can  eliminate $\rho_{\rm shock} $ by observing that on average the particle position $u$ follows the spring, i.e.
\be\label{w-w'}
w = \overline {u(w'+w)- u(w') } = \left< S\right>  \rho_{\rm shock} w \punkt
\ee
This yields 
\be
\rho_{\rm shock} = \frac1{\left< S\right>}.
\ee
Expanding \Eq{369} in $w$, and retaining only the term linear in $w$    yields
\be\label{Delta'(0+)}
\highlight{ \rule[-2.7ex]{0mm}{6.5ex}-\Delta'(0^+) = {m^4} \frac{\left< S^2\right>}{2\left< S\right>}\punkt}
\ee
We just showed that having a cusp non-analyticity in $\Delta(w)$ is a necessity if the system under consideration has shocks or avalanches.  The latter are a consequence of metastability (i.e.\ existence of local minima), thus metastability implies a cusp in $\Delta(w)$.

\subsection{The effective disorder correlator in the field-theory}\label{measurecusp}

The above toy model can   be generalized to the field theory \cite{LeDoussal2006b}. 
Consider an interface in a random potential, as given by \Eqs{Hel}-\eq{HDO}
\be
{\cal H}_{\mathrm{tot}}^{w}[u] =  \int_{x}\frac{m^2}{2} [u(x)-w]^2 + {\cal
H}_{\mathrm{el}}[u] + {\cal H}_{{\rm dis}}[u]\punkt
\ee
Physically, the role of the well is to forbid the interface to wander
off to infinity. This avoids that observables are dominated by rare events. In each sample (i.e.\ disorder configuration), and given $w$, one
finds the minimum-energy configuration. The corresponding 
ground-state energy, or effective potential, is
\begin{eqnarray}
\hat V(w) := \min_{u(x)} {\cal H}_{\mathrm{tot}}^{w}[u]\punkt
\end{eqnarray}
Let us call $u_w^{\rm min}(x)$ this configuration. Its center-of-mass position is
\be\label{uCOM}
u_{w}:=\frac{1}{L^{d}} \int_{x} u_w^{\rm min}(x)\punkt
\ee
Both $\hat V(w)$ and  $u_{w}$ vary with $w$ as well as from sample to sample. Let us now look at their second cumulants. The effective potential $\hat V(w)$  defines a function $R(w)$, 
\be  
\highlight{R(w-w') := L^{-d}\,\overline{ \hat V(w)  \hat V(w') }^{\,\rm c}} . \label{defR}
\ee
This is the same function as computed in the field theory, defined there from the zero-momentum action. 
 The factor of volume $L^d$ is necessary.  The interface is correlated over a  length  $\xi=1/m$, while its width
$\overline{u^2}$  is bounded by the confining well. This means that the interface is made of roughly
$(L/\xi)^d$  independent pieces of length $\xi$:
Eq.~(\ref{defR}) expresses the central-limit theorem and $R(w)$
measures the second cumulant of the disorder seen by any one of the
independent pieces.

\begin{figure}\setlength{\unitlength}{1mm}
\centerline{\mbox{\begin{picture}(82,57)
\put(-2,2){\fig{8.5cm}{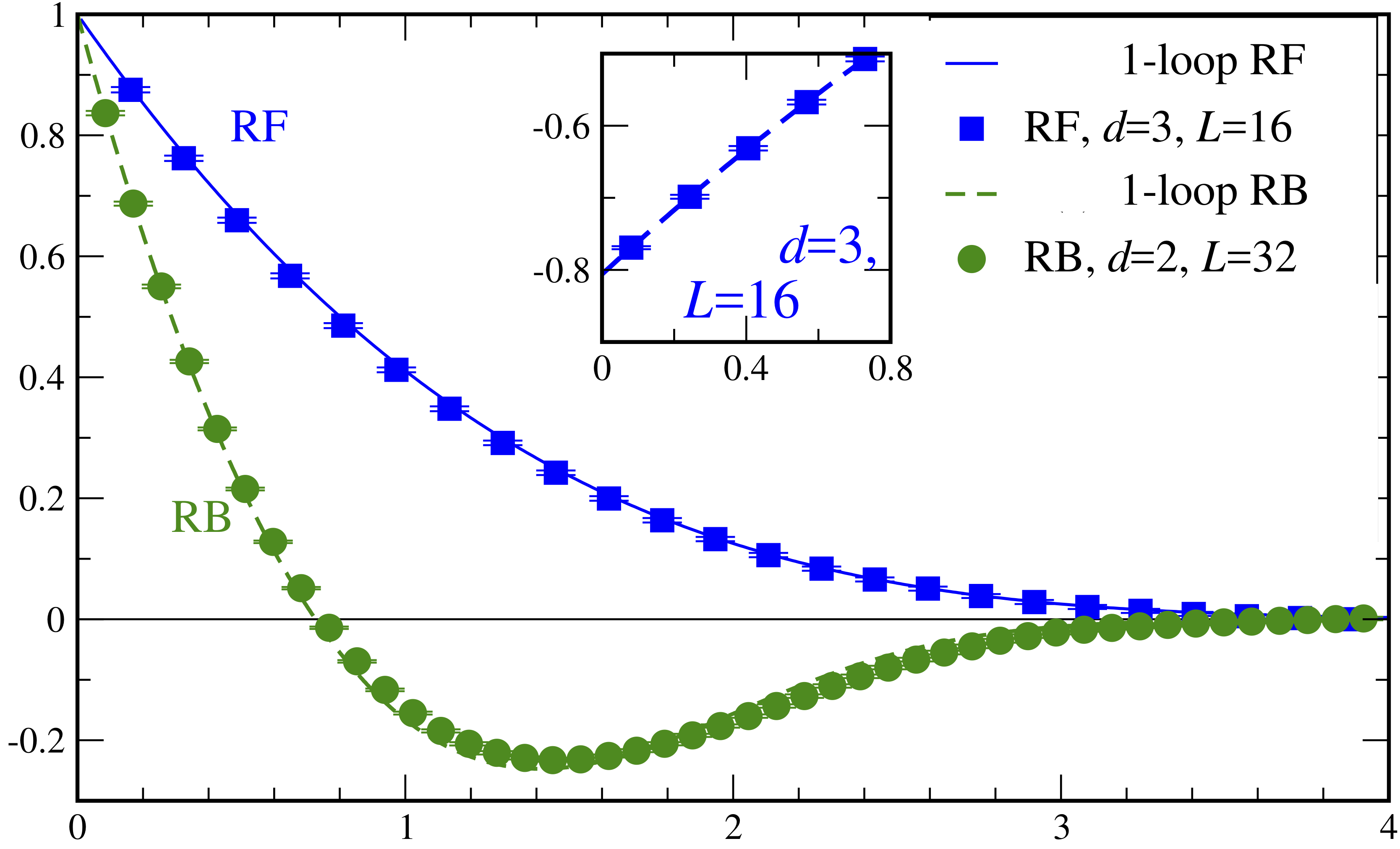}}
\put(63,-1){$u$, $u/4$ for RB}
\put(0,55){$\Delta(u)$}
\put(27,49){$\scriptstyle\Delta'(u)$}
\end{picture}}}
\caption{Filled symbols show numerical results for $\Delta(u)$, a
normalized form of the interface displacement correlator $-R''(u)$
[Eq.\ (\ref{defDe})], for $D=2+1$ random field (RF) and $D=3+1$ random
bond (RB) disorders. These suggest a linear cusp. The inset plots the
numerical derivative $\Delta'(u)$, with intercept $\Delta'(0^+)\approx -0.807$
from a quadratic fit (dashed line).  The
points are for confining wells with width given by $m^2=0.02$.
Comparisons to 1-loop FRG predictions (curves) are made with no
adjustable parameters. Reprinted from \protect\cite{MiddletonLeDoussalWiese2006}.
}  
\label{f:Alan1}
\end{figure}
The nice thing about \Eq{defR} is that it can be measured. One
varies $w$ and computes (numerically) the new ground-state energy,
finallying averaging over disorder  realizations. 
In fact, what is even easier to measure are   
the fluctuations of the center-of-mass position $u_{w}$, related to the total force acting on the interface. 
To see this, write the
  condition for the interface to be in a minimum-energy configuration, 
\bea\label{eq-cond}
0 = -\frac{\delta {\cal H}[u]}{\delta u(x)} = \nabla^{2} u(x) - m^{2 }[u(x)-w]  + {F} \big(x,u(x)\big)\komma  \nn\\
 F(x,u)=-\partial_{u}V(x,u)\punkt
\eea  
Integrating over space, and  using periodic boundary conditions, the term $\sim \nabla^{2} u(x)$   vanishes. At the minimum-energy configuration $u_{w}^{{\rm min}}(x)$, this yields 
\bea\label{Fhat}
m^{2}  (u_{w}-w ) = \frac{m^{2}}{L^{d}}\int_{x}  u_{w}^{\min}(x)-w  \nn\\
=\frac1{L^{d}} \int_{x} {F} \big(x,u_{w}^{{\rm min}}(x)\big)=: \hat F(w)\punkt
\eea
The last equation defines the effective force $\hat F(w)$.
Its second cumulant reads
\bea 
\highlight{
\overline{\hat F(w) \hat F(w')}^{\,\rm c} \equiv m^{4} \overline{[w-u_{w}] [w'-u_{w'}] }^{\,\rm c} \nn\\
= L^{-d}   \Delta(w-w')\punkt}
\label{defDe}
\eea
Taking two derivatives of \Eq{defR}, one verifies that  the effective correlators for potential and force are related by
$\Delta(u)=-R''(u)$, as in the microscopic relation \eq{Delta-R-rel}.    \Eq{Delta'(0+)} remains valid (without an additional factor of $L^d$).

\subsection{$\Delta(u)$ and the cusp in simulations}
A numerical check has been performed
  in Ref.~\cite{MiddletonLeDoussalWiese2006}, using a powerful
exact-minimization algorithm, which finds the ground state in a time
polynomial in the system size. 
The result of these measurements is presented in figure \ref{f:Alan1}. The  function $\Delta(u)$ is normalized   to $1$ at $u=0$,  and 
the $u$-axis is rescaled (to yield integral 1) to eliminate all non-universal scales.
As a result, the plot is parameter free, thus what one compares is purely the shape. It has
several remarkable features. Firstly, it   shows that a linear
cusp exists in all dimensions. Next it is very close to the 1-loop
prediction. Even more remarkably the statistics is good enough to reliably estimate the deviations from the 2-loop predictions of \cite{ChauveLeDoussalWiese2000a}, see figure \ref{f:RF-dep}.

While we vary the position $w$ of the
center of the well, it is not a real motion.  Rather it means to find the
new ground state given $w$.  Literally {\em moving} $w$ is another
  interesting possibility: It measures the universal properties of the so-called {\em depinning transition}, see section \ref{s:dynamics}.

\paragraph{A technical point.} A field theory is usually defined by its partition function $ Z[J]$ in presence of an applied field $J$.
To obtain the effective action $\Gamma(u)$, one evaluates the free energy $\ca F[J]:=-kT \ln  Z[J] $, and then performs a Legendre transform from $\ca F[J]$ to $\Gamma[u]$. The effective action, solution of the FRG flow equation, is the 2-replica term in $\Gamma[u]$, and not $\ca F[J]$ itself.  
  When measuring the force-force correlations in \Eq{defDe}, these are technically part of $\ca F[J = m^2 w]$. Passing from $\ca F[J]$ to $\Gamma[u]$ is achieved by amputating the   correlation function.   Due to the statistical tilt symmetry discussed in section \ref{s:tilt-symmetry}, the latter does not renormalize. For  the zero-mode (zero momentum) we consider, this amounts to   multiplying twice with $m^2$, resulting in the prefactor of $m^4$ in \Eq{defDe}. In the dynamics, this remains   true in the limit of a vanishing driving velocity. 
These points are further discussed in Refs.~\cite{LeDoussal2006b,MiddletonLeDoussalWiese2006,LeDoussalWiese2006a,WieseLeDoussal2007,LeDoussal2008,terBurgWiese2020}.

\subsection{Beyond 1-loop order}\label{beyond1loop} 
We have successfully applied 
functional renormalization  at 1-loop
order. From a field theory, we  demand more. Namely that
it
\begin{enumerate}

\item be renormalizable\footnote{\new Renormalizability  is a key concept of (perturbative) field theory about the organization of the leading divergences in perturbation theory, which imposes   constraints on higher-order diagrams. These constraints were historically important to derive the RG equation at 2-loop order \cite{ChauveLeDoussalWiese2000a,LeDoussalWieseChauve2003}.}, 

\item allows for systematic corrections beyond 1-loop order, 

\item and thus allows us to make universal predictions. 

\end{enumerate}\begin{widefigure}
\setlength{\unitlength}{.37mm}
{\begin{picture}(462,140)
\put(-1,0){\fig{0.48\textwidth}{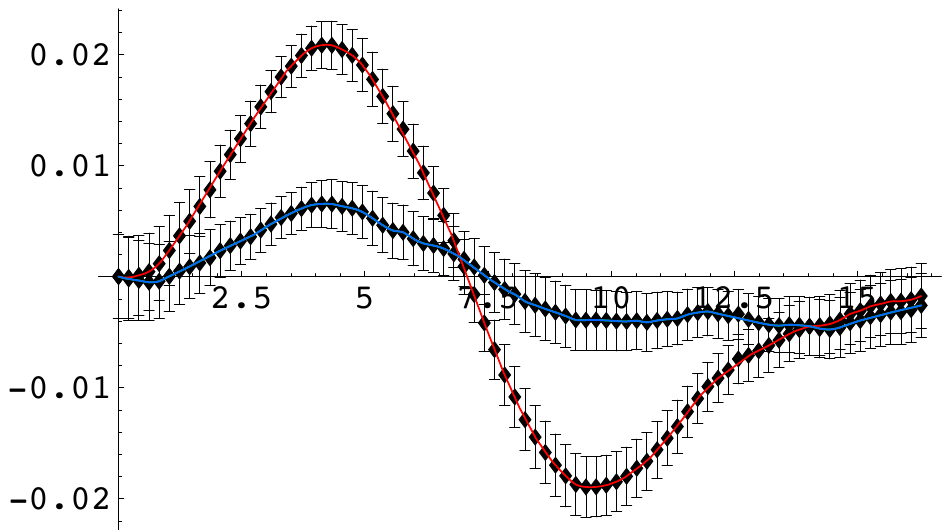}}
\put(0,130){ $ \delta \Delta (u) $}
\put(90,110){\rot $\Delta_{\rm RB}^{\rm sim}(u)$ (1-loop)}
\put(120,65){\blue $\Delta_{\rm RB}^{\rm sim}(u)$ (2-loop)}
\put(340,110){\rot $\Delta_{\rm RF}^{\rm sim}(u)$ (1-loop)}
\put(310,60){\blue $\Delta_{\rm RF}^{\rm sim}(u)$ (2-loop)}
\put(220,55){ $u$} 
\put(232,0){\fig{0.48\textwidth}{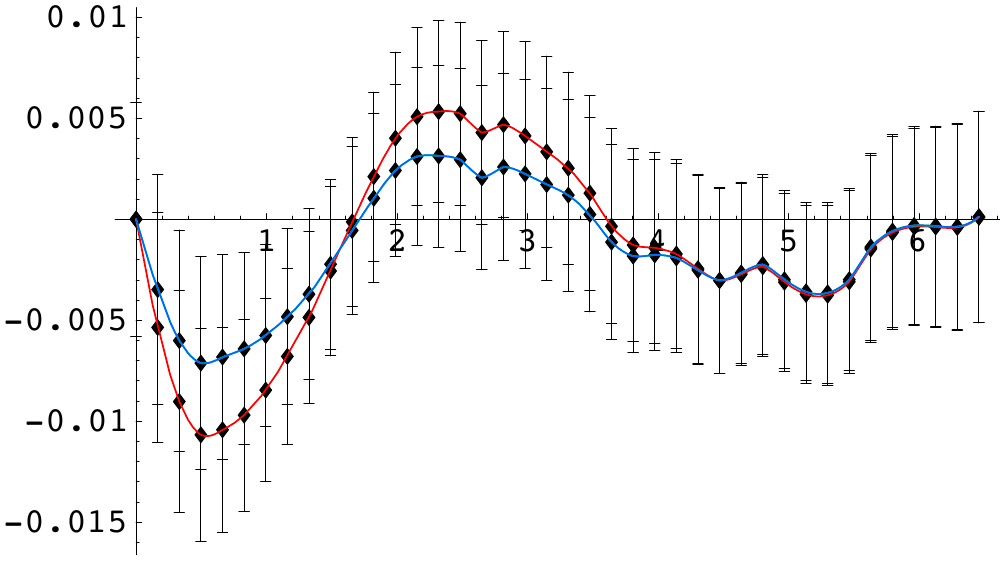}}
\put(236,130){ $ \delta \Delta (u) $}
\put(455,70){$u$}
\end{picture}}
\caption{The measured $\Delta(u)$ in equilibrium with the 1-loop (red) and 2-loop corrections (blue) subtracted. Left: RB-disorder $d=2$. Right: RF-disorder $d=3$. One sees that the 2-loop corrections improve the precision, and that the second-order correction is stronger in $d=2$ than in $d=3$.}
\label{f:Delta-loops}
\end{widefigure}%
\noindent This has been a puzzle since 1986, and it was even
  suggested that the theory is not renormalizable due to the
appearance of terms of order $\epsilon ^{\frac{3}{2}}$
\cite{BalentsDSFisher1993}. Why is the next order so complicated? The
reason is that it involves terms proportional to $R''' (0)$. A look at
figure \ref{fig:cusp} or \ref{f:Alan1} explains the puzzle. Shall we use the symmetry of $R (u)$ to
conclude that $R''' (0)$ is 0? Or shall we take the left-hand or
right-hand derivatives, related by
\begin{eqnarray}
R''' (0^{+}) := \lim_{{u>0}\atop {u\to 0}} R ''' (u) = -
\lim_{{u<0}\atop {u\to 0}} R ''' (u) =:- R''' (0^{-}) .\nn\\
\end{eqnarray}
Below,  we   present the solution of this puzzle, obtained
at 2- and 3-loop order. This is then extended to finite $N$ (section \ref{s:finiteN}),    compared to large
$N$ (section \ref{largeN}), and   the driven dynamics (section \ref{s:dynamics}).

The flow equation
  was first calculated at 2-loop order without the anomalous terms $\sim R'''(0^+)^2$   \cite{WagnerGeshkenbeinLarkinBlatter1999}. The full result with  the necessary anomalous terms was first obtained at 2-loop order \cite{LeDoussalWieseChauve2003,ChauveLeDoussalWiese2000a,Scheidl2loopPrivate,DincerDiplom,ChauveLeDoussal2001},  and later extended to 3-loop order  
\cite{HusemannWiese2017,WieseHusemannLeDoussal2018}.\numparts
\begin{eqnarray}\label{3loopRGR}
\partial _{\ell} \tilde R (u) = \left(\epsilon -4 \zeta  \right) \tilde R (u) +
\zeta u \tilde R' (u) \nn\\
+ \frac{1}{2} \tilde R'' (u)^{2}-\tilde R'' (u)\tilde R'' (0) \nn \\
 +  ({\textstyle \frac{1}{2} {+}\ca{C}_1 \epsilon} ) \Big[ \left(\tilde R'' (u){-}\tilde R'' (0) \right)\tilde R'''
(u)^{2}{-} \tilde R''' (0^{+})^{2 } \tilde R'' (u) \Big]\nn\\
+ \ca{C}_4 \Big\{ \tilde{R}''(u) \big[ \tilde{R}'''(u)^2 \tilde{R}''''(u) -\tilde{R}'''(0^+)^2\tilde{R}''''(0^+)\big] \nn\\
\qquad-\tilde{R}''(0^+)  \tilde{R}'''(u)^2 \tilde{R}''''(u)\Big\} \nn \\
 + \ca{C}_3 \big[\tilde{R}''(u){-}\tilde{R}''(0^+)\big]^2 \tilde{R}''''(u)^2 \nn\\
 + {\ca{C}_2 \big[ \tilde{R}'''(u)^4 - 2\tilde{R}'''(u)^2\tilde{R}'''(0^+)^2\big]}
\\
{\cal C}_{1} =  \frac1 {36} \textstyle \big[{9+4\pi ^{2}-6
\psi' (\frac{1}{3})}\big] =-0.335976 \komma \\
{\cal C}_{2}=  \frac{3}{4}\zeta (3)+\frac{\pi ^{2}}{18}-\frac{\psi'
(\frac{1}{3})}{12} = 0.608554\komma   \\
{\cal C}_{3}=  \frac{\psi'
({\textstyle\frac{1}{3})}}{6} -\frac{\pi ^{2}}{9} =  0.585977\komma \\
{\cal C}_{4} =  2+\frac{\pi ^{2}}{9}-\frac{\psi' (\frac{1}{3})}{6}= 1.414023\punkt
\end{eqnarray}\endnumparts
The first line contains the rescaling terms, the second line the result at 1-loop order, already given in
Eq.~(\ref{RG1loop}). The third line is new; setting there $\epsilon=0$ is the 2-loop result. All remaining terms (proportional to ${\cal C}_{1},...,{\cal C}_{{4}}$) are 3-loop contributions, which we put here   for completeness. 

Consider now 
the last term of the third line, which involves $R''' (0^{+})^{2}$ and which we call {\em anomalous}.  The hard task is to fix the prefactor
$-1$.   There are  different prescriptions to
do this: The sloop-algorithm, recursive construction,
reparametrization invariance, renormalizability, potentiality and exact RG
\cite{ChauveLeDoussalWiese2000a,LeDoussalWieseChauve2003,WieseHusemannLeDoussal2018}. For lack
of space, let us consider only renormalizability, a {\em necessary} property for a field theory. The following  2-loop
diagram leads to the   anomalous term
\bea\label{IA-subdivergence}
{\parbox{2.4cm}{{\begin{tikzpicture}
\coordinate (x1) at  (0.35,0) ; 
\coordinate (x2) at  (1.35,0) ; 
\coordinate (x3) at  (0.85,0.866025) ; 
\node (x) at  (-.3,0)    {$\!\!\!\parbox{0mm}{$\raisebox{-3.5mm}[2.5mm][2.5mm]{$\scriptstyle R'''$}$}$};
\node (x) at  (1.7,0)    {$\!\!\!\parbox{0mm}{$\raisebox{-3.5mm}[2.5mm][2.5mm]{$\scriptstyle R'''$}$}$};
\node (x) at  (0.6,1)    {$\!\!\!\parbox{0mm}{$\raisebox{0mm}[1.5mm][0mm]{$\scriptstyle R'''$}$}$};
\fill (x1) circle (2pt);
\fill (x2) circle (2pt);
\fill (x3) circle (2pt);
\draw  (x2) arc(60:120:1);
\draw  (x2) arc(-60:-120:1);
\draw  (x1)--(x3)--(x2);
\draw[red,thick,dashed](0.1,-.25)--(1.6,-.25)--(1.6,0.25)--(0.1,0.25)--(0.1,-.25);
\end{tikzpicture}}}}
\label{rebi}
\\
\longrightarrow\ \frac{1}{2}\Big[ \big(R'' (u)-R'' (0)
\big)R'''
(u)^{2} -  R'' (u)R''' (0^{+})^{2} \Big]\punkt \nn
\eea
The momentum integral is
\bea
\LH = \int_k\int_p \frac1{(k^2+m^2)^2}\frac{1}{p^2+m^2}
\frac1{ (k+p)^2+m^2} \punkt \nn\\
\eea
In units where the 1-loop integral \eq{I1} is $1/\epsilon$, it reads
\be\label{IA}
\frac{\LH}{\left[ \epsilon \;\Ione\right]^2} 
=  
\frac{1}{2 {\epsilon}^2}+\frac{1}{4 \epsilon}+\ca{O}(\epsilon^0)  \punkt
\ee
The integral \eq{IA-subdivergence}
contains a sub-divergence, which is indicated by the red dashed
box, and which yields the leading  $ 1/\epsilon^{2}$ term in \Eq{IA}. Renormalizability demands that this term be canceled by a 1-loop counter-term. The
latter is unique; it is obtained by replacing $R(u)$ in 
the 1-loop correction $\delta R(u)=\half R''(u)^{2}-R''(u)R''(0)$ by $\delta R(u)$ itself; the last term then yields
\be
\delta R''(0) := \lim_{u\to 0 } \delta R''(u) = \lim_{u\to 0 }   R'''(u)^{2} = R'''(0^{+})^{2}\punkt
\ee 
This fixes the prefactor of the last (anomalous) term in the third line of \Eq{3loopRGR}. 

A physical requirement is that the disorder correlations remain potential,
i.e.\ that forces are derivatives of a potential. The force-force
correlations   being $-R'' (u)$, this means that the flow of
$R' (0^{+})$ has to vanish. (The simplest way to see this is to
study a periodic potential.) From Eq.~(\ref{3loopRGR}) one can check that
this does not remain true if one changes the prefactor of the last
term in the third line of Eq.~(\ref{3loopRGR}); thus fixing it.

\begin{widefigure}
\begin{center}
\begin{tabular}{|c|c|c|c|c|c|}
\hline
$\zeta _{\rm eq}$ & 1-loop & 2-loop & 3-loop & Pad\'e-(2,1) &
simulation and exact\\
\hline \hline $d=3$  & 0.208 &  0.215  & 0.204  &
0.211 &
$0.22\pm 0.01$ \cite{Middleton1995}  \\
\hline
$d=2$ &0.417 &0.444 & 0.358 & 0.423 & $0.41\pm 0.01$ \cite{Middleton1995},  $0.42$ \cite{AlavaDuxbury1996} \\
\hline
$d=1$ & 0.625 & 0.687 & 0.396 & 0.636 & 2/3 \cite{KardarHuseHenleyFisher1985} \\
\hline
\end{tabular}
\vspace{1mm}
\caption{Roughness exponent for random bond disorder obtained by an $\epsilon$-expansion in comparison with exact results and numerical simulations. In the fourth column is an estimate value using a (2,1)-Pad\'e approximant of the 3-loop result.
}
\label{fig:numstat}
\end{center}
\end{widefigure}

\subsubsection*{RP disorder.}
Let us give   results for cases of physical interest. First of all,
for  a periodic potential (RP), which is relevant for
charge-density waves, the fixed-point function can be calculated
analytically. With the notations of \Eqs{RP-fixed-point}-\eq{RP-fixed-point-2} this reads (with the choice of period 1,   $u\in
\left[0,1 \right]$)\numparts
\bea\label{RP-3-loop}
R_{\rm RP} (u) = - u^{2}(1-u)^{2} \frac g {24}+\mbox{const.}, \\
\Delta_{\rm RP}(u) = \frac g {12}- \frac g 2 u(1-u),\\
 g=  \frac{\epsilon }{3}+ 
 \frac{2 \epsilon ^2}{9}+\frac{\epsilon ^3}{81} \ \Big[9 + 2 \pi ^2-18 \zeta
   (3)-3 \psi ' ({\textstyle\frac{1}{3} })\Big]\nn\\
\quad~~  +\ca O(\epsilon ^4).
\eea\endnumparts
This gives a universal amplitude for the 2-point function at 2-loop \cite{LeDoussalWieseChauve2003} and 3-loop order \cite{HusemannWiese2017}, 
\be\label{universal-amplitude-1}
 \overline{\tilde u(q) \tilde u(-q) }\Big|_{q=0} = \frac{g  }{6 m^d}.
\ee
This in turn leads to a logarithmic growth of the  2-point function in position space.
The amplitude is  more complicated to extract, as one needs to extract the asymptotic behavior of scaling functions involved in this transformation. 
Using  Eqs.~(4.13)-(4.18) of \cite{HusemannWiese2017}\footnote{Eq.~(4.15) of \cite{HusemannWiese2017} should read $F_d(0)=1$.}, it can be written   as\numparts
\bea\label{universal-amplitude-2}
\half \overline{[u(x)-u(0)]^2 } = \frac{g B(d)}{6 (4 \pi)^{\frac d 2}\Gamma(\frac d 2)} \ln\!\left(\frac{|x|}{L}\right), \\
 B(d) = \frac{1+0.134567 \epsilon}{1+1.134567\epsilon} + \ca O(\epsilon^3).
 \label{universal-amplitude-2b}
\eea\endnumparts

\subsubsection*{RF disorder.}
For  random-field
disorder, the argument  given in \Eq{357} is still valid, and  $\zeta
=\frac{\epsilon }{3}$ remains valid, equivalent to the Flory estimate (\ref{a3}). The fixed-point function $\Delta(u)$ changes, and can up to 3-loop order be given analytically \cite{HusemannWiese2017}. 

\subsubsection*{RB disorder.}
\label{s:2-loop-RB}
For random-bond disorder (short-ranged
potential-potential correlation function) we have to solve
\Eq{3loopRGR} numerically, order by order in $\epsilon$. The result is \cite{HusemannWiese2017}
\bea
\zeta_{\rm RB} = 0.208 298 04
\epsilon +0.006858 \epsilon ^{2}  
 -0.01075  \epsilon ^{3}+ {\cal O}(\epsilon^{4})\punkt\nn\\
\eea This compares well with numerical
simulations, see figure \ref{fig:numstat}. It is also surprisingly close to, but distinct from, the Flory estimate (\ref{a2}), $\zeta=\epsilon/5$. For $d=1$ ($\epsilon=3$) it gets close to the exact value  \cite{Kardar1987}
\be\label{zeta=2/3}
\zeta_{\rm RB}^{d=1}=\frac23.
\ee 
The fixed-point function $\Delta(u)$ can be obtained up to 3-loop order numerically  \cite{HusemannWiese2017}.

\subsection{Stability of the fixed point}
\label{s:Stability of the fixed point}

Having found a fixed point, 
\be\label{86}
\partial_\ell \Delta(u) = \beta[\Delta](u)=0, 
\ee
one has to ascertain that it is stable. Linear stability is analyzed by considering infinitesimal perturbations of the fixed point 
\be
\delta \beta[\Delta,z](u):= \frac{\rmd }{\rmd \kappa } \beta[\Delta +\kappa z ](u)\Big|_{\kappa=0}  \punkt
\ee
Assuming that $\Delta(u)$ is a   solution of \Eq{86},   the eigenvalue equation reads
\be\label{omega4FRG}
\delta \beta[\Delta, z](u) = -\omega z(u).
\ee
The exponent $\omega$, if it exists, is the standard {\em correction-to-scaling exponent} \cite{Zinn} {\new associated to the eigen-mode $z(u)$. In contrast to standard RG, more than one eigen-mode may exist.}
The solutions to \Eq{omega4FRG} depend on the  universality class. 

First, for the periodic fixed point \eq{RP-3-loop}, there is a discrete spectrum of solutions\footnote{As in \cite{KompanietsWiese2019} we use the high-energy-physics conventions with $\omega>0$ for an IR-attractive fixed point. This is opposite to some earlier work, as  the leading solution   given in \cite{HusemannWiese2017}.}, 
\numparts
\bea
\omega_{-1} = -\epsilon \komma\quad z_{-1}(u) =1. \\
\label{omega-1-CDW}
\omega_1 =\epsilon-\frac23 \epsilon^2 +\frac{5+12 \zeta(3)}9 \epsilon^3+\ca O(\epsilon^4)\komma \\
  z_1(u) = 1-6u(1-u). \nn\\
\omega_2 = 4 \epsilon -5 \epsilon^2 +\frac56 [13+12\zeta(3)]\epsilon^3 +\ca O(\epsilon^4)\komma\\
z_2(u) = 1 \nn\\
{-} \Big\{ 15 {+}5 \epsilon {-}\frac56 \epsilon ^2 \Big[ 12 \zeta
   (3){+}5{+}2 \pi ^2-3 \psi'({\textstyle\frac{1}{3}})\Big]\Big\} u(1{-}u) \nn\\
 + \Big\{ 45{+}25 \epsilon {-}\frac{25}{6} \epsilon ^2
   \Big[ 12 \zeta (3){+}5{+}2 \pi ^2{-}3 \psi'(\textstyle\frac{1}{3})\Big] \Big\} [u(1{-}u)]^2 \nn\\
\omega_3 =\frac{25 \epsilon }{3}-\frac{140 \epsilon
   ^2}{9}+\frac{70}{9} \big[4 \zeta (3){+}7\big]
   \epsilon ^3    + \ca O(\epsilon^4)\\
\hspace{4cm}\vdots \nn
\eea\endnumparts
The first solution $\omega_{-1}=-\epsilon$ is relevant, and comes with a constant perturbation for $\Delta(u)$. It is inadmissible in equilibrium, where $\int_u \Delta(u)=0$, but shows up at depinning, see section \ref{s:dynamics}. 
Thus the {\em leading} perturbation is $z_1(u)$, proportional to the fixed-point solution $\Delta^*(u)$ itself. As the flow in this subspace can be represented by the flow of a  single coupling constant $g$, the $\beta$-function must be a polynomial in $g$, and at leading order it is a parabola. This parabola has two fixed points, with slope $-\epsilon$ at $g=0$ and consequently slope $\epsilon$ at the non-trivial fixed point. This explains why the eigenvalue $\omega_1$ starts with $\epsilon$, making the fixed point stable, exactly as in scalar $\phi^4$ theory \cite{Zinn}. 
The following solutions $z_n(u)$ can be classified by their maximal order in $[u(1-u)]^n$. One sees that the larger $n$, the larger $\omega_n$. Thus this fixed point is perturbatively stable. 

The analysis is more difficult for the non-periodic fixed points, i.e.\ those which allow for a non-trivial exponent $\zeta>0$. The random-bond and random-field fixed points above belong to this class. 
While a proof of stability   even for the 1-loop fixed point is 
still lacking, there are two analytical solutions which can be given  (\cite{LeDoussalWiese2003a}, section VII):
\bea
\omega_0 = 0 \\
z_0(u) = u \Delta'(u) - 2 \Delta(u)\nn\\
\omega_1 = \epsilon \\
z_1(u) = \zeta u \Delta'(u) + (\epsilon-2 \zeta)\Delta(u).\nn
\eea
The first one is a  redundant perturbation {\new in the sense of Wegner \cite{Wegner1974}: it is a consequence of the invariance of the $\beta$-function under the rescaling $\Delta(u) \to \kappa^{-2}\Delta(\kappa u)$. 
In   conformal field theory, redundant operators are associated to null states \cite{PolandRychkovVichi2019}. Their eigenvalues have no physical meaning.}
The dominant solution thus is $\omega_1$, $z_1(u)$, which for $\zeta=0$ reduces (at leading order) to \Eq{omega-1-CDW}. 
Subleading solutions can be constructed numerically \cite{HusemannWiese2017}. 

To conclude, we believe that all the FRG fixed points discussed above are perturbatively stable, and that the leading eigenvalue, i.e.\ correction-to-scaling exponent is $\omega_1 = \epsilon + \ca O(\epsilon^2)$. Order-$\epsilon^2$ corrections depend on the universality class \cite{HusemannWiese2017}.

\subsection{Thermal rounding of the cusp}\label{s:Rounding the cusp} 
\subsubsection*{Generalities.}
As we have seen, a cusp non-analyticity necessarily arises at zero temperature, due
to the jumps between   metastable states.  Interestingly,
this cusp can be rounded by several effects: By {\new a non-zero temperature
$T>0$ (see below), {\em disorder chaos} as defined in section \ref{s:Disorder chaos}, or a non-zero driving velocity in the dynamics}
(section \ref{s:The effective disorder, and  rounding of the cusp by a finite driving velocity}). 

{\new Let us start by a finite temperature, which is 
  easy to include   in}
the FRG equation   \cite{ChauveGiamarchiLeDoussal2000}. The additional 1-loop correction to $R(u)$ is 
\bea
\delta R(u) =    T \left[\, {\parbox{1.5cm}{{\begin{tikzpicture}
\coordinate (x1t1) at  (0,0) ; 
\coordinate (x1t2) at  (1,0) ;
\fill (x1t1) circle (2pt);
\fill (x1t2) circle (2pt);
\draw  (x1t2) arc(-90:270:.4);
\draw [dashed,thick] (x1t1) -- (x1t2);
\end{tikzpicture}}}} \;\right] =  T R''(u) \times  \ITP, \\
\label{ITP}
 I_{\rm TP}:= \ITP = \int_{k}\frac{ 1}{k^2+m^2}.
\eea
The combinatorial factor is $2$ for the two ends of the interaction, and $1/2$ accompanying the second derivative; this can be checked for $R(u_a-u_b)=(u_a-u_b)^2$.
The RG flow of the tadpole diagram is 
\be
-m \partial_m \left[ \ITP \right] = 2 m^2\Big[ \Ione\Big]= 2 m^2 I_1, 
\ee with $I_1$ given in  \Eq{I1}. 
This leads to the $\beta$-function
\bea\label{tempRG}
\partial _{\ell} \tilde R (u) = \left(\epsilon -4 \zeta  \right) \tilde R (u) +
\zeta u \tilde R' (u) \nn\\
+ \frac{1}{2} \tilde R'' (u)^{2}-\tilde R'' (u) \tilde R'' (0) + \tilde T_\ell
\tilde R''(u) \punkt
\eea
The  dimensionless temperature $\tilde T_\ell$ is
\be
\tilde T_\ell := \frac{2T}\epsilon \left( \epsilon I_1\big|_{m=1}\right)  m^{\theta} = \frac{2T}\epsilon \left( \epsilon I_1\big|_{m=1}\right)  \rme^{- \theta \ell} .
\ee  
The power of $m$ is obtained from   \Eq{Rtillde-def} as 
  the scaling of $m^2 \tilde R''(u) $, i.e.\ $m^{2-\epsilon+2\zeta} = m^{d-2+2\zeta}= m^\theta $.
Although $\tilde T_{\ell}$ finally flows to zero since $\theta >0$ (see \Eq{a8}),
in \Eq{tempRG} it acts as a ``diffusion''
term smoothening the cusp. In fact, at non-zero temperature there
is no cusp, and $R(u)$ remains analytic. The convergence to the
fixed point is non-uniform. For $u$
fixed, $\tilde R(u)$ rapidly converges to the zero-temperature fixed point, except
near $u=0$, or more precisely in a boundary layer of size $u \sim
\tilde T_\ell$, which shrinks to zero in the large-scale limit $\ell \to \infty $, i.e.\ $m\to 0$.
Non-trivial consequences are: The curvature blows up as $R''''(0) \sim
\rme^{\theta \ell}/T \sim L^\theta/T$. We  show in section \ref{s:droplet} that this is related
to the existence of thermal excitations, or {\em droplets} in the statics
\cite{BalentsLeDoussal2004}, and of {\em barriers} in the dynamics, which
grow as $L^\theta$ \cite{BalentsLeDoussal2003}.

\subsubsection*{An analytic solution for the thermal boundary layer.}
\begin{widefigure}
{\fig{0.315\textwidth}{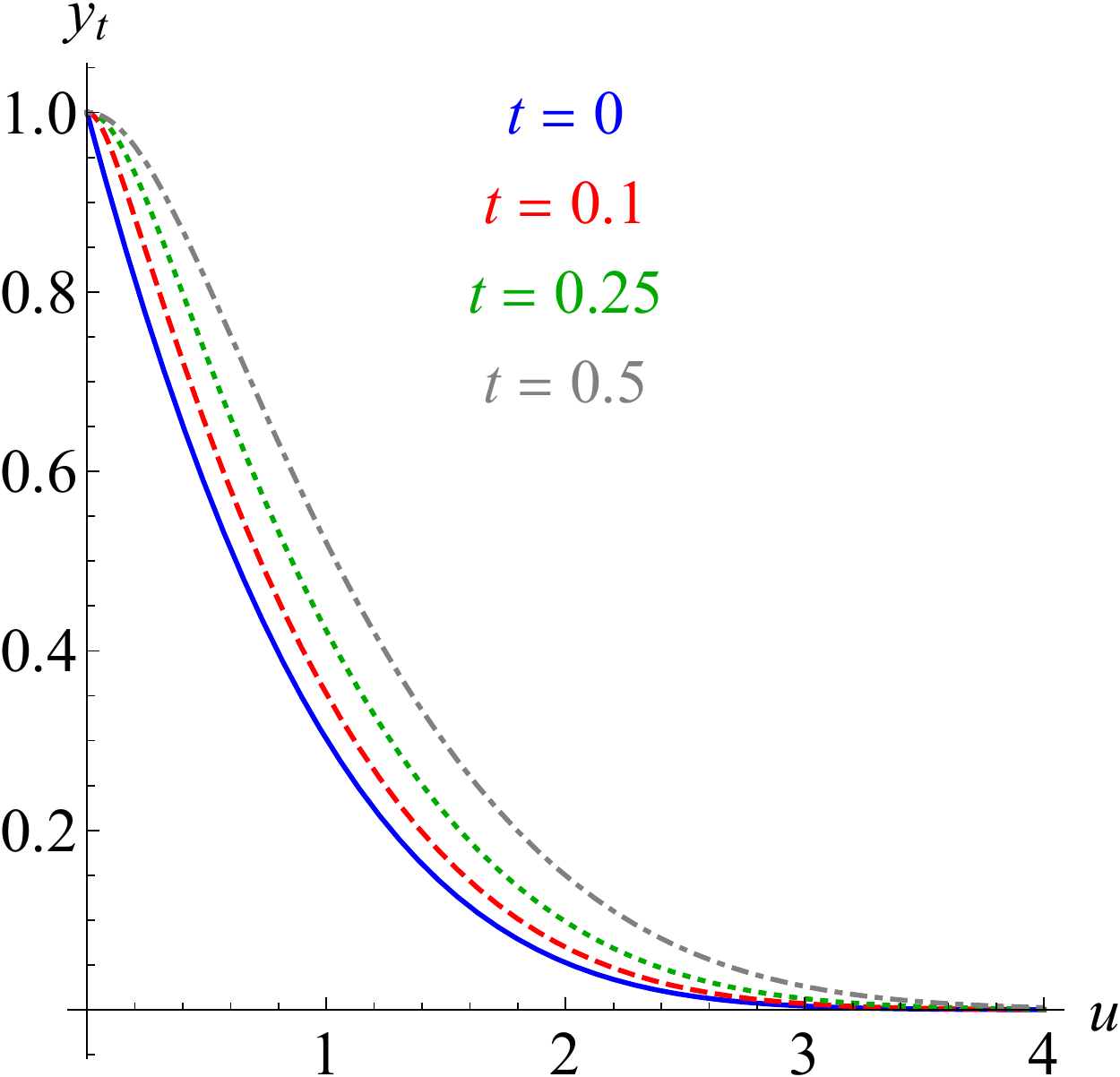}}\hfill\fig{0.315\textwidth}{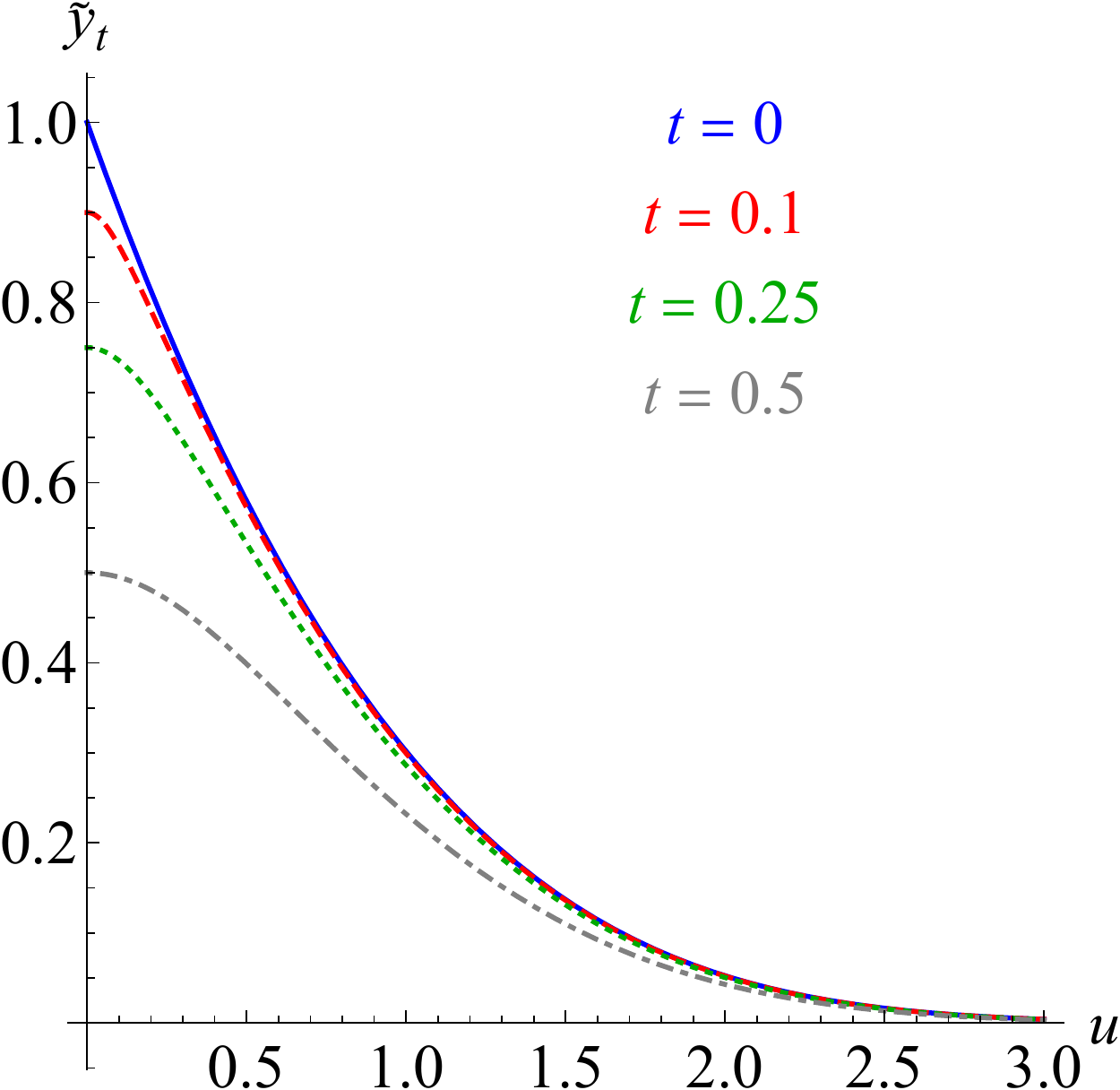}\hfill~~{\fig{0.33\textwidth}{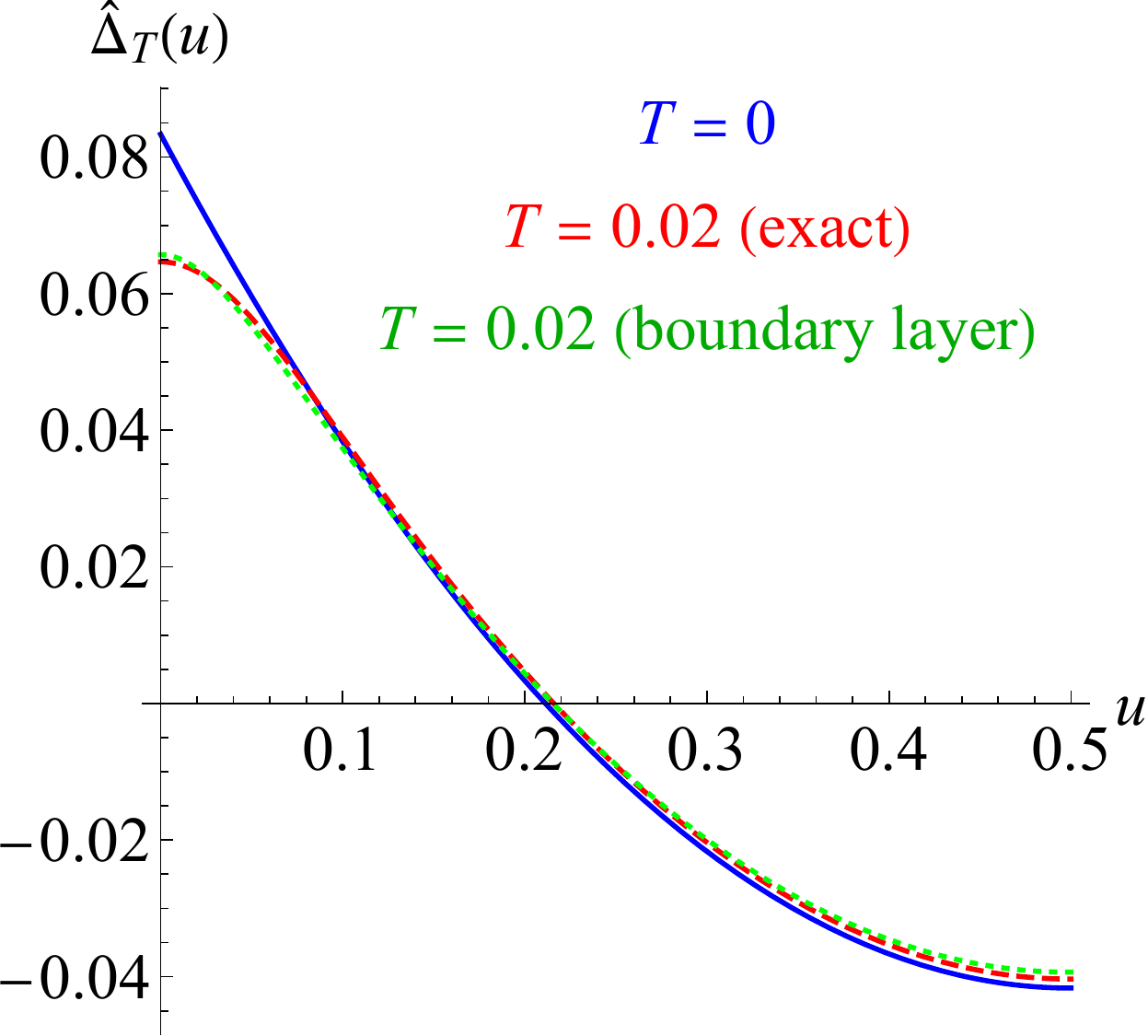}}
\caption{Left: The RF-solution $y_t$ given in \Eq{B13}. Middle:  the rescaled solution $\tilde y_t$ given in  \Eq{B16}. Right: Solution for the toy model. The blue line is the exact result at $t=T=0$; the red dashed line is the numerical integral \eq{CDW-Delta-rounded} for $t=T|_{m=1}=0.02$; the green dotted line is the boundary-layer approximation \eq{CDW-FP-finite-T-approx}.}
\label{f:boundary-layer}
\end{widefigure}
Consider the flow equation \eq{tempRG} for RF disorder. Following section \ref{s:FRG-fixed-points}, we solve it analytically. Setting 
\bea\label{B.10}
-\tilde R''(u) \equiv \tilde \Delta(u)=\frac\epsilon3 \kappa ^{-2} y_t(\kappa u) \\
 y_t(0) = 1\komma  \qquad \tilde T_\ell = \frac{\epsilon}3\kappa^{-2} t \komma  
\eea
 and taking two derivatives of \Eq{tempRG} yields in generalization of  \Eq{2.33}    
\be\label{2.33+T}
\partial_u \left[ u y_t(u) -\half \partial_u \big(y_t(u)-1\big)^2 + t y_t'(u)\right]=0\punkt
\ee
The expression in the square brackets is a constant, fixed to $0$ by considering the limit of $u\to \infty$.
Simplifying gives
\be
u y_t(u) + \big[t +1-y_t(u)\big] y_t'(u) =0\punkt
\ee
Dividing by $y_t(u)$ and integrating once more we arrive at
\be\label{B13}
\frac {u^2}2  + \big(t +1 \big) \ln\!\big(y_t(u)\big) - y_t(u) = -1\punkt
\ee
The integration constant was fixed by considering the limit of $u\to 0$,  $y_t\to 1$. 
This is an explicit analytic solution, plotted on figure 
\ref{f:boundary-layer}. 

It is instructive to relate this to the solution at $t=0$, which will guide us to a general finite-$T$ approximation. To this aim, rewrite \Eq{B13} as 
\be
\frac {u^2} {2(1+t)}  = - \ln\big(y_t (u)\big)  - \frac{1-y_t(u)}{1+t}\punkt
\ee
It can be reduced to  the solution $y_0$ at $t=0$, by  setting $y_t\to (1+t)y_0$, $u^2\to u^2 (1 + t) - 2 \ln (1 + t) (1 + t) + 2 t $.
As a consequence, 
\bea\label{B15}
y_t(u) =  (1+t)\,y_0\!\left( \sqrt{\frac{u^2-2 t }{1 + t}  + 2 \ln(1 +t) }\right).
\eea
Finally using the rescaling invariance \eq{B.10}, we find yet another solution of the flow equation,  \numparts
\bea\label{B16}
\tilde y_t (u) := \frac{1}{1+t'}\, y_{t'} \big(u \sqrt{1+t'}\big)\\
 t = \frac{t'}{1+t'} ~~\Leftrightarrow~~ t'= \frac t{1-t}\punkt
\eea\endnumparts
Using this and the r.h.s.\ of \Eq{B15} yields  
\bea\label{boundary-layer}
\tilde y_t (u)  &=& y_0 \left( \sqrt{u^2-\frac{2 t}{t+1}+2 \ln
   (t+1)} \right) \nn\\
& \approx& y_0 \big( \sqrt{u^2 + t^2} \big)\punkt
\eea
This solution, often in the approximate form of the second line, is commonly used in a {\em boundary-layer} analysis. The idea of the latter is to match a solution in one range, say at small $u$, for which $T\tilde \Delta''(u)$ is large but the non-linear terms in $\partial_\ell \tilde \Delta(u)$ can be neglected, to a solution at large $u$, 
where the former can be neglected. 
As a consequence, 
\be
\lim_{t\to 0} \lim_{u\to 0} t \partial_u^2 \tilde y_t(u)  =  \lim_{u\to 0}   \lim_{t\to 0}  \partial_u \tilde y_t(u).
\ee
To rewrite this in terms of $\tilde \Delta(u)$ is not immediate as we do not know the scale $\kappa$.
However, we can derive a relation directly from the 1-loop flow equation for $\tilde \Delta(u)$, obtained from \Eq{tempRG} after taking two derivatives
\bea\label{tempRG-Delta}
\partial _{\ell} \tilde \Delta (u) = \left(\epsilon -2 \zeta  \right) \tilde \Delta (u) +
\zeta u \tilde \Delta' (u) \nn\\
- \partial_u^2 \frac{1}{2}\left[  \tilde \Delta (u) -\tilde \Delta (0) \right]^2 + \tilde T_\ell
\tilde \Delta''(u) \punkt
\eea
Suppose that    the fixed point is attained and the l.h.s.\ vanishes. Evaluating \Eq{tempRG-Delta} once for   $\tilde T_\ell=0 $, i.e.\ $\tilde T_\ell \to 0 $ and then  $u\to 0$, and once for finite $\tilde T_\ell$, where the limit $u \to 0$ is taken first, we obtain
\be
\left(\epsilon -2 \zeta  \right) \tilde \Delta (0) = \left \{ \begin{array}{ll}
 \tilde \Delta'(0^+)^2, &  \tilde T_\ell=0\\
-\lim_{\tilde T_\ell \to 0} \tilde T_\ell \tilde \Delta''(0), ~ & \tilde T_\ell>0
\end{array}\right. .
\ee
This implies
\be
\tilde \Delta'(0^+)^2\Big|_{T_\ell=0} = - \lim_{\tilde T_\ell\to 0} \tilde T_{\ell} \tilde \Delta''(0)\Big|_{\tilde T_\ell>0}.
\ee
There is a large mathematics and physics literature on the subject. Relevant keywords are  {\em boundary layer} (physics literature) or {\em singular perturbation theory} (mathematics literature); a few references to start with are \cite{Wasow1965,Bogolyubov2011,Smith1985,HairerWanner1996}.

\subsubsection*{Check for a toy model.}
Consider a particle subject to periodic disorder (CDW, RP universality class). 
Suppose that the minimum of the random potential is at $u=u_0+ n i $, $i \in \mathbb Z$, and that this minimum is rather sharp. Then for small $m$, the effective potential $\hat V(w)$ is 
\be
\hat V(w) = - T \ln\! \left( \sum_{i\in \mathbb Z} \exp\Big({-}\frac{(w{-}i{-}u_0)^2m^2}{2T}\Big)\right) \punkt
\ee 
The effective force is 
\be
\hat F(w) = - \partial _w \hat V(w)\punkt 
\ee
We need the force-force correlator, which is obtained as 
\be\label{CDW-Delta-rounded}
  \Delta_T (w) = \left< \hat F(w) \hat F(0)\right>^{\!\!\rm c} = \int_0^1\rmd u_0 \,\hat F(w) \hat F(0)\punkt 
\ee\Review{\begin{widefigure}
\centerline{\mbox{\parbox{0.55\textwidth}{\unitlength1mm
\begin{picture} (90,58)
\put(0,0){\fig{90mm}{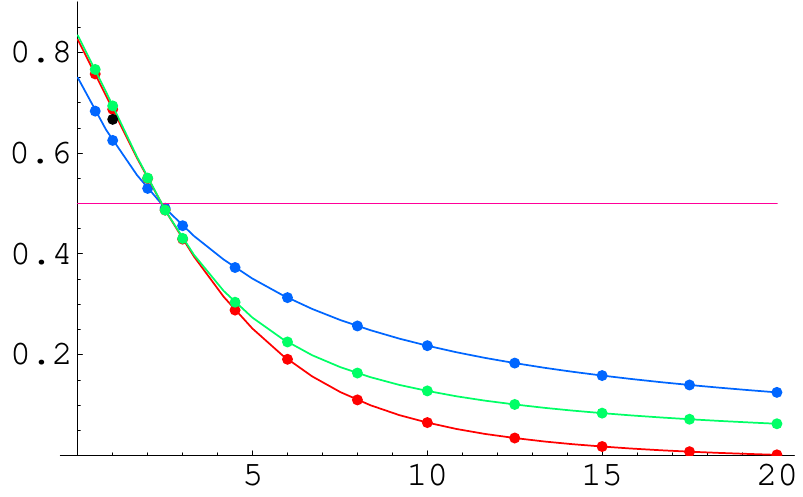}}
\put(5,55){$\zeta $}
\put(90,3){$N$}
\put(70,15){1-loop}
\put(30,7){2-loop}
\end{picture}}}~~~~~~~~~\mbox{\parbox{0.315\textwidth}{\begin{tabular}{|c||c|c|}
\hline
$N$ & $\zeta_{1}$ & $\zeta_{2}$  \\
\hline\hline 1 &0.2082980 &0.0068573 \\\hline 
2 & 0.1765564 &0.17655636 \\\hline
2.5 & 0.1634803 &-0.000417 \\\hline
3 &0.1519065  & -0.0029563  \\\hline
4.5 & 0.1242642 & -0.009386 \\\hline
6 & 0.1043517 & -0.0135901 \\\hline
8 &  0.0856120 & -0.0162957 \\\hline
10 & 0.0725621 &-0.016942 \\\hline
15 &0.0528216 &-0.01564 \\\hline
20 &0.0417 &-0.0138  \\
\hline
\end{tabular}}}
}
\caption{The roughness exponent $\zeta$ as
a function of the number of components $N$:  1 loop (blue),  2 loops  (red), and a 2-loop Pad\'e-(1,1) (green). Reprinted from \cite{LeDoussalWiese2005a}.}  \label{f:Ncomp}
\end{widefigure}}
For $T=0$ we find at $m=1$
\be
  \Delta_0 (w) = \frac1{12} - \frac 12 w (1 - w)\punkt
\ee
This solution is shown in blue in Fig.~\ref{f:boundary-layer} (right). There is also the numerically evaluated integral \eq{CDW-Delta-rounded} (red, dashed). Let us finally consider the FRG-equation for the rescaled disorder $\tilde \Delta_T(w) := m^4   \Delta(w)$, 
\bea
\partial_\ell \tilde \Delta_T (w) &=& 4 \tilde \Delta_T (w) - 4  \partial_u^2 \frac12 \left[\tilde  \Delta_T(w) -\tilde  \Delta_T(0) \right]^2 \nn\\
&&+ 2 T m^2 \tilde \Delta_T ''(w)\punkt
\eea
The prefactors in their order of appearance are: $\epsilon = 4$, $4=-  m \partial_m \ln I_1$ from the 1-loop diagram  $I_1$, and $2 = -  m \partial_m \ln I_{\rm TP}$ from the tadpole $I_{\rm TP}$ defined in \Eq{ITP}. 
Thus the FP equation   at $m=1$ is as above 
\be
0 =  \tilde \Delta_T (w) -  \partial_u^2 \frac12 \left[\tilde  \Delta_T(w) -\tilde  \Delta_T(0) \right]^2 + \frac T 2 \tilde \Delta_T ''(w).~~~~
\ee
At $m=1$,  $\tilde \Delta_T$ and $   \Delta$ coincide, resulting in
\be\label{CDW-FP-finite-T-approx}
\tilde \Delta_T (w) =   \Delta_T (w) \approx   \Delta_0 \left(\sqrt{w^2 + \frac{T^2}4}\right) + \mbox{const}\punkt 
\ee
The constant is chosen s.t.\ $\int_0^1 \rmd w\,   \Delta_T (w) = 0$. This approximation works quite well, see Fig.~\ref{f:boundary-layer}, right.

For complementary descriptions of the high-temperature regime we refer to \cite{BustingorryLeDoussalRosso2010}.

\subsection{Disorder chaos}
\label{s:Disorder chaos}

When  changing the disorder slightly, e.g.\ by  varying    the magnetic
field in a superconductor, the new
ground state may change macroscopically, a phenomenon  termed {\em disorder
chaos} 
\cite{LeDoussal2006a,MiddletonLeDoussalWiese2006,DuemmerLeDoussal2007}. Not all types of disorder exhibit
chaos. Using FRG, one studies a model with two copies, $i=1,2$,
each seeing a slightly different   potential $V_{i} (x,u (x))$ in
Eq.~(\ref{HDO}).  The latter are mutually correlated Gaussian random
potentials with correlation matrix
\begin{eqnarray}\label{ViVj}
 \overline{V_i(x,u) V_j(x',u')} =  \delta^d(x-x') R_{ij}(u-u')\punkt
\end{eqnarray}
At zero temperature, the FRG equations for $R_{11} (u) =R_{22} (u)$
are the same as in \Eq{RG1loop}. The one for the cross-correlator
$R_{12}(u)$ satisfies    equation   (\ref{tempRG}), with
$\tilde T_\ell$  replaced by $\hat T:=R''_{12}(0)-R''_{11}(0)$. The flow of this fictitious temperature
 must be determined
self-consistently from the FRG equations.
As for a real
temperature the cusp is  rounded, leading to a non-trivial cross-correlation function.

\begin{reviewKay}
\subsection{Finite $N$}\label{s:finiteN} Up to now, we have studied the
functional RG for one component $N=1$. The general case of
 $N\neq 1$, here termed {\em finite $N$}, is more difficult to handle, since derivatives of the
renormalized disorder now depend on the direction in which this
derivative is taken. Define amplitude $u:=|\vec u|$ and direction
$\hat u:= \vec u/|\vec u|$ of the field. Then deriving the latter
variable leads to terms proportional to $1/u$, which are diverging in
the limit of $u\to 0$. This poses   problems in the
calculation, and it is a priori not clear that the theory at $N\neq1$
exists, supposed this is the case for $N=1$. At 1-loop order
everything is well-defined \cite{BalentsDSFisher1993}. A 
consistent FRG-equation at 2-loop order  is \cite{LeDoussalWiese2005a}
\begin{eqnarray}\label{2loopFPENcomp}
\partial_{\ell } \tilde R(u) = (\epsilon - 4 \zeta) \tilde R(u) + \zeta u \tilde R'(u)
\nn\\
+\frac{1}{2} \tilde R''(u)^2 - \tilde R''(0) \tilde R''(u) \nn\\
+\frac{N-1}{2} \frac{\tilde R'(u)}{u}
\left[\frac{\tilde R'(u)}{u} - 2 \tilde R''(0)\right]
\nn \\
+\frac{1}{2} \Big[ \tilde R''(u) - \tilde R''(0) \Big] \,{\tilde R''' (u)}^2
\nn\\
+\frac{N{-}1}{2} \frac{{\Big[ \tilde R'(u) {-} u\tilde R''(u) \Big] }^2\, \Big[ 2
\tilde R'(u) {+} u(\tilde R''(u) {-}3 \tilde R''(0) ) \Big]}{u^5}
\nonumber \\
  -\tilde R'''(0^{+})^{2} \left[\frac{N+3}{8}\tilde R''(u)+\frac{N-1}{4}\frac{\tilde R'(r)}{u} \right]
\punkt
\end{eqnarray}
The first line is from   rescaling, the next two lines are the 1-loop contribution given in
\cite{BalentsDSFisher1993}, with the third line containing additional contributions for $N\neq 1$ as compared to \Eq{RG1loop}. The last three lines represent the
2-loop contributions, with   the new anomalous terms proportional to $R'''
(0^{+})^{2}$ in the last line.

The fixed-point equation (\ref{2loopFPENcomp}) can be integrated
numerically, order by order in $\epsilon$. The result, specialized to
directed polymers, i.e.\ $\epsilon =3$ is plotted on figure
\ref{f:Ncomp}.  We see that the 2-loop corrections are rather big at
large $N$, so some doubt on the applicability of the latter down to
$\epsilon=3$ is advised. However both 1- and 2-loop results reproduce
well the two known points on the curve: $\zeta =2/3$ for $N=1$ and
$\zeta =0$ for $N=\infty$. The latter result will be given in section
\ref{largeN}. As   discussed in section \ref{s:KPZ, Burgers, and the directed polymer}, the directed polymer
  in $N$ dimensions treated here, and the KPZ-equation of
non-linear surface growth in $N$ dimensions are related, identifying $z_{\rm KPZ} = 1/\zeta$,  see \Eq{zKPZ=1zeta}. Using the analytic solution for the latter in dimension $N=1$, $\zeta_{\rm KPZ}^{N=1}=1/2$ (\Eq{zeta-KPZ-d=1}), and the scaling relation $z_{\rm KPZ}+\zeta_{\rm KPZ}=2$ (\Eq{Galileo-exp-relation})  leads to 
\be
\zeta_{d=1}^{N=1}=\frac 2 3. 
\ee
The line $\zeta =1/2$ given on figure \ref{f:Ncomp} plays a special
role: In the presence of thermal fluctuations, we expect the
roughness-exponent of the directed polymer to be bounded by $\zeta \ge
1/2$. In the KPZ-equation, this corresponds to a dynamic exponent
$z_{\mathrm{KPZ}}=1/\zeta\le 2$,  which due to the   scaling relation
$z_{\mathrm{KPZ}}+\zeta_{\mathrm{KPZ}}=2$ is an upper bound in the
strong-coupling phase. The results above    suggest that there
exists an upper critical dimension in the KPZ-problem, with
$d_{\mathrm{uc}}\approx 2.4$. Even though the latter value might be
an underestimation, it is hard to imagine what can go wrong {\em
qualitatively} with this scenario. The debate in the literature is far from settled, and we summarize it in section \ref{s:An upper critical dimension for KPZ?}. 

\end{reviewKay}

\subsection{Large $N$}\label{largeN} 
In the last sections we  
discussed renormalization in a loop expansion, i.e.\ an expansion in
$\E=4-d$. In order to   check consistency, we now turn to a 
non-perturbative approach which can be solved analytically in the large-$N$ limit. 
The starting point is a straightforward generalization of \Eq{H}, 
\begin{eqnarray}\label{HlargeN}
{\cal H}[\vec u,\vec j ] &=& \frac{1}{2T} \sum _{a=1}^{n}\int_{x}
[  \vec u_{a} (x)-  \vec w]  (m^{2} -\nabla^{2}  )[ \vec u_{a} (x)- \vec w]\nn\\
&&   -\frac{1}{2 T^{2}}  \sum
_{a,b=1}^{n} \int_x B \left((\vec u_{a} (x)-\vec u_{b} (x))^{2} \right), \\
B(u^2)& =& R(|u|).
\end{eqnarray}
For large $N$ the saddle-point equation reads
\cite{LeDoussalWiese2001,LeDoussalWiese2003b}
\begin{equation}\label{saddlepointequation}
\tilde B' (w_{ab}^{2}) = B' \!\Big(w_{ab}^{2}{+}2 T I_{\rm TP} {+} 4
I_{1} [\tilde B' (w_{ab}^{2}){-}\tilde B' (0)] \Big).
\end{equation}
This equation gives the derivative of the effective (renormalized)
disorder $\tilde B$ as a function of the (constant) background field
$w_{ab}^{2}= (\vec w_{a}-\vec w_{b})^{2}$ in terms of: the derivative of the
microscopic (bare) disorder $B$, the temperature $T$ and the integrals $I_1$ and $I_{\rm TP}$ defined in \Eqs{I1} and \eq{ITP}.
The saddle-point equation can   be turned into a closed functional
renormalization group equation for $\tilde B$ by taking a derivative
w.r.t.\ $m$. In analogy to \Eq{RG1loop}, and with the same notation used there, one obtains \cite{LeDoussalWiese2001,LeDoussalWiese2003b}
\bea
\partial _{\ell}\tilde B (x)&:=& -\frac{m \partial }{\partial m}\tilde
B (x) =\left(\epsilon -4\zeta \right)\! \tilde B (x) + 2 \zeta x
\tilde B' (x) \nn\\
&&+\frac{1}{2}\tilde B' (x)^{2}-\tilde B' (x) \tilde B'
(0)+ \frac{\epsilon\, T \tilde B' (x)}{\epsilon {+}\tilde B'' (0)}.\,\,\,
\eea
This is a complicated nonlinear partial differential equation. It is
  surprising that one can find an analytic solution: The
trick (reminding   the RF-solution \eq{RF-FP-y}) is to examine the flow equation for the inverse function of
$y(x):=   - \tilde B' (x)$,  which  is the dominant term at large $N$ for the  force-force correlator\footnote{The sign of the last term in Eq.~(11) of \cite{LeDoussalWiese2001} must be reversed.},
\bea
m \partial_m x(y) &=& (\epsilon - 2 \zeta) y x'(y) + 2 \zeta x(y) +  y_0 - y
 \nn\\
 &&+\frac{ T_m }{ 1 -(\epsilon x'_0)^{-1}} . \label{linear} 
\eea
 Let us only give the results of
this analytic solution: First of all, for long-range correlated
disorder of the form $\tilde B' (x)\sim x^{-\gamma }$, the exponent
$\zeta $ can be calculated analytically as $\zeta =\frac{\epsilon }{2
(1+\gamma )}\punkt $ It agrees with the replica-treatment in
\cite{MezardParisi1991},  the 1-loop treatment in
\cite{BalentsDSFisher1993}, and the Flory estimate (\ref{a4}). For short-range correlated disorder,
$\zeta =0$.  Second, it demonstrates that before reaching the Larkin-length,
$\tilde B (x)$ is analytic and  dimensional reduction
holds. Beyond the Larkin length, $\tilde B'' (0)=\infty $, a cusp
appears and dimensional reduction is broken. This shows again that
the cusp is not an artifact of the perturbative expansion, but an
important property of the exact solution of the problem (here for large $N$).

\begin{reviewKay}
\subsection{Corrections at order $1/N$}\label{sec:1overN}
In a graphical notation, we find \cite{LeDoussalWiese2004a}
\begin{eqnarray}
\delta B^{(1)}=
\!\!\diagram{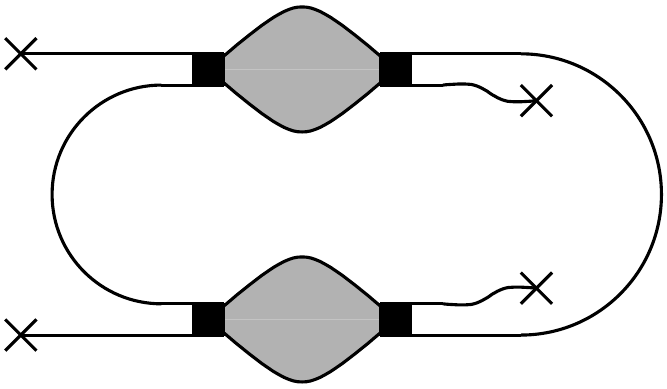}\!\!+\!\!\!\diagram{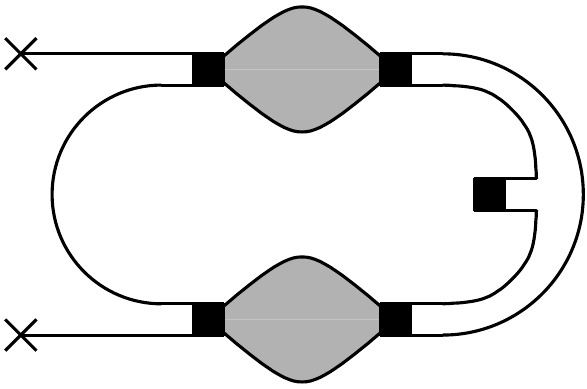}\nn\\
+\!\!\diagram{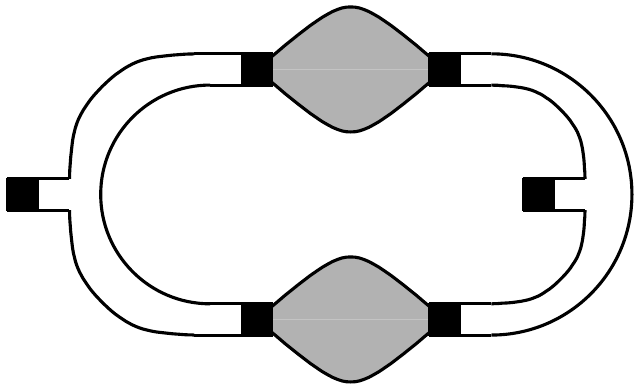}\!\!+\!\!\!\diagram{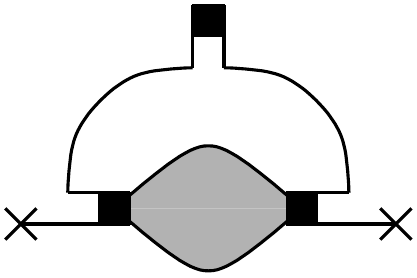}\!\!+\!\!\diagram{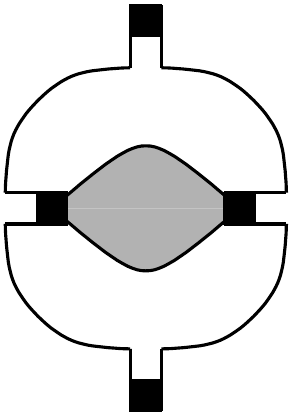}\!\! \nonumber
\\
 +T\Bigg[ \!\!\diagram{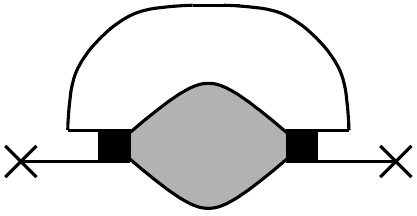} \!\!+ \!\!\diagram{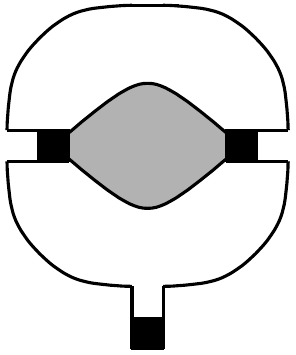} \!\!+
\!\!\diagram{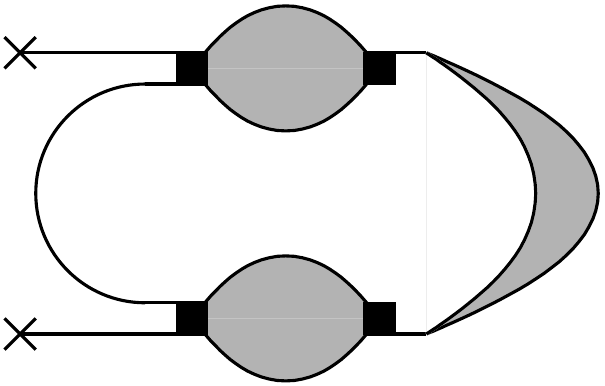} \nn\\
+ \!\!\diagram{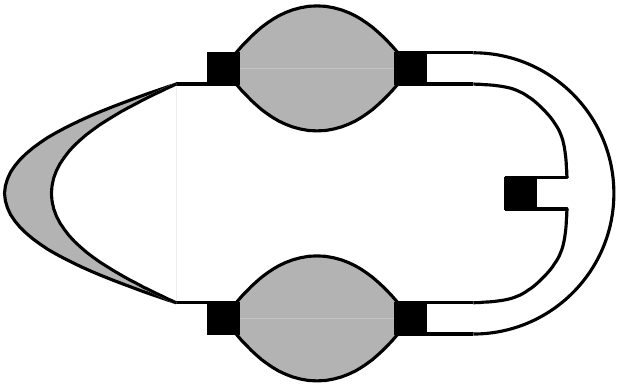}\!\!+
\!\!\diagram{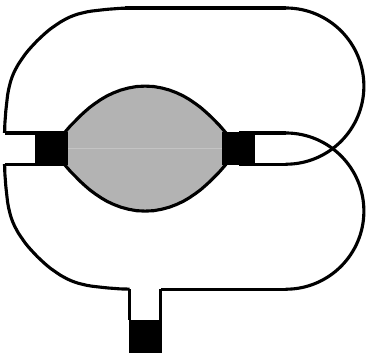} \!\!+ \!\!\diagram{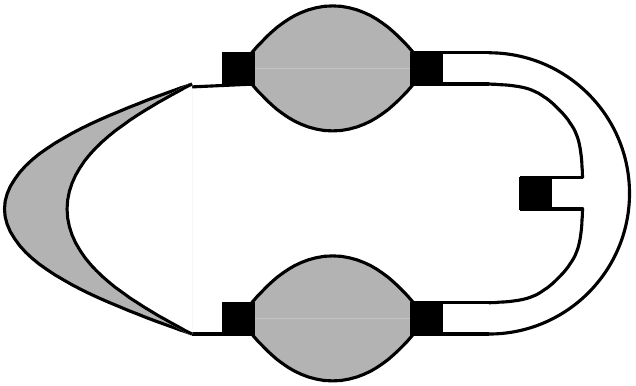}\!\! \Bigg]\nonumber \\
 + T^{2}\Bigg[ \!\!\diagram{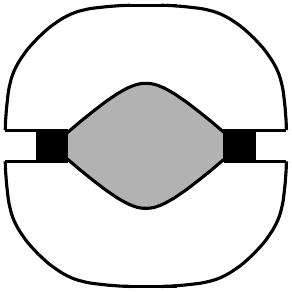} \!\!+ \!\!\diagram{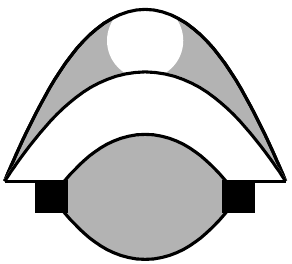}
\!\!+
\!\!\diagram{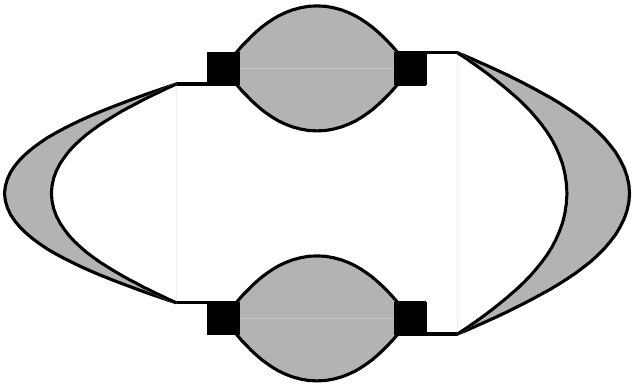} + {\cal A}^{T^{2}}\Bigg]\\
\diagram{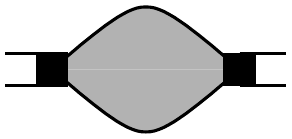}=B'' (\chi _{ab})\left(1-4A_{d} I_{2} (p)B'' (\chi
_{ab}) \right)^{-1}\komma  \nn\\ \diagram{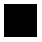}=B(\chi_{ab})\komma 
\end{eqnarray}
where $\chi_{ab}$ is the argument of the r.h.s.~of \Eq{saddlepointequation}.
More  explicit expressions are given in Ref.~\cite{LeDoussalWiese2004a}.

By varying the IR-regulator, one can derive a $\beta$-function at
order $1/N$  \cite{LeDoussalWiese2004a}. At $T=0$, it is
UV-convergent, and should allow one to find a non-perturbative fixed point. This goal has currently only been achieved to 1-loop order \cite{LeDoussalWiese2004a}. 
Another open problem is the behavior at finite $T$.

\begin{reviewKay}
\subsection{Relation to Replica Symmetry Breaking (RSB)}\label{s:RSB} 
One of the key methods employed in  disordered   systems is a method termed {\em replica-symmetry breaking} (RSB) \cite{Parisi1979,Parisi1980c,Parisi1980b,Parisi1980d,ParisiToulouse1980,ParisiToulouse1980b,MezardParisiSourlasToulouseVirasoro1984,MezardParisiSourlasToulouseVirasoro1984b,MezardParisiVirasoro1985,MezardParisiVirasoroBook}, sometimes  referred to as {\em Gaussian variational ansatz}, or simply   {\em mean field}, since there is no tractable scheme to go beyond that limit. It is an interesting task to confront this alternative approach to the FRG. As we saw above, FRG works very well for the experimentally most relevant case of $N=1$, whereas the RSB ansatz only holds in the limit of $N\to \infty$ \cite{MezardParisi1991}. So what is the idea? Ref.~\cite{MezardParisi1991} starts from \Eq{HlargeN} but
{\em without}\/ a source-term $w$, i.e.\ without an applied field, a relevant difference. In the limit of large $N$, a
Gaussian variational energy of the form
\begin{eqnarray}\label{HlargeNMP}
{\cal H}_{\mathrm g}[\vec u] &=& \frac{1}{2T} \sum _{a=1}^{n}\int_{x}
 \vec u_{a} (x)\left(-\nabla^{2}{+}m^{2} \right) \vec u_{a} (x)
\nn\\&&   -\frac{1}{2 T^{2}}  \sum
_{a,b=1}^{n}  \int_x\sigma_{ab} \,\vec u_{a} (x)\vec u_{b} (x)
\end{eqnarray}
becomes exact. The art is to make an appropriate ansatz for
$\sigma_{ab}$. The simplest possibility, $\sigma _{ab}=\sigma $ for
all $a\neq b$ reproduces the dimensional-reduction result, which we know to
break down at the Larkin length. Beyond that scale, a replica
symmetry broken (RSB) ansatz for $\sigma _{ab}$ is suggestive. To this
aim, one   breaks $\sigma _{ab} $ into four blocks of equal size, and
chooses two (variationally optimized) values for the   diagonal and off-diagonal 
blocks. This is termed {\em 1-step RSB}. One  then iterates the procedure on the diagonal blocks, proceeding via a {\em 2-step} to an {\em infinite-step} RSB. The final result has the form
\begin{equation}\label{RSB}
\sigma_{ab} =
\left(\,\parbox{.25\textwidth}{\fig{.25\textwidth}{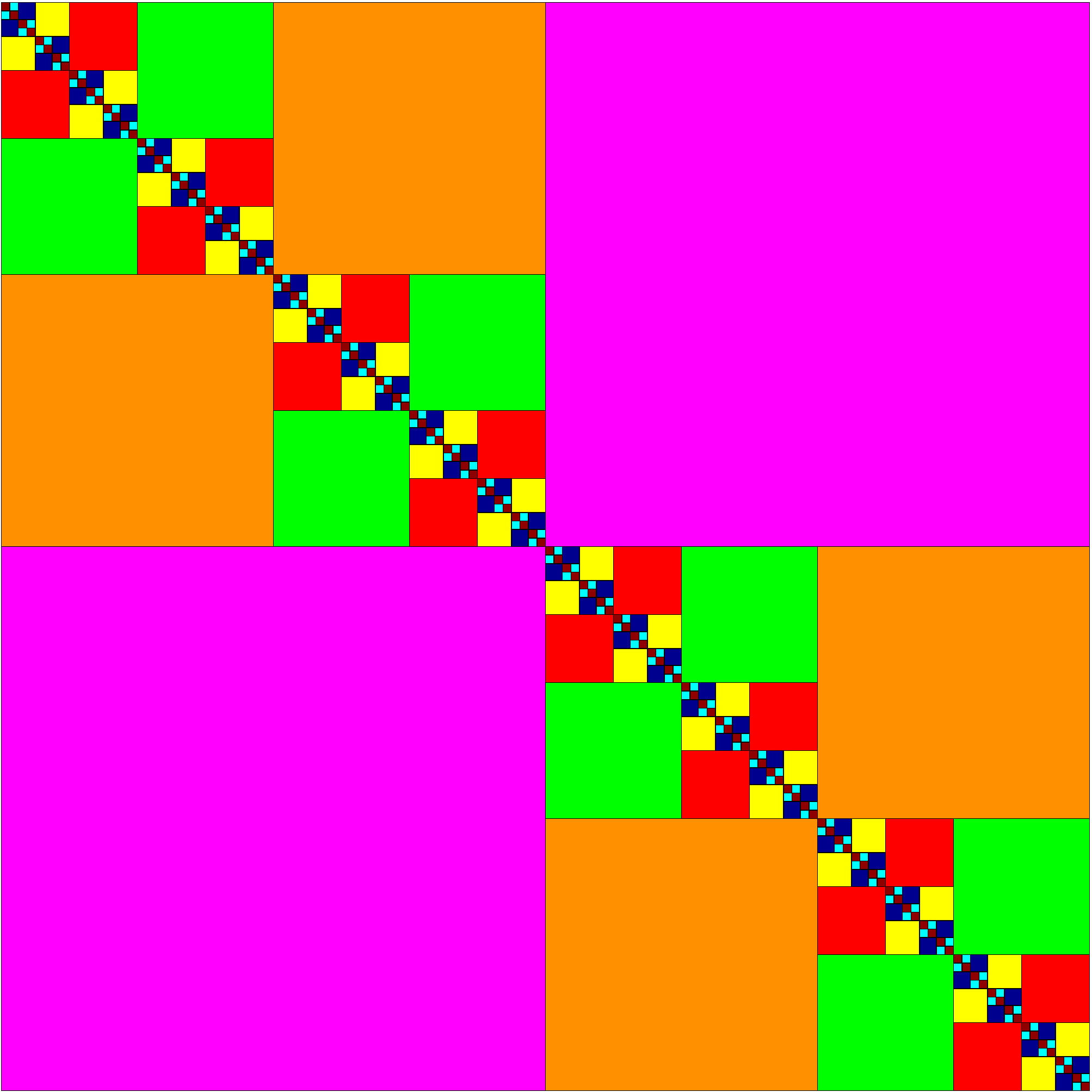}}\,\right)\punkt
\end{equation}
One finds that the more often one iterates, the more stable (to perturbations) and precise the result becomes. In
fact, one has to repeat this procedure infinitely many times. This seems
like a hopeless endeavor, but Parisi has shown \cite{Parisi1979} that the infinitely
often replica-symmetry broken matrix can be parameterized by a
function $[\sigma] (z)$ with $z\in \left[0,1 \right]$. In the
SK-model, $z$ has the interpretation of an overlap between
replicas. While there is no such simple interpretation for the model
(\ref{HlargeNMP}), we retain that $z=0$ describes distant states,
whereas $z=1$ describes nearby states. The solution of the large-$N$
saddle-point equations leads to the curve depicted in figure 6.
Knowing it, the 2-point function is given by
\bea\label{RSBformula}
\left< \tilde u_{k} \tilde u_{-k} \right>=\frac{1}{k^{2}+m^{2}}\left(1+\int_{0}^{1}
\frac{\rmd z}{z^{2}} \frac{\left[\sigma \right]
(z)}{k^{2}+\left[\sigma \right] (z)+m^{2}} \right)\punkt\nn\\
\eea
The important question is: What is the relation between the two
approaches, which both declare to calculate the same 2-point function?
Comparing the analytical solutions, we find that the 2-point function
given by FRG is the same as that of RSB, if in the latter expression
we only take into account the contribution from the most distant
states, i.e.\ those for $z$ between 0 and $z_{m}$ (see figure
\ref{fig:MP-function}). \begin{figure}[t]
\centerline{\fig{8cm}{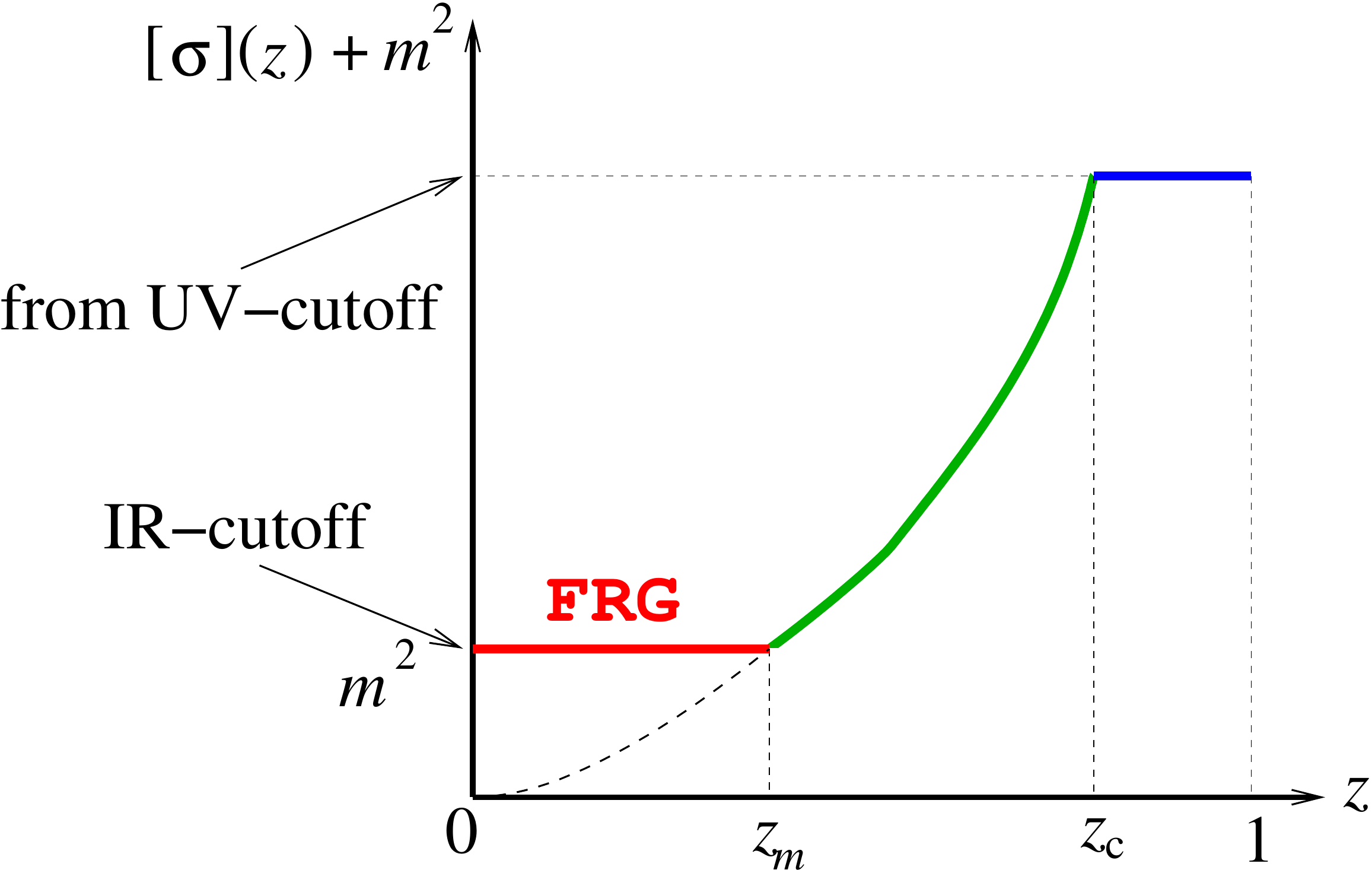}}
\caption{The function $\left[\sigma \right] (u)+m^{2}$ as given in
\protect\cite{MezardParisi1991}.} \vspace{-0.1cm}\label{fig:MP-function}
\end{figure}To understand why this is so, we have to
remember that the two calculations are done under quite different
assumptions: In contrast to the RSB-calculation, the FRG-approach
calculates the partition function in presence of an external field
$w$, which is then used to give via a Legendre transform the
effective action. Even if the field $w$ is finally tuned to 0, the
system   remembers its preparation, as does a magnet:
Preparing the system in presence of a magnetic field   results in a
magnetization which aligns with this field. The magnetization  
remains, even if finally the field is turned off. The same phenomenon
happens here: By explicitly breaking the replica-symmetry through an
applied field, all replicas   settle into distant states, and the
close states from the Parisi-function $\left[\sigma \right] (z)+m^{2}$
(which represent {\em spontaneous} RSB) will not contribute.  However,
 the full RSB-result can be reconstructed by remarking
that the part of the curve between $z_{m}$ and $z_{c}$ is independent
of the infrared cutoff $m$. Integrating over $m$
\cite{LeDoussalWiese2001} then yields ($m_{c}$ is the mass corresponding to
$z_{c}$)
\begin{equation}\label{RSB=intFRG}
\left< \tilde u_{k} \tilde u_{-k} \right>\Big|^{\mathrm{RSB}}_{k=0} =\frac{\tilde
B'_{m}(0)}{m^{4}} +\int_{m}^{m_{c}} \frac{\rmd \tilde
B'_{\mu}(0)}{\mu^{4}} + \frac{1}{m_{c}^{2}}-\frac{1}{m^{2}}\punkt
\end{equation}
We also note that a similar effective action  has been proposed in
\cite{BalentsBouchaudMezard1996}. While it agrees qualitatively, it
does not reproduce the correct FRG 2-point function, as
it should.

To go further, one needs to redo the analysis of \REF{MezardParisi1991} in presence of an applied field, a formidable task. 
A first step in this direction was taken in  \REF{BalentsBouchaudMezard1996}, building on the technique developed in \cite{Duplantier1992}. However, the function $R(u)$
defined in that work does not coincide with the one usually studied in field theory (there is an additional Legendre transform), making a precise comparison
difficult. This goal was finally achieved in \REF{LeDoussalMuellerWiese2007}. 
In summary, there are two distinct scaling regimes,
\begin{equation}\label{summary}
\tilde B(w^2) - \tilde B(0) = \left\{\begin{array}{cccc} L^{-d} \tilde b(w^2 L^d) &\mbox{for}&  w^2 \sim L^{-d}, & \mbox{(i)} \\
 N b(w^2/N) &\mbox{for}&   w^2 \sim N, & \mbox{(ii)}
\end{array}\right. 
\end{equation}
 (i) a ``single shock'' regime, $w^2 \sim L^{-d}$ where $L^d$ is
the system's volume,  and (ii) a ``thermodynamic'' regime, with $w^2 \sim   N$, independent of $L$. In regime (i) all the equivalent  RSB saddle points within the Gaussian variational approximation contribute, while in regime
(ii) the effect of RSB enters only through a single anomaly. When   RSB is continuous (e.g., for short-ranged
disorder, in dimension $2 \leq d \leq 4$),   regime (ii) yields the large-$N$ FRG function shown above. In that case, the disorder correlator exhibits a cusp in both regimes, though with different
amplitudes and of different physical origin. When the RSB solution is 1-step and non-marginal (e.g.\ in $d < 2$ for SR
disorder), the correlator $\tilde R(w)=\tilde B(w^2)$ in regime (ii) is considerably reduced, and exhibits no cusp.

\subsubsection*{RSB at finite $N$.} 
The Gaussian variational ansatz with an infinite-step RSB is possible also at finite $N$, {\new and termed the Gaussian variational model (GVM)}.
For $m=0$, 
\be
[\sigma](u) \sim u^\alpha, \quad \alpha = \frac{4 + N}{d -  N (1- \frac d2)}.
\ee
As a consequence,  
\be
\zeta _{\rm GVM } = \frac{4-d}{4+N} \equiv \zeta_{\rm Flory}^{\rm RB}, 
\ee
where $\zeta_{\rm Flory}^{\rm RB}$ is the Flory estimate \eq{a2}. 
How can this be explained? In  the GVM solution \cite{AgoritsasLecomte2017}, all power-laws can be deduced from dimensional considerations, leaving no room for a deviation from the Flory estimate \eq{a2}. Deviations are possible with additional scales \cite{AgoritsasLecomteUnpublished}.

\end{reviewKay}

\subsection{Droplet picture}
\label{s:droplet}
The droplet picture was   proposed  \cite{BovierFrohlich1986,FisherHuse1986} for Ising spin glasses (more in \cite{FisherHuse1987,HuseFisher1987a,FisherHuse1988,FisherHuse1988a}), as a conceptual alternative to the Parisi solution \cite{Parisi1979,Parisi1980c,Parisi1980b,Parisi1980d} (more in \cite{ParisiToulouse1980,ParisiToulouse1980b,MezardParisiSourlasToulouseVirasoro1984,MezardParisiSourlasToulouseVirasoro1984b,MezardParisiVirasoro1985,MezardParisiVirasoroBook}) of the Sherrington-Kirkpatrick (SK) model \cite{KirkpatrickSherrington1978,SherringtonKirkpatrick1975}. Using the concept of replica-symmetry breaking (RSB), the latter yields  infinitely many extremal Gibbs states at very low temperature, organized within an ultrametric topology, {\new i.e.\ }arrangeable  as a tree. As   temperature is raised, states at increasing distance  coalesce until a unique state remains at $T=T_{\rm c}$.
Appropriate for the infinite-range SK model, its validity for short-ranged spin glasses as the Edwards-Anderson (EA) model \cite{EdwardsAnderson1975} 
is disputed. 
The latter assumes an energy
\be\label{H-EA}
\ca H_{\rm EA}=\sum _{\left<i,j\right>} J_{ij} S_i S_j
\ee
with uncorrelated random couplings $J_{ij}$, drawn from a probability distribution $P(J)$.

Existence of the spin-glass phase is detected by the {\em Edwards-Anderson order parameter}
\be\label{qEA}
q_{\rm EA} (T):= \overline{ \left< S_i\right>_t^2 }, 
\ee
where the overline denotes (as above) the   disorder average (over $J$), and $\left< ...\right>_t$ the temporal average over an infinite time. One expects $q_{\rm EA}(T)= 0 $ for $T>T_{\rm c}$, and $q_{\rm EA}(T)> 0 $ for $T<T_{\rm c}$.

In contrast to the RSB scenario with a finite density of states at $q=0$, the droplet picture proposes that the low-lying excitations which dominate the long-distance and long-time correlations are clusters of (nested) droplets of coherently flipped spins. 
Let us denote by $\ca F_0$ the ground-state free energy, i.e.\ the infimum of all free energies $\ca  F_i$, 
\be
\ca F_{0}:= \inf_i (\ca  F_i).
\ee
 In a pure system at $T=0$, 
the energy of a domain wall can be measured by imposing anti-periodic boundary conditions, which force a domain wall of size $L^{d-1}$ and energy $  \ca E_L \simeq  \Upsilon  L^{d-1}$. In a disordered system at $T>0$ this generalizes to  
\be\label{delta F with theta}
 \ca F_L \simeq  \Upsilon L^\theta,
\ee 
with $\theta <d-1$, where $\ca F_L$ is the   free energy at scale $L$. If one supposes that the free energies $\ca F_L$ in the domain wall are uncorrelated, using the central-limit theorem improves the bound\footnote{One might argue that if a domain wall is rough, then its size scales as $L^{d_{\rm f}}$ with $d-1\le d_{\rm f}\le d$, and we get a weaker bound for $\theta$. This is incorrect. While the minimum-energy domain wall may be larger, its energy is lower; otherwise the minimal-energy interface would be flat.} to $\theta \le \frac{d-1}2$. 
The probability of droplet excitations with free-energy differences $\ca F_L$, given size $L$, has the scaling form 
\bea\label{theta-bound}
\rho(\ca F_L| L) = \frac{\rho(\ca F_L/\Upsilon L^\theta)}{ \Upsilon L^\theta}, \quad \rho(0)>0, \quad 0 < \theta \le\frac{d-1}2. \nn\\
\eea
Values of $\theta$ satisfying the bound \eq{theta-bound} have   been reported \cite{BrayMoore1985}.

Next consider an ensemble of independent (but possibly nested) droplets of size between $L^d$ and $(L+\delta L)^d$. The probability that a spin is inside such a droplet is independent of $L$: while the probability that a spin is inside a droplet scales as $L^d$, the number of droplets scales as $L^{-d}$. 
Thus the   measure for integration  of $\rho(\ca F_L| L)$ over $L$ is 
\be\label{droplet measure}
\rho(L) \rmd L = \frac{\rmd L}{L} \rho(\ca F_L| L).
\ee
Stated differently, a given spin has a finite, $L$-independent, probability to be inside a droplet of size $L$.
 
 Only if both points $i$ and $i+r$ are inside a droplet,
the connected spin-spin correlation function $ \overline{ \left< S_i
\right>_t \left<S_{i+r}\right>_t^{\rm }  }^{\rm c}$,  is non-zero. To give a non-vanishing contribution to the space-dependent version of the Edwards-Anderson order parameter \eq{qEA}, the droplet has to be bigger than $r$, leading to 
\be\label{SS-droplet}
q_{\rm EA}(r) := \overline{ \left< S_i
\right>_t \left<S_{i+r}\right>_t^{\rm }  }^{\rm c} \sim  r^{-\theta}, ~ \mbox{ as }r\to \infty.
\ee
For higher-energy excitations, it is natural to suppose that 
their activation barriers  scale with  $L$ as
\be
\ca   F_{\rm b} \simeq \Delta L^\psi, \quad \theta \le \psi \le d-1.
\ee
The upper bound is given by the maximal energy needed to flip a flat wall. The lower bound    insures that $ \ca F_L$ from \Eq{delta F with theta} satisfies $ \ca F_L< \ca F_{\rm b}$.

To address dynamical properties,  suppose   a droplet persists for a time $t = t_0 \rme^{\ca F_{\rm b}/T}$. The line of  arguments used above   implies that the autocorrelation function decays as 
\bea
C(t):= \overline{ \left< S_i(0) S_i(t)\right>_t^{\rm }  } \simeq \frac{q_{\rm EA} T}{\Upsilon} \left[ \frac{\Delta}{T \ln(t/t_0) }\right]^{\frac \theta \psi}.
 \eea
To conclude our excursion into the droplet picture for the EA-spin glass (for further reading see \cite{FisherHuse1986,HuseFisher1987a,FisherHuse1988,FisherHuse1988a,NewmanStein1997}), let us mention an interesting result due to 
Bovier and Fr\"ohlich
 \cite{BovierFrohlich1986}, who analyze EA-spin glasses with concepts from gauge theory. They  state that  for $d= 2$ the Gibbs state is unique at all temperatures. In the language used here, this implies $\theta<0$. 
The debate whether   EA-spin glasses are better described by the droplet picture or RSB is still raging \cite{MooreBokilDrossel1998,MarinariParisiRuizLorenzoRitort1996,MarinariParisiRicci-TersenghiRuiz-LorenzoZuliani2000,AspelmeierWangMooreKatzgraber2016,CharbonneauYaida2017,Moore2019,YeoMoore2020,HollerRead2020}. It is  possible that depending on the dimension, the distance to $T_{\rm c}$, and small modifications of the EA spin-glass energy \eq{H-EA}, both phases are realized in some domains of the phase diagram, while in other domains, none of them is appropriate.

Let us finally  
apply the droplet ideas to the directed polymer   \cite{HwaFisher1994b}, and more generally disordered elastic manifolds 
\cite{BalentsLeDoussal2002,BalentsLeDoussal2003,BalentsLeDoussal2004,LeDoussal2006b}. 
First of all, the droplet exponent $\theta$ as defined in \Eq{delta F with theta} is the one  given in \Eq{a8}
\be
\theta = d-2 + 2 \zeta. 
\ee
The bound $\theta\le d/2$ translates into
\be
\zeta\le \zeta_{\rm droplet~bound} = \frac{4-d}4.
\ee
It is less clear what the exponent $\psi$ is, but there is some evidence \cite{DrosselKardar1995,MikheevDrosselKardar1995} that 
\be
\psi = \theta,  
\ee
up to   logarithmic corrections, shown to exist at least in one case \cite{MikheevDrosselKardar1995}. 
As an immediate generalization to \Eq{SS-droplet}, we expect that at finite temperature \cite{BalentsLeDoussal2002}
\be
\left< [u(x)-u(y)]^{2n}\right> \simeq T |x-y|^{2n \zeta-\theta} .
\ee
To build a consistent field theory is a challenge. Technically, one has to connect the thermal boundary layer of $\Delta(w)$ (section \ref{s:Rounding the cusp}) with the outer region. Physically,  one needs to make the connection to the    droplet picture. This problem is 
considered in \cite{BalentsLeDoussal2002,BalentsLeDoussal2003,BalentsLeDoussal2004,LeDoussal2006b}.

\subsection{Kida model}
\label{s:Kida}
\begin{figure}[t]
{\setlength{\unitlength}{1mm}\begin{picture}(83,51)
\put(0,0){\Fig{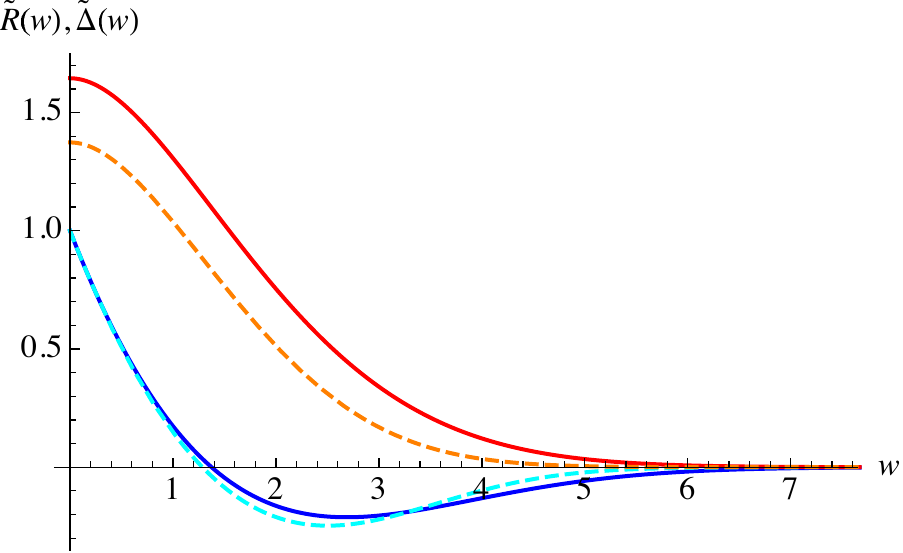}}
\put(40,15){\includegraphics[width=4.3cm]{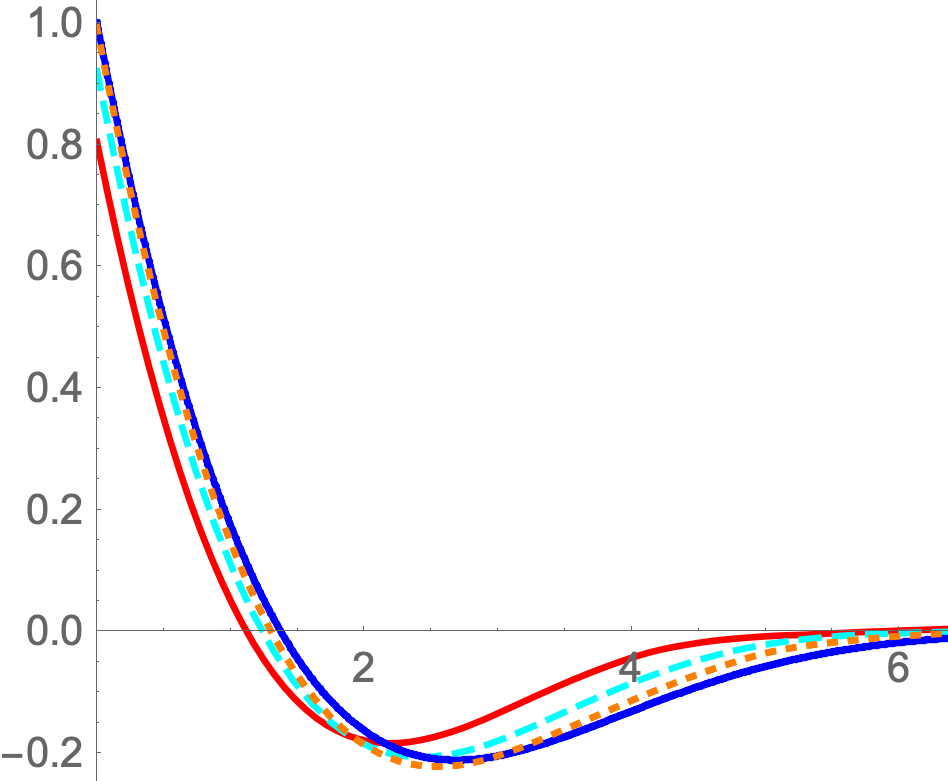}}
\end{picture}}
\caption{The force-force correlator $\Delta(u)$ for the Kida model (blue solid), compared to the RB 1-loop FRG result (cyan, dashed) already shown in Fig.~\ref{f:RF-FP} (right), rescaled to have the same value and slopes as \Eq{Delta-Kida}. In solid red the potential-potential correlator $\tilde R(w)$, in dashed the corresponding 1-loop FRG result. Inset: Numerical simulations \cite{terBurgPhD,terBurgWieseUnpublished} for $m^2=10^{-1}$ (red solid), $m^2=10^{-2}$ (cyan, dashed), and $m^2=10^{-4}$ (orange dotted), compared to the theory curve (blue). Convergence is slow. Statistical errors are within the line thickness.}
\label{f:Kida}
\end{figure}
In Ref.~\cite{Kida1979} Shigeo Kida solved the problem how a random short-ranged correlated   velocity field   decays under action of the Burgers equation. As we discuss in sections \ref{s:Burgers}--\ref{s:Decaying KPZ, and shocks}, this is equivalent to   minimizing the energy 
\be\label{Kida-minimization-problem}
\ca H_w(u) := \frac{m^2}2(u-w)^2 + V(u).
\ee
The random potential $V(u)$ is defined for $u\in \mathbb Z$, and each $V$ is drawn from a probability distribution $P(V)$. 
The effective potential is defined as in \Eq{5.37}.
Kida's solution for the $\hat V(w)$ correlations, rephrased in Refs.~\cite{BouchaudMezard1997,LeDoussal2008} for the minimization problem \eq{Kida-minimization-problem}, is constructed in several steps: 
Define 
\bea
P_<(V):= \int_{-\infty }^V  P(V') \rmd V' \simeq  \rme^{-A(-V)^\gamma} \mbox{ for } V\to -\infty.\nn\\
\eea
The characteristic $u$-scale is 
\be
\rho_m  = \left[ m^2\gamma \ln(m^{-1})^{1-\frac 1 \gamma} A^{ \frac 1 \gamma}\right]^{-\frac12}.
\ee
For a standard Gaussian distribution, $A=1/2$, $\gamma=2$, this can be simplified to
\be
\rho_m^{\rm Gauss} = \frac1 m \left[ \ln(m^{-2})\right]^{-\frac14}, \quad \mbox{i.e.}\quad \zeta_{\rm Kida} = 1^-.
\ee
The $u$-distribution  minimizing the energy \eq{Kida-minimization-problem} is
\be
P(u) \approx \frac1{\rho_m \sqrt{2\pi}}\rme^{-\half (u/\rho_m)^2}.
\ee
In order to proceed, define the auxiliary function 
\be
\Phi(x) := \int_0^\infty \rmd y \,\rme^{- \frac{y^2}2 + x y }=\sqrt{\frac{\pi }{2}}\, \rme^{\frac{x^2}{2}}
   \Big[ \text{erf}\big(\textstyle \frac{x}{\sqrt{2}}
   \big){+}1\Big].
\ee
The effective disorder force-force correlator $\Delta(w)$ and $ R(w)$ are then given by  
\be
\Delta(u) = m^4 \rho_m^2 \tilde \Delta(w/\rho_m), ~~ R(u) = m^4 \rho_m^4 \tilde R(w/\rho_m),
\ee
\be
\tilde \Delta(w) =  \frac{\rmd }{\rmd w}  \int_0^\infty 
\, \frac {2w}{\Phi(\frac w 2+x)+\Phi(\frac w 2-x)}\,\rmd x,
\label{Delta-Kida}
\ee
\be\label{R-Kida}
\tilde R(w) = \frac{\pi^2}6- \int_0^\infty \frac{2 w^2x \,\Phi(\frac w2-x)}{\Phi(\frac w 2+x)+\Phi(\frac w 2-x)}\, \rmd x.
\ee
The solutions $\tilde \Delta(w)$ and  $\tilde R(w)$ are compared in Fig.~\ref{f:Kida} to numerical simulations, and to the fixed point obtained by solving the 1-loop FRG equation 
\eq{RF-FP}, for $\zeta=\zeta_{\rm RB}$, \Eq{zeta-RB}.
For reference we give for the Kida-solution
\bea
\tilde \Delta(0)=1 , \quad 
\tilde \Delta'(0^+) = -\frac{2}{\sqrt{\pi}}, \quad
\tilde \Delta''(0^+)= \frac{3^{\frac32}}{\pi }{-}1,\nn\\
\frac{\Delta(0)\Delta''(0)}{\Delta'(0^+)^2} \approx 0.5136.
\label{Delta-Kida-derivatives}
\eea
Using \Eq{Delta'(0+)}    the universal avalanche scale is
\be
S_m:=\frac{\left< S^2\right> }{2\left< S\right> }=\frac2{\sqrt{\pi}} \rho_m .
\ee

\subsection{Sinai model}
\label{s:Sinai}
\begin{figure}[t]
{\setlength{\unitlength}{1mm}\begin{picture}(83,49.2)
\put(0,0){\Fig{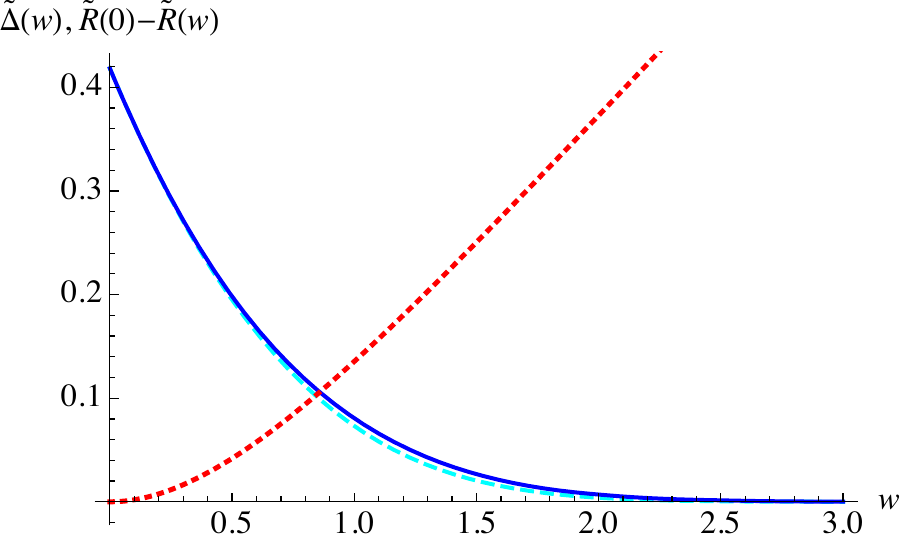}}
\put(40,14){\includegraphics[width=4.3cm]{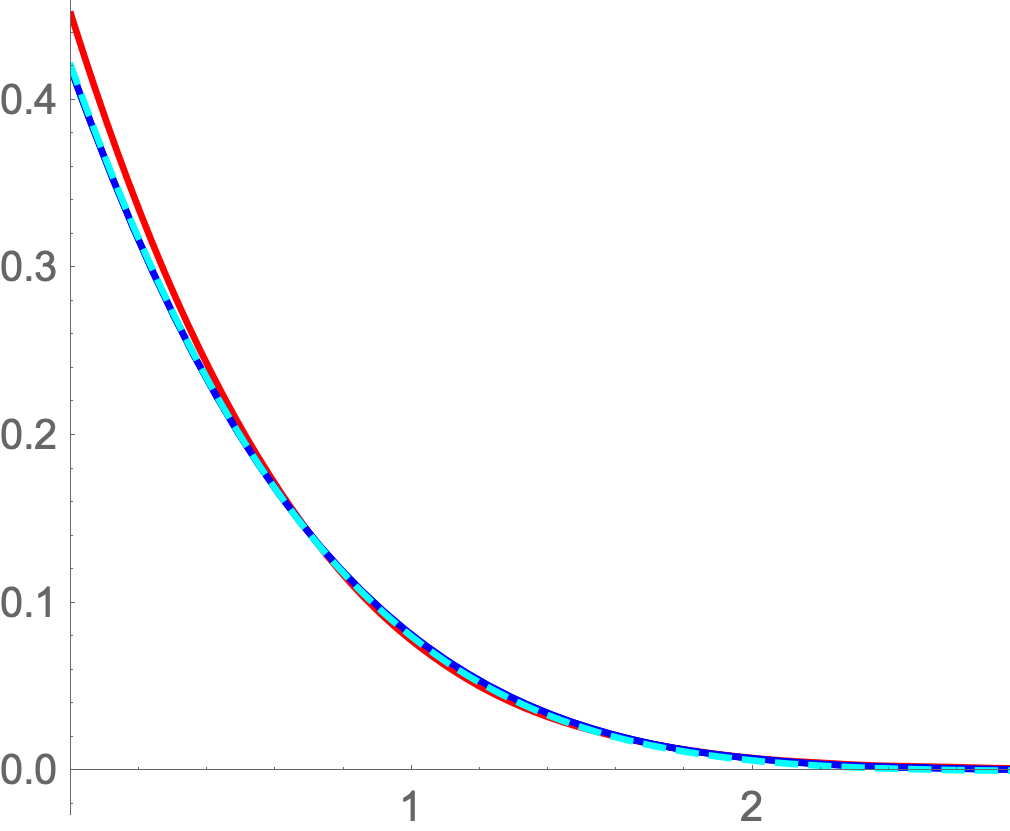}}
\end{picture}}

\caption{$\tilde \Delta(w)$ for the Sinai model (blue), obtained by numerical integration of \Eq{tDelta-Sinai}. In cyan dashed the solution \eq{Delta=y...}-\eq{RF-FP-y} of the 1-loop FRG equation, rescaled to the same value and slope at $w=0$. In red, dotted $\tilde R(0)-\tilde R(w)$. Inset: numerical simulation \cite{terBurgPhD,terBurgWieseUnpublished} for $m^2=10^{-1}$ (red, solid), and $m^2=10^{-2}$ (cyan, dashed),   indistinguishable from the theory   (blue solid).   Statistical errors are within the line thickness.}
\label{f:Sinai}
\end{figure}
In 1983  Y.G.~Sinai asked: Consider a  random walk $X_n$, which with probability $p_n\equiv p_{X_n}$   increases by 1 in step $n$, and with probability $1-p_n$   decreases by 1, assuming  the $p_X\in [0,1]$ themselves to be (quenched)  {\em independent random variables}. 
Sinai showed \cite{Sinai1983} that contrary to a normal random walk, which has variance $n$ after $n$ steps,  the process $X_n$ defined above has a variance which grows as $\ln^2(n)$. Interpreting the $p_n$ as random field disorder, Sinai's theorem shows that the walk is localized even at finite temperature. 

Let us consider again the model defined in \Eq{Kida-minimization-problem}, but with a potential which itself is   a random walk, 
\bea\label{Sinai-minimization-problem}
\ca H_w(u) := \frac{m^2}2(u-w)^2 + V(u),\\
V(u) = - \int_0^ u  F(u')\, \rmd u', \\
\overline{F(u) F (u') } = \sigma \delta(u-u'). 
\eea
Thus 
\be\label{176}
\frac 12 \overline{[V(u)-V(u')]^2} = \sigma |u-u'|.
\ee
Using the methods developed in Ref.~\cite{LeDoussalMonthus2003}, 
the renormalized disorder correlator $\Delta(w)$ for the Sinai model has been obtained in   Ref.~\cite{LeDoussal2008}. Here we give a simplified version\footnote{We    simplify the result of  \cite{LeDoussal2008} such that the only appearance of $a=2^{-1/3}$ or $b= 2^{  2/ 3}$ is in the scale $\rho_m$ of \Eq{rho-m-Sinai}. We further correct several misprints: First, the formulas given for $a$ and $b$ in \cite{LeDoussal2008} can only be used for $m=\sigma=1$, or one would have to rescale the term  $\sim w^3$ in the exponential as well. The factor of $w$ in the innermost integral for $\Delta$ is missing in Eq.~(304) of \cite{LeDoussal2008}, while it is there in Eq.~(293)  for $R$.}:
\bea
\Delta(w) = m^4 \rho_m^2 \tilde \Delta(w/\rho_m )  ,\\
R(w) =  m^4 \rho_m^4 \tilde R(w/\rho_m ), \\
\rho_m =    2^{\frac 2 3}  m^{-\frac 43}\sigma^{\frac 1 3} .
\label{rho-m-Sinai}
\eea
The effective disorder correlator   reads
\bea\label{tDelta-Sinai}
\tilde \Delta(w) &=& -\frac{  \rme^{-\frac {w^3}{12}}}{ 4\pi^{\frac 3 2}\sqrt w}\int\limits _{-\infty}^{\infty} \rmd \lambda_1 \int\limits_{-\infty}^{\infty} \rmd \lambda_2 \,\rme^{-\frac{(\lambda_1{-}\lambda_2)^2}{4 w  }} \nn\\
&& \times \rme^{ i \frac w {2} (\lambda_1+\lambda_2)} \frac{\mbox{Ai}'(i \lambda_1)}{\mbox{Ai}(i \lambda_1)^2}
\frac{\mbox{Ai}'(i \lambda_2)}{\mbox{Ai}(i \lambda_2)^2}\nn\\
&&\times \!\bigg[1{+} 2 w\frac{\int_0^\infty \rmd V \rme^{w V}  \mbox{Ai}(i \lambda_1{+} V) \mbox{Ai}(i \lambda_2{+} V) }{\mbox{Ai}(i \lambda_1) \mbox{Ai}(i \lambda_2)} \bigg]\!.\nn\\
\eea
For reference we give \bea
\tilde \Delta(0)\approx 0.418375 , \quad 
\tilde \Delta'(0^+) \approx -0.566, \nn\\
\tilde \Delta''(0^+)\approx 0.52, \quad \frac{\Delta(0)\Delta''(0)}{\Delta'(0^+)^2} \approx 0.68.
\label{Delta-tilde-Sinai}
\eea
Using \Eq{Delta'(0+)}, this predicts, among others,  the universal avalanche scale, 
\be
S_m:=\frac{\left< S^2\right> }{2\left< S\right> }=0.566  \rho_m .
\ee 
One can also give an explicit formula for the potential-potential correlator $\tilde R(w)$
\bea\label{tildeR-Sinai}
\tilde R(w) &=& - \frac{\sqrt{   w}\, \rme^{-\frac {w^3}{12}}}{16 \pi^{\frac32}} \int\limits _{-\infty}^{\infty} {\rmd \lambda_1} \int\limits_{-\infty}^{\infty} {\rmd \lambda_2} \,\rme^{-\frac{(\lambda_1{-}\lambda_2)^2}{4w}} \nn\\
&& \times \frac{\rme^{ i \frac w 2 (\lambda_1+\lambda_2)}}{\mbox{Ai}(i \lambda_1) \mbox{Ai}(i \lambda_2)}  \bigg[1- \frac{ (\lambda_1{-}\lambda_2)^2}{2w} \bigg] 
 \nn\\
&&\times \bigg[1+2 w\frac{\int_0^\infty \rmd V \rme^{\frac{w}2 V}  \mbox{Ai}(i \lambda_1{+}V) \mbox{Ai}(i \lambda_2{+}V) }{\mbox{Ai}(i \lambda_1) \mbox{Ai}(i \lambda_2)} \bigg]\nn\\
&&+\tilde R(0).
\eea
We checked numerically that $\tilde \Delta(w) = -\tilde R''(w)$.
We find that 
\bea\label{Sinai-200}
\lim_{w\to \infty} \tilde R(0) - \tilde R(w) -\frac w 4  = 0.127689 ,  \\
\lim_{w\to \infty} -R'(w) = \frac{m^4 \rho_m^3}4   = \sigma .
\eea
The latter is a consequence of the FRG equation:   it cannot renormalize the tail of $R(w)$, given for the microscopic disorder in \Eq{176}. 
Finally note that if in the square brackets of the second line of \Eq{tildeR-Sinai} one only retains the ``$1$'', then  the dominant term  $w/4$ of \Eq{Sinai-200} is obtained.
A plot, a comparison to the 1-loop FRG fixed point \eq{RF-FP}, and a numerical verification are presented in Fig.~\ref{f:Sinai}.

\subsection{Random-energy model (REM)}
\label{s:REM}
The random-energy model (REM) was introduced by B.~Derrida in 1980 \cite{Derrida1980,Derrida1981} as an exactly soluble, albeit extreme simplification of a spin glass. It was further studied in \cite{DerridaToulouse1985,Ruelle1987,Mukaida2015}.
Consider an Ising model with $N$ spins. It has   $2^N$ distinct configurations, labeled by $i=1,...,2^N$. 
Suppose that the energy $E_i$  of each state $i$ is taken from a Gaussian distribution  
\be\label{P-REM}
P(E) = \frac1{\sqrt{\pi N}} \rme^{-E^2/N}.
\ee
Thus $\left< E\right> =0$, and $\left< E^2\right> = N/2$.
The factor of $N$ is chosen to obtain a non-trivial limit for $N\to \infty$ below. 

A sample of the REM consists of $2^N$ random energies $E_i$ drawn from \Eq{P-REM}. The partition function at temperature $T=1/\beta$,  and the occupation probabilities are 
\be
 Z_0= \sum_{i=1}^{2^N} \rme^{-\beta E_i}, \quad p_i = \frac{\rme^{-\beta E_i}}{ Z_0}.
\ee
\begin{figure}
\Fig{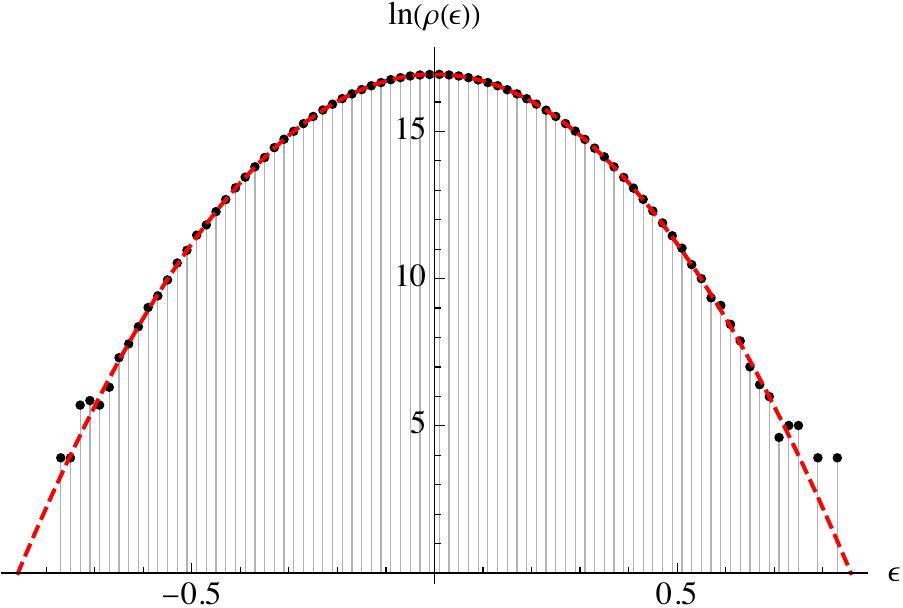}
\caption{The log of the density of states $\ln \rho(\epsilon)$ for the REM, $n=23$. The maximal and minimal energies in this sample are $0.8376$ and $-0.7736$, compared to $\epsilon_*=0.8582$: The histogram vanishes for $|\epsilon|\ge \epsilon_*$. }
\label{f:REM}
\end{figure}
Consider the number $\ca N(\epsilon, \epsilon+\delta)$ of states in the interval $[N\epsilon, N(\epsilon+\delta)]$.
Setting  
\be
\ca I_\epsilon:= \int_{N\epsilon}^{ N(\epsilon+\delta)} P(x) \,\rmd x \equiv \sqrt{ \frac N{\pi}} \int_{\epsilon}^{\epsilon+\delta} \rme^{-y^2}\rmd y, 
\ee
the expectation and variance of $\ca N(\epsilon, \epsilon+\delta)$ are 
\bea\label{REM:197}
\left< \ca N(\epsilon, \epsilon+\delta) \right> = 2^N \ca I_\epsilon, \\
\label{REM:198}
\left< \ca N(\epsilon, \epsilon+\delta)^2 \right>^{\rm c} = 2^N \ca I_\epsilon(1-\ca I_\epsilon).
\eea
This allows us to write the density of states $ \rho(\epsilon) \simeq \frac 1\delta \ca N(\epsilon, \epsilon+\delta)$ as 
\be
\ln \rho(\epsilon) \simeq N  \left[ \ln (2) -  {\epsilon^2} \right] + \frac12 \ln\!\left(\frac N\pi\right).
\ee
Define $\epsilon_*$ s.t.
\be
\ln \rho(\epsilon_*) =0 ~~ \Longleftrightarrow ~~ \epsilon_* 
=\sqrt{\ln 2}+\frac{\ln (N/\pi)}{4 N
   \sqrt{\ln 2}}+ ...
\ee
It is instructive to run a simulation: as can be seen in Fig.~\ref{f:REM}, there is not a single state for $|\epsilon|\ge \epsilon_*$. Second, according to \Eqs{REM:197}-\eq{REM:198}, relative fluctuations are suppressed, 
\be
\frac{\left< \ca N(\epsilon, \epsilon+\delta)^2 \right>^{\rm c}}{\left< \ca N(\epsilon, \epsilon+\delta) \right>^2}= 2^{-N}\frac{1-\ca I_\epsilon}{\ca I_\epsilon}.
\ee
To good precision one can therefore approximate at leading order in $1/N$
\be\label{REM:202}
\ln \rho(\epsilon) = N s(\epsilon), \quad s(\epsilon) = \left\{ 
\begin{array}{ccl}
 \ln(2)- \epsilon^2 &,& |\epsilon|\le \epsilon_* \\
  -\infty  &,& |\epsilon|> \epsilon_* 
\end{array}
\right..
\ee
The quantity $s(\epsilon)$ is interpreted as the entropy of the system; $s(\epsilon)=-\infty$ means that the corresponding density  $\rho(\epsilon)$ vanishes. Note that the  factor of $N$ in \Eq{P-REM} is introduced in order to render thermodynamic quantities as \Eq{REM:202} extensive, i.e. $\sim N$. 

We now proceed to  other thermodynamic properties, notably the free energy
\be\label{REM:s(epsilon)}
\rme^{-\beta N f(\epsilon)} =  Z_0 \simeq \int_{-\epsilon_*}^{\epsilon_*} \rmd \epsilon\, \rme^{-N [ \beta \epsilon - s(\epsilon) ] },
\ee
equivalent to 
\be
f(\epsilon) \simeq \min_{\epsilon\in [-\epsilon_*,\epsilon_*]} \Big[ \epsilon - \frac{ s(\epsilon)}\beta\Big].
\ee
This is a Legendre transform, typical of thermodynamic arguments; the restriction of $\epsilon$ to $ [-\epsilon_*,\epsilon_*]$ is implicit in the definition \eq{REM:s(epsilon)}, and allows us to use the parabolic form valid in that domain. 
An explicit calculation yields
\be\label{REM:213}
f(\epsilon) = \left\{ 
\begin{array}{ccl}
 -\frac\beta4 - \frac {\ln (2) }\beta &,& \beta \le \beta_c   \\
  -\sqrt {\ln (2)}   &,&  \beta > \beta_c
\end{array}
\right., \quad \beta_c = 2 \sqrt {\ln (2)}.
\ee
Next define the {\em participation ratio} inspired by spin glasses \cite{GrossMezard1984,MezardParisiVirasoro1985}
\be
Y\equiv Y(\beta) = \frac{\sum_i \rme^{-2 \beta  E_i}}{ \Big[\sum_i \rme^{-\beta  E_i} \Big]^2}.
\ee
It was shown \cite{DerridaToulouse1985} that all moments can be calculated analytically (see Eqs.~(10)--(11) of \cite{DerridaToulouse1985})
\bea
g(\mu) = \int_0^\infty (1-\rme^{-u -\mu u^2}) u^{-\frac T{T_{\rm c}} -1}\rmd u,\\
\left< Y^n\right> = \frac{(-1)^{n+1}}{\Gamma(2n)}\frac {T_c}T \frac{\rmd^n \ln g(\mu)}{\rmd \mu^n}\Big|_{\mu=0}.
\eea
The first  moments read
\be
\left< Y \right> = 1-\frac T {T_c}, \quad 
\left< Y^2 \right>^{\rm c} = \frac13 \frac T {T_c}\left(1-\frac T {T_c}\right).
\ee
Thus $Y$ is a random variable, with rather large,  non-selfaveraging fluctuations. 

Finally, one can calculate the partition function in presence of a magnetic field \cite{Mukaida2015}, by generalizing its definition to 
\be
 Z(h) = \sum_{i=1}^{2^N} \rme^{-\beta E_i -\beta  M_i h},
\ee 
where $M_i$ is the total magnetization of the sample (the number of up spins minus the number of down spins).
As the energies $E_i$ are independent of the spin configuration $\sigma_i$, and its total magnetization $M_i$, the expectation value of $ Z(h)$  factorizes,
\be
\left<  Z(h)\right>  = \left<  Z _0 \right>\times \left< \rme^{-\beta h M_1} \right>_{\sigma_1} .
\ee
Using this factorization property (which also holds for SK), the partition function for two copies reads
\bea
\frac{\left<  Z(h_1)  Z(h_2) \right>}{ Z_0^2} = \sum_{i=j}\frac 1{ Z_0^2} \left< \rme^{-\beta [ 2 E_i + (h_1+h_2) M_i ]} \right>\nn\\
\hphantom{\frac{\left<  Z(h_1)  Z(h_2) \right>}{ Z_0^2}}+ \frac1{ Z_0^2}\sum_{i\neq j}\left< \rme^{-\beta [   E_i +E_j +  h_1 M_i+ h_2 M_j ]} \right>\nn \\
= \left< Y \right> \left< \rme^{-\beta (h_1+h_2)M_1 } \right>_{\sigma_1} \nn\\
+ \left(1-2^{-N}\right)\left< \rme^{-\beta h_1 M_1 } \right>_{\sigma_1}\left< \rme^{-\beta h_2 M_1 } \right>_{\sigma_1}.
\eea
The average over   spin configurations factorizes\footnote{We refer to \cite{Mukaida2015} 
for  details.}, 
\be
\left< \rme^{-\beta h M_1}\right>_{\sigma_1} =  \cosh(\beta h)^N.
\ee
In the high-temperature phase where $\left< Y\right>$ vanishes the partition function factorizes.
In the low-temperature phase, the first term dominates for $h_1 h_2>0$, whereas the second one does for $h_1 h_2<0$. 
It   leads to a non-analyticity  of the effective action for $T<T_c$. 
This behavior is comparable to that of $\Delta(w)$, and different from that of $R(w)$ (section \ref{s:cusp}). How can we understand this? The reason is that the disorder is {\em very strong}: flipping a single spin changes the energy as much as flipping a finite fraction of the spins. Thus there is no continuity in energy as for random manifolds, and shocks appear   in the energy, rather than in the force. A cusp in the correlations of energy is   
a rather natural consequence.

\subsection{Complex disorder and localization}
\begin{figure}[t]\setlength\unitlength{1mm}
\mbox{\begin{picture}(83,68)
\put(15,5){\includegraphics[width=.8\columnwidth]{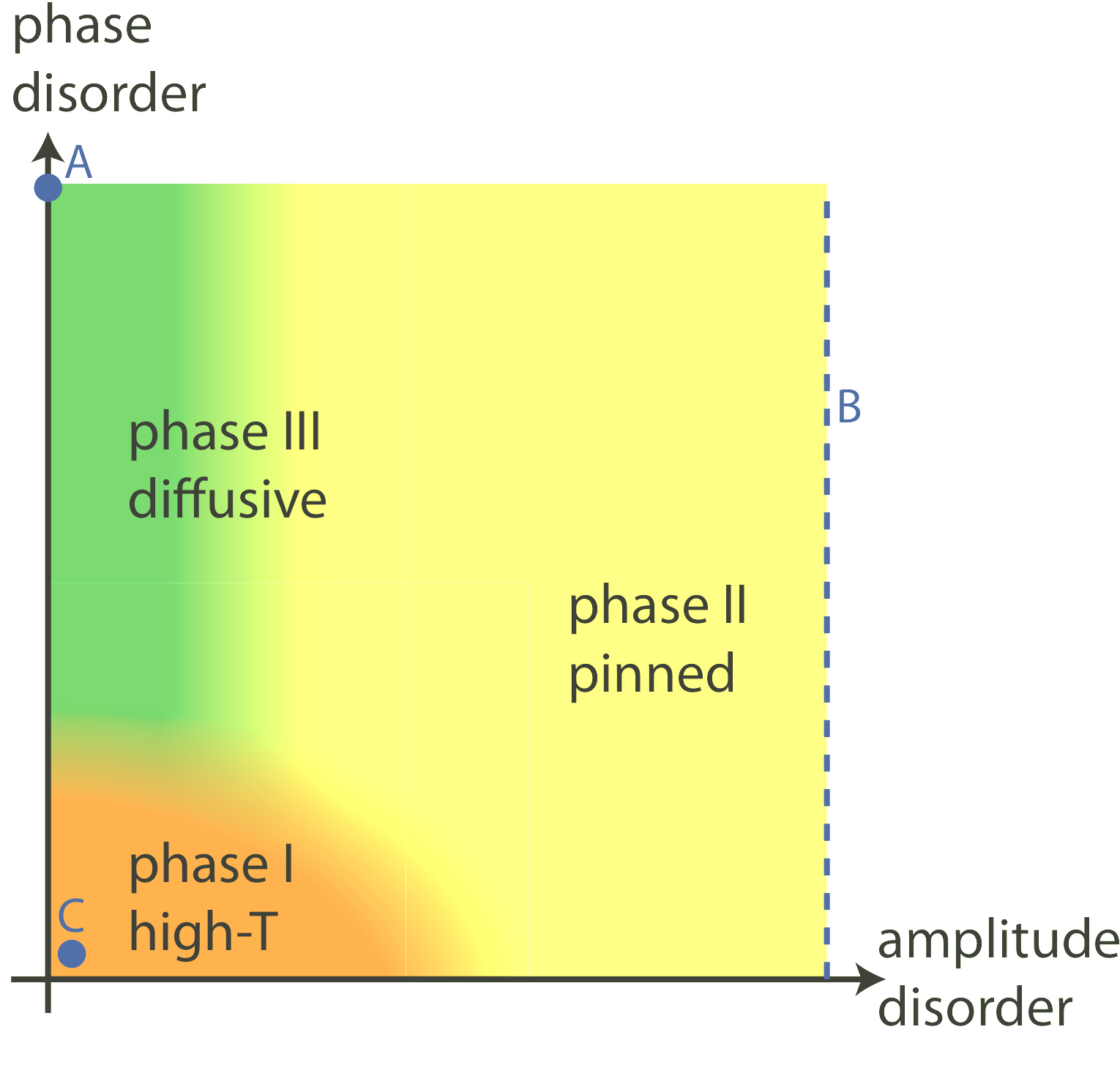}}
\put(0,40){\includegraphics[width=4cm]{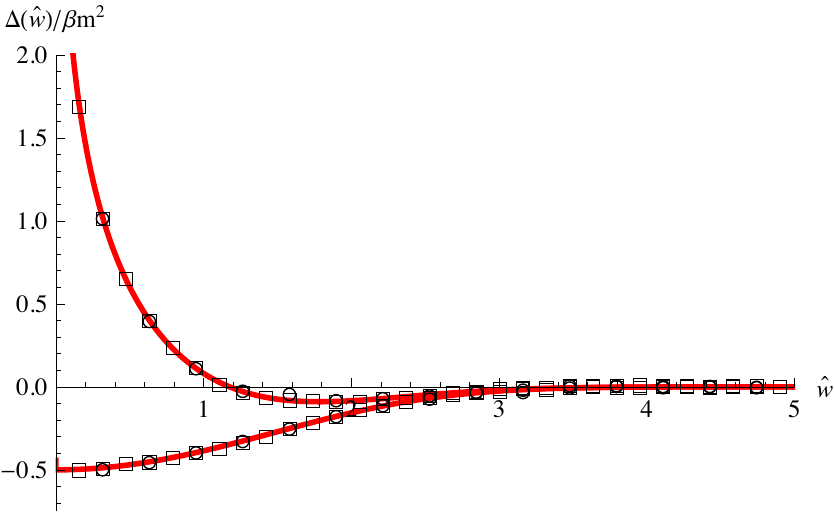}}
\put(5,60){$\scriptstyle\Delta_{ZZ^*}$}
\put(3,39){$\scriptstyle\Delta_{ZZ}$}
\put(38,47){$\scriptstyle w$}
\put(43,38){\includegraphics[width=4cm]{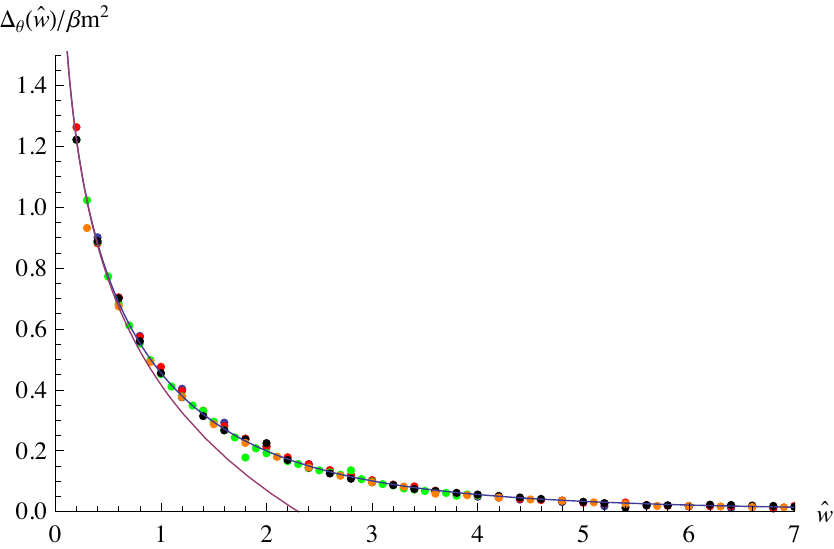}}
\put(44,64){$\scriptstyle\Delta_{\theta}$}
\put(81,41){$\scriptstyle w$}
\put(0,0){\includegraphics[width=4cm]{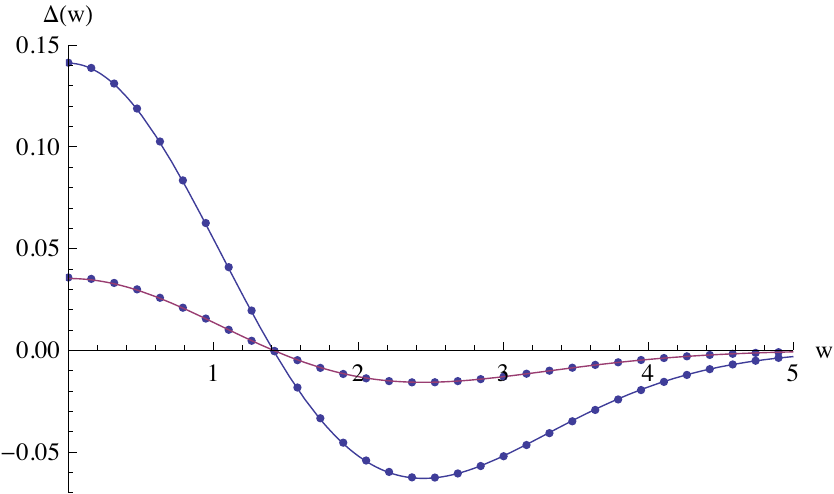}}
\put(2,24){$\scriptstyle\Delta$}
\put(38,8){$\scriptstyle w$}
\end{picture}}
\caption{Phase portrait of the model {\new defined in \Eq{eq:IntroZ}}, following the conventions of Ref.~\cite{Derrida1991}.
 The horizontal
axis is the strength of $V$,  the vertical axis the strength of
$\theta$. The effective disorder correlators for points A (deep in the
diffusive phase), B (deep in the
pinned phase), C
(infinitesimally small disorder) are shown as well. Symbols are from numerical simulations.\figinfo{Figure adopted from \cite{DobrinevskiLeDoussalWiese2011}.}}
\label{fig:CxPhases}
\end{figure}
\subsubsection*{Motivation.} 
FRG is used with success to describe the statistics of elastic objects (this section \ref{s:The field-theoretic treatment}) and depinning (next section \ref{s:dynamics}), subjected to quenched {\em real} disorder. An interesting question is whether it can also be applied to systems with complex disorder {\new (complex random potential or force)}, relevant in quantum mechanics. There is one study for quantum creep \cite{GorokhovFisherBlatter2002}, but what we are   after is a situation where interference becomes important. 

Let us give a specific example. The hopping conductivity of electrons in
disordered insulators in the strongly localized regime is described by the Nguyen-Spivak-Shklovskii
(NSS) model \cite{NguenSpivakShklovskii1985}. The probability amplitude $J(a,b)$ for a transition from $a$ to $b$ 
 is  the sum over interfering directed paths $\Gamma$ from  $a$ to $b$ \cite{MedinaKardar1991,MedinaKardar1992,MedinaKardarShapirWang1989,MedinaKardarShapirWang1990,SomozaOrtunoPrior2007,PriorSomozaOrtuno2009}
\begin{equation}
\label{eq:IntroNSS}
J(a,b) := \sum_{\Gamma} \prod_{j\in \Gamma} \eta_j .
\end{equation}The conductivity between sites $a$ and $b$  is   given by $g(a,b)=|J(a,b)|^2$. Each lattice site $j$ contributes a random sign $\eta_j=\pm 1$ (or, more generally a complex phase $\eta_j=e^{i\theta_j}$). 

Another example is the Chalker-Coddington model \cite{ChalkerCoddington1988} for the quantum Hall
(and spin quantum Hall) effect, where the
transmission matrix $J$ between
two contacts $a$ and $b$ is given by
\cite{Cardy2010,BeamondCardyChalker2002}
\begin{equation} 
\label{eq:IntroCC}
J (a,b) =\sum_{\Gamma} \prod_{(i,j)\in \Gamma} S_{(i,j)}.
\end{equation}
The  random variables $S_{(i,j)}$ on every bond $(i,j)$ are  $U(N)$
matrices,  with
$N=1$ for the charge quantum Hall effect and $N=2$ for the spin
quantum Hall effect.   $\Gamma$ are paths subject to some rules imposed at the vertices. 
 The conductance  is   given by $g (a,b)=
\text{tr}\,\big( J (a,b)^{\dagger}J (a,b) \big)$.

In both models, one would like to understand the dominating
contributions to the sum $Z$ over  paths with random weights  
  $J (a,b)$. In contrast to the thermodynamics of classical models,
where all contributions are positive, contributions between paths with
different phases can   cancel. One is   interested in the
expected phase transitions.

\subsubsection*{Definitions.}
This is a  complicated problem. In Refs.~ \cite{CookDerrida1990,Derrida1991b,DerridaEvansSpeer1993,DobrinevskiLeDoussalWiese2011}  simplified models  were considered. Here we consider the toy model of Ref.~\cite{DobrinevskiLeDoussalWiese2011}, which allows one to define the central objects of the FRG. The partition function at finite $T=\beta^{-1}$ reads in generalization of \Eq{5.37} 
\bea
\label{eq:IntroZ}
 Z (w) &=&  \sqrt{\frac{\beta m^2}{2\pi}} \int\limits_{-\infty}^{\infty}\rmd x\, e^{-\beta \big[V(x)+\frac{m^2}{2}(x-w)^{2}\big] -i \theta(x)}\nn\\
&=:& \label{eq:DefEffPot}
\rme^{- \beta \hat{V}(w) - i\hat{\theta}(w) }.
\eea
One wishes to  study the correlations
\begin{eqnarray}\label{eq:DefCorrVV}
\Delta_{V}(w_1-w_2)&:=& \overline{\hat{V}'(w_1)\hat{V}'(w_2)}, \\
\label{eq:DefDelta}
\Delta_{\theta}(w_1-w_2)&:=&\overline{\hat{\theta}'(w_1)\hat{\theta}'(w_2)}.
\end{eqnarray}
They are related to the correlations of $\partial_w  Z(w)$ and $\partial_w { Z^*}(w)$ by
\begin{eqnarray}
\label{eq:DefCorrZZs}
\Delta_{ Z  Z^*}(w) =\Delta_{V}(w) + \beta^{-2} \Delta_{\theta}(w),\\
\label{eq:DefCorrZZ}
\Delta_{ Z  Z}(w) = \Delta_{V}(w) - \beta^{-2} \Delta_{\theta}(w) .
\end{eqnarray}

\subsubsection*{Results.}

As established by Derrida   \cite{Derrida1991}, 
there are  three phases:  high-temperature phase I, frozen phase II, and strong-interference phase III,   depicted in the center of  Fig.~\ref{fig:CxPhases}, accompanied by their   correlation functions  \cite{DobrinevskiLeDoussalWiese2011}.

\paragraph{Phase I:} For weak disorder (perturbative regime) one can evaluate the integral \eq{eq:IntroZ} by Taylor expanding to leading order in both $V$ and $\theta$, to obtain 
\be
\Delta_V(w) \sim \Delta_\theta(w) \sim - \partial_w^2 \rme^{-\frac{m^2}4 w^2}.
\ee

\paragraph{Phase II:}
This phase may be seen as a deformation of the localized phase,  encountered for  short-ranged disorder in the Kida model (section \ref{s:Kida}), or for long-ranged disorder in the Sinai model (section \ref{s:Sinai}). The key change is a deformation of the shocks, which shows up in an additional logarithmic deformation of the force-force correlator inside a boundary lawyer of size $w\sim T$, see Fig.~\ref{fig:CxPhases}.

\paragraph{Phase III:}
Here $V(u)=0$.
This phase is the one most closely     related to the NSS or  Chalker-Coddington models. Contrary to the Kida or Sinai models which lead to a localization of the path (the partition function is dominated by a minimizing path) fluctuations of $Z$ are important, and zero-crossings are observed. They seem to be rather independent of the nature of the $\theta$-disorder, which we attribute to   $\theta$ being a compact variable.
The zero crossings    lead to  a logarithmic divergence of the correlation functions
\eq{eq:DefCorrVV} to \eq{eq:DefCorrZZ}, see Fig.~\ref{fig:CxPhases}.

\subsection{Bragg glass and vortex glass}
\label{s:Bragg glass and vortex glass}
\begin{figure}
\Fig{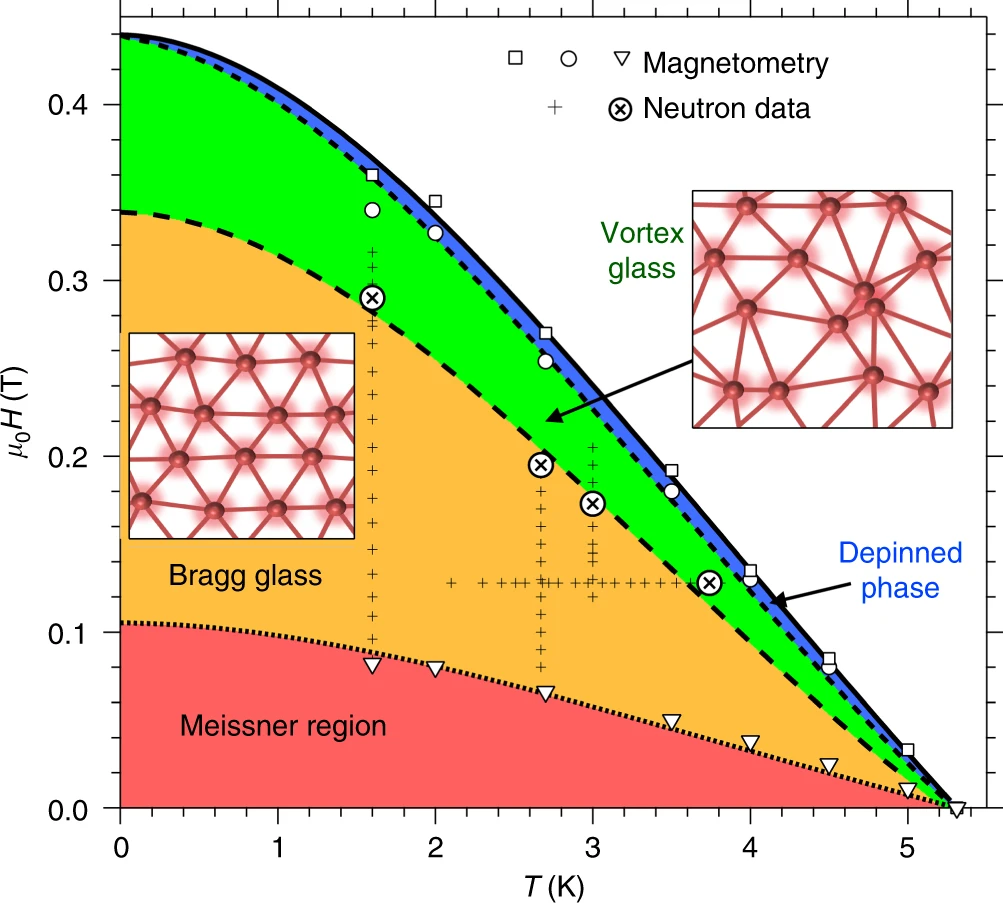}
\caption{Phase diagram for vortices in a type-II superconductor \cite{Toft-PetersenAbrahamsenBalogPorcarLaver2018}. (Figure reproduced   with kind permission from the authors). From bottom to top these are the Meissner, vortex-free region (red), followed by the Bragg glass (orange) and a vortex glass (green and blue), above which super-conductivity vanishes (white).}
\label{f:Toft-PetersenAbrahamsenBalogPorcarLaver2018-fig1}
\end{figure}
When the magnetic field $H$ is increased in a pure {\new type-II} superconductor, there are two phase transitions: For low magnetic fields,  $H<H^{\rm c}_{1}(T)$,  
vortices are  expelled   due to the  Meissner effect \cite{MeissnerOchsenfeld1933} (red region in Fig.~\ref{f:Toft-PetersenAbrahamsenBalogPorcarLaver2018-fig1}).
For $H>H^{\rm c}_{1}(T)$ flux-vortices appear. Increasing the magnetic field further, superconductivity breaks down for  $H> H^{\rm c}_{2}(T)$ (white region in Fig.~\ref{f:Toft-PetersenAbrahamsenBalogPorcarLaver2018-fig1}). 

Let us now consider a dirty magnet. There has been a long debate whether the vortex lattice present for $H^{\rm c}_{1}(T)<H <H^{\rm c}_{2}(T)$ can be described by a Bragg glass, or a vortex glass. In a Bragg glass favored by \cite{GiamarchiLeDoussal1994,GiamarchiLeDoussal1996b,CarpentierLedoussalGiamarchi1996,KleinJoumardBlanchardMarcusCubittGiamarchiLeDoussal2001}, the Abrikosov lattice of vortices (see sketch in Fig.~\ref{sketch-Bragg-glass}) is elastically deformed, but there are no topological defects and each vortex line retains six nearest neighbors. According to the theory of disordered elastic manifolds, the correlation function given in \Eq{universal-amplitude-2} grows logarithmically with distance, preserving enough coherence to show up in a Bragg peak in neutron scattering experiments (hence the name). The alternative theory, termed vortex glass and favored (for sometimes quite different reasons) in \cite{Fisher1989,KochFogliettiGallagherKorenGuptaFisher1989,RegerTokuyasuYoungFisher1991,MoserHugParashikovStiefelFritzThomasBaratoffGuntherodtChaudhari1995,BalentsMarchettiRadzihovsky1998,AransonScheidlVinokur1998,ScheidlVinokur1998,PfeifferReiger1999}, assumes that topological defects destroy the order, and as a consequence  the Bragg peak in neutron scattering experiments.

The current experimental situation \cite{Toft-PetersenAbrahamsenBalogPorcarLaver2018} shown in Fig.~\ref{f:Toft-PetersenAbrahamsenBalogPorcarLaver2018-fig1}   favors, in agreement with intuition, a Bragg glass for smaller fields, and a vortex glass for larger ones, with a transition at $H^{\rm c}_{\rm B/V}(T)$, with $H^{\rm c}_{1}(T)<H^{\rm c}_{\rm B/V}(T)<H^{\rm c}_{2}(T)$.

If the disorder contains  columnar defects, it may  exhibit \cite{Balents1993,Fedorenko2008}  a transverse Meissner effect, 
for which the tilt $\vartheta$ to an applied field $H$ vanishes up to $H^{\rm c}_{\rm tilt}$, after which it grows as $\vartheta \sim |H-H_{\rm tilt}^{\rm c}|^{\phi}$, with $\phi<1$. 
One-dimensional quantum systems (Luttinger liquid) map to two-dimensional classical systems with columnar disorder, thus can be understood in the same framework  \cite{Fedorenko2008}, or alternatively \cite{Dupuis2019,DupuisDaviet2020,Dupuis2020,DavietDupuis2020,DavietDupuis2021} within the NPFRG approach  (section \ref{s:NPRG}). One further encounters a rounding of the cusp through quantum fluctuations, similar to the quantum creep regime of Ref.~\cite{GorokhovFisherBlatter2002}.

There are also situations where one of the two phases is absent, as  defects can destabilize the Bragg-glass phase 
\cite{EmigNattermann2006}.
Similar physics may be observed in charge-density waves \cite{EmigNattermann1997}. 
For   details we refer the reader to \cite{NattermannScheidl2000,LeDoussal2010Book}. 
Discussion of a moving vortex lattice is referred to section \ref{s:Depinning of vortex lines or charge-density waves}.

\subsection{Bosons and fermions in $d=2$, bosonization}

To proceed, we need to establish connections between theories in  two dimensions, including relations between fermions and bosons, known as bosonization.   
Dimension $d=2$ is special as the Gaussian free field is dimensionless, allowing for a number of constructions of which we show some below.
Our account is very condensed, and we refer   to  \cite{CardyBook,DiFrancescoMathieuSenechal,vonDelftSchoeller1998,GiamarchiBook} for background reading. 

\subsubsection*{Bosons.}
\begin{equation} \label{bosonen}
{\cal H}_{\rm boson}=\frac1{8\pi} \int\rmd^2 \vec z\, \left[\nabla \Phi(\vec z)\right]^2 
\ .
\end{equation}
Appendix \ref{inv lap} implies
\begin{equation}
\left< \Phi(\vec z) \Phi(0) \right> = - \ln |\vec z|^{2} =- \ln z - \ln \bar z
\ .
\end{equation}
This suggests that one can decompose $\Phi(\vec z)$ into a holomorphic and antiholomorphic part, $\Phi(\vec z)\equiv \Phi(z,\bar z)= {\phi}(z) + \bar {\phi}(\bar z)$, with 
\begin{eqnarray}
\left< {\phi}(z) {\phi}(w) \right> &=& -\ln (z-w) , \nn \\
\left< \bar{\phi}(\bar z) \bar{\phi}(\bar w) \right> &=& -\ln (\bar z-\bar w),\\
\left< {\phi}(z) {\bar \phi}(\bar w) \right> &=& 0.\nn
\end{eqnarray}
Since $\phi(z)$ has logarithmic correlations, an infinity of power-law correlated {\em vertex operators} can be constructed (the dots indicate normal ordering, i.e.\ exclusion of self-contractions at the vertex), 
\bea
V_\alpha(z)= \; \bn\rme^{\alpha \phi(z)}\en\\
\left<V_\alpha(z) V_\beta(w)\right> = \rme^{-\alpha \beta \ln(z-w)} = (z-w)^{-\alpha \beta}.
\eea

\subsubsection*{Majorana fermion.}\label{majorana}
Consider a Majorana (real) fermion, constructed from anti-commuting Grassman fields (section \ref{s:Grassmann})
\begin{equation}\label{fermion-action2}
{\cal H}_{\rm Majorana} = \frac{1}{2\pi }\int \rmd ^{2}\vec z \, \big[\bar \psi(\bar z) \partial
\bar \psi(\bar z) +\psi (z) \bar \partial \psi (z) \big].
\end{equation}
Its correlation-functions are obtained from \Eq{partial inv}
as 
\begin{eqnarray}\label{psipsi}
\left< \psi (z)\psi (w) \right> &=& \frac{1}{z-w},\\
\left< \bar \psi (\bar z) \bar \psi (\bar w) \right> &=& \frac{1}{\bar z-\bar w}.
\label{psipsibar}
\end{eqnarray}
As for the bosons above, the theory can be split into a holomorphic and an antiholomorphic part. 

\subsubsection*{Dirac fermion.}\label{dirac}
A  Dirac (complex) fermion is made out of two  Majorana-fermions
$\psi _{1}=\psi _{1}^{*}$ and $\psi _{2}=\psi _{2}^{*}$, 
\begin{equation}\label{psi}
\psi =  \psi _{1}+ i \psi _{2}, \quad \psi^* =  \psi _{1}- i \psi _{2}.
\end{equation}
Corresponding rules hold for the antiholomorphic fields. 
Choosing \begin{equation}\label{H-Dirac}
{\cal H}_{\rm Dirac} = \frac{1}{\pi } \int_{z} \bar \psi ^{*}(\bar z)  \partial \bar \psi(\bar z)  
+\psi ^{*} (z)  \bar \partial \psi(z)   , 
\end{equation}
  the correlation functions  for the components $\psi_1$ and $\psi_2$ have an additional factor of $1/2$ as compared to \Eqs{fermion-action2}-\eq{psipsi}, resulting in 
\begin{eqnarray}\label{fermion-correlators}
\left< \psi^{*} (z)\psi (w) \right> =\left< \psi (z)\psi^{*}  (w)
\right> = \frac{1}{z-w}, \\
\left< \psi (z)\psi (w) \right> =\left< \psi^{*} (z)\psi^{*} (w) \right> = 0  .
\end{eqnarray}
Similar relations hold for the anti-holomorphic parts $\bar \psi $, 
and  correlations vanish between $\psi $ and $\bar \psi$.

\subsubsection*{Bosonization.}\label{bosonize}
Since a Dirac-fermion and a free boson have both central charge\footnote{\new The conformal charge $c$ is the amplitude of the leading term in the OPE of the stress-energy tensor with itself. It can be measured from the finite-size corrections of the free energy of a system \cite{DotsenkoCFT,DiFrancescoMathieuSenechal,HenkelCFT,CardyBook}. The central charge is often used to identify or distinguish systems. This has to be taken with some precaution, as the total central charge of non-interacting systems is   the sum of the central charges of its components.}  $c=1$, we may expect
a closer relation between objects in these theories. 
Indeed setting 
\begin{eqnarray}\label{bosrule1}  
\psi (z) &\hat =& \bn \rme^{i{\phi} (z)}\en \ \qquad  \psi^{*} (z)\hat =  \bn
\rme^{-i{\phi}  (z)}\en\\
\label{bosrule2} 
\bar \psi (\bar z) &\hat =& \bn \rme^{i\bar {\phi} (\bar z)}\en \ \qquad  \bar 
\psi^{*} (\bar z) \hat =   \bn \rme^{-i\bar {\phi} (\bar z)}\en 
\end{eqnarray}
the (diverging part of the) fermion correlation functions  are
reproduced within the bosonic theory. 
Products of fermion operators are obtained from the point-splitted product
\bea\label{bosruleinv}
\left[\psi ^{*}\psi \right]  (z)&:=&\lim_{z\to w} \psi ^{*} (w) \psi (z)  =\ \bn
\rme^{-i{\phi}(w) }  \rme^{i{\phi}(z) }\en
\frac{1}{w-z} \nn\\ 
&=&\frac{1}{i}\partial {\phi} (z).
\eea 
This rule allows one to decouple the 4-fermion interaction as 
\be\label{4-psi-bos}
\bar \psi^*(\bar z) \bar \psi(\bar z)   \psi^*(  z)   \psi (z)  \quad  \hat= \quad \bar \partial   \phi(\bar z)   \partial   \phi(  z) .
\ee
Note that since there are two complex fermions, the only non-vanishing combination one can write down is the one  given in \Eq{4-psi-bos}.
Introductory texts on bosonization techniques can be found in \cite{vonDelftSchoeller1998,GiamarchiBook}.

\subsubsection*{Thirring and sine-Gordon model.}
The {\em Thirring model}   introduced in Ref.~\cite{Thirring1958} is the most general 2-dimensional model with two independent families of fermions, and a kinetic term with a single derivative. 
\bea\label{Thirring}
{\cal S}_{\rm Thirring} = \frac{1}{\pi } \int_{z} \Big\{ \bar \psi ^{*} (\bar z)\partial \bar \psi(\bar z)
+\psi ^{*} (z)   \bar \partial \psi(z)  \nn\\
\hp{{\cal H}_{\rm Thirring} = \frac{1}{\pi } \int_{z}}
+ \frac\lambda2 \Big[ \bar \psi (\bar z)   \psi (  z) + \bar \psi^* (\bar z)   \psi^* (  z)\Big]  \nn\\
\hp{{\cal H}_{\rm Thirring} = \frac{1}{\pi } \int_{z}}
+ \frac g2 \bar \psi^*(\bar z) \bar \psi(\bar z)   \psi^*(  z)   \psi (z) \Big\}.
\eea
Using the dictionary provided by \Eqs{bosrule1}, \eq{bosrule2} and \eq{4-psi-bos} shows equivalence to the {\em sine-Gordon model} \cite{Coleman1975} 
\bea\label{H-sine-Gordon}
{\cal H}_{\rm SG}=\int\rmd^2 \vec z\; \frac{1+g}{8\pi} \big[\nabla \Phi(\vec z)\big]^2 + \frac\lambda\pi  \cos \!\big( \Phi(\vec z)\big) . 
\eea
\subsection{Sine-Gordon model, Kosterlitz-Thouless transition}
\label{s:Sine-Gordon model, Kosterlitz-Thouless transition}
The sine-Gordon model can be treated in a perturbative expansion in $\lambda$. The leading-order correction comes at second order, and corrects $g$. Noting 
\be
T:= \frac1{1+g}, 
\ee
it can be written as\footnote{Combinatorial factors are obtained from $\cos( \Phi ) = \half (\rme^{i\Phi} +\rme^{-i\Phi}) $, leaving two combinations with overall charge neutrality, which cancel against the $1/2!$ from the expansion of $\rme^{-\ca H_{\rm SG} }$. The dots denote normal-ordering, the change in $\lambda$ being absorbed therein. Vector notation is suppressed for simplicity. For an introduction into the technique see \cite{WieseHabil}.}
\bea
\left(\frac{\lambda}{2\pi}\right)^{\!2} \int_{x,y} \bn \rme^{i\Phi(x)}\en\; \bn\rme^{-i\Phi(y)}\en \nn\\
= \left(\frac{\lambda}{2\pi}\right)^{\!2} \int_{x,y}  \bn \rme^{i[\Phi(x)-\Phi(y)]}\en |x-y|^{-2T} \nn\\
\simeq \left(\frac{\lambda}{2\pi}\right)^{\!2} \int_{x,y} \Big\{ 1-  \half \bn [\Phi(x)-\Phi(y)]^2\en + ...\Big\} |x-y|^{-2T} \nn\\
\simeq \left(\frac{\lambda}{2\pi}\right)^{\!2} \int_{x,y} \Big\{ 1-  \frac1{2} \bn  \Big[(x{-}y) \nabla\Phi(\textstyle{\frac{x+y}2})\Big]^2\en\,  + ...\Big\} \nn\\
\hp{\left(\frac{\lambda}{2\pi}\right)^{\!2} \int_{x,y} \Big\{1} \times |x-y|^{-2T} \nn\\
\simeq \left(\frac{\lambda}{2\pi}\right)^{\!2} \int_{x,y} \Big\{ 1-  \frac{(x{-}y)^2}{4} \bn \Big[ \nabla\Phi(\textstyle{\frac{x+y}2})\Big]^2\en  + ...\Big\} \nn\\
\hp{\left(\frac{\lambda}{2\pi}\right)^{\!2} \int_{x,y} \Big\{1} \times |x-y|^{-2T} .
\eea\begin{figure}[tb]
\Fig{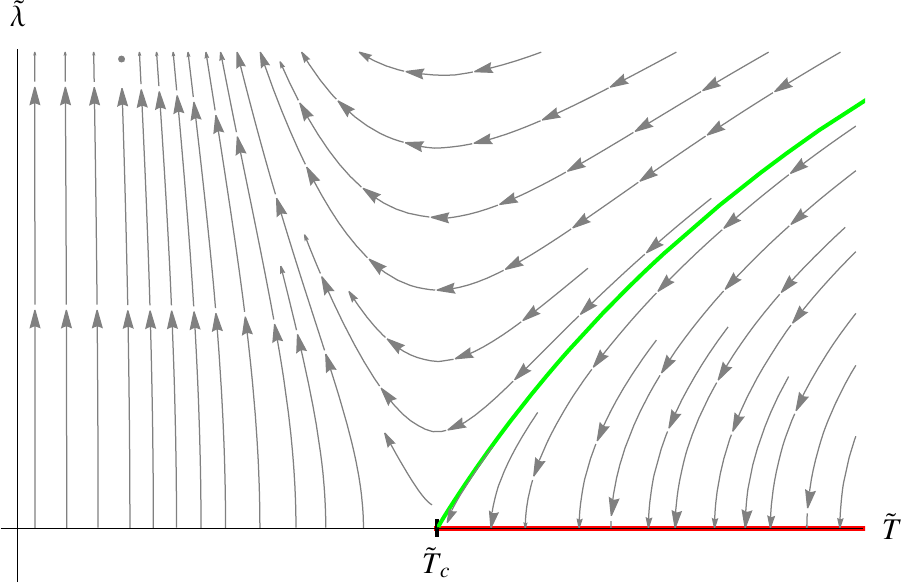}
\caption{Flow diagram of the Kosterlitz-Thouless transition: All trajectories starting from the green line are attracted to $\tilde \lambda = 0$, and $\tilde T>\tilde T_{\rm c}$ (line of fixed points),  the remaining ones to the strong-coupling regime with $\tilde \lambda \gg 1$.}
\label{f:KT-flow}
\end{figure}The first term yields a correction to the free energy; it necessitates a relevant counter term (UV divergent, IR finite), but does not enter into IR properties of the theory.
The second term is a correction to $g$, 
\be
\delta g = \lambda^2 \int_0^L \frac {\rmd z}z {z}^{4-2T} = \lambda^2 \frac {L^{4-2 T}}{4-2T}.
\ee
Defining the dimensionless couplings as $\tilde g \equiv    g_{\rm eff} = g+ \delta g$,  $\tilde \lambda : = \lambda_{\rm eff} L^{2-T}$, with $\lambda_{\rm eff} = \lambda + \ca O(\lambda^3)$, one obtains the $\beta$ functions\footnote{Definition of the $\beta$ functions are as in section \ref{s:deriveRG}. \Eqs{Thirring} and \eq{H-sine-Gordon} were tuned  to have the   simplest coefficients later.} \numparts
\bea
\beta_{\tilde g}(\tilde \lambda, \tilde g) = \tilde \lambda^2 + ... \\
\beta_{\tilde \lambda}(\tilde \lambda, \tilde g) = (2-T) \tilde \lambda +  ...
=  (2-\textstyle\frac{1}{1{+}\tilde g} ) \tilde \lambda +  ...
\eea\endnumparts
Subdominant corrections are down by a factor of $\tilde \lambda^2$. Rewritten in terms of $\tilde \lambda$ and $\tilde T= 1/(1+\tilde g)$, this yields  \numparts
\bea
\beta_{\tilde T}(\tilde T, \tilde \lambda) = - \tilde T ^2 \tilde \lambda ^2  + ... = -   \tilde T_{\rm c} ^2 \tilde\lambda ^2 + ... \\
\beta_{\tilde  \lambda} (\tilde T, \tilde \lambda) = (\tilde T_{\rm c} - \tilde T) \tilde \lambda+ ...
\eea\endnumparts
The reader will mostly see these equations  expanded around $\tilde T_{\rm c}= 2$, as done above. The schematic flow chart is shown in Fig.~\ref{f:KT-flow}. 

There is a line of fixed points for $\tilde T>\tilde T_{\rm c}$ (red in Fig.~\ref{f:KT-flow}). All these fixed points have $\tilde \lambda=0$, thus are Gaussian theories. Below $\tilde T_{\rm c}$, the flow is to strong coupling. Physically, $\rme^{i \Phi}$ and  $\rme^{-i \Phi}$ are interpreted as vortices and anti-vortices,  topological defects with charge $\pm 1$, chemical potential $\lambda$,  interacting via Coulomb interactions. For $\tilde T>\tilde T_{\rm c}$ they are bound, and only a finite number is present. For $\tilde T<\tilde T_{\rm c}$ they are unbound, gaining enough entropy to overcome the energetic costs for  their core. 
The transition at $\tilde T= \tilde T_{\rm c}$ is known as the Kosterlitz-Thouless transition \cite{KosterlitzThouless1973}. 

Higher-loop calculations can be performed, both in the Thirring model \eq{fermion-action2}, as in the sine-Gordon model \eq{H-sine-Gordon}.
As our   treatment shows, they are rather straight-forward, and one should be able to go at least to 3-loop order in sine-Gordon, and to even higher order in the fermionic model. It is therefore surprising to read about massive contradictions in the literature   at 2-loop order \cite{BalogHegedus2000}, confirming the original 1980 result of \cite{AmitGoldschmidtGrinstein1980}, but declaring later calculations in \cite{Lovelace1986,Boyanovsky1989,Naik1993} 
as well as \cite{KonikLeClair1996} to be incorrect.

What the RG approach cannot reach is the strong-disorder fixed point $\tilde \lambda \gg 1$. The latter has been studied in the 
Wegner flow-equation approach 
\cite{Kehrein2001}, and via NPFRG \cite{DavietDupuis2019}.

\subsection{Random-phase sine-Gordon model}
\label{s:The random-phase sine-Gordon model}
The sine-Gordon model \eq{H-sine-Gordon} with quenched disorder coupling to $\rme^{i \Phi}$ reads (writing $\vec z\to z$)
\bea\label{H-rp-sine-Gordon}
{\cal H}_{\rm rpSG} =\int_{ z} \frac{\big[\nabla \Phi( z)\big]^2}{8\pi T}  +   \xi(z)\bn\rme^{ i\Phi( z)}\en +  \xi^*(z)\bn\rme^{- i\Phi( z)}\en   \nn\\ 
\overline{\xi(z)\xi^*(z') }= \frac \lambda{2\pi} \delta^2(z-z') .
\eea
After replication (see section \ref{s:replicas}), the effective action reads
\bea\label{rpSG}
{\cal S}_{\rm rpSG}= \!\int_z \sum_\alpha  \frac{[\nabla \Phi_\alpha(  z)]^2}{8\pi T}  - \frac{\lambda}{2\pi}\int_z {\sum_{\alpha \neq \beta}} \bn\rme^{i[ \Phi_\alpha(z){-}\Phi_\beta(z)]}\!\en  \nn\\
\hp{{\cal S}_{\rm rpSG}=}- \frac{\sigma}{4\pi} \int_z{\sum_{\alpha \neq \beta}}\nabla \Phi_\alpha(  z) \nabla \Phi_\beta(  z).
\eea
We   added an additional off-diagonal term in the last line since it is generated under RG; we will see this shortly.

\begin{figure}[t]
\Fig{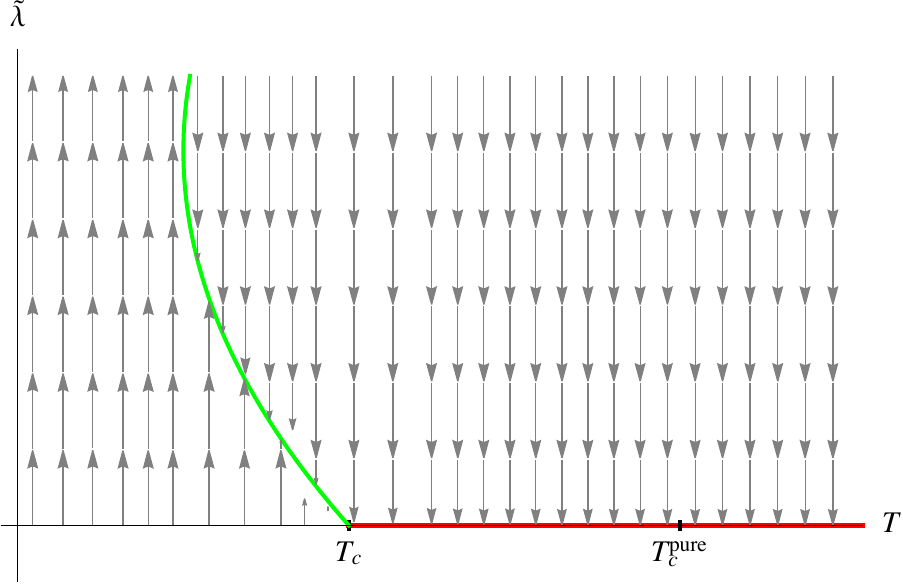}
\caption{The RG flow for $\tilde \lambda$ as a function of $T$.}
\label{figdiagram}
\end{figure}

Perturbation theory is performed with 
\begin{equation}\label{K:a2}
\left< \Phi_{\alpha} (x)\Phi_{\beta} (y) \right>_0 = -  T\delta_{\alpha
\beta}   \ln\big(|x-y|^2\big)\ .
\end{equation}

\subsubsection*{First diagram (one loop).}
We use the graphical notation
\begin{equation}\label{a6}
{_{\alpha }}\!\!\diagram{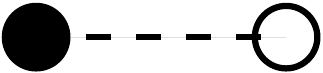}\!\!{_{\beta}} =\; \bn\rme^{i[ \Phi_{\alpha} (x) -\Phi_{\beta} (x)]}\en 
\end{equation}
 The contributions to the effective action  are 
 $\delta {S}_{i} \equiv \int_x \delta { s}_{i}$. 
The first one is (an  ellipse encloses the same replica)
\begin{eqnarray}\label{a7}
-\delta { s}_{1} = \diagram{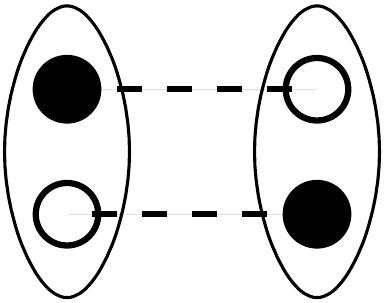}^{x}_{y} \nn\\
= \frac{1}{2!} \left(
\frac{ \lambda }{2\pi} \right)^{2}{\sum_{\alpha\neq\beta}} \int_{y}
:\!  \rme^{i [\Phi_{\alpha} (x) -\Phi_{\beta} (x) - \Phi_{\alpha} (y)
+\Phi_{\beta} (y)] }\!:\nn\\
 \qquad \qquad  \qquad \quad\times  \rme^{-4 T\ln |x-y|}.
\end{eqnarray}
This term contains a strongly UV-divergent contribution to the free
energy (which we do not need) and the   sub-dominant term
\begin{eqnarray}\label{rpSG-a8}
- \delta { s}_{1}& \approx&  -\frac{1}{4} \left(
\frac{ \lambda}{2\pi} \right)^{2}{\sum_{\alpha \neq \beta}}\int_{y} |x-y|^{-4 T  } \times \nn\\
&&
\times :\! \left[(x-y)\cdot \nabla \Phi_{\alpha} (x) - (x-y)\cdot \nabla \Phi_{\beta} (x) \right]^{2} \!:
  \nn\\
&=&  -\frac{1}{4} \left(
\frac{ \lambda}{2\pi} \right)^{2}{\sum_{\alpha \neq \beta}}\int_{y} |x-y|^{2-4 T  } \times \nn\\
&&\times:\! \left[ \nabla \Phi_{\alpha} (x)- \nabla \Phi_{\beta} (x)  \right]^{2}\!:
\end{eqnarray}
It corrects $\sigma$, \numparts
\bea
\delta \sigma=\frac{1}{2}\lambda^{2}   \times {I}_1,\\
{I}_1 = \frac{1}{2\pi} \int \rmd y^{2}
|y|^{2-2T}\Theta(|y|<L) = \frac{L^{4\tau}}{4\tau},
\qquad\\
\tau:= 1-T.
\eea
\endnumparts

\subsubsection*{Second diagram (one loop).}
\begin{eqnarray}\label{a10}
-\delta {s}_{2}=\diagram{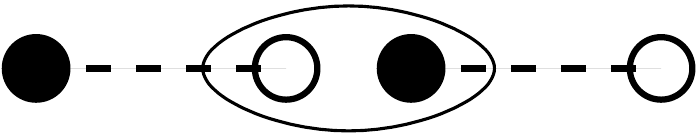} \nn\\
= \frac{2}{2!} \Big(
\frac{ \lambda}{2\pi} \Big)^{\!2}\sum_{\alpha \neq \beta \neq \gamma} \int_{y}  \bn\rme^{i [\Phi_{\alpha} (x)- \Phi_{\gamma} (y)]}\en
:\!  \rme^{ - i [\Phi_{\beta} (x)
-\theta_{\beta} (y)] }\!: \nn\\
 \qquad \qquad \qquad\qquad  \times\rme^{- 2T\ln |x-y|}
 .
\eea 
Projecting onto the interaction yields
\bea\label{rpSG-a11}
-\delta { s}_{2}\approx  \Big(
\frac{ \lambda}{2\pi} \Big)^{\!2}\times (n{-}2) \sum_{\alpha\neq  \gamma} \bn \rme^{i [\Phi_{\alpha} (x)- \Phi_{\gamma} (x)]}\en \,   {I}_2\ , \\
{I}_2 = \frac{1}{2\pi} \int \rmd^2 y\, |y|^{-T} \Theta(|y|<L)=\frac{L^{2\tau}}{2\tau}\ .
\eea
Setting the number of  replicas $n \to 0$, we get    
\begin{eqnarray}\label{a12}
 \delta  \lambda&=& -2 \lambda^{2}  \times {I}_2 \ .
\end{eqnarray}
Defining the $\beta$-functions  as the variation with respect to the large-scale cutoff $L$, keeping
  the bare coupling $\lambda$, one obtains after some algebra\begin{figure}[t]
\mbox{\setlength{\unitlength}{1mm}\begin{picture}(83,60)
\put(0,1){\Fig{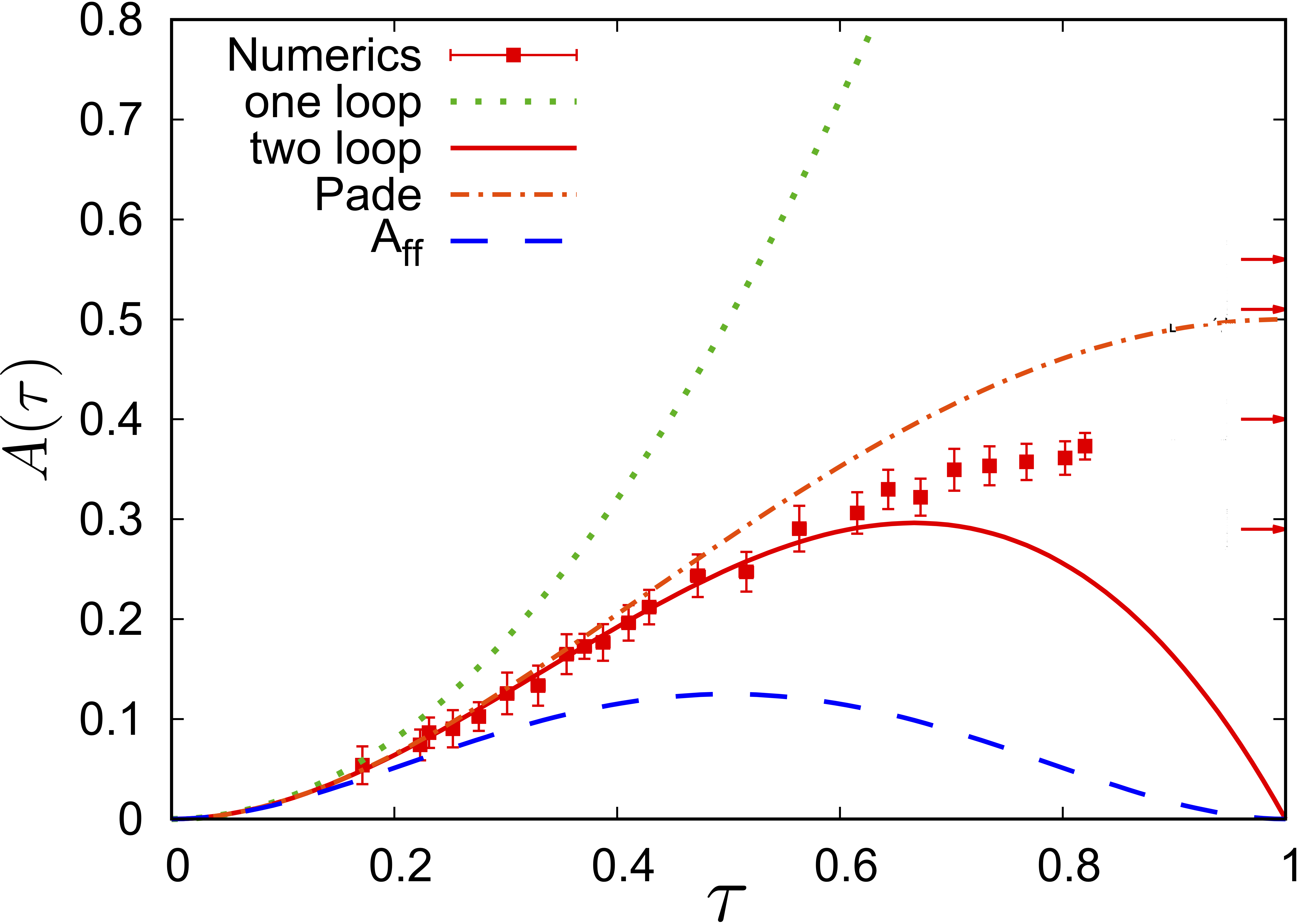}}
\put(72.5,43){\small\protect\cite{RiegerBlasum1997}}
\put(72.5,39.5){\small\protect\cite{ZengMiddletonShapir1996}}
\put(72.5,32.5){\small\protect\cite{Kenyon2001}}
\put(72.5,25.5){\small\protect\cite{LeDoussalSchehr2007}}
\put(74,0){$T=0$}
\put(6,0){$T=T_{\rm c}$}
\end{picture}}
\caption{The amplitude ${A}(\tau)$, characterizing the super-rough phase (Fig.~from \cite{PerretRistivojevicLeDoussalSchehrWiese2012}). The squares are   numerical estimates  using  the  algorithm of \cite{Propp2003}.  ``one loop''   indicates the 1-loop result $A(\tau)=2 \tau^2$ while   ``two loop'' refers to \Eq{amplitude},   (A Pad{\'e} resummation of it is shown as well). $A_{\rm ff}$ is the result of Ref.~\cite{GuruswamyLeClairLudwig2000}. We also show   values obtained numerically at $T=0$ in the corresponding references.}\label{fig_amplitude}
\end{figure}\numparts
\begin{eqnarray}
\beta_{\tilde \lambda}(\tilde \lambda) &:=& L\frac{\partial }{\partial L} \tilde \lambda  = 2\tau \tilde \lambda  -2 \tilde \lambda ^2 +\tilde \lambda ^3+\mathcal{O}(\tilde \lambda^4),\qquad\\
\beta_\sigma(\tilde \lambda) &:=& L\frac{\partial }{\partial L} \sigma   = \frac 12 \tilde \lambda^2+\mathcal{O}(\tilde \lambda ^4).
\end{eqnarray}\endnumparts
The additional 2-loop coefficients are obtained in Ref.~\cite{RistivojevicLeDoussalWiese2012}.
What are the physical consequences of these $\beta$-functions?
First of all, 
there is a non-trivial fixed point for $\tilde \lambda$ at  
\be
\tilde \lambda_{\rm c} = \tau + \frac12 \tau^2 + \ca O(\tau^3),
\ee
see Fig.~\ref{figdiagram}. 
Second, integrating the   $\beta$-function for $\sigma$, starting at a microscopic scale $a$, yields
\be
\sigma = \half \tilde \lambda_{\rm c}^2 \ln \!\Big(\frac L a\Big) + ... = \left[ \frac{\tau^2}2 {+} \frac{\tau^3}2{+}\ca O(\tau^4)\right] \ln \!\Big(\frac L a\Big),
\ee
equivalent to 
\be
 \sigma(k ) \simeq - \left[ \frac{\tau^2}2 {+} \frac{\tau^3}2{+}\ca O(\tau^4)\right] \ln ( a k).
\ee
This allows us to obtain the $k$ dependent 2-point function as 
\be
\overline{ \left < \tilde \Phi_1(k) \tilde \Phi_2(-k) \right> } = \left( \frac{4 \pi T}{k^2}\right)^{\!2}   \frac{\sigma(k) k^2}{2\pi}.  
\ee
As a consequence\footnote{Intermediate steps are
\begin{eqnarray}\nn
\overline {\left< \Phi_1(x) \Phi_1(0)\right>}-\overline {\left< \Phi_1(x) \Phi_2(0)\right>} = \int\frac{\rmd^2 k}{(2\pi)^2}\,
\frac{4 \pi T}{k^2}   \rme^{i k x} \nn\\
= -2T \ln|x/a|. \nn \\
\overline {\left< \Phi_1(x) \Phi_2(0)\right>} = \int\frac{\rmd^2 k}{(2\pi)^2}\,
 \left( \frac{4 \pi T}{k^2}\right)^{\!2}   \frac{\sigma(k) k^2}{2\pi} \rme^{i k x}\nn\\
\nn
=-\big[ \tau^2+\tau^3 +\ca O(\tau^4) \big] T^2 \ln^2|x/a| .
\end{eqnarray}},
\bea\label{eq4amplitude}
\overline{ \left< \Phi(x) - \Phi(0)\right>^2 } = \ca A \ln(x/a)^2 +\ca O\big(\ln(x/a)\big),\\
\ca A =2 (1-\tau)^2 \big[ \tau^2+\tau^3 +\ca O(\tau^4) \big] \nn\\
=  2 \tau^2 -2\tau^3 + \ca O(\tau^4). 
\label{amplitude}
\eea
A numerical test \cite{PerretRistivojevicLeDoussalSchehrWiese2012} using a combinatorial algorithm growing   polynomial in system size (the concept behind this achievement is discussed in section \ref{s:Simulations in equilibrium}) are shown in Fig.~\ref{fig_amplitude}.

The result \eq{amplitude} was obtained in a perturbative expansion in $T-T_{\rm c}$. Even if one could calculate the following orders, and resum them properly, one   wonders whether the expansion remains correct down to $T=0$. This is unlikely: we know from the $\epsilon$-expansion, see \Eq{RP-fixed-point}, that the fixed-point at $T=0$ has to all orders in $\epsilon$ the form (with $\ca C$ a constant)
\be
\Delta(\Phi) = \ca C \left[ \frac{2\pi^2}3- \Phi(2\pi-\Phi) \right] \equiv  4 {\ca C} \sum_{q=1}^\infty \frac{\cos(q \Phi)}{q^2}.
\ee
It contains an infinity of subdominant modes indexed by $q$, which one can try to incorporate into the perturbative result. Naively one expects the mode $q$ to show up at $T_{\rm c}(q)=T_{\rm c}/q^2$, and it is likely to increase the perturbative result.
Attempts to do so have been undertaken in Refs.~\cite{CarpentierLeDoussal1998,CarpentierLeDoussal2000,CarpentierLeDoussal2001,CarpentierLeDoussal2007}.

\subsection{Multifractality}
Consider an observable $\ca O(\ell)$ such as the field difference between two   points  a distance $\ell$ apart.  Its $n$-th moment  reads
\be
\left < \ca O(\ell)^n \right> \sim \ell^{\zeta_n} .
\ee
Generically there are two possibilities 
\begin{itemize}
\item[(i)] $\zeta_n = n \zeta $: fractal 
\item[(ii)]  $\zeta_n \neq n \zeta $: multifractal  
\end{itemize}
In some cases, e.g.\ in the critical dimension, one has 
\be
\left < \ca O(\ell)^n \right> \sim \ell^{\zeta_n} = \rme^{\zeta_n \ln(\ell)} \to {\zeta_n \ln(\ell)}, 
\ee
i.e.\ the universal anomalous dimension appears as the amplitude of the log, see e.g.\ \Eq{universal-amplitude-2} and section \ref{s:Behavior at the upper critical dimension}. We still think of these systems as  fractal or multifractal, depending on which of the two choices above applies. 
 
Most pure critical systems are fractals. Famous multifractal systems are Navier-Stokes  turbulence
\cite{FrischBook,LesieurBook,Gawedzki2008},
or   the more tractable passive advection of particles, the  {\em passive scalar},
\cite{KraichnanYakhotChen1995,GawedzkiKupiainen1995,BernardGawedzkiKupiainen1998,Antonov1999,AdzhemyanAntonovVasilev1998,AdzhemyanAntonovBarinovKabritsVasilev2001,AdzhemyanAntonovBarinovKabritsVasilev2001b}, or its  generalization to  the advection of extended   elastic objects \cite{Wiese1999}. 

An important question is whether  systems with {\em quenched disorder} show multifractality. A prominent example is the wave-function statistics at a delocalization transition, such as the Anderson metal-insulator transition at the mobility edge in three spatial dimensions, or the integer quantum-Hall plateau transition in two dimensions. Here one considers
(see \cite{FosterRyuLudwig2009} for a concise introduction{ \new or the classic Ref.~\cite{HalseyJensenKadanoffProcacciaShraiman1986}})
\be
P_q(\epsilon_i) := \int_{L^d} \rmd^dr \,|\psi_i(r)|^{2q}  \sim L^{-\tau(q)},
\ee
where $\epsilon_i$ is the energy of the state $i$, and $\psi_i(r)$ its wave function. 
Note that normalization imposes $\tau(1)=0$.
For  extended states inside a band $\tau(q) =d(q-1)$, while for localized states $\tau(q)=0$. At the band edge $\tau(q)$ is non-trivial. Define by $f(\alpha)$ its Legendre transform, 
\be
\mbox{Legendre}_{\alpha\leftrightarrow q} \quad f(\alpha) + \tau(q) = \alpha q.
\ee
Then the set of points at which  an eigenfunction takes the value $|\psi(r)|^2= \ca A\,L^{-\alpha}$ has weight $L^{f(\alpha)}$.
Both $f(\alpha)$ and $\tau(q)$ are convex.

The question relevant for this review is whether disordered elastic manifolds show   multifractality. As long as the roughness exponent $\zeta>0$, this does not seem to be the case. The situation is different for $\zeta=0$, i.e.\ charge-density waves or  vortex lattices. Technically, it can be accessed either via a $4-\epsilon$ expansion \cite{FedorenkoLeDoussalWiese2014} 
or directly in two dimensions \cite{LeDoussalRistivojevicWiese2013}.

\subsubsection*{Multifractality of the random-phase sine-Gordon model in dimension $d=2$.}
The random-phase sine-Gordon model was introduced above in section \ref{s:The random-phase sine-Gordon model}. 
The object  to be considered is 
\begin{equation}\label{Cdef}
C(q,r):=\overline{\langle e^{iq\left[\Phi(r)-\Phi(0)\right]}\rangle}.
\end{equation}
It was shown in Ref.~\cite{LeDoussalRistivojevicWiese2013} that with $\ca A$ given in \Eq{amplitude}, 
\bea\label{Cfinalform}
C(q,r)\simeq\left(\frac{a}{r}\right)^{\eta(q)}
\exp\left(-\frac{1}{2}\mathcal{A}q^2 \ln^2(r/a)\right).
\eea
The anomalous exponent $\eta$ defined in Eq.~(\ref{Cfinalform}) reads
\bea\label{etaFINAL}
\eta(q)=2q^2(1-\tau)[1+2(1-\tau)\sigma'] +\tau^2\eta_g(q)+\mathcal{O}(\tau^3). \nn\\
\eea
Its nontrivial part $\eta_g$ is \cite{LeDoussalRistivojevicWiese2013}  
\bea\label{etagfinal}
\eta_g(q)=\left\{ 
\begin{array}{cc}
q^2[1-2\gamma_{\mathrm{E}}-\psi(q)-\psi(-q)], &q<1\\
-2,& q=1
\end{array}
\right.
\eea
Here $\gamma_{\mathrm{E}}$ is   Euler's constant. 
The result for the correlation function (\ref{Cdef}) enables one to calculate the leading large-distance behavior of all higher powers of the connected correlation functions in the super-rough phase, i.e., for $T<T_{c}$ ($\tau>0$, see Fig.~\ref{figdiagram}). Using 
\be
\eta_g(q) = q^2 + 2 \sum_{n=2}^\infty \zeta(2n-1) q^{2n}, 
\ee
 one sees that odd moments vanish, the second moment is given by \Eqs{eq4amplitude}-\eq{amplitude}, and higher even moments by
\bea
\frac{(-1)^{n}}{(2n)!}\overline{\left\langle\left[\Phi(r){-}\Phi(0)\right]^{2n}\right\rangle}_{\rm c}
= -2\tau^2\zeta(2n{-}1)\ln\left(\frac{r}{a}\right)\nn +...\\
\eea
This system is multifractal.

\subsubsection*{FRG in dimension $4-\epsilon$.}
Following 
\cite{FedorenkoLeDoussalWiese2014}, define
\bea
\mathcal{G}[\lambda]:=\overline{\left\langle{ \rme^{  \int_x \lambda({x}) u ({x}) }}  \right\rangle}=\lim\limits_{n\to 0}\left< \rme^{\int_x \lambda({x}) u_1({x}   )}\right>_{\cal S},\\
\lambda(x) = i q[ \delta(x)-\delta(x+r)].
\eea
This can be calculated with methods similar to \cite{FedorenkoLeDoussalWiese2006}, using from the action \eq{H} only the  cubic vertex,  
\bea
{\frac{1}{2T^2} \int_x  \sum_{a,b}  R\big(u_a(x) {-} u_b(x)\big)  }   \nn\\
 \to 
 \frac{\sigma} {12 T^2}\sum_{a,b}|u_a(x){-}u_b(x)|^3, \quad \sigma = R'''(0^+).
\eea
The connected part of $\ca G[\lambda]$ reads (the correlation function $C(x)$ is defined in \Eq{m:cor1bis})
\bea
\ln ( \ca G[\lambda] )=~ \parbox{35mm}{\includegraphics[width=35mm]{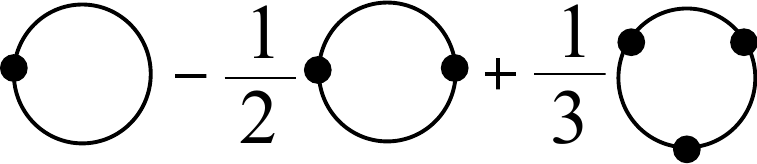}}+...\nn\\
=\ln\left(\frac{\det ( - \nabla^2 +\sigma U(x) + m^2 ) }{
  \det ( - \nabla^2 + m^2 )}\right)
\label{formidable-determinant}  
  \\
U(x) =   \int_y C(x-y)  \lambda(y)  .
\eea
Calculating the determinant \eq{formidable-determinant} is a formidable task, usually only possible in perturbation theory in $\sigma$. 
Here we give analytical results, using three consecutive tricks: 
\begin{enumerate}
\item   Solve the problem for a spherically symmetric source $\lambda(y)$, assuming  a uniformly distributed positive unit charge on a circle of radius $a$, and a compensating negative   charge on a circle of radius $L\gg a$. The potential is $U(x)=  (r^{-2}-L^{-2})/(2\pi)^2$ for $a\le r\le L$, and constant beyond.

\item   Write the Laplacian in distance and angle variables,
\begin{equation}
-\nabla^2 \to {\mathcal H}_{l}:=
-\frac{\rmd^2}{\rmd r^2}+\frac{\left(l+\frac{d-3}{2}\right)
\left(l+\frac{d-1}{2}\right)}{r^2}.
\end{equation}
\item
$\ln ( \ca G[\lambda] )$ is
written as a sum of the logarithms of
the 1-dimensional determinant ratios ${\cal B}_l$ for partial waves,
weighted with the degeneracy of
angular momenta $l$,
\begin{equation}
\ln ( \ca G[\lambda] )
=\sum_{l=0}^\infty \frac{(2l{+}d{-}2)(l{+}d{-}3)!}{l!(d{-}2)!}
\ln ({\cal B}_l). \label{eq-dunne}
\end{equation}
\item
The  Gel'fand-Yaglom method \cite{GelfandYaglom1960}  explained in appendix \ref{s:Gelfand-Yaglom} gives the ratio of the
1-dimensional functional determinants for each partial wave $l$ as
\begin{equation}{\cal B}_l:=\frac{\det \left[{\mathcal H}_{l}+\sigma U(r) + m^2\right]}{\det
\left[{\mathcal H}_{l}+m^2\right]}=
\frac{\psi_l(L)}{\tilde{\psi}_l(L)}. \label{eq-det-l}
\end{equation}
Here ${\psi}_l(r)$ is the solution of
\begin{equation} \label{eq-Schodinger}
\left[{\mathcal H_l}+\sigma U(r)+m^2\right]{\psi}_l(r)=0,
\end{equation}
satisfying
$ {\psi}_{l}(r) \sim r^{l+(d-1)/2}$ for $r\to 0$; $\tilde {\psi}_l(r)$ is the solution for $\sigma=0$.
\item After some pages of algebra,  one finds (modulo odd terms which cancel below) 
\bea
 \ln ( \ca G[\lambda] )  = \ca F\Big(\frac{\sigma q}{(2\pi)^2} \Big) +\mbox{terms odd in $\sigma$}
\\
\label{16}
\ca F(s)=-\sum_{l=0}^{\infty} (l+1)^2 \Big( \sqrt{(l+1)^2+s} -(l+1)\nn  \\
\hp{\ca F(s)=-\sum_{l=0}^{\infty} }  -\frac{s}{2(l+1)}+\frac{s^2}{8 (l+1)^3}\Big)\ .
\end{eqnarray}  
Resummation yields (dropping   odd terms)
\begin{eqnarray}\label{eq-F-fun}
\ca F(s) = \sum _{n=2}^{\infty }\frac{ s^{2n}\Gamma(2n-\frac12) \zeta(4n-3)
  }{ 2 \sqrt{\pi}(2n)! }.
\end{eqnarray}
\item In a last step one proves perturbatively that all $n$-point functions remain unchanged if one moves the charge on the sphere at $|r|=L$ to a single point at distance $L$ from the origin.  The combinatorial analysis yields an additional factor of $2$. 
\end{enumerate} 
Using the FRG fixed point \eq{RP-fixed-point-2}, $s=\frac\epsilon 3 q $,   the    \(2n\)-th   cumulant of relative displacements is obtained as   
\bea
\overline{\left<[u(r)-u(0)]^{2n}\right>^{\rm c} } \simeq {\cal A}_{2n}  \ln (r/a)\\
{\cal A}_{2n} = - \Big(\frac\epsilon 3\Big)^{2n} 
 \frac{\Gamma(\textstyle 2n-\frac12) \zeta(4n-3)
  }{  \sqrt{\pi} }, \quad n\ge 2 .
\eea
This correlation function is multifractal.

\end{reviewKay}

\subsection{Simulations in equilibrium: polynomial versus NP-hard}
\label{s:Simulations in equilibrium}
Finding the ground state of a disordered system is in general a very difficult problem, often even NP-hard, meaning no algorithm exists which is guaranteed to find the ground state in polynomial time, i.e.\ within a time which does not grow faster than $N^p$, with $N$ the system size, and $p$ a finite number. This statement should be viewed as ``state of the art'': E.g.\ we know of no algorithm to find the ground state of the SK model in polynomial time; but this does not imply that no such algorithm can exist. What computer scientists have proven is that if one day an algorithm is found to solve one NP-hard problem, all other NP-hard problems can be solved as well. 
We refer the reader to the textbooks \cite{HartmannRiegerBook,KrauthBook} and collection  \cite{HartmannRiegerBook2}    for a more precise definition and  further information on the subject.  

While many ground-state calculations are considered NP-hard, there are some notable exceptions: For the RNA-folding problem \cite{TinocoBustamante1999} a polynomial algorithm exists \cite{McCaskill1990}, which evaluates the partition function at all temperatures  in a time growing as $N^3$, allowing one not only to find the ground state but even the phase transition from a frozen to a molten phase \cite{BundschuhHwa1999,Higgs2000,BaezWieseBundschuh2019}.

Another notable exception are disordered elastic manifolds, or more specifically the ground state of an Ising ferromagnet coupled to either random-bond or random-field disorder. 
It can be solved by the {\em minimum-cut algorithm} \cite{Sedgewick1990}. This has been used in numerous publications: To find the roughness exponent $\zeta$ in dimensions 2 and 3 \cite{Middleton1995}, see Fig.~\ref{fig:numstat}; to measure the FRG-fixed point function \cite{MiddletonLeDoussalWiese2006}, see Fig.~\ref{f:Alan1}.  or avalanche-size distributions in equilibrium \cite{LeDoussalMiddletonWiese2008}, see Fig.~\ref{f:2}. Further 
for flux lines in a disordered environment \cite{Rieger1998,NohRieger2001}; or  solid-on-solid models with disordered substrates \cite{RiegerBlasum1997}.

\subsection{Experiments in equilibrium}
\label{s:Experiments at equilibrium}
There are   few experiments which  really are in equilibrium.  
The main reason is that in most cases the exponent $\theta $  defined in \Eq{a8} is positive, restricting the energy fluctuations which  according to \Eq{free-energy-scaling} grow as $L^\theta$. As a consequence, there is a maximal length $L^{\rm max}_{T}$ up to which the system can equilibrate, i.e.\ find the minimum-energy configuration. Let us give a list of notable exceptions.
\begin{enumerate}
\item {\em domain walls in   thin magnetic films} with random-bond disorder have long been interpreted \cite{LemerleFerreChappertMathetGiamarchiLeDoussal1998,Diaz-PardoMoisanAlbornozLemaitreCurialeJeudy2019,Diaz-PardoPhD}  as showing the roughness exponent of $\zeta_{\rm RB}^{d=1}=2/3$ given in  \Eq{zeta=2/3}.  This interpretation probably holds only on small scales, see the discussion in section \ref{s:Experiments on thin magnetic films}.

\item {\em hairpin unzipping} reported in \cite{HuguetFornsRitort2009} is consistent with a roughness exponent $\zeta=4/3$, in agreement with the value predicted for a single degree of freedom, i.e.\ $d=0$ or $\epsilon=4$ in \Eq{zeta-RF-1loop}. We discuss this experiment in   detail in section \ref{s:Experiment on RNA-DNA unzipping}. There it is confronted with an experiment using the much softer peeling mode, placing it in the different depinning universality class with $\zeta=2^-$.

\item{Vortex lattices (section \ref{s:Bragg glass and vortex glass}).}

\end{enumerate}

\section{Dynamics, and the depinning transition}\label{s:dynamics}
\begin{widefigure}
\begin{center}
\href{http://www.phys.ens.fr/~wiese/masterENS/movies/contact-line.mov}{\fig{0.58\textwidth}{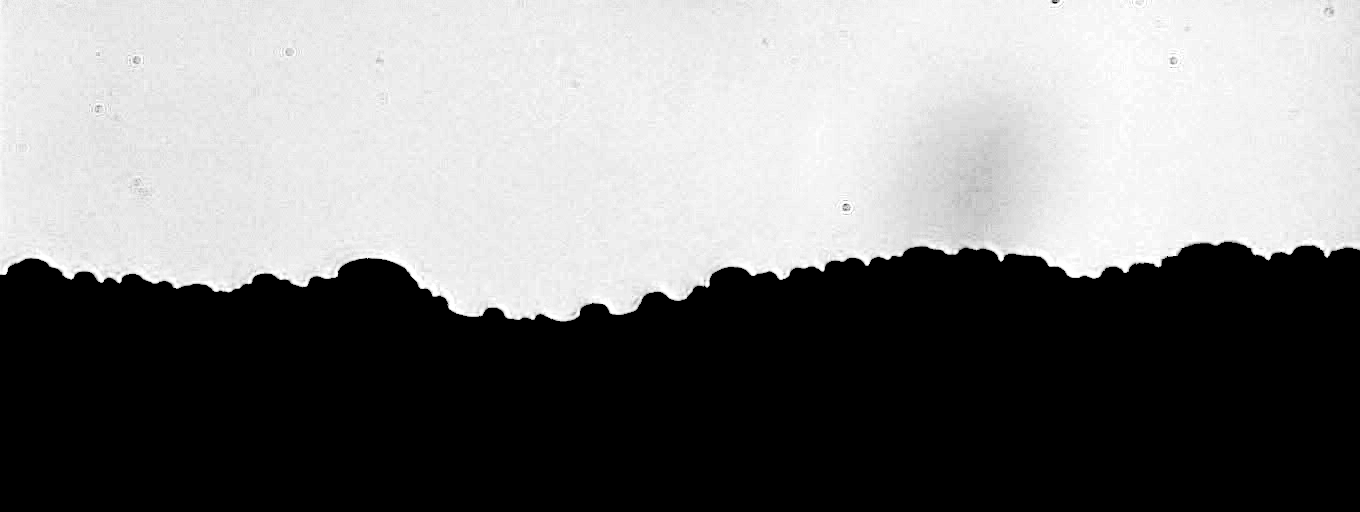}}\hfill
\fig{0.34\textwidth}{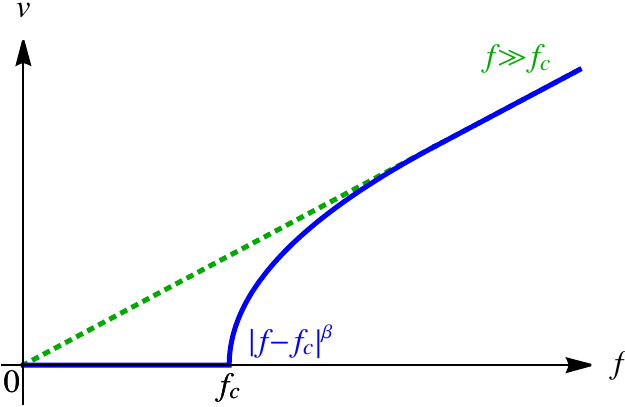}\end{center}
\caption{Left: Snapshot of a contact-line at depinning, courtesy E.~Rolley [\href{http://www.phys.ens.fr/~wiese/masterENS/movies/contact-line.mov}{movie}].
Observables derived from this system are shown in Figs.~\ref{f:Delta3} and \ref{f:avalanche-obs}.
Right: Velocity of a pinned interface as a function of the applied
force. $f=0$: equilibrium. $f=f_{c}$: depinning. For an experimental confirmation of the $v(f)$ curve in a thin magnetic film, see Fig.~\ref{fig-RF-dep-experiment}.}
\label{f:vel-force}
\end{widefigure}

\subsection{Phenomenology}
\label{s:Phenomenology}
Another important class of phenomena for elastic manifolds in disorder
is the so-called {\em depinning transition}: Applying a constant force
to the elastic manifold, e.g.\ a constant magnetic field to the
ferromagnet mentioned in the introduction, the latter will   move
if, and only if, a certain critical threshold force $f_{c}$ is surpassed, see
figure \ref{f:vel-force}. (This is fortunate, since otherwise the
magnetic domain walls in the hard-disc drive onto which this article
is stored would move,  with the effect of deleting all information,
depriving you from your reading.)  At $f=f_{c}$, the so-called
depinning transition, the manifold has a  roughness exponent $\zeta$ (see Eq.~(\ref{roughness})), distinct from the
equilibrium ($f=0$). 
The equation describing the movement of the interface is
\bea\label{eq-motion-f}
\highlight{\partial_{t} u (x,t) = (\nabla^{2}{-}m^{2}) u (x,t) + F \big(x,u (x,t)\big) + f(x,t) \komma \nn\\  \\
F (x,u)=-\partial_{u} V (x,u)\punkt}
\eea
There are two main driving protocols, depending on whether one controls the applied force, or the mean driving velocity.
\subsubsection*{Force-controlled depinning.}
Let us   impose a driving force $f(x,t)=f$, and set   $m\to 0$.
For $f>f_{\rm c}$, the manifold then moves with velocity $v$.  Close to
the transition, new critical exponents appear:
\begin{itemize}\itemsep0mm
\item
a velocity-force relation given by (see figure \ref{f:vel-force})
\be\label{v=|f-fc|^beta}
\highlight{ v \sim |f-f_{\rm c}|^{\beta} \mbox{~~ for ~~} f>f_{\rm c}\komma }
\ee
\item  a {\em dynamic} exponent $z$ relating correlation functions in space and time\be\label{exponent-z}
\highlight{ t\sim x^{z}\punkt}
\ee
Thus if one has a correlation or response function $R(x,t)$, it will be for short times and distances be a function of $t/x^z$ only,
\be
R(x,t) \simeq R(t/x^z)\punkt
\ee
\item a correlation length $\xi$ set by the distance to $f_{c}$
\be\label{def:nu}
\highlight{ \xi = \xi_f \sim |f-f_{\rm c}|^{-\nu }}\punkt
\ee
Remarkably,  this relation holds on both sides of the transition: For $f<f_{\rm c}$, it describes how starting from   a flat or equilibrated configuration,  the correlation length $\xi$, which can be interpreted as the avalanche extension (defined below in \Eq{ava-extension-def}),  increases as one approaches $f_{\rm c}$. Arriving at $f_{\rm c}$, each segment of the interface has moved. Above $f_{\rm c}$, the interface is always moving, and the correlation length $\xi$ (which now decreases upon an increase in $f$) gives the size of coherently  moving pieces.

\item 
The  new exponents $z$, $\beta$ and $\nu$ are not   independent, but related
\cite{NattermannStepanowTangLeschhorn1992}.  Suppose that   $f>f_{\rm c}$, and we witness  an avalanche of extension $\xi$. Then its mean velocity scales as
\bea\label{beta-rel}
v \sim \frac u t \sim \frac{\xi^\zeta}{\xi^z}\sim |f-f_{{\rm c}}|^{-\nu (\zeta-z)}\nn\\ \Longrightarrow \quad\highlight{ \beta = \nu (z-\zeta)}\punkt
\eea
One can   make the same argument below $f_{\rm c}$, by slowing increasing $f$ to $f_{\rm c}$.
\item 
Suppose that below $f_{c}$ the manifold is in a pinned configuration. Increasing $f$   leads to an avalanche, of extension $\xi$, and a change of elastic force (per site) $\sim  \xi^{\zeta-2}$. This has to be balanced by the driving force, i.e.
\be\label{nu-rel}
\xi^{\zeta-2} \sim |f-f_{\rm c}| \quad \Longrightarrow\quad \highlight{ \nu =\frac{1}{2-\zeta }}\punkt
\end{equation}
\end{itemize}
\subsubsection*{Velocity-controlled depinning.}If $m>0$, then we can rewrite the equation of motion \eq{eq-motion-f} as 
\bea\label{eq-motion-w}
\highlight{\partial_{t} u (x,t) = (\nabla^{2}-m^{2}) [u (x,t)-w] + F \big(x,u (x,t)\big)\komma  \nn\\
    w = vt\punkt}
\eea
The phenomenology   changes:
\begin{itemize}
\item
The driving force acting on the interface is fluctuating as well as the velocity,  while the mean driving velocity is fixed 
\be
\overline {\dot u(x,t)}= v\komma  \quad f = \frac{1}{L^d}\int_x\overline {F\big(x,u(x,t)\big)}\punkt
\ee
\item The correlation length $\xi$ is set by the confining potential, 
\be\label{xi-m}
\highlight{ \xi = \xi_m=\frac 1 m}\punkt
\ee
\end{itemize}

\subsection{Field theory of the depinning transition, response function}
\label{s:Field theory of the depinning transition}
\Review{We can enforce the equation of motion \eq{eq-motion-w} with an auxiliary field $\tilde u(x,t)$\footnote{This trick is known as the MSR formalism \new\protect\cite{MSR,Janssen1976,DeDominicis1976,Janssen1985,Tauber2012}. It is the generalization to a field of the relation 
$\int_k \rme^{i k x} = \delta (x)$: the {\em response field} $\tilde u(x,t)$ enforces the Langevin equation \eq{eq-motion-w} for each $x$ and $t$. A short introduction is given in appendix \ref{MSR-formalism}.}}
\bea\label{S[u,utilde,F]}
{\cal S}[u,\tilde u,F] = \!\int_{x,t} \!\tilde u(x,t) \Big[& \big( \partial_{t}  - \nabla^{2}+m^{2}\big) \big( u (x,t)-w\big) \nn\\
 &-F \big(x,u (x,t)\big) - f(x,t) \Big] \punkt
\eea
We   need to average over disorder,  to obtain the disorder-averaged action 
$
\rme^{-{\cal S}[u,\tilde u]}  := \overline {\rme^{-S[u,\tilde u,F]}}
$, with 
\bea\label{dyn-action}
\highlight{{\cal S}[u,\tilde u] =\! \int_{x,t} \!\tilde u (x,t)\Big[ (\partial_{t}{-}\nabla^{2}{+}m^{2}) [ u
(x,t){-}w] {-}f(x,t)\Big]\nn\\
  -
\frac12 \int_{x,t,t'} \tilde u (x,t)\Delta \big(u (x,t){-}u (x,t')\big)\tilde u
(x,t')}\punkt
\eea
We remind the definition of the force-force correlator given in \Eq{Delta-def}.

\subsubsection*{Response function and the free theory:}
The response of a system is defined as the answer of the system given a perturbation $f(x,t)$. The response can be any observable, as the avalanche-size distribution defined below in \Eq{rhof}, but the simplest one is the response of the field $u(x',t')$ itself,
\bea\label{170}
R_f(x',t'|x,t) := \frac{\delta}{\delta f(x,t)} {\overline{u(x',t')} } = \left<   u(x',t') \tilde u(x,t)  \right>  \punkt \nn\\
\eea
In a translationally invariant system, $R_f(x',t'|x,t) $ does only depend on $x'-x$ and $t'-t$, and is denoted  
\be
R(x'-x,t'-t) := R_f(x',t'|x,t)  .
\ee
In the second equality of \Eq{170} we used the average provided by the action \eq{dyn-action}.
The formalism is explained in appendix \ref{MSR-formalism}, see \Eq{response} and following.  
The most convenient representation  is the spatial  Fourier transform calculated for the free theory in  \Eq{R(k,t)}, 
\be
R(k,t) = \left< {u(k,t{+}t') \tilde u(-k,t')} \right> = \rme^{-(k^2+m^2)t} \Theta(t). 
\ee
We could introduce   response functions as the answer to different perturbations, e.g. increasing $w$ instead of $f$, 
\be\label{Rw}
R_w(k{=}0,t) := \frac{\rmd }{\rmd w} \! \left< u(k{=}0,t) \right> = m^2 R(k{=}0,t) . 
\ee
This changes the normalization,  \be
\label{Rwint}
\int_{t} R_w(k=0,t) = 1.
\ee 
While \Eq{Rw} is the  free-theory result, corrected in perturbation theory, \Eq{Rwint} is    by construction exact.

\subsection{Middleton theorem} 
\label{s:Middleton}
We now state the famous  \cite{Middleton1992} \smallskip

\noindent{{\bf Middleton Theorem}:}
If $F(x,u)$ is continuous in $u$, and $\dot u(x,t)\ge 0$, then $\dot u(x,t')\ge 0$ for all $t'\ge t$. Moreover, if two configurations are ordered, $u_2(x,t)\ge u_1(x,t)$, then they remain ordered for all times, i.e.\ $u_2(x,t'')\ge u_1(x,t')$ for all $t''\ge t'>t$.

\smallskip
\noindent{{\bf \em Proof}:} Consider an interface discretized in $x$. 
The trajectories $
u(x,t)
$ are a function of time.  Suppose that there exists $x$ and $t'>t$ s.t.\ $\dot u(x,t')<0$. Define $t_0$ as the first time when this happens, 
$t_0:=\inf_x\inf_{t'>t}\{ \dot u(x,t')<0\}$, and $x_0$ the corresponding position $x$. By continuity of $F$ in $u$, the velocity $\dot u$ is   continuous in time, and  $\dot u(x_0,t_0)=0$. 
This implies that the disorder force acting on $x_0$  does not change in the next (infinitesimal) time step, and the only changes in force can come from a change in the elastic terms. Since by assumption no other point has a negative velocity, this change in force can not be negative,  contradicting   the assumption. 

To prove the second part of the theorem, 
consider the following configuration at time $t_0$:
$$
{\parbox{7.5cm}{{\begin{tikzpicture}[scale=1.2]
\coordinate (v1) at  (0,0) ; 
\coordinate (v2) at  (1,-.25) ; 
\coordinate (v3) at  (2,.25) ; 
\coordinate (v4) at  (3,.75) ; 
\coordinate (v5) at  (4,.25) ; 
\coordinate (v6) at  (5,.5) ;  
\coordinate (u1) at  (0,1) ; 
\coordinate (u2) at  (1,.25) ; 
\coordinate (u3) at  (2,1.) ; 
\coordinate (u4) at  (3,.75) ; 
\coordinate (u5) at  (4,1) ; 
\coordinate (u6) at  (5,.75) ;  
\node (x) at  (5.4,-.5)    {$\!\!\!\parbox{0mm}{$\raisebox{-3mm}[0mm][0mm]{$ x$}$}$};
\node (x0) at  (3,0.7)    {$\!\!\!\parbox{0mm}{$\raisebox{-3mm}[0mm][0mm]{$ x_{\rm 0}$}$}$};
\node (uu1) at  (0,0)    {$\!\!\!\parbox{0mm}{$\raisebox{1mm}[0mm][0mm]{$ u_1$}$}$};
\node (uu2) at  (0,1)    {$\!\!\!\parbox{0mm}{$\raisebox{1mm}[0mm][0mm]{$ u_2$}$}$};
\node (u) at  (-.7,1.2)    {$\!\!\!\parbox{0mm}{$\raisebox{1mm}[0mm][0mm]{$ u$}$}$};
\draw [blue,thick] (v1) -- (v2)--(v3)--(v4)--(v5)--(v6);
\draw [red,thick] (u1) -- (u2)--(u3)--(u4)--(u5)--(u6);
\draw[enddirected](-.5,-.5)--(5.5,-.5);
\draw[enddirected](-.5,-.5)--(-.5,1.5);
\fill (v1) circle (1.5pt);
\fill (v2) circle (1.5pt);
\fill (v3) circle (1.5pt);
\fill (v4) circle (1.5pt);
\fill (v5) circle (1.5pt);
\fill (v6) circle (1.5pt);
\fill (u1) circle (1.5pt);
\fill (u2) circle (1.5pt);
\fill (u3) circle (1.5pt);
\fill (u4) circle (1.5pt);
\fill (u5) circle (1.5pt);
\fill (u6) circle (1.5pt);
\end{tikzpicture}}}}
$$
Here the red configuration is ahead of the blue one, except at position $x_0$, where they coincide. 
As in the first part of the proof, we wish to bring to a contradiction the hypothesis that at some later time  $u_1(x_0)$ (blue) is ahead of $u_2(x_0)$ (red). For this reason, we have chosen $t_0$ the infimum of   times contradicting the theorem, $t_0:=\ \inf_{t'>t}\{  u_1(x_0,t')>u_2(x_0,t')\}$.
 Consider  
the    equation of motion  \Eq{eq-motion-w} for the difference between $u_1$ and $u_2$, 
\bea
\partial_t \left[ u_2(x_0,t) - u_1(x_0,t)\right]\big|_{t=t_0}\nn\\
 = \nabla^2  \left[ u_2(x_0,t_0) - u_1(x_0,t_0)\right]\punkt
\eea
The disorder force terms have canceled as well as the term of order $m^2$, since by assumption $u_2(x_0,t_0) = u_1(x_0,t_0)$. 
By construction, the r.h.s.\ is positive, leading to the desired contradiction. 
\smallskip

\paragraph{Remark: Uniqueness of perturbation theory.}
 As in the statics, one encounters terms
proportional to $\Delta' (0^{+})\equiv -R''' (0^{+})$. Here the
sign problem can  be solved unambigously  by observing that due to Middleton's theorem the manifold
only moves forward,
\begin{equation}\label{jump-ahead}
t'>t\quad \Longrightarrow \quad u (x,t')- u (x,t) \ge 0 \punkt\end{equation}
Thus the argument of $\Delta\big (u (x,t')- u (x,t)\big)$ has a well-defined sign, allowing us to interpret derivatives at vanishing arguments correctly. 
Practically this means that when evaluating diagrams containing
$\Delta (u (x,t)-u (x,t'))$, one splits them into two pieces, one with
$t<t'$ and one with $t>t'$. Both pieces are well defined, even in the
limit of $t\to t'$.

\subsection{Loop expansion}
\label{s:dep-loops}
Consider the field theory defined by the action \eq{dyn-action}. \Review{
To appreciate the problem, let us remind that in equilibrium a model is defined by its Boltzmann weight. As long as the system is ergodic, it can be sampled  with the help of a Langevin equation, and equilibrium expectations can be evaluated as expectations in the dynamic field theory. This goes hand in hand with  identical renormalizations, as is e.g.\ known  for the effective coupling in $\phi^4$ theory. On the other hand, it does  not fix the dynamics. It is indeed well-known that a different dynamics leads to a different {\em dynamic universality class}, as exemplified by the Hohenberg-Halperin classification of dynamical critical phenomena \cite{HohenbergHalperin1977}, leading to the zoo of  models A, B, C, ..., F, and J.} We might therefore not be   surprised if below we find the  same renormalization for the disorder     in the driven dynamics. 
On the other hand, equilibrium and  out-of-equilibrium     are two distinct phenomena, and may have  distinct critical exponents. As we will see below, at 1-loop order all comparable observables are identical, whereas differences are manifest at 2-loop order.

Let us start by rederiving the corrections to the  renormalized disorder correlator at 1-loop order. 
The replica diagram in \Eq{80cis} is one of the two contributions to the effective potential-potential correlator  $R(u)$
 given in \Eq{351}. In the dynamics, the disorder term in \Eq{dyn-action} is   the bare (microscopic) {\em force-force correlator}  $\Delta_0(u)$, which we note graphically as
 \bea
 \int_{x,t_1,t_2}\tilde u(x,t_1)\tilde u(x,t_2)\,\Delta_0\big(u(x,t_1)-u(x,t_2) \big)\nn\\
 = ~\;
 {\parbox{0.8cm}{{\begin{tikzpicture}
\coordinate (x1t1) at  (0,0) ; 
\coordinate (x1t2) at  (0,.5) ; 
\node (x) at  (0,0)    {$\!\!\!\parbox{0mm}{$\raisebox{-3.5mm}[2.5mm][2.5mm]{$\scriptstyle x$}$}$};
\node (t1) at  (-.25,0)    {$\!\!\!\parbox{0mm}{$\raisebox{-1mm}[0mm][0mm]{$\scriptstyle t_1$}$}$};
\node (t2) at  (-.25,0.5)    {$\!\!\!\parbox{0mm}{$\raisebox{-1mm}[0mm][0mm]{$\scriptstyle t_2$}$}$};
\fill (x1t1) circle (2pt);
\fill (x1t2) circle (2pt);
\draw [dashed,thick] (x1t1) -- (x1t2);
\draw [enddirected]  (x1t1)--(0.5,0);
\draw [enddirected]  (x1t2)--(0.5,0.5);
\end{tikzpicture}}}}
\eea 
The arrows are the response fields $\tilde u(x,t_1)\tilde u(x,t_2)$; some authors   represent them   by a wiggly line. 
Since the   response function has a  direction in time, the static diagram \eq{80cis} has two {\em descendants} in the dynamic formulation, 
 \be\label{7.16}
{\parbox{1.65cm}{{\begin{tikzpicture}
\coordinate (x1t1) at  (0,0) ; 
\coordinate (x1t2) at  (0,.5) ; 
\coordinate (x2t3) at  (1.5,0) ; 
\coordinate (x2t4) at  (1.5,0.5) ; 
\node (x) at  (0,0)    {$\!\!\!\parbox{0mm}{$\raisebox{-3.5mm}[2.5mm][2.5mm]{$\scriptstyle x$}$}$};
\node (y) at  (1.5,0)    {$\!\!\!\parbox{0mm}{$\raisebox{-3.5mm}[2.5mm][2.5mm]{$\scriptstyle y$}$}$};
\fill (x1t1) circle (2pt);
\fill (x1t2) circle (2pt);
\fill (x2t3) circle (2pt);
\fill (x2t4) circle (2pt);
\draw  (x1t1) -- (x2t3);
\draw  (x1t2) -- (x2t4);
\draw [dashed,thick] (x1t1) -- (x1t2);
\draw [dashed,thick] (x2t3) -- (x2t4);
\end{tikzpicture}}}}
~\longrightarrow~
{\parbox{2.1cm}{{\begin{tikzpicture}
\coordinate (x1t1) at  (0,0) ; 
\coordinate (x1t2) at  (0,.5) ; 
\coordinate (x2t3) at  (1.5,0) ; 
\coordinate (x2t4) at  (1.5,0.5) ; 
\node (x) at  (0,0)    {$\!\!\!\parbox{0mm}{$\raisebox{-3.5mm}[2.5mm][2.5mm]{$\scriptstyle x$}$}$};
\node (y) at  (1.5,0)    {$\!\!\!\parbox{0mm}{$\raisebox{-3.5mm}[2.5mm][2.5mm]{$\scriptstyle y$}$}$};
\fill (x1t1) circle (2pt);
\fill (x1t2) circle (2pt);
\fill (x2t3) circle (2pt);
\fill (x2t4) circle (2pt);
\draw [directed] (x1t1) -- (x2t3);
\draw [directed] (x1t2) -- (x2t4);
\draw [dashed,thick] (x1t1) -- (x1t2);
\draw [dashed,thick] (x2t3) -- (x2t4);
\draw [enddirected]  (x2t3)--(2,0);
\draw [enddirected]  (x2t4)--(2,0.5);
\end{tikzpicture}}}} ~+~  {\parbox{2.5cm}{{\begin{tikzpicture}
\coordinate (x1t1) at  (0,0) ; 
\coordinate (x1t2) at  (0,.5) ; 
\coordinate (x2t3) at  (1.5,0) ; 
\coordinate (x2t4) at  (1.5,0.5) ; 
\node (x) at  (0,0)    {$\!\!\!\parbox{0mm}{$\raisebox{-3.5mm}[2.5mm][2.5mm]{$\scriptstyle x$}$}$};
\node (y) at  (1.5,0)    {$\!\!\!\parbox{0mm}{$\raisebox{-3.5mm}[2.5mm][2.5mm]{$\scriptstyle y$}$}$};
\fill (x1t1) circle (2pt);
\fill (x1t2) circle (2pt);
\fill (x2t3) circle (2pt);
\fill (x2t4) circle (2pt);
\draw [directed]  (x2t3)--(x1t1) ;
\draw [directed] (x1t2) -- (x2t4);
\draw [dashed,thick] (x1t1) -- (x1t2);
\draw [dashed,thick] (x2t3) -- (x2t4);
\draw [enddirected]  (x1t1)--(-.5,0);
\draw [enddirected]  (x2t4)--(2,0.5);
\end{tikzpicture}}}}
\ee
The first descendant with the corresponding times is
\bea
{\parbox{2.3cm}{{\begin{tikzpicture}
\coordinate (x1t1) at  (0,0) ; 
\coordinate (x1t2) at  (0,.5) ; 
\coordinate (x2t3) at  (1.5,0) ; 
\coordinate (x2t4) at  (1.5,0.5) ; 
\node (x) at  (0,0)    {$\!\!\!\parbox{0mm}{$\raisebox{-3.5mm}[2.5mm][2.5mm]{$\scriptstyle x$}$}$};
\node (y) at  (1.5,0)    {$\!\!\!\parbox{0mm}{$\raisebox{-3.5mm}[2.5mm][2.5mm]{$\scriptstyle y$}$}$};
\node (t1) at  (-.25,0)    {$\!\!\!\parbox{0mm}{$\raisebox{-1mm}[0mm][0mm]{$\scriptstyle t_1$}$}$};
\node (t2) at  (-.25,0.5)    {$\!\!\!\parbox{0mm}{$\raisebox{-1mm}[0mm][0mm]{$\scriptstyle t_2$}$}$};
\node (t3) at  (1.75,-0.2)    {$\!\!\!\parbox{0mm}{$\raisebox{-1mm}[0mm][0mm]{$\scriptstyle t_3$}$}$};
\node (t4) at  (1.75,0.7)    {$\!\!\!\parbox{0mm}{$\raisebox{-1mm}[0mm][0mm]{$\scriptstyle t_4$}$}$};
\fill (x1t1) circle (2pt);
\fill (x1t2) circle (2pt);
\fill (x2t3) circle (2pt);
\fill (x2t4) circle (2pt);
\draw [directed] (x1t1) -- (x2t3);
\draw [directed] (x1t2) -- (x2t4);
\draw [dashed,thick] (x1t1) -- (x1t2);
\draw [dashed,thick] (x2t3) -- (x2t4);
\draw [enddirected]  (x2t3)--(2,0);
\draw [enddirected]  (x2t4)--(2,0.5);
\end{tikzpicture}}}} \nn\\
=
-\int_{t_1,t_2,x} R(x-y,t_4-t_2) R(x-y,t_3-t_1)  \nn\\
~~~~~~~~ \times\Delta_0\big(u(x,t_2)-u(x,t_1)\big)\Delta_0''\big(u(y,t_4)-u(y,t_3)\big) \nn\\
~~~~~~~~ \times \tilde u(y,t_3)\tilde u(y,t_4)\nn\\
\simeq -\int_k\int_{t_1<t_3,t_2<t_4} \rme^{-(k^2+m^2)(t_4-t_2)} \rme^{-(k^2+m^2)(t_3-t_1)}   \nn\\
~~~~~~~~ \times\Delta_0\big(u(y,t_4)-u(y,t_3)\big)\Delta_0''\big(u(y,t_4)-u(y,t_3)\big)\nn\\
~~~~~~~~ \times \tilde u(y,t_3)\tilde u(y,t_4)\nn \\
=-\int_k\frac{1}{(k^2+m^2)^2} \; \Delta_0\big(u(y,t_4)-u(y,t_3)\big) \nn\\
 ~~~~~~~~\times \Delta_0''\big(u(y,t_4)-u(y,t_3)\big)\tilde u(y,t_3)\tilde u(y,t_4)\punkt~~\qquad ~~
\eea
Some remarks are in order: This is a correction to $\Delta$, and we have not   written the integrations over $t_3$, $t_4$ and $y$. The derivatives of $\Delta$   come from the Wick contractions as in \Eq{45}. The global minus sign in the first line originates from the derivatives  acting once on the field at time $t_3$, and once at time $t_4$. Going to the second line, we have in the argument of $\Delta$ replaced fields at time $t_2$ by those at time $t_4$, and fields at time $t_1$ by those at time $t_3$; this is justified since the response function $R$ decays rapidly in time. In the argument of $\Delta$ we have also replaced  $x$ by $y$, as we did in the statics after arriving at \Eq{81}. The remaining two times $t_3$ and $t_4$ can be taken arbitrarily far apart, thus this diagram encodes a contribution to the effective disorder.

The second descendant   gives after the same steps 
\bea
{\parbox{2.5cm}{{\begin{tikzpicture}
\coordinate (x1t1) at  (0,0) ; 
\coordinate (x1t2) at  (0,.5) ; 
\coordinate (x2t3) at  (1.5,0) ; 
\coordinate (x2t4) at  (1.5,0.5) ; 
\node (x) at  (0,0)    {$\!\!\!\parbox{0mm}{$\raisebox{-3.5mm}[2.5mm][2.5mm]{$\scriptstyle x$}$}$};
\node (y) at  (1.5,0)    {$\!\!\!\parbox{0mm}{$\raisebox{-3.5mm}[2.5mm][2.5mm]{$\scriptstyle y$}$}$};
\node (t1) at  (-.25,-.2)    {$\!\!\!\parbox{0mm}{$\raisebox{-1mm}[0mm][0mm]{$\scriptstyle t_1$}$}$};
\node (t2) at  (-.25,0.5)    {$\!\!\!\parbox{0mm}{$\raisebox{-1mm}[0mm][0mm]{$\scriptstyle t_2$}$}$};
\node (t3) at  (1.75,0)    {$\!\!\!\parbox{0mm}{$\raisebox{-1mm}[0mm][0mm]{$\scriptstyle t_3$}$}$};
\node (t4) at  (1.75,0.7)    {$\!\!\!\parbox{0mm}{$\raisebox{-1mm}[0mm][0mm]{$\scriptstyle t_4$}$}$};
\fill (x1t1) circle (2pt);
\fill (x1t2) circle (2pt);
\fill (x2t3) circle (2pt);
\fill (x2t4) circle (2pt);
\draw [directed]  (x2t3)--(x1t1) ;
\draw [directed] (x1t2) -- (x2t4);
\draw [dashed,thick] (x1t1) -- (x1t2);
\draw [dashed,thick] (x2t3) -- (x2t4);
\draw [enddirected]  (x1t1)--(-.5,0);
\draw [enddirected]  (x2t4)--(2,0.5);
\end{tikzpicture}}}}\nn\\
 \simeq -\int_k\frac{1}{(k^2+m^2)^2} \;  \Delta_0'\big(u(y,t_4){-}u(y,t_3)\big) ^2 \,\nn\\
 \qquad \times \tilde u(y,t_3)\tilde u(y,t_4) .
\eea
We used that $\Delta'(u)$ is odd in $u$.
Together, these two diagrams give with $I_1$ defined in \Eq{I1}
\bea\label{7.19}
{\parbox{2.1cm}{{\begin{tikzpicture}
\coordinate (x1t1) at  (0,0) ; 
\coordinate (x1t2) at  (0,.5) ; 
\coordinate (x2t3) at  (1.5,0) ; 
\coordinate (x2t4) at  (1.5,0.5) ; 
\fill (x1t1) circle (2pt);
\fill (x1t2) circle (2pt);
\fill (x2t3) circle (2pt);
\fill (x2t4) circle (2pt);
\draw [directed] (x1t1) -- (x2t3);
\draw [directed] (x1t2) -- (x2t4);
\draw [dashed,thick] (x1t1) -- (x1t2);
\draw [dashed,thick] (x2t3) -- (x2t4);
\draw [enddirected]  (x2t3)--(2,0);
\draw [enddirected]  (x2t4)--(2,0.5);
\end{tikzpicture}}}} ~+~  {\parbox{2.5cm}{{\begin{tikzpicture}
\coordinate (x1t1) at  (0,0) ; 
\coordinate (x1t2) at  (0,.5) ; 
\coordinate (x2t3) at  (1.5,0) ; 
\coordinate (x2t4) at  (1.5,0.5) ; 
\fill (x1t1) circle (2pt);
\fill (x1t2) circle (2pt);
\fill (x2t3) circle (2pt);
\fill (x2t4) circle (2pt);
\draw [directed]  (x2t3)--(x1t1) ;
\draw [directed] (x1t2) -- (x2t4);
\draw [dashed,thick] (x1t1) -- (x1t2);
\draw [dashed,thick] (x2t3) -- (x2t4);
\draw [enddirected]  (x1t1)--(-.5,0);
\draw [enddirected]  (x2t4)--(2,0.5);
\end{tikzpicture}}}} \\
 \simeq -I_1\;  \Big[ \Delta_0\big(u(y,t_4)-u(y,t_3)\big)\Delta_0''\big(u(y,t_4)-u(y,t_3)\big) \nn\\
~~~~~~~~~~~~+ \Delta_0'\big(u(y,t_4)-u(y,t_3)\big)^2\Big] \tilde u(y,t_3)\tilde u(y,t_4). \nn
\eea
Taking care of the combinatorial factors, and the factors of $1/2$ in the action, 
we read off their contribution to the effective disorder $  \Delta(u)$, 
\bea
\delta_1   \Delta(u) = -  \left[ \Delta_0(u)\Delta_0''(u) + \Delta_0'(u)^2\right] I_1 \nn\\
=- \partial_u^2 \half  \Delta_0(u)^2 I_1.
\eea
This is the same contribution as given by the diagram in \Eq{80cis}, noting that $\Delta_0(u)=-R_0''(u)$, and using  the combinatorial factor $1/2$ reported in \Eq{351}.

To complete our analysis, consider the 
  second diagram; it also has two descendants, 
\bea
{\parbox{1.6cm}{{\begin{tikzpicture}
\coordinate (x1t1) at  (0,0) ; 
\coordinate (x1t2) at  (0,.5) ; 
\coordinate (x2t3) at  (1.5,0) ; 
\coordinate (x2t4) at  (1.5,0.5) ; 
\node (x) at  (0,0)    {$\!\!\!\parbox{0mm}{$\raisebox{-3.5mm}[2.5mm][2.5mm]{$\scriptstyle x$}$}$};
\node (y) at  (1.5,0)    {$\!\!\!\parbox{0mm}{$\raisebox{-3.5mm}[2.5mm][2.5mm]{$\scriptstyle y$}$}$};
\fill (x1t1) circle (2pt);
\fill (x1t2) circle (2pt);
\fill (x2t3) circle (2pt);
\fill (x2t4) circle (2pt);
\draw  (x1t1) -- (x2t4);
\draw  (x1t2) -- (x2t4);
\draw [dashed,thick] (x1t1) -- (x1t2);
\draw [dashed,thick] (x2t3) -- (x2t4);
\end{tikzpicture}}}}
~~\longrightarrow~~
{\parbox{2.1cm}{{\begin{tikzpicture}
\coordinate (x1t1) at  (0,0) ; 
\coordinate (x1t2) at  (0,.5) ; 
\coordinate (x2t3) at  (1.5,0) ; 
\coordinate (x2t4) at  (1.5,0.5) ; 
\node (x) at  (0,0)    {$\!\!\!\parbox{0mm}{$\raisebox{-3.5mm}[2.5mm][2.5mm]{$\scriptstyle x$}$}$};
\node (y) at  (1.5,0)    {$\!\!\!\parbox{0mm}{$\raisebox{-3.5mm}[2.5mm][2.5mm]{$\scriptstyle y$}$}$};
\fill (x1t1) circle (2pt);
\fill (x1t2) circle (2pt);
\fill (x2t3) circle (2pt);
\fill (x2t4) circle (2pt);
\draw [directed] (x1t1) -- (x2t4);
\draw [directed] (x1t2) -- (x2t4);
\draw [dashed,thick] (x1t1) -- (x1t2);
\draw [dashed,thick] (x2t3) -- (x2t4);
\draw [enddirected]  (x2t3)--(2,0);
\draw [enddirected]  (x2t4)--(2,0.5);
\end{tikzpicture}}}} ~+~  {\parbox{2.5cm}{{\begin{tikzpicture}
\coordinate (x1t1) at  (0,0) ; 
\coordinate (x1t2) at  (0,.5) ; 
\coordinate (x2t3) at  (1.5,0) ; 
\coordinate (x2t4) at  (1.5,0.5) ; 
\node (x) at  (0,0)    {$\!\!\!\parbox{0mm}{$\raisebox{-3.5mm}[2.5mm][2.5mm]{$\scriptstyle x$}$}$};
\node (y) at  (1.5,0)    {$\!\!\!\parbox{0mm}{$\raisebox{-3.5mm}[2.5mm][2.5mm]{$\scriptstyle y$}$}$};
\fill (x1t1) circle (2pt);
\fill (x1t2) circle (2pt);
\fill (x2t3) circle (2pt);
\fill (x2t4) circle (2pt);
\draw [directed]  (x2t4)--(x1t1) ;
\draw [directed] (x1t2) -- (x2t4);
\draw [dashed,thick] (x1t1) -- (x1t2);
\draw [dashed,thick] (x2t3) -- (x2t4);
\draw [enddirected]  (x1t1)--(-.5,0);
\draw [enddirected]  (x2t3)--(2,0.);
\end{tikzpicture}}}}\nn~.\\
\eea
After time-integration this yields
\bea
{\parbox{2.3cm}{{\begin{tikzpicture}
\coordinate (x1t1) at  (0,0) ; 
\coordinate (x1t2) at  (0,.5) ; 
\coordinate (x2t3) at  (1.5,0) ; 
\coordinate (x2t4) at  (1.5,0.5) ; 
\node (x) at  (0,0)    {$\!\!\!\parbox{0mm}{$\raisebox{-3.5mm}[2.5mm][2.5mm]{$\scriptstyle x$}$}$};
\node (y) at  (1.5,0)    {$\!\!\!\parbox{0mm}{$\raisebox{-3.5mm}[2.5mm][2.5mm]{$\scriptstyle y$}$}$};
\node (t1) at  (-.25,0)    {$\!\!\!\parbox{0mm}{$\raisebox{-1mm}[0mm][0mm]{$\scriptstyle t_1$}$}$};
\node (t2) at  (-.25,0.5)    {$\!\!\!\parbox{0mm}{$\raisebox{-1mm}[0mm][0mm]{$\scriptstyle t_2$}$}$};
\node (t3) at  (1.75,-0.2)    {$\!\!\!\parbox{0mm}{$\raisebox{-1mm}[0mm][0mm]{$\scriptstyle t_3$}$}$};
\node (t4) at  (1.75,0.7)    {$\!\!\!\parbox{0mm}{$\raisebox{-1mm}[0mm][0mm]{$\scriptstyle t_4$}$}$};
\fill (x1t1) circle (2pt);
\fill (x1t2) circle (2pt);
\fill (x2t3) circle (2pt);
\fill (x2t4) circle (2pt);
\draw [directed] (x1t1) -- (x2t4);
\draw [directed] (x1t2) -- (x2t4);
\draw [dashed,thick] (x1t1) -- (x1t2);
\draw [dashed,thick] (x2t3) -- (x2t4);
\draw [enddirected]  (x2t3)--(2,0);
\draw [enddirected]  (x2t4)--(2,0.5);
\end{tikzpicture}}}}\nn\\
\simeq \int_k\frac{1}{(k^2+m^2)^2} \; \Delta_0(0)\Delta_0''\big(u(y,t_4)-u(y,t_3)\big).
\eea
\bea
{\parbox{2.5cm}{{\begin{tikzpicture}
\coordinate (x1t1) at  (0,0) ; 
\coordinate (x1t2) at  (0,.5) ; 
\coordinate (x2t3) at  (1.5,0) ; 
\coordinate (x2t4) at  (1.5,0.5) ; 
\node (x) at  (0,0)    {$\!\!\!\parbox{0mm}{$\raisebox{-3.5mm}[2.5mm][2.5mm]{$\scriptstyle x$}$}$};
\node (y) at  (1.5,0)    {$\!\!\!\parbox{0mm}{$\raisebox{-3.5mm}[2.5mm][2.5mm]{$\scriptstyle y$}$}$};
\node (t1) at  (-.25,-.2)    {$\!\!\!\parbox{0mm}{$\raisebox{-1mm}[0mm][0mm]{$\scriptstyle t_1$}$}$};
\node (t2) at  (-.25,0.5)    {$\!\!\!\parbox{0mm}{$\raisebox{-1mm}[0mm][0mm]{$\scriptstyle t_2$}$}$};
\node (t3) at  (1.75,-0.2)    {$\!\!\!\parbox{0mm}{$\raisebox{-1mm}[0mm][0mm]{$\scriptstyle t_3$}$}$};
\node (t4) at  (1.75,0.5)    {$\!\!\!\parbox{0mm}{$\raisebox{-1mm}[0mm][0mm]{$\scriptstyle t_4$}$}$};
\fill (x1t1) circle (2pt);
\fill (x1t2) circle (2pt);
\fill (x2t3) circle (2pt);
\fill (x2t4) circle (2pt);
\draw [directed]  (x2t4)--(x1t1) ;
\draw [directed] (x1t2) -- (x2t4);
\draw [dashed,thick] (x1t1) -- (x1t2);
\draw [dashed,thick] (x2t3) -- (x2t4);
\draw [enddirected]  (x1t1)--(-.5,0);
\draw [enddirected]  (x2t3)--(2,0.);
\end{tikzpicture}}}}\nn\\
 \simeq \int_k\frac{1}{(k^2+m^2)^2} \; \Delta_0'(0^+)\Delta_0'\big(u(y,t_4)-u(y,t_3)\big).
\eea
The last diagram contains a first factor of $\Delta'(0^+)$; the definite sign results from  the causality of the response functions 
ensuring $t_2<t_1$. It 
is asymmetric under ex\-change of $t_3$ and $t_4$, thus vanishes after integrating  over these times.
(B.t.w., inserted into a 2-loop diagram, it is this diagram which is responsible for the differences seen there, especially for the 2-loop contribution to $\zeta$.)
Together, they give a second contribution to the effective disorder
\be\label{7.24}
\delta_2   \Delta(u) =  \Delta_0(0)\Delta_0''(u) \; I_1= \partial_u^2 \left[ \Delta_0(0)\Delta_0(u)\right] I_1\punkt
\ee
This is the same contribution as given by \Eq{81}.

The last diagram we   drew for the equilibrium was given in \Eq{339}. Its descendant reads
\be\label{7.25}
{\parbox{1.65cm}{{\begin{tikzpicture}
\coordinate (x1t1) at  (0,0) ; 
\coordinate (x1t2) at  (0,.5) ; 
\coordinate (x2t3) at  (1.5,0) ; 
\coordinate (x2t4) at  (1.5,0.5) ; 
\fill (x1t1) circle (2pt);
\fill (x1t2) circle (2pt);
\fill (x2t3) circle (2pt);
\fill (x2t4) circle (2pt);
\draw  (x2t4) arc(60:120:1.5);
\draw  (x2t4) arc(-60:-120:1.5);
\draw [dashed,thick] (x1t1) -- (x1t2);
\draw [dashed,thick] (x2t3) -- (x2t4);
\end{tikzpicture}}}} 
~~~\longrightarrow~~~{\parbox{2.2cm}{{\begin{tikzpicture}
\coordinate (x1t1) at  (0,0) ; 
\coordinate (x1t2) at  (0,.5) ; 
\coordinate (x2t3) at  (1.5,0) ; 
\coordinate (x2t4) at  (1.5,0.5) ; 
\fill (x1t1) circle (2pt);
\fill (x1t2) circle (2pt);
\fill (x2t3) circle (2pt);
\fill (x2t4) circle (2pt);
\draw [directed] (x2t4) arc(60:120:1.5);
\draw [directed] (x1t2) arc(-120:-60:1.5);
\draw [dashed,thick] (x1t1) -- (x1t2);
\draw [dashed,thick] (x2t3) -- (x2t4);
\draw [enddirected]  (x1t1)--(-.5,0);
\draw [enddirected]  (x2t3)--(2,0.);
\end{tikzpicture}}}}\qquad .
\ee
While the static diagram on the l.h.s.\ does not contribute to the effective disorder since it is a 3-replica term (three independent sums over replicas), the dynamic diagram on the r.h.s.\ does not contribute due to the acausal loop, as it does not allow for any time integration, thus vanishes.

For completeness, we write the   effective disorder-force correlator at 1-loop order,
\be
      \Delta(u) = \Delta_0(u) -\partial_u^2 \left[ \half  \Delta_0(u)^2 - \Delta_0(0)\Delta_0(u) \right] I_1.
\ee
This result is the same as when 
applying $-\partial_u^2$ to \Eq{351}. 
We thus recover the same flow equation for the renormalized dimensionless force-force correlator as given in \Eq{flow-Delta} and first derived in \cite{NattermannStepanowTangLeschhorn1992,NarayanDSFisher1992a,NarayanDSFisher1992b,LeschhornNattermannStepanowTang1997}
\bea\label{beta-dep-1loop}
\highlight{\partial_\ell \tilde \Delta({ u}) =  (\epsilon -2 \zeta) \tilde\Delta({ u}) + \zeta
{  u} \tilde\Delta'({ u}) \nn\\
\qquad~~~~~~~ - \partial_u^2 \,\half \big[ \tilde\Delta ({ u})^{2} - \tilde\Delta(0) \big]^2\punkt} 
\eea
While this might not be surprising on a formal level, it is   {\em very surprising} on a physical level: The effective disorder \eq{351} is for the minimum energy state, while the derivation given above is for a state  at depinning. We will see in the next section \ref{s:dyn-2loop} that there are indeed corrections at 2-loop order which account for this difference, and which are important to reconcile the physically observed differences in exponents and other observables with the theoretical prediction. 

Before going there, let us complete our analysis with two additional contributions not present in the statics, and which we will interpret as the critical force at depinning, and a renormalization of friction, leading to a non-trivial dynamical exponent $z$,  as defined in \Eq{exponent-z}.   The diagram in question is
\bea
{\parbox{1.1cm}{{\begin{tikzpicture}
\coordinate (x1t1) at  (0,0) ; 
\coordinate (x1t2) at  (0,1) ; 
\node (x) at  (0,0)    {$\!\!\!\parbox{0mm}{$\raisebox{-3.5mm}[2.5mm][2.5mm]{$\scriptstyle x$}$}$};
\node (t1) at  (-.25,0)    {$\!\!\!\parbox{0mm}{$\raisebox{-1mm}[0mm][0mm]{$\scriptstyle t_1$}$}$};
\node (t2) at  (-.25,1.2)    {$\!\!\!\parbox{0mm}{$\raisebox{-1mm}[0mm][0mm]{$\scriptstyle t_2$}$}$};
\fill (x1t1) circle (2pt);
\fill (x1t2) circle (2pt);
\draw [directed] (x1t1) arc(-90:90:0.5);
\draw [dashed,thick] (x1t1) -- (x1t2);
\draw [enddirected]  (x1t2)--(-.5,1);
\end{tikzpicture}}}} \nn\\
= \tilde u(x,t_2)\,\int_{t_1,k}\Delta_0'\big(u(x,t_2)-u(x,t_1)\big)\rme^{-(t_2-t_1)(k^2+m^2)}\nn\\
\hp{= \tilde u(x,t_2)\,\int_{t_1,k}}  \times \Theta(t_1<t_2)\nn\\
\simeq \tilde u(x,t_2)\,\int_{t_1,k}\Big[\Delta_0'(0^+)+\Delta_0''(0^+) (t_2{-}t_1)\dot u(x,t_2)\,{+} ...\Big]\nn\\
\hp{\simeq \tilde u(x,t_2)\,\int_{t_1,k}\Big[}\times\rme^{-(t_2-t_1)(k^2+m^2)}\Theta(t_1<t_2) \nn\\
= \tilde u(x,t_2)\,\int_{k}\frac{ \Delta_0'(0^+) }{k^2+m^2}+\frac{\Delta_0''(0^+)}{(k^2+m^2)^2}  \dot u(x,t_2)+ ...
\label{7.32}
\eea
The first term  corresponds to a constant driving force $f$ in \Eq{eq-motion-f}, and can be interpreted as the threshold force below which the manifold will not move. In terms of the renormalized disorder, it reads
\bea\label{fc}
\highlight{f_{\rm c} = - \Delta'(0^+)  I_{\rm TP}\komma  \\
 I_{\rm TP}= \ITP = \int_{k}\frac{ 1}{k^2+m^2}\punkt}
\eea 
Its value is non-universal, but gives us a pretty good idea how strong we have to drive. In the driving protocol with a parabola centered at $w$ as given in \Eq{eq-motion-w}, it gives us the size of the {\em hysteresis loop}, illustrated in Fig.~\ref{figgraphicalA}, 
\be
m^2\left[ \overline { u_w-w}^{\rm forward} - \overline { u_w-w}^{\rm backward}\right] = 2 f_{\rm c}\punkt
\ee
Let us now turn to the second term in \Eq{7.32}. Restoring the friction coefficient $\eta$ in front of $\partial_t u(x,t)$ in the equation of motion \eq{eq-motion-w} yields
\be
\eta_{\rm eff} = 1 - \Delta_0''(0^+)I_1 + ...
\ee
The dynamical exponent $z$, expressed in terms of the renormalized disorder,  is then obtained as 
\be
\highlight{z= 2   - m \partial_m \ln \eta_{\rm eff}  = 2  - \tilde \Delta''(0^+)+...}
\ee
Taking one derivative of \Eq{RF-FP}, or equivalently of \Eq{beta-dep-1loop} at the fixed point $\partial_\ell \tilde \Delta(u)=0$,  and evaluating it in the limit of $u\to 0$ allows us to conclude that
\be\label{Delta''(0)-FP}
\tilde \Delta''(0^+) = \frac{\epsilon -\zeta }{3}\punkt
\ee
This depends on the universality class, 
\be
z= 2- \frac{\epsilon -\zeta }{3} + ...= \left\{ \begin{array}{cl}
\displaystyle
2-\frac{\epsilon}3+ ...& \mbox{RP disorder}\\
\displaystyle  \rule{0mm}{4.2ex}2-\frac{2\epsilon}9+ ... &\mbox{RF disorder}
\end{array}\right.
\ee
We do not give a value of  $z$ for the RB fixed point, as the latter is unstable under RG, as we will see in the next section. 

\begin{figure}
\setlength{\unitlength}{.34mm}
\centerline{{\begin{picture}(235,140)
\put(0,0){\fig{0.45\textwidth}{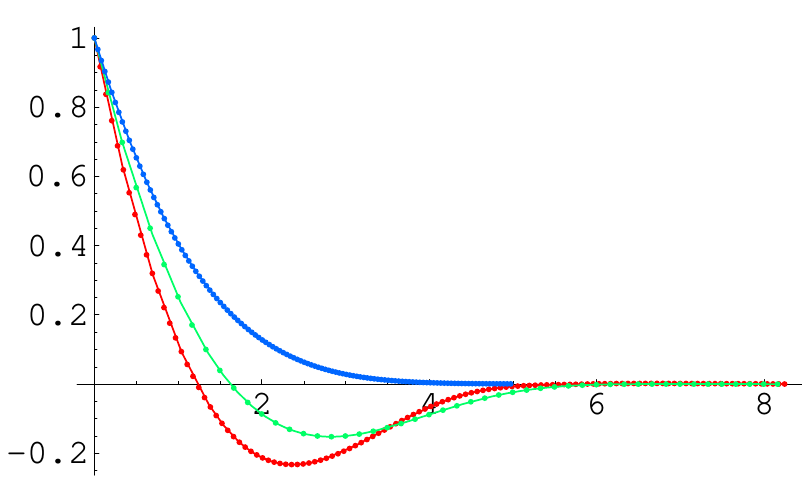}}
\put(-13,130){ $ \Delta (u) $}
\put(235,25){ $u$} 
\put(35,6){\color{red} $m^{2}=1$}
\put(130,15){\color{green} $m^{2}=0.5$}
\put(92,45){\color{blue} $m^{2}=0.003$}
\end{picture}}}
\caption{{Doing RG in a simulation: Crossover from RB disorder to RF for a driven particle   \protect\cite{MiddletonLeDoussalWiese2006}.}}
\label{f:RF-dep-toy}
\end{figure}
\subsection{Depinning beyond leading order}\label{s:dyn-2loop}

Renormalization at the depinning transition was first treated at 1-loop order by Natterman et
al.~\cite{NattermannStepanowTangLeschhorn1992}, soon followed by Narayan and
Fisher \cite{NarayanDSFisher1993a}. As we have seen, the 1-loop flow equations are
identical to those of the statics. This is surprising, since
   equilibrium and   depinning are quite
different phenomena. There was even a claim by \cite{NarayanDSFisher1993a},
that the roughness exponent in the random field universality class is
  $\zeta =\epsilon /3$ also at depinning. After a long debate among numerical physicists, the issue is
now resolved: The roughness is significantly larger, and reads e.g.\
for the driven polymer $\zeta =1.25\pm 0.005$ \cite{RossoKrauth2001b,FerreroBustingorryKolton2012}, and possibly exactly $\zeta=\frac54$ \cite{GrassbergerDharMohanty2016}; this should be contrasted to $\zeta=1$  at equilibrium, see \Eq{zeta-RF-1loop}. Clearly, a 2-loop analysis is necessary to resolve these
issues. The latter was performed in Refs.~\cite{ChauveLeDoussalWiese2000a,LeDoussalWieseChauve2002}. 
\begin{widefigure}
\setlength{\unitlength}{.37mm}
\mbox{\begin{picture}(463,165)
\put(0,0){\fig{0.45\textwidth}{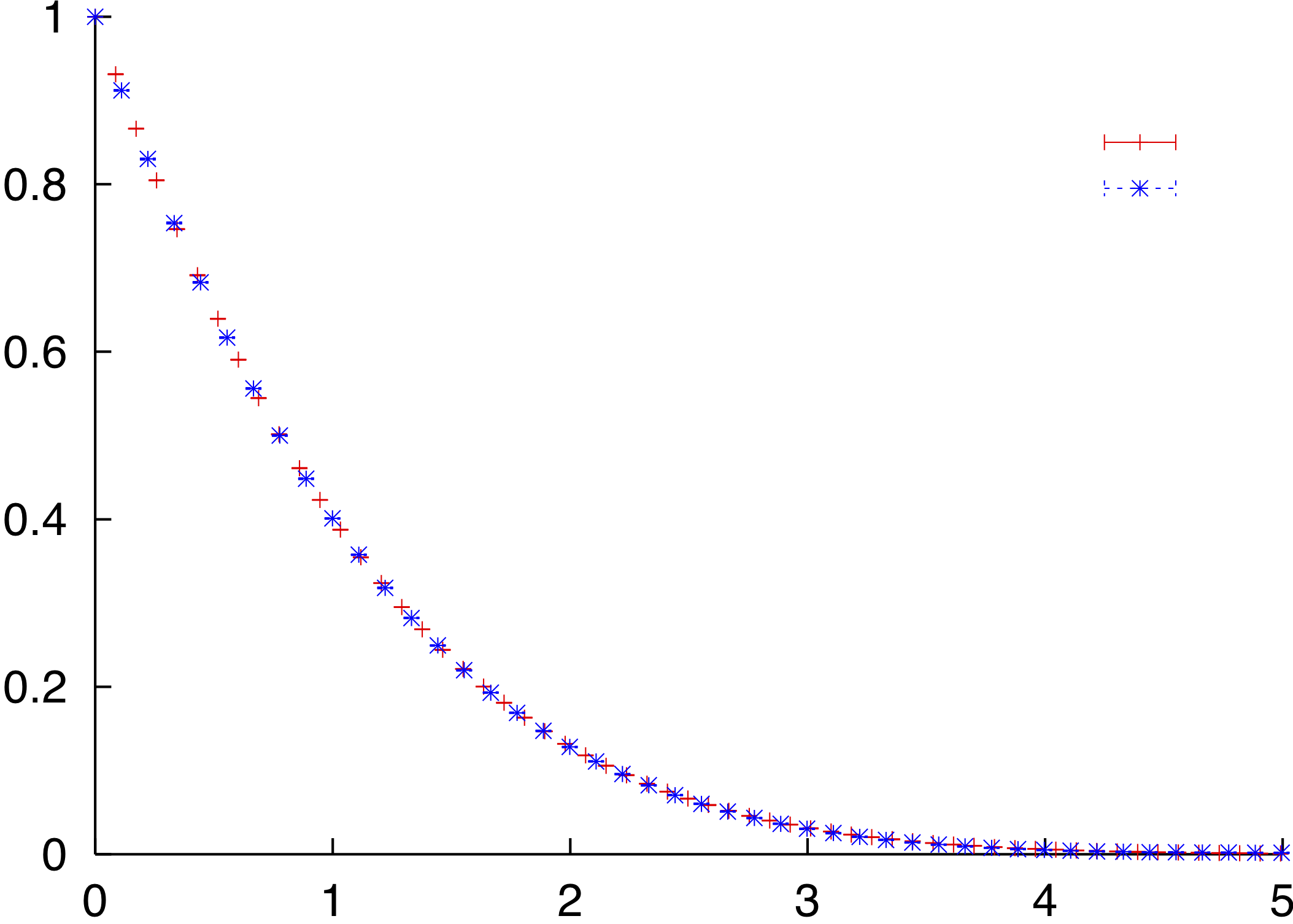}}
\put(0,155){ $ \Delta (w) $}
\put(210,10){ $w$} 
\put(85,125){{\small \rot RF, $m=0.071$, $L=512$}}
\put(85,115){{\small \blue  RB, $m=0.071$, $L=512$}}
\put(245,0){\fig{0.45\textwidth}{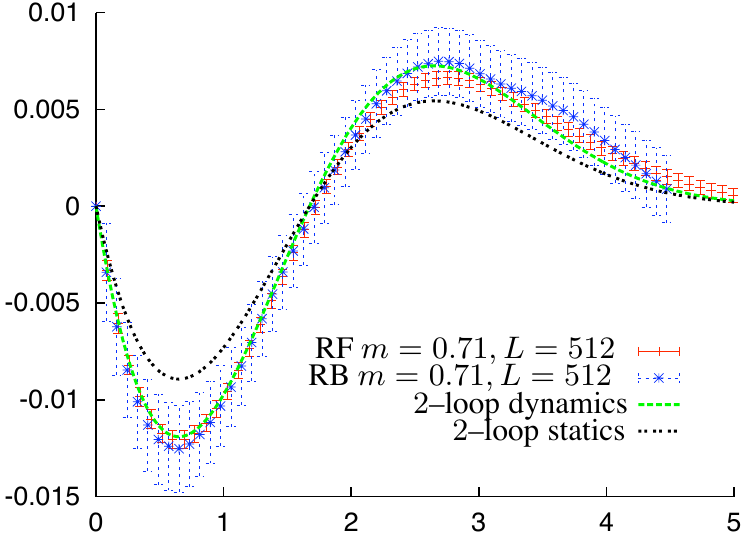}}
\put(245,155){ $ \delta \Delta (w) $}
\put(455,10){$ w $}
\end{picture}}
\caption{{Left: The   fixed point  $\Delta(w)$  for the force-force correlations in $d=1$, rescaled s.t. $\Delta(0)=1$, and  $\int_w \Delta(w)=1$, starting both from   RB and RF initial condition \protect\cite{RossoLeDoussalWiese2006a}. Right: Residual error   $\delta \Delta(w)$   after subtracting the 1-loop correction. The measured difference is consistent with the depinning fixed point, but not the static one.}}
\label{f:RF-dep}
\end{widefigure}
At the depinning transition, the 2-loop FRG flow equation reads
\cite{ChauveLeDoussalWiese2000a,LeDoussalWieseChauve2002}
\begin{eqnarray}\label{two-loop-FRG-dyn}
\partial_{\ell}\tilde  \Delta (u) \nn\\
= (\epsilon -2 \zeta)\tilde \Delta (u)+\zeta u \tilde \Delta' (u)
  -\frac{1}{2} \partial_u^2 \left[ \tilde \Delta(u)-\tilde \Delta(0)\right]^2 \nonumber \\
 +\frac{1}{2}\,\partial_u^2 \left\{ \left[\tilde \Delta(u)-\tilde \Delta (0) \right] \tilde\Delta' (u)^{2}
~{\mbox{\rot\bf +}}~ \tilde\Delta' (0^{+})^{2}\tilde \Delta
(u) \right\}\punkt
\end{eqnarray}
Compared to the FRG-equation \eq{3loopRGR} for the statics,  
 the only change is in the last sign on the second
line of \Eq{two-loop-FRG-dyn}, given in red (bold). This ``small change'' has important consequences for
the physics.
 First of all, the roughness exponent $\zeta$ for the
random-field universality class changes:
Integrating \Eq{two-loop-FRG-dyn} the last term yields a boundary term at $u=0$, and due to the different sign it no longer cancels with the preceding one, resulting in 
\be\label{integrated-two-loop-FRG-dyn}
0 =    (\epsilon - 3 \zeta) \int_0^\infty  
\tilde  \Delta(u) \,\rmd u -  \tilde \Delta'(0^+)^3+ \ca O(\epsilon^3).
\ee
Inserting the 1-loop fixed point \eq{Delta=y...}-\eq{RF-FP-y} leads to\footnote{For details see \cite{LeDoussalWieseChauve2002}, section IV.A.}
\begin{equation}\label{zetaRFdyn}
\zeta^{\rm dep}_{\rm RF} =\frac{\epsilon}{3} (1 +0.143317 \epsilon +\ldots )\punkt
\end{equation}
Other critical exponents mentioned above can also be
calculated. The dynamical exponent $z$   reads
\cite{ChauveLeDoussalWiese2000a,LeDoussalWieseChauve2002}
\begin{equation}\label{zdyn}
\zeta^{\rm dep}_{\rm RF}=2-\frac{2}{9}\epsilon -0.04321\epsilon^{2} + \dotsb
\end{equation}
The remaining exponents are related via the scaling relations (\ref{beta-rel}) and (\ref{nu-rel}).
That the method works well   quantitatively can be inferred from
table \ref{dyn-data}.

\begin{table}[b]
{\begin{tabular}{|c|c|c|c|c|c|}
\hline
 & $d$ & $\epsilon$ & $\epsilon^2$ & estimate & simulation~~~\\
\hline
\hline
        & $3$ & 0.33 & 0.38 & 0.38$\pm$0.02 & 0.355$\pm$0.01 \cite{RossoHartmannKrauth2002} \\
\hline
$\zeta$ & $2$ & 0.67 & 0.86 & 0.82$\pm$0.1 & 0.753$\pm$0.002  \cite{RossoHartmannKrauth2002}$\!$ \\
\hline
        & $1$ & 1.00 & 1.43 & 1.2$\pm$0.2 & 5/4 \cite{GrassbergerDharMohanty2016} \\
\hline
\hline
        & $3$ & 1.78  & 1.73  &  1.74$ \pm$0.02  & 1.75$\pm$0.15 \cite{NattermannStepanowTangLeschhorn1992}\\
\hline
$z$ & $2$ & 1.56 & 1.38  &  1.45$ \pm$0.15   &  1.56 $\pm$0.06 \\
\hline
        & $1$ & 1.33 & 0.94  & 1.35$ \pm$0.2    & 10/7  \cite{GrassbergerDharMohanty2016}\\
\hline
\hline
        & $3$ & 0.89 & 0.85 & 0.84$\pm$0.01 & 0.84$\pm$0.02 \cite{NattermannStepanowTangLeschhorn1992} \\
\hline
$\beta$ & $2$ & 0.78 & 0.62 & 0.53$\pm$0.15  & 0.64$\pm$0.02
\\
\hline
        & $1$ & 0.67 & 0.31 & 0.2$\pm$0.2 & 5/21 \cite{GrassbergerDharMohanty2016}  \\
\hline
\hline
        & $3$ & 0.58 & 0.61 &   0.62$\pm$0.01  & \\
\hline
$\nu$ & $2$ & 0.67 & 0.77 &  0.85$\pm$0.1   & 0.77$\pm$0.04 \cite{RotersHuchtLubeckNowakUsadel1999} \\
\hline
        & $1$ & 0.75 & 0.98 &   1.25$\pm$0.3  & 4/3  \cite{GrassbergerDharMohanty2016}\\
\hline

\end{tabular}}\\
\caption{Critical exponents at the depinning transition  for short-ranged elasticity  ($\alpha=2$). 1-loop and 2-loop results compared to estimates  based on three Pad\'e approximants, scaling relations and common sense.}
\label{dyn-data}
\end{table}

The random-bond fixed point is unstable and  
renormalizes to the random-field universality class. This might  physically be
expected: Since the manifold only moves   forward, each time it advances it experiences a
new disorder configuration, and it has no way to ``know'' whether this
disorder is derived from a potential or not. 
This can   be seen from the integrated FRG equation \eq{integrated-two-loop-FRG-dyn}:
According to \Eq{rel-RB}, an RB fixed point is characterized by a vanishing of the integral in \Eq{integrated-two-loop-FRG-dyn}, but this does not solve \Eq{integrated-two-loop-FRG-dyn}.
The instability of the RB fixed point can already be seen for a toy model with a single particle, measuring the renormalized disorder correlator at a scale $\ell=1/m$ set by the confining potential, see figure \ref{f:RF-dep-toy}. 
Generalizing the arguments of section \ref{measurecusp} one shows \cite{LeDoussalWiese2006a} that \Eq{defDe} remains valid in the limit of $w=vt$, $v \to 0$. It was      confirmed numerically for  a string that both RB and RF disorder flow to the RF fixed point \cite{RossoLeDoussalWiese2006a}, and that this fixed point is   close to the analytic solution of \Eq{two-loop-FRG-dyn}, see figure \ref{f:RF-dep}.

The non-potentiality of the depinning fixed point
is also observed in the random periodic universality class,
 relevant   for charge-density waves. The fixed point
for a periodic disorder of period one reads (remember $\tilde \Delta (u)=-\tilde R''
(u)$)
\begin{equation}\label{rand-per-fp}
\tilde\Delta  (u) =\frac{\epsilon}{36}+\frac{\epsilon^{2}}{108}
-\left(\frac{\epsilon}{6}+\frac{\epsilon^{2}}{9} \right) u (1{-}u)+ \ca O(\epsilon^3).
\end{equation}
Integrating over a period, we find 
\begin{equation}\label{period}
\int_{0}^{1}\rmd u \, \tilde \Delta  (u) \equiv \int_{0}^{1}\rmd u\
\overline{\tilde F (u) \tilde F (u')}= -\frac{\epsilon^{2}}{108}\punkt
\end{equation}
In   equilibrium, this correlator  vanishes since
potentiality requires $\int_0^{1}\rmd u\,\tilde  F (u)\equiv 0$. Here, there
are non-trivial contributions at 2-loop order,  $\ca O(\epsilon^{2})$,
violating this condition and rendering the system non-potential. 

If an additional constant term $\tilde \Delta_0$ cannot be excluded as is the case in equilibrium, then according to \Eq{two-loop-FRG-dyn}  it flows as 
\be
\partial_{\ell}\tilde \Delta_0 
= (\epsilon -2 \zeta) \tilde\Delta_0.
\ee 
It acts as a {\em Larkin term} leading to a  roughness exponent  \cite{NarayanDSFisher1992a,LeDoussalWieseChauve2002,KasparMungan2013}
\be
\zeta^{\rm CDW}_{\rm obs} =\zeta_{\rm Larkin} =\frac{4-d}2.
\ee
For the dynamic exponent $z$,  one can go     further \cite{WieseFedorenko2018,WieseFedorenko2019,KompanietsWiese2019,ShapiraWiese2020}, using the equivalence to $\phi^4$-theory discussed in section \ref{s:Loop-erased random walks, and other models equivalent to CDWs}, 
\begin{eqnarray} \nn
z &=  2-\frac{\epsilon }{3}- \frac{\epsilon^2}{9}
+ \bigg[\frac{2 \zeta (3)}{9}-\frac{1}{18}\bigg]\epsilon^3 \nn\\
&- \bigg[\frac{70 \zeta (5)}{81} -\frac{\zeta(4)}{6} -\frac{17 \zeta (3) }{162}
   +\frac{7}{324} \bigg]\epsilon^4 \nn \\
&- \bigg[ \frac{541 \zeta (3)^2}{162} +\frac{37 \zeta (3)}{36}+\frac{29 \zeta (4)}{648}+
  \frac{703
   \zeta (5)}{243}\nn\\
  &  ~\quad +\frac{175 \zeta (6)}{162}-\frac{833 \zeta (7)}{216}+\frac{17}{1944}\bigg]\epsilon^5\nn\\
  & -11.7939\epsilon^6+ {\cal O}(\epsilon^7).  \label{eq:On-24} \ \ \ \ \ \ \ 
\end{eqnarray}

\subsection{Stability of the depinning fixed points} 
The stability analysis at depinning is done as in section \ref{s:Stability of the fixed point} for the equilibrium. 
\paragraph{RP fixed point.}
The RP fixed point at depinning is stable perturbatively,  
(appendix I of   \cite{LeDoussalWieseChauve2002}). The    leading three modes are 
\numparts
\bea
\omega_{-1} = -\epsilon \komma\quad z_{-1}(u) =1. \\
\omega_1 =\epsilon+\frac73 \epsilon^2 +\ca O(\epsilon^3)\komma \\
  z_1(u) = 1-(6+4\epsilon)u(1-u). \nn\\
\omega_2 = 2 \epsilon +4 \epsilon^2 + \ca O(\epsilon^3)\komma\\
z_2(u) = 1-(15+20 \epsilon)u(1{-}u) + (45 + 85 \epsilon) [u(1{-}u)]^2, \nn\\
\omega_3 = \frac{25}3\epsilon + \frac{140}{9} \epsilon^2  + \ca O(\epsilon^3). \eea
\endnumparts
\paragraph{RF fixed point.}\numparts
\bea
\omega =  \epsilon + 0.0186 \epsilon^2 + \ca O(\epsilon^3) .
\eea
The fixed-point function is 
\bea
z(u,\epsilon) = \epsilon z_1(u) + \epsilon^2 z_2(u) +   \ca O(\epsilon^3), \\
z_1(u) = \zeta u \Delta'(u) + (\epsilon- 2 \zeta) \Delta(u) \Big|_{\epsilon=1}.
\eea\endnumparts
While the first-order term can rather instructively  be expressed in terms of the fixed point $\Delta(u)$ itself, the higher-order terms are more complicated (section 6.5 of \cite{HusemannWiese2017}).

 \subsection{Non-perturbative FRG} The fixed points discussed above are also present in the non-perturbative functional renormalization group (NP-FRG) approach \cite{BalogTarjusTissier2019}, leading to slightly varying numerical predictions in the values of the critical exponents. 
 The result of    NP-FRG  for the RF class is  $z = 1.69$ ($d=3$), $z= 1.33$ ($d=2$), and $z=0.97$ ($d=1$). 
For the roughness at depinning this yields $\zeta = 0.37$ ($d=3$), $\zeta = 0.76$ ($d=2$), and $\zeta= 1.15$ ($d=1$).

\subsection{Behavior at the upper critical dimension}
\label{s:Behavior at the upper critical dimension}

\cite{FedorenkoStepanow2002,LeDoussalWiese2003a} consider depinning at the upper critical dimension. 
To derive this,  note that the integral $I_1$ defined in \Eq{I1} has a well-defined limit for $\epsilon\to 0$, if one introduces as in \Eq{2.8} an UV-cutoff $\Lambda\sim 1/a$, 
\bea\label{I1-bis}
\highlight{ I_{1}:=\Ione=  \int^{\Lambda} \frac{\rmd ^d k}{(2\pi)^d}\frac{1} {(
k^{2}+m^{2})^{2}} \nn\\
~~~~~= \frac {m^{-\epsilon}-\Lambda^{-\epsilon}}\epsilon  \frac{2\Gamma(1+\frac\epsilon2)}{(4 \pi)^{{d/2}}}}
\to \frac {\ln ( \Lambda/m)} {8 \pi^2} .
\eea
This suggests as scale for the RG flow 
\be
  \ell: = \ln( \Lambda/m) .
\ee
Let us make in generalization of \Eq{Rtillde-def} the ansatz
\bea
  \Delta(u) = {8\pi^2} \ell^{2 \zeta_1-1}\tilde \Delta_\ell(u \ell^{-\zeta_1}) ,\\
\zeta = \zeta_1 \epsilon + \zeta_2 \epsilon ^2 + ... 
\eea
Then $\tilde \Delta_\ell(u)$ satisfies the flow equation  \cite{FedorenkoStepanow2002,LeDoussalWiese2003a}
\bea
\partial_\ell \tilde \Delta_\ell(u) = (1- 2 \zeta_1) \tilde \Delta_\ell(u) + \zeta_1 u \tilde \Delta_\ell'(u)\nn\\
 - \half \partial_u^2 \left[ \tilde \Delta_\ell(u)-\tilde \Delta_\ell(0) \right]^2 + \sum_{n>1} \ell^{1-n} \beta_n(u),
 \eea
 where $ \beta_n(u)$ are the $n$-loop contributions to the $\beta$-function. 
As a consequence, 
\numparts
\bea\label{224}
\overline{\tilde u_q \tilde u_{-q}}\Big|_{q\ll m} \simeq   \frac {  \Delta(0)}{m^4} \left[ 1 + \ca O(\ell^{-1}) \right] \nn\\ 
\simeq \frac{ 8 \pi^2 \ln(\Lambda/m)^{2 \zeta_1-1} }{ m^4}+ ...~.
\eea
This formula is     valid  both in equilibrium and at depinning. 
For RF disorder,     $\zeta_1= 1/3$, leading to an additional factor of $\ln( \Lambda/m)^{-1/3}$ in the 2-point function \eq{224} as compared to naive expectations. 
In position space this reduces the expected $\ln x$ behavior to 
\be\nn
\overline{[u(x)-u(0)]^2} \sim ( \ln   x)^{2/3}.
\ee
\endnumparts
Thus mean field is   invalid at the upper critical dimension.

\subsection{Extreme-value statistics: The Discretized Particle Model (DPM)}
\label{s:DPM}
\begin{widefigure}
\centerline{\fig{9cm}{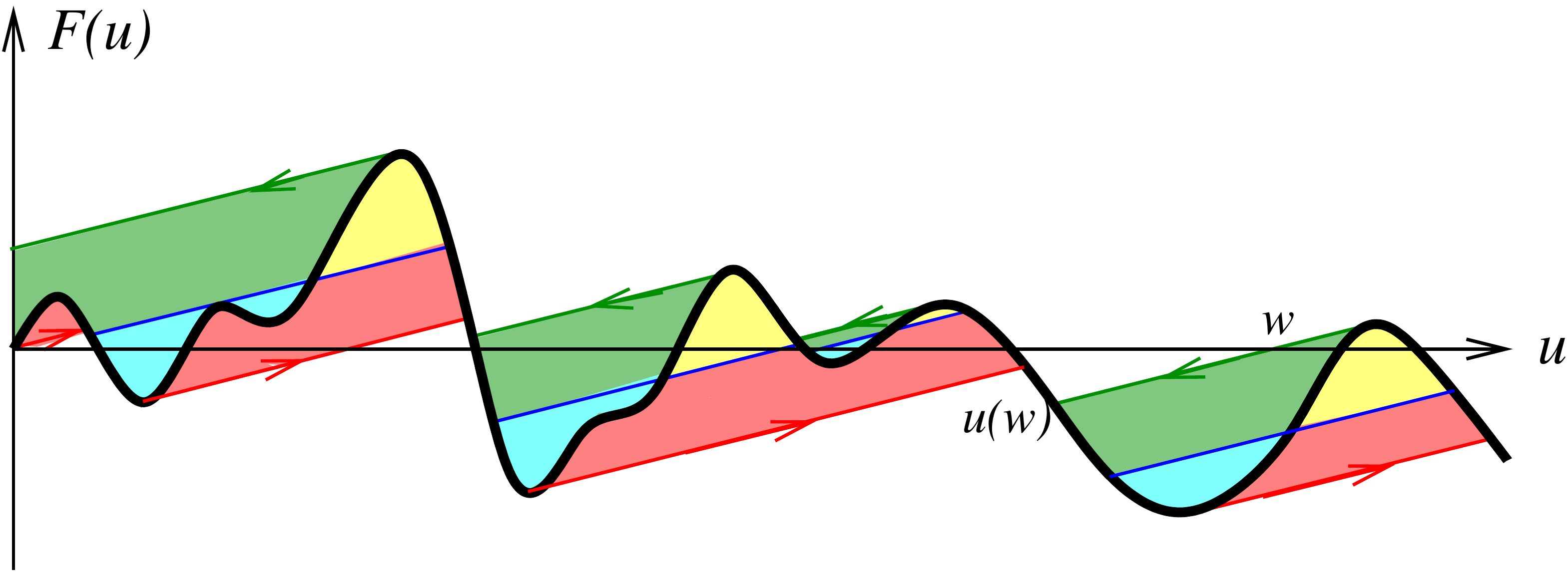}~\hfill~\raisebox{4mm}[0mm][0mm]{\fig{7.5cm}{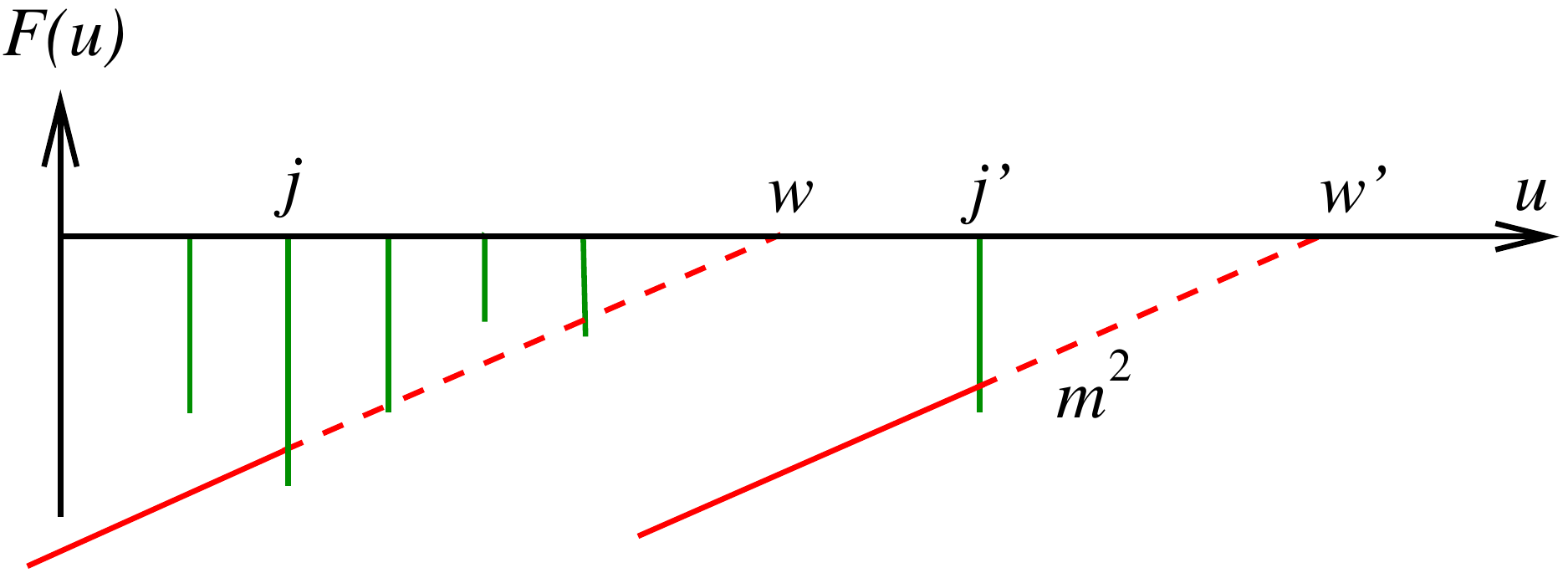}}}
\caption{Construction of $u(w)$ in $d=0$, for the pinning force $F(u)$ (bold black line). The two quasi-static
motions driven to the right and to the left are indicated by red and green arrows, and exhibit jumps
(``dynamical shocks''). The position of the shocks in the statics is shown for comparison, based on the Maxwell
construction (equivalence of light blue and yellow areas, both bright in black and white). The critical force is
$1/(2 m^2)$ times the area bounded by the hull of the construction. Right: The needles of the discretized particle model (DPM)  \cite{LeDoussalWiese2008a}.  $u_w$ as a function of $w$ is given by the left-most intersection of $m^2 (u-w)$ with a needle, here $u_w=j$, and $u_{w'} = j'$.\figinfo{Figures from \cite{LeDoussalWiese2008a}.}} \label{figgraphicalA}
\end{widefigure}For a single particle, 
there is   a   nice geometrical construction to obtain the particle trajectories,   indicated in figure \ref{figgraphicalA}: For   given $w$, draw a line   $m^2 (u-w)$. For forward driving, $u(w)$ is the leftmost intersection with the   pinning force $F(u)$, while for backward driving it is the rightmost such intersection. As indicated by the arrows, this is equivalent to shining light with slope $m^2$, either from the left for forward driving, or from the right for backward driving. Parts in the shadow are never visited, while   illuminated ones are. The jumps are the avalanches of section \ref{s:shocks} and are further discussed in section \ref{s:Avalanches}.

\begin{figure}
{\setlength{\unitlength}{1mm}\begin{picture}(83,52)
\put(0,0){\Fig{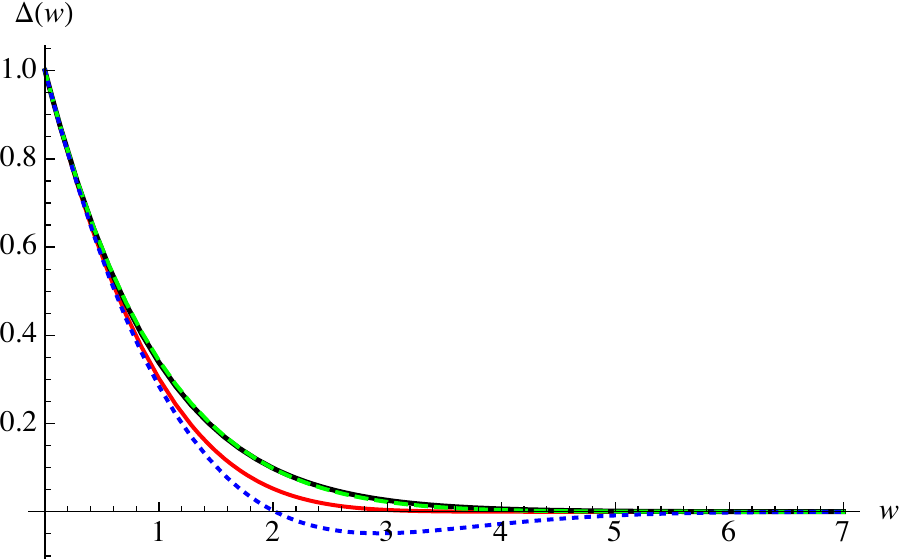}}
\put(25,16.5){\fig{5.8cm}{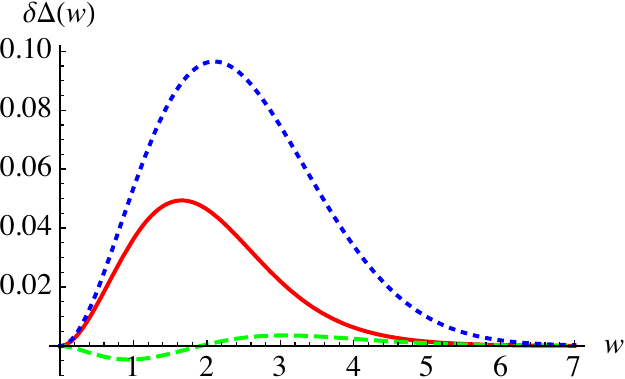}}
\end{picture}}
\caption{Main plot: The function \eq{DeltaGumbel} rescaled s.t.\ $\Delta(0)=-\Delta'(0^+)=1$ (in black). This is compared to the similarly rescaled  1-loop prediction (\ref{RF-FP-y}) (in red), and the straightforward  2-loop prediction obtained from \Eq{two-loop-FRG-dyn} (blue dashed). To improve convergence, we have   used a Pad\'e-(1,1) resummation (green, dashed), defined  in \Eq{Gumbel-from-Pade}. Inset: The same functions with the black curve subtracted. We see that the 1-loop result is decent, and that the Pad\'e-resummed 2-loop result strongly improves on it.}
\label{f:Gumbel-test}
\end{figure}

Using this construction, one can obtain both the distribution of critical forces, as well as the renormalized disorder force-force correlator $\Delta(w)$ analytically
\cite{LeDoussalWiese2008a}. The distribution of threshold forces corresponds to the three main classes of extreme-value statistics. 
Let us according to \Eq{defDe}   define
\begin{equation}
  \Delta(w-w') := m^4 \overline{[w-u(w)][w'-u(w')]}^{\,\rm c}  \punkt
 \label{65}
\end{equation}
Each class (discussed below), has its own exponent $\zeta$, setting a scale $\rho_m \sim m^{- \zeta}$. At  small $m$, force-force correlations are universal,   given by
\begin{equation}
  \Delta(w) = m^4 \rho_m^2 \tilde \Delta (w/\rho_m) \punkt  \label{formdelta}
\end{equation}
The
fixed-point function $\tilde \Delta (w)$   depends on the universality class. The three classes are distinguished by the distribution of the random forces $F$ for  the {\em most blocking} forces.
\paragraph{Gumbel class:}
\be
{  P}(F)\simeq \rme^{-A(-F)^\gamma},  \mbox{ as } F\to -\infty\punkt
\ee
The  threshold forces $f_{\rm c}$ are distributed according to a {\em Gumbel} distribution (tested in \cite{terBurgWiese2020}),
\bea
P_{\rm G}(a) =   \exp(- a) \Theta(a),\\
f_{\rm c} = \left[\frac{-\ln  (m^2 a) }A \right] ^{\frac 1 \gamma}
 = f_{\rm c}^0 - \ln (a)\, m^2 \rho_m  + ... 
\eea
The constant $f_{\rm c}^0$, the scale $\rho_m$, and the exponent $\zeta$ are
\bea\label{zeta-DPM}
f_{\rm c}^0  = A^{-\frac 1\gamma} (\ln m^{-2})^{\frac{1}{\gamma}}\komma  \nn \\
 {\rho_m^{-1}} =   {\gamma A^{\frac 1 \gamma} m^\zeta (\ln m^{-2})^{1-\frac{1}{\gamma}}}\komma  \  \zeta=2^-  \punkt
\eea
The effective disorder correlator reads
\bea\label{DeltaGumbel}
\tilde \Delta_{\rm G}(w) &=& \frac{w^2}2+\mbox{Li}_2(1-\rme^{w})+\frac{\pi^2}6\nn\\
&=&\mbox{Li}_2(\rme^{-w})- w \ln(1-\rme^{-w}).
\eea
Its first derivatives are 
\bea\label{DPM-Gumbel}
\tilde\Delta_{\rm G}(0)=\frac{\pi^2}6, \quad \tilde\Delta_{\rm G}'(0)=-1, \quad \tilde\Delta_{\rm G}''(0)=\frac12, \nn\\
\frac{ \tilde\Delta_{\rm G}(0) \tilde\Delta_{\rm G}''(0)}{ \tilde\Delta_{\rm G}'(0)^2}=\frac{\pi^2}{12}=0.822467.
\eea
It can be compared to the FRG fixed point for the RF class at depinning, see Fig.~\ref{f:Gumbel-test}, and discussed below around \Eq{Gumbel-from-Pade}. 
\paragraph{Fr\'echet class:}
\bea
P(F) \simeq   A \,  
\alpha (  \alpha+1) (-F)^{- 2- \alpha}\,\Theta(-F) \  \mbox{as}\  F\to -\infty\punkt\nn\\
\eea
The  threshold forces $f_{\rm c}$ are distributed according to a {\em Fr\'echet} distribution ($\alpha>0$), 
\be
f_{\rm c} =  x\, m^2 \rho_m\komma \   P_{\rm F}(x) = \alpha\, x^{-\alpha-1} \rme^{-x^{-\alpha}} \Theta(x)\punkt
\ee
The mean pinning force $f_{\rm c}$, the scale $\rho_m$, and the exponent $\zeta$ are
\bea
\overline {f_{\rm c}} = \Gamma(1- {\textstyle \frac 1\alpha}) \, m^2 \rho_m\komma  \  {\rho_m} =     A^{\frac 1 \alpha} m^{-\zeta} \komma  \nn\\
\zeta = 2 + \frac{2}{\alpha}\punkt
\eea
The effective disorder correlator can   be written as an integral, and is ill-defined   for $\alpha<2$, where the second moment of the force-force fluctuations vanishes. For $\alpha>2$ it  has a cusp at small $w$, and decays algebraically at large $w$, 
\bea\label{DeltaFrechet}
\tilde \Delta_{\rm F}(w)&\simeq & \Gamma \left({\textstyle \frac{\alpha -2}{\alpha }}\right)- \Gamma(1-{\textstyle
\frac{1}{\alpha}})^2 + {\textstyle \frac1 {\alpha
   ^2}  \Gamma
   \left(-\frac{1}{\alpha }\right) w} \nn\\
    &&+\frac{\alpha  w^2}{4 \alpha +2} + ... \  \mbox{for } w\to 0, \nn\\
\tilde \Delta_{\rm F}(w)&\simeq & \frac{w^{2-\alpha } \alpha }{(\alpha -2) (\alpha -1)}+ ...  \  \mbox{for } w\to \infty.
\eea
\paragraph{Weibull class:}
In this class, the random forces are bounded from below, growing as a power law above the threshold, here chosen to be zero,
\be
P(F) = A\,\alpha (\alpha -1)    F^{  \alpha-2} \theta(F), \  \alpha>1\punkt
\ee
The  threshold forces are distributed according to a {\em Weibull} distribution  
\bea
f_{\rm c} =  x\, m^2 \rho_m\komma \nn\\
 P_{\rm W}(x) = \alpha\, (-x)^{\alpha-1} \rme^{-(-x)^{\alpha}} \Theta(-x)\punkt
\eea
The mean pinning force $\overline {f_{\rm c}}$, the scale $\rho_m$, and the exponent $\zeta$ are
\bea
\overline {f_{\rm c}} =-A^{-\frac 1\alpha} \, m^{\frac 2\alpha} \,\Gamma(1+ {\textstyle \frac 1\alpha}) \  ,\nn\\
 {\rho_m} =     A^{-\frac 1 \alpha} m^{-\zeta}\, , \quad  \zeta = 2 - \frac{2}{\alpha}\punkt
\eea
The most important class is the box distribution with minimum at 0 ($\alpha=2$). Its force-force correlator is
\bea \label{DeltaWeibullalpha=2}
\tilde \Delta_{\rm W}^{\alpha=2}(w)
&=&\frac{\rme^{-w^2}} {4
   w}{ \left[2 w{-}e^{w^2} \sqrt{\pi } \left(2 w^2{+}1\right) \mbox{erfc}(w){+}\sqrt{\pi }\right]} \nn\\
   && +\frac{1}{2}  \sqrt{\pi } \left[w\, e^{-w^2} - \Gamma \left(\textstyle\frac{3}{2},w^2\right)\right]\punkt
\eea

\paragraph{Comparison to the $\epsilon$-expansion.}
An interesting question is whether one of the cases discussed above can be related to the $\epsilon$-expansion. 
The most natural candidate is the Gumbel class with  $\gamma=2$, as field theory assumes bare Gaussian disorder.  In that case $\zeta=2^-$, close to the 2-loop result (\ref{zetaRFdyn}), i.e.\ $\zeta(\epsilon=4)=2.098$. 
While a straightforward $\epsilon$-expansion for  $\tilde \Delta(w)$ is not  satisfactory at 2-loop order, we can use a Pad\'e approximant, 
\be\label{Gumbel-from-Pade}
\tilde \Delta^{\mbox{\scriptsize Pad\'e}}(w) = \frac{\tilde \Delta_1(w)+\alpha \epsilon \tilde \Delta_2(w)}{1+(\alpha-1)\epsilon \tilde \Delta_2(w)/\tilde \Delta_1(w)}.
\ee
A comparison  between \Eqs{DeltaGumbel} and \eq{Gumbel-from-Pade}, rescaled s.t.\ $\tilde \Delta(0)=1$, and $\int_0^\infty \rmd w\,\tilde \Delta(w)=1$, is shown on figure \ref{f:Gumbel-test}. 
As   can be seen there,  $\alpha=0.35$ yields a good approximation, making it a strong candidate to compare to numerical simulations or experiments in $d=1$ and $d=2$. Note however, that the functions \eq{DeltaGumbel}, \eq{DeltaFrechet} and \eq{DeltaWeibullalpha=2}    
 are close, so that the $\epsilon$ expansion might not be able to discriminate  between  them.

\paragraph{Avalanches and waiting times.} 
For simplicity we restrict ourselves to the   Gumbel class, where both the {\em waiting distances} $w$ between jumps as well as the avalanches size $S$ have a pure exponential distribution, (for definitions see section \ref{s:Avalanches}),
\bea
P(w) = \rho_m^{-1} \exp(-w/\rho_m),\\
P(S) = \rho_m^{-1} \exp(-S/\rho_m). \label{PofS-discrete}
\eea

\paragraph{Dynamics.} The model defined in Ref.~\cite{LeDoussalWiese2008a} and discussed in this section advances instantaneously. The easiest way to endow it with a   dynamics is to consider a Langevin equation \cite{terBurgWiese2020}, 
\be\label{216}
\eta \partial_t u(t) = m^2 [w-u(t)] + F\big(u(t)\big).
\ee
If the disorder is needle-like as on the right of Fig.~\ref{figgraphicalA} (the original construction of \cite{LeDoussalWiese2008a}), then either the particle is at rest, blocked by a needle, or it moves, and the only force acting on it  comes from the spring. Neglecting that the spring gets shorter during the movement, the response-function is then given by 
$
R(t) \sim  P(S/v)
$, where $\eta v = f_{\rm c}$, resulting for Gaussian disorder (Gumbel class with $\gamma=2$, $A=1/2$) into (check in \cite{terBurgWiese2020})
\be
R(t) = \tau_m^{-1} \rme^{-t/\tau_m}, \quad \tau_m = \frac\eta{2 m^2 \ln(m^{-2})} .
\ee

\subsection{Mean-field theories}
\label{s:Mean-field theories}
The framework of disordered elastic manifolds covers many experiments, from contact-line depinning over magnetic domain walls to earthquakes. Many of these experiments, or at least aspects thereof, are successfully described by {\em mean-field theory}. For driven disordered systems the first question to pose is: What is meant by mean-field (MF)? Let us define {\em 
mean-field theory as a theory which reduces an extended system to a single degree of freedom}\footnote{\new  In the literature, the term MF is used with varying meanings: It was coined for magnetic systems, when each spin interacts with all the other spins, the {\em mean fielld}. This   approximation is valid in  Ginzburg-Landau theory above a critical dimension and  MF theory is   often equated with the minimum of the Ginzburg-Landau free energy. In the bootstrap approach to CFT, Gaussian theories with long-range interactions are termed MF. In the context of avalanches, MF equates with the ABBM model introduced below. The term MF is further used for the Gaussian variational ansatz to replica-symmetry breaking (section \ref{s:RSB}), and  in dynamical systems (section \ref{s:Manna:MF}).}, in general its center of mass $u$. For depinning,  $u$ then follows the  equation of motion \eq{eq-motion-f}, reduced to a single degree of freedom, 
\be\label{218}
\partial_t   u(t) = m^2 \left[ w-   u(t) \right] +     F \big(u (t)\big) .
\ee
Specifying the correlations of $F(u)$   selects one   mean-field theory. However, when the reader   encounters the term  ``mean-field theory''   in the literature, it is quite generally employed for a model where the forces perform a random walk, 
\bea
\label{219}
 \partial_u F \big(u  \big) =   \xi(u) \komma  \\
 \label{220}
  \left< \xi(u)\xi(u') \right> = 2 \sigma \delta (u-u').
\end{eqnarray}
This model was introduced in 1990 by Alessandro, Beatrice, Bertotti and Montorsi (ABBM) \cite{AlessandroBeatriceBertottiMontorsi1990,AlessandroBeatriceBertottiMontorsi1990b} to describe magnetic domain walls. There $F(u)$ are the ``coercive magnetic fields'' pinning the domain wall, which were observed experimentally \cite{VergneCotillardPorteseil1981} to change with a  {\em seemingly} uncorrelated function $\xi(u)$. 
The decision of ABBM \cite{AlessandroBeatriceBertottiMontorsi1990} to model $\xi(u)$  in \Eq{220} as a white noise  is a  strong assumption, a posteriori justified by the applicability to experiments \cite{AlessandroBeatriceBertottiMontorsi1990b}. 
It means that $F(u)$ has the statistics of  a random walk. 
\begin{figure}
\Fig{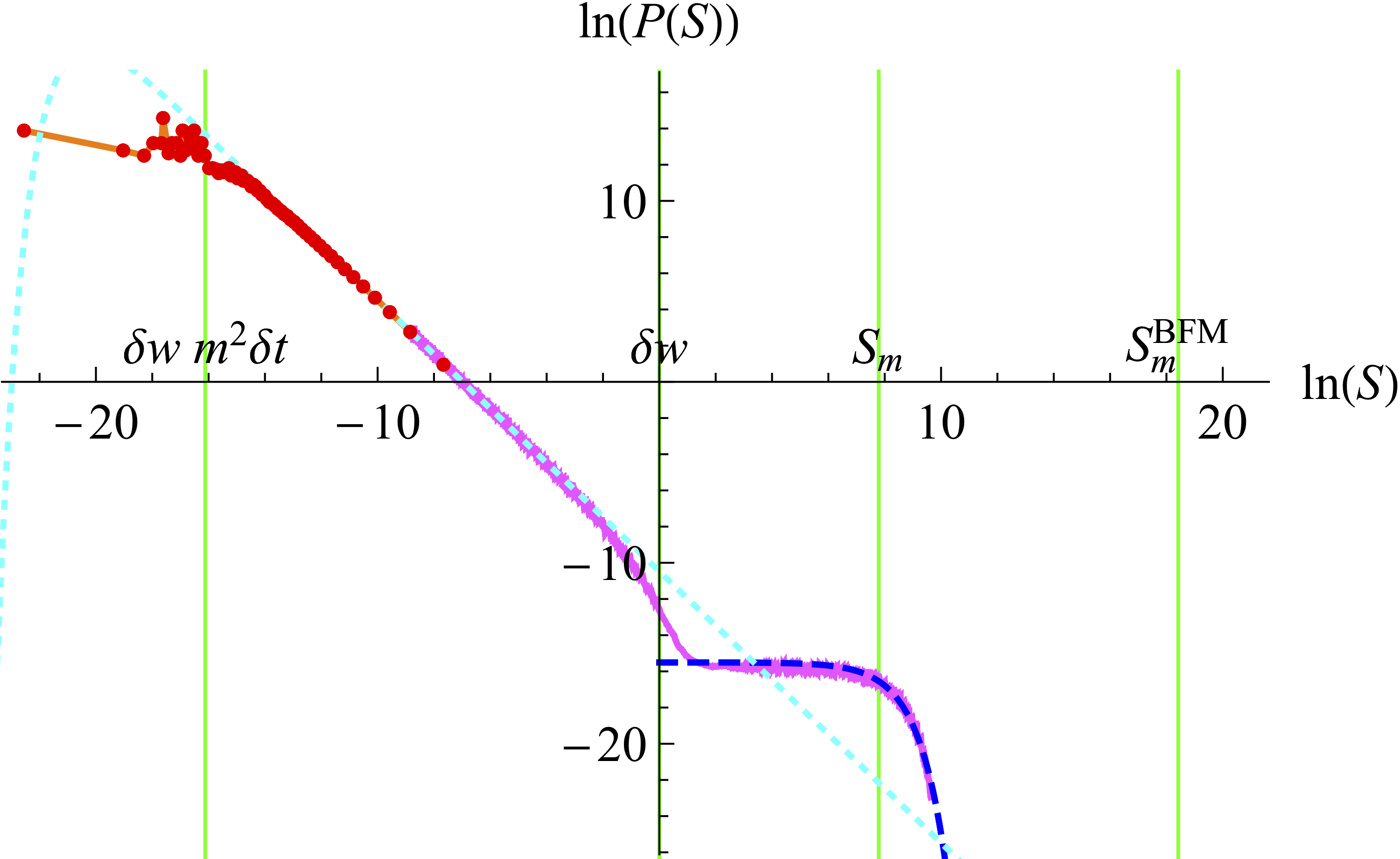}
\caption{Avalanche-size distribution $P(S)$ for a particle evolving due to \Eq{216}, with forces $F(u)$ modeled by the   Ornstein-Uhlenbeck process \eq{Ornstein-Uhlenbeck}.  The theoretical curves are  the  kicked ABBM model as given by \Eq{PwS(S)} (cyan dotted), and the discrete particle model as given by \Eq{PofS-discrete} (blue, dashed). $m^2=10^{-4}$, $\delta w=\ca A = \rho = 1$, $\delta t=10^{-4}$, $S_m=\left<S^2\right>/(2\left< S\right>)=2408.89$, $\rho_m=2329.95$, $10^8$ samples. Reprinted from \cite{terBurgWiese2020}.}
\label{ABBM-2-DPM-crossover4P(S)}
\end{figure}

Field theory \cite{LeDoussalWieseChauve2003,LeDoussalWieseChauve2002,ChauveLeDoussalWiese2000a}  gives a more differentiated view:  First of all, mean-field theory should be applicable (with additional logarithmic corrections, see section \ref{s:Behavior at the upper critical dimension}) in $d=d_{\rm c}$ \cite{FedorenkoStepanow2002,LeDoussalWiese2003a}, which contains magnets with strong dipolar interactions \cite{DurinZapperi2006b}, earthquakes \cite{DSFisher1998}, and micro-pillar shear experiments  \cite{CsikorMotzWeygandZaiserZapperi2007}. 
As $F(u)$ has the statistics of a random walk, the  (microscopic) force-force correlator of \Eqs{219}-\eq{220} is    
\bea\label{DeltaABBM}
\Delta(0) - \Delta(u-u') =  
\half \left < \left[ F(u)-F(u')\right]^2\right> = \sigma |u-u'| .\nn\\
\eea
Our argument for RF-disorder in section \ref{model}, the strongest microscopic disorder at our disposal, predicts such a behavior for the   correlator $R(0)-R(u)= \half \left <  [ V(u){-}V(0) ]^2\right> $ of the potential, but not of the force. 
On the other hand, the effective (renormalized) force-force correlator $\Delta(u)$ has a cusp, so \Eq{DeltaABBM} with $\sigma = |\Delta'(0^+)|$ is an approximation, valid for small $u$. The ABBM model \eq{218}-\eq{220} should then be viewed as  an effective  theory, arriving {\em  after renormalization}.

If indeed the microscopic disorder has the statistics of a random walk, then the force-force correlator \eq{DeltaABBM} does not change under renormalization, as is easily checked by inserting it into the 1-loop  \eq{beta-dep-1loop} or 2-loop \eq{two-loop-FRG-dyn} flow equation. Counting of derivatives for higher-order corrections proves that this statement persists to all orders in perturbation theory. Thus even an extended (non-MF) system where each degree of freedom sees a random force which performs a random walk, the Brownian-force model (BFM), introduced in \cite{LeDoussalWiese2012a} and  further discussed in section \ref{s:BFM}, is stable under renormalization, and has a  a roughness exponent\footnote{This was indirectly   numerically verified in  Ref.~\cite{ZhuWiese2017}.} 
\be
\zeta_{\rm ABBM} =\epsilon .
\ee
Our discussion shows that the ABBM model \eq{219}-\eq{220} is   adequate only at small distances, but   fails at larger ones, where the force-force correlator decorrelates. We therefore expect that at large distances it crosses over to the DPM of section \ref{s:DPM}. Ref.~\cite{terBurgWiese2020} proposed to model the crossover by replacing the random walk \Eq{219} by an Ornstein-Uhlenbeck process,  
\numparts
\be\label{Ornstein-Uhlenbeck}
 \partial_u F \big(u  \big) =  -   F(u)+ \xi(u) .
\ee
This equation is solved by 
\be
F(u) = \int_{-\infty}^u \rmd u_1 \, \rme^{-\ca  (u-u_1)}\xi(u_1) .
\ee
\endnumparts
It leads to microscopic correlations 
\bea\label{FFOU}
\Delta(u-u') = \overline {F(u) F(u')} \nn\\
= \int_{-\infty}^u \rmd u_1 \int_{-\infty}^{u'} \rmd u_2\, \rme^{- (u+u' -u_1-u_2)} \overline{\xi(u_1)\xi(u_2)}
\nn
\\
= 2\sigma  \int_{-\infty}^{{\rm min}(u,u')} \rmd \tilde u \,\rme^{- (u+u' - 2\tilde u)}  \nn
\\
=  \sigma \,\rme^{- |u-u'|}.
\eea
The small-distance behavior of $\Delta(u-u')$ is as in \Eq{DeltaABBM}. The crossover was confirmed numerically   \cite{terBurgWiese2020}, and in  experiments on magnetic domain walls \cite{terBurgBohnDurinSommerWiese2021} and knitting \cite{DouinterBurgLechenaultWieseUnpublished}. On Fig.~\ref{ABBM-2-DPM-crossover4P(S)} we show a simulation for the crossover in the avalanche-size distribution (section \ref{s:Avalanches}) from $\tau=3/2$ for ABBM, given in \Eq{PwS(S)}, to $\tau=0$ as given by \Eq{PofS-discrete}.

\Eq{Ornstein-Uhlenbeck} also serves as an  effective theory for the crossover observed in systems of linear size $L$, from a regime with $mL\gg 1$ described by an extended elastic manifold (section \ref{s:dep-loops}),  to a single-particle regime as described by the DPM model (section \ref{s:DPM}). This crossover has indeed be seen in numerical simulations for a line with periodic disorder \cite{BustingorryKoltonGiamarchi2010,BustingorryKolton2010,KoltonBustingorryFerreroRosso2013}.

\subsection{Effective disorder, and  rounding of the cusp by a finite driving velocity}
\label{s:The effective disorder, and  rounding of the cusp by a finite driving velocity}
Suppose the system is driven quasi-statically, such that whenever we measure,  almost surely $\partial_t u(x,t)=0$.  Then the condition \eq{eq-cond} derived for equilibrium is valid too. As is illustrated in Fig.~\ref{figgraphicalA}, the chosen minimum is not the global minimum, but the leftmost stable one (driving from left to right), as obtained by the construction contained in Middleton's theorem of section \ref{s:Field theory of the depinning transition}. Thus there are three relevant local minima: From left to right these are the (local) depinning minimum, the equilibrium one, and finally the (local) depinning minimum for driving the system in the opposite direction. The arguments in the construction entering \Eqs{eq-cond}, \eq{Fhat} and \eq{defDe} remain valid, and \Eq{defDe} is the prescription to measure $\Delta(w)$ at depinning. 
Defining  with $w=vt$, $w'=vt'$
\bea 
\Delta_v(w-w'):= L^d m^{4} \overline{[w-u_{w}] [w'-u_{w'}] }^{\,\rm c} , 
\label{De-u}
\eea
the renormalized force-force correlator is 
\be
\Delta(w-w')= \lim_{v\to 0} \Delta_v(w-w').
\ee
In an experiment, the driving velocity $v$ is finite, and  it is impossible to take the limit of $v\to 0$. However, the observable \eq{De-u} can be calculated as \cite{terBurgWiese2020}
\bea\label{Delta-u-theory}
\Delta_v(w) =  \int\limits_0^\infty\!\rmd t \! \int\limits_0^\infty\!\rmd t' \, \Delta(w{-}vt{+}v t' )R_w(t) R_w(t'), 
\eea
where $R_w(t)$ is the response \eq{Rw} of the center of mass to  an increase in $w$, and $\int_t R_w(t) = 1$. This implies that  the integral of $\Delta_v(w)$ is independent of $v$.

As an illustration, consider $\Delta(w) = \Delta(0) \rme^{-|w|/\xi}$, and $R_w(t)=\tau^{-1}\rme^{-t/\tau}$.
Then 
\be
\Delta_v(w) = \Delta(0) \, \frac{  
   \rme^{-{|w|/\xi }}-\frac{\tau  v}{\xi}
   \rme^{-{|w|/(\tau 
   v)}}}{1- \big(\frac{\tau  v}{\xi} \big)^2}.
\ee
This is a superposition of two exponentials, with the natural scales $\xi$ and $\tau v$. Since
\be
\Delta_v'(0^+)=0,
\ee
the cusp is rounded. This
  can   be proven   in general  from \Eq{Delta-u-theory}. However, $\Delta_v(w)$ is {\em not} analytic, contrary to the thermal rounding discussed in section \ref{s:Rounding the cusp}. 
As long as  $\tau v\ll \xi$, the second term decays much faster than the first, allowing us to perform a boundary-layer analysis, already encountered for the thermal rounding of the cusp in \Eq{boundary-layer}.
 \Eq{Delta-u-theory} is  approximated  by   the   boundary-layer ansatz 
\bea\label{BL}
\Delta_v(w) \simeq \ca A_v\, \Delta \Big(\sqrt{w^2+(\delta^{\rm BL} _w)^2}\Big)\komma \\
\delta^{\rm BL} _w =v \tau,\quad \tau :=   \int_0^\infty \rmd t \, R_w(t)  t \ ,\\
\ca A_v = \frac{\int_0^\infty\rmd w\, \Delta(w)}{\int_0^\infty\rmd w\, \Delta(\sqrt{w^2+(\delta^{\rm BL}_w)^2})}\punkt
\eea
The   amplitude $\ca A_v$ ensures   normalization. 
While the response function  $R_w(t)$ (and possibly $\Delta(w)$) in \Eq{Delta-u-theory} may   depend   on $v$,  our considerations using the zero-velocity expressions in \Eq{Delta-u-theory} yield at least the correct small-velocity behavior \cite{terBurgWiese2020}. 

If in an experiment the response function is unavailable, its characteristic time scale $\tau$ can  be reconstructed approximatively from $\Delta_v(w)$ as 
\be
\tau \simeq \frac1v \frac{\lim_{w\to 0 }\Delta'(w)}{\Delta_v''(0)}.
\ee
In the numerator we have written  $\lim _{w\to 0}\Delta'(w)$, which  is obtained by extrapolating $\Delta_v'(w)$ from outside the boundary layer, i.e $w\ge \delta^{\rm BL} _w = v\tau$, to $w=0$.
\begin{figure} 
\Fig{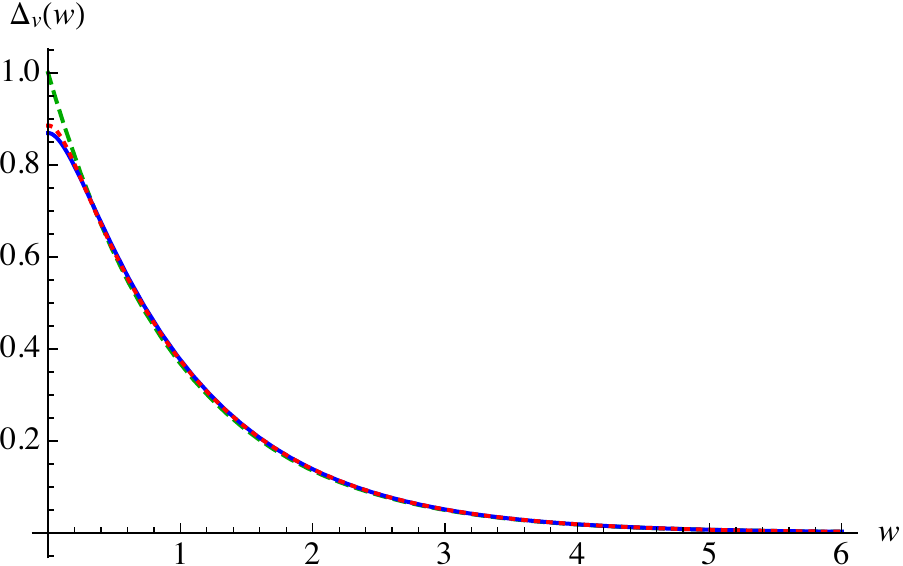}
\caption{Rounding of $\Delta(w)$ (green, dashed) at finite $v$ to  $\Delta_v(w)$ (blue solid)  given by \Eq{Delta-u-theory}   and the boundary-layer approximation \eq{BL} (red dotted).  }
\end{figure}

There are two other, and more precise,    strategies  to obtain $\tau$,  and at the same time reconstruct the zero-velocity correlator:
\begin{enumerate}
\item use the boundary-layer formula \eq{BL} to plot the measured $\Delta_v(w)$ against $\sqrt{w^2+(v\tau )^2}$; find the best  $\tau$ which removes the curvature of $\Delta_v(w)$. This yields $\tau$, and by extrapolation to $w=0$ the full $\Delta(w)$.
\item use that $(\tau \partial_t +1)  R_w(t) = \delta(t)$  to remove  the response functions in \Eq{Delta-u-theory}, 
\be\label{242}
\Delta_{v=0}(w) = \left[1-\left( \tau {v } \frac{\rmd}{\rmd w}\right)^{\!2} \right]\Delta_v(w).
\ee 
\end{enumerate}
In both approaches, 
the fitting parameter $\tau$ can rather precisely be    obtained by plotting  $-\Delta_{v=0}'(w)/\Delta_{v=0}(w)$, and optimizing to render the plot as straight as possible for small $w$, i.e.\ inside the boundary layer $w\le \delta^{\rm BL} _w = v\tau$. While \Eq{242} is more precise, and reconstructs $\Delta_{v=0}(w)$ down to $w=0$, the boundary analysis is more robust for noisy data \cite{terBurgWiese2020,terBurgBohnDurinSommerWiese2021}.

\subsection{Simulation strategies}
\label{s:Simulation strategies}
In order to test the predictions of the field theory, one needs efficient simulation algorithms. There are 
three categories.
\begin{enumerate}
\item 
{\em Cellular automata} are simple to implement, either directly for the elastic manifold, or for one of the related sandpile models (section \ref{s:sandpiles}). Direct implementations for the elastic manifold are tricky, as extended moves are necessary \cite{RossoKrauth2001a}. 

\item {\em Langevin dynamics} is the most realistic approach, and the best approach  to access the dynamics \cite{FerreroBustingorryKolton2012}; even though dynamical simulations are sometimes   performed in cellular automata.

\item {\em Critical configurations} in a continuous setting can be sampled most efficiently by the Rosso-Krauth algorithm   \cite{RossoKrauth2002,Rosso2002}, termed by the authors variant Monte Carle (VMC). The idea is simple: When updating the position of a single site, the latter site can be moved as far ahead as the equation of motion permits, without updating at the same time its neighbors. Since Middleton's theorem guarantees  that the such generated configuration cannot surpass the next pinning configuration, the algorithm converges to the latter. This fictitious dynamics   is much faster than the Langevin one, and gives precise estimates for the roughness $\zeta$ in dimension $d=1$  \cite{RossoKrauth2001b}, and higher  \cite{RossoHartmannKrauth2002}. 
   
\end{enumerate}

\begin{figure}[t]
\begin{center}
\mbox{{{\begin{tikzpicture}
\coordinate (0) at  (0,0)  ; 
\coordinate (h) at (1,0);
\coordinate (1) at  (2,0) ;
\coordinate (y) at (2,1.7) ;
\coordinate (y) at (2,2.7) ;
\coordinate (yp) at (0.7,0.95) ;
\node at (-.2,-.2) {$0$};
\node at (2.2,-.2) {$1$};
\node at (2.2,2.7) {$z$};
\node at (0.7,1.2) {$z'$};
\node at (1.15,0.25) {$\phi$};
\fill (0) circle (1.5pt);
\fill (h) circle (1.5pt);
\fill (1) circle (1.5pt);
\fill (y) circle (1.5pt);
\fill (yp) circle (1.5pt);
\draw [dashed](0) circle (2);
\draw [thick] (h) circle (1);
\draw  (1.5,0) arc (0:108:0.5);
\draw [thick](2,-2) -- (2,3);
\draw [](-2.5,0) -- (2.5,0);
\draw [dashed] (0) -- (y);
\draw [dashed] (h) -- (yp);
\draw [red,thick] (yp) -- (1);
\end{tikzpicture}}}}
\end{center}
\caption{The conformal transformation $z\to z'$ as given in \Eq{CFT-1}.}
\label{f:conformal}
\end{figure}

\subsection{Characterization of the  1-dimensional string}
\label{s:Characterisation of the  1-dimensional string}
\subsection*{Scaling variables}
To keep a system translationally invariant, 
simulations are usually performed with periodic boundary conditions. 
Trying to extract critical exponents by 
plotting simulation results against distance   yields poor results. 
There are two natural scaling variables:

\begin{enumerate}
\item {\em Polymer Scaling}: For a non-interacting polymer of size $x$, or   random walk of time $x$, the probability that monomers $0$ and $x$ come close  in $d$-dimensional space is
$P(x) = \ca A x^{-d/2}$. The probability that a ring polymer of size $L$ has monomers 0 and $x$ close together equals the probability to have two rings of sizes $x$ and   $L-x$,
\be
P(x|L) = P(x) P(L-x) =\ca A^2 \big[ x(L-x) \big] ^{-\frac d2}.
\ee
This identifies the natural scaling variables
\be
x^{(1)}_{ p} := \frac{4x(L-x)}{L^2}\komma \quad x^{(2)}_{ p}:= \big( x^{(1)}_{ p} \big)^2 .
\ee

\item
{\em Conformal Invariance}:
The conformal mapping from  the line $z= 1+iy$ to the circle  of diameter 1 as shown in Fig.~\ref{f:conformal} and known as an {\em inversion at the circle}, maps
\bea\label{CFT-1}
z= 1+i y ~\longrightarrow z' =   \frac 1{1-i y}  = \frac12 \left( 1 + \rme^{i \phi} \right) .
\eea
This implies (for details see \cite{SparfelWiese2021})
\be
y = \tan(\phi/2).
\ee
Suppose the 2-point function on the infinite axis is
\be
\left< \ca O(y_1) \ca O(y_2) \right> = \frac1{|y_1-y_2|^{2\Delta}}.
\ee
Conformal invariance  \cite{GinspargLH1988,DiFrancescoMathieuSenechal,CardyInDombGreen} implies that on the circle
\bea\label9
\left< \ca O(\phi_1) \ca O(\phi_2) \right> = \left(\frac{\partial y_1}{\partial \phi_1 } \right)^{\!\!\Delta}
 \left(\frac{\partial y_2}{\partial \phi_2 } \right)^{\!\!\Delta} \left< \ca O(y_1) \ca O(y_2) \right> \nn\\
 = t_{12}^{-2 \Delta} .
\eea{\begin{widefigure}[t]\fig{0.32\textwidth}{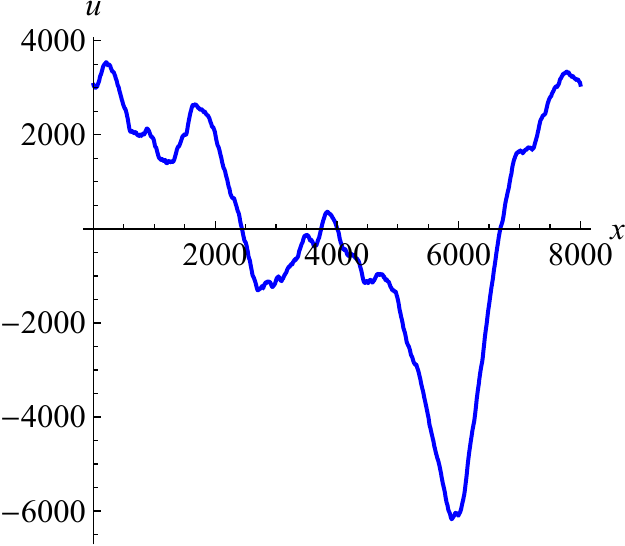}\hspace{0.02\textwidth}\fig{0.32\textwidth}{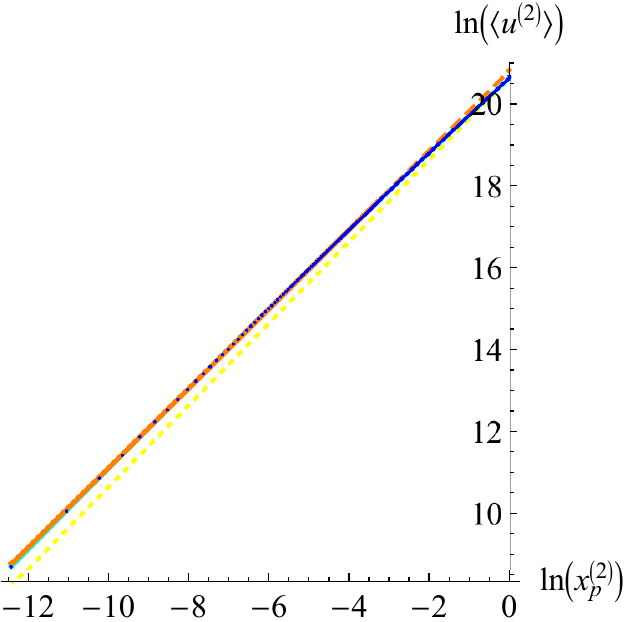}\hspace{0.02\textwidth}\fig{0.32\textwidth}{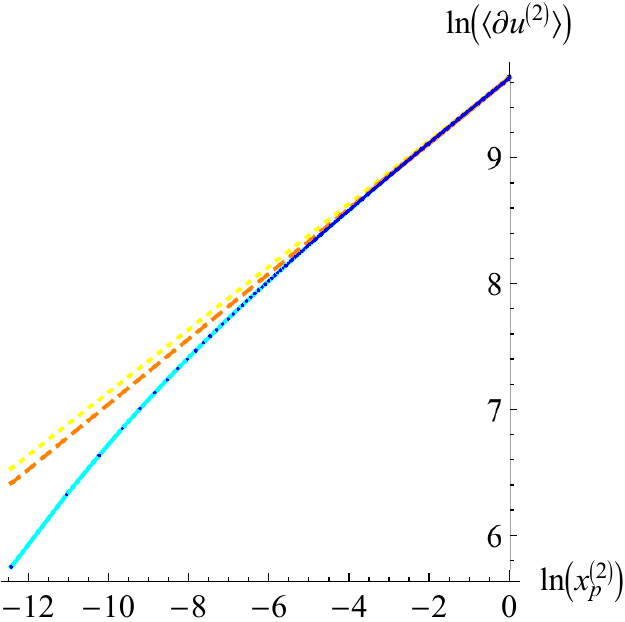}\caption{Left: A critical string at depinning, $L=8000$, $mL=1$. Middle and right: The 2-point function for  $L=2000$, $mL=1$. The measured slopes (in orange) are $0.970$, and $0.259$,  confirming the  theoretically expected $1$, and  $0.25$ (in yellow)}\label{f:dep-d=1}\end{widefigure}}\begin{widefigure}{\fig{0.32\textwidth}{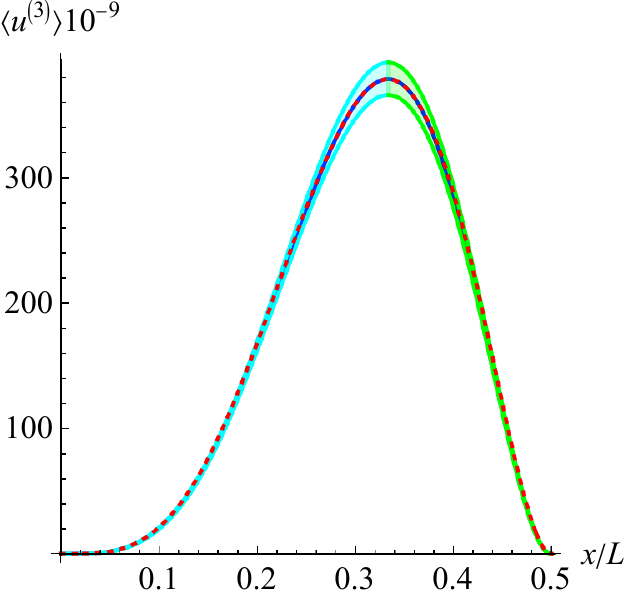}\hspace{0.02\textwidth}\fig{0.32\textwidth}{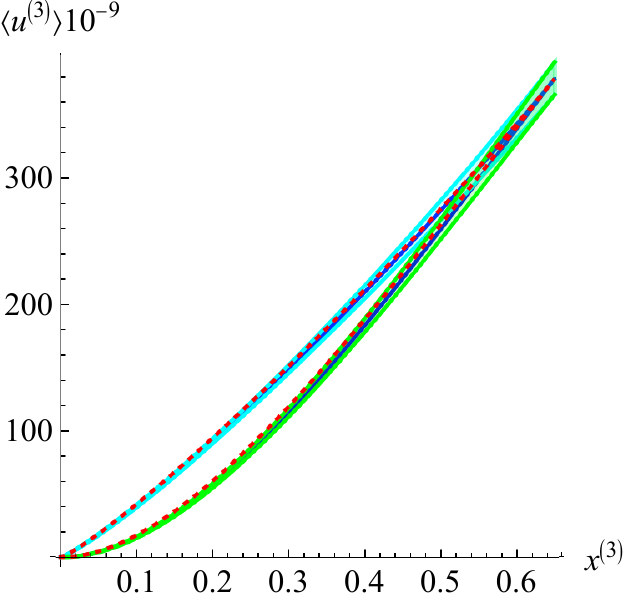}\hspace{0.02\textwidth}\fig{0.32\textwidth}{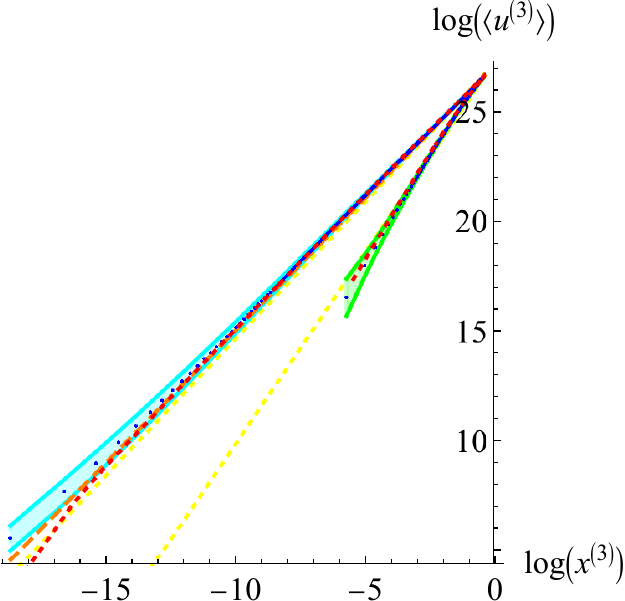}}
\caption{The 3-point function $\left< u^{(3)} \right>$ for $L=2000$, $mL=1$. The measured slope is $1.21\pm 0.04$ (orange, dashed), as compared to the expected   $\zeta=1.25$ (in yellow).  The descending branch of the last curve has slope $3\zeta-2=1.75$. The red dotted curves are the theoretical prediction from \cite{SparfelWiese2021}.}
\label{f:dep-d=1bis}\end{widefigure}\hspace*{-1.5ex}
Here $t_{12}$ is the chordal distance between the two points parameterized by $\phi_1$ and $\phi_2$, i.e.
\bea\label{202}
t_{12}= t(\phi_1{-}\phi_2) = |\rme^{i \phi_1}{-}\rme^{i \phi_2}| = 2 \left| \sin \left( \frac{\phi_1-\phi_2}{2} \right)\right|\punkt \nn\\
\eea 

Similarly, for the 3-point function of   scalar operators of dimension $\Delta$, conformal invariance implies  
\be\label{203}
\left< \ca O(\phi_1) \ca O(\phi_2)  \ca O(\phi_3)  \right> =\ca A \; \Big( t_{12} t_{13} t_{23}\Big)^{\!\!-\Delta}.
\ee
\end{enumerate}

\subsubsection*{An anomalously large roughness, $\zeta=5/4$:}
Consider the standard definition of the 2-point function
\be
\left< u^{(2)} (x) \right> := \frac 12 \left< [u(x) -u(0)]^2 \right> .
\ee
We expect that $\left< u^{(2)} (x) \right> \sim |x|^{2\zeta}$. This is not possible for $\zeta>1$, as is shown by the following simple argument \cite{LeschhornTang1993}:
\bea
\half \left< [u(x) -u(0)]^2 \right> \nn\\
= \half\sum_{i=1}^x \sum_{j=1}^x \left<  [ u(i)-u(i-1)] [u(j)-u(j-1)] \right>.
\eea
The expression inside the expectation value  depends on $i-j$, and is maximal for $i=j$. Thus
\be
\half \left< [u(x) -u(0)]^2 \right>  \le \frac{x^2}{2}   \left<  [ u(1)-u(0)]^2   \right>.
\ee
As can be seen on the middle of Fig.~\ref{f:dep-d=1}, the bound is almost saturated. 
Thus, we expect \bea\label{199}
\half \left< [u(x) -u(0)]^2 \right> \approx \ca A x^{(2)}_p L^{2\zeta}
\simeq 4 {\ca A}\, x^2 L^{2\zeta-2},  ~x\ll L.\nn\\
\eea
The roughness exponent $\zeta$ can be observed in the overall scaling, evaluating 
$\left< u^{(2)} (x) \right> $ at its maximum $x=L/2$, in Fourier space, or by measuring correlations of the discrete derivative of $u(x)$. 
We expect that 
\bea
\left< \partial u^{(2)} (x) \right> &:=& \frac 12 \left< \Big[\big( u(x{+}1){-}u(x)\big) {-} \big(u(1) {-}u(0)\big) \Big]^2 \right> \nn\\
&=&  \ca B \big( x^{(2)} \big)^{\zeta-1}   
\simeq 4 {\ca B}\, x^{2\zeta-2},  ~x\ll L.
\eea
This is indeed satisfied, see the middle of Fig.~\ref{f:dep-d=1}.
The exponent is consistent with $\zeta=5/4$, as conjectured in Ref.~\cite{GrassbergerDharMohanty2016}, see section \ref{zeta=5/4}.

\subsubsection*{Skewness:}

In equilibrium the connected three-point function $\left< u(x) u(y) u(z) \right>^{\rm c}$   vanishes, due to the symmetry $u\to -u$. 
At  depinning  this symmetry may be  broken,  but no signs   were found  yet \cite{RossoKrauthPrivate,KoltonPrivate}. 

On  figure \ref{f:dep-d=1bis} we show non-vanishing  simulation results for the 3-point function \cite{SparfelWiese2021}
\bea\label{3-pint-nonvanishing}
\big\langle u^{(3)}(x)\big\rangle := { \left<[u(x)-u(0)]^2 [u(-x)-u(0)]\right>}.
\eea
Note that the  more symmetric-looking variant 
\be
\left< [u(x)-u(y)][u(y)-u(z)] [u(z)-u(x)] \right>^{\rm c} = 0 
\ee
vanishes as indicated,  which can be shown by expansion. 
The simplest  symmetric, non-vanishing combination  is 
\bea
\big\langle u^{(3)}(x,y,z)\big\rangle :=\frac16 {\left< \Delta u(x,y,z) \Delta u(y,z,x) \Delta u(z,x,y)\right>^{\rm c}},
\nn\\
\Delta u(x,y,z):= u(x)+u(y)-2 u(z)\punkt
\eea
It is related to the combination in \Eq{3-pint-nonvanishing} by
\be
\big\langle u^{(3)}(x,0,-x)\big\rangle \equiv \big\langle u^{(3)}(x)\big\rangle .
\ee
From scaling, we expect that 
\be
\big\langle u^{(3)}(x)\big\rangle \sim |x|^{3 \zeta}\komma\quad x\ll L\ , 
\ee
as long as we escape the argument that leads to \Eq{199}. Figure \ref{f:dep-d=1bis} shows that this is the case. 
What is tested there in addition is whether conformal invariance holds. According to \Eq{203}, conformal invariance implies that 
\bea
\big\langle u^{(3)}(x)\big\rangle  = \ca A (x^{(3)})^{\zeta},\\
x^{(3)} =  t(x)^2t(2x).
\eea
with $t(x)$ introduced in \Eq{202}. While this does not hold conformal symmetry may be  present in  a different observable.

\begin{figure}
\parbox{0.48\textwidth}{\newlength{\inset}
\setlength{\inset}{6.15cm}
\fig{0.48\textwidth}{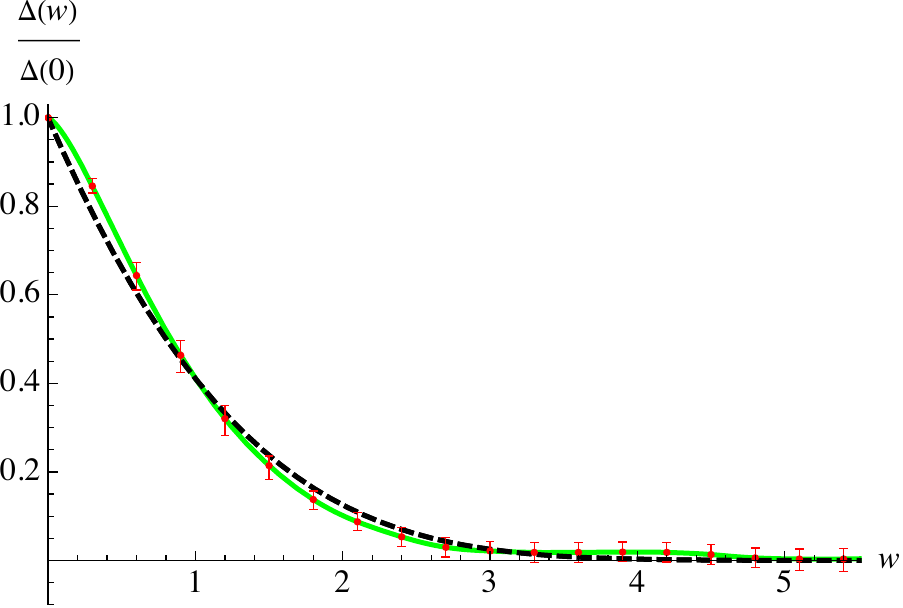}\hspace{-\inset}\raisebox{21.7mm}{\parbox[b]{\inset}{\fig{\inset}{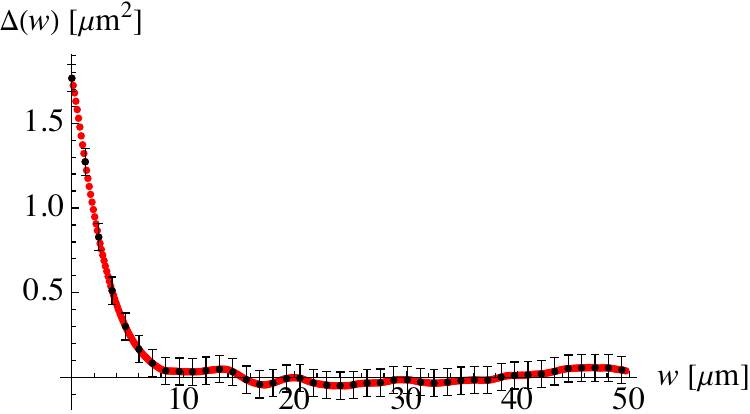}}}}\caption{Inset: The disorder correlator $\Delta (w)$ for contact line depinning of $\mathrm{H}_2$/Cs, with
error bars estimated from the experiment. Main plot:
The rescaled disorder correlator   (green/solid) with error bars (red).
The dashed line is the 1-loop result; figure from \protect\cite{LeDoussalWieseMoulinetRolley2009}. 
Note that   the boundary layer due to the finite driving velocity (section \ref{s:The effective disorder, and  rounding of the cusp by a finite driving velocity}) is not deconvoluted.} \label{f:Delta3}
\end{figure}

\subsection{Theory and numerics for long-range elasticity: contact-line depinning and  fracture}
\label{Theory for contact-line depinning}
\begin{figure}
\centerline{\begin{tabular}{|c|c|c|c|c|c|}
\hline
 & $\epsilon $ & $\epsilon ^{2}$ & estimate & simulation \\
\hline
\hline
$\zeta $& 0.33     &  0.47    & $0.47\pm0.1$ & $0.388 \pm 0.002$~\cite{RossoKrauth2002}  \\
 \hline
 $\beta $  & 0.78 & 0.59 & $0.6\pm 0.2$ & $0.625\pm
0.005$~\cite{DummerKrauth2007}   \\
\hline
$z$  &0.78 &0.66 &$0.7\pm 0.1$ & $0.770 \pm
0.005$~\cite{DummerKrauth2007} \\
\hline $\nu $ & 1.33 & 1.58 & $2\pm0.4$ & $1.634 \pm
0.005$~\cite{RossoKrauth2002} \\
\hline
\end{tabular}}
\caption{Exponents for the depinning of a line with long-range elasticity ($\alpha=1)$, relevant for contact-line depinning and fracture. The last exponent $\nu$ was obtained from $\nu = 1/(1-\zeta)$.}
\label{tab:LR-d=1}
\end{figure}
Contact-line depinning  can be treated by a modification of the theory for disordered elastic manifolds, using the long-range elasticity introduced in section \ref{s:LR-elasticity}, \Eq{Hel-contact-line}. The theory was developed to $\ca O(\epsilon)$ in Ref.~\cite{ErtasKardar1994b} and to $\ca O(\epsilon^2)$ in Ref.~\cite{ChauveLeDoussalWiese2000a,LeDoussalWieseChauve2002}. Key predictions for $\alpha=1$ are \cite{LeDoussalWieseChauve2002}
\bea
\epsilon = 2-d ,\\
\zeta = \frac{\epsilon}3 \left(1+ 0.39735  \epsilon^2\right) +\ca O(\epsilon^3) ,\\
z = 1- \frac{2}{9 }\epsilon -0.1132997 \epsilon^2 +\ca O(\epsilon^3) .
\eea
The other exponents are obtained via scaling, $\nu=1/(1-\zeta)$, and $\beta = \nu (z-\zeta)$.
Simulation  results are  
\bea
\label{zeta-LR-num}
\zeta &=& 
0.388
\pm
0.002 \quad \cite{RossoKrauth2002} ,\\
\label{z-LR-num}
z &=& 0.770 \pm
0.005 \quad \cite{DummerKrauth2007}, \\
\label{beta-LR-num}
\beta &=& 0.625\pm
0.005 \quad \cite{DummerKrauth2007}.
\eea
For arbitrary $\alpha$ the roughness reads \cite{WieseOriginalReview}
\be
\zeta(\alpha) =\frac\epsilon3 +  \frac{\psi(\alpha )-2 \,\psi\!\left(\frac{\alpha}{2}\right)-\gamma_{\rm E} }{20.9332}\epsilon^2 + \ca O(\epsilon^3).
\ee
For the exponent $z$  expressions are more involved, and we only give an additional value for $\alpha=3/2$ \cite{WieseOriginalReview}, 
\be
z(\alpha=3/2)=  \frac32 - \frac{2}{9} \epsilon  -0.0679005  \epsilon^2.
\ee
Numerical values both for the $\epsilon$-expansion and for simulations are collected on table \ref{tab:LR-d=1}.

\subsection{Experiments on contact-line depinning}
\label{s:Experiments on contact-line depinning}
Contact lines are a nice experimental realization of depinning, as one can watch and film them to extract  not only their roughness, but also dynamical properties.  The value of the roughness exponent, given as $\zeta = 0.51$ in \cite{MoulinetGuthmannRolley2002}, but observed smaller $\zeta\approx 1/3$ in earlier work \cite{RolleyGuthmannGombrowiczRepain1998}, is still   debated  \cite{IlievPeshevaIliev2018}, and many effective exponents are found in the literature. Our own theoretical work \cite{LeDoussalWieseRaphaelGolestanian2004,BachasLeDoussalWiese2006} does not allow to exclude an exponent of $\zeta > \zeta_{\rm dep}^{\rm LR} = 0.38$, but   we do not believe  this to be likely.

Contact-line depinning is also the first system where the renormalized disorder correlator $\Delta(w)$ was measured, both for liquid hydrogen on a disordered Cesium substrate, and for isobutanol on a randomly silanized silicon wafer  \cite{LeDoussalWieseMoulinetRolley2009}. Earlier experiments with water on a glass plate with randomly deposited Chromium islands \cite{MoulinetGuthmannRolley2002} turned out to have long-range correlated correlations, both due to the impurity of water as of the inhomogeneity of glass.
Measurements of the renormalized force-force correlator $\Delta(w)$ as defined in \Eq{defDe} are shown in Fig.~\ref{f:Delta3}, using the cleaner of the two systems, liquid hydrogen on a disordered Cesium substrate. The agreement is satisfactory.

\subsection{Fracture}
\label{s:fracture}
\begin{figure}[b]
\setlength{\unitlength}{1mm}
\Fig{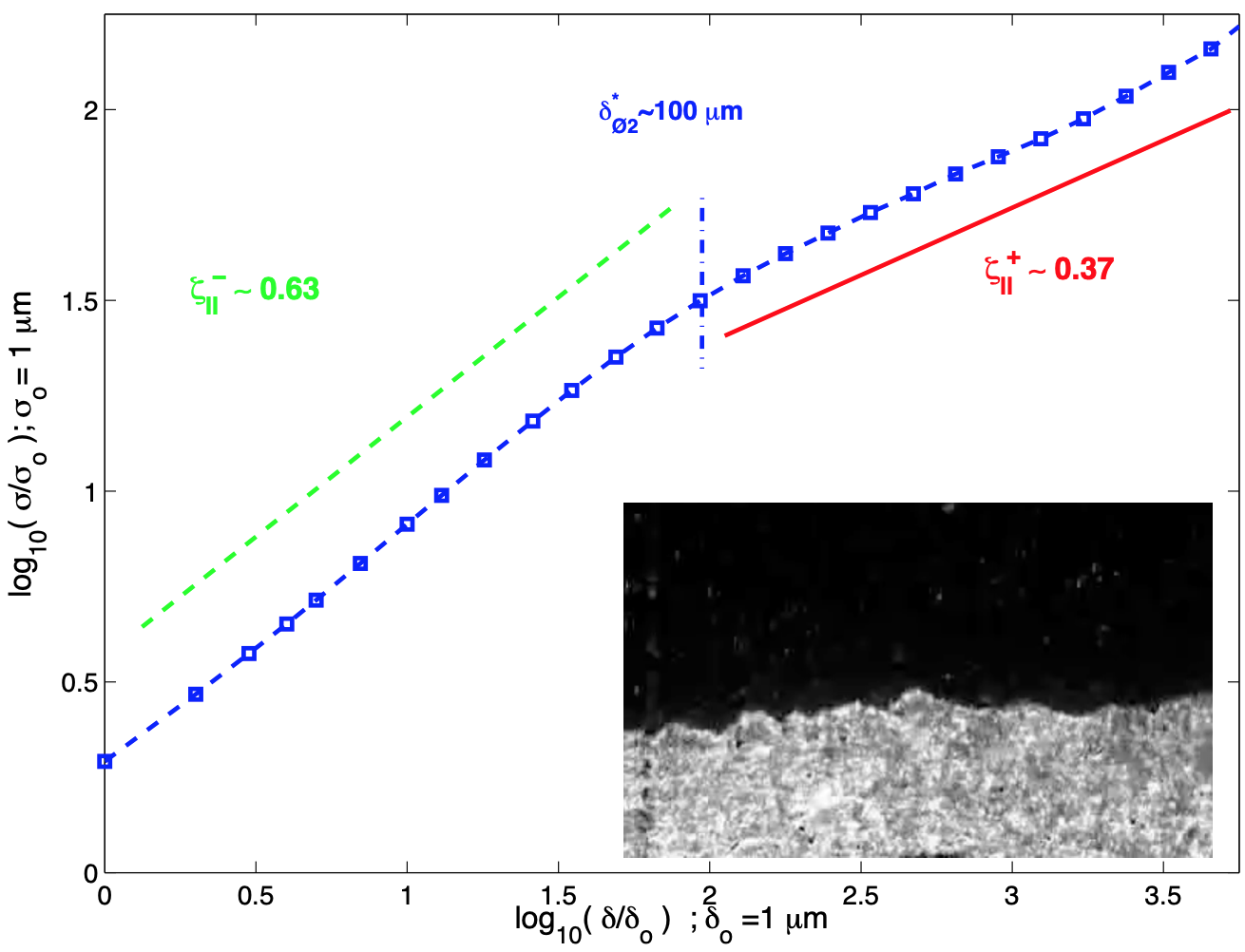}
\caption{Scaling behavior of the height-height correlations  with two different roughness exponents $\zeta_{\parallel }\approx 0.63$ below the critical scale $\delta^*=100 \mu \rm m$ and $\zeta_{\parallel }\approx 0.37$   above   \cite{SantucciGrobToussaintSchmittbuhlHansenMaloy2010}. The inset shows the fracture front, moving from bottom to top.}
\label{f:Santucci}
\end{figure}

There are two main types of fracture experiments: Fracture along a fault plane \cite{RamanathanFisher1997,RossoKrauth2002,KatzavAdda-BediaBenAmarBoudaoud2007,SantucciMaaloyDelaplaceMathiesenHansenHaavigBakkeSchmittbuhlVanelRay2007,SantucciGrobToussaintSchmittbuhlHansenMaloy2010}, and fracture of a bulk material \cite{BouchaudLapassetPlanes1990,LawnBook1993,ParisiCaldarelliPiotronero2000,ArndtNattermann2001,BonamyPonsonPradesBouchaudGuillot2006,PonsonBonamyBouchaud2006,Ponson2007,Ponson2008}.

\subsubsection*{Fracture along a fault plane.}
Let us start with the conceptionally simpler fracture along a fault plane. It is characterized by a   roughness exponent $\zeta \equiv \zeta_{\parallel}$  in the propagation direction.  A beautiful example is the {\em Oslo experiment} \cite{SantucciMaaloyDelaplaceMathiesenHansenHaavigBakkeSchmittbuhlVanelRay2007,SantucciGrobToussaintSchmittbuhlHansenMaloy2010}, where two transparent plexiglas plates are sandblasted rendering them opaque.   Sintered together the sandwich becomes   transparent. Breaking the crack open along the fault plane between the two plates, the damaged parts become again opaque, allowing one to observe and film the advancing crack, see the inset of Fig.~\ref{f:Santucci}. Below a characteristic scale $\delta_0 \approx 100\mu$m, which also is the correlation length of the disorder, the roughness exponent is $\zeta _\parallel\approx 0.63$, which is interpreted  \cite{SantucciGrobToussaintSchmittbuhlHansenMaloy2010} as the roughness exponent in directed percolation $\zeta =  0.632613 (3)$, see \Eq{exponents-DP-d=1}. (For reasons discussed in section   \ref{s:qKPZ}, directed percolation is also relevant for anisotropic depinning.) For larger scales, the roughness crosses over to a smaller exponent of $\zeta _\parallel \approx 0.37$, consistent with the roughness exponent for    depinning of a line with long-ranged elasticity, $\zeta =
0.388
\pm
0.002$  see \Eq{zeta-LR-num} \cite{RossoKrauth2002}. (Long-range elasticity  is explained in section \ref{s:LR-elasticity}.) For  fracture    it was introduced in Ref.~\cite{Rice1985}. Finally,  interface configurations are non-Gaussian \cite{TallakstadToussaintSantucciMaaloy2013}.

\subsubsection*{Fracture of bulk material.}
Fracture of a bulk material is more complicated. To get the notations straight, we show the coordinate system favored in the fracture community (drawing of \cite{Ponson2007}):

\noindent
\Fig{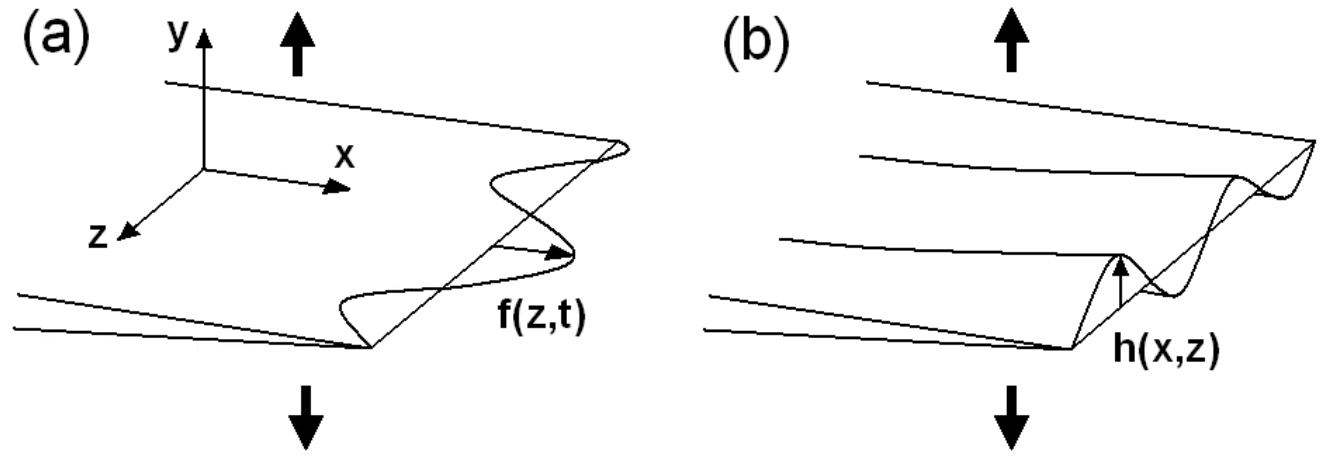}

\noindent
Applying stress in the direction of the fat arrows, the crack advances on average in the $x$ direction. The crack front  as a function of time is parameterized by 
\bea
x(z,t) = w + f(z,t),\\
y(z,t) = \hat h(z,t ) = h\big( x(z,t) , z\big).
\eea
The quantity $w= v t $ is the external control parameter   (used throughout this review).
Several critical exponents can be defined. 
Denote
\be\label{276}
\delta h(\delta x,\delta z) ^2:= \left< [ h(x+\delta x,z+\delta z)-h(x,z)]^2  \right>,
\ee 
where the average is taken over all $x$ and $z$ (and samples, if possible).  
The critical exponents ${\hat\beta}$ and $\zeta$ defined in the literature are (we changed $\beta\to \hat \beta$ in order to avoid confusion with the exponent $\beta$ defined in \Eq{v=|f-fc|^beta})
\bea\label{277}
\delta h(\delta x,0)  \sim \delta x^{\hat\beta}\ , \quad 
\delta h(0,\delta z) \sim \delta z^\zeta .
\eea
Scaling implies that \Eq{276} can be written as
\bea
\delta h(\delta x,\delta z) = \delta x^{\hat\beta} f\!\left( \frac{\delta z}{\delta x^{1/z}}\right),\\
 f(u) \sim u^ \zeta, \mbox {~~ for $u$  large, and } \zeta =   {\hat\beta} z.
\eea
A third exponent $\zeta_\parallel$ can be defined by the fluctuations of $f$, 
\be
\delta f \sim \delta z ^{\zeta_\parallel}.
\ee
As   measurements  are in general   {\em post-mortem}, $\zeta_\parallel$ is  inaccessible, except if the broken material is transparent, and one can observe the crack front advancing. It  has been measured in a clever experiment where a crack was filled with color, and broken open after the color had dried, confirming the small-scale regime $\zeta_\parallel\approx 0.6$ \cite{Bouchaud1997}. Numerical simulations suggest \cite{PonsonPrivate} that a smaller exponent $\zeta_{\parallel}\approx 0.38$ should hold at larger scales; an experimental confirmation is outstanding \cite{PonsonPrivate}.
One imaging option is to use a synchrotron; this challenging experiment has to  our knowledge not been attempted. 
Common values for materials as diverse as  silica glass, 
aluminum, mortar or wood give ${\hat\beta} \approx 0.6$, and $\zeta  \approx 0.75 $ to $0.8$ \cite{Bouchaud1997,Ponson2007}. 
The question arises whether there is a connection to fracture along a fault plane, and depinning.

To make contact to the latter, we first observe that fracture is irreversible, a crucial point  for  depinning. 
To proceed, note the 2-d vector
\be
\vec u(z,t) = \big( f(z,t), h(z,t) \big).
\ee
Knowing the elastic kernel \eq{Hel-alpha=1-pos} for long-range elasticity with $\alpha=1$ \cite{Rice1985}, 
the only Langevin equation linear in $u(z,t)$ one can write down is (see e.g.\ \cite{RamanathanErtasFisher1997,BonamySantucciPonson2008})
\noindent
\bea\label{EOM-crack-effective}
\partial_t \vec u(z,t) = \frac{\gamma}{2\pi}\int_{z'} \frac{ \vec u(z',t){-}\vec u (z,t)}{(z{-}z')^2} 
+\eta\big(\vec u(z,t),z\big),\\
\overline{\eta(x,y,z) \eta(x',y',z')} = \sigma\delta (x{-}x')\delta(y{-}y')\delta(z{-}z').
\eea
Assuming that the scenario of Refs.~\cite{ErtasKardar1994,ErtasKardar1996} holds   for long-ranged elasticity, 
  {\em at large scales} the longitudinal exponent $\zeta_{\parallel}$ should be that of a contact line, with according to  \Eqs{zeta-LR-num}-\eq{beta-LR-num} an exponent $\zeta_\parallel =0.388$. The 
transversal roughness $\zeta_\perp$ should   be {\em thermal}, which for $\alpha=1$ means logarithmically rough ($\int_k \rme^{ikz}/|k| \sim \ln z$). This agrees with \cite{RamanathanErtasFisher1997}, and was  experimentally verified in \cite{DalmasLelargeVandembroucq2008}.

The question arises whether this LR universality class, and especially the roughness exponent 
$\zeta=0.38$ can be seen in an experiment. The first  such experiment is in  Ref.~\cite{BonamyPonsonPradesBouchaudGuillot2006}, using a very brittle material. 
The authors of this study conjecture that \cite{BonamyPonsonPradesBouchaudGuillot2006}
``both critical scaling regimes can be observed in all   heterogeneous materials:'' For length scales smaller  than the process  zone, the larger exponents ($\zeta \approx 0.75$, $\hat \beta \approx 0.6$, $z = \zeta /\hat \beta \approx 1.2$) should be relevant, and the fracture surface was reported to be multi-fractal \cite{VernedePonsonBouchaud2015}. 
For larger scales the exponents are those of depinning ($\zeta \approx 0.4$, $\hat \beta =0.5$, $z = \zeta/\beta\approx 0.8$).
However, these observations were made for the transversal roughness, accessible post-mortem, whereas according to the scenario proposed above it should hold for the longitudinal roughness which is   less   accessible in experiments. The emerging consensus \cite{PonsonPrivate} of the  community seems to be that the large-scale roughness in the longitudinal direction is $\zeta_\parallel \approx 0.38$, whereas the transversal roughness $\zeta_\perp =0$ (logarithmic rough), as observed in \cite{DalmasLelargeVandembroucq2008}; and that whenever a roughness of $\zeta_\perp \approx 0.38$ has been observed,  it has to do with physics related to short scales (damage zone). 

What is the {\em process-zone} mentioned above? 
The standard theory for fracture is based on work by Griffith \cite{Griffith1921}, with a crucial improvement by Irwin \cite{Irwin1957}. The idea of Griffith \cite{Griffith1921} was to write an energy balance between the stress released by the crack, and the surface energy necessary to create it. Irwin \cite{Irwin1957} realized that for ductile material, part of the released energy  goes into a plastic deformation, i.e.\ heat,   at the crack front. The size of the zone affected is the {\em process zone}. 
It  ranges from $\xi = 50\pm 9\mu $m  for ceramics, over $\xi = 170\pm 12\mu $m for aluminum  to $\xi = 450 \pm 35 \mu $m for mortar  \cite{VernedePonsonBouchaud2015}.

\begin{figure}[b]
{\scalebox{0.96}{\fboxsep0mm\mbox{\begin{tikzpicture}
\node at (4.3,1.5) {\includegraphics[width=7.5cm]{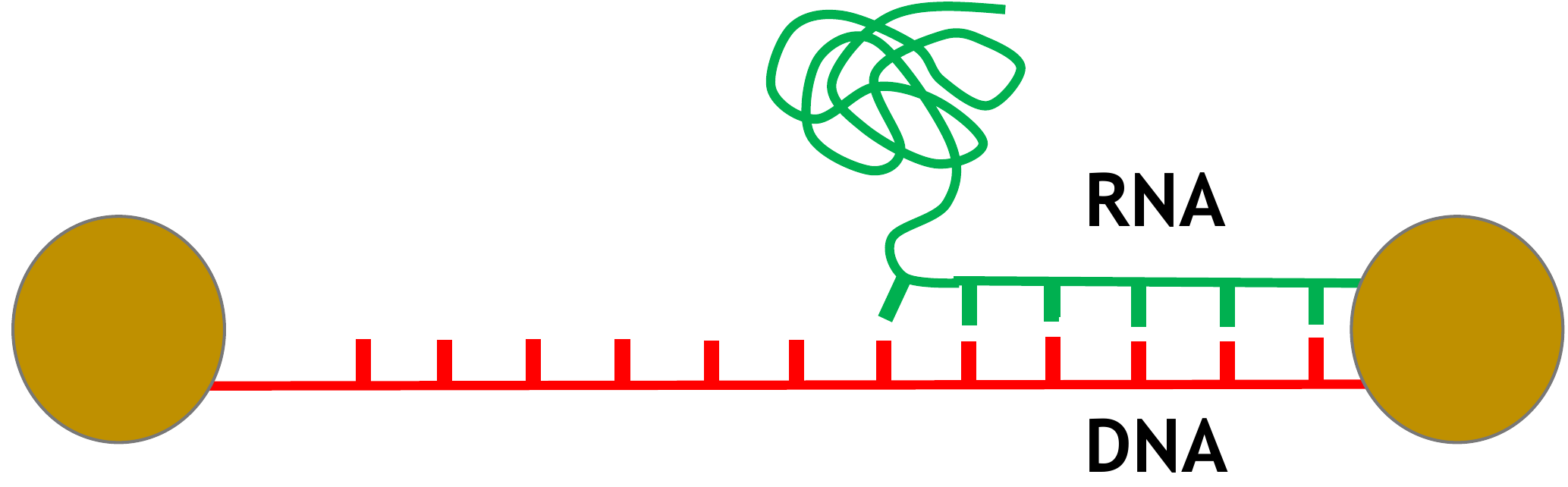}};
\draw [blue,thick] (1,0) parabola (2,1.5) ;
\draw [blue,thick] (1,0) parabola (0,1.5) ;
\draw [blue,thick] (7.6,0) parabola (8.6,1.5) ;
\draw [blue,thick] (7.6,0) parabola (6.6,1.5) ;
\draw [thick,dashed] (1,0) -- (1,-0.35);
\draw [thick,dashed] (7.6,0) -- (7.6,-0.35);
\draw [thick,->] (4,-0.25) -- (1,-0.25) ;
\draw [thick,->] (4.6,-0.25) -- (7.6,-0.25) ;
\node at (4.3,-0.25) {$w$};
\draw [thick,dashed]  (1.1,3.05) -- (1.1,1);
\draw [thick,dashed] (7.5,3.05) -- (7.5,1) ;
\draw [thick,->] (4,3) -- (1.1,3) ;
\draw [thick,->] (4.6,3) -- (7.5,3) ;
\node at (4.3,3) {$u$};
\end{tikzpicture}}}}
\caption{Peeling of an RNA-DNA double strand. The RNA sequence  is from subunit 23S of the ribosome in E.~Coli, prolonged to attach the beads (brown circles, with a much larger radius than drawn here). The DNA sequence is its complement. The beads sit in an optical trap (blue), at a distance $w$. (Drawing not to scale.) Fig.~reprinted from \cite{WieseBercyMelkonyanBizebard2019}.}
\label{f:peeling}
\end{figure}

\begin{figure}[t]
{\fboxsep0mm
\mbox{\setlength{\unitlength}{0.95cm}\begin{picture}(8.6,5.83)
\put(0,0){\fig{0.45\textwidth}{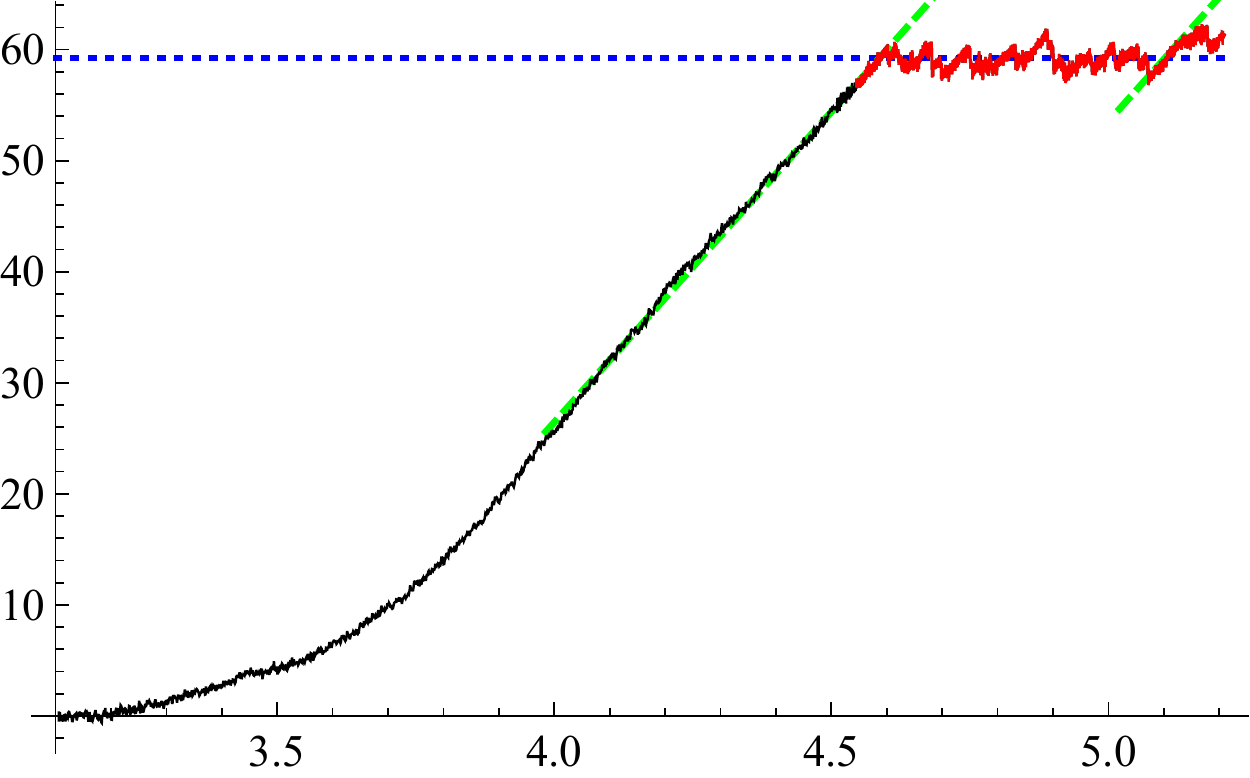}}
\put(7.3,0.6){\scalebox{0.8}{$w[\mu\rm m]$}}
\put(0,5.5){\scalebox{0.8}{$F[\rm pN]$}}
\end{picture}}}\caption{Left: A sample force-extension curve. For the data-analysis  only  the last plateau part of the curve  is used (in red).
The effective stiffness  $m^{2}$ in \Eq{eq-motion-f} is estimated from the slope  of the green dashed lines as $m^{2} = 55\pm5  \rm pN/\mu m$  at the beginning of the plateau, which remains at least approximately correct at the end of the plateau.
The driving velocity is about    $7 \rm n m/s$. Fig.~reprinted from \cite{WieseBercyMelkonyanBizebard2019}.
}
\label{f:RNA-unzipping}
\end{figure}

\subsubsection*{Fracture in thin sheets.}
In thin sheets, very different roughness exponents have been reported: $\zeta=0.48 \pm 0.05$ for polysterene, and $\zeta = 0.67 \pm 0.05$ for paper \cite{Ponson2016}. We may speculate that the larger one is related to directed percolation (section \ref{s:A short summary on directed percolation}).

\subsubsection*{Random fuse models.}
Random fuse models, a.k.a.\ {\em damage percolation}, have been proposed 
\cite{deArcangelisRednerHerrmann1985}
as a model for fracture: Consider a regular lattice, where on each bond is placed a fuse of unit resistance, and a random maximum carrying capacity $i_c$ (maximal current), 
in most studies drawn from a uniform distribution,  $i_c \in [0,1]$. The system may be 2 or 3-dimensional, with a voltage applied in one direction. To avoid finite-size effects due to the electrodes, it is advantages to use periodic boundary conditions \cite{BatrouniHansen1998}, with an additional voltage gain $V$ in one dimension. The voltage is then ramped up from 0, until one of the fuses exceeds its carrying capacity, at which point it is considered broken, i.e.\ having an infinite resistance. One then recalculates the current distribution and checks whether another fuse   breaks. If not, one increases the applied voltage. 

This is an interesting model for fracture: (i) by solving the Laplace equation to find the current distribution, it incorporates the  elasticity of the bulk of the material, providing an effective long-range elasticity; (ii) when a part of the material is broken, it is removed. 
It incorporates ingredients found in Laplacian walks   (section \ref{s:Loop-erased random walks, and other models equivalent to CDWs}) and  DLA (solving in both cases the Laplace equation to determine the most likely point of action), and cellular automata as {\bf TL92} (section \ref{s:qKPZ}).

A roughness exponent of the fracture surface in $d=2+1$ was reported to be $\zeta = 0.62\pm 0.05$ \cite{BatrouniHansen1998}, apparently not too different from some experiments \cite{BatrouniHansen1998}.
Other authors focused on the distribution of strength, or broken fuses upon failure 
\cite{NukalaimunoviZapperi2004,ZapperiNukalaSimunovic2005,ZapperiNukala2006}. A variant is the fiber-bundle model \cite{GjerdenStormoHansen2014,StormoLenglineSchmittbuhlHansen2016}.

\subsection{Experiments for peeling and unzipping}
\label{s:Experiment on RNA-DNA unzipping}
\begin{figure}[t]\setlength{\unitlength}{0.1\columnwidth}
\fboxsep0mm
{\begin{picture}(10,8.7)
\put(3,2.5){\fig{5.82cm}{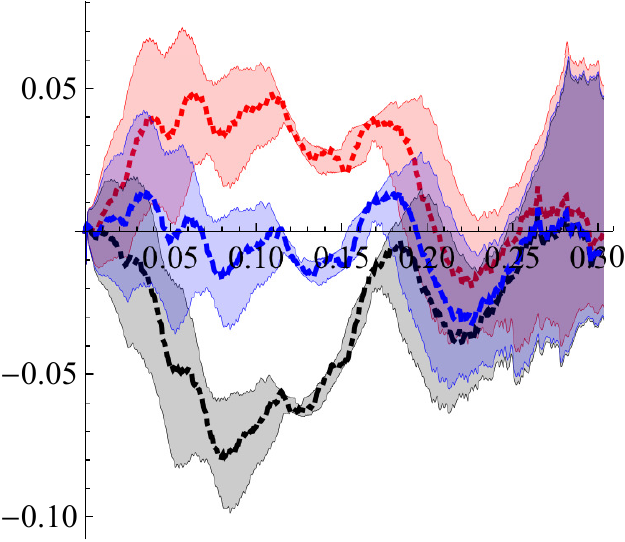}}
\put(0,0){\Fig{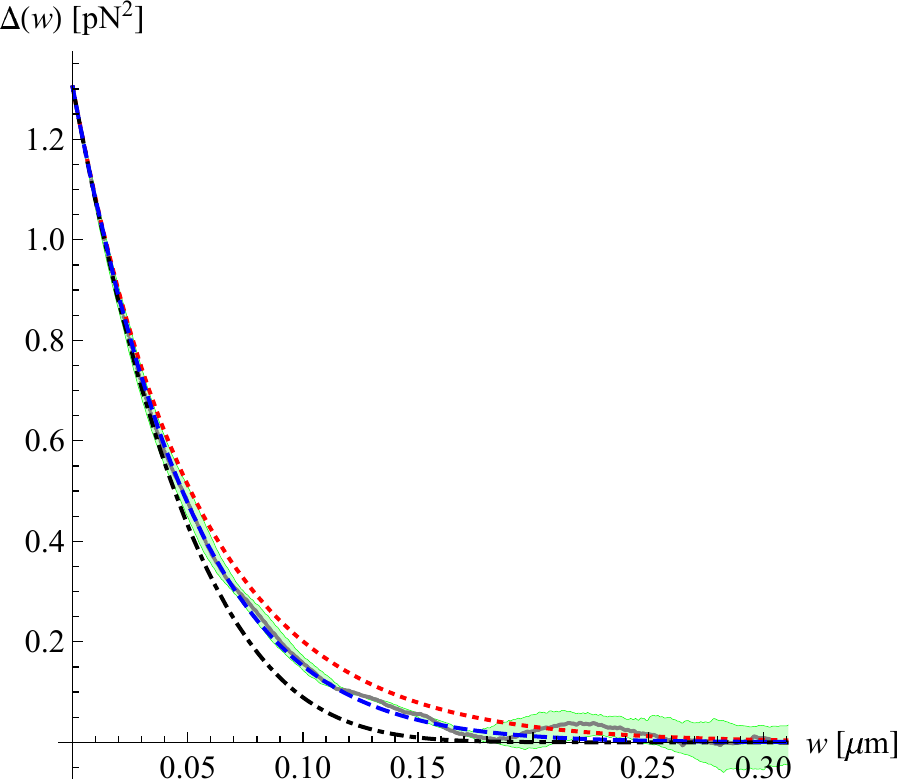}}
\end{picture}}
\caption{Measurements of $\Delta(w)$ (in grey), with 1-$\sigma$ error bars (green shaded), compared to three theoretical curves: pure exponential decay (dotted red), 1-loop FRG, \Eq{Delta=y...} (black dot-dashed), and  DPM,  \Eq{DeltaGumbel} (blue dashed), all rescaled to have the same value and slope at $u=0$. Inset:  theoretical curves with the data subtracted (same color code). The blue curve is   the closest to the data. The correlation length estimated from  $\Delta(w)$  is $\xi = 0.055\pm0.005 \mu \rm m \simeq 186$ base pairs. Fig.~reprinted from \cite{WieseBercyMelkonyanBizebard2019}.
}
\label{f:our-choice}
\end{figure}

There are two ways to open a double helix made out of two complementary RNA or DNA strands, or one RNA and its complementary DNA strand: peeling and unzipping. In both cases beads are fixed to the molecules, and then pulled in an optical or magnetic trap. 
 In the literature, the word {\em peeling}  is used for the setup of Fig.~\ref{f:peeling},  where forces act along the helical axis from opposite  extremities of a duplex, and one of the two strands peels off.  {\em Unzipping} denotes an alternative setup where the right bead of Fig.~\ref{f:peeling}
 is attached to the free end of the upper strand. 
 As the reader can easily verify with a twisted thread, unzipping is much easier to accomplish than peeling.
Let us start with peeling \cite{WieseBercyMelkonyanBizebard2019}, for which a typical force-extension curve is shown in Fig.~\ref{f:RNA-unzipping}. The stationary regime is the plateau part (in red). 
Averaging over about 400 samples, the effective disorder $\Delta(w)$ defined in \Eq{defDe} is measured. 
The resulting curve, including error bars for the shape \cite{WieseBercyMelkonyanBizebard2019}, is shown in grey in Fig.~\ref{f:our-choice}, where it is compared to three theoretical curves: an exponentially decaying function (red, dotted, top curve), the DPM solution \eq{DeltaGumbel} for the Gumbel class (blue, dashed, middle curve), and the 1-loop FRG solution given by \Eqs{Delta=y...} and \eq{RF-FP-y}, all rescaled to have the same value and slope at $u=0$. The experiment clearly favors the DPM solution, best seen in the inset of Fig.~\ref{f:RNA-unzipping}. While this is expected, it is a nice confirmation of the theory in a delicate experiment.

 One should be able to extract $\Delta(w)$ also from the unzipping   of a hairpin. Interestingly, experiments report that the  scaling of  \Eq{zeta-DPM} is   replaced by \cite{HuguetFornsRitort2009}
\be
\rho_m \sim m^{-4/3}, \quad \mbox{i.e.}\quad
\zeta = \frac43.
\label{271}
\ee 
This is  a clear signature of a different universality class, namely   ``random-field'' disorder in equilibrium, for which the roughness exponent \eq{zeta-RF-1loop}  to all orders in $\epsilon$ reads $\zeta=\epsilon/3$; setting $\epsilon =4$ leads to \Eq{271}. An analytic solution is given in section \ref{s:Sinai}.
This scenario is possible through the much larger effective stiffness $m^2$ there, which manifests itself in correlation lengths of $\xi = 1$ to $35$ base pairs, as compared to $\xi = 186$ base pairs for peeling. Equilibrium is observed experimentally \cite{HuguetFornsRitort2009}
 through a vanishing hysteresis curve.

\subsection{Creep, depinning and flow regime}
\begin{figure}[t!]
\centerline{\fig{7.5cm}{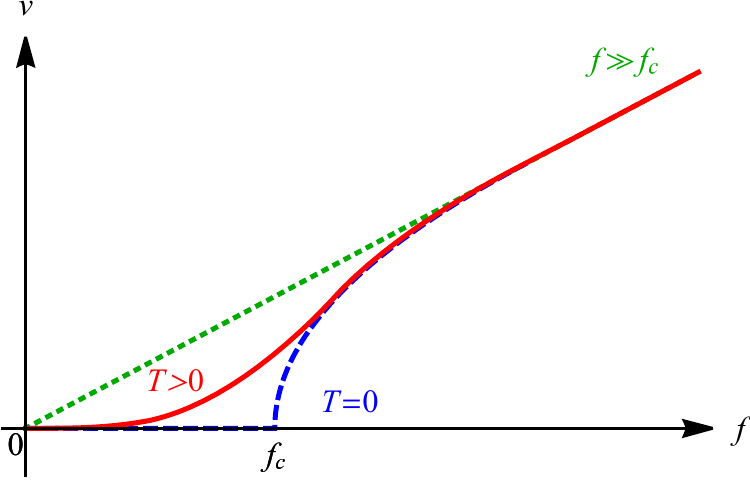}}
\caption{Sketch of velocity force curve at vanishing ($T=0$, depinning) and finite temperature ($T>0$, creep). For an experimental test see Fig.~\ref{fig-RF-dep-experiment}.}
\label{f:creep1}
\end{figure}

In section \ref{Larkin},  \Eqs{a8}-\eq{free-energy-scaling}, we had argued that in equilibrium  the elastic energy scales as
\be\label{187}
\ca E_{\rm el} (\ell)\sim  \ell^\theta\komma \quad \theta = 2 \zeta_{\rm eq} + d-\alpha \komma
\ee
and as long as $\theta>0$ the temperature
$T$ is irrelevant at large scales. On the other hand, if the driving velocity $v=0$, and leaving the system enough time to equilibrate,  it is in equilibrium. As sketched in Fig.~\ref{f:creep1}, there are  three different fixed points: equilibrium ($v=f=0$, $T\to 0$), depinning ($T=0$, $v\to 0$ or $f\to f_{\rm c}$), and large $v$ or $f$, for which we expect 
$
\eta v = f
$.
Let us now consider perturbations of the equilibrium fixed point, i.e.\ $T$ small, and $f\ll f_{\rm c}$, commonly referred to as the {\em creep regime}. Scaling arguments   first proposed by Ioffe and Vinokur \cite{IoffeVinokur1987}, and 
Nattermann \cite{Nattermann1990}, were later put on  more solid ground   via FRG  \cite{ChauveGiamarchiLeDoussal1998,ChauveGiamarchiLeDoussal2000}. Scaling arguments compare the elastic energy \eq{187} with the energy gained through the advance of the interface, i.e.\ an avalanche of size $S$, 
\be
\ca E_{f}(\ell ) =- f \int _x \delta u(x) \equiv- f S \sim - f   \ell^{d+\zeta_{\rm eq}} \punkt
\ee
As $\zeta_{\rm eq} < \alpha$, the   energy $\ca E_{f}(\ell )$ dominates over  $\ca E_{\rm el}(\ell )$ for large $\ell$, and the {\em optimal fluctuation}
is obtained for $\partial_\ell [ \ca E_{\rm el}(\ell) +\ca E_f (\ell)] \stackrel !=0$,  
resulting in
$
\ell_{\rm opt}^{\zeta_{\rm eq}-\alpha} \sim  f, 
$ or 
\bea
\ell_{\rm opt} \sim f^{-\nu_{\rm eq}}\ , \quad \nu_{\rm eq} = \frac1{\alpha-\zeta_{\rm eq}},\\
\ca E_{\rm opt} \sim f^ {- \mu_{\rm eq}}, \ \mu_{\rm eq} = \nu_{\rm eq}\theta = \frac{2 \zeta_{\rm eq} + d-\alpha}{\alpha-
\zeta_{\rm eq}}.
\eea
This identifies the {\em creep law} as 
\be\label{creep-law}
v(f,T) = v_0 \,\rme^{-\frac{T_*}T \left(\frac {f_{\rm c}}f\right)^{\mu_{\rm eq}}}, \quad f\ll f_{\rm c}.
\ee
We remind that for depinning   (see \Eqs{v=|f-fc|^beta} and \eq{beta-rel})
\be
v \sim (f-f_{\rm c})^\beta, \quad f\ge f_{\rm c},
\ee
and that for large $f$
\be
v \simeq \frac f \eta,\quad f\gg f_{\rm c}.
\ee
There are thus three regimes, sketched in Fig.~\ref{f:creep1}: $f\ll f_{\rm c}$, the creep regime discussed above, governed by the $T=0$ equilibrium fixed point; $T=0$, and $f\approx f_{\rm c}$, the depinning fixed point; and the large-$f$ and large-$v$ regime, where the disorder resembles a thermal white noise, with amplitude proportional to $1/v$. The latter can be understood from the relation 
\numparts
\bea
\Delta^{\rm RF} (w) \simeq \delta(w) = \delta (v t) = \frac 1 v \delta(t),\\
\Delta^{\rm RB} (w) \simeq - \delta''(w) = - \delta'' (v t) = -\frac 1 {v^3} \delta''(t).
\eea
\endnumparts
More precisely, for RF it looks like a thermal noise with temperature 
\be
T^{\rm RF}=\frac1 v \int_0^\infty \rmd w \,  \Delta(w).
\ee
For RB disorder, the    noise decays as $T^{\rm RB}\sim 1/v^3$ \cite{EliasKoltonWiese2022}.

\subsubsection*{Creep in simulations.}
\begin{figure}
\Fig{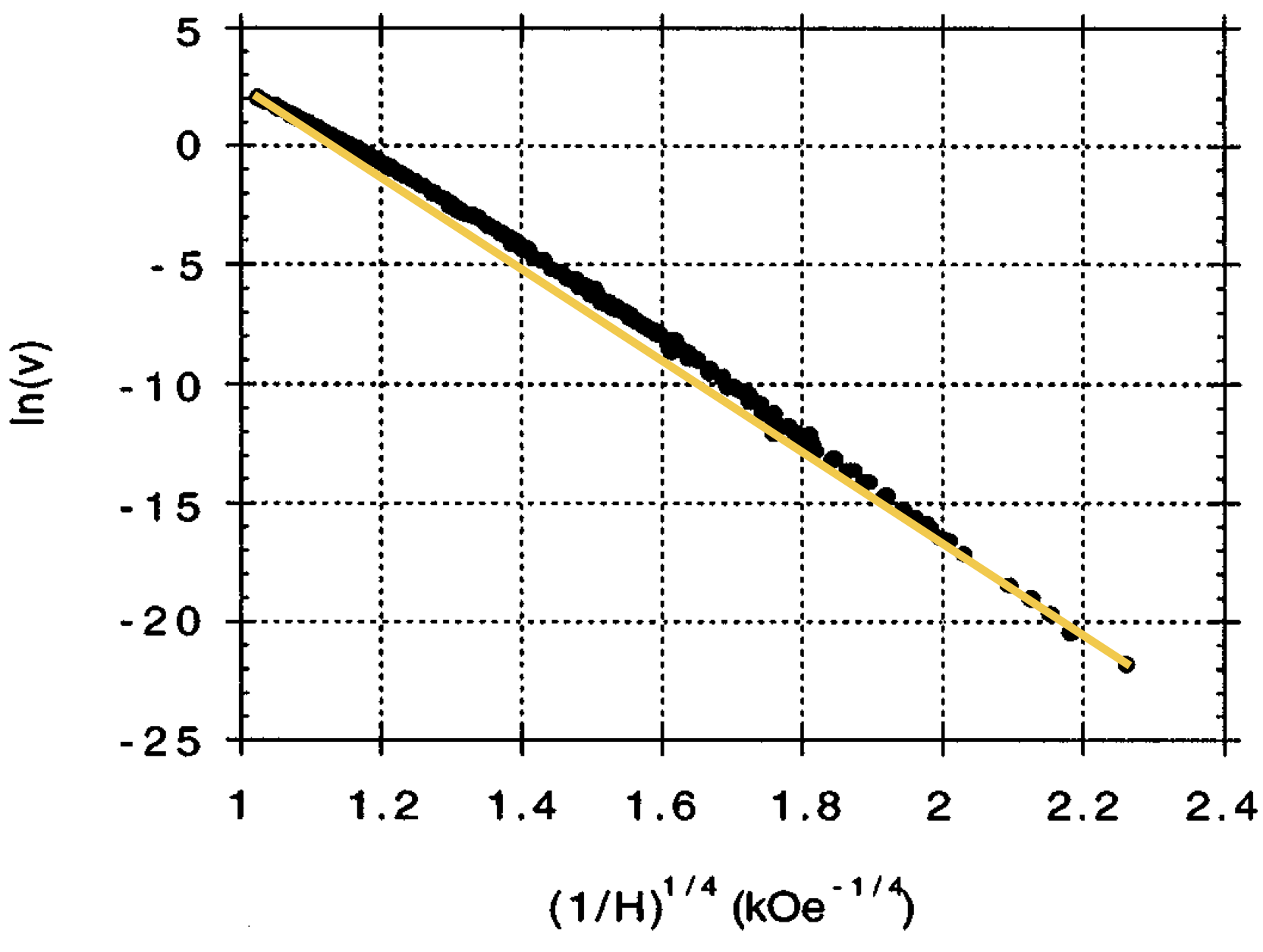}\caption{Experimental confirmation  of the creep-law $\ln (v) \sim f^{-\mu}$ 
in an ultrathin PtCoPt film  
\cite{LemerleFerreChappertMathetGiamarchiLeDoussal1998}. 
Tested is the hypothesis  $\mu_{\rm eq}^{\rm RB}=1/4$; 
as the added orange line shows, a larger value of $\mu_{\rm eq} \approx 1/3$ 
should improve the fit, consistent with a value of $\zeta>2/3$. One might see the beginning of the large-scale regime with $\zeta = \zeta_{\rm dep}=5/4$, see e.g.\ \cite{FerreroFoiniGiamarchiKoltonRosso2020}. A recent experiment is shown in Fig.~\ref{fig-RF-dep-experiment}.}
\end{figure}
In $d=1$, the creep law \eq{creep-law} was verified numerically \cite{KoltonRossoGiamarchi2005,KoltonRossoGiamarchiKrauth2006,KoltonRossoGiamarchiKrauth2009,KoltonBustingorryFerreroRosso2013,FerreroBustingorryKoltonRosso2013,FerreroFoiniGiamarchiKoltonRosso2017} both for random-bond and random-field disorder, and short-ranged elasticity ($\alpha=2$):
\bea
\zeta_{\rm eq}^{\rm RB}=\frac23~~&\Longrightarrow&~~ \mu_{\rm eq}^{\rm RB} = \frac14,\\
\zeta_{\rm eq}^{\rm RF}=1~~&\Longrightarrow&~~ \mu_{\rm eq}^{\rm RF} = 1.
\eea
The numerical work \cite{KoltonRossoGiamarchi2005,KoltonRossoGiamarchiKrauth2009,KoltonBustingorryFerreroRosso2013,FerreroBustingorryKoltonRosso2013,FerreroFoiniGiamarchiKoltonRosso2017} was possible through the realization that for $T\to
 0$ the sequence of states, in which the interface rests, becomes deterministic, and can be found by a clever enumeration of all possible saddle points.

\subsubsection*{Creep in Experiments.} 
The exponent $\mu_{\rm eq}^{\rm RB} =1/4$ was first found experimentally in Ref.~\cite{LemerleFerreChappertMathetGiamarchiLeDoussal1998}, 
and later confirmed in numerous other magnetic domain-wall experiments \cite{RepainBauerJametFerreMouginChappertBernas2004,MetaxasJametMouginCormierFerreBaltzRodmacqDienyStamps2007,GorchonBustingorryFerreJeudyKoltonGiamarchi2014,JeudyMouginBustingorrySavero-TorresGorchonKoltonLemaitreJamet2016,Diaz-PardoSavero-TorresKoltonBustingorryJeudy2017}. These experiments use a Kerr microscopic to image the domain wall; a sample image is given in Fig.~\ref{exp:Magnet}. At large scales, the domain-wall roughness is expected to  cross over to $\zeta_{\rm qKPZ}=0.63$ or $\zeta_{\rm dep}=1.25$,  see the discussion in section \ref{s:Experiments on thin magnetic films}.

Creep motion was in less depth     studied in vortex lattices \cite{TroyanovskiAartsKes1999}, 
fracture experiments \cite{TallakstadToussaintSantucciSchmittbuhlMaaloy2011,Vincent-DospitalCochardSantucciMaloyToussaint2020}, and quantum systems 
\cite{GorokhovFisherBlatter2002,NattermannGiamarchiLeDoussal2003,AndreanovFedorenko2014}.

For $f=f_{\rm c}$ (critical driving), Ref.~\cite{KoltonJagla2020} claims that for periodic disorder
\be
v(T,f=f_{\rm c}) \sim T^{\chi}, \quad \chi = \frac{d+2}{6-d}.
\ee
In the fixed-velocity ensemble, the scaling \eq{v=|f-fc|^beta} suggests
\be
 \lim_{v\to 0} \left[ \left< f(0,v)\right> - \left< f(T,v)\right> \right]\sim T^{\chi/\beta}. 
\ee

\subsection{Quench}
\label{s:Quench}

In most of this review we studied situations where the system is equilibrated, either in its ground state, or in the steady state. One may   ask how it reacts to a quench. This question was first considered for model A (Langevin dynamics for   $\phi^4$-theory, classification of \cite{HohenbergHalperin1977}) in Ref.~\cite{JanssenSchaubSchmittmann1989}. There one starts with a system at $T\gg T_{\rm c}$, where   correlations vanish. At $t=0$ one {\em quenches} it to $T=T_{\rm c}$. The response function $R(q,t_w,t)$ 
then depends on $t$ where one measures the field, and a {\em waiting time} $t_w<t$ at which a small kick was performed.
For disordered elastic manifolds, the state with vanishing correlations is a flat interface. It can be obtained   by imposing $u(x,t=0)=0$,   by moving the interface with a very large velocity $v\gg 1$ up to $t=0$,  or by switching on the disorder at time $t=0$. 
Scaling implies that $R$  takes the   form\footnote{The dimension is $R(q,t)\sim t^{\frac{2-z}z}\sim  1/({q^2 t})$, s.t.\ $\int_tR(q,t)\sim  q^{-2}$, reflecting the non-renormalization of the elasticity (STS, section \ref{s:tilt-symmetry}).}
\bea\label{343}
R(q,t_{\rm w},t) {=} \Big( \frac t {t_{\rm w}}\Big)^{\!\theta_{\rm R}} (t{-}t_{\rm w})^{\frac{2-z}z} f_{\rm\scriptscriptstyle R}(q^z(t{-}t_{\rm w}), t/ t_{\rm w}),  \nn \\
f_{\rm \scriptscriptstyle R}(x,y)\to \mbox{const} ~~\forall x\to 0, ~\mbox{or} ~  y\to \infty. 
\eea
A similar ansatz holds for the correlation function. 
\Eq{343} would simplify if 
\be\label{thetarel?}
\theta_{\rm \scriptscriptstyle R} \stackrel ? = \frac{z-2} z.
\ee
In model A,    $\theta_{\rm \scriptscriptstyle R}$ violates   \Eq{thetarel?} at 2-loop order \cite{JanssenSchaubSchmittmann1989}\footnote{Ref.~\cite{JanssenSchaubSchmittmann1989} does not state the $\epsilon$ expansion for $\theta_{\rm \scriptscriptstyle R}$, but for a related object $\eta_0$. The missing relation is 
$
\theta_{\rm \scriptscriptstyle R} =-  \frac{\eta_0}{2z}
$,
 confirmed in \cite{ChenGuoLiMarculescuSchulke2000}.}. For disordered elastic manifolds at depinning, \Eq{thetarel?} is satisfied at 2-loop order \cite{SchehrLeDoussal2005,KoltonSchehrLeDoussal2009}, but   may  be violated in simulations in $d=1$ \cite{KoltonSchehrLeDoussal2009}; to decide the matter,   larger systems   need to be simulated  \cite{KoltonPrivate}.

Technically, the two calculations are rather different: Imposing $\phi(x,t=0)=0$ in model A amounts to using Dirichlet boundary conditions. New divergences then appear between fields $\phi(x,t)$, and their mirror images at $t<0$. 
For disordered elastic manifolds no mirror images appear,  and one can simply switch on the disorder at $t=0$, reducing the possibility for independent divergences; it may thus well be that relation \eq{thetarel?} remains valid at all orders.

The situation simplifies in the limit of  $t_{\rm w}\to 0$. 
Standard power counting then implies that  (see section \ref{s:Phenomenology})
\be
\left< \dot u(t) \right> \sim t^{-\frac \beta{\nu z}} \equiv t^{\frac \zeta z -1} .
\ee
Similarly, the squared interface width    grows as 
\be
 {\left< L^{-2d}\textstyle \int_{x,y}[u(x)- u(y)]^2 \right>}  \to t^{\frac {2\zeta} z }\, .
\ee
In Ref.~\cite{FerreroBustingorryKolton2012} these relations  were used in simulations of system sizes up to $L=2^{25}$ to give the most precise (direct) estimation of the two independent exponents $\zeta$ and $z$ in dimension $d=1$, yielding $\zeta =1.25\pm 0.005$, $\nu = 1.333 \pm 0.007$, $\beta = 0.245 \pm 0.006,$ and $z = 1.433 \pm 0.007$. This should be compared to the values conjectured to be exact reported in table~\ref{dyn-data}.

A quench has also been studied in the Manna model \cite{DickmanAlavaMunozPeltolaVespignaniZapperi2001,KwonKim2016,TapaderPradhanDhar2020}, and interpreted as a dependence of the dynamical exponent $z$ on the initial condition. As $z$ is a bulk property, this is hard to believe. It seems \cite{KoltonPrivate}
that the systems   used in the simulations are too small to be in the asymptotic regime.

\begin{figure}[t]
\includegraphics[width=\columnwidth]{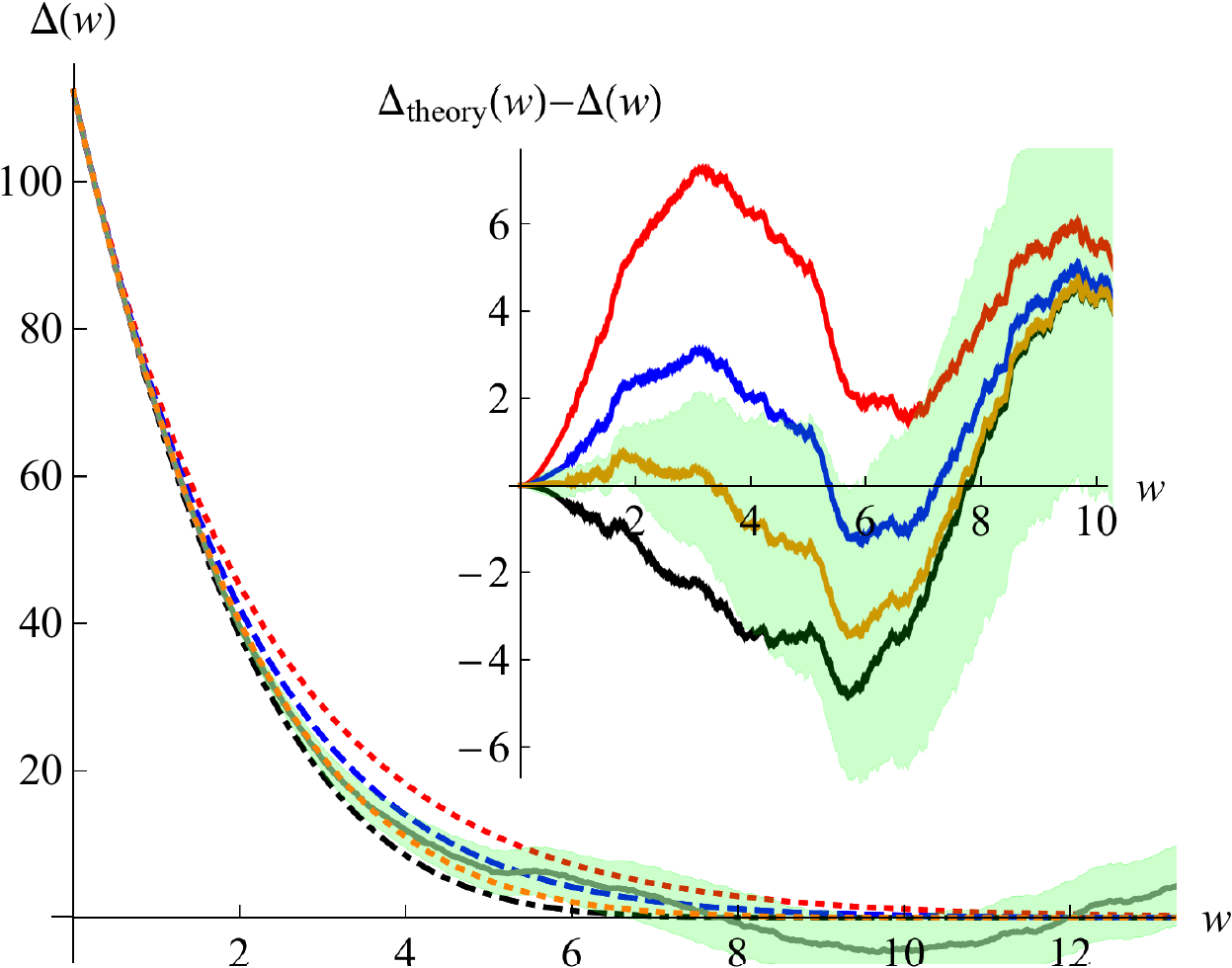}
\caption{Measured force-force correlator $\Delta(w)$ for a 200nm ribbon of FeSiB, a bulk   magnet with SR elasticity \cite{terBurgBohnDurinSommerWiese2021} (in grey, with error bars in shaded green, arbitrary units), after correcting for the finite driving velocity  (section \ref{s:The effective disorder, and  rounding of the cusp by a finite driving velocity}). This is compared to several theoretical curves (from top to bottom): an exponential function (red, dotted), the DPM correlator \eq{DeltaGumbel} (blue, dashed), FRG resummation \eq{Gumbel-from-Pade} for $\epsilon=2$  (orange, dashed), and 1-loop (black, dot-dashed). 
Error bars are at $68\percent$ confidence level.
The inset shows   theory minus   measurement, favoring the FRG fixed point at $\epsilon=2$  (with error bars  for this curve only).}
\label{FigSRNEC200nmFixedPoint} 
\end{figure}

\subsection{Barkhausen noise in magnets ($d=2$)}
\begin{figure}[t]
\includegraphics[width=\columnwidth]{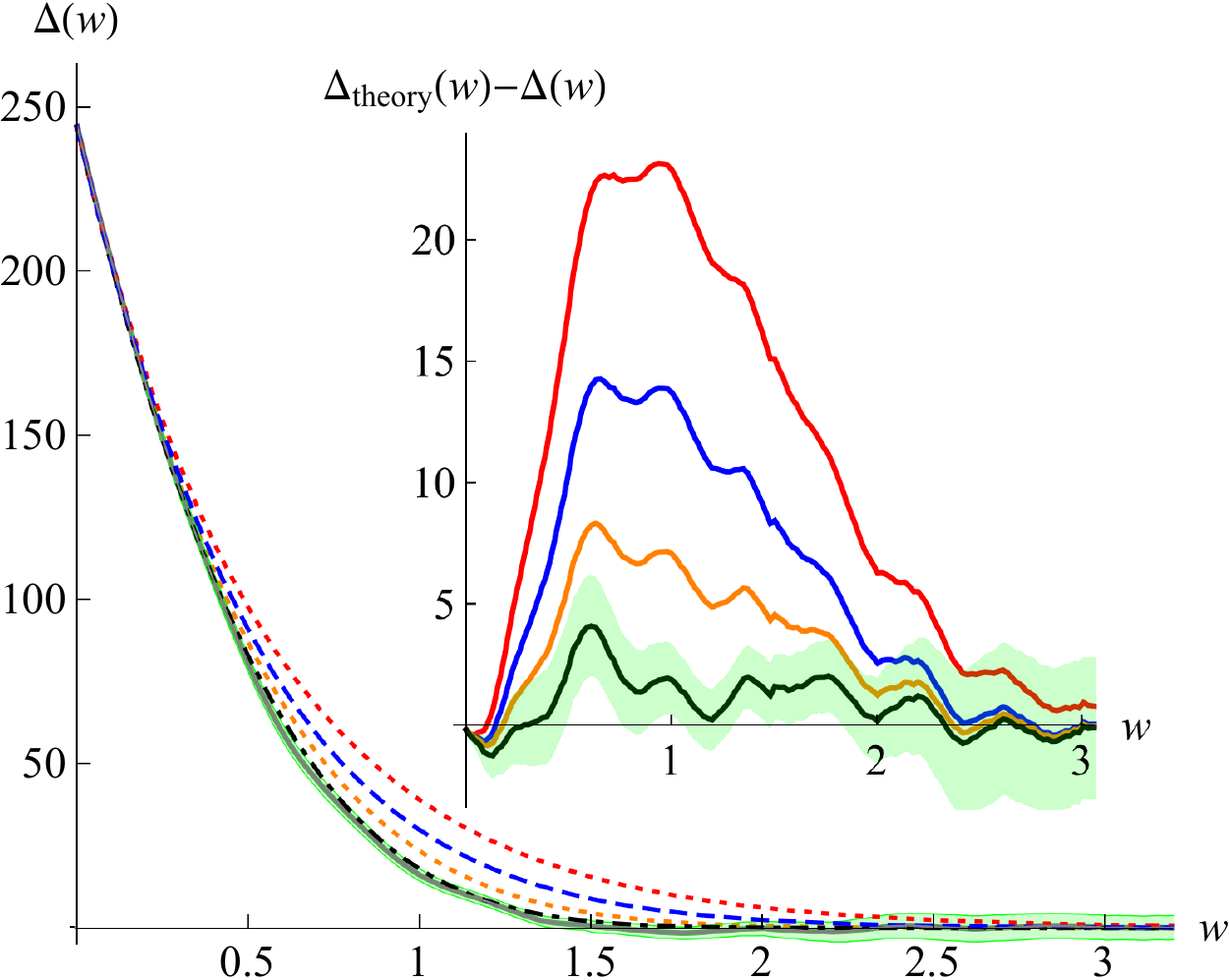}
\caption{Measured force-force correlator $\Delta(w)$  as in  figure \ref{FigSRNEC200nmFixedPoint},    for FeSi 7.8$\percent$, a bulk magnet   with strong dipolar interactions, making the elasticity long-ranged  \cite{terBurgBohnDurinSommerWiese2021}. The FRG prediction for $\Delta(w)$ is the 1-loop fixed point (section \ref{s:Behavior at the upper critical dimension}). For better visibility, error bars are at $90\percent$ confidence level, not accounting for a remaining oscillation from the power grid with period in $w$ of about $0.4$.}
\label{LRWEC78gr001timefixedpointinset} 
\end{figure}

To our knowledge, {\new domain walls in} bulk magnets are the only system     to realize depinning of a  2-dimensional manifold. Two universality classes need to be distinguished  \cite{DurinZapperi2006b,TheScienceOfHysteresis}: 
\begin{itemize}
\item
magnets with short-ranged elastic interactions, as the Ising model, for which $\alpha=2$ (notations as in section \ref{s:LR-elasticity}), and $\epsilon=2$. 
\item
 magnets with strong dipolar interactions,   which have long-range elasticity with $\alpha=1$, thus  $d_{\rm c}=2$ is the upper critical dimension (see section \ref{Theory for contact-line depinning}). 
\end{itemize}
For dynamic properties the influence of eddy currents, which varies from sample to sample,
  needs to be taken into account. A simple model is discussed in section \ref{s:Avalanches with retardation}. 
Here we consider the renormalized disorder correlator, for which eddy currents are less important.  The signal obtained  experimentally is the  current induced in a pickup coil, which we identify as $\dot u(t)$, the velocity of the center of mass of the interface. Integrating once yields $u(w=vt)$. $\Delta_v(w)$ is its auto-correlation function defined in \Eq{De-u}. 
Using the deconvolution procedure  of \Eq{242} (section \ref{s:The effective disorder, and  rounding of the cusp by a finite driving velocity}), 
allows one to extract the zero-velocity limit $\Delta(w)$. The latter is plotted in Fig.~\ref{FigSRNEC200nmFixedPoint} for the SR sample, and in Fig.~\ref{LRWEC78gr001timefixedpointinset} for the LR sample. 
For the SR sample, we expect $\Delta(w)$ to be closest to the resummed loop expansion at $\epsilon=2$, as given in \Eq{Gumbel-from-Pade}. For the LR sample, we expect the FRG-fixed point at the upper critical dimension (section \ref{s:Behavior at the upper critical dimension}), equivalent to the 1-loop FRG fixed point \eq{Delta=y...}--\eq{RF-FP-y}. In both cases, the agreement is very good, and clearly allows us to distinguish between the different universality classes. Let us stress that while the LR sample has critical exponents consistent with the ABBM model \cite{DurinZapperi2006b}, the disorder correlator $\Delta(w)$ is clearly distinct\footnote{As $\Delta(w)$ for ABBM is not renormalized, we  should measure it if  the microscopic disorder   were of the ABBM type.} from the one in ABBM, given in  \Eq{51}. 
 
The measured $\Delta(w)$ can  be compared to the correlator for RNA-DNA peeling in Fig.~\ref{f:our-choice} ($d=0$), and to   contact-line depinning ($d=1$, $\alpha=1$) in Fig~\ref{f:Delta3}. Note that for the latter the boundary layer due to the finite driving velocity (section \ref{s:The effective disorder, and  rounding of the cusp by a finite driving velocity}) was not deconvoluted.

\subsection{Experiments on thin magnetic films ($d=1$)}
\label{s:Experiments on thin magnetic films}
Thin magnetic films are appealing, as the position of the domain wall can be visualized using Kerr microscopy, see Fig.~\ref{exp:Magnet}.
There is a long and somewhat controversial interpretation of the results, which we summarize:

\begin{figure}[t]
\Fig{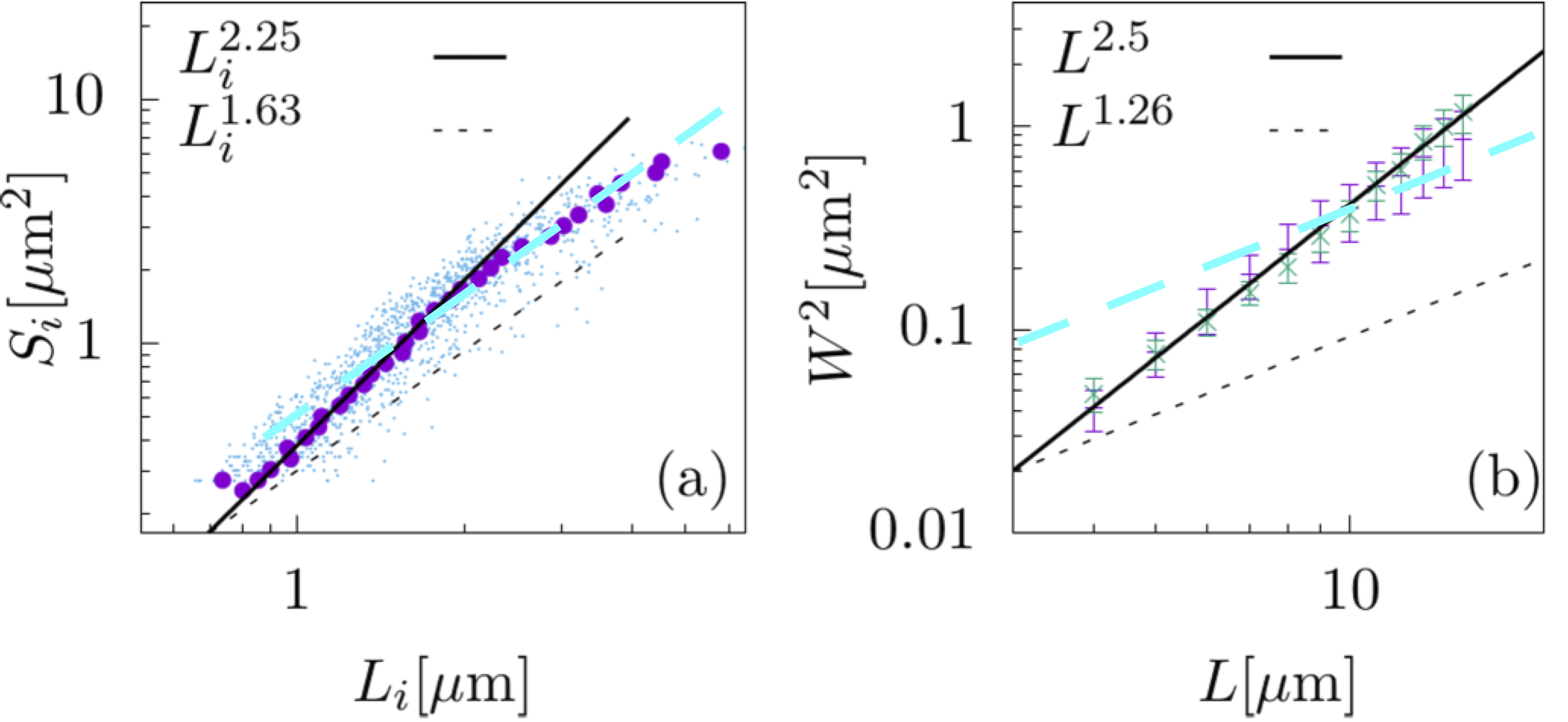}
\caption{Experimentally observed scaling of avalanche sizes (left) and squared width (right) for magnetic domain walls in ferromagnetic Pt/Co/Pt thin films. The solid lines indicate the exponent expected from depinning ($\zeta=1.25$), whereas dashed lines are for qKPZ ($\zeta=0.63$). From \cite{GrassiKoltonJeudyMouginBustingorryCuriale2018}, with kind permission.}
\label{f:qEW2qKPZ}
\end{figure}

\begin{figure}
\Fig{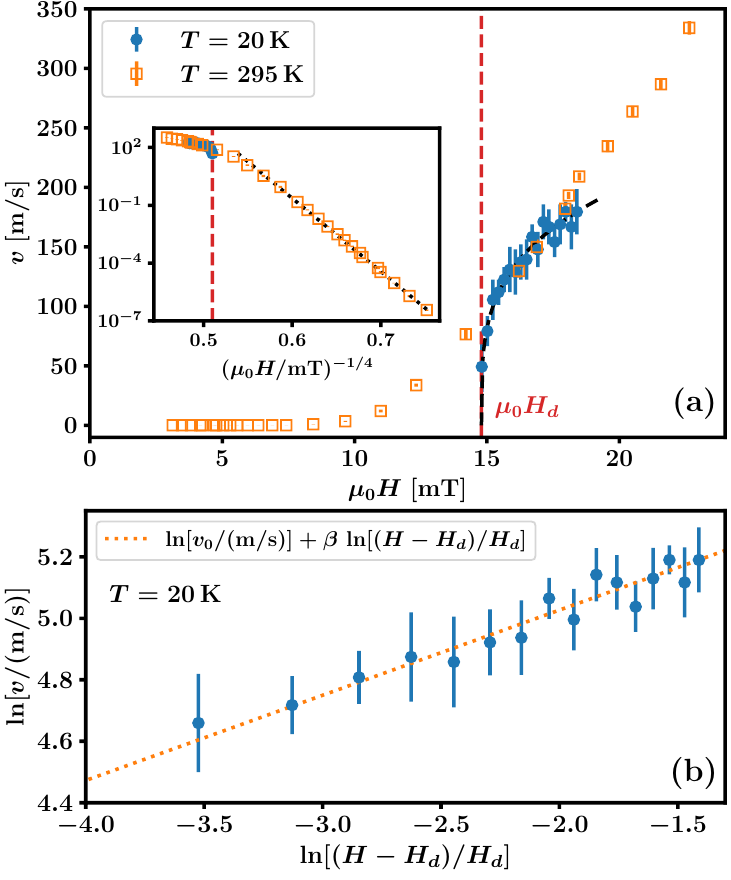}
\caption{The velocity as a function of applied force $f=\mu_0 H$ for a thin GdFeCo film (top). The sharp transition at $T=20$K is rounded at $T=295$K. The bottom plot shows the quality of the determination of $\beta=0.3\pm0.03$. The inset shows the creep law \eq{creep-law} with $\mu_{\rm eq}=1/4$.  Figure from \cite{AlbornozFerreroKoltonJeudyBustingorryCuriale2021}, with kind permission.}
\label{fig-RF-dep-experiment}
\end{figure}

\begin{enumerate}
\item 
Guided by a theoretical framework based on creep, the first experiments were interpreted as a perturbation of the RB equilibrium fixed point with $\zeta=2/3$ 
\cite{LemerleFerreChappertMathetGiamarchiLeDoussal1998}, even though configurations seem to be frozen. 

\item Creep exponents are reported   \cite{GorchonBustingorryFerreJeudyKoltonGiamarchi2014,Diaz-PardoSavero-TorresKoltonBustingorryJeudy2017,JeudyDiaz-PardoSavero-TorresBustingorryKolton2018} without a measurement of the roughness.

\item A   roughness exponent of $\zeta\approx 0.6$ was reported \cite{ShibauchiKrusin-ElbaumVinokurArgyleWellerTerris2001} together with a plateau of the 2-point function for large distances due to the confining potential, here a  consequence of dipolar interactions. 
(The plateau was interpreted as a smaller roughness $\zeta \approx 0.17$, a conclusion we do not share.)

\item Bound pairs of domain walls are reported \cite{BauerMouginJametRepainFerreStampsBernasChappert2005}, with no clear theoretical interpretation in terms of the phenomena discussed  here.

\item Ref.~\cite{MoonKimYooChoHwangKahngMinShinChoe2013} shows clear evidence for the  negative qKPZ class for current-induced depinning, and for the positive qKPZ class for field-induced depinning. We discuss this experiment   in section \ref{s:negative-QKZ}.

\item Refs.~\cite{Diaz-PardoMoisanAlbornozLemaitreCurialeJeudy2019,Diaz-PardoPhD} give the most complete analysis of the roughness exponent  to date. 
 The   beautiful image shown in Fig.~\ref{f:time-evolution-DW-shape} qualitatively confirms the findings of \cite{MoonKimYooChoHwangKahngMinShinChoe2013}. Note the strong up-down asymmetry which is inconsistent with an equilibrated systems (see section \ref{s:Characterisation of the  1-dimensional string}). 
 Nevertheless, the equilibrium RB fixed point with $\zeta_{\rm eq} = 2/3$ is still the key theoretical class the experiments are compared to. In the case of faceting as in Fig.~\ref{f:time-evolution-DW-shape}, the analysis is applied to fluctuations of the facets itself.

\item For a thin antiferromagnetic GdFeCo film, exponents consistent with the 1-dimensinal RF depinning class were found:   $\beta=0.30\pm 0.03$ and $\nu=1.3 \pm 0.3$     \cite{AlbornozFerreroKoltonJeudyBustingorryCuriale2021}, see Fig.~\ref{fig-RF-dep-experiment}. In particular the determination of $\beta$ is remarkable. The sample in question has a vanishing mass, $m^2\approx 0$.  Direct confirmation of a roughness $\zeta> \zeta_{\rm qKPZ}$ is   more tentative \cite{GrassiKoltonJeudyMouginBustingorryCuriale2018}, see Fig.~\ref{f:qEW2qKPZ}.

\item
 Experiments for alternating drive 
\cite{DomenichiniQuinterosGranadaCollinGeorgeCurialeBustingorryCapelutoPasquini2019}.

\item More experimental results can be found in Ref.~\cite{FerreMetaxasMouginJametGorchonJeudy2013}.

\end{enumerate}

\begin{figure}
\Fig{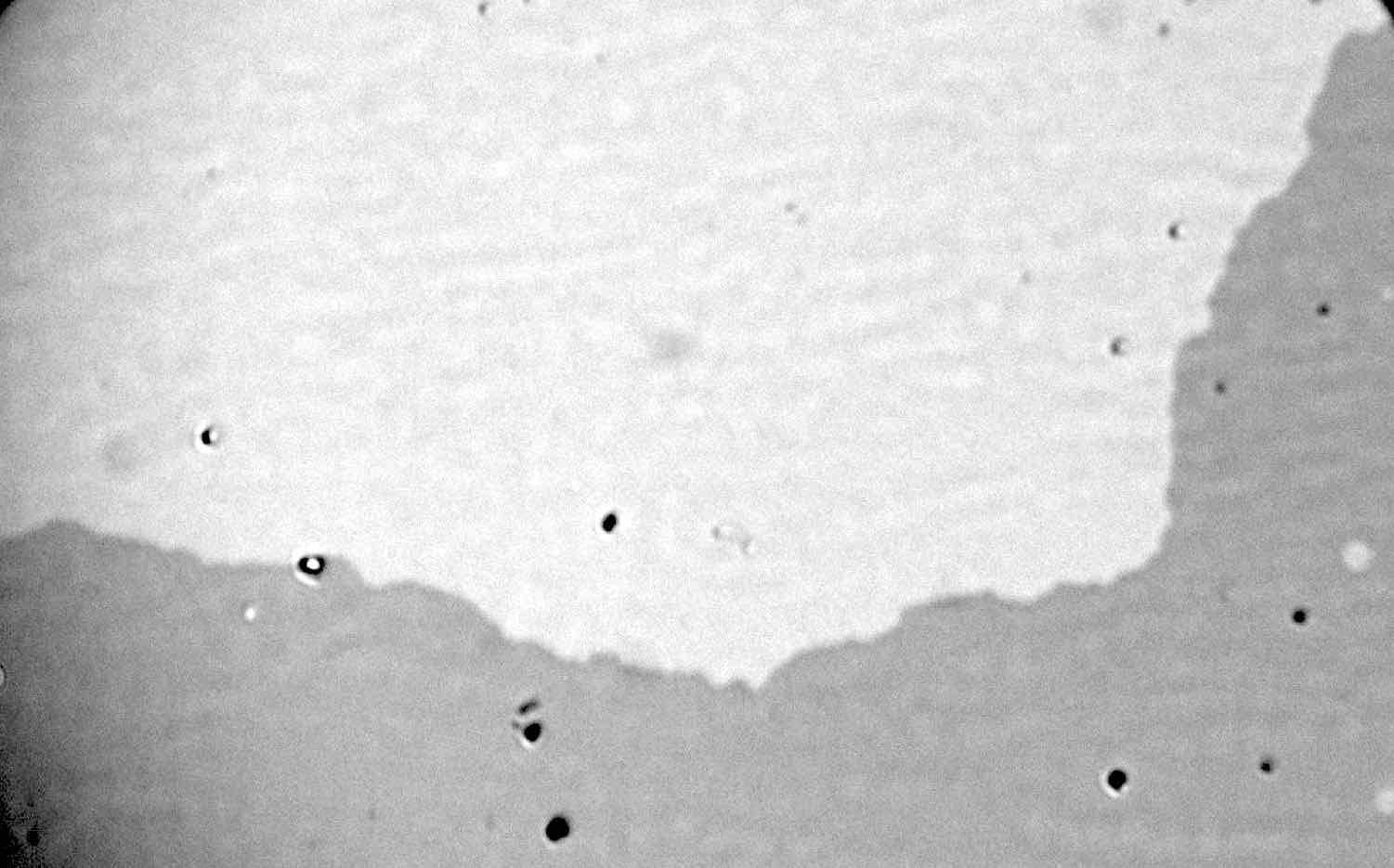}
\caption{PMOKE image of a domain wall in the GdFeCo sample used for Fig.~\ref{fig-RF-dep-experiment}. From \cite{Albornoz2021} with kind permission.}
\end{figure}

\noindent
To conclude:
Evidence for the equilibrium RB universality class with 
$\zeta=2/3$ (sections \ref{s:FRG-fixed-points} and \ref{beyond1loop}) seems to evaporate in favor of the quenched KPZ class with  $\zeta=0.63$   (section \ref{s:qKPZ}). At small scales depinning without KPZ-terms is visible \cite{GrassiKoltonJeudyMouginBustingorryCuriale2018}, but remains to be confirmed. Our conclusion is that at short scales KPZ terms are absent, leading to the RF-depinning class with $\zeta=1.25$. At larger scales, the KPZ term becomes relevant, and one crosses over to one of the qKPZ classes: positive qKPZ for field-induced driving, and negative qKPZ for current-induced driving, see Figs.~\ref{f:time-evolution-DW-shape} and \ref{f:mag-domain-walls-PRL-110-107203} (page \pageref{f:mag-domain-walls-PRL-110-107203}).

\subsection{Hysteresis}
\label{s:Hysteresis}
As we have seen in sections \ref{s:Phenomenology}, \ref{s:dep-loops} and \ref{s:DPM}, 
hysteresis in a driven disordered system is a sign of a non-vanishing force at depinning. In a real magnet the overall magnetization is bounded, thus the critical force depends on where one is on the hysteresis loop. This allows one to invent a plethora of protocols: One can try to get as close as possible into equilibrium by ramping up and down the magnetic field, while reducing the amplitude of the field in each cycle. One can also study {\em sub-loops}, by varying the applied field in a much smaller range than necessary for a full magnetization reversal. The reader wishing to enter the {\em Science of Hysteresis} can find a book with this title \cite{TheScienceOfHysteresis}, or one of the many original research articles \cite{LyuksyutovNattermannPokrovsky1999,NattermannPokrovskyVinokur2001,GlatzNattermannPokrovsky2002,KleemannRhensiusPetracicFerreJametBernas2007}. 
\begin{figure}[t]
\Fig{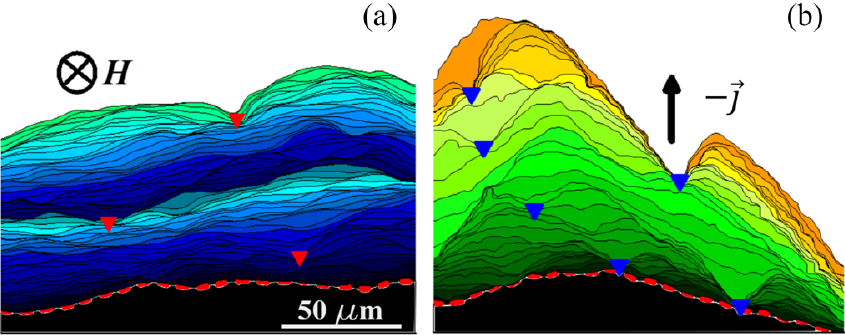}
\caption{Fig.~1 from \cite{Diaz-PardoMoisanAlbornozLemaitreCurialeJeudy2019} (with kind permission), showing the 
time evolution of the domain-wall  shape. (a) Successive positions   at $T = 28$K, 
driven by a magnetic field ($H = 0.16$mT, delay between images $t = 0.5 $s, total duration $60\rm s$).  
(b) ibid.~current driven ($j \approx 0.5 \rm GA/m^2$, $t= 0.5 \rm s$, total duration $16 s$.) 
observed at the same sample location. The DW moves in the direction opposite to the current density, which is indicated by the arrow. The initial DW position is underlined by a thick dashed line. The triangles indicate the strongest DW pinning positions.}
\label{f:time-evolution-DW-shape}
\end{figure}
There are few analytical result, a notable exception being the hysteresis curve in the ABBM model 
\cite{DobrinevskiPhD}.

\subsection{Inertia, and   a large-deviation function}
\label{s:Inertia}
Inertia plays an important role in everyday-life  with depinning: When we were children, we   pulled a block or a cart with the help of an elastic string, observing stick-slip motion, with a certain periodicity. At a higher frequency stick-slip motion may be  observed when breaking a bike or opening a door, often amplified by a resonance excited in the surrounding  medium. 
Despite its ubiquity in everyday life, stick-slip motion is absent in the avalanche phenomenology discussed above. The reason for this absence is the modeling of the equation of motion \eq{eq-motion-f} or \eq{eq-motion-w} via an overdamped Langevin equation, neglecting inertia. 

This is a good occasion to remind us how the overdamped Langevin equation \eq{eq-motion-w} is derived from 
Newton's equation of motion for depinning  ($w = vt$):
\bea\label{eq-motion-w+inertia}
M \partial_t^2 u(x,t) &=&
-\eta \partial_{t} u (x,t) + (\nabla^{2}-m^{2}) [u (x,t)-w] \nn\\
&& + F \big(x,u (x,t)\big)\punkt
\eea
Inertia $M$ times acceleration $\partial_t^2 u(x,t)$ are balanced by friction $-\eta \partial_t u(x,t)$,   forces exerted by the elasticity of the interface ($\sim \nabla^2 u(x,t)$), the confining well $m^2[ u(x,t)-w]$ and disorder $F(x,u)$. Neglecting inertia, i.e.\ setting $M\to 0$ yields back the standard equation of motion \eq{eq-motion-w}, written there with $\eta=1$. 

Assuming an exponential behavior $u(t) \sim \rme^{-t/\tau}$, the 
time scale $\tau$ satisfies
\be
 M \tau^{-2} -\eta \tau^{-1}+m^2  = 0.
\ee
The solution for $M\to 0$ starts with $\tau =\eta/ m^{2} $; when $M$ reaches 
\be
M_{\rm c} = \left(\frac{\eta}{2 m}\right)^{\!2}
\ee
this solutions splits into two complex ones, and movement becomes oscillatory. This can be interpreted as a dynamical phase transition.  
One may conjecture that this remains valid for an extended elastic system. This was indeed observed in MF theory \cite{SchwarzFisher2001}. A careful scaling analysis shows that this extends to systems below the upper critical dimension, where MF theory is no longer valid. 
Analytic progress was made 
\cite{LeDoussalPetkovicWiese2012} for various toy models generalizing ABBM (section \ref{s:ABBM}), i.e.\ quenched forces which have the statistics of a random walk. All these models share a common {\em large-deviation function} (a concept discussed below) for a large driving velocity $v$. They differ in how returns, which are difficult to incorporate in the field theory, are treated. If instead of returning on the same quenched disorder,  new random forces are generated with the same statistics of a random walk, then despite dissipation one finds a new active  steady state in the limit of a vanishing driving velocity $v\to 0$.

\paragraph{Large-deviation function.}
An interesting concept, well studied in the literature of driven systems, is the large-deviation function, see e.g.\ \cite{LebowitzSpohn1999,LeDoussalPetkovicWiese2012,MajumdarSchehr2014,KrapivskyMallickSadhu2014,SadhuDerrida2015}. 
Set $F_v(x)$ and $Z_v(\lambda)$ to be 
\be\label{LDF1}
\ca F_v(x):= - \frac{\ln\big(P(x v)\big)} v \ , \quad Z_v(\lambda) := \frac{\ln\big( \overline{ \rme^{\lambda \dot u}} \big)}{v} .
\ee
Define $\ca F(x)$ and $Z(\lambda)$ as the large-$v$ limits, if they exist,  of the above functions,
\be\label{LDF2}
\ca F(x):= \lim_{v\to \infty} \ca F_v(x), \quad Z(\lambda):= \lim_{v\to \infty}  Z_v(\lambda).
\ee
The existence of the second limit can be shown in field theory. In simple models,  $Z_v(\lambda)$ is even independent of $v$, see \Eq{central} in a simpler setting. Supposing the latter, we obtain
\bea
 \overline{ \rme^{\lambda \dot u}}  = \rme^{v Z(\lambda) } = \int_0^\infty \rmd \dot u  \, \rme^{\lambda \dot u} P(\dot u) \nn\\
=v \int_0^\infty \rmd x\, \rme^{v[\lambda x - \ca F_v(x) ]} .
\eea
If $v$ is large, the latter integral can be approximated by its    saddle point with $\lambda= \partial_x F_v(x)$. We recognize  a Legendre transform, 
\be
Z(\lambda) {+} \ca F(x ) = \lambda x, \quad   \lambda = \partial_x \ca F(x), \quad x = \partial_\lambda Z(\lambda). 
\ee
As by assumption $Z (\lambda)$ does not depend on $v$, this   shows that also the first limit in \Eq{LDF2} exists, and  $\ca F_v(x)\equiv \ca F(x)$, independent of $v$. The large-deviation function for the FBM model defined in section \ref{s:BFM}
with an additional inertia term as in \Eq{eq-motion-w+inertia} can then be constructed. Setting for simplicity   $m=\eta=1$, it reads \cite{LeDoussalPetkovicWiese2012} 
\bea
\ca F(x,M) = x {-} \ln(x){-}1 + M \!\Big[ \frac x2 {-}\frac 1{2x} {-} \ln (x)\Big]\! +\ca O(M^2), \nn\\
= \frac {(1{-}x)^2}2 + \frac {(1{-}x)^3}3  (1+  M)  +...  ,
\eea
where on the second line terms of order $(1{-}x)^4$ have been dropped. 
Units are restored by \cite{LeDoussalPetkovicWiese2012} 
\be
P_{v\gg 1}(\dot u) \simeq \rme^{-v \frac{ \eta m^2}{\sigma} \ca F(\frac{\dot u}v , \frac{M m^2}{\eta^2})}. 
\ee
A similar form holds for the joint distribution of velocities and   accelerations. 

\subsection{Plasticity}
Most systems and their deformations discussed so far  are {\em elastic}, indicating that conformational changes are reversible, and   nearest-neighbor relations   fixed. When the experiment is repeated, it passes through the same configurations, as given by Middleton's theorem (section \ref{s:Middleton}).
There are numerous systems where this is not the case, as   in sheared colloidal systems, termed {\em plastic} for their irreversible deformations. The question relevant for us is  how much of the phenomenology and methodology developed for disordered elastic manifolds carries over to plastic systems. The gap has not been bridged yet, but efforts have been undertaken starting from disordered elastic manifolds, 
\cite{VinokurNattermann1997,MarchettiMiddletonPrellberg2000,MarchettiSaunders2002,MarchettiDahmen2002,SaundersSchwarzMarchettiMiddleton2004,Marchetti2005,Marchetti2006,LeDoussalMarchettiWiese2008,FerreroJagla2019}, and plastic (mostly sheared colloidal) systems
\cite{NicolasMartensBocquetBarrat2014,AgoritsasBertinMartensBarrat2015,VasishtGoffMartensBarrat2018,TyukodiPatinetRouxVandembroucq2016}. A good starting point for the latter is the  review  \cite{NicolasFerreroMartensBarrat2018}. 

\subsection{Depinning of vortex lines or charge-density waves, columnar defects, and non-potentiality}
\label{s:Depinning of vortex lines or charge-density waves}
A   vortex line in 3-dimensional space,  driven through quenched disorder, has   the statistics of a depinning line in the driving direction ($\zeta_{\parallel}=5/4$, section \ref{s:Characterisation of the  1-dimensional string}) and   Gaussian fluctuations ($\zeta_{\perp}=1/2$) in the transversal direction \cite{ErtasKardar1994,ErtasKardar1996,EliasKoltonWiese2022}.  While the dynamical exponent $z_{\parallel}= 10/7$ in the driving direction is unchanged, the perpendicular dynamical exponent is   larger, 
\be
z_\perp = z_\parallel + 2-\zeta = \frac{61}{28} = 2.17857...
\ee 
We   updated the values for the exponents of  \cite{ErtasKardar1994,ErtasKardar1996} to today's best estimates (section \ref{s:Characterisation of the  1-dimensional string}).

A defect line binds a vortex line more strongly than point disorder. 
 Tilting the sample such that the  columnar defect no longer aligns with the magnetic field,  one observes an unbinding transition of the vortex line,
  known as the  {\em transverse Meissner effect} \cite{Balents1993,BalentsKardar1993,ChauveLeDoussalGiamarchi2000,Fedorenko2008,OliveSoretDoussalGiamarchi2003}. This is also  observed as an effective model for   sliding charge-density waves \cite{ChenBalentsFisherMarchetti1996}.

In the above setting, forces are assumed to be derivatives of a potential, i.e.\ conservative. If they are non-conservative, as e.g.\ in presence of stable advecting  currents, then a new universality class is reached,   accessible  perturbatively 
\cite{LeDoussalCugliandoloPeliti1997,LeDoussalWiese1997,WieseLedoussal1998}.

In section \ref{s:Bragg glass and vortex glass} we had shown experimental and theoretical evidence for the existence of an ordered phase in vortex lattices (Bragg glass) at weak disorder. Refs.~\cite{GiamarchiLeDoussal1996,LeDoussalGiamarchi1997}
argue that  the Bragg-glass phase is stable w.r.t.\ slow driving, with the lattice responding by flowing  through well-defined, elastically coupled, static channels. If the lattice is preserved, then after it has moved by a full lattice constant, it comes back   to its original configuration. In this case, one expects the velocity to be periodic in time 
\cite{BalentsFisher1995}. 

In \cite{BalentsMarchettiRadzihovsky1997} it was found that  translational order  in the driving direction can be destroyed.

\begin{reviewKay}
\subsection{Other universal distributions}\label{s:distribution}
\begin{figure} \centerline{\fig{0.5\textwidth}{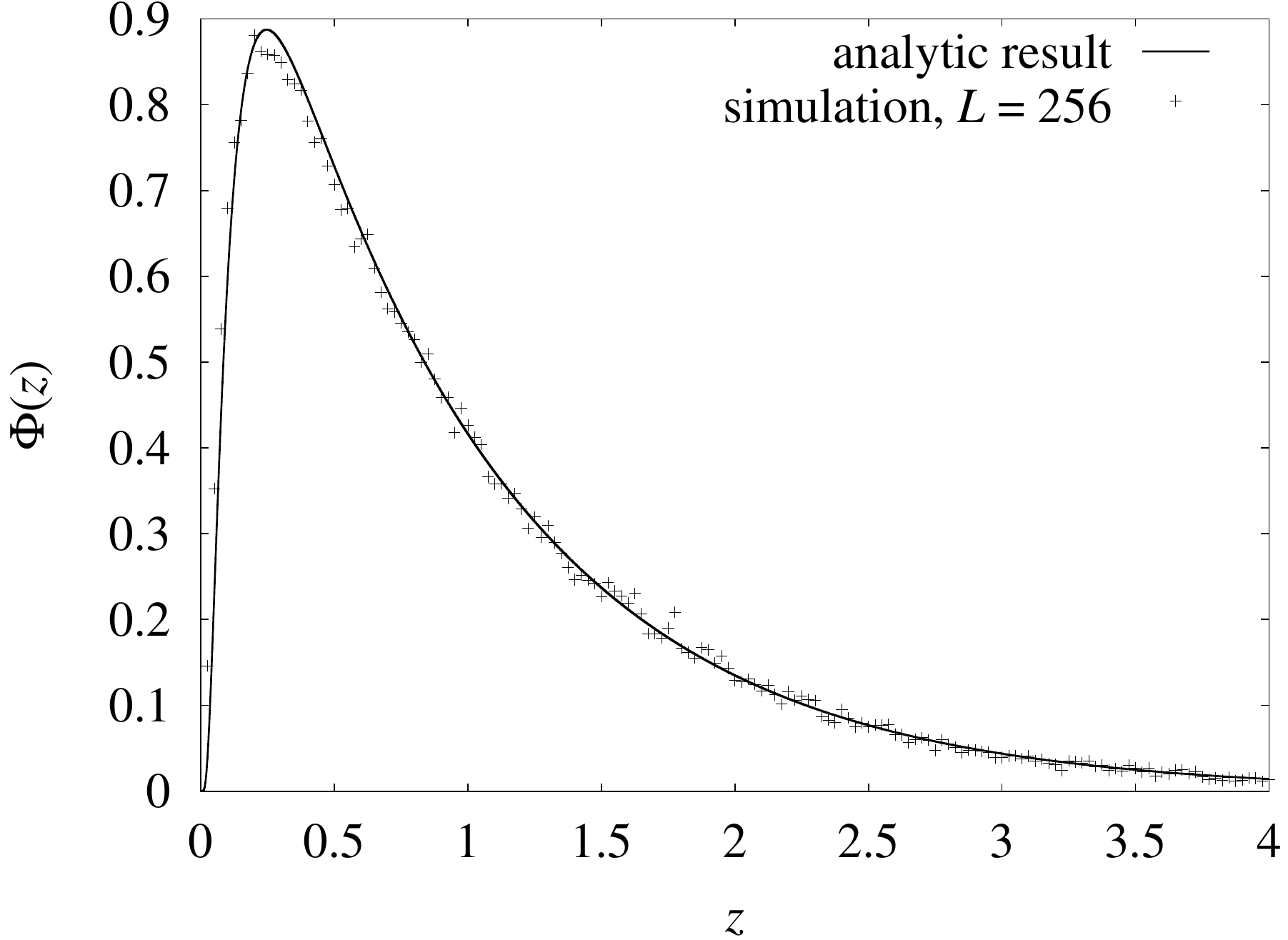}}
\caption{Scaling function $\Phi(z)$ for the ($1+1$)--dimensional
harmonic model, compared to the Gaussian approximation for
$\zeta=1.25$. Data and Fig.~from
\protect\cite{RossoKrauthLeDoussalVannimenusWiese2003}.}
\label{f:Dhm}
\end{figure}
Exponents are not the only interesting observables: In experiments and
simulations, often whole distributions can be measured, as e.g.\ the
 width distribution of an interface   at
depinning
\cite{RossoKrauthLeDoussalVannimenusWiese2003,LeDoussalWiese2003a,MoulinetRossoKrauthRolley2004}. Be
$\left< u \right>$ its spatial average   for a {\em
given} disorder configuration, then the   width
\begin{eqnarray}\label{w2}
w^{2}:= \frac{1}{L^{d}}\int_{x}\left(u (x)-\left< u \right> \right)^{2}
\end{eqnarray}
is a random variable, with
distribution $P (w^{2})$. The rescaled function $\Phi (z)$, defined by
\begin{equation}\label{width:Phi}
P (w^{2})={1}/{\overline{w^{2}}}\,\Phi
\left({w^{2}}/{\overline{w^{2}}} \right)
\end{equation}
is expected to be  universal, i.e.~independent of microscopic details and the
size of the system.

Supposing $u(x)$ to be Gaussian, $\Phi (z)$ was 
calculated analytically to leading order. It depends on two parameters, the roughness
exponent $\zeta$ and the dimension $d$. Numerical simulations \cite{RossoKrauthLeDoussalVannimenusWiese2003,LeDoussalWiese2003a}
displayed in Fig.~\ref{f:Dhm} show   agreement between
analytical and numerical results. The  distribution is  distinct from a Gaussian.

There are more observables of which distributions have been calculated
within FRG, or measured in simulations. Let us mention fluctuations of
the elastic energy \cite{FedorenkoStepanow2003}, and of the depinning
force \cite{FedorenkoLeDoussalWiese2006,BolechRosso2004}.
\end{reviewKay}

\begin{figure}
\Fig{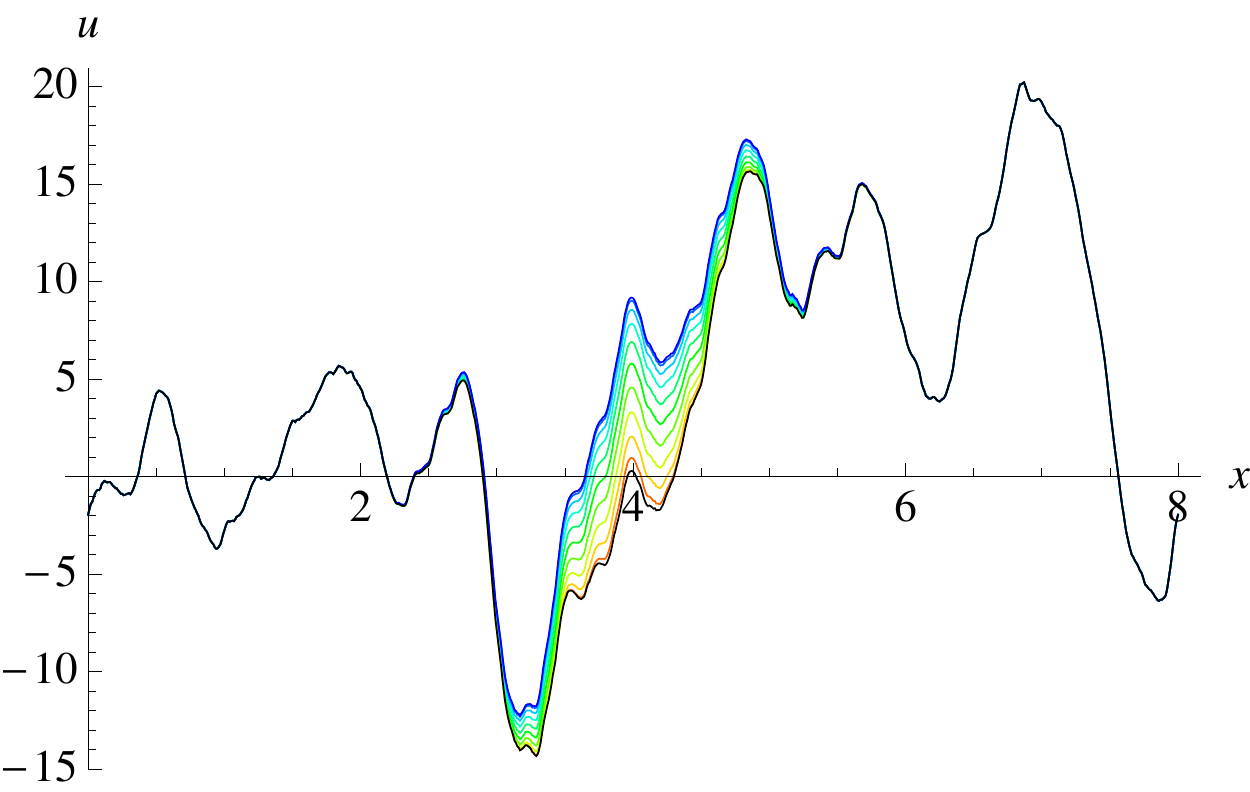}
\caption{Temporal evolution of an avalanche starting at $x=4$ at $t=0$, evolving to the top until time $T$. Interface positions at intermediate times $t=i T/10$ are shown for $i=1,2,...,10$. Fig.~reprinted from \cite{LeDoussalWiese2012a}.}
\label{f:one-avalanche}
\end{figure}

\section{Shocks and avalanches}
\label{s:Avalanches}
\begin{widefigure}
\centerline{{\parbox{0.38\textwidth}{\fig{0.38\textwidth}{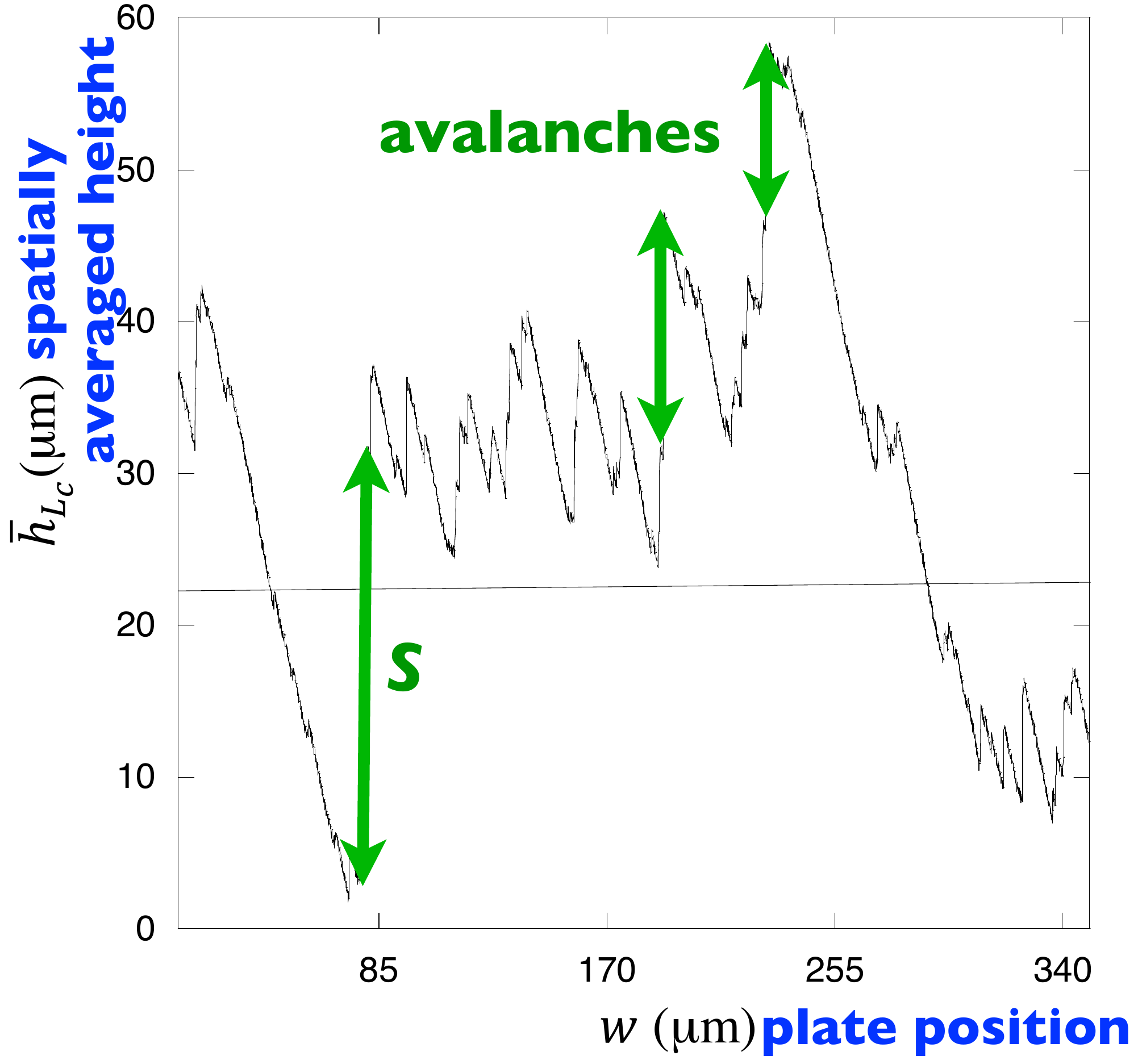}}~~\parbox{0.6\textwidth}{\fig{0.6\textwidth}{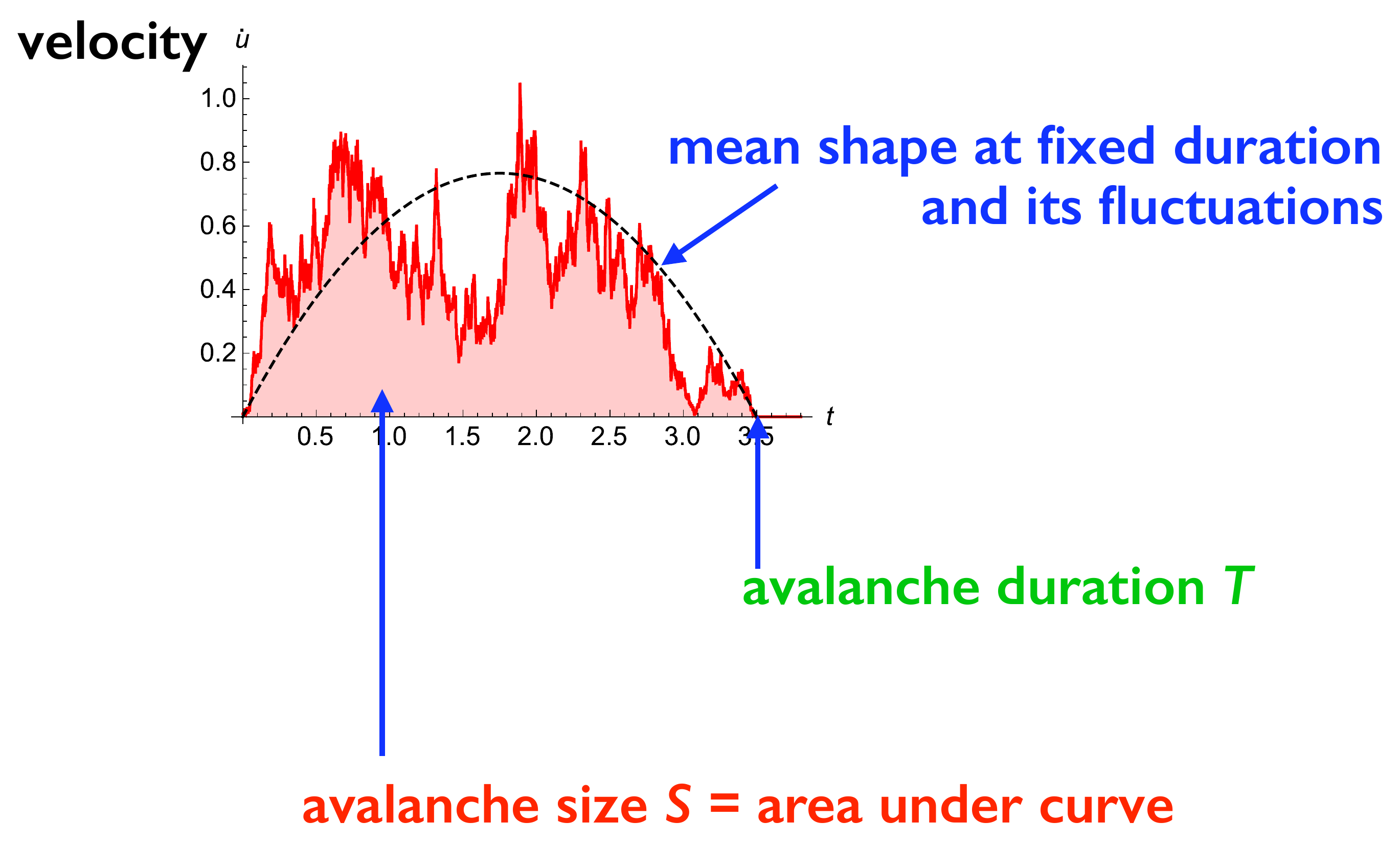}}}}
\caption{Left: The mean spatial height in the contact-line depinning experiment of Fig.~\ref{f:vel-force}. Right: Increasing the time resolution to resolve a single avalanche (jump, marked by the arrow), the velocity inside a single avalanche can be viewed as a random walk with absorbing boundary conditions at vanishing velocity. This allows us to define observables as the mean avalanche shape, the size $S$ (area under the curve), or its duration $T$.}
\label{f:avalanche-obs}
\end{widefigure}

\subsection{Observables and scaling relations}
\label{s:Observables and scaling relations}
When slowly driving a system,   according to \Eq{eq-motion-w}  long times of inactivity are followed by bursts of activity, which on these long time scales look  instantaneous. To be specific, we start with the system at rest, and instantaneously increase $w\to w+\delta w$. Observables of interest are 
\begin{itemize}
\item
 center-of-mass   position 
 \be
 u(t) := \frac1{L^d} \int_x u(x,t),
\ee 
\item
center-of-mass  velocity   
\be
 \dot u(t) := \partial_t u(t),
\ee  
\item
duration $T$, see Fig.~\ref{f:avalanche-obs}. This quantity is well-defined, as  every  avalanche stops at some point when it is no longer driven (see section \ref{s:END}), so 
\be
T:= \inf_t \{t, \dot u(t)>0 \}  < \infty.
\ee  
\item 
 shape $\left< \dot u(t)\right>$.
\item
avalanche size $S$,
\be\label{S-def}
S:=\int_x \delta u(x) = \int_x u_2(x)-u_1(x), 
\ee
where   $u_1(x)$ is the interface position before, and $u_2(x)$  after an avalanche, see Figs.~\ref{f:one-avalanche} and~\ref{f:avalanche-obs}. 
\item
the avalanche extension $\ell$,
\be\label{ava-extension-def}
\ell:= \sup_{x,y}  \{|x-y|, \delta u(x)>0,  \delta u(y)>0 \} . 
\ee
\end{itemize}
These are the main  observables. 
Setting and language here are for depinning. Some observables, as the avalanche-size distribution can also be formulated in the statics: These {\em static avalanches}, also termed {\em shocks}, are the changes in the ground-state configuration upon a change in the applied field, i.e.\ the position $w$ of the confining potential. We will comment on differences between these two concepts at the appropriate positions. 
Key points are
\begin{itemize}
\item
avalanches are the response of the system  to an increase in force. They have a typical size 
\be\label{Sm}
\highlight{ S_m:=\frac{\left< S^2\right>}{2\left< S\right>}\sim \xi^{d+\zeta} \sim m^{-(d+\zeta)}}\punkt
\ee
In the literature one sometimes finds the notation $D=d+\zeta$ for the fractal dimension of an avalanche.
\item the {\em avalanche-size distribution} per unit force $\delta f=m^2 \delta w$, 
\bea\label{rhof}
\rho_f(S) := \frac{\delta N(S)}{\delta f} \simeq S^{-\tau } f_S(S/S_m) g_S(S/S_0)\komma \nn\\S_0 \ll S_m,
\eea
has a large-scale cutoff $S_m$ defined in \Eq{Sm} due to the confining potential, and a small-scale cutoff $S_0$ due to the size of the kick or   discretization effects (as in a spin system). The scaling functions 
are expected to have a finite limit when $m\to 0$, i.e.\ $\lim_{x\to 0} f_S(x) = \mbox{const}$, and  $\lim_{x\to \infty}g_S(x)=1$.

\item
An increase $\delta f$ in the total integrated force is then on average given by an increase $\delta u(x)$ in $u$, which integrated over space gives $S$.
On the other hand, we can  integrate \Eq{rhof} over $S$. Together, these relations give
\bea
\delta f &=&m^2\int_x \left< \delta u(x)   \right>
= m^2 \left< S\right> \nn\\
&=& \delta f\,m^2 \int_0^\infty \rmd S\, S\rho_f(S) \nn\\
&\sim& \delta f\, m^2 \left[ S_m^{2-\tau}-   {\cal O}(S_0^{2-\tau}) \right] .\eea
As we will see below    $\tau<2$, and the last term can be dropped due to the  assumption of $S_0\ll S_m$ in \Eq{rhof}.
This gives
\be\label{m-Sm}
m^2 \sim S_m^{ \tau-2}.
\ee
Inserting $S_m$ from \Eq{Sm}   yields
\be\label{tau-relation}
\highlight{ \tau_{\rm SR} = 2 -\frac2{d+\zeta}}\punkt
\ee
In $d=1$ and with $\zeta=5/4$ this gives 
\be
\tau_{\rm SR}^{d=1} = \frac{10}9 = 1.111...  
\ee
\item  {\em LR-elasticity}: Here one   replaces $m^2 \to m^\alpha$, leading to 
\be\label{tau-relation-alpha}
\tau_\alpha=2-\frac{\alpha}{d+\zeta}.
\ee
\begin{table*}[t]{\setlength{\tabcolsep}{5.3pt}
\renewcommand{\arraystretch}{1.5}
\scalebox{1}{\begin{tabular}{|c|c|c|c|c|c|c|}
\hline 
 & $\rho(S) $ & $\rho(S_\phi)$ & $\rho(T)$ & $\rho(\dot u)$ & $\rho(\dot u_\phi)$   & $\rho(\ell )$\\
\hline 
\hline 
    & $S^{-\tau}$ & $ S_\phi^{-\tau_\phi}$ & $T^{-\tilde \alpha}$ & $\dot u^{-\sf a}$ & $\dot u_\phi^{{-\sf a}_\phi}$ & $\ell^{-\sf k}$ \\
\hline 
  SR elasticity        & $\tau = 2 - \frac2{d+\zeta}$ &$\tau_\phi = 2 - \frac2{d_\phi+\zeta}$ & $\tilde\alpha = 1+ \frac{d-2+\zeta}{z}$ & ${\sf a} = 2- \frac2{d+\zeta-z}$ & ${\sf a}_{\phi} = 2- \frac2{d_\phi+\zeta-z}$ & ${\sf k}=d{+}\zeta{-}1$ \\
\hline 
LR elasticity       & $\tau = 2 - \frac1{d+\zeta}$ &$\tau_\phi = 2 - \frac1{d_\phi+\zeta}$ & $\tilde \alpha = 1+ \frac{d-1+\zeta}{z}$ & ${\sf a} = 2- \frac1{d+\zeta-z}$ & ${\sf a}_{\phi} = 2- \frac1{d_\phi+\zeta-z}$ & ${\sf k} = d+\zeta $ \\
\hline 
\end{tabular}}}
\caption{Scaling relations discussed in the main text, specifying to SR elasticity ($\alpha=2$), and standard LR elasticity $\alpha=1$. }
\label{tab1}
\end{table*}\item{\em Alternative scaling argument}:
In Ref.~\cite{KadanoffNagelWuZhou1989} it was suggested in the context of sandpile models (see section \ref{s:Manna}) that a single grain performs a random walk which has to reach the boundary, implying  that $\left< S \right> \sim L^2$. Using $\left< S \right>\sim S_m^{2-\tau}$, and $L\sim m^{-1}$ leads to $(2-\tau) (d+\zeta)=2$, equivalent to \Eq{tau-relation}. 

\item{\em Boundary driving}: when a (SR-elastic) system is driven at the boundary (tip driving \cite{AragonKoltonDoussalWieseJagla2016}) there is a drift (advection) away from this boundary\footnote{We do not understand the first-return-to-the-origin argument of Ref.~\cite{PaczuskiBoettcher1996}.}, leading to a linear scaling,   $\left< S \right> \sim L$, and as a consequence   $(2-\tau) (d+\zeta)=1$ \cite{PaczuskiBoettcher1996,NakanishiSneppen1997,AragonKoltonDoussalWieseJagla2016},  
\be
\tau^{\rm tip}_{\rm SR} = 2-\frac1{d+\zeta}. 
\ee
In $d=1$ and for $\zeta=5/4$ this gives 
\be
\tau^{{\rm tip}, d=1}_{\rm SR} = \frac{14}9 = 1.555...  >\frac32.
\ee

\item the distribution of avalanche sizes in a submanifold $\phi$ of dimension $d_\phi$, 
\bea\label{tau-phi-relation}
\highlight{\rho_f^{\phi} (S^{\phi})\sim S^{-\tau _\phi}}, \quad S_m^{\phi}\gg S^{\phi} \gg S_0^{\phi},\nn\\
\highlight{\tau_{\phi} = 2-\frac{\alpha}{d_\phi+\zeta}}\punkt
\eea
The derivation proceeds as for \Eqs{tau-relation} and \eq{tau-relation-alpha}.

\item avalanche size and duration are related via
\be\label{gamma}
\hl{ S_m \sim T_m^{\gamma}} \komma  \quad \hl{\gamma = \frac{d+\zeta}z} \punkt
\ee
This is obtained from the scaling relations $S_m\sim m^{-d-\zeta}$, and $T_m \sim m^{-z}$.

\item the (unnormalized) duration distribution per unit force is\footnote{We note the exponent as $\tilde \alpha$ instead of the standard notation $\alpha$   to avoid confusion with  $\alpha$ for the exponent of LR elasticity in \Eq{Hel-alpha-k}, and here present in \Eqs{tau-phi-relation}, \eq{tilde-alpha-LR}, \eq{a-in-P(udot)}, \eq{486}, and \eq{487}.}
\bea\label{rho-T}
\hl{\rho_f(T) \sim    T^{-\tilde\alpha}   }, \quad T_m \gg T\gg T_0, \quad   
\hl{T_m =\frac{  \left< T^3 \right>}{ \left<T^2\right>  }}. \nn\\ 
\eea
The   integral relation  $  \rho(S) \rmd S = \rho(T)\rmd T $ implies  $S_m^{1-\tau}\sim T_m^{1-\tilde \alpha}$. Using $   S_m\sim m^{-(d+\zeta)}$, and $ T_m\sim m^{-z}$ yields with the help of \Eq{tau-relation-alpha}
\be\label{tilde-alpha-LR}
\highlight{ \tilde\alpha = 1+ \frac{d+\zeta-\alpha}{z}} \punkt
\ee

\item the (unnormalized) velocity distribution
\begin{equation}\label{rho-dotu}
\hl{\rho_f (\dot u)  \sim   
 {\dot u}^{- {\sf a}} } \komma  \quad \dot u_m\gg \dot u \gg \dot u_0 \komma  \quad \hl{ \dot u_m =  \frac{S_m}{T_m} }\punkt
\end{equation}
The exponent ${\sf a}$ is obtained from arguments similar to those used in the derivation of Eqs.~\eq{tau-relation} and \eq{tau-phi-relation}, with the result that in the denominator the dimension of the obeservable in question appears. For the velocity distribution it yields
\be\label{a-in-P(udot)}
\hl{{\sf a} = 2-\frac{\alpha}{d+\zeta-z}} \punkt
\ee

\item avalanche extension: In general, 
avalanches have a well-defined spatial extension $\ell$, allowing us to define their distribution $\rho_f(\ell)$. If $\ell\ll \xi=1/m$, then $\ell$, and not $\xi$ is the relevant scale, and  $S \sim \ell^{d+\zeta}$. Writing $\rho_f(S)\rmd S = \rho_f(\ell)\rmd \ell $ allows us to conclude \cite{DelormeLeDoussalWiese2016} that for extensions between the lattice cutoff $a$ and $\xi=1/m$, 
\bea\label{486}
\hl{\rho_f(\ell) \sim \ell^{-{\sf k}}}, \quad a\ll \ell \ll \frac1m , \quad \nn\\
\hl{{\sf k}= d+\zeta+1-\alpha} \punkt 
\eea

\item avalanche volume: in higher dimensions, it is difficult to define the spatial extension of an avalanche, while its volume is well-defined. Using $\rho_f(V)\rmd V = \rho_f(S)\rmd S$, and $S\sim\ell^{d+\zeta}$, $V\sim \ell^{{d}}$, we arrive at 
\bea\label{487}
\hl{\rho_f(V) \sim V^{-{\sf k}_{V}}}, \quad a^{d}\ll V \ll \frac1{m^{d}}, \nn\\
\hl{{\sf k}_{V}= 2- \frac{\alpha-\zeta}d\punkt}
\eea

\end{itemize}
\medskip

\noindent{\em Differences between static avalanches (shocks) and avalanches at depinning:} 
A conceptually and practically  important question is whether static avalanches and avalanches  at depinning are in the same universality class. As the roughness exponent $\zeta$   differs from one class to the other, 
 \Eq{tau-relation} implies that they also have a different avalanche-size exponent $\tau$, and thus must be different.
We will see below that this difference is not visible at 1-loop order, but  shows up at 2-loop order. 

\medskip

\noindent {\em Phenomenology, and a warning:} For magnetic domain walls, where avalanche phenomena were first observed as {\em  Barkhausen noise} \cite{Barkhausen1919}, one distinguishes in general SR samples ($\alpha=2$) from LR samples ($\alpha=1$), and samples with noticeable  eddy currents from those without. A good review is
\cite{DurinZapperi2006b}. The reader should   realize that the ABBM model (section \ref{s:ABBM}) is often equated with the LR class or {\em mean field}, even though this is not true \cite{terBurgWiese2020,terBurgBohnDurinSommerWiese2021}. The line of theory we develop below   starts with the ABBM model, generalizes it to the Brownian Force Model (BFM) (section \ref{s:BFM}), and then proceeds to short-range correlated disorder (section \ref{s:Short-ranged rough disorder}).

\subsection{A theory for the velocity}

Up to now, our modeling of depinning was based on the equation of motion (\ref{eq-motion-w}) for the position of the interface. This formulation makes it difficult to extract observables involving the velocity. For this purpose it is better to take a time derivative of \Eq{eq-motion-w}, to get an equation of motion for the velocity $\dot u(x,t)$, 
\bea\label{eq-motion-udot1}
\hl{ 
\partial_{t} \dot u (x,t) = \displaystyle(\nabla^{2}{-}m^2) \left[\dot u (x,t){-} \dot w(t) \right] + \partial_t F \big(x,u (x,t)\big)}
\punkt
\nn\\
\eea

\subsection{ABBM model}
\label{s:ABBM}
The field theory to be constructed below gives a quantitative description of avalanches in a force field $F(x,u)$, with  short-ranged correlations  in both the $x$ and $u$-directions. 
We   start with a toy model for a single degree of freedom, and then proceed in two steps to short-range correlated forces for an interface.  

The toy model in question is the ABBM model, introduced in 1990 by Alessandro, Beatrice, Bertotti and Montorsi \cite{AlessandroBeatriceBertottiMontorsi1990,AlessandroBeatriceBertottiMontorsi1990b}, see also \cite{Colaiori2008,DobrinevskiLeDoussalWiese2011b}. Setting $w(t) = v t$, it reads
\begin{eqnarray}\label{48-2}
\partial_t \dot u(t) = m^2 \left[ v- \dot u(t) \right] +    \partial_t F \big(u (t)\big), \\
\label{49bis}
 \partial_t F \big(u (t)\big) = \sqrt{\dot u(t)} \xi(t) \komma  \\
  \left< \xi(t)\xi(t') \right> = 2 \sigma \delta (t-t') .\label{49noise}
\end{eqnarray}
The last equation implies that $F(u)$ is a random walk, as can be seen as follows: As $\dot u$ is non-negative, $t$ is an increasing function of $u$, and we can  change variables from $t$ to $u$, 
\beq
\partial_u F (u) = \bar \xi(u)\komma  \quad \left< \bar\xi(u)\bar\xi(u') \right> =2 \sigma \delta (u-u')\punkt
\eeq
As a random walk, $F(u)$ has  correlations   
\bea \label{51}
   \Delta(0)-\Delta(u-u')    \equiv \frac12 \left < \left[ F(u)-F(u') \right]^2 \right> = \sigma |u-u'|\punkt \nn\\
\eea
In the language introduced above,  the (bare) disorder  has a cusp, with amplitude $|\Delta'(0^+)|=\sigma $. 

The ABBM model is traditionally treated \cite{AlessandroBeatriceBertottiMontorsi1990,AlessandroBeatriceBertottiMontorsi1990b,Colaiori2008} via the associated Fokker-Planck equation \eq{forwardFP} (for a derivation see appendix \ref{s:FP}),
\bea
\partial_t P(\dot u ,t) = \sigma \frac{\partial^2}{\partial \dot u^2}\Big[    \dot u P(\dot u,t)\Big] +{  m^{2}} \frac{\partial}{\partial \dot u} \Big[(\dot u-v) P(\dot u,t) \Big]\punkt\nn\\
\eea
This approach is difficult for time-dependent quantities, but efficient for observables in the steady state. As an example, consider the steady-state distribution of velocities, obtained by solving $\partial_t P(\dot u ,t) =0$,
\be\label{P-udot-ABBM}
P(\dot u) = \frac{\dot u ^{\frac{m^{2}v}{\sigma}-1}\rme^{-\dot u \frac{m^{2}}\sigma}}{\Gamma\left(\frac{m^{2 }v}{\sigma}\right)}  \left( \frac{m^{2}}{\sigma} \right)^{{\frac{m^2 v}\sigma} }\punkt
\ee
Setting $\sigma=m=1$ to simplify the expressions yields 
\be\label{P-udot-ABBM-parameters=1}
\hl{ P(\dot u) = \frac{\dot u ^{v-1}\rme^{-\dot u}}{\Gamma(v)}}\punkt
\ee

\subsection{End of an avalanche, and an efficient simulation algorithm}
\label{s:END}
It is important to remark that an avalanche stops at a given well-defined time. To see this, we solve \Eqs{48-2}-\eq{49bis} for $m=0$, given that at time $t=0$ the velocity is $\dot u_0$. 
The associated Fokker-Planck equation is 
\be\label{25bis}
\partial _t P(\dot u, t) = \partial_{\dot u}^2 \left[ \sigma \dot u P(\dot u,t)\right] .
\ee
It can be solved analytically, for   given initial distribution $P(\dot u,0) =  \delta(\dot u-\dot u_0)$, as  
\be\label{57-1}
\begin{array}{rcl}

\displaystyle P(\dot u,t) &=& \displaystyle \delta(\dot u) \exp\left(-\frac{\dot u_0}{\sigma t}\right) \nn\\
&& + \frac{\exp\left(-\frac{\dot u_0+\dot u}{\sigma t}\right) }{\sigma t}\sqrt{\frac{\dot u_0}{\dot u}}\, I_1\!\left(\frac{2 \sqrt{\dot u_0 \dot u}}{\sigma  t}\right)\punkt 
\end{array}
\ee
($I_1$ is the Bessel-function of the first kind.)
This can be checked by inserting the solution into the differential equation (\ref{25bis}). 
\Eq{57-1} teaches us that for an  initial velocity   $\dot u_0$,   with a finite probability $\exp(-\frac{\dot u_0}{\sigma t})$ the velocity will be (strictly) zero after time $t$. It also means that the end of an avalanche is well defined in time, which is crucial to define its duration. This would not be the case for a particle in a smooth potential: Linearizing the potential close to the endpoint of the avalanche assumed to be $u(t=\infty)=0$  yields
\be
\partial_t u(t) \simeq - \alpha u(t) \quad \Longrightarrow \quad u(t) \simeq u_0 \rme^{-\alpha t}\punkt
\ee
\Eq{57-1} can serve as an efficient simulation algorithm, replacing $t$ by the time-discretization step $\delta t$, and alternately  integrating the forcing term $\delta \dot u(t) = m^2[v-\dot u(t)]\delta t$ and the stochastic process according to \Eq{57-1}. 
This is not straightforward, due to the appearance of multiplicative noise \cite{Munoz2004}.
We can use an additional trick \cite{DornicChateMunoz2005}: 
Observe that the solution \eq{57-1} can be written as
\bea
P(\dot u,t) = \delta(\dot u) \exp\left(-\frac{\dot u_0}{\sigma t}\right)\nn\\
  + \sum_{n=1}^\infty \frac{\left(\frac{\dot u_0}{\sigma  t}\right)^n \exp\left(-\frac{\dot u_0}{\sigma  t}\right)}{n!}\times \frac1{\sigma t}\frac{\left(\frac{\dot u}{\sigma  t}\right)^{n-1} \exp\left(-\frac{\dot u}{\sigma  t}\right)}{(n-1)!}
\nn\\
= \sum_{n=0}^{\infty} p_n \,\frac{1}{\sigma t}P_n \!\left(\frac{\dot u}{\sigma  t}\right).
\eea
Here, $p_n$ is a normalized probability vector, i.e.\ $\sum_{n=0}^\infty p_n =1$, and each probability $P_n(x)$ is normalized, $\int_0^\infty \rmd x\ P_n(x)=1$.  Explicitly, we have 
\bea
p_n = \frac{\left(\frac{\dot u_0}{\sigma  t}\right)^n \exp\left(-\frac{\dot u_0}{\sigma  t}\right)}{n!},\\
P_0(x) = \delta(x), \\
P_n(x)= \frac{x^{n-1} \exp\left(-x\right)}{(n-1)!}\komma  \quad n\ge 1\punkt
\eea
Given $\dot u_0$, one obtains $\dot u$ with probability $P(\dot u,t)$ as follows:
\begin{enumerate} \item[(i)]
 draw an integer random number $n$, from the Poisson distribution $p_n$; the latter has parameter $\dot u_0/(\sigma t)$,
 \item[(ii)] if $n=0$, return $\dot u=0$,
 \item[(iii)]
  else draw a positive real random number $x$, from the Gamma distribution with parameter $n$. 
\item[(iv)] return $\dot u = \sigma t x$, 
\end{enumerate} Contrary to a naive integration of the stochastic differential equation which yields $\partial_t F\big(u(t)\big)\delta t = \xi_t \sqrt{\delta t}$, $\left< \xi_t \xi_{t'}\right> = \delta_{t ,t'}$,  this algorithm is linear in $\delta t$.

This allows us to define the distribution of durations, given below in  \Eq{P-duration}, and the mean temporal shape \eq{t-shape}, without introducing an (arbitrary) small-velocity cutoff.

\subsection{Brownian Force Model (BFM)}
\label{s:BFM}
The model defined in \Eqs{48-2} and \eq{49bis} is a model for a single degree of freedom, not for an interface. A model for an interface can be defined by \cite{LeDoussalWiese2012a}
\begin{eqnarray}\label{BFM1}
\partial_t \dot u(x,t) = \nabla^2 u(x,t) + m^2 \left[ v- \dot u(x,t) \right]\nn \\
\hphantom{\partial_t \dot u(x,t) = }+    \partial_t F \big(x,u (x,t)\big) ,\\
\label{BFM1bis}
 \partial_t F \big(x,u (x,t)\big) = \sqrt{\dot u(x,t)} \xi(x,t) \komma   \\
  \left< \xi(x,t)\xi(x',t') \right> = 2 \sigma \delta (t-t') \delta^d(x-x').
  \label{BFM2}
\end{eqnarray}
Since each degree of freedom sees a force which is a random walk, this model is termed the {\em Brownian Force Model} (BFM) \cite{LeDoussalWiese2012a}. 

\subsection{Short-ranged rough disorder}
\label{s:Short-ranged rough disorder}

Both the ABBM model as its spatial generalization, the BFM model, are pathologic in the sense that the force-force correlator grows for all distances instead of saturating as   expected in short-ranged correlated systems, and as is reflected in the FRG fixed points discussed in section \ref{s:FRG-fixed-points}. To remedy this, one can keep the equation of motion \eq{BFM1}, but add an additional damping term in the evolution equation \eq{BFM1bis} of the force, 
\begin{eqnarray}\label{SR1}
\label{SR2}
 \partial_t F \big(x, u (x,t)\big) = - \gamma \dot u(x,t) F\big(x, u(x,t)\big) \nn\\
~~~~~~~~~~~~~~~~~~~~~~~~~~~~~  +\sqrt{\dot u(x,t)} \xi(x,t) \komma\\
  \left< \xi(x,t)\xi(x',t') \right> = 2 \sigma \delta (t-t') \delta^d(x-x')\punkt
\end{eqnarray}
As $\dot u(x,t) \ge 0$, the equation of motion for the force is equivalent to 
\begin{eqnarray}\label{eq1}
 \partial_u F(x,u) = - \gamma   F(x,u) + \tilde \xi(x,u)\komma \\
  \left< \tilde \xi(x,u) \tilde \xi(x',u') \right> = 2 \sigma \delta (u-u') \delta^d(x-x')\punkt
\end{eqnarray}
This system has the force-force correlator \be\label{71}
\left < F(x,u) F(x',u') \right>^c = \sigma \delta^d(x-x') \frac{\rme^{-\gamma |u-u'|}}{  \gamma}\komma
\ee
derived in \Eq{FFOU}. 
For $u\ll \gamma$, we recover the correlations \eq{51}.

\subsection{Field theory}
Consider the equation of motion \eq{eq-motion-udot1} with generic short-ranged force-force correlators.
The dynamical action is obtained by multiplying the equation of motion with $\tilde u(x,t)$, and averaging over disorder\footnote{The response field $\tilde u(x,t)$ is different from that in \Eqs{S[u,utilde,F]}-\eq{dyn-action}. One can derive \Eq{dyn-action2} by substituting  in \Eq{dyn-action} $\tilde u(x,t)\to -\partial_t \tilde u(x,t)$, and then integrating by parts in time.},
\bea\label{dyn-action2}
\hl{ {\cal S} = \int_{x,t} \tilde u (x,t)\Big[ (\partial_{t}{-}\nabla^{2}) \dot u
(x,t) + m^2 \big( \dot u (x,t){-}\dot w \big) \Big] \nn\\
 -\half \int_{x,t,t'} \tilde u (x,t)\tilde u
(x,t') \partial_t \partial_{t'}\Delta \big(u (x,t){-}u (x,t')\big)} \punkt
\eea
The second line contains
\bea
\partial_t \partial_{t'}\Delta \big(u (x,t)-u (x,t')\big) \nn\\
= \dot u(x,t) \partial_{t'}  \Delta' \big(u (x,t)-u (x,t')\big)  \nn\\
 = \dot u(x,t)   \left[ \Delta' (0^+) \partial_{t'} \mbox{sign}(t-t') -\Delta''(0^+) \dot u(x,t') + ... \right]\nn\\
 = -2 \dot u(x,t)    \Delta' (0^+)  \delta (t-t') \nn\\
\hphantom{ = } - \Delta''(0^+) \dot u(x,t) \dot u(x,t') + ...
 \label{73}
\eea
The terms dropped in this expansion are higher derivatives of $\Delta(u)$, and they come with higher powers of $\dot u(x,t)$, and its time-integral  $ u(x,t)-u(x,t') = \int _{t'}^{t} \rmd \tau\, \dot u(x,\tau)$, reminding that $\dot u(x,t)$ and not $u(x,t)$ is the variable for which we wrote down the equation of motion. 

This expression is quite remarkable: The leading term is proportional to $\delta (t-t')$, rendering the last term in \Eq{dyn-action2} {\em local} in time. 
It is therefore appropriate to    start our analysis of the theory with this term only. 
The action we obtain is 
\begin{eqnarray}\label{dyn-action3}
&& {\cal S}_{{\rm BFM}} [\dot u,\tilde u]\nn\\
&&= \int_{x,t} \tilde u (x,t)\Big[ (\partial_{t}-\nabla^{2}) \dot u
(x,t) + m^2 \big(\dot u(x,t)- \dot w (x,t) \big) \Big] \nn\\
&& \qquad +\Delta'(0^{+})  \tilde u (x,t)^{2 }\dot u(x,t)\punkt
\end{eqnarray}
This is  the action of the BFM introduced in \Eqs{BFM1}-\eq{BFM2}: as $\Delta(u)$ only has a first non-vanishing derivative,   all subsequent terms in \Eq{73} vanish. 
Corrections are obtained by adding the omitted terms perturbatively. The subleading term  is
\bea\label{corrections2BFM}
\hl{{\cal S}  [\dot u,\tilde u] = {\cal S}_{{\rm BFM}} [\dot u,\tilde u] \\
+\frac{\Delta'' (0^+)}2 \int_{x,t,t'} \tilde u (x,t)\tilde u
(x,t') \dot u(x,t) \dot u(x,t') + ... }\punkt\nn
\eea
We   show below  on page \pageref{theorem2} several  theorems, indicating that at the upper critical dimension  the action \eq{dyn-action2} leads to the results of the BFM model, and that the additional  term in \Eq{corrections2BFM}   is sufficient for the 1-loop corrections, of order $\epsilon=d_{\rm c}-d$.

\subsection{FRG and scaling}
The FRG equation \eq{two-loop-FRG-dyn} for the disorder has the following   structure
\bea\label{78}
\partial_{\ell} \tilde \Delta (u) &=& (\epsilon -2 \zeta)\tilde\Delta (u)+\zeta u \tilde\Delta' (u) \nn\\
&&+ \sum_{n=1}^\infty \partial_u^{2n} \left[ \tilde\Delta(u)-\tilde\Delta(0)\right] ^{n+1}\punkt
\eea
The $n$-loop terms are highly symbolic, since the derivatives can be distributed in different ways on the $n+1$ disorder correlators, and we have dropped all prefactors. 
We now assume that the microscopic disorder has the form (\ref{51}), thus $\tilde \Delta(0)-\tilde \Delta(u)$ has only a  term linear in $u$. 
This implies that the term of order $n=1$ may contribute a constant to \Eq{78}, while   terms with $n\ge 2$ vanish. Thus 
to all orders, the roughness exponent is given by
\be\label{zeta-BFM}
\hl{\zeta_{\rm BFM} = \epsilon = 4-d} \punkt
\ee
As a consequence, the unrescaled disorder   is scale independent (does not renormalize),   and 
\be
\hl{\Delta_{\rm BFM}(0)-\Delta_{\rm BFM}(u) \equiv  \sigma |u| }\punkt
\ee
Note   that $\Delta(0)$ is not well-defined, since random forces grow unboundedly. 

Similarly, the dynamical exponent $z$ has corrections proportional to $\Delta''(0)$, which vanish. As a consequence, $z=2$, and   all  exponents can be obtained analytically,  
\bea
z_{\rm BFM} = 2 \komma   \quad \beta_{\rm BFM} = {\sf a}_{\rm BFM} =1\komma  \quad \gamma_{\rm BFM}=\alpha_{\rm BFM}=2\komma  \nn \\
 \tau_{\rm BFM} = \frac32\komma  \quad \kappa_{\rm BFM}=3\  .~~
\eea

\subsection{Instanton equation}
We want to construct observables for the BFM such as
\bea\label{75}
 Z[\lambda,w]:=\left< \rme^{{\int_{{x,t}} \lambda(x,t)\dot u(x,t)}}\right> \nn\\
= \int {\cal D}[\dot u]{\cal D}[\tilde u] \, \rme^{{\int_{{x,t}} \lambda(x,t)\dot u(x,t)} -{\cal S}_{\rm BFM} [\dot u,\tilde u] }\punkt\eea
This includes the avalanche-size distribution with  $\lambda(x,t)=\lambda$, the velocity distribution with $\lambda(x,t)=\lambda \delta (t)$, the local avalanche-size distribution with  $\lambda(x,t)=\lambda\delta (x)$, a.s.o.

The key observation is that $\dot u(x,t)$ appears   linearly in the exponent, thus the path integral over $\dot u$ can be performed  {\em exactly}, enforcing an {\em instanton equation} for $\tilde u(x,t)$, 
\bea\label{inst-equation}
\hl{ \left( -\partial_{t     }-\nabla^{2} + m^{2}\right) \tilde u_{\rm inst}(x,t)-\sigma \tilde u_{\rm inst} (x,t)^{2 } = \lambda(x,t)} \punkt \nn\\
\eea
Here $\sigma \equiv - \Delta'(0^{+})>0$, see e.g.\ \Eqs{51} and \eq{71}.
The expectation \eq{75} simplifies to
\bea\label{central}
\hl{ Z[\lambda,w]  = \rme^{\int_{x,t} m^{2} \dot w(x,t) \tilde u(x,t)} \Big|_{\tilde u = \tilde u_{{\rm inst}}}} \punkt
\eea
The term with $\dot w(x,t)$ in the exponential is the only one not proportional to $\dot u(x,t)$; the latter vanish due to the instanton equation \eq{inst-equation}.
Let us   consider some examples.

\subsection{Avalanche-size distribution}\label{s:Avalanche-size distribution in the BFM}
The simplest example is the   avalanche-size distribution. Noting that 
$
S = \int _{x,t} \dot u(x,t)
$, we have to solve the instanton equation \eq{inst-equation} for $\lambda(x,t) = \lambda$. The solution for $\tilde u = \tilde u_{\rm inst}$ will   be   constant in space and time,  thus the instanton equation \eq{inst-equation} reduces to  
\be
m^2 \tilde u -\sigma \tilde u^2 = \lambda\punkt
\ee
This quadratic equation has two solutions. The relevant   one   vanishing at $\lambda=0$ reads
\be\label{80bis}
\tilde u = \frac{m^2-\sqrt{m^4-4 \lambda 
   \sigma }}{2 \sigma }\punkt
\ee
We now insert this solution into \Eq{central}. As $\tilde u(x,t)$ is constant in space and time, the integral on the r.h.s.\ of  \Eq{central} is $\tilde u$ times 
\be
w:= \int_{x,t} \dot w(x,t) >0 \punkt
\ee
This yields with $\tilde u$ given in \Eq{80bis}
\be
\left< \rme^{\lambda S}\right> =  \rme^{m^2 w \tilde u}\punkt
\ee
Taking the inverse Laplace transform gives
\be\label{PwS(S)}
\hl {P_w^S(S) = m^2 w  \frac{ e^{-\frac{m^4 (S-w)^2}{4
   \sigma  S}}}{2 \sqrt{\pi  \sigma } S^{3/2}} }\punkt
\ee
One checks that $P_w^S(S)$ is normalized, $\left< 1\right>_{w}=\int_0^\infty \rmd S\, P_w^S(S) =1$, and that the first avalanche-size moment is  $ \left< S\right>_{w}= \int_0^\infty \rmd S\, S P_w(S) =w$. If $w(x,t)$ is constant in $x$, then $ \left< S\right>_{w} =w$ is nothing but the displacement of the confining parabola.

  $P_w(S)$ is the response of the system to a displacement $w$, or equivalently to a  {\em force kick} $\delta f = m^2 w$. We now  take the limit of an infinitesimally small   displacement $w$, and to this purpose define 
\bea\label{83}
P^S(S):=  \lim _{w\to 0} \frac{\left< S\right>_{w}}{w} {P^S_w(S)}   \nn\\
= \left< S\right> m^2\frac{ e^{-\frac{m^4 S}{4
   \sigma  }}}{2 \sqrt{\pi  \sigma } S^{3/2}} 
   \equiv  \left< S\right>\frac{ e^{-\frac{ S}{4
   S_m  }}}{2 \sqrt{\pi  S_m } S^{\tau}}\komma  \\
 \tau= \tau_{\rm ABBM} = \frac32,\\
S_m:= \frac{\left< S^2\right> }{2\left< S\right> } = \frac{\sigma }{m^4}   \punkt
\eea
Since $\left< S\right> _{w} =w$, by construction all moments which do not necessitate a small-$S$ cutoff, i.e.\ $\left< S^n\right>_{w}$ with $n\ge 1$ have a well-defined small-$w$ limit, given by \Eq{83}. What one looses when taking the limit of $w\to 0$ is  normalizability, as formally $\left< 1\right> = \lim_{w\to 0} w^{-1}=\infty$.

\subsection{Watson-Galton process, and first-return probability}
\label{s:Watson-Galton process}
The avalanche size-exponent $\tau=3/2$ observed in the ABBM model appears in many contexts: It was first studied in the survival probability of a noble man's  name (male descendents) 
\cite{WatsonGalton1875}. The latter has the equations of motion \eq{48-2}-\eq{49noise}, where $\dot u(t)$ is the number of descendants in a generation, $v=0$, and $m^2$ the mean relative decrease in male descendants in a generation (which could be negative),
\bea\label{EOM-WG}
\partial_t u(t) = - m^2 \dot u(t) + \sqrt{\dot u(t)} \xi(t), \\
\left<   \xi(t)   \xi(t') \right> = 2 \sigma \delta(t-t'). 
\eea
 As $\dot u(t)>0$, the total number of descendants $u(t)$ is monotonically increasing, allowing us to write $\dot u(t) \equiv \dot u(u(t))$. The equation of motion \eq{EOM-WG} then becomes
\bea
\partial_t \dot u(u(t)) = \dot u (u)   \partial_u \dot u(u  )  =  - m^2 \dot u(u)  {+} \sqrt{\dot u (u) }\, \xi(t). 
\eea
As long as $\dot u>0$, we can divide both sides by $\dot u$ to arrive at
\bea
\partial_u \dot u(u) = - m^2 +\tilde \xi(u), \\
 \left< \tilde \xi(u) \tilde \xi(u') \right> = 2 \sigma \delta(u-u'). 
\eea
Note the change of argument in the noise.
Thus $\dot u(u)$ performs a random walk in ``time'' $u$ with drift $-m^2$ and  absorbing boundary conditions at $\dot u=0$. 
In absence of an absorbing wall, the probability that $\dot u(u)$ starts close to zero at $u=0$, and returns to zero behaves for large $u$ as 
\be\label{P-return}
 P_{\rm return}(u) \sim \frac{\rme^{-m^2 u}}{\sqrt{u}} . 
\ee
In presence of an absorbing wall, we need the  probability to arrive at zero for the first ``time'' $u$. The latter   is obtained by taking  a ``time'', i.e.\ $u$-derivative of $ P_{\rm return}(u)$, 
\be\label{P-first-return}
P_{\rm first}(u) =  -\partial_u P_{\rm return}(u) \sim \frac{\rme^{-m^2 u}}{u^{\frac 32}} . 
\ee
This is again the ABBM avalanche-size distribution \eq{83} with $\tau=3/2$, interpreted as first-return probability of a random walk. 

\subsection{Velocity distribution}
To simplify further considerations, we   set \be
m^{2}\to 1 \komma  \quad -\Delta '(0^{+}) \equiv \sigma \to 1\punkt
\ee
To obtain the instantaneous velocity distribution, we  evaluate \Eqs{75}-\eq{central} for $\lambda(x,t) = \lambda \delta (t)$, setting $  \dot w(x,t) = v$ (uniform driving). 
The instanton equation to be solved is 
\be\label{v-inst-eq}
-\partial_t \tilde u(t) +\tilde u(t) -\tilde u(t)^2= \lambda \delta (t)\punkt
\ee
To impose proper boundary conditions, look at the r.h.s.\ of \Eq{central}: Driving at times $t>0$ does not affect the velocity distribution at  $t=0$, thus the instanton solution $\tilde u(t)$ must   vanish for positive times. \Eq{v-inst-eq}  with this constraint is solved by
\be\label{inst88}
\tilde u(t) = \frac{\lambda \Theta(-t)}{\lambda +(1-\lambda) \rme^{-t} } \punkt
\ee
With the above solution \Eq{central} reduces to 
\bea
\left< \rme^{{\lambda \int_{{x}} \dot u(x,0)}}\right> = \rme^{  v L^{d}\int_{t}    \tilde u(t)} = \rme^{ - v L^{d}\ln(1-\lambda) } \nn\\
= (1-\lambda)^{-v L^{d}} \punkt
\eea
The inverse Laplace transform for the integrated velocity $\dot u=\int_x\dot u(x,0)$ is \be
\hl {P_{v,L}^{\dot u}(\dot u) = \frac{\dot u^{vL^{d}-1}\rme^{-\dot u}}{\Gamma(v L^{d})} }\punkt
\ee
Apart from  the {\em source} $vL^{d}$, this result is independent of the dimension $d$, and agrees with the ABBM result \eq{P-udot-ABBM}, there derived for a single degree of freedom.
We can take the limit of $v\to 0$, and define 
\be
P^{\dot u}(\dot u) := \lim_{v\to 0} \frac{P_{v,L}^{\dot u}(\dot u) }{v L^{d}} = \frac{\rme^{-\dot u}}{\dot u} \punkt
\ee

\subsection{Duration distribution}
The probability that the avalanche has velocity zero at time $0$ after a kick of size $w$ at time $t=-T$, with $T>0$, can be obtained from the central result \eq{central} with the instanton \eq{inst88} as
\bea
P(\dot u(x,0)=0~\forall x) = \lim _{\lambda \to -\infty} \left< \rme^{\lambda \int_x \dot u(x,0)}\right>\nn\\
= \lim _{\lambda \to -\infty} \rme^{w\, \tilde u_\lambda(-T)}  
 = \exp\left(-\frac{w}{\rme^{T}-1} \right)\punkt
\eea
This is also the probability that the duration following a kick of size $w$ is smaller than $T$. The distribution of durations is   obtained by taking a derivative w.r.t.\ $T$, 
\bea\label{P-duration1}
P^{\rm duration}_w(T) = \partial_T \exp\left(-\frac{w}{\rme^{T}-1} \right)\nn\\
 =w  \exp\left(-\frac{w}{\rme^{T}-1} \right) \frac{e^{-T}}{\left(e^{-T}-1\right)^2} \nn\\
= w  \exp\left(-\frac{w}{\rme^{T}-1} \right)  \frac1{[2\sinh(T/2)]^2}\punkt
\eea
This distribution is normalized. As at the end of section \ref{s:Avalanche-size distribution in the BFM}, let us define the (unnormalized) probability density in the limit of $w\to 0$, 
\bea\label{P-duration}
P^{\rm duration} (T):= \lim_{w\to 0} \frac{P^{\rm duration}_w(T) }{w} =  \frac1{[2\sinh(T/2)]^2}\punkt\nn\\
\eea

\subsection{Temporal shape of an avalanche}
In order to obtain the temporal shape of an avalanche, we need to solve the instanton equation 
\be\label{v-inst-eq2}
-\partial_t \tilde u(t) +\tilde u(t) -\tilde u(t)^2= \lambda \delta (t_{\rm f}-t) +\eta \delta(t-t_{m}) \komma 
\ee
where $t_{\rm f}$ is the final time (where the avalanche stops) and $\tm$ the time at which the velocity is measured. 
Since we only need its first moment, we can construct $\tilde u(t)$ perturbatively in $\eta$. To that purpose write 
\bea\label{96}
\tilde u(t) = \tilde u_0(t) + \eta \tilde u_1(t)+ \eta^2 \tilde u_2(t) + {\cal O}(\eta^3)\komma  \nn\\
 \tilde u_0(t) = \frac{  \Theta(t_{\rm f}-t)}{1- \rme^{t_{\rm f}-t} } \punkt
\eea
The solution $\tilde u_0(t)$ is the solution (\ref{inst88}), translated to stop at $t=t_{\rm f}$, in the limit of $\lambda\to -\infty$.
Inserting \Eq{96} into \Eq{v-inst-eq2} and collecting terms of order $\eta$ yields 
\be
-\partial_t \tilde u_1(t) +\tilde u_1(t) -2\tilde u_1(t)\tilde u_0(t)= \delta(t-t_{m})\punkt
\ee
The solution is 
\be
\tilde u_1(t) = \frac{ \sinh^2
   (\frac{\tf-\tm}2)}{\sinh^2 (\frac{\tf-t}{2})}\, \theta (\tm-t)\punkt
\ee
Performing a kick of size $w$ at $t=0$, and constraining the avalanche to stop at time $\tf=T$, the shape can   be written as 
\bea
\left< \dot u(\tm) \right> = \partial_\eta\Big|_{\eta=0} \ln \left( \partial_{\tf}\rme^{w \tilde u(t)} \right)\Big|_{T=\tf}
\nn\\
= 4 w \frac{\sinh ^2 (\frac{T-\tm}{2} )}{\sinh ^2 (\frac{T }{2} )} +4 \frac{\sinh (\frac{T-\tm}{2} ) \sinh (\frac{\tm}{2} )}{\sinh  (\frac{T }{2} )}\punkt
\eea
Consider now the limit of $w\to 0$, for which the first term vanishes: 
For short durations $T$, $\left< \dot u(t_{\rm m})\right>$ converges to a parabola, 
\be\label{t-shape}
\left< \dot u(t_{\rm m})\right>  = 2 \frac{\tm (T-\tm)}{T}\punkt
\ee
For long durations, it settles on a plateau at $\left< \dot u(t)\right>=2$, see figure \ref{f:udot}. 

Pursuing to  the next order, one finds for the connected average
\bea
\left< \dot u(\tm)^2 \right>^{\rm c} &=& 4 w \frac{\sinh ^3 (\frac{T-\tm}{2} ) \sinh (\frac{\tm}{2} )}{\sinh ^3 (\frac{T }{2} )} \nn\\
&& +8 \frac{\sinh^2 (\frac{T-\tm}{2} ) \sinh^2 (\frac{\tm}{2} )}{\sinh^2  (\frac{T }{2} )}\punkt
\eea
At $w=0$, quite remarkably the ratio 
\be
\frac{\left< \dot u(\tm)^2 \right> }{ \left< \dot u(\tm) \right> ^2} = \frac32
\ee
is   time independent. 

Further observables, as well as loop corrections are obtained in \cite{DobrinevskiLeDoussalWiese2014a,DobrinevskiPhD}, and compared to experiments in \cite{DurinBohnCorreaSommerDoussalWiese2016}.
\begin{widefigure}[bt]
\centerline{
\fig{8cm}{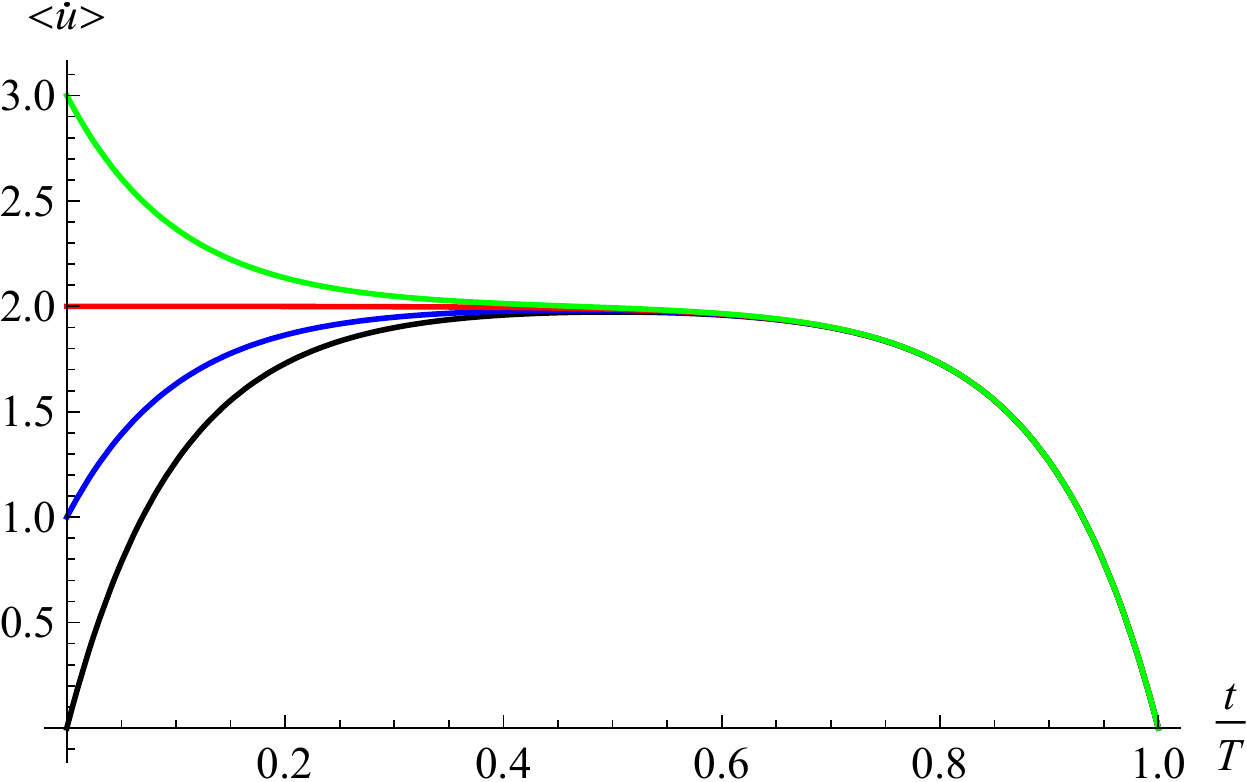}~~~~~~~ \fig{8cm}{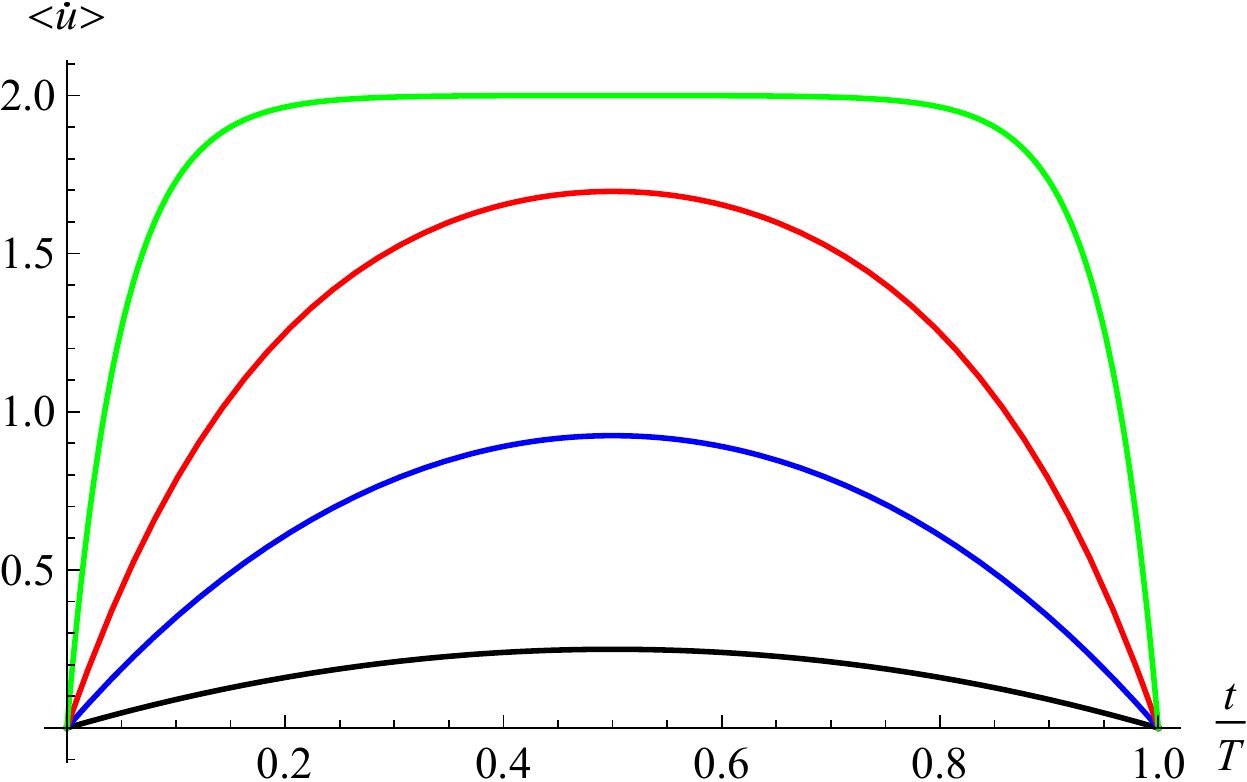}
}
\caption{Expectation $\left< \dot u(t)\right> $. Left: $T=10$, and from bottom to top $w=0$, $1$, $2$, and $3$. Right: $w=0$ (infinitesimal kick), and from bottom to top $T=\half$, $2$, $5$, and $20$. Fig.~reprinted from \cite{LeDoussalWiese2012a} }
\label{f:udot}
\end{widefigure}

\subsection{Local avalanche-size distribution}
\label{s:loc-size}
We now consider avalanches on a codimension-1 hyperplane, i.e.\ at a point for a line, or on a line for a 2d interface, a.s.o., by choosing
\be
\lambda (x,\vec {x}_{\perp}) = \lambda \delta (x)\punkt
\ee 
As a consequence, $\vec {x}_{\perp}$ drops from the instanton equation. 
Setting again $\sigma=m^2=1$, one arrives at 
\be\label{90}
\tilde u(x) -\tilde u''(x)- \tilde u(x)^2 = \lambda \delta(x)\punkt
\ee
The only solution which vanishes at infinity and satisfies the instanton equation at $\lambda=0$ is  
\be
\tilde u(x) = \frac{3}{1+\cosh(x+x_0)}  \punkt
\ee
It can be promoted to a solution at $\lambda \neq0$ by setting $\tilde u(-x) = \tilde u(x)$. 
The parameter $ x_0 = x_0(\lambda)$ has to be chosen to satisfy the instanton equation at $x=0$. Integrating \Eq{90} within a small domain around $x=0$ yields
\be\label{92}
\lambda = -2 \tilde u'(0^+) = \frac{6 \sinh(x_0)}{[1+\cosh(x_0)]^2}\punkt
\ee
On the other hand, the generating function is  
\be\label{93}
Z  :=  \int_{-\infty}^\infty \rmd x\, \tilde u(x) = \frac{12}{1+\rme^{x_0}}.
\ee
Solving \Eq{93} for $x_0$ and inserting into \Eq{92} yields 
\be
\lambda= \frac{Z (Z-6)(Z-12)}{72}\punkt
\ee
The inverse Laplace transform is a priori difficult to perform, as $Z(\lambda)$ is a complicated function of $\lambda$. The trick is to write
\bea
P_w(S_0)&:=&\left< \rme^{\lambda S_0}\right> = \int_{-i \infty}^{i \infty} \frac{\rmd \lambda}{2\pi i}\,\rme^{-\lambda S_0}\rme^{w Z(\lambda)} \nn\\
 & =& \int_{-i \infty}^{i \infty} \frac{\rmd Z}{2\pi i}\frac{\rmd \lambda(Z)}{\rmd Z}\rme^{-\lambda(Z)S_0+w Z}  \nn\\
&=& -\frac1{S_0}\int_{-i \infty}^{i \infty} \frac{\rmd Z}{2\pi i} \rme^{wZ}\frac{\rmd }{\rmd Z}\rme^{-\lambda(Z)S_0}  \nn\\
&=&\frac w{S_0}\int_{-i \infty}^{i \infty} \frac{\rmd Z}{2\pi i} \rme^{wZ} \rme^{-\lambda(Z)S_0}  \nn\\ 
&=& \frac{6 e^{6 w} w}{\pi  S_0} \int_0^\infty \rmd x \,\cos\! \big(3 x (S_0 x^2+S_0+2 w)\big) \nn\\
&=&\frac{2 e^{6 w} w \sqrt{S_0+2 w}
  }{\pi 
   S_0^{3/2}} K_{\frac{1}{3}} \!\!\left(\frac{2 (S_0+2
   w)^{3/2}}{\sqrt{3 S_0}}\right)\punkt ~~~~ \nn\\
\eea
This can also be written in terms of the Airy function (formula~(21) of Ref.~\cite{DelormeLeDoussalWiese2016}). In the limit of $w\to 0$, it reduces to 
\be
P(S_0) = \frac{2 }{\pi  S_0} K_{\frac{1}{3}}\!\left(\frac{2 S_0}{\sqrt{3}}\right)\punkt
\ee
One can also give analytical expressions for the joint distribution of avalanche size $S$ and local size $S_0$, as well as of size $S$ and spatial extension $\ell$. The interested reader   finds this in \cite{DelormeLeDoussalWiese2016}.

\subsection{Spatial shape of avalanches} 
We now turn to the spatial shape of avalanches \cite{ZhuWiese2017}.
We  remind that avalanches have a well-defined extension $\ell$, beyond which there is no movement.  For a given avalanche, denote its advance by $S(x)$, and  its size by $S=\int_x S(x)$. We call avalanche extension $\ell$ the size of the smallest ball into which we can fit the avalanche. 
As long as  $\ell \ll m^{-1}$,  \be\label{ansatz}
\left< S(x)\right>_\ell =  \ell^\zeta g(x/\ell)
\komma 
\ee
where $g(x)$ is non-vanishing in the unit ball. Integrating over space yields 
$
S\sim \ell^{d+\zeta}, 
$
the canonical scaling relation between size and extension of avalanches, confirming the ansatz (\ref{ansatz}).

We now want to deduce how $g(x)$ behaves close to the boundary. For simplicity of notations, we write our argument for the left boundary in $d=1$, which we place at $x=-\ell/2$. Imagine the avalanche dynamics for a discretized system. The avalanche starts at some point, which in turn may trigger an advance  of its neighbors, a.s.o. This   leads to a shock front   propagating outwards from the seed to the left and to the right. As long as the elasticity is local,  the dynamics of these two shock fronts is local: If one conditions on the position of the $i$-th point away from the front, with $i$ being much smaller than the total extension $\ell$ of the avalanche (in fact, we only need that the avalanche started right of this point), then we expect that the joint probability distribution for the advance of points $1$ to $i-1$  depends on $i$, but is independent of the size $\ell$. 
Thus we expect that   {\em in this discretized model} the shape $\left< S(x-r_1)\right> $ close to the left boundary $r_1$ is  independent of $\ell$. Let us now turn to avalanches of large size $\ell$, so that we are in the continuum limit studied in the field theory. 
Our argument then implies that the shape $\left< S(x-r_1)\right>$ measured from the left boundary $r_1=-\ell/2$, is independent of $\ell$. In order to cancel the $\ell$-dependence in Eq.~(\ref{ansatz}) this in turn implies that \cite{ZhuWiese2017}
\be \label{gofx}
g(x ) = {\cal B} \times  (x-1/2)^\zeta\komma  
\ee
with some  amplitude ${\cal B}$. For the Brownian force model in $d=1$, the roughness exponent is 
$
\zeta_{\rm BFM} = 4-d = 3
$, and one can further show that ${\cal B}=\sigma/21$ \cite{ZhuWiese2017}.

On the left of figure \ref{f:samples}, we show twenty  realizations of avalanches, with mean given by the thick black line. On the right   we compare numerical averages with the theory sketched below. Note that the latter indeed has a cubic behavior close to the boundary, as predicted by \Eq{gofx}.

\begin{widefigure}
\setlength{\unitlength}{0.995cm}
{$\!\!$\begin{picture}(8.4,5.7)
\put(0.5,0){\fig{8\unitlength}{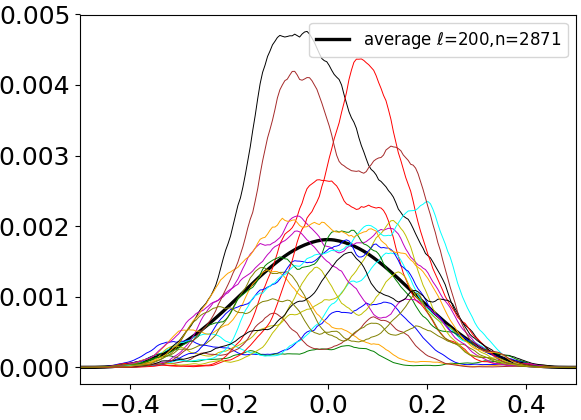}}
\put(0,5){\rotatebox{90}{$S(x)$}}
\put(8.2,0.1){$x$}
\end{picture}}~~~~~~
{$\!\!$\begin{picture}(8.4,6.05)
\put(0.5,0.){\fig{8\unitlength}{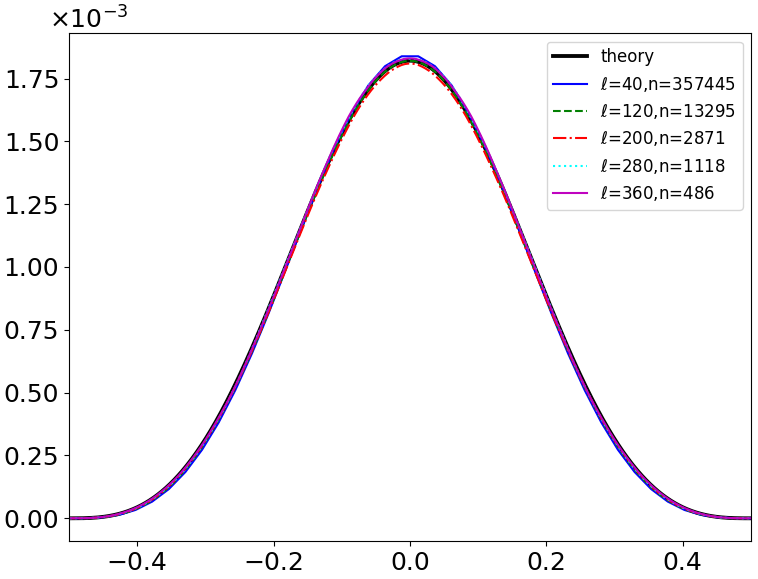}}
\put(0,4.9){\rotatebox{90}{$\left<{S(x)}\right>$}}
\put(8.2,0.1){$x$}
\end{picture}}
\caption{Left: 20 avalanches with extension $\ell=200$, rescaled to $\ell=1$.  $n=2871$ is the number of samples used for the average.
Right: The shape $\left< S(x)\right>\equiv \left< S(x/\ell)\right>_{\ell } /\ell^3$ averaged for all avalanches with a given $\ell$  between $40$ and $360$. To reduce  statistical errors, we have symmetrized this function. Fig.~reprinted from \cite{ZhuWiese2017}.}
\label{f:samples}
\end{widefigure}

We now turn to the theory:  
In order to get the spatial avalanche shape, one needs to construct a solution of the instanton equation 
\eq{inst-equation}, with a source
\bea \label{30-1}
\lambda(x) = -\lambda_1 \delta(x-r_1) -\lambda_2 \delta(x-r_2) + \eta \delta(x-x_{\rm c})\komma \nn\\
  \lambda_1,\lambda_2 \to \infty\punkt
\eea
Looking at our central result \eq{central}, this choice implies that the avalanche kicked at $x=x_0$ does not extend to $x=r_{1,2}$. To simplify matters further, one replaces $m^2 w(x)\to f(x)$, and considers the response to a kick in the force. This allows us to take the limit of $m\to 0$. Setting further  $\sigma=1$, the instanton equation  to be solved is 
\be\label{34}  
\tilde u''(x)  +\tilde u(x)^2 =  -\lambda(x)\komma  
\ee 
The source $\eta$ generates moments of the avalanche size at $x_{\rm c}$. 
While   unsolvable for arbitrary $\eta$, \Eq{34} can be solved perturbatively in $\eta$, allowing us to construct moments of the spatial avalanche shape. This   solution has the   form  
\bea\label{31}  
\tilde u(x) &=& \tilde u_0(x) + \eta \tilde u_1(x)+ \eta^2 \tilde u_2(x)+ ...,
\\
\label{inst-8}
\tilde u_0(x) &=& \frac{1}{(r_2-r_1)^2}\, f\!\left(\frac{2x-r_1-r_2}{2(r_2-r_1)} \right) ,
\\
f(x) &=& - 6 \,{\cal P}\!\left(x+1/2; g_2=0,g_3 = \frac{\Gamma \left(\frac{1}{3}\right)^{18}}{
   (2\pi) ^6}\right) \label{50}
\punkt
\eea
Here ${\cal P}$ is the Weierstrass-P function, diverging at $x=0$ and $x=1$. The subdominant terms in $\eta$ are obtained by realizing that if $f(x)$ is solution of the instanton equation \eq{34}, so is $\kappa^2 f(\kappa x +c)$. 
The  details of this calculation are  cumbersome, and can be found in Ref.~\cite{ZhuWiese2017}. On the right of figure \ref{f:samples} we show $\left< S(X)\right>$ predicted by the theory, and its numerical test. 

Note that  the shape of very large avalanches does not scale as $\ell^3$ but $\ell^4$; it also has a different shape \cite{ThieryLeDoussalWiese2015}.

\subsection{Some theorems}
Inspired by the calculations done so far, 
one can show the following   theorems \cite{LeDoussalWiese2012a}: 
\paragraph{Theorem 1:}
The zero-mode $\dot u(t):=\frac1{L^d} \int_x \dot u(x,t)$ of the BFM field theory \eq{dyn-action3} is the same random process as in the ABBM model, \Eq{48-2}.

\paragraph{Theorem 2:}
\label{theorem2} The field theory of this process is the sum of all tree diagrams, involving   $\Delta'(0^+)$ as a vertex.

\paragraph{Theorem 3:}
\label{theorem3}
Tree diagrams are relevant at the upper critical dimension $d_{c}$. Corrections   involve loops and can be constructed in a controlled  $\epsilon$, i.e.\ loop, expansion around the upper critical dimension $d_{\rm c}$.  

\paragraph{Sketch of Proof:}
One first constructs the generating function for a spatially constant observable, as the velocity or the size in the BFM model. As we saw, these generating functions involve instanton solutions constant in space, thus independent of the dimension. Graphically this can be understood by constructing $\tilde u$ perturbatively, with  
 vertices  proportional to $ \sigma = -\Delta'(0^+)$, and lines which are  response functions, possibly integrated over time. Since by assumption external observable vertices are at zero momentum, all response functions are at zero momentum. This proves theorems 1 and 2. 
 
 We now consider models with one of the fixed points studied above, be it RB, RF or periodic  disorder,  at equilibrium or at depinning. Since each vertex is proportional to $\epsilon$, the leading order is again   given by trees constructed from $\Delta'(0^+)$. 
 The only thing which can be added are loops. Each loop  comes with an additional factor of $\epsilon$ from the additional vertex, of which  the leading one is  given in \Eq{corrections2BFM}. As long as the ensuing momentum integrals are finite, thus do not yield a factor of $1/\epsilon$, these additional contributions are of order $\epsilon^n$, where $n$ is the number of loops. 
 That the momentum integrals are finite can be checked; it reflects the fact that the theory is renormalizable, i.e.\ that all divergences which can possibly appear have already been taken care of by the counter  terms introduced earlier, see section \ref{s:dep-loops} for depinning.

\subsection{Loop corrections}
Loop corrections are cumbersome to obtain, and prone to  errors. To avoid the latter, one should check the obtained results by explicitly constructing  them perturbatively in $\Delta(u)$, and $\lambda$. This is done in the relevant research literature \cite{LeDoussalWiese2012a,LeDoussalWiese2008c,LeDoussalWiese2011b,DobrinevskiPhD}.
Here we sketch the generally applicable method of Ref.~\cite{LeDoussalWiese2012a}, to which we refer for   details. 

\paragraph{Simplified model:}\label{sec:simplified}

Consider the action \eq{corrections2BFM}. To leading order, we can decouple the  term in addition to the BFM   via 
\bea\label{corrections2BFM1}
\rme^{-S[\dot u,\tilde u]} = \left<\rme^{-S_\eta[\dot u,\tilde u]} \right> _\eta\\
\hl{{\cal S}_\eta  [\dot u,\tilde u] = {\cal S}_{{\rm BFM}} [\dot u,\tilde u] + \int_{x,t} \eta(x)\tilde u(x,t) \dot u(x,t)}\punkt
\eea
 $\eta (x) $ is an (imaginary) 
Gaussian disorder to be  averaged over,  with correlations
\be 
\langle \eta(x) \eta(x') \rangle_\eta =- \Delta''(0) \delta^d(x-x') \label{etadis}
\punkt
\ee  
For each realization   $\eta(x)$, the theory has the same form as  in the preceding sections. In particular, the total action (including the sources) is linear in the velocity field, and the only change is an additional term in the instanton equation \eq{inst-equation}, 
\bea\label{inst-equation-eta}
&&\left( -\partial_{t     }-\nabla^{2} + m^{2}\right) \tilde u(x,t)-\sigma \tilde u (x,t)^{2 } \nn\\
&&\qquad = \lambda(x,t) + \eta(x) \tilde u(x,t)\punkt 
\eea
Our central result \eq{central} remains unchanged. 
\paragraph{Perturbative solution:}\label{q1}
To simplify notations,  we   set $m= \sigma=1$.
We expand the solution of \Eq{inst-equation-eta} in powers of $\eta(x)$, denoting by   $\tilde  u^{(n)}(x,t)$ the term of order $\eta^n$, 
\begin{equation}\label{q4}
\tilde u (x,t) = \tilde u^{(0)}(x,t) + \tilde u^{(1)}(x,t) + \tilde u^{(2)}(x,t) + ... 
\punkt
\end{equation} 
The hierarchy of equations to be solved is \begin{eqnarray}\label{3eq}
 \left[- \partial_t - \nabla_x^2 + 1 \right] \tilde u^{(0)}(x,t) =  \lambda(x,t)  +  \tilde u^{(0)}(x,t)^2 \komma \quad \\
\left[- \partial_t - \nabla_x^2 + 1 - 2 \tilde u^{(0)}(x,t)  \right] \tilde u^{(1)}(x,t) \nn\\
=  \eta_{x} \tilde u^{(0)}(x,t) \komma  \\
\left[- \partial_t - \nabla_x^2 + 1 - 2 \tilde u^{(0)}(x,t)  \right] \tilde u^{(2)}(x,t) \nn\\
=  \tilde u^{(1)}(x,t)^2 + \eta(x) \tilde u^{(1)}(x,t)\punkt \quad
\label{3.20}
\end{eqnarray}
The first line is the usual   instanton equation (\ref{inst-equation}). 
Let us introduce   the dressed response kernel 
\bea\label{3.21new}
\left[- \partial_t - \nabla_x^2 + 1 - 2 \tilde u^{(0)}(x,t)  \right]
\mathbb{R}_{x't',xt} \nn\\
=  \delta^{d} (x-x') \delta (t-t')
\punkt
\eea
It has the usual causal structure of a response function, and
obeys a backward evolution equation. It allows us to rewrite the solution of the system of equations (\ref{3eq}) to (\ref{3.20}) as
\begin{eqnarray}\label{q5}
\tilde  u^{1}(x,t) &=& \int_{x'}\int_{t'>t} \eta (x')\,\tilde u^{0}(x',t')  \mathbb{R}_{x't',xt} \label{u1} \komma \\
\tilde u^{(2)}(x,t) &=&  \int_{x'}\int_{t'>t} \left[ \tilde u^{1}(x',t')^2 + \eta(x') \tilde u^{(1)}(x',t') \right] \nn\\
&&\hphantom{ \int_{x'}\int_{t'>t} }\times \mathbb{R}_{x't',xt}~.~~~~~~~~
\label{u2}
\end{eqnarray}
Consider now the average (\ref{etadis}) over $\eta(x)$.
Since $ \langle\tilde  u^{(1)}(x,t) \rangle_\eta = 0$, the lowest-order correction is given by the average of $ \tilde u^2(x,t)$,
\be 
Z[\lambda] = Z_{\mathrm{tree}}[\lambda] + \int_{xt} \langle\tilde  u^{(2)}(x,t) \rangle_\eta +... 
\punkt
\ee  
Inserting Eq.~(\ref{u1}) into Eq.~(\ref{u2}), and performing the average over $\eta$, one  finds
\begin{eqnarray}\label{q6}
\langle  \tilde u^{(2)}(x,t) \rangle_\eta \nn\\
=  -\Delta''(0) \int_{t<t_{1}<t_{2},t_{3}} \int_{x_{1},x'} \tilde  u^{(0)}(x',t_{2}) \tilde u^{(0)}(x',t_{3})\, \times \nn\\
\qquad\qquad\qquad\qquad \quad\times \mathbb{R}_{x' t_{2},x_{1}t_{1}}
 \mathbb{R}_{x't_{3},x_{1}t_{1}} \mathbb{R}_{x_{1}t_{1}, xt} \nonumber \\
-\Delta''(0) \int_{t<t_{1}<t_{2}} \int_{x'}\tilde u^{(0)}(x',t_{2})\,
\mathbb{ R}_{x't_{2},x' t_{1}} \mathbb{R}_{x' t_{1}, xt} \punkt
\end{eqnarray}
It admits the following graphical representation
\begin{equation}\label{q7}
\langle \tilde u^{(2)}(x,t) \rangle_\eta = \diagram{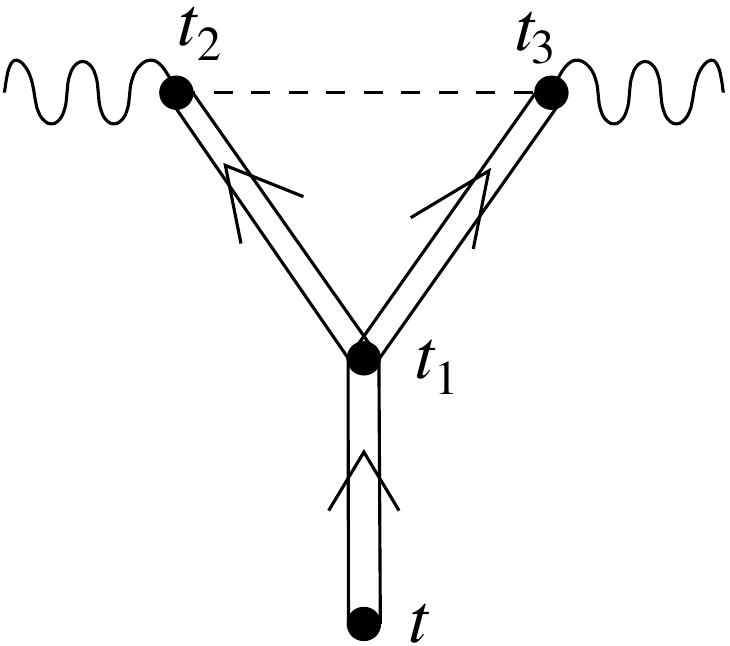}+\diagram{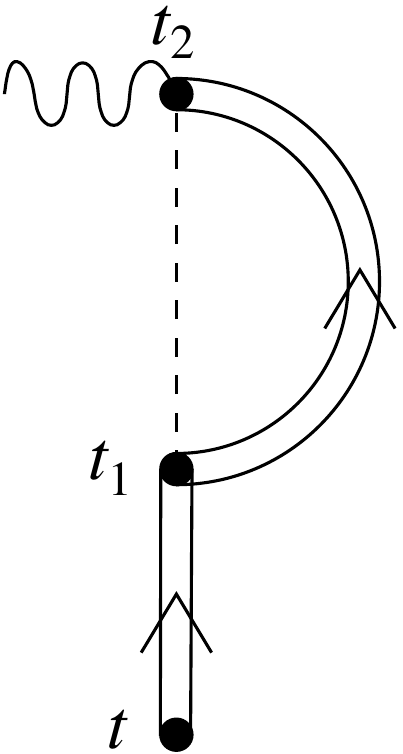}\punkt 
\end{equation}
The symbols are as follows: (\textit{i}) a wiggly line represents $\tilde u^{(0)}(x,t)$, the mean-field solution;
(\textit{ii}) a double solid line is a dressed response function $\mathbb{R}$,
advancing in time following the arrow (upwards), thus times are
ordered from bottom to top. 
We now define the combination 
\begin{equation}\label{Phi}
\Phi (x',x,t):= \int_{t'>t} \tilde u^{(0)}(x',t') \,\mathbb{R}_{x't',xt}\komma 
\end{equation}
in terms of which \bea\label{tilde-u2}
\langle  \tilde u^{(2)}(x,t) \rangle_\eta \nn \\
= \int_{t',x'} \left[\int_{y} \Phi
(y,x',t')^{2}+ \Phi (x',x',t') \right] \mathbb{R}_{x't',xt}
\punkt 
\eea
There are  several additional terms: (i)  a counter-term for the disorder, showing up in a change of $\Delta'(0^+)$ to its renormalized value.  (ii)  a counter-term   to  friction. (iii) a missed boundary term, due to the replacement of $\Delta''(u_t-u_{t'})$ which decays to zero for large times by $\Delta''(0)$, which does not.

\paragraph{1-loop corrections to the avalanche-size distribution:}
Let us construct the 1-loop corrections to the avalanche-size distribution, following the formalism developed above. 
For $\lambda(x,t)=\lambda$,  the solution of the unperturbed instanton equation was given in \Eq{80bis}. For $m=\sigma=1$, it reads
\be\label{Z-S-MF}
Z_{\rm MF}(\lambda) \equiv\tilde u^{(0)} = \frac{1}{2} \left(1- \sqrt{1-4 \lambda}\right) 
\punkt
\ee
The dressed response kernel in Fourier representation  becomes
\be 
   \mathbb{R}_{k,t_{2},t_{1}} =  e^{ - (k^2+1- 2 \tilde u^{(0)}) (t_2-t_1)   } \theta (t_2-t_1) 
   . 
\ee 
It is   the bare response function up to the replacement $m^2 \to m^2 - 2 \tilde u^{(0)}(\lambda)$.
The combination in the exponential simplifies, 
\be
k^2+1- 2 \tilde u^{(0)} = k^2 +\sqrt{1-4\lambda}\punkt
\ee
Formula (\ref{Phi}) then gives\be
\Phi(k,t_1) = \tilde u^{(0)}\int_{t_1<t_2} {\mathbb R}_{k,t_2,t_1} = \frac{\tilde u^{(0)}}{k^2+1-2 \tilde u^{(0)}} 
\punkt
\ee
\begin{widefigure}
\begin{center}
\mbox{\fig{0.47\textwidth}{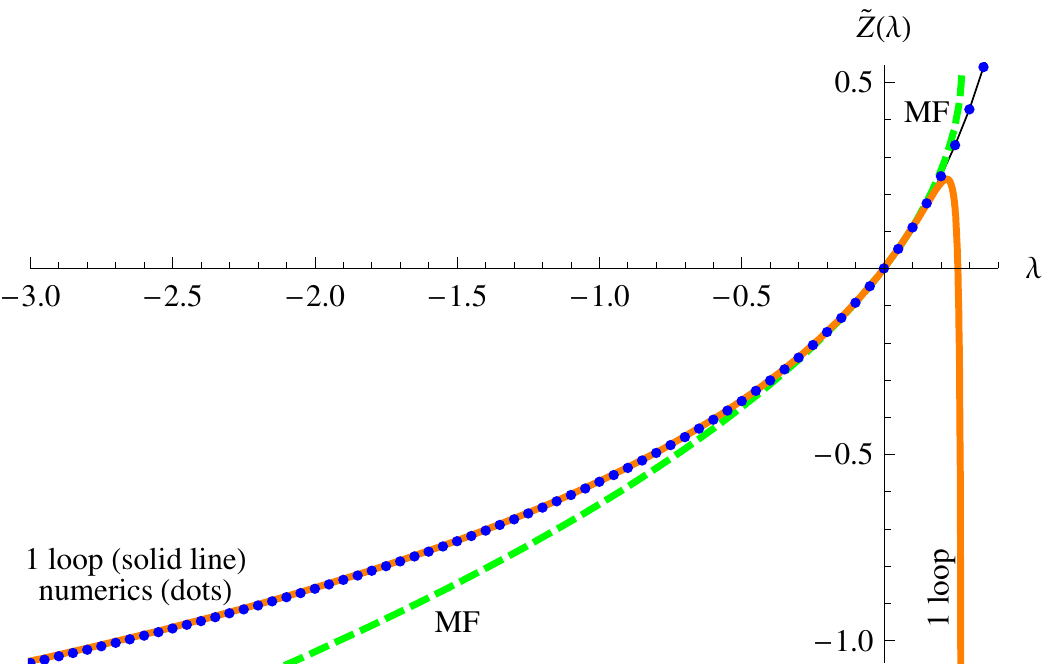}}~~\mbox{\fig{0.51\textwidth}{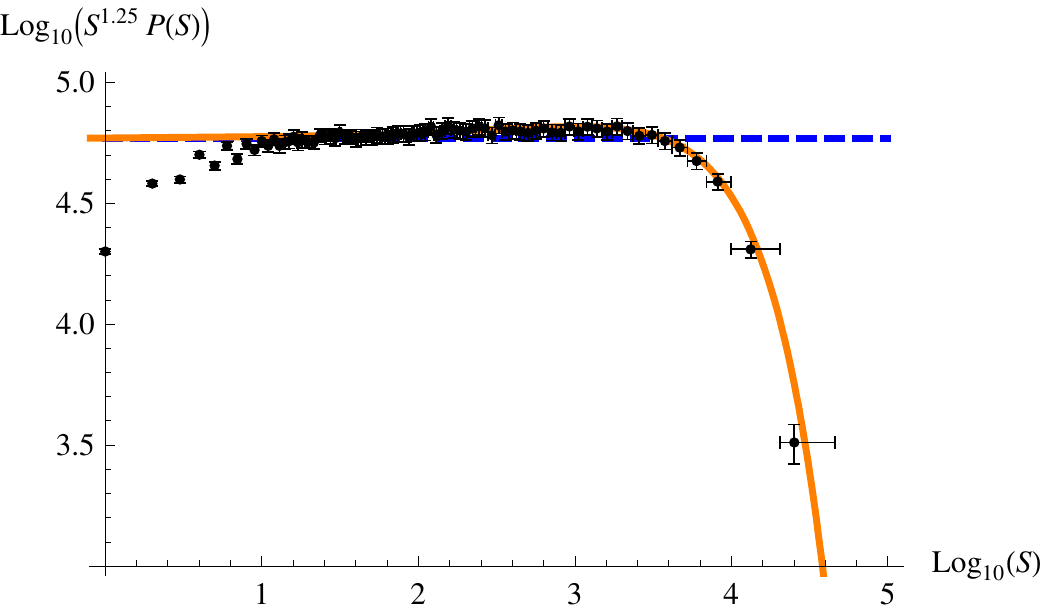}\hspace{-7.5cm}\raisebox{20mm}[0mm][0mm]{\parbox{0mm}{\fig{5cm}{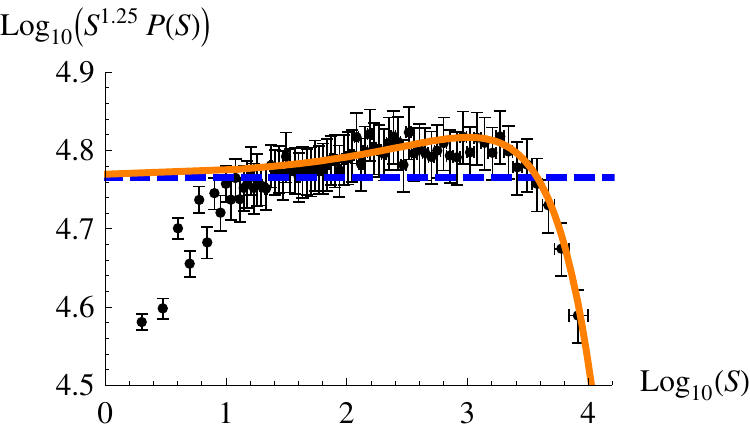}}}\hspace{7.5cm}}
\end{center}
\caption{Results of Ref.~\cite{LeDoussalMiddletonWiese2008} for RF disorder, $d=2$. Left: Numerically measured $\tilde Z(\lambda)$  (blue dots).   MF result \eq{Z-S-MF} (green dashed),     1-loop result \eq{Z-aval-1-loop} (orange solid). The latter is rather precise, almost up to the singularity at $\lambda=1/4$.
Right: Avalanche-size distribution $P(S)$, multiplied by $S^{\tau}$ with $\tau=1.25$ from \Eq{tau-relation} (dots).
The orange solid curve is the prediction from \Eq{finalS}.
The dashed line is a constant (guide to the eye). 
Inset: blow-up of main plot.
}
\label{f:2}
\end{widefigure}With the additional integral over $\mathbb R$ in \Eq{tilde-u2}, the latter    becomes
\bea\label{Jt2}
\langle  \tilde u^{(2)}  \rangle_\eta
= -\frac{\Delta''(0^+)}{ 1-2\tilde u^{(0)}}\nn \\
\times \int_k 
 \bigg(\frac{\tilde u^{(0)}}{k^2+1-2 \tilde u^{(0)}}\bigg)^{\!2}  + \frac{\tilde u^{(0)}}{k^2+1-2\tilde u^{(0)}} \punkt
\eea
Adding the proper counter-terms, and replacing the bare disorder by the renormalized one \cite{LeDoussalWiese2012a}, 
the full generating function is
\bea\label{Z-aval-1-loop}
Z(\lambda)   \equiv \tilde u
= \tilde u^{(0)} -\frac{\tilde \Delta''(0^+)}{ 1-2\tilde u^{(0)}}\frac1{\epsilon I_1} \times\\
  \int_k \bigg[ 
 \bigg(\frac{\tilde u^{(0)}}{k^2{+}1{-}2 \tilde u^{(0)}}\!\bigg)^{\!\!2}  {+} \frac{\tilde u^{(0)}}{k^2{+}1{-}2\tilde u^{(0)}} 
 {-} \frac{\tilde u^{(0)}}{k^2{+}1} {-} \frac{3 (\tilde u^{(0)})^2 }{(k^2{+}1)^2} \bigg]. \nn
\eea

\paragraph{Avalanche-size distribution at 1-loop order:}
The generating function \eq{Z-aval-1-loop} can be inverted analytically \cite{LeDoussalWiese2008c}. The result for avalanches larger than a microscopic cutoff $S_0$ is  to $\ca O(\epsilon)$
\bea\label{finalS}
P(S) = \frac{\left<S\right>}{2 \sqrt{\pi}} S_m^{\tau-2} A S^{-\tau} \exp\!\left(C \sqrt{\frac{S}{S_m}} -
\frac{B}{4} \left[\frac{S}{S_m}\right]^\delta\right)\punkt\nn\\
\eea
The  coefficients are to $\ca O(\epsilon)$
\bea \label{amp} 
 A = 1+ \frac{1}{8} (2-3 \gamma_{\mathrm{E}} ) \alpha    , \quad B = 1-\alpha \left(1+\frac{\gamma_{\mathrm{E}}}{4}\right) \komma  \nn \\
  C=- \frac{\sqrt{\pi}}{2}  \alpha   , \quad \alpha=  \frac{\zeta-\epsilon }{3}\komma 
\eea
and 
$\gamma_{\mathrm{E}}=0.577216$ is Euler's number. The exponent $\tau $ is consistent with the scaling relation \eq{tau-relation}, while the new exponent $\delta$ reads
\be\label{a64}
 \delta= 1 + \frac{\epsilon - \zeta}{12}   \punkt
\ee

\begin{figure*}
 \centerline{{\includegraphics[width=0.98\columnwidth]{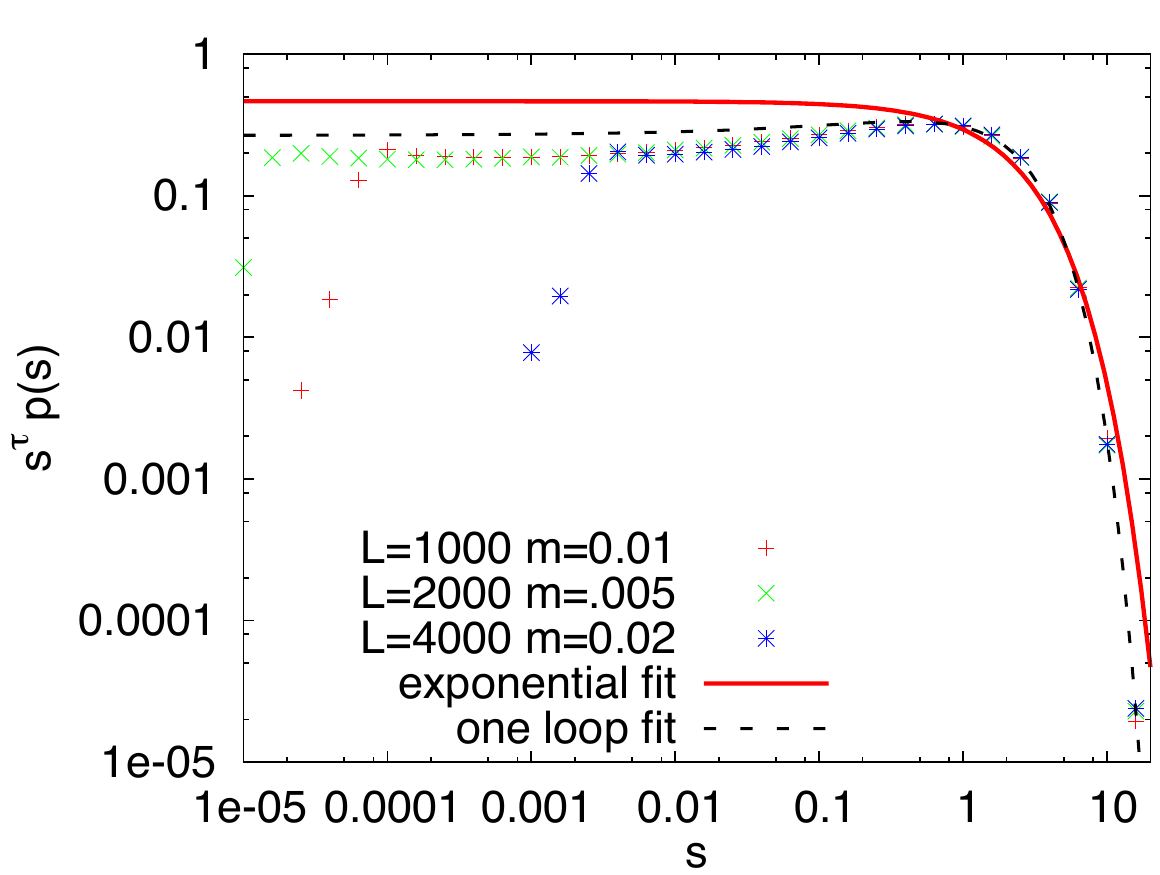}}~~\hfill~~\includegraphics[width=1\columnwidth]{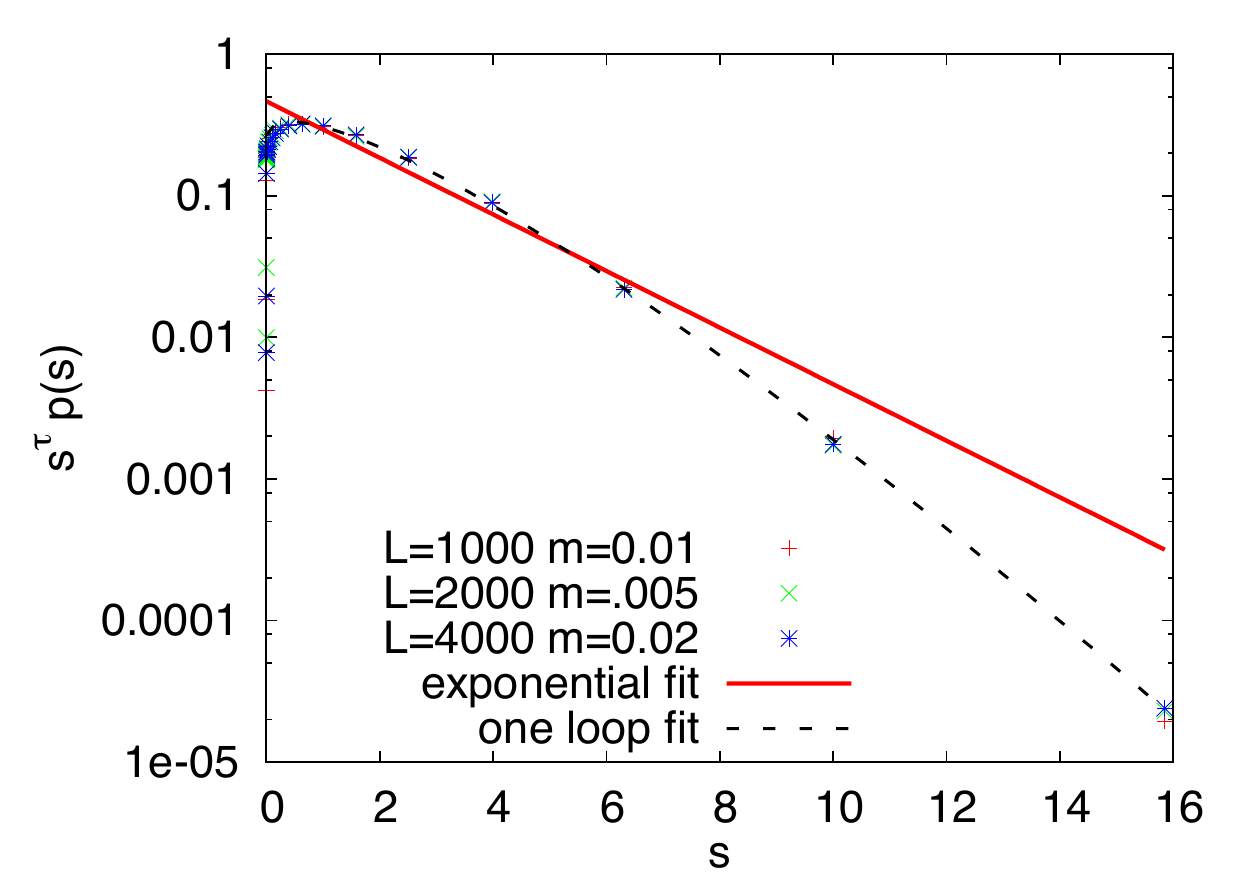}}
\caption{Left: Avalanche-size distribution for Random Field in $d=1$ at depinning. The variable $s=S/S_m$.  Blow up of the power-law
region. The red solid curve is given by the MF result \Eq{83},  the black dashed line  by
\Eq{finalS}, with $A=0.947$, $B=1.871$ and $C=0.606$. Right: the same for the tail. Data and Figs.~from \cite{RossoLeDoussalWiese2009a}.}
\label{f:RF-aval-dep}
\end{figure*}

\subsection{Simulation results and experiments}
\subsubsection*{Avalanche-size distribution:}
The result for the avalanche-size distribution has been verified numerically, both for the statics \cite{LeDoussalMiddletonWiese2008} as for depinning \cite{RossoLeDoussalWiese2009a}. 

For the statics (equilibrium) \cite{LeDoussalMiddletonWiese2008}, we show plots on figure \ref{f:2}. The simulations are for a 3-dimensional RF magnet, with weak disorder s.t.\ only a single domain wall appears, yielding $d=2$, $\epsilon=2$, and $\zeta=\zeta_{\rm RF}=2/3$. The generating function $Z(\lambda)$ is verified with high precision. For the avalanche-size distribution, the agreement is   good, even though there is appreciable noise due to binning, which is absent from the generating function $Z(\lambda)$.

At depinning,
 avalanches  are simulated for an elastic string in $d=1$ \cite{RossoLeDoussalWiese2009a}. The results for system sizes   up to $L=4000$ are shown on figure \ref{f:RF-aval-dep}. The statistics is good, allowing to verify  \Eq{finalS}  in the tail region, with $\delta=7/6$.

\subsubsection*{The temporal avalanche shape at fixed duration $T$:}
The temporal shape at fixed duration $T$  is predicted by the theory \cite{DobrinevskiLeDoussalWiese2014a,DobrinevskiPhD}
as 
\begin{eqnarray}\label{T-shape}
\left< \dot  u\left(t = \vartheta T \right) \right>_T =2 {\cal N}\Big[ T\vartheta (1-\vartheta )\Big]^{\gamma-1} \\
\times\exp\!\bigg(\!- \frac{16 \epsilon}{9 d_{\rm c}} \bigg[ \mbox{Li}_2(1-\vartheta )-\mbox{Li}_2\Big(\frac{1-\vartheta }{2}\Big)+\frac{\vartheta  \ln (2 \vartheta )}{\vartheta -1} \nn\\
~~~\qquad +\frac{(\vartheta +1) \ln
   (\vartheta +1)}{2(1- \vartheta )}\bigg] \bigg)\punkt \nn
\end{eqnarray}
The exponent $\gamma$ is given in \Eq{gamma}.
The temporal shape is  well approximated by 
\be
\left<\dot u(t = \vartheta T)\right>_T \simeq  \left[T \vartheta(1-\vartheta)\right] ^{\gamma-1} \exp\left({\cal A}[ \textstyle\frac12-\vartheta]\right) \punkt
\ee
The asymmetry ${\cal A}\approx - 0.336 (1-d/d_{\rm c})$ is negative for $d$ close to $d_{\rm c}$, skewing the avalanche towards its end, as observed in numerical simulations in $d=2$ and $3$ \cite{LaursonPrivate}. For $d=1$ the asymmetry is positive 
in numerical simulations \cite{LaursonIllaSantucciTallakstadyAlava2013}. In experiments on  magnetic avalanches (Bark\-hausen noise), and in fracture experiments, the asymmetry is difficult to see \cite{LaursonIllaSantucciTallakstadyAlava2013}.

\begin{figure}
\includegraphics[width=8.3cm]{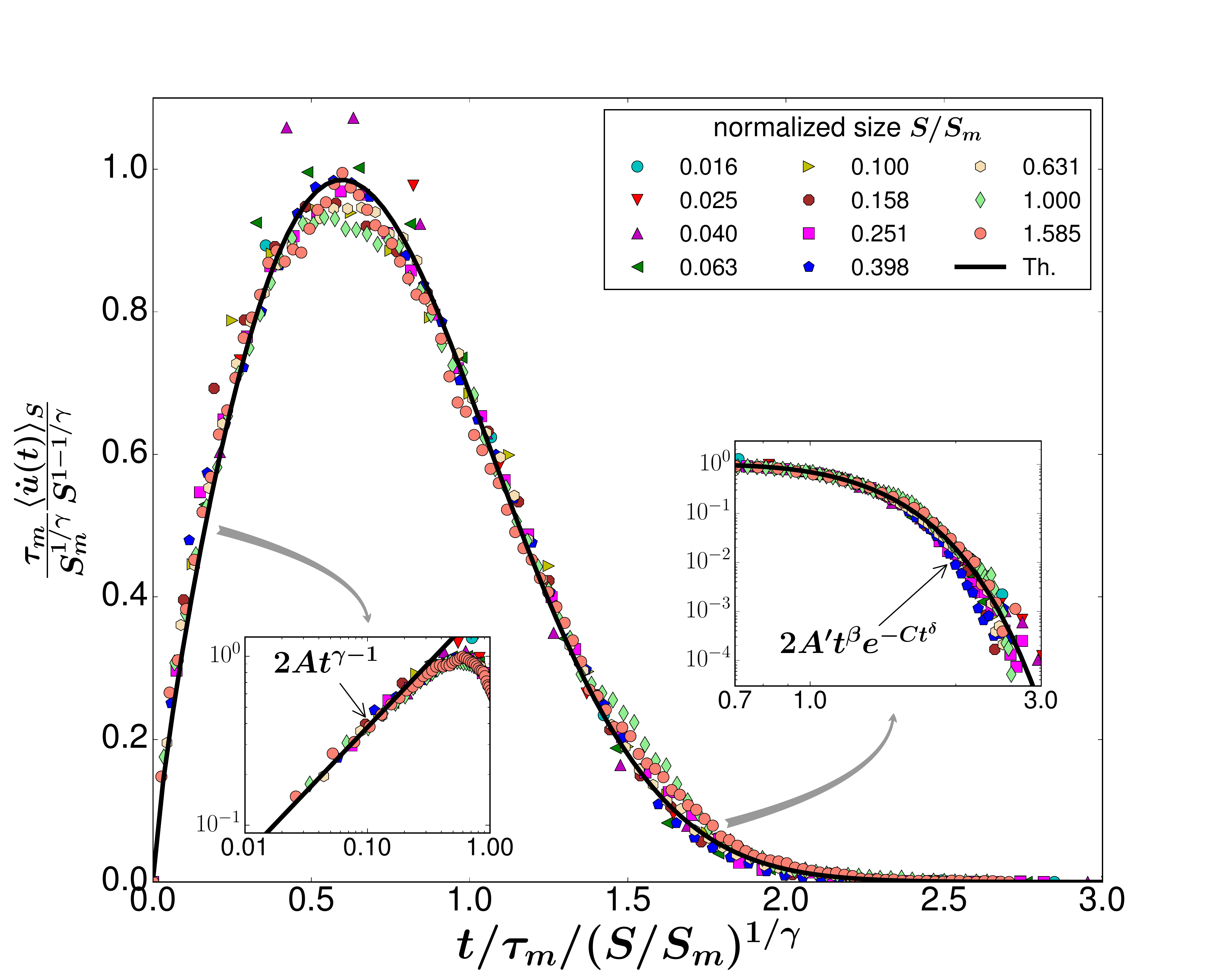}\caption{Scaling collapse of the average shape at fixed avalanche sizes 
$\langle \dot {\sf u}(t) \rangle_S$, according to Eq.\ (\ref{shapeS}), in the FeSiB thin film. 
The continuous line is the prediction for the universal SR scaling function of 
Eq.~(\ref{S-shape}). 
The insets show comparisons of the tails of the data
with the predicted asymptotic behaviors of 
Eqs.~(\ref{asympt}) and (\ref{asympt2}), setting $\epsilon=2$, with 
$A=1.094, A'=1.1, \beta=0.89, C=1.15$, and $\delta=2.22$. Consistent with scaling relations, the  measured   $\gamma=1.76$. Fig.~from \cite{DurinBohnCorreaSommerDoussalWiese2016}.
}\label{fig:aveShape_FeSiB}
\end{figure}

\subsubsection*{The temporal avalanche shape at fixed size $S$:}
The temporal shape can also be calculated at fixed size $S$.
Scaling suggests
that 
\be \label{shapeS} 
\langle \dot u(t) \rangle_S = \frac{S}{\tau_m} \Big(\frac{S}{S_m}\Big)^{\!-\frac{1}{\gamma}} f\bigg( \frac{t}{\tau_m}\Big(\frac{S_{m}}{S}\Big)^{\!\frac{1}{\gamma}} \bigg)\komma 
\ee 
with $\int_0^{\infty} \rmd t\, f(t) = 1$, where $f(t)$ may depend on $S/S_m$. In mean field, the scaling function $f(t)$
is independent of $S/S_m$ \cite{DobrinevskiLeDoussalWiese2013},
and reads\bea
f_0(t) = 2 t e^{- t^2}    , \quad \gamma=2\punkt
\eea 
To one loop one obtains
$
f(t) = f_0(t) - \frac{\epsilon}{9} \delta f(t) . $ 
Expressions for arbitrary $S/S_m$ are lengthy. The universal small-$S$ limit reads
\bea
\label{228} \nn
 \delta f(t) =  \frac{f_0(t)}{4} \bigg[&\pi  \left(2
   t^2+1\right) \mbox{erfi}(t)   +2 \gamma_{\rm E} 
   \left(1-t^2\right)-4
\\  & -2 t^2 \left(2 t^2+1\right) \,
   _2F_2\left(1,1;\frac{3}{2},2;t^2\right)\nn\\
   & -2 e^{t^2}
   \Big(\sqrt{\pi } t\,
   \mbox{erfc}(t)-\mbox{Ei}\left(-t^2\right)\Big)\bigg]\punkt ~~~~~~~
\eea 
It satisfies $\int_0^{\infty} \rmd t\, \delta f(t) =0$. The asymptotic behaviors  are 
\begin{eqnarray} \label{asympt} 
f(t) \simeq_{t \to 0} 2 A t^{\gamma-1}, \quad A=1 + \frac{\epsilon}{9} (1-\gamma_{\rm E}), \\
 f(t) \simeq_{t \to \infty}  2 A' t^{\beta} e^{- C t^\delta} \label{asympt2}, \quad  \delta = 2 + \frac{\epsilon}{9} , \quad \beta = 1- \frac{\epsilon}{18}, \nn \\
A' = 1+ \frac{\epsilon}{36}  (5 - 3 \gamma_{\rm E} - \ln 4), \quad 
C = 1+ \frac{\epsilon}{9} \ln 2. 
\end{eqnarray}
The amplitude $A$
leads to the same universal short-time behavior as in \Eq{T-shape}. To properly extrapolate to larger values of $\epsilon$, we use
\beq \label{S-shape}
f(t) \approx 2 t e^{-C t^\delta} {\cal N} \exp\!\left(- \frac{\epsilon}{9} \!\left[\frac{\delta f(t)}{f_0(t)}{-}t^2 \ln (2t)\right]  \right),
\eeq
with the normalization ${\cal N}$ chosen s.t.\ $\int_0^{\infty} \rmd t f(t) = 1$.  \Eq{S-shape} is exact  to ${\cal O}(\epsilon)$
and satisfies the asymptotic expansions  (\ref{asympt}) and  (\ref{asympt2}).

This result has beautifully been measured in the Barkhausen noise experiment of  Ref.~\cite{DurinBohnCorreaSommerDoussalWiese2016}, see figure \ref{fig:aveShape_FeSiB}.

\subsubsection*{The spatial avalanche shape (in $d=1$):} The spatial avalanche shape for the BFM was shown on figure \ref{f:samples}. 
For systems with SR-correlated disorder, it was measured for two different driving protocols: tip driven (driving at a single point), and spatially homogeneous driving by the parabola, the protocol used above. For tip-driven avalanches at the non-driven end, as well as for homogeneously driven avalanches, \Eq{gofx} predicts that the avalanche shape at fixed extension $\ell$ grows close to the boundary point $b$ as 
\be
\left< S(x)\right> _\ell \sim |x-b|^\zeta\punkt
\ee
For $\zeta=1.25$ one thus expects this curve to have a slightly positive curvature at these points, consistent with   plots 3 and 5 of Ref.~\cite{AragonKoltonDoussalWieseJagla2016}. 

Let us also mention the studies of 
\cite{ThieryLeDoussalWiese2015} for avalanches with a large aspect ratio in the BFM which are rare, and with fixed seed position \cite{ThieryLeDoussal2016} which are difficult to realize in an experiment.

\subsubsection*{The velocity distribution.}
\begin{figure}[t]
\Fig{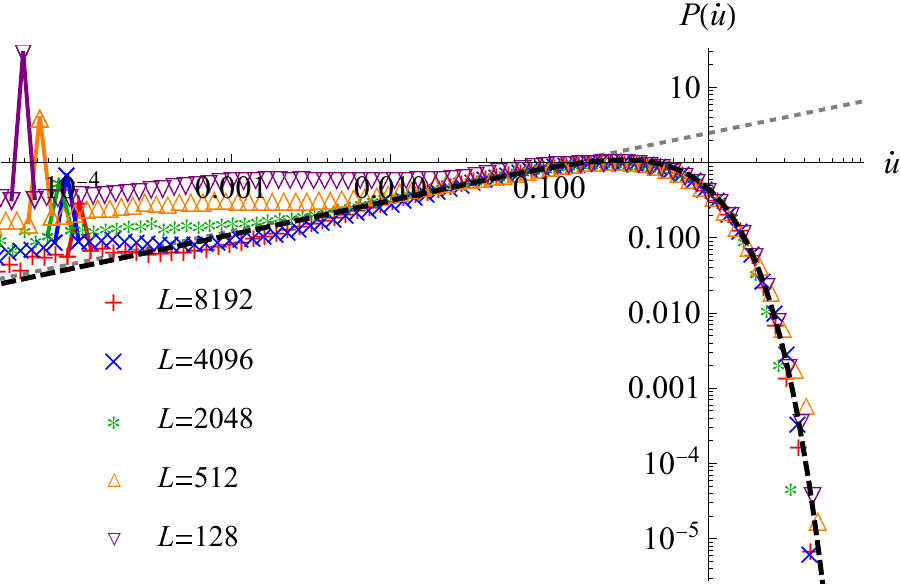}
\caption{The center-of mass velocity distribution $P(\dot u)$. The  weight of the peak at $\dot u=v_{\rm kick}$  is $\frac{\delta t}{\left< T\right>}\sim L^{-z}\sim m^z$, where $T$ is the duration of an avalanche and $\delta t$ the time discretization step.  
The analytic result (black dashed line) is from Eq.~(385) of Ref.~\cite{LeDoussalWiese2012a}, the dotted gray line the pure power law  $P(\dot u)\sim {\dot u}^{{-\sf a}}$, with   ${\sf a} = -\frac{10}{23} =-0.435$ as  given in \Eq{a-in-P(udot)}. There is no adjustable (fitting) parameter, thus  convergence to the theory including all scales is 
  read off from the plot. Plot from \REF{KoltonLeDoussalWiese2019}.}
\label{f:Pofv}
\end{figure}The velocity distribution was analytically obtained in Refs.~\cite{LeDoussalWiese2011a,LeDoussalWiese2012a}, and numerically checked in  \REF{KoltonLeDoussalWiese2019}. The   scaling relation of \Eq{a-in-P(udot)} actually predicts a negative exponent 
$a=-10/23$, implying $P(\dot u) \sim \dot u^{10/23}$. Despite the change in sign, this is beautifully verified in Fig.~\ref{f:Pofv}.

\subsection{Correlations between avalanches}
In section \ref{s:shocks}, we had asked how avalanche moments are encoded in $\Delta(w)$, and found the key relation  \eq{Delta'(0+)}.
We can further ask how avalanches at $w_1$ and $w_2$ are correlated. This can be evaluated along the same lines \cite{ThieryLeDoussalWiese2016}: On one hand, 
\bea\label{576}
\overline{[u_{w_1+\delta w_1} - u_{w_1}]  [u_{w_2+\delta w_2} - u_{w_2}]}\nn\\
= \left< S_{w_1} S_{w_2} \right> \delta w_1 \, \delta w_2 \,\rho_2(w_1-w_2)  + \ca O(\delta w^3), 
\eea
where $\rho_2(w)$ is the probability density to have two shocks a distance $w$ apart. 
On the other hand, 
\bea\label{577}
 \overline{[u_{w_1+\delta w_1} - u_{w_1}]  [u_{w_2+\delta w_2} - u_{w_2}]} -\delta w_1 \delta w_2 \nn\\
= \frac1{m^{4}L^{d}}\big[ 
 \Delta(w_1{+}\delta w_1 {-}w_2 {-}\delta w_2) -\Delta(w_1 {-}w_2 {-}\delta w_2)\nn\\
\hp{= m^{-4}L^{-d}\big[ } -\Delta(w_1{+}\delta w_1 {-}w_2 )+ \Delta(w_1  {-}w_2  ) \big]   \nn\\
= -\delta w_1 \delta w_2   \frac{  \Delta''(w_1{-}w_2)}{m^4L^d} + \ca O(\delta w^3).
\eea
Using \Eq{rho-shock} in \Eq{576}, and comparing to \Eq{577} for small $\delta w$ implies
\be\label{e:anticorrelations}
\frac {\left< S_{w_1} S_{w_2} \right>^{\!\rm c}}{\left< S\right> ^2}\equiv \frac {\left< S_{w_1} S_{w_2} \right>}{\left< S\right> ^2} -1 = -\frac{\Delta''(w_1-w_2)}{m^{4}L^d}.
\ee
\begin{figure}[t]
\centerline{(a)\hspace{.24\textwidth}(b)}
\vspace*{-0.2cm}
\centerline{     
    \includegraphics[width=.25\textwidth]{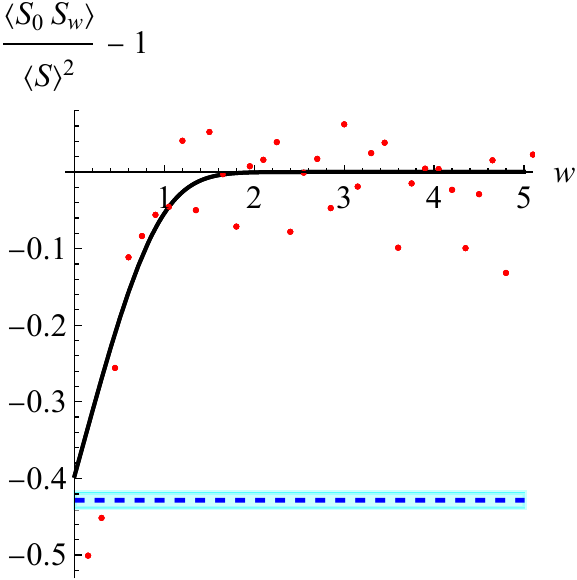}\hfill
    \includegraphics[width=.25\textwidth]{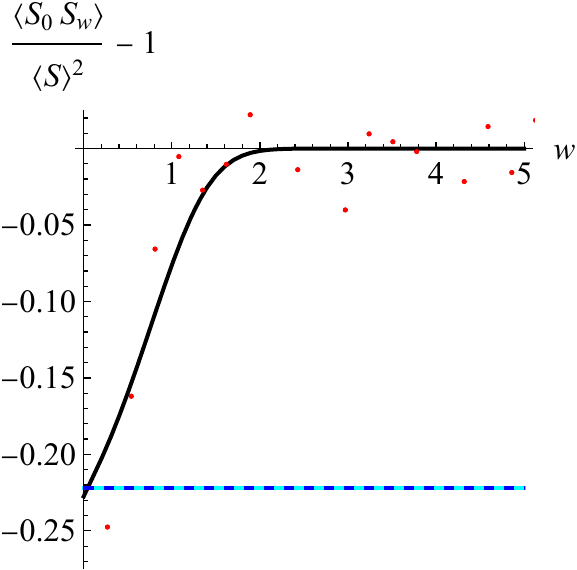}}
\caption{Anticorrelation of avalanches as a function of $w$, for   two samples with eddy currents, SR (a), and LR (b). The solid line is the prediction for $-\Delta''(w)/(m^4L^d)$ from \Eq{e:anticorrelations}, as obtained from the experiment. The dashed lines are bounds on the maximally achievable reduction from the $\epsilon$-expansion \eq{bound}, with error bars in cyan for SR. There are no fitting parameters.}
\label{Fig4}
\end{figure}Since $\left< S_{w_1} S_{w_2} \right> \ge0$, the r.h.s.\ is bounded from below by $-1$, or
\be
\frac{\Delta''(w)}{m^{4}L^d} \le \frac{\Delta''(0^+)}{m^{4}L^d} \le 1.
\ee
For the Kida and Sinai models, this yields the bounds
$
{\tilde \Delta''(0^+)} \le 1, 
$ which are indeed satisfied by \Eqs{Delta-Kida} and \eq{Delta-tilde-Sinai}. At depinning, the DPM has $\tilde \Delta''(0^+)=0.5$, see \Eq{DPM-Gumbel}.
In the perturbative FRG,  \Eqs{Rtillde-def} and \eq{Delta''(0)-FP}, extended by the 2-loop results of \cite{LeDoussalWieseChauve2002}, imply \be\label{bound}
 \frac{  \Delta''(0^+) } {m^4 L^d }\simeq  \left(  \frac{2}{9} +0.107533 \epsilon + \ca O(\epsilon^2) \right)   \frac1{ m^4 L^d  I_1 } .
\ee
The diagram $I_1$ defined in \Eq{I1} as an integral, here depends both on $m$ and $L$, and  is evaluated as a discrete sum over momenta $k_i=n_i 2\pi/L$, $n_i\in \mathbb Z$.  
One shows that  $ m^4 L^d I_1\ge 1$,   the bound is saturated for $mL\to 0$, and deviations from the bound remain smaller than $10\percent$ for  $mL\le 3.2$ in $d=1$, $mL\le 2.4$ in $d=2$, $mL\le 1.8$ in $d=3$, and $mL\le 0.6$ in $d=4$,   indicating optimal choices for the sample size.  The experiments \cite{terBurgBohnDurinSommerWiese2021} shown on Fig.~\ref{Fig4} satisfy \eq {e:anticorrelations}, and  almost saturate the bound \eq{bound}.  
Further relations  are studied in Refs.~\cite{ThieryLeDoussalWiese2016,LeDoussalThiery2020}.

\subsection{Avalanches with retardation}
\label{s:Avalanches with retardation}
In magnetic systems, a change in the magnetization induces an {\em eddy current}, which in turn can reignite an avalanche which  had already   stopped  \cite{ZapperiCastellanoColaioriDurin2005}. The   simplest model exhibiting this phenomenon, and which remains analytically solvable \cite{DobrinevskiLeDoussalWiese2013} reads
\begin{eqnarray}
  \partial_t u(t) &= F\big(u(t)\big) + m^2 \big[w(t) - u(t)\big] -  a   h(t),\\
 \tau \partial_t h(t) &= \partial_t u(t) - h(t) .
\label{eq:EOMExp}
\end{eqnarray}
While many observables can be obtained analytically \cite{DobrinevskiLeDoussalWiese2013}  and  measured, e.g.\ the temporal shape given $S$, other ones are not well-defined, as the duration of an avalanche. As due to the eddy current $h(t)$, an avalanche can restart, this complicates the data-analysis in real magnets \cite{DurinZapperi2006b}.

\subsection{Power-law correlated random forces, relation to fractional Brownian motion}
Fractional Brownian motion (fBm) is the unique Gaussian process $X_t$ which is  scale and translationally invariant, see e.g.~\cite{PiterbargBook1995,PiterbargBook2015,Michna2009,DelormeWiese2015}. It is uniquely characterized by its 2-point function 
\be
\left< X_t X_s \right> = \sigma \left( t^{2H} + s^{2H} - |s-t|^{2H}\right) .
\ee
The Hurst exponent $H$ may take values between 0 and 1, 
\be
0< H \le 1. 
\ee 
Note that $X_t$ is non-Markovian, since the 2-time correlations of increments at times $t\neq s$
\be
\left< \partial_t X_t \partial_s X_s \right> =  \, 2H (2H-1) \sigma |s-t|^{2H-2}
\ee
do not vanish, except for $H=1/2$, for which the fBm reduces to standard Brownian motion. 

Since $X_t$ is a Gaussian process, many observables can be calculated analytically. This is   interesting, since one can access 
analytically, in an expansion in $H-1/2$,  most variables of interest for extremal statistics   \cite{WieseMajumdarRosso2010,DelormeWiese2015,DelormeWiese2016b,DelormeWiese2016,DelormeRossoWiese2017,Wiese2018,SadhuDelormeWiese2017,BenigniCoscoShapiraWiese2017,WalterWiese2019a,WalterWiese2019b,ArutkinWalterWiese2020}. An example of such an observable is the maximum relative height of elastic interfaces in a random medium 
\cite{RambeauBustingorryKoltonSchehr2011}.
Fractional Brownian motion is also the   simplest choice if one only knows the scaling dimension $H$ of a process, without further insight into higher correlation functions. 
\begin{figure}[t]
\fig{4.1cm}{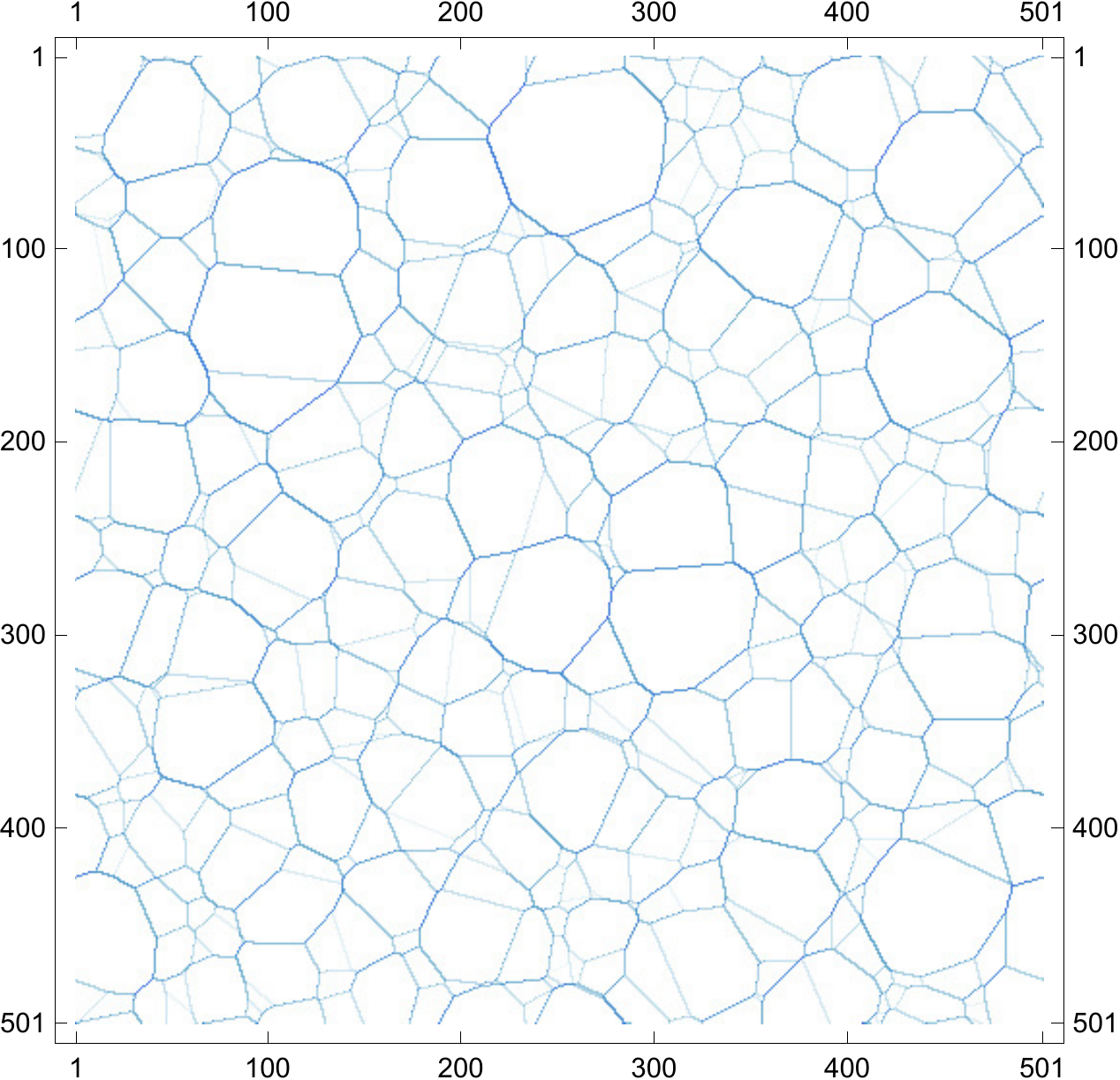}\hfill\fig{4.12cm}{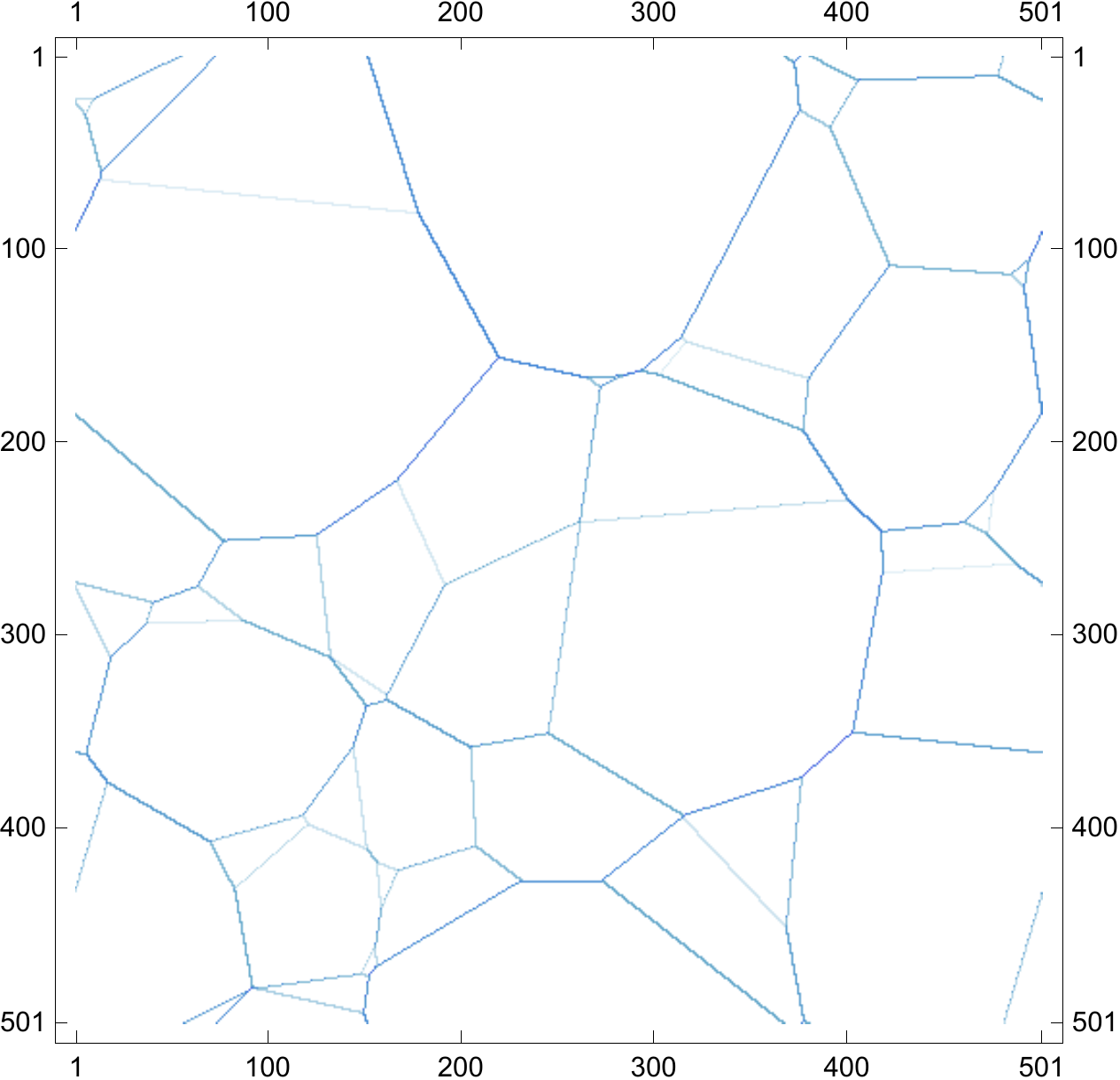}
\caption{Shocks in a 2-dimensional system with short-ranged correlated disorder,  size $L=500$, and periodic boundary conditions, for two different masses $m^2=10^{-3}$ (left) and $m^2=10^{-4}$ (right). Decreasing $m^2$, shocks merge. Shock fronts are almost straight.}
\label{f:2D-shocks}
\end{figure}

Returning to depinning, 
suppose that   random forces are Gaussian and correlated as a fractional Brownian motion  
\be
\Delta(0)-\Delta(u) = \sigma |u|^{2H}.
\ee
Solving \Eq{two-loop-FRG-dyn}, and realizing that  loop corrections are subdominant\footnote{Corrections in the FRG equation \eq{flow-Delta} are $\delta[ \Delta(0)- \Delta(u)] \sim u^{4H-2} \ll u^{2H}$ for $u\to \infty$.} in the tail for   $H< 1$, we obtain 
similar to  the derivation of \Eq{zeta-BFM}
\be\label{615}
\zeta = \frac{\epsilon}{2(1-H)}.
\ee
As a consequence of \Eq{tau-relation}, the avalanche-size exponent is 
\be\label{taufBm}
\tau = 2-\frac2{d+\zeta} = 2-\frac{4(1- H)}{4+d(1-2H)}.
\ee
Interestingly, in $d=0$, i.e.\ for a particle, this reduces to 
\be\label{taufBm-d=0}
\tau \big|_{d=0} = 1+H.
\ee
This is consistent with the first-return probability derived in Refs.~\cite{DelormeWiese2015,DelormeWiese2016b}. 
Indeed, the probability to return to the origin of a fBm  $X_t$ with Hurst exponent $H$ is $P(t) = \left< \delta (X_t)\right>\sim t^{-H}$,  equivalent to Eq.~(40) of \cite{DelormeWiese2016b}. 
The probability to return for the first time is $\partial_t P(t)\sim t^{-(1+H)}$, equivalent to \Eq{taufBm-d=0}. These considerations generalize those leading to \Eqs{P-return} and \eq{P-first-return}, and in $d=0$ confirm \Eqs{615} and \eq{taufBm}.

\subsection{Higher-dimensional shocks}
Little is known  about higher-dimensional shocks or avalanches. As in our understanding of the cusp, the 2-dimensional toy model \eq{5.37} is helpful here, with $V(u)$ drawn as uncorrelated Gaussian random variables with variance $1$ on a unit grid. On Fig.~\ref{f:2D-shocks} we show the shocks, i.e.\ the locations where the minimizer $u$ in \Eq{5.37} changes discontinuously. 
Principle properties are 
\begin{enumerate} 
\item $\hat V(u)$ can be interpreted as a decaying KPZ height field, and $\hat F(u):=-\nabla \hat V(u)$ as a decaying Burgers velocity, see section \ref{s:Decaying KPZ, and shocks}.
\item decreasing $m^2$, i.e.\ increasing {\em time} $t \sim m^{-2}$ in the KPZ/Burgers formulation, shocks merge and annihilate.
\item shock fronts are straight lines. 
\item when crossing a shock line, the minimizer $u$ of \Eq{5.37} jumps perpendicular to the shock.
\end{enumerate}
Properties (iii) and (iv) suggest to write (with $S=|\vec S|)$ \cite{LeDoussalWieseUnpublished}
\be
P(\vec S) \rmd S_{1}\rmd S_{2}= \mathbb P(S) \rmd S\, \cos \theta \rmd \theta .
\ee
Using $
\rmd S_{1}\,\rmd S_{2} = S \,\rmd S\, \rmd \theta
$
yields 
\be
P(S_{1},S_{2}) = \frac{\mathbb P(S) }{S} \cos \theta.
\ee
On the other hand, one can again solve the problem in the mean-field limit \cite{LeDoussalRossoWiese2011,LeDoussalWieseUnpublished}, valid if the microscopic disorder  $R(0)-R(u)\sim |u|^3$. 
In this limit, shocks are  {\em an infinitely divisible process} \cite{LeDoussalWiese2011b}. As a consequence, 
\be
\overline{\rme^{\vec \lambda [\vec u (\vec w)-u (\vec 0)-\vec w]}} =
\rme^{w Z (\vec \lambda)} = \int \rmd^{N}\vec S\, \rme ^{\vec \lambda \vec S} P
(\vec S,\vec w).\label{250}
\ee
As in section \ref{s:Inertia}, the 
{\em large-deviation function} $F (\vec x)$ can be  defined as
\begin{equation}\label{k4}
F (\vec x):=-\lim_{w\to\infty} \frac{ \ln P(\vec x w,\vec w)}{w}.
\end{equation}
Inserting this expression into Eq.~(\ref{250}) yields
\begin{equation}\label{k5}
\rme^{ w Z (\vec \lambda)}=  w^{N} \int \rmd^{N} \vec x\, \rme^{w [\vec \lambda \vec x - F (\vec
x )]\  }
.\end{equation}
This shows that 
the generating function 
$Z (\vec \lambda)$ and the large-deviation function $F(\vec w)$ are 
Legendre-transforms of each other, 
\bea\label{k6}
Z (\vec\lambda) +F (\vec x) =\vec  \lambda \vec x, \\
\lambda_{i} = \frac{\partial}{\partial x_{i}} F (\vec x), \quad 
x_i = \frac{\partial}{\partial \lambda_{i}} Z (\vec \lambda ).
\eea
It is non-trivial to show \cite{LeDoussalRossoWiese2011,LeDoussalWieseUnpublished} that 
\begin{equation}\label{k9}
F ( x_{1},x_{2}) = \frac{2 x_2^2+\left[x_2^2+\left(x_1-1\right) x_1\right]^2}{4
   \big(x_1^2+x_2^2\big)^{3/2}}\ .
\end{equation}
Measuring only  a single component, equivalent to setting $x_{2}=\lambda_2=0$, this reduces  to 
\begin{equation}\label{k10}
F (x,0) = \frac{(1{-}x)^2}{4 x} , \quad 
Z (\lambda,0) = \frac{1}{2}\left (1{-}\sqrt{1{-}4\lambda} \right).
\end{equation}
This is the same generating function as in  \Eq{80bis}, thus the probability distribution for the longitudinal component $S_1$ is as given in \Eq{PwS(S)} (standard Watson-Galton process  \cite{WatsonGalton1875,LeDoussalWiese2008a}).
The transversal avalanche-size distribution is more involved, but a parametric representation for $\tilde Z_2(\lambda):= Z(0,\lambda) $ can be given \cite{LeDoussalRossoWiese2011}, 
\begin{figure}
\includegraphics[width=4.23cm]{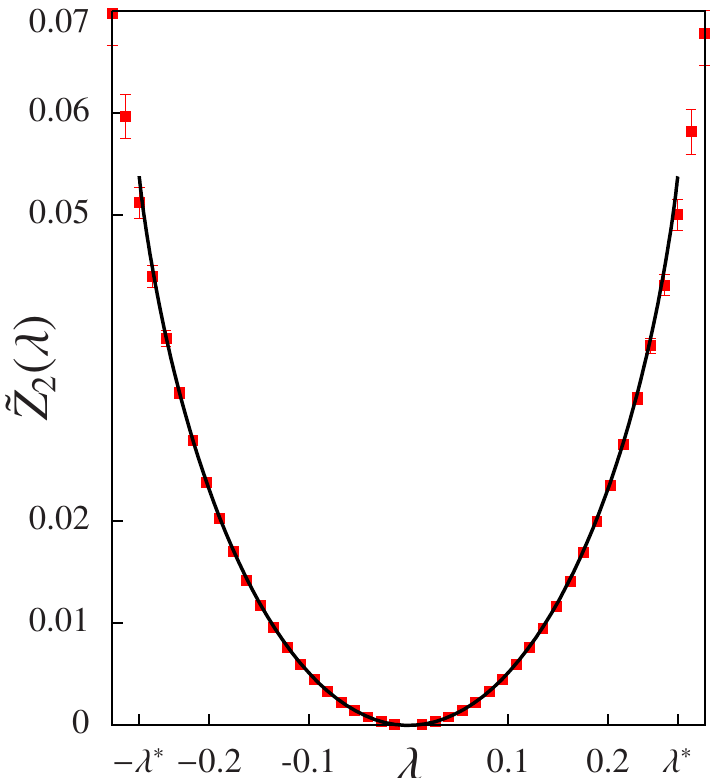}\hfill\includegraphics[width=4.0cm]{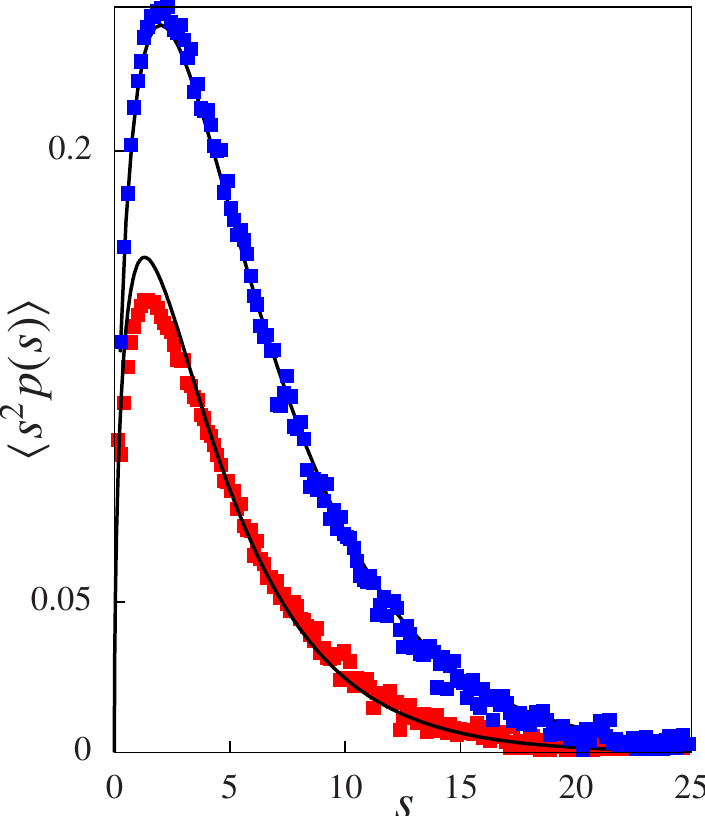}
\caption{Left:  measured $ \tilde Z_2(\lambda)$  (squares) compared to the prediction  (\ref{z2}) (solid line).
Right: Plot of $s_x^2 p_1(s_x)$ (top curve) and $s_\perp^2 [p_2(s_\perp)+ p_2(-s_\perp)]$ (bottom curve).  
Solid lines represents the analytical predictions. Results and Figs.~from \cite{LeDoussalRossoWiese2011}.
}\label{Z+P-N=2}
\end{figure}\begin{eqnarray} \label{z2}
\begin{array}{rcl}
 \lambda (\theta) &=& \ds \sin( \theta )\frac{ \sqrt{5-\cos (4 \theta)}+2}{\big[1-\cos (2 \theta)+\sqrt{5-\cos (4
   \theta)}\big]^2},  \qquad \\
\tilde Z_2(\theta) &=& \ds \frac{\cos (\theta)}{2} \frac{ \sqrt{5-\cos (4 \theta)}-2}{1-\cos (2 \theta)+\sqrt{5-\cos (4
   \theta)}} .
\end{array}
\end{eqnarray}
This allows one to obtain the graph of $\tilde Z_2(\lambda)$, and even to Laplace-invert it. The results and numerical tests  are shown in Fig.~\ref{Z+P-N=2}.
 
\subsection{Clusters of avalanches in systems with long-range elasticity}
\label{s:avalanches-LR-elasticity}
When   elasticity is long-ranged, avalanches can nucleate away from the   part of the   avalanche   including the first point to have moved. 
This is an old problem, with many references, see e.g.\ \cite{MaaloySantucciSchmittbuhlToussaint2006,LaursonSantucciZapperi2010,BudrikisZapperi2013,LePriolChopinLeDoussalPonsonRosso2020,LePriolLeDoussalRosso2020,LePriolThesis}.

Suppose that 
each  avalanche of size $S$ is composed of $N_c(S)$ clusters, distributed as 
\be
P(S_c|S) \sim S_c^{-\tau_c } \Theta(S_c< S), \quad \tau_c<2.
\ee
Then the  typical size of clusters, given  avalanche size $S$, is 
\be
\left<S_c\right>_S = \int_0^\infty \rmd S_c \, S_c P(S_c|S) \sim S^{2-\tau_c}.
\ee
There are 
\be
N_c(S) \simeq  \frac {S}{\left< S_c\right> } \sim S^{\tau_c -1}
\ee
clusters. Suppose that the number of clusters scales as 
\be
 P(N_{\rm c})  \sim  N_c^{-\mu} \Theta(N_c < N_c(S)).
\ee
On dimensional grounds, $P(S)\rmd S \sim P(N_c)\rmd N_c$. Inserting the above relations yields
\be\label{tau_c}
\tau_c -1 = \frac {\tau -1}{\mu-1}.
\ee 
If one further supposes \cite{LePriolChopinLeDoussalPonsonRosso2020,LePriolLeDoussalRosso2020,LePriolThesis} that the generation of a new cluster is a Galton-Watson process (section \ref{s:Watson-Galton process}), then 
\bea
\mu = 3/2, 
\eea
simplifying \Eq{tau_c} to \cite{LePriolChopinLeDoussalPonsonRosso2020,LePriolLeDoussalRosso2020,LePriolThesis}
\be
\tau_{\rm cluster}= 2\tau -1.
\ee
Numerically it was checked \cite{LePriolChopinLeDoussalPonsonRosso2020,LePriolLeDoussalRosso2020} that this scaling relation works for all $0\le \alpha< 2$; it might actually continue to work for $\alpha=2$, if one keeps a finite value for $\ca A_d^\alpha$ in \Eq{A-d-alpha}  avoiding to  reduce the power-law kernel to short-ranged correlations in that limit (see section \ref{s:LR-elasticity}). These results have recently been questioned \cite{Terrot2021}.

\subsection{Earthquakes}

Gutenberg and Richter 
\cite{GutenbergRichter1944,GutenbergRichter1956}  first reported that the magnitude of earthquakes in California follows a power-law, equivalent to an avalanche-size exponent\footnote{Geophysicist usually consider the cumulative distribution of magnitude. The magnitude was originally defined as ``proportional to the log of the maximum amplitude on a standard torsion seismometer'' \cite{Nature1945}.} of  $\tau=3/2$. 
Due to its enormous impact on society, much research is done in the domain, both by geophysicists with the aim of predicting the next big earthquake, and by theoretical physicists, trying to put earthquakes into the framework of disordered elastic manifolds.  
The latter is successful to a certain extend:
\begin{itemize}
\item
 the elastic object depinning is a 2-dimensional fault plane, to which the relative movement is confined, often with sub-mm precision ({\em localization}),
\item 
driving is through the tectonic plates, equivalent to the  parabolic confining potential of \Eq{Hconf},
\item the elastic interactions on the fault plane are long-ranged  since elasticity is mediated by the bulk. The calculation is essentially the same as for contact lines in section \ref{s:LR-elasticity}, and yields $\alpha=1$ in \Eq{Hel-alpha-k},
\item
the critical dimension  $d_c(\alpha)= 2\alpha = 2$ is the dimension of the fault plane. The system is in its critical dimension. As a consequence $\zeta=0$, $z=2$, and $\tau = 3/2$, which correctly predicts the Gutenberg-Richter law.
\end{itemize} 
 But there is an additional element: After an earthquake, the fault is {\em damaged}, rendering it less resistant to further movement before the damage is {\em healed}, which happens on   a much  longer time scale (see e.g.\ \cite{Dieterich1992}). As a consequence, immediately after a big earthquake, the likelihood of another earthquake is increased. It is indeed found that the probability for an {\em aftershock} to appear decays (roughly) as $1/t$ in time $t$, known today as  Omori's law \cite{Omori1894}. 

For further reading,  we refer  to the original literature 
\cite{BurridgeKnopoff1967,Dieterich1992,BenZionRice1993,Ruina1983,CarlsonLangerShaw1994,BenZionRice1997,FisherDahmenRamanathanBenZion1997,DSFisher1998,Scholz1998,ShomeCornellBazzurroCarballo1998,Monte-MorenoHernandez-Pajares2014,Kagan2002,SchwarzFisher2003,JaglaKolton2009}
and to some of the  relevant concepts discussed in this review,
long-range correlated elasticity (section \ref{s:avalanches-LR-elasticity}), and inertia (section \ref{s:Inertia}).

\subsection{Avalanches in the SK model}
The ABBM model, the BFM, or any other approach based on a random walk, and commonly summarized as ``mean field'' gives an avalanche-size exponent $\tau =3/2$ (defined in \Eq{rhof}), bounding from above all experiments and simulations  on disordered elastic manifolds
\bea
\zeta_{\rm ABBM}= \zeta_{\rm BFM} = \zeta_{\rm MF} = \frac 32 \ge \zeta^d_{\rm dep} \equiv 2-\frac2{d + \zeta}>1.\nn\\
\eea
On the other hand, since $d+\zeta>2$, even in dimension $d=1$, there seems to be a lower bound on $\tau$ as well, indicated above.
Note that for $\tau\le 1$    the avalanche-size distribution becomes non-integrable at large $S$ in absence of an IR cutoff. 

It is thus quite surprising to learn that in the SK model \cite{SherringtonKirkpatrick1975} the exponent $\tau $ is smaller \cite{LeDoussalMuellerWiese2010,LeDoussalMuellerWiese2011}, 
\be
\tau^{\rm equilibrium}_{\rm SK} = \tau^{\rm dynamic}_{\rm SK} = 1.
\ee
This result for the equilibrium was obtained \cite{LeDoussalMuellerWiese2010,LeDoussalMuellerWiese2011} within a full-RSB scheme, relevant for SK. 
Curiously, exactly the same exponent is found in numerical simulations \cite{PazmandiZarandZimanyi1999} out of equilibrium, where one simply increases the magnetic field until one spin becomes unstable, which is then flipped. While finding the ground state is an NP-hard problem, this dynamic algorithm is trivial to implement. Still, the exponent $\tau$ is the same.
It is also counterintuitive to learn that avalanches in the SK model involve a finite fraction of its total of $N$ spins, changing the magnetization on  average by  $\sqrt{N}$ for an increase in external field by  order  $1/\sqrt{N}$, i.e.
\be
S_{\rm typ} := \frac{\left< S^2\right> }{\left< 2S\right> } = \sqrt{N}.
\ee
 This means that on the complete hysteresis curve  each spin flips on average an order of   $\sqrt{N}$ times. This is very different from magnetic domain walls, where each spin flips exactly once. It is compatible with the non-integrable tail in the size distribution, $P(S)\sim 1/S$, knowing that there is no natural IR cutoff other than the system size.

\section{Sandpile Models, and Anisotropic Depinning}
\label{s:sandpiles}
\subsection{From charge-density waves to sandpiles}
While nowadays sandpile models  constitute a domain of statistical physics and mathematics on their own, 
it is worth reminding that they originated in the study of charge-density waves. In the seminal paper \cite{BakTangWiesenfeld1987}, the authors considered an array of rotating pendula elastically coupled to their neighbors via weak torsion springs, a mechanical analogue of a charge-density wave. In any equilibrium state the pendulum will almost point down. 
Consider  a decomposition of the  positions $u_i$ of the pendula, into their integer part  
$ \bar u(i)$ and a rest $\delta u(i)$, 
\be
u(i) = \bar u(i) + \delta u(i)  \ , \quad \bar u(i) \in \mathbb{Z}.
\ee
 The limit considered in 
\cite{BakTangWiesenfeld1987} is that of weak springs as compared to the gravitational forces, implying that $\delta u(i)$ is small. 
The forces acting on pendulum $i$ are 
\bea\label{630}
z(i) &=&g \cos \big(2\pi u(i)\big)+\sum_{j\in {\rm nn}(i)} u(j)-u(i) +F(i) \nn\\
&=&g \cos (2\pi \delta u(i))+\sum_{j\in {\rm nn}(i)} \bar u(j){-}\bar u(i) \nn\\
&&+ \sum_{j\in {\rm nn}(i)} \delta u(j) {-}\delta u(i)+F(i)\punkt
\eea
$F(i)$ are $u$-independent applied forces, and $g$ is the gravitational constant (with mass and length of the pendula set to 1).
The sum runs over the nearest neighbors $j$ of $i$, denoted ${\rm nn}(i)$.
If a pendulum becomes unstable, $\bar u(i)\to \bar u(i)+1$. 
The model \eq{630} can also be viewed as a charge-density wave at depinning (sections \ref{model}, \ref{s:Charge-density wave (CDW) fixed point} and \ref{s:dyn-2loop}).

Supposing that the $\delta u(i)$ are small,   the update rule for 
$z(i)$  can be written as  
\bea\label{ASM-toppling}
z(i)  \to z(i) - 2d , \nn \\
z(j) \to z(j) +1\komma \quad \mbox{for }j \in {\rm nn}(i)\punkt
\eea
Again neglecting $\delta u(i)$, the  condition for the event \eq{ASM-toppling} is  
\be
z(i)>z_{\rm c},
\ee
with $ z_{\rm c} = g$.

\subsection{Bak-Tang-Wiesenfeld, or  Abelian sandpile model}
The Bak-Tang-Wiesenfeld (BTW) model \cite{BakTangWiesenfeld1987}, also known as the Abelian sandpile model (ASM), uses the update rules 
\eq{ASM-toppling}
combined with 
\be
z_{\rm c}=2d .
\ee 
It is   interpreted as a sandpile of height $z(i)$. A site {\em topples}, i.e.\ the rule \eq{ASM-toppling} is performed, when its height exceeds $z_{\rm c}$.
If several sites become unstable at the same time, one has to choose an order of the topplings. Considering the original model in terms of the $u(i)$, and using Middleton's theorem,  it is clear that the final state is independent of the order of   updates. Stated differently, the topplings commute. For this reason the model is also referred to as the Abelian sandpile model (ASM). Its algebra was studied in detail, especially by D.~Dhar \cite{MajumdarDhar1992,Dhar1999,Dhar1999b,Dhar2006}.

In this model, one   starts from $z(i)=0$ for all $i$, chosen to belong to a finite lattice with open boundaries, as a chess board.  At randomly chosen sites $i$ grains are  added, $z(i)\to z(i)+1$. If a site becomes unstable, it topples. If this toppling renders   one of its neighbors  unstable, it   topples in turn.  Grains  fall off at the boundary.  When   topplings have stopped,  a new grain is added.

In the interface formulation,   grains falling off at the boundary  correspond to an interface where $u(i)=0$ outside the finite lattice (``on the boundary'').  
As a result, the system is automatically in a critical state. This phenomenon called {\em self-organized criticality} (SOC), made the BTW model \cite{BakTangWiesenfeld1987} popular. It is now recognized that if a system can become critical, slowly driving it   achieves criticality. 
In the language developed in this review, it is   {\em velocity-controlled depinning}, instead of {\em force-controlled depinning} (section \ref{s:Phenomenology}).
Many natural phenomena are self-organized critical, and a large literature exists on the topic \cite{PruessnerBook,BonachelaAlavaMunoz2008,BonachelaPhD,BonamySantucciPonson2008,BonachelaChateDornicMunoz2007,UritskyPaczuskiDavilaJones2007,Dhar2006,Jeng2005,StapletonChristensen2005,Dhar2004,Alava2003,Alava2002,BasslerPaczuskiAltshuler2001,DickmanMunozVespignaniZapperi2000,PazmandiZarandZimanyi1999,Dhar1999,Dhar1999b,DickmanVespignaniZapperi1998,BasslerPaczuski1998,CuleHwa1998,Jensen1998,TanguyGounelleRoux1998,ChristensenCorralFretteFederJossang1996,FretteChristensenMalthe-SorenssenFederJossangMeakin1996,UrbachMadisonMarkert1995,Sneppen1992,Manna1991,DharMajumdar1990,DharRamaswamy1989,TangBak1988,BakTangWiesenfeld1987}.

 Configurations in the ASM can be classified as recurrent or not. Recurrent configurations can   be realized in the steady state, while non-recurrent ones can not. An example for a non-recurrent configuration is the initial state $u(i)=0$. Recurrent configurations can be mapped one-to-one onto   uniform spanning trees, and the $q$-states Potts model in the limit of $q\to 0 $. There further is an injection onto  loop-erased random walks. We discuss this in more depth in section \ref{s:Loop-erased random walks, and other models equivalent to CDWs}.
 We refer the reader to the cited literature and especially \cite{PruessnerBook,Dhar1999} for  details.

\subsection{Oslo model}
\begin{figure}[t]
\Fig{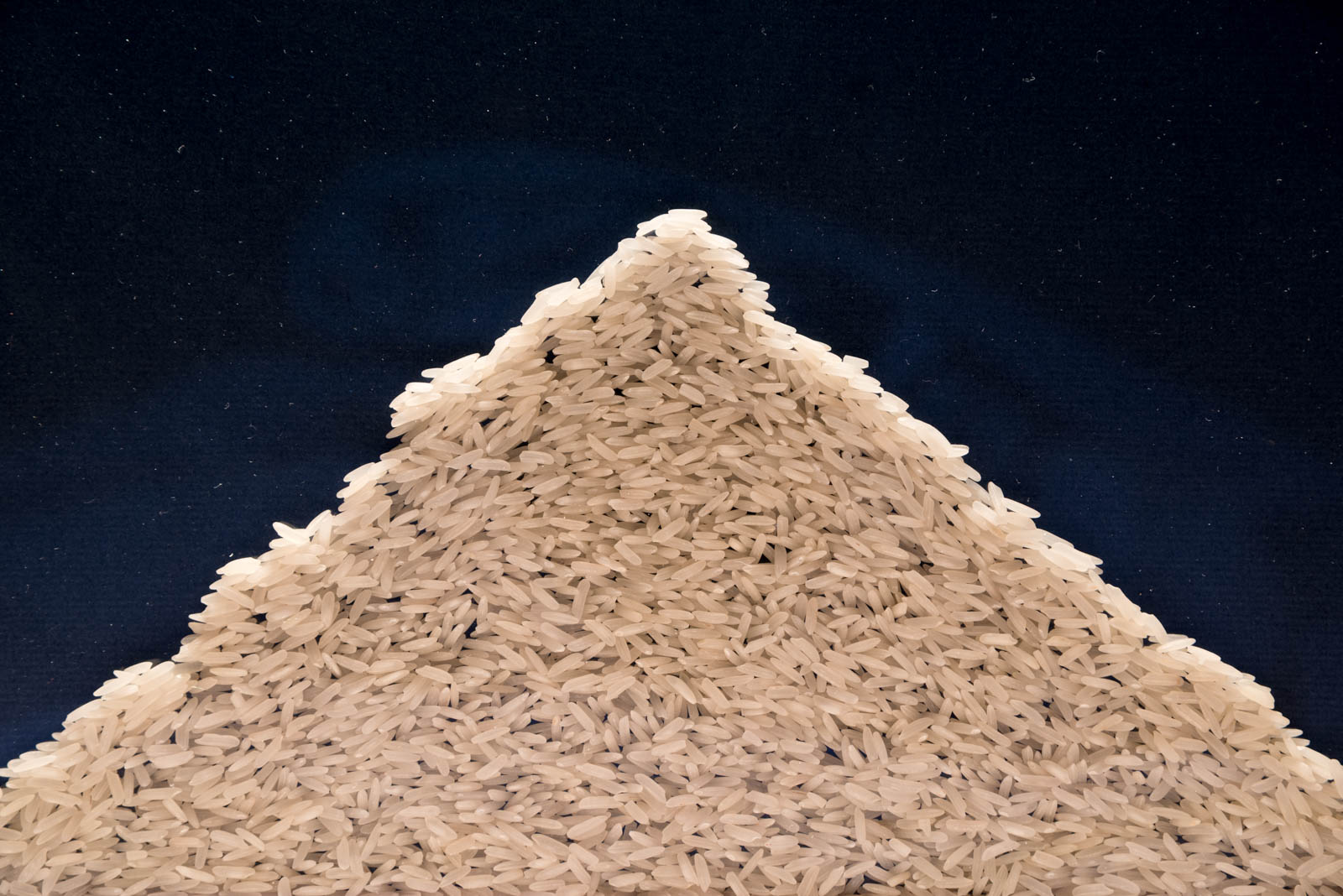}
\caption{Stable configuration in a rice pile experiment. (Photo by the author). The grains are between two glass plates 5mm apart. The pile was prepared by slowly increasing the inclination of the plates from horizontal to vertical. Brighter grains sit at the top and are more likely to topple.}
\label{f:Kay-sandpile-experiment}
\end{figure}
\begin{figure}
\begin{center}
\fig{7cm}{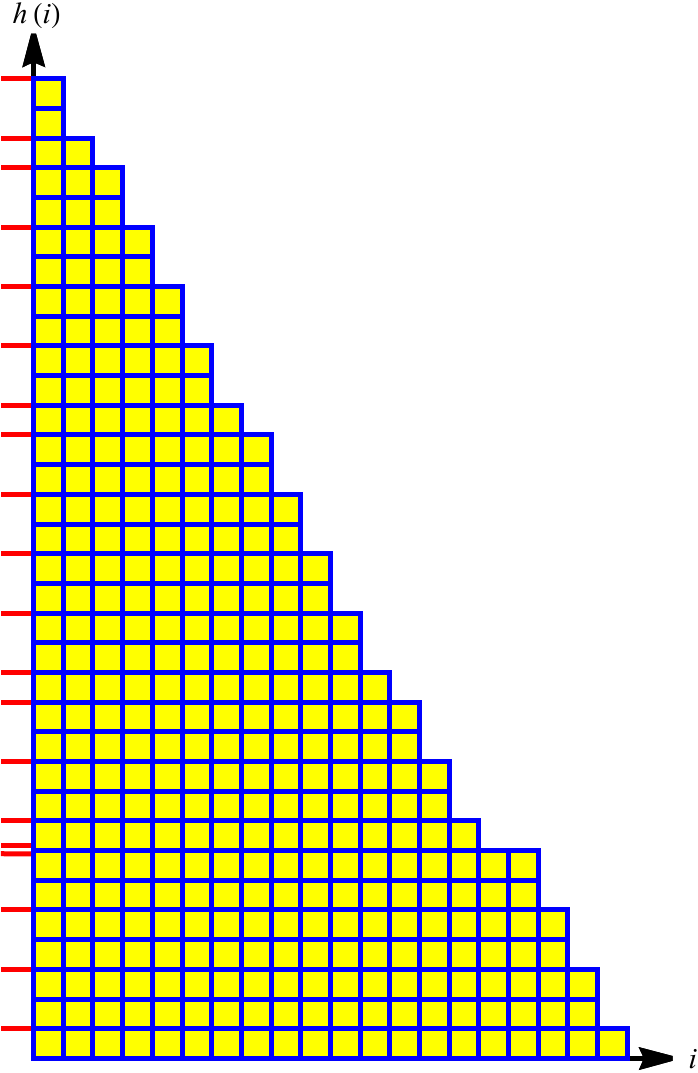}
\fig{7cm}{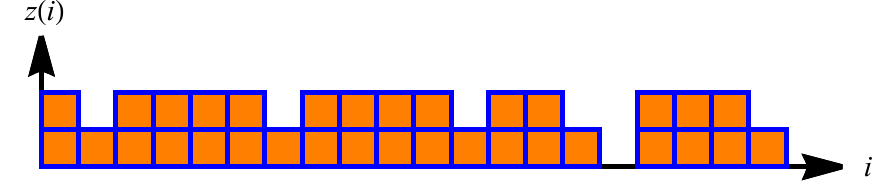}
\fig{7cm}{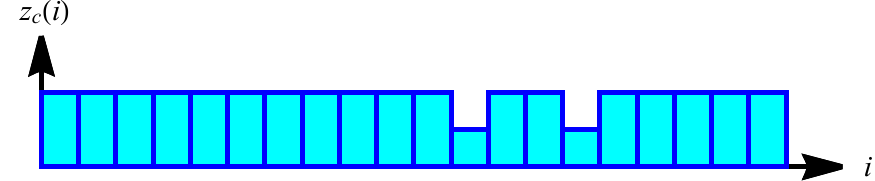}
\end{center}
\caption{A stable configuration of the Oslo model. The latter is a cellular automaton version of the right half of the rice pile in Fig.~\ref{f:Kay-sandpile-experiment}. The red lines indicate the particle positions of particles used in section \ref{zeta=5/4}. Note that there is one plateau where two particles sit on top of each other, drawn here slightly apart.}
\label{f:Oslo-model}
\end{figure}
Albeit we used the term ``sandpile'', we did not yet  motivate its use. To this aim, consider Fig.~\ref{f:Kay-sandpile-experiment}. The system is in a stable configuration, characterized by a mean slope, plus fluctuations. A grain may start to slide, 
depending on the local slope, the friction  between the neighbors, and its orientation. The ASM does not contain any randomness, but instead is deterministic. Randomness enters only through the driving, i.e.\ the order in which grains are added. 
Any realistic model for a   sandpile must   contain some randomness. A simple 1-dimensional model to accomplish this is the Oslo model.

It is defined as follows
\cite{Frette1993,ChristensenCorralFretteFederJossang1996}: 
Consider the height function $h(i)$ of the sand  or rice pile as shown in Fig.~\ref{f:Oslo-model}.
To each height profile $h(i)$ is associated a stress field $z(i)$ defined by 
\be
z(i) := h(i)-h(i+1)\punkt
\ee
A toppling 
is invoked if $z(i)>z_{\rm c}(i)$, $i>1$.
The toppling rules are equivalent to those of \Eq{ASM-toppling}, 
\bea
z(i) \to z(i)-2 \komma \quad
z(i\pm 1)\to z(i\pm 1)+1 \punkt 
\eea
They can be interpreted as moving a grain from the top of the pile at site $i$ to the top of the pile at site $i+1$, 
\be
h(i)\to h(i)-1\komma\quad h(i+1)\to h(i+1)+1\punkt
\ee
After such a move, the threshold $z_{\rm c}(i)$ for site $i$  is updated,
\bea
 z_{\rm c}(i) \to \mbox{new random number} .
\eea
In its original version, the random number is $1$ or $2$ with probability $1/2$. To obtain Fig.~\ref{f:Oslo-model} we used a random number drawn uniformly from the interval $[0,2]$. This reduces the critical slope, and the result looks closer to the experiment in Fig.~\ref{f:Kay-sandpile-experiment}. 

The function $h(i)$ is not one of the usual random-manifold coordinates. If we use the interpretation in \Eq{ASM-toppling} that  $z(i)$ is the discrete Laplacian of the interface position $u(i)$, then the interface position $u(i)$ is given by 
\be\label{h=du}
h(i) = u(i-1)-u(i).
\ee
The random force sits in the threshold $z_{\rm c}$.
The variable $u(0)$ can be identified as  the total number of grains added to the pile. The Oslo model can thus be viewed as an elastic string, pulled at $i=0$. Its average profile is parabolic,  
\bea
\left< u(i) \right> \approx \frac{\left< z\right>}{2} (L-i)^2 +  \# \{ \mbox{grains fallen off at the right} \}\punkt\nn\\
\eea
As the disorder is renewed after each displacement, it falls into the random-field universality class. 

Is this model realistic for the rice pile of Fig.~\ref{f:Kay-sandpile-experiment}? According to \cite{FretteChristensenMalthe-SorenssenFederJossangMeakin1996}, 
this depends on the shape of the grains and their friction. If the grains are round,  the system goes into a self-organized critical state, described by the Oslo model. On the other hand, if the grains are longish (as on our photo), this does not work.
It appears that the direction of the grains is a relevant variable, to be incorporated.

For further reading on the Oslo model, we refer to Refs.~\cite{Dhar2004,HuynhPruessnerChew2011,GrassbergerDharMohanty2016,Hinrichsen2000,HenkelHinrichsenLubeck2008}.

\subsection{Single-file diffusion, and $\zeta^{\rm dep}_{d=1}=5/4$}
\label{zeta=5/4}
Let us consider the heights $h(i)$ of the plateaus in Fig.~\ref{f:Oslo-model}. They are marked on the left as red lines, which we interpret as particles. If a plateau has length   $n\ge2$, then $n$ particles  are at the same position.  (In the figure there is one plateau of length 2, for which we have drawn the two particle positions slightly apart.) As $h(i)$ is a monotonically decreasing function, $h(i)\ge h(i+1)$. This induces a half-order $h(i)\succeq h(i+1)$ on the particle positions. We can extend this to an order by the convention that if $i<i+1$, and $h(i) \ge h(i+1)$, then  $h(i)\succ h(i+1)$. Topplings preserve this order. If we      identify this process as single-file diffusion \cite{WeiBechingerLeiderer2000,KrapivskyMallickSadhu2015,KrapivskyMallickSadhu2015a}, then its Hurst exponent is 
$H_{\rm SFD}=1/4$. 
The additional advection term (grains always topple to the right) converts the temporal correlations into spatial ones,  resulting into $\left< [h(i) -h(j)]^2\right>^{\rm c}\sim |i-j|^{2H_{\rm SFD}}$.
Using that according to \Eq{h=du} $h$ is the discrete gradient of $u$, we conclude that \cite{ShapiraWieseUnpublished}
\bea
\left< [u(i){ -}u(j)]^2\right>^{\rm c}\sim |i-j|^{2 \zeta}\komma \\
 \zeta^{\rm dep}_{d=1} = 1+H_{\rm SFD} = \frac54. ~~
\eea
A roughness exponent $\zeta^{\rm dep}_{d=1}=5/4$ is indeed conjectured   in Ref.~\cite{GrassbergerDharMohanty2016}.

\subsection{Manna model}
\label{ss:Manna}
We introduced the Abelian sandpile model with toppling rules \eq{ASM-toppling}, i.e.\ if $z(i)\ge 2d$, then one grain is moved to each of the $2d$ neighbors of site $i$. In 1991, S.S.~Manna \cite{Manna1991} proposed a stochastic variant\footnote{The original version moves {\em all} the grains to randomly selected neighbors. This version is not Abelian, whereas \Eq{zstab} is. Some authors call it the Abelian Manna model.}
\bea\label{zstab}
z(i)\ge 2 :\mbox{ move 2 grains to randomly chosen neighbors}.\nn\\
\eea
The chosen neighbors may be identical. 
Again, we wish   to  introduce a random-manifold variable $u(i)$, s.t.\ a toppling on site $i$ corresponds to $u(i)\to u(i)+1$, while the remaining $u(j)$ remain unchanged. 
To do so, let us define the discrete Laplacian of $u(i)$ as 
\be
\nabla^2 u(i) :=   \sum_{j\in {\rm nn}(i)} u(j)-u(i) \punkt
\ee
Write 
\be\label{idea}
z(i) := \frac1d\left[ \nabla^2 u(i) + F(i)\right].
\ee
Suppose two grains from site $i$ go to sites $i_1$ and $i_2$, possibly identical.  
Then  choose for the site $i$ and its nearest neighbors $j$
\bea
u(i) \to u(i)+1\komma\quad F(i) \mbox{ unchanged}\komma \\
F(j) \to F(j) + \delta F(j), \\
 \delta F(j)=  d (\delta_{j,i_1}+\delta_{j,i_2}) -1\punkt
\eea
The total random force remains constant, 
$
\sum_{j\in {\rm nn}(i)} \delta F(j)  = 0
$. We may think of this process as distributing $2d$ grains onto the $2d$ neighbors, but instead of doing this  uniformly as in the ASM, twice $d$ grains are moved collectively to a randomly chosen neighbor. Eqs.\ \eq{zstab} and \eq{idea} imply that the interface position $u(i)$ increases by $1$ if the r.h.s.\ of  \Eq{idea} is larger than $2$. This can be interpreted as a cellular automaton   for the equation of motion \eq{eq-motion-f},  if  $F(i)$ has the  statistics of a random force. 
One can show that in any dimension $d$
\be\label{Manna-noise}
\delta F(i) =  \nabla \left[  \dot u(i)    \vec \eta(i) \right], 
\ee
where $\vec \eta(i)$ is a white noise in space and time\footnote{In $d=1$ it is uncorrelated in space and time, whereas in $d=2$ it has a non-trivial spatial structure, but remains short-ranged correlated.}.

As a result, for each $i$ the variable $F(i)$ performs a random walk, which due to \Eq{Manna-noise}, and the equation of motion, cannot grow unboundedly. 
This suggests that the Manna model is in the same universality class as disordered elastic manifolds. 
In section \ref{s:Manna} we  give a more formal 2-step mapping of the Manna model onto disordered elastic manifolds.

\subsection{Hyperuniformity}
Consider a   stationary random point process on the line. It is said to be {\em hyperuniform}  \cite{GrassbergerDharMohanty2016}, if the number $n_L$ of points in an interval of size $L$ has a variance which scales with $L$ as 
\be
\mbox{var}(n_L) \sim L^{\zeta_{\rm h}}\ , \qquad 0 \le \zeta_{\rm h} \le 1\punkt
\ee 
A Poisson process has   $\zeta = 1$, a periodic function $\zeta = 0$.

For sandpile models, this  property was first observed in \cite{BasuBasuBondyopadhyayMohantyHinrichsen2012}, and later verified in 
\cite{HexnerLevine2015,Lee2014,DickmanCunha2015}. Recent references analyzing or using hyperuniformity include \cite{GrassbergerDharMohanty2016,Garcia-MillanPruessnerPickeringChristensen2018,BerthierChaudhuriCoulaisDauchotSollich2011}. 
Hyperuniformity renders simulations much better convergent, allowing for results from the Manna or Oslo model to exceed in size those obtained directly for the depinning of a disordered elastic manifold.

\subsection{A cellular automaton for  fluid invasion, and related models}\label{s:qKPZ}
There are intriguing  connections between invasion of porous media, directed percolation (DP), and depinning of disordered elastic manifolds when the nearest-neighbor interactions grow   stronger than linearly. Let us start our considerations with the cellular automaton   proposed 1992 by Tang and Leschhorn in  Ref.~\cite{TangLeschhorn1992}. Variants of this model can be found in \cite{BuldyrevBarabasiCasertaHavlinStanleyVicsek1992}, where it is applied to experiments on fluid invasion, both numerically and experimentally; see also \cite{GlotzerGyureSciortinoConiglioStanley1994}. 

\begin{figure}[t]
\Fig{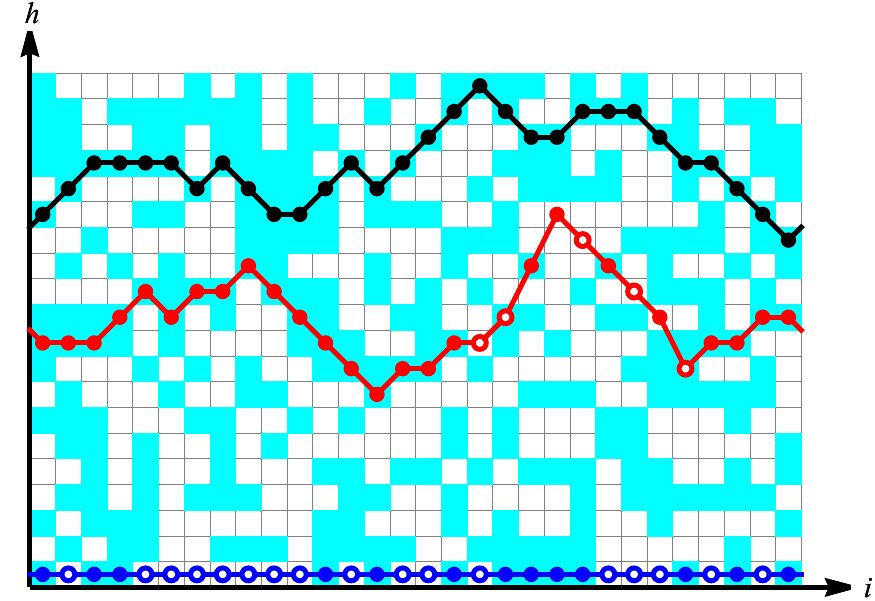}
\caption{
The cellular automaton model {\bf TL92}. Blocking cells, i.e.\ cells above the threshold are drawn in  cyan; those  below in white. The initial configuration is the string at height 1 (dark blue). The interface moves up.  An intermediate configuration is shown in red,   the final configuration  in black. Open circles represent unstable points, i.e.\ points which can move forward; closed circles are stable.}
\label{f:qKPZcellur}
\end{figure}
The model {\bf TL92} proposed in Ref.~\cite{TangLeschhorn1992} uses a square lattice as shown in Fig.~\ref{f:qKPZcellur}. To each cell $(i,j)$ is assigned a random variable $f(i,j) \in [0,1]$. If $f(i,j)<f_{\rm c}$, the cell is considered closed (blocking), drawn in cyan. Open cells (not blocking) are drawn in white. The interface starts as a flat configuration at the bottom (dark blue in Fig.~\ref{f:qKPZcellur}). A point $(i,h(i))$ on this interface is unstable and can move forward by 1, $h(i)\to h(i)+1$, according to the  following rule in meta code:

\begin{figure}[t]
\Fig{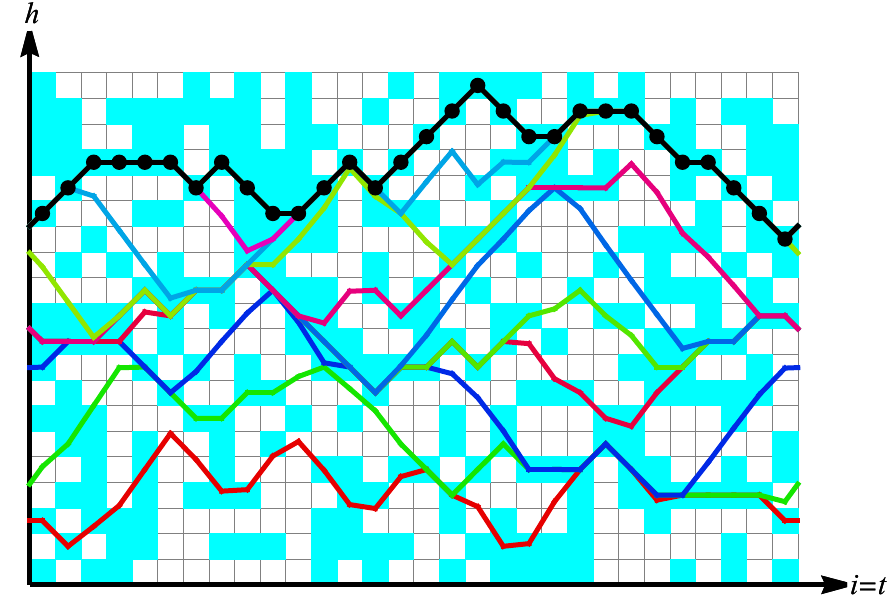}
\caption{
Simulation of the continuous version of the cellular automaton model {\bf TL92}. The continuous configurations (in color) converge reliably against the directed-percolation solution (black, with filled circles).}
\label{f:qKPZsmooth}
\end{figure}

\medskip

\noindent\begin{minipage}{\textwidth}
\noindent  {\bf unstable}$(i)$

\quad {\sl \# links cannot be longer than 2}
 
\quad {\bf if} $h(i)-h(\mbox{neighbor}) \ge 2$ {\bf return} false  

\quad {\sl \# move forward if open}   

\quad {\bf if} $f\big(i,h(i)\big) >f_{\rm c}$ {\bf return} true

\quad {\sl \# move forward if a neighbor is 2 ahead}
 
\quad {\bf if} $h(\mbox{neighbor})-h(i) \ge 2$ {\bf return} true
   
{\bf end}
\end{minipage}
\medskip

\noindent
This cellular automaton models a fluid invading a porous medium. Invasion takes place if a cell is open (second ``if'' above), or can be invaded from the side (third ``if''). The  process stops if all points $(i,h(i))$ are stable. As is illustrated in Fig.~\ref{f:qKPZcellur}, this stopped configuration is a directed path from left to right passing only through blocked sites, commonly referred to as  a {\em directed percolation} path. One can convince oneself that upon stopping the algorithm yields the lowest-lying directed percolation path. This can be implemented both for open and periodic  boundary conditions. The latter are chosen in Fig.~\ref{f:qKPZcellur}.
The automaton {\bf TL92} can straightforwardly 
 be generalized to higher dimensions \cite{BarabasiGrinsteinMunoz1996}, but there is a priori no directed percolation process in the orthogonal direction. 

\begin{table}
\begin{tabular}{|c|c|c|c|}
\hline
& $d=1$ & $d=2$ & $d=3$ \\
\hline
 $\zeta$ & 0.63  \cite{TangLeschhorn1992} &  0.45 & ? \\
\hline
 $z$ & 1 \cite{TangLeschhorn1992} & $1.15 \pm 0.05$ \cite{HavlinAmaralBuldyrevHarringtonStanley1995} & $1.36 \pm 0.05$ \cite{HavlinAmaralBuldyrevHarringtonStanley1995}\\
\hline
\end{tabular}
\caption{The exponents of qKPZ.}
\label{qKPZ-table}
\end{table}
Two continuous equations of motion may be associated with this surface growth. The  first is the (massive) quenched KPZ equation, 
\bea\label{lf28}
 \partial_t u (x,t) &=& c \nabla^2 u (x,t) + \lambda \left[ \nabla u
(x,t)\right]^2 +m^2 [w-u (x,t)] \nn\\
&& + F(x,u (x,t) ). 
\eea
This is almost the equation of motion \eq{eq-motion-f} for a disordered elastic interface; the additional non-linear term 
 proportional to $\lambda$  is  referred to as a
KPZ-term, due to its appearance in the famous KPZ equation of non-linear surface growth \cite{KPZ}. 
The latter accounts for the surface growing in its normal direction, and not in the direction of $h$. For a derivation see section \ref{s:KPZ}. 
For an early reference see \cite{LeeKim2005}.
{\new In the present context it was first observed in   
simulations \cite{TangKardarDhar1995}, where an increase in   
the drift-velocity was found upon tilting the interface.}

The second model one can associate with the automaton {\bf TL92} is depinning of an elastic interface. As  {\bf TL92} makes no distinction between nearest-neighbor distances $0$ or $\pm 1$, has strong interactions at distance $2$, and forbids larger distances, the corresponding elastic energy ${\cal H}_{\mathrm{el}}[u] $ must be strongly anharmonic.
Our choice is (with $u(L+1)=u(1)$)
\bea
{\cal H}_{\mathrm{el}}[u]  = \sum_{i=1}^L \ca E_{\rm el} \big( u(i)-u(i+1) \big), \\
 \ca E_{\rm el} (u) = \left\{ 
 \begin{array}{ccc}
 0 &, & |u| \le 1 \\ 
 \frac1{24} (u^2-1)^2 &, & | u|>1.
\end{array}
 \right.
\eea
This implies an elastic nearest-neighbor force 
\bea
f_{\rm el}(u):=-\partial_u \ca E_{\rm el} (u) = \left\{ 
 \begin{array}{ccc}
 0 &, & |u| \le 1, \\ 
 -\frac1{6} u (u^2-1) &, & | u|>1.
\end{array}
 \right.
\eea
It evaluates to $-1$ at $u=2$, which is sufficient to overcome any obstacle; and to $-4$ at $u=3$, making the latter unattainable. 
The full equation of motion for site $i$ then reads
\bea
\partial_t u(i,t)
&=& f_{\rm el} \big(u({i},t){-}u(i{+}1,t)\big) {+}f_{\rm el}  \big(u({i},t){-}u(i{-}1,t)\big)\nn\\
 && {+} F(i,u(i,t))\punkt
\eea
The last term is the disorder force, which we choose to be $f(i,j)-f_{\rm c}$ if $u$ is within  $\delta$ close to $j$. Thus disorder acts as an obstacle close to an integer $h$. To mimic {\bf TL92}, we wish the manifold to advance freely  between obstacles, setting there      $F=f_+$.  Formally
\be
F(i,u):= \left\{
\begin{array}{ccc}
f(i,j)-f_{\rm c} &,& ~\exists j, |u-j|<\delta\komma  \\
f_+ &,& \mbox{else}\punkt
\end{array}
 \right.
\ee
The $f(i,j)$ are the threshold forces of {\bf TL92}.
The parameter $\delta$ is a regulator. One checks that $\delta=10^{-3}$, and $f_+=2$    reproduces the time evolution of {\bf TL92}, if movement is restricted to a single degree of freedom $i$, and one stops when $u(i)$ hits the next barrier. 
This is not how Langevin evolution works: the latter being parallel, we cannot expect   trajectories to go through the same states. However, due to Middleton's theorem (see section \ref{s:Field theory of the depinning transition}), the blocking configurations of both algorithms are the same. We have verified with numerical simulations that the Langevin equation of motion finds exactly the same blocking configurations as the cellular automaton {\bf TL92}. 
This proves that the critical configurations of the former are states of directed percolation.

While this statement was proven above for a specific non-linearity, we expect that it is more generic, and applies to any convex elastic energy which at large distances grows stronger than a parabola. This was numerically verified for several anharmonicities in  \cite{RossoKrauth2001b}.

\begin{figure}
\Fig{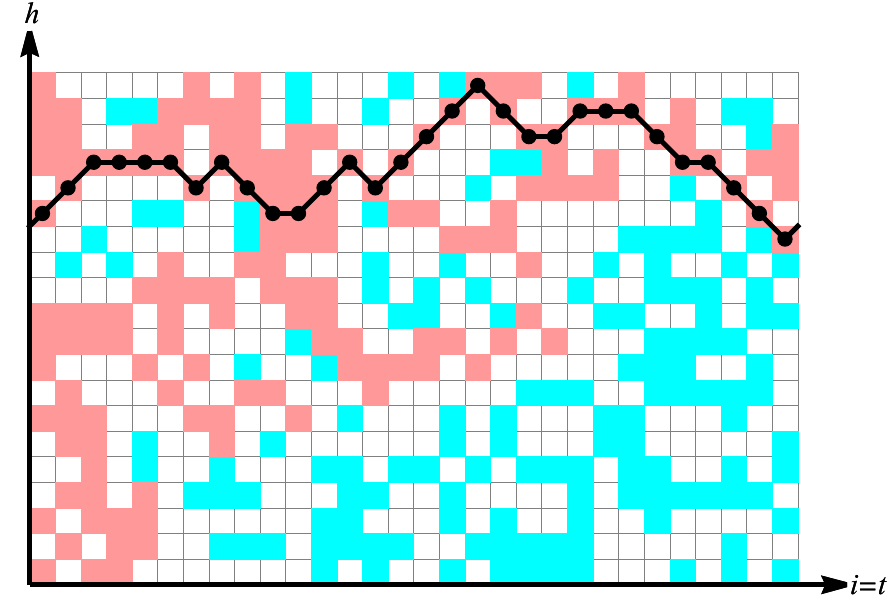}
\caption{Directed percolation from left to right. A site $(i,h)$  is  defined to be {\em connected} if it is occupied, and at least one of its left neighbors $(i-1,h)$,  $(i-1,h\pm 1)$ is connected. 
The index $i$ takes the role of time $t$. }
\label{f:dir-percol2}
\end{figure}

\subsection{Brief summary of directed percolation}
\label{s:A short summary on directed percolation}
Directed percolation (DP) is a mature domain of statistical physics
\cite{Hinrichsen2000,AraujoGrassbergerKahngSchrenkZiff2014,Dhar2017}. Consider Fig.~\ref{f:dir-percol2}. 
Sites are empty or full with probability $p$, which in our discussion above equals $p=f_{\rm c}$. 
A site $(i,h)$ is said  to be connected to the left boundary, if it is occupied, and at least one of its three  left neighbors $(i-1,h)$,  $(i-1,h\pm 1)$ is connected to the left boundary. The system is said to percolate, if at least one point on the right boundary is connected to the left boundary. 
For small $p$, this is unlikely, whereas for large $p$ this is likely. There is a transition at $p=p_{\rm c}$.  
What is commonly considered are the three independent  exponents $\beta$, $\nu_{\parallel}$, and $\nu_{\perp}$, defined via
\bea
\rho(t) := \left< \frac1H \sum_h s_h(t)    \right> ~~ \stackrel{t\to \infty}{\longrightarrow} ~~ \rho^{\rm stat},
\eea
where $s_h(t)=1$  if site $(t,h)$ is occupied, and 0 else.  
\bea
 \rho^{\rm stat} \sim (p-p_{\rm c})^\beta\komma \quad p>p_{\rm c}, \\
\xi_{\parallel} = |p-p_{\rm c}|^{- \nu_\parallel}, \\
\xi_{\perp} = |p-p_{\rm c}|^{- \nu_\perp}.
\eea
The last two relations imply
\be
  \xi_{\perp} \sim \xi_{\parallel}^\zeta\ , \quad \zeta := \frac{\nu_\perp}{\nu_\parallel}.
\ee
Hinrichsen \cite{Hinrichsen2000} gives   in $d=1$:
\bea\label{exponents-DP-d=1}
\nu_{\parallel} &=& 1.733 847 (6), \quad 
\nu_{\perp} = 1.096 854 (4), \nn \\
\beta &=& 0.276 486 (8), \quad 
\Rightarrow ~\zeta = 0.632613 (3). 
\eea
In $d=2$:
\bea\label{exponents-DP-d=2}
\nu_{\parallel} =  1.295(6), \quad 
\nu_{\perp} =   0.734(4), \quad 
\beta =  0.584(4) .  
\eea
In $d=3$:
\bea\label{exponents-DP-d=3}
\nu_{\parallel} = 1.105(5) , \quad 
\nu_{\perp} = 0.581(5), \quad   
\beta = 0.81(1).  
\eea
For {\bf TL92} ($d=1$), the exponent $\zeta$ is interpreted as the roughness exponent. 
Simulations in dimensions $d=1$ to $4$  yield  \cite{Hinrichsen2000}:
\be
\zeta = \frac{\nu_{\perp}}{\nu_{\parallel}} = \left\{ \begin{array}{ccc} 
0.632613 (3) &, & \quad d=1\\ 
0.566(7) &, & \quad d=2\\ 
0.526(7)  &, & \quad d=3\\ 
 0.5  &, &\quad  d\ge 4 
\end{array}
\right.
\label{222}
\ee
Field-theory for directed percolation is derived in section
\ref{s:Field theory for directed percolation}. 
At 2-loop order  \cite{Janssen1981,BronzanDash1974,CardySugar1980,JanssenTaeuber2005} it reads\footnote{Note that the notations in these papers are somehow contradictory. The dynamical critical exponent $z$ is related to our roughness $\zeta$ via $z=1/\zeta$. The $z$ defined in \cite{BronzanDash1974,AbarbanelBronzanSugarWhite1975,CardySugar1980} is  in Reggeon field theory, and equals $z_{\rm Reggeon} = 2 \zeta$. Partial results at 3-loop order are reported in   \cite{AdzhemyanHnaticKompanietsLucivjanskyMizisin2019}.}
\bea
 \nu_{\parallel}=  1+\frac{\epsilon }{12}+\frac{\epsilon
   ^2 \left[109-110 \ln
    (\frac{4}{3} )\right]}{34
   56}+\ca O(\epsilon ^3), \\
\nu_{\perp}= \frac{1}{2}+\frac{\epsilon
   }{16}+\frac{\epsilon ^2
   \left[107-34 \ln
    (\frac{4}{3} )\right]}{46
   08}+\ca O(\epsilon ^3) , \\
\beta=
1-\frac{\epsilon }{6}+\frac{\epsilon
   ^2 \left[11-106 \ln
    (\frac{4}{3} )\right]}{17
   28}+\ca O(\epsilon ^3) .
\eea
This yields
\be
\zeta:= \frac{\nu_{\perp}}{\nu_{\parallel}}  = \frac{1}{2}+\frac{\epsilon
   }{48}+\frac{\epsilon ^2
   \left[79{+}118 \ln
   \left(\frac{4}{3}\right)\right]}{13
   824}+\ca O(\epsilon ^3).
\ee
In $d=1$ ($\epsilon=3)$, these values are in decent agreement with those of \Eq{exponents-DP-d=1}.

\subsection{Fluid invasion fronts from directed percolation}
\label{s:Scaling for anisotropic depinning and qKPZ}

To avoid confusion, let us define 
\bea\label{2-point-function-qKPZ}
\left< [u(x,t) - u(0,t)]^2 \right> \sim 
\left\{ 
\begin{array}{ccc}
{|x-x'|^{2\zeta }} &\mbox{~for~} &|x-x'|\ll \xi_m  \\
\\
m^{-2 \zeta_m} &\mbox{~for~} & |x-x'|\gg \xi_m 
\end{array} \right.\nn\\
\eea
In $d=1$, 
the scaling of $x$ and $u$ as a function of $p-p_{\rm c}$ is
\bea
x &\sim& \xi_\parallel \sim |p-p_{\rm c}|^{-\nu_{\parallel}} , \\
\label{nu-par}
u  &\sim& \xi_\perp \sim |p-p_{\rm c}|^{-\nu_{\perp}} .
\eea
This implies
\be\label{qKPZ:zeta}
u \sim x^{\zeta}, \quad \zeta = \frac{\nu_\perp}{\nu_\parallel} .
\ee
The exponent $\nu$ defined for depinning in \Eq{def:nu} is identified from \Eq{nu-par} as 
\be
\nu\equiv \nu_{\rm dep} = \nu_\parallel.
\ee
If we drive with a parabolic  confining potential, 
\be
m^2 u \simeq |p-p_{\rm c}|.
\ee
This yields 
\be
u \sim m^{- \zeta_m}, \quad \zeta_m = \frac{2 \nu_\perp}{1+\nu_\perp}.
\ee
Let us define the correlation length $\xi_m$ as the  $x$-scale at which the crossover between the two regimes of \Eq{2-point-function-qKPZ} takes place. This yields
\be
\xi_m \sim  m^{-\frac{\zeta_m}{\zeta }}, \qquad \frac{\zeta_m}{\zeta }= \frac{2\nu_{\parallel}}{1+\nu_\perp}. 
\ee
(We remind that in contrast for qEW $\zeta = \zeta_m$, and $\xi_m = 1/m$, see \Eq{xi-m}.)
The avalanche-size exponent also changes. An avalanche scales as   
\be
S_m = \xi_m^{d+\zeta} \equiv \xi_m^{d } m^{-\zeta_m}.
\ee
Since 
\Eq{m-Sm} was derived under the sole assumption that the avalanche density has a finite IR-independent limit for $m\to 0$, it remains valid, implying
\be
\tau = 2 -\frac{2   }{d+\zeta }  \frac{\zeta }{\zeta_m}.
\ee
In dimension $d=1$, the dynamical exponent $z=1$  (see below).
The depinning scaling relation  \eq{beta-rel} 
can be rewritten with \Eq{qKPZ:zeta} and $z=1$ as 
\be
\beta_{\rm dep} = \nu_{\parallel} (z-\zeta) \equiv \nu_{\parallel} -\nu_\perp.
\ee 
Using the   values of Ref.~\cite{Hinrichsen2000} combined with the above scaling relations, the  
numerical values in $d=1$ are  
\bea
\nu_{\rm dep}\equiv\nu_{\parallel} = 1.733 847 (6) \\
\nu_{\perp} = 1.096 854 (4) \\
\zeta = 0.632613 (3), \\
\zeta_m =  1.04619  (2) , \\
\frac{\zeta_m}{\zeta} =1.47955(3),  \\
\tau = 1.259246(2), \\
\beta_{\rm dep} = 0.636993(7) ,\\
z=1 .
\label{qKPZ:z=1}
\eea

\subsubsection*{The dynamic exponent $z$.}
\label{s:The dynamic exponent and relation to percolation}
In Ref.~\cite{HavlinAmaralBuldyrevHarringtonStanley1995} it was proposed that the dynamical exponent $z$ is related to the fractal dimension $d_{\rm min}$ of the shortest path connecting two points a distance $r$ apart in a percolation cluster. Denoting its length by $\ell \sim r^{d_{\rm min}}$,  the conjecture is 
\be
z = d_{\rm min}\; .
\ee
This relation was confirmed numerically, and yielded the dynamical exponents $z$ reported in table \ref{qKPZ-table}. Curiously, the upper critical dimension of percolation is $d=6$, whereas the theory for directed percolation used in the preceding section has an upper critical dimension of $d=4$. As a consequence, constructing a field theory encompassing both seems challenging.

\subsection{Anharmonic depinning and FRG}
Let us finally study anharmonic  depinning within FRG, and to this purpose consider the standard elastic energy \eq{Hel}, supplemented by an 
additional anharmonic (quartic) term, 
\begin{equation}\label{Eanharm}
\ca H_{\mathrm{el}}[u]= \int_{x} \frac{1}{2}\left[\nabla
u (x)\right]^{2}+\frac{c_{4}}{4} \left[\left(\nabla u (x)\right)^2\right]^{2}\punkt 
\end{equation}
The corresponding equation of motion reads
\bea\label{lf28m}
 \partial_t u (x,t) &=&  \nabla^2 u (x,t) +c_4 \nabla \left\{\nabla u(x,t)\left[ \nabla u
(x,t)\right]^2 \right\}\nn\\
&& + F(x,u (x,t) ) + f\punkt 
\eea
Since the r.h.s.\ of \Eq{lf28m} is a total derivative, it is surprising 
that a KPZ-term can be generated  in the limit of a {\em vanishing}
driving-velocity. This puzzle was solved in Ref.~\cite{LeDoussalWiese2002a}, where the KPZ term  arises by contracting the non-linearity with one disorder, following the rules of section \ref{s:dep-loops} (setting $m=0$):
\begin{widefigure}[t]
\parbox{8.cm}{\includegraphics[width=8.cm]{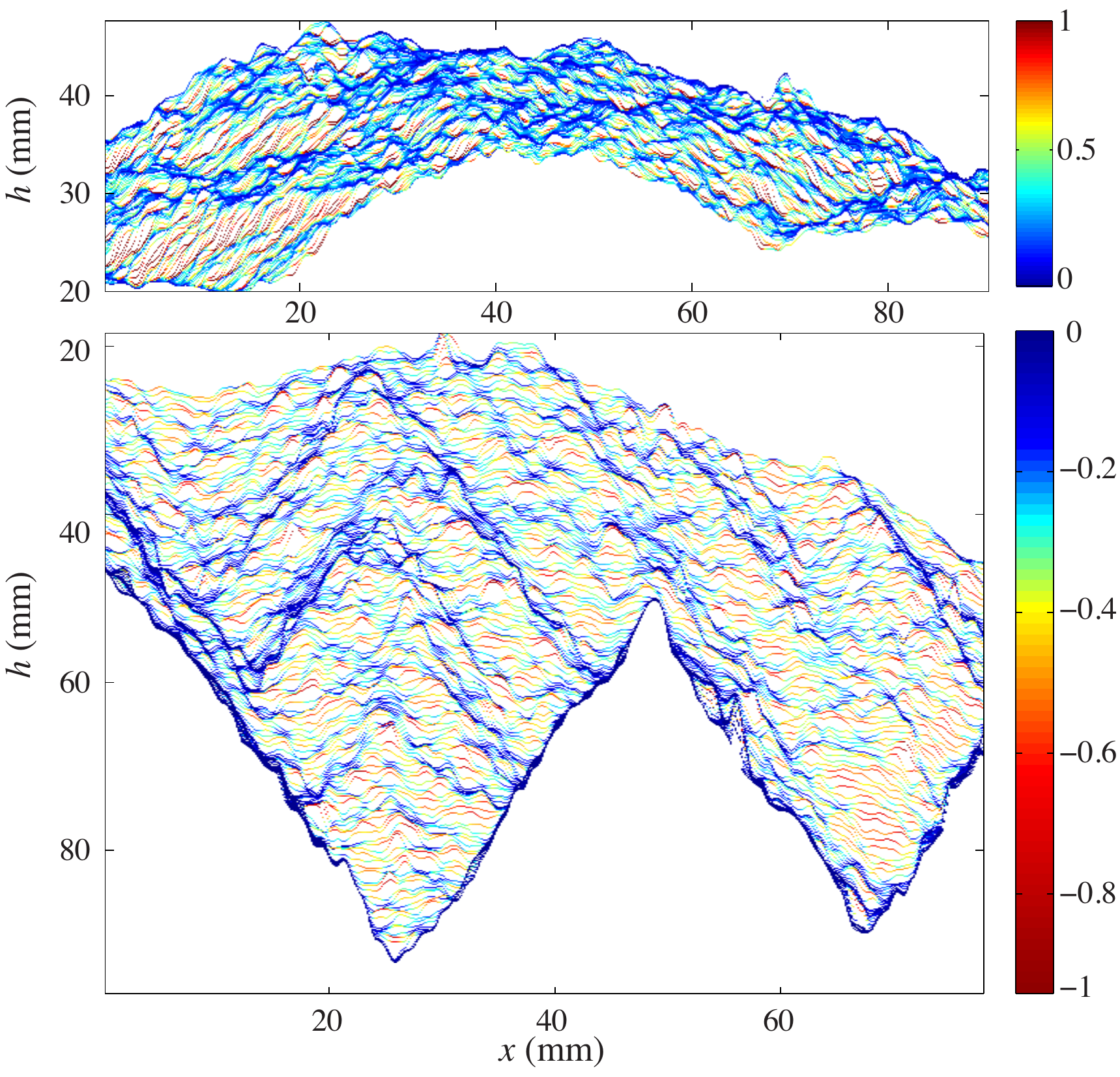}}~~~\parbox{9.5cm}{\includegraphics[width=9.5cm]{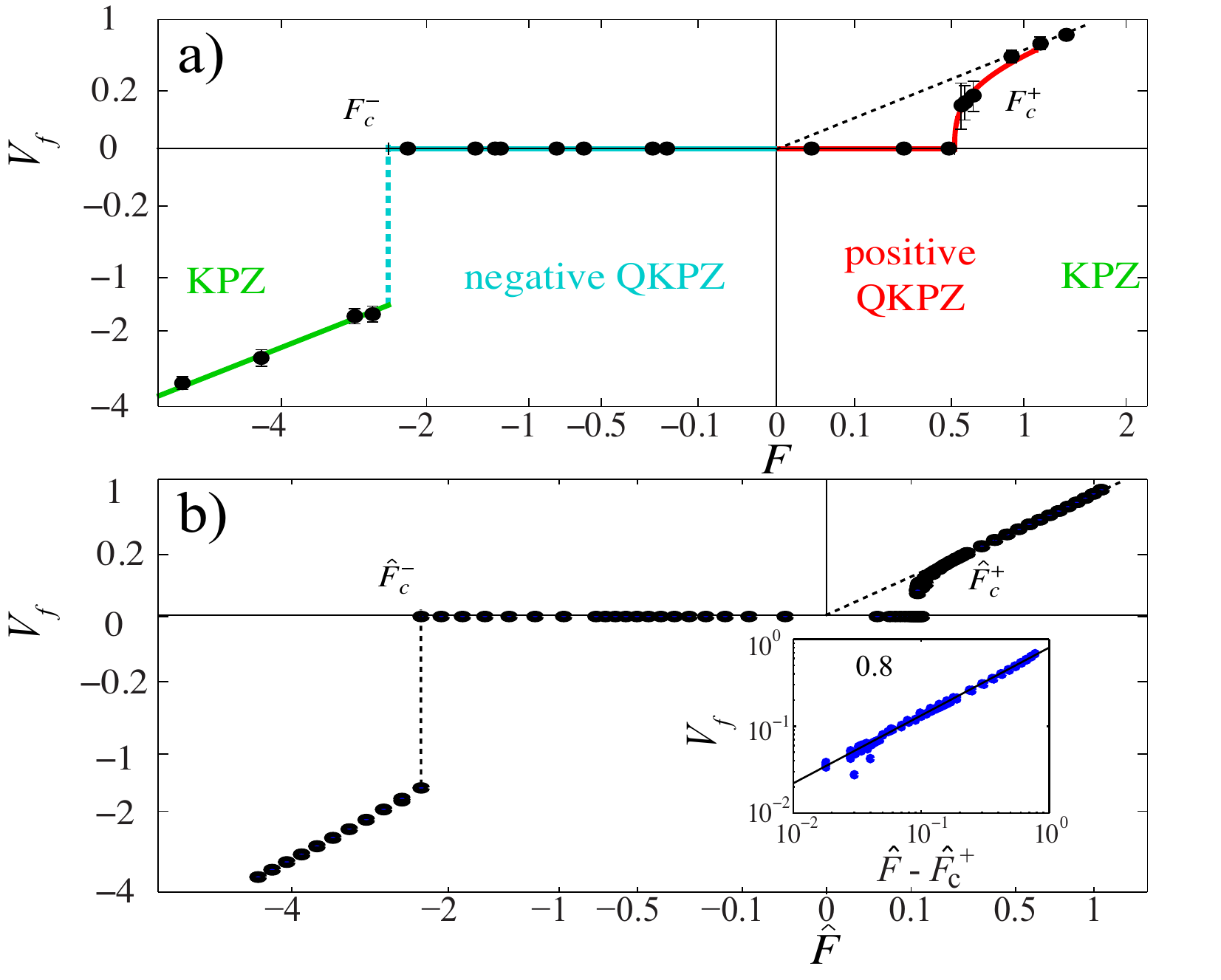}}
\caption{Left: Successive experimental fronts at constant time intervals in a self-sustained reaction,     propagating in a disordered environment made by polydisperse beads  \cite{AtisDubeySalinTalonLeDoussalWiese2014}. Color represents local front velocity. Top left:
upward propagating front near $F_{\rm c}^+$. Bottom left: backward propagating front near
$F_{\rm c}^-$.  
\newline
Right: Front velocity $V_{\rm f}$ versus the applied force $F$, in adverse flow.
a) experiments (black dots with error bars), b) numerics. Dashed lines are a linear extrapolation of the advancing branch.
To put all data on one plot, axes are rescaled according to $F\to F/|F|^{1/2}$,  $V_{\rm f}\to V_{\rm f}/|V_{\rm f}|^{1/2}$.
Insert: log-log plot of front velocity versus $\hat F - \hat{F}_{c^+} $. The continuous line corresponds to  
$v(\hat F) \propto ( \hat F - \hat F_{c^{+}} )^{ 0.8 \pm 0.05}$. Fig.~from \cite{AtisDubeySalinTalonLeDoussalWiese2014}.}
\label{VF} 
\end{widefigure}
\begin{eqnarray}
\delta \lambda = \parbox{2cm}{\fig{2cm}{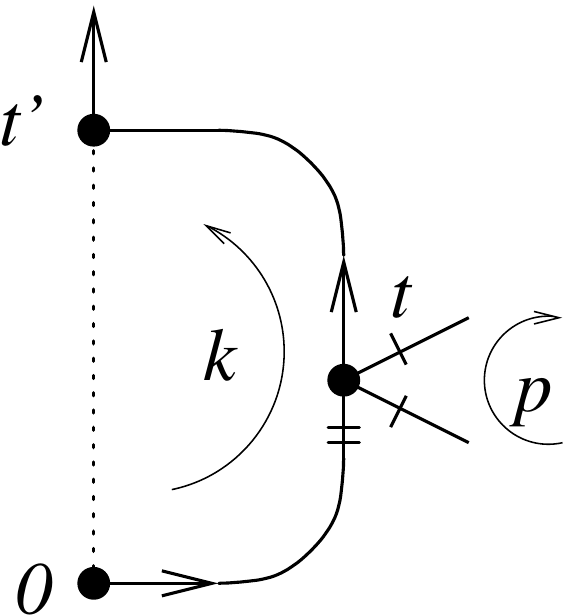} } \nn \\
= -\frac{c_{4}}{p^{2}} 
\int_{t>0} \int_{{t'>0}} \int_{k} \rme^{-(t+t')k^2} \left(k^{2}p^{2}+2
(kp)^{2}\right) \nonumber \\
\qquad \qquad \qquad \quad 
\times  \Delta'(u_{x,t+t'}-u_{x,0}) \punkt
\label{k1}
\end{eqnarray}
As $u(x,t+t')- u(x,0)\ge 0$,  \Eq{k1} can be written as 
\bea\label{lf2a}
\delta\lambda = - \frac{c_{4}}{p^{2}} \int_{t}\int_{t'}\int_{k}
\rme^{-(t+t')k^2} 
\left(k^{2}p^{2}+2 
(kp)^{2}\right)  \Delta' (0^{+}). \nn\\
\eea
Integrating over $t,t'$ and
using the radial symmetry in $k$ yields
\begin{equation}\label{lf4}
\delta \lambda = - c_{4} \left(1+\frac{2}{d}
\right)\int_{k}\frac{\Delta' (0^{+})}{k^{2}}  . 
\end{equation}
This shows that in the FRG a KPZ term is generated from the non-linearity. Field theory does not yet permit to calculate the ensuing roughness exponent, nor explain the mapping onto directed percolation, even though a mechanism for the generation of a branching-like process was found \cite{LeDoussalWiese2002a}.

\subsection{Other models in the same universality class}
Many models nowadays are recognized as being in the universality class of Directed Percolation (DP). 
This started with   work by Janssen \cite{Janssen1981} and Grassberger \cite{Grassberger1982}, who conjectured that the findings 
``suggest another type of universality, comprising all critical points with an absorbing state and a single order parameter in one universality class''  \cite{Grassberger1982}. As a general rule, a model belongs to the universality  class of directed percolation, as long as it has no additional symmetry. A notable exception is the Manna model (see section \ref{s:Manna}). Note that   additional conserved quantities are not enough, as exemplified by   directed percolation   with many absorbing states  \cite{Jensen1993,MunozGrinsteinDickmanLivi1996,MunozGrinsteinDickmanLivi1997,MunozGrinsteinDickman1998,BonachelaAlavaMunoz2008}.
The reader wishing to explore the large literature further can find a lot of  material  in the context of {\em Phase Transitions into Absorbing States}, see  \cite{Hinrichsen2000,Tauber2012} for review, as well as \cite{MunozGrinsteinDickmanLivi1996,Munoz1998,VespignaniDickmanMunozZapperi1998,DickmanMunozVespignaniZapperi2000,VespignaniDickmanMunozZapperi2000,AlavaMunoz2002,BonachelaAlavaMunoz2008}.

\begin{widefigure}[t]
{\fig{9cm}{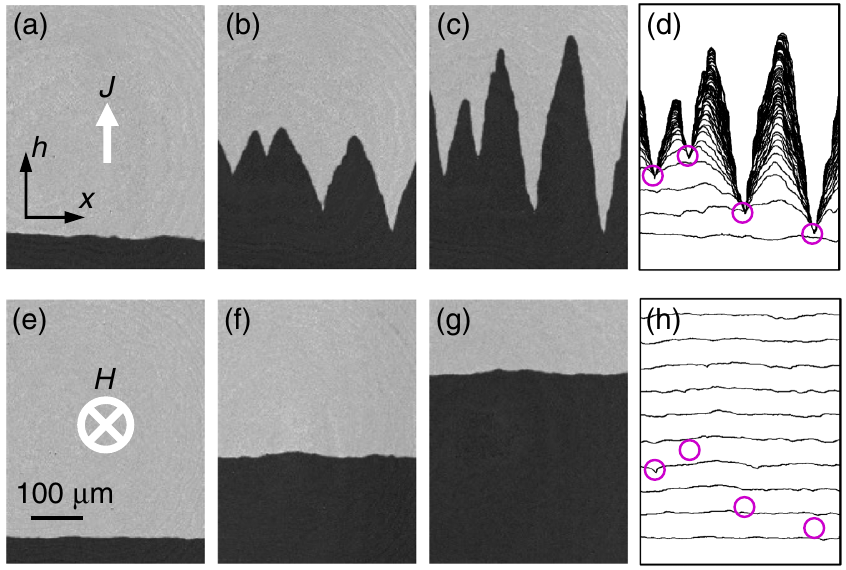}}\hfill{\fig{7.96cm}{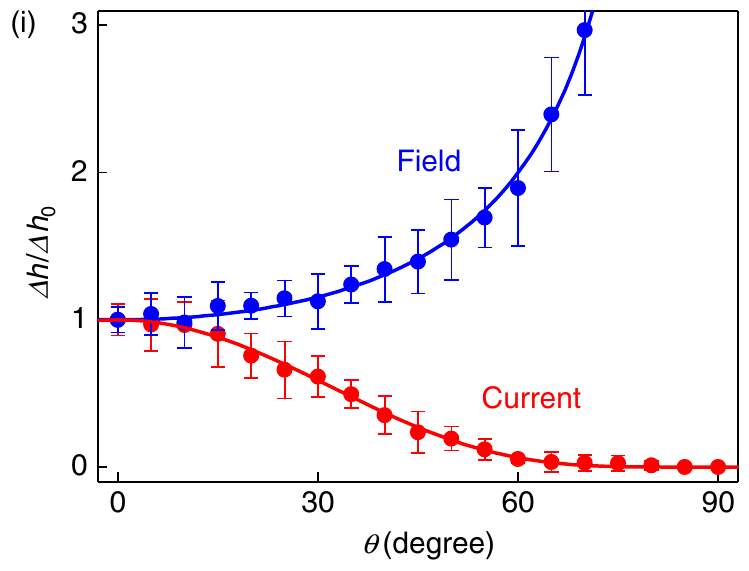}}
\caption{Magnetic domain walls in a two-dimensional Pt-Co-Pt thin film. (a-d) images for current-driven walls at increasing times. (e)-(h) ibid.~field-driven. (i) measurement of the angle-dependent force as extracted from an analysis of the creep laws. From \cite{MoonKimYooChoHwangKahngMinShinChoe2013}, with kind permission.}
\label{f:mag-domain-walls-PRL-110-107203}
\end{widefigure}

\subsection{Quenched KPZ with a reversed sign for the non-linearity}
\label{s:negative-QKZ}
As long as the disorder $F(x,u)$ is statistically invariant under $u\to -u$, 
the quenched KPZ (qKPZ) equation \eq{lf28} is invariant under $u\to -u$, $\lambda \to - \lambda$, and $f\to -f$. This leads to two distinct cases: $\lambda f>0$ the {\em positive qKPZ} class, and $\lambda f<0$ the {\em negative qKPZ} class. 
Consider $f>0$, and $\lambda>0$, then the KPZ term facilitates depinning. In the opposite case, assuming a tilted configuration allows the interface to remain pinned for larger applied forces. It then assumes a sawtooth shape, with the bottom kinks at the strongest pinning centers, and the slope given by $f\approx (-\lambda) (\nabla u)^2$. This was first observed numerically
\cite{JeongKahngKim1996,JeongKahngKim1999}, and later confirmed     experimentally \cite{AtisDubeySalinTalonLeDoussalWiese2014,MoonKimYooChoHwangKahngMinShinChoe2013}, as beautifully  illustrated in Figs.~\ref{VF} and \ref{f:mag-domain-walls-PRL-110-107203}, and discussed in the next section.

\subsection{Experiments for directed Percolation and quenched KPZ}
\label{s:Directed Percolation and qKPZ experiments}

Experiments for directed percolation seem to be scarce \cite{Hinrichsen2000}. A notable exception is Refs.~\cite{TakeuchiKurodaChateSano2007,TakeuchiKurodaChateSano2009}, where 
 a transition between two topologically different turbulent states, called dynamic scattering modes 1 and 2 (DSM1 and DSM2), is observed upon an increase in the applied voltage. This allows them to measure directly the exponent $\beta$ as 
 \be
 \beta_{\rm DP}^{d=2} = 0.59(4)\\
 \ee
The remaining exponents are obtained from a quench. Citing only the most precise values, 
\bea
\nu_{\parallel} &=& 1.18_{-(21)}^{+(14)} ,\\
\nu_{\perp} &=& 0.77(7).
\eea 
The theory values are given in \Eq{exponents-DP-d=2}.

More experiments have been done for the qKPZ class. A particularly nice example are self-sustained reaction fronts   propagating in a disordered environment made by polydisperse beads  \cite{AtisDubeySalinTalonLeDoussalWiese2014}, as depicted on figure \ref{VF}. The measured spatial and temporal fluctuations   are consistent with three distinct universality classes in dimension $d=1+1$, controlled by a single parameter, the mean (imposed) flow velocity. The three classes are
\begin{enumerate}
\item the Kardar-Parisi-Zhang (KPZ) class for fast advancing or receding fronts, with a roughness exponent of $\zeta\approx 0.5$, see \Eq{zeta-KPZ-d=1}. (Purely diffusive motion with the same roughness exponent is excluded by  the   temporal correlations.) 
\item the quenched Kardar-Parisi-Zhang class (positive-qKPZ) when the mean-flow velocity almost cancels the reaction rate. It has  a roughness of $\zeta \approx 0.63$, in agreement with our discussion in section \ref{s:Scaling for anisotropic depinning and qKPZ}.  A depinning transition with a non-linear velocity-force characteristics, $v\sim |F-F_{\rm c}|^{\beta} $ is observed, see figure \ref{VF}.
\item  the negative-qKPZ class for   receding fronts, close to the lower depinning threshold $\hat{F}_{c^-} $. One observes characteristic saw-tooth shapes, see Fig.~\ref{VF}, bottom left. 
\end{enumerate}
To our knowledge, this system is the only one where all three KPZ universality classes have been observed in a single experiments. 

The qKPZ phenomenology is also   observed in    domain walls in thin magnetic films \cite{MoonKimYooChoHwangKahngMinShinChoe2013} (see section \ref{s:Experiments on thin magnetic films}), either driven by an applied field (positive qKPZ)
or a current (neagtive qKPZ).  The experiment performed in Ref.~\cite{MoonKimYooChoHwangKahngMinShinChoe2013} cleverly extracts the slope-dependent mean force as a function of the angle, see Fig.~\ref{f:mag-domain-walls-PRL-110-107203} (right). This firmly establishes the relevance of the two qKPZ classes for domain wall experiments. 
It would be interesting to  drive the system both with a magnetic field and   a current, chosen s.t.\ the two effects cancel.

\section{Modeling Discrete Stochastic Systems}
\label{s:CSPI}

\subsection{Introduction}
In {\em discrete stochastic processes}   the elementary degrees of freedom are discrete variables. This can be the number of colloids in a suspension, the number of bacteria, fishes and their predators in the ocean, or the grains in sandpile models. 
There are two powerful methods to treat these systems (for a pedagogical introduction see \cite{Wiese2015})
\begin{itemize}
\item[(i)] the coherent-state path integral {\cite{Doi1976a,Doi1976b,Peliti1985,Cardy2006,Wiese2015}},
\item[(ii)] effective stochastic equations of motion.
\end{itemize}
The first method, 
the coherent-state path integral, is an exact method, and as such a natural starting point  in a field-theoretic setting, i.e.\ to construct a dynamic action, similar to the  Martin-Siggia-Rose action (section \ref{MSR-formalism}). As we will see in the next section \ref{s:summary CSPI}, despite the fact that it is an exact method, or maybe due to it, it has its problems. \Review{They arrive when decoupling the non-linear terms via an auxiliary noise.} This noise is in general   imaginary, leading to problems both in the interpretation, as  in simulations. 
As a caveat to the reader, let us mention that things sometimes get messed up in the literature: Starting with the coherent-state path integral, one sees emerging   an effective stochastic equation of motion with real noise. We   show below why this is in general {\em not possible}. 

Real noise appears in a different modeling of stochastic systems, via {\em effective stochastic equations of motion}. Here the noise   stems from the fact that one tries to approximate a {\em discrete}
 random process by a {\em continuous} one, and one has to add back the appropriate {\em shot noise}. 
 
 Another important question we need to deal with is the notion of the {\em Mean-Field} approximation in stochastic equations. We will give a simple and precise  definition of the latter.  To our astonishment, we have not found a discussion of this in the literature pior to \cite{Wiese2015}.

\subsection{Coherent-state path integral, imaginary noise and its interpretation}
\label{s:summary CSPI}
\begin{widefigure}
\fig{0.4\textwidth}{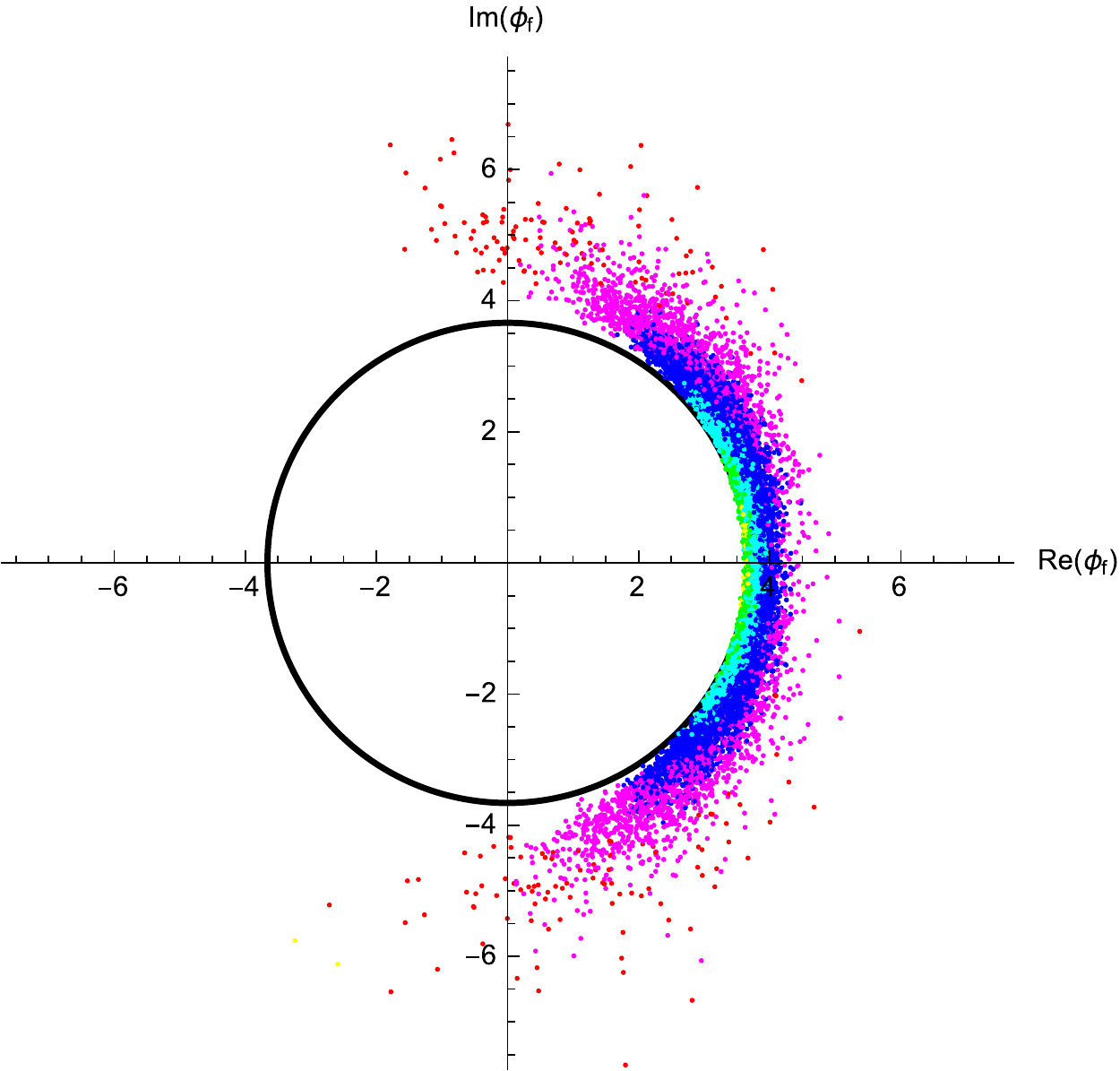} 
\fig{0.6\textwidth}{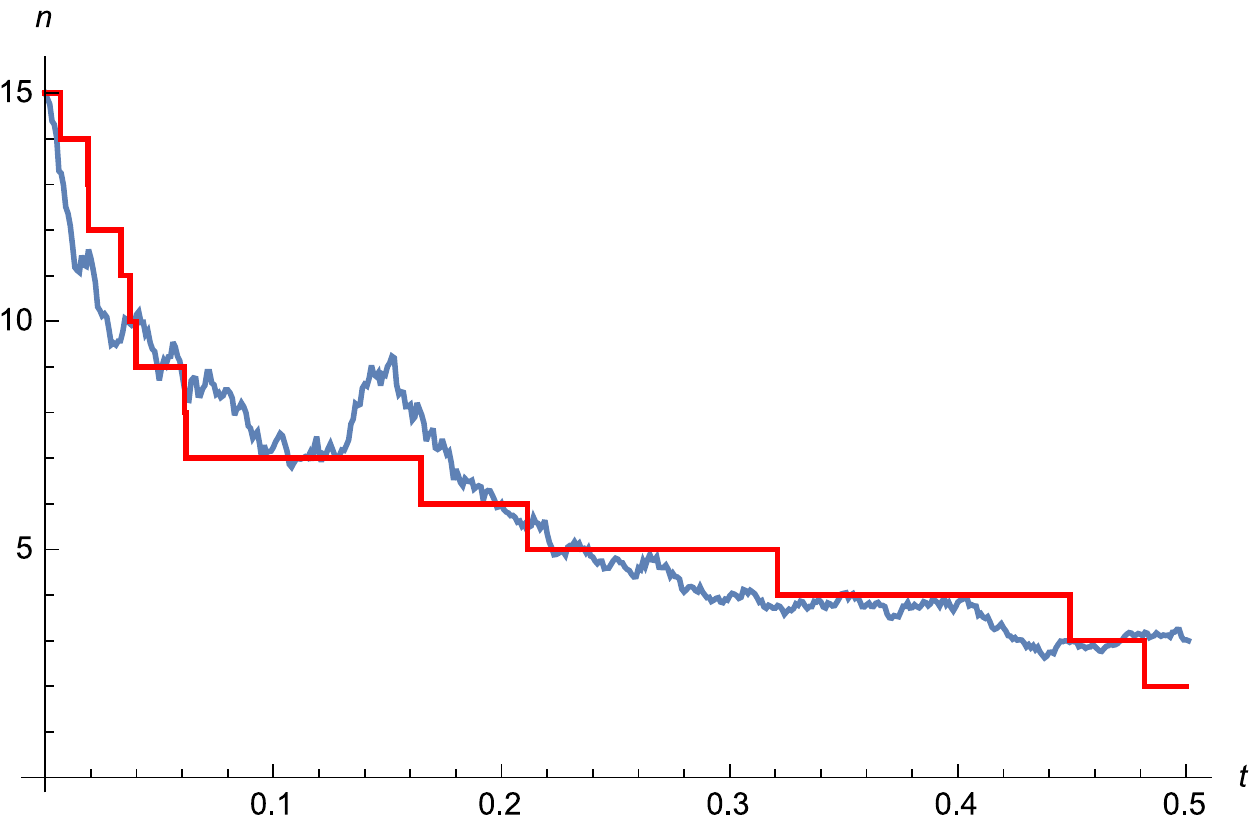}
\caption{Left: Result  of the integration of \Eq{92},
with $\nu=1$, total time $t_{\rm f}-t_\rmi=0.5$, and initial state $\phi_\rmi=15$. The black circle has radius
 $\phi_{\rm f}=3.6614$, obtained by integrating the drift term $\partial_t \phi_t = -\phi_t^2+\phi_{t}+t/2$. Using an algorithm which splits points which are likely to contribute more to the final result, the color codes less probable values,  from yellow over green, cyan, blue, magenta to red. (Thus a red point has $2^{-5}$ times the weight of a yellow point.)
Right: One trajectory each for process $n_t$, i.e.\ a direct numerical simulation of $A+A\to A$ (red, with jumps), and ${\hat n}_t$, Eq.~(\ref{a-process}) (blue-grey, continuous, rough). The rate is $\nu=1$. We have chosen two trajectories which look ``similar''. Note that $\hat n_{t}$ is not monotonically decreasing. Figs.~from \cite{Wiese2015}.}\label{f:cute-algo}
\end{widefigure}
The coherent-state path-integral \Review{\cite{Doi1976a,Doi1976b,Peliti1985,Cardy2006,Wiese2015}} is constructed by using   creation   and annihilation operators familiar from quantum mechanics, 
\beq\label{commutator2}
\left[ \ah , \ad  \right] = 1 
\komma  \qquad \left| n \right>:= (\ad)^n \left|0 \right>\punkt
\eeq
The state $ \left| n \right>$ is interpreted as   $n$-times occupied. 
Eigenstates of $\ah$ are  
coherent states. They are the building blocks of the formalism, giving it its name
\beq\label{CS2}
\left| \phi \right>:= \rme^{\phi \ad } \left| 0 \right> \qquad \Rightarrow \qquad \ha \left| \phi \right> =\phi \left| \phi \right>\punkt
\eeq
Taylor expanding $\rme^{\phi \ad } \left| 0 \right>$, one sees that 
coherent  states are Poisson distributions with $n$-fold occupation probability given by 
\beq \label{25}
p  (n) = \rme^{-\phi} \frac{\phi^n}{n!} \punkt
\eeq
Note that $\left<n\right> = \left<n^2\right>^{\rm c} =\phi$, thus the parameter $\phi$ characterizing a coherent state is both  its mean and variance.

Consider   the reaction-diffusion process with diffusion constant $D$ and reaction rate $ A+ A \stackrel{\nu}\longrightarrow A $. 
The action \Review{, see (Eq.~(112) of Ref.~\cite{Wiese2015}}) reads
\bea
{\cal S}'[\phi^*,\phi] = \int_{x,t}   \phi^{*}({x,t})\Big[\partial_{t} \phi({x,t}) - D   \nabla^2\phi({x,t} )\Big] \nn\\
  +\int_{x,t}  \frac{\nu}2 \Big[\phi^{*}({x,t}) \phi({x,t})^2+\phi^{*}({x,t})^2 \phi({x,t})^2\Big] \punkt 
\label{92aBIS}
\eea
The first two terms are similar to those appearing in the MSR formalism for diffusion, identifying the tilde fields there with star fields here. The next term 
$\phi^{*}({x,t}) \phi({x,t})^2$ is also intuitive: Two particles are destroyed, and one is created. The only surprising term is the last one. It appears in the formalism to ensure that the probability is conserved, and can be interpreted as a first-passage time problem \Review{ \cite{Wiese2015}}.
If the last term were not there, then we could interpret the action as an equation of motion for $\phi(x,t)$. To include the latter, let us decouple  the quartic term by introducing an auxiliary field $\xi(x,t)$, to be integrated over in the path integral, 
\bea
{\cal S}'[\phi^*,\phi,\xi] = \int_{x,t}   \phi^{*}({x,t})\Big[\partial_{t} \phi({x,t}) - D   \nabla^2\phi({x,t} ) \nn\\
 + \frac{\nu}2 \phi({x,t})^2 - i \sqrt{\nu} \xi(x,t)   \phi({x,t})\Big]+ \frac12 \xi(x,t)^2 \punkt 
\label{92aBIS2}
\eea
The corresponding equation of motion and noise correlations are
\begin{eqnarray}\label{SEOMApAgA}
\partial_{t} \phi({x,t}) = -\frac\nu2 \phi({x,t})^{2}+D\nabla^{2} \phi({x,t})\nn\\
~~~~~~~~~~~~~~~~~~~+i \sqrt{\nu}\phi({x,t})\xi({x,t})\komma  \\
\left<\xi({x,t})  \xi({x',t'}) \right>= \delta(t-t')\delta(x-x')\punkt
\end{eqnarray}
This noise is imaginary. It has puzzled many researchers whether this is unavoidable \cite{Munoz1998,AndreanovBiroliBouchaudLefevre2006,GredatDornicLuck2011,TaeuberBook},  or could even be beneficial \cite{DeloubriereFrachebourgHilhorstKitahara2002}. 

For the moment, let us restrict our considerations to a single site, starting at time $t=t_{\rm i}$ with the initial state, $\phi_{t_{\rm i}} =\phi_{\rm i}$, 
\bea\label{92bis}
\partial_{t} \phi({t}) = -\frac \nu 2 \phi({t})^{2}+i \sqrt{\nu} \phi({t}) \xi({t}) \komma  \nn\\
 \left< \xi({t})  \xi(t')\right>= \delta(t-t')\punkt
\eea
This equation is integrated from $t=t_{\rm i}$ to $t_{\rm f}$.
On the left of figure \ref{f:cute-algo} we show the result for $\phi_{t_{\rm f}}$ for different realizations of the noise $ \xi({t}) $. Since $\phi_{t_{\rm f}}$ is complex, the question  is how to interpret these states.  The answer is that the probability distribution is given, in generalization of Eq.~(\ref{25}), by \cite{Wiese2015}
\beq\label{magicBIS}
p^{\rm SEM}_{t} (n) := \left<\ \!\rme^{-\phi_{t}}\frac{\phi_{t}^n}{n!} \right>_{\!\xi}\punkt
\eeq
A complex $\phi({t})$ is  necessary, since the final distribution is {\em narrower} than a Poissonian\footnote{First, large particle numbers have a higher probability to annihilate, thus the tail of the distribution is suppressed, making it decay faster than an exponential. Second, it is  impossible to construct a probability distribution which is narrower than a Poissonian  by a superposition of Poissonians with positive coefficients.}. 
The problem with the stochastic average (\ref{magicBIS}) is that when $\arg (\phi_{t})$ grows in time, it is {\em dominated} by those $\phi_{t}$ with the smallest real part, and the estimate (\ref{magicBIS}) breaks down.  A stochastic equation of motion for the coherent-state path integral is thus not a valid simulation tool. This is similar to what happens in the REM model (section \ref{s:REM}): in both cases rare events give a substantial contribution to the observable we want to calculate, which is missed in  any finite sample or simulation.

In the next sections, we   follow a different strategy: We give up on the discreteness of the number $n(t)$ of particles, and replace it by a continuous variable $\hat n(t)$. In exchange we need to introduce a stochastic noise. 
  
\subsection{Stochastic noise as a consequence of the discreteness of the state space}
\label{s:discreteness} 
We   want to derive a stochastic differential equation with real noise. 
To this aim let us simulate directly the random process $A+A\stackrel{\nu}\longrightarrow A$. Each simulation run gives one possible realization of the process, in the form of  an integer-valued monotonically decreasing function $n(t)$.  Averaging over these runs, one samples the final distribution $P_\f(n)$, or, equivalently,  moments of $n_\f$. 
We   ask the question: Is there a continuous random process $\hat n(t)$ which has the same statistics as $n(t)$?  

\begin{figure}
\centerline{\Fig{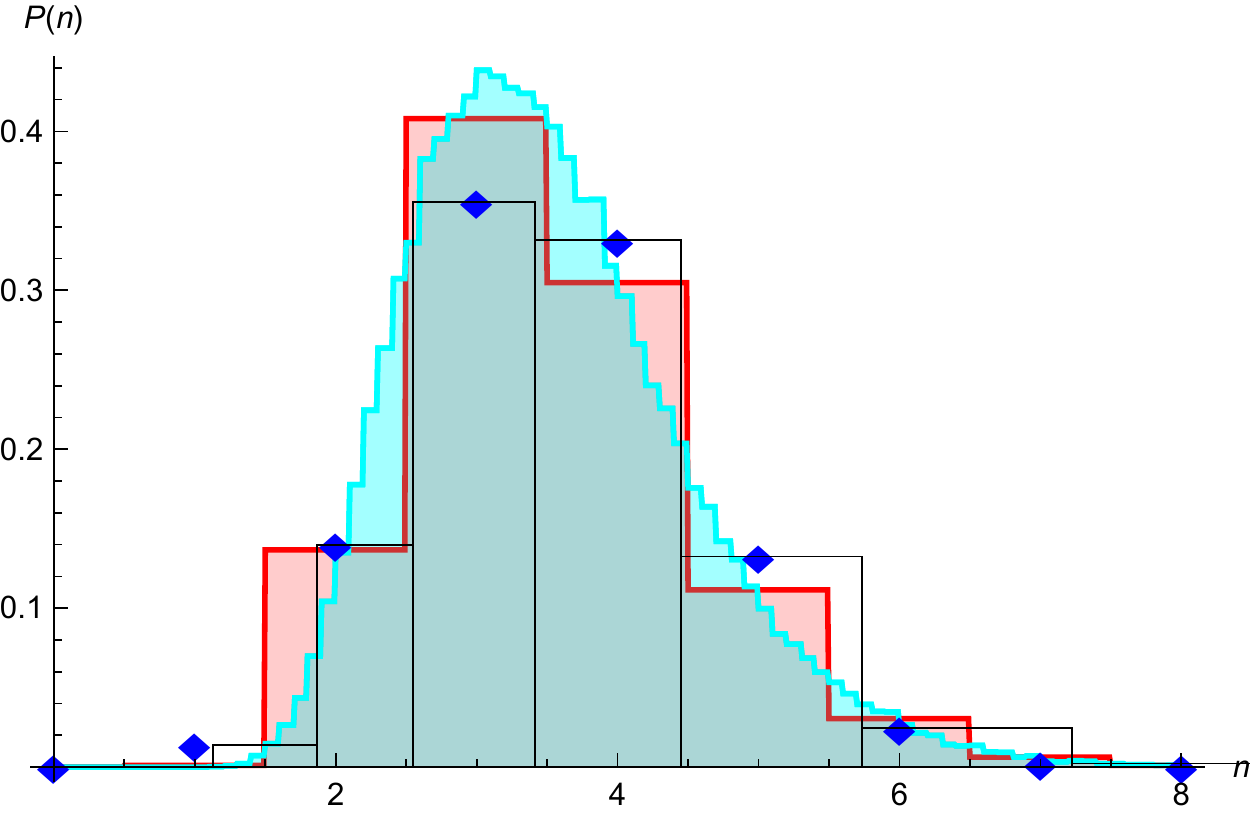}}
\caption{Result of a numerical simulation, starting with $n_{\rmi}=15$ particles, and evolving for $t_{\f}-t_{\rmi}=0.025$. Blue diamonds: Direct numerical simulation of the process $A+A\to A$ with rate $\nu=1$.  Cyan: Distribution of the continuous random walk (\ref{a-process}). Red: The latter distribution, when rounding $n_{\f}$ to the nearest integer. Black boxes: The size of the boxes in $n$-direction to obtain the result of the direct numerical simulation of the process $A+A\to A$. Both processes have first moment $3.511\pm0.001$, and second connected moment $1\pm 0.05$; the third connected moments already differ quite substantially, $0.75$ versus $0.2$.  Fig.~from \cite{Wiese2015}.}
\label{f:2-distributions}
\end{figure}

Let us consider a   more general problem: 
Be $n(t)$ the number of particles at time $t$. With rate $r_+$ the number of particles increases by one,  and with rate $r_-$ it decreases by one. This implies that after one time step,  as long as  $r_\pm \delta t$ are small,  
\bea \label{151}
\left< n{(t+\delta t)}-n(t)\right> = (r_+-r_-)\delta t\komma \\
\left< [n{(t+\delta t)}-n(t)]^2\right> = (r_++r_-) \delta t\punkt 
\eea 
The following {\em continuous random process} ${\hat n}(t)$ has the same  first two moments  as $n(t)$,  
\footnote{Despite our best efforts, we have not been able to locate a source prior to \cite{Wiese2015} for this simple argument in the literature. It is applied in \cite{Janssen1981}, but the cited source \cite{GardinerMcNeilWallsMatheson1976} drily states ``Our whole work depends on the use of stochastic master equations, which, we believe, have a better conceptual and intuitive basis than the fluctuating force formalism of Langevin equations.''}
\bea\label{215}
&& \hl{ \rmd {\hat n}(t) = (r_+-r_-) \rmd t +  \sqrt{ r_++r_-} \, \xi({t})\rmd t } \komma \qquad \\
&& \left< \xi(t) \xi(t')\right> = \delta(t-t')\punkt
\label{215b}
\eea 
This procedure can be modified to include higher cumulants of $n{(t+\delta t)}-n(t)$, leading to more complicated noise correlations.  Results along these lines were obtained in Ref.~\cite{JanssenTaeuber2005} by considering cumulants generated in the effective field theory.

\subsection{Reaction-annihilation process}
For the reaction-annihilation process, the rate $r_+=0$, and $r_-=\frac\nu 2 \hat n{(t)}(\hat n{(t)}-1)$; the latter, in principle, is only defined on integer $\hat n{(t)}$, but we will use it for all $\hat n{(t)}$.
Thus the best we can do to replace the discrete stochastic process with a continuous one is to write
\bea\label{a-process}
\frac{\rmd {\hat n}({t})}{\rmd t} = - \frac\nu2 {\hat n}({t}) ({\hat n}({t})-1)+  \sqrt{\frac{\nu}2 {\hat n}({t}) ({\hat n}({t})-1)} \,  \xi({t}) \komma  \nn\\
\left< \xi({t})  \xi(t')\right>= \delta(t-t')\punkt
\eea
Using $n_{\rmi}=15$, and $\nu=1$, we have shown two typical trajectories on figure \ref{f:cute-algo} (right), one for the process $n(t)$ (red, with jumps), and one for the process ${\hat n}_t$ (blue-grey, rough). While by construction both processes  have (almost) the same  first two moments, clearly ${\hat n}(t)$ looks different: It is continuous, which $n(t)$ is not, and it can  increase in time, which $n(t)$ can not. 
One can also compare the distribution for $t_\f-t_\rmi=0.5$, see figure \ref{f:2-distributions}.
While the distribution of $n_\f$ is discrete (blue diamonds), the one for $\hat n_\f$ is continuous (cyan). Rounding $n_\f$ to the nearest integer gives a different distribution (red). We have also drawn (black lines) the size of the boxes which would produce $p(n)$ from $p(\hat n)$.
Clearly, there are differences. On the other hand, it is also evident that these differences   diminish when increasing $n_{\rmi}$.

\begin{figure}[tb]
\centerline{\Fig{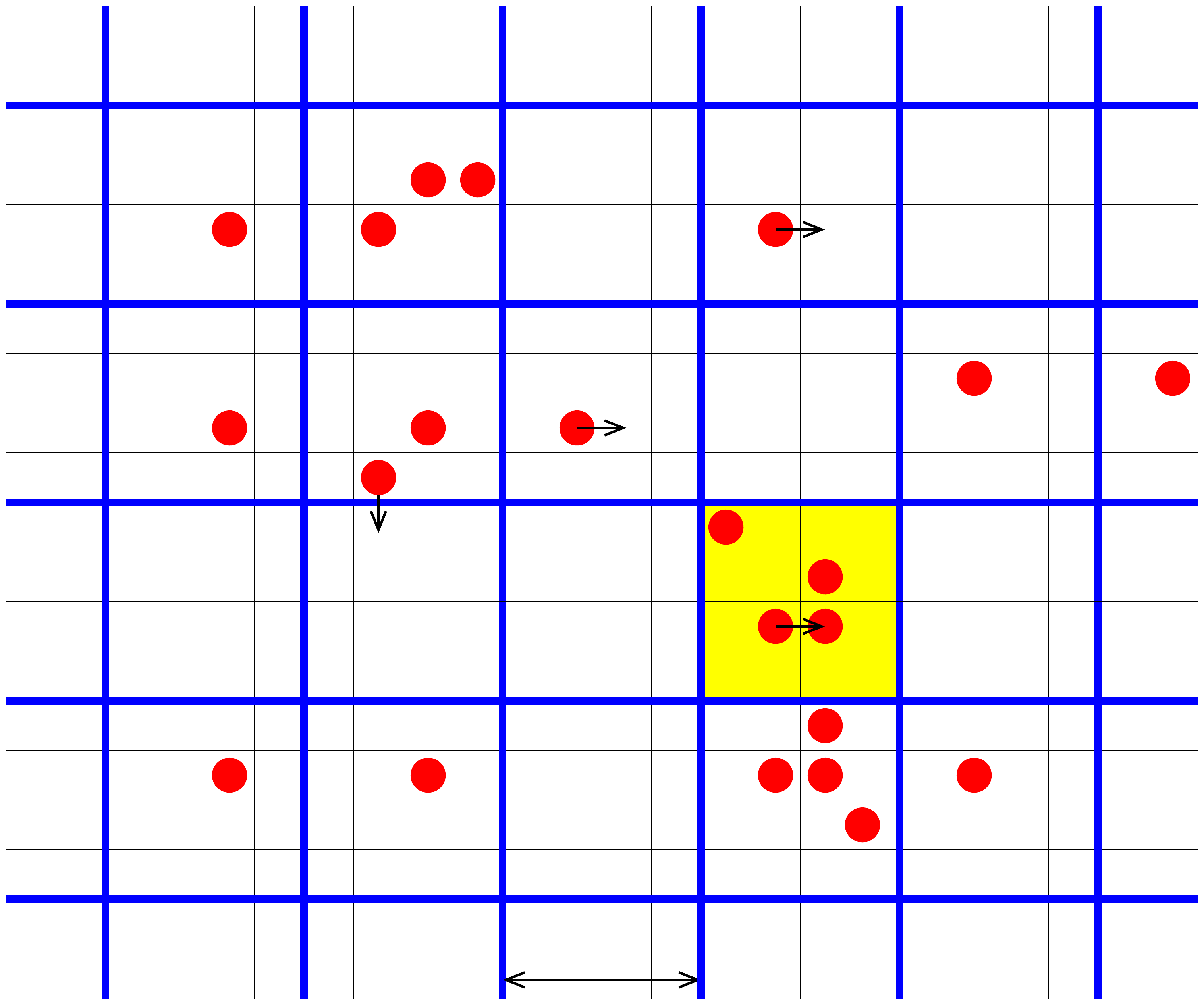}}

\vspace*{-1mm}
\centerline{$\ell$}
\caption{A coarse-grained lattice with box-size $\ell=4$. The yellow box
contains $n=4$ particles.  Fig.~from \cite{Wiese2015}.}
\label{f:7}
\end{figure}

\subsection{Field theory for directed percolation}
\label{s:Field theory for directed percolation}
There are several paths to   a field theory for reaction-diffusion or directed percolation. A beautiful derivation is given by Cardy and 
Sugar~\cite{CardySugar1980}. The authors start from an exact microscopic modelization, before introducing an auxiliary field resulting into the action given below in \Eq{Regge-action}. 
They then use perturbative results obtained for the equivalent action   in Reggeon field theory. The latter is an effective theory for deep-inelastic scattering \cite{AbarbanelBronzanSugarWhite1975}, quantum gravity (simplicial gravity) \cite{WilliamsTuckey1992}, vortices in He-II \cite{RasettiRegge1975}, and many more\footnote{\label{footnote-Regge}Note however, that different theories are associated with the name ``Regge'': Sometimes the cubic vertex is  an antisymmetrized combination of a field with two field derivatives, as in Ref.~\cite{RasettiRegge1975}.}. 

For pedagogic reasons, we   apply the formalism developed in section \ref{s:discreteness} \cite{WieseOriginalReview}: 
Denote $n \equiv n(x,t)$ the number of  particles inside a box located around $(x,t)$ with size $\ell^d$, see Fig.~\ref{f:7}.
There we could draw time as coming out of the plane. 

 In figure \ref{f:dir-percol2} a different view is taken: Here $n(x,t)$ is the coarse-grained  number of occupied sites connected to the left border, there drawn in red. Going one step in $t$ to the right,  $n$ grows with rate  
\bea
n \stackrel {r_+}\longrightarrow n+1\komma \quad r_+ = \alpha_+ n \komma\\
   \alpha_+\approx 3 p\punkt
\eea
Here $3$ is the number of left neighbors per site, and $p$ is the probability that the site itself is not empty, and thus can be connected.
The rate to reduce $n$ by one is given by  
\bea
n \stackrel {r_-}\longrightarrow n-1\komma \quad r_- = \alpha_- n+\beta n^2\komma  \\
\alpha_-= 1-p \komma \quad  \beta\approx \frac{1}{\ell^d} \punkt
\eea
The first term $\alpha_-$  takes into account that if the site itself is empty, it cannot be connected. The second term
proportional to $n^2$ ensures that the fraction of connected sites cannot grow beyond  $1$.  
According to \Eq{215}, this leads to the stochastic equation of motion 
\bea\label{reac-diff-eq=DP}
\partial_t \hat n(x,t) &=& \nabla^2 \hat n(x,t)+ (\alpha_+ - \alpha_-)\hat n(x,t) - \beta \hat n(x,t)^2 \nn\\
&+& \sqrt{\hat n(x,t)} \sqrt{\alpha_+ + \alpha_-  + \beta \hat n(x,t)}\xi(x,t).
\eea
Note that we have added a diffusive term (rescaling $x$ if necessary to set its prefactor to 1). In directed percolation (figure \ref{f:dir-percol2}) it arises since the left neighbor to which a site is connected  can be one up or down on the lattice. 

 Multiplying \Eq{reac-diff-eq=DP} with a response field $\tilde   n(x,t)$, and averaging over the noise $\xi(x,t)$ yields the dynamic action 
\bea\label{action-dir-percol-raw}
S[\tilde n, \hat n] \nn \\
=\int_{x,t} \tilde n(x,t) \left[ \partial_t \hat n(x,t) - \nabla^2 \hat n(x,t) + (\alpha_- {-} \alpha_+)\hat n(x,t)\right] \nn\\
+ \int_{x,t} \beta \tilde n(x,t)   \hat n(x,t)^2  \nn\\
- \int_{x,t}  \frac12 {   \left[\alpha_+ + \alpha_- + \beta \hat n(x,t) \right]} \tilde n(x,t) ^2 \hat n(x,t).
\eea
Let us rewrite this action. As one can see from the equation of motion, 
the combination $m^2 :=\alpha_--\alpha_+$ measures the distance to criticality (without perturbative corrections).
The term proportional to $\beta$ in the last line gives a quartic term, which is irrelevant. Finally, one can change normalization of the fields, setting $\hat n\to \lambda \phi  $, $\tilde n \to \lambda^{-1}\tilde \phi$, which leaves the quadratic terms invariant, but changes the relative magnitude of the two cubic terms. As a result, we obtain the action with coupling const $g = \sqrt{\beta (\alpha_+ + \alpha_-)/2}$, 
\bea\label{Regge-action}
S[\tilde n, \hat n]  
=\int_{x,t} \tilde \phi (x,t) \left[ \partial_t \phi(x,t) - \nabla^2 \phi(x,t) + m^2 \phi(x,t)\right] \nn\\
~~~~~~~~~~~~+   \int_{x,t}    g \left[  \tilde \phi(x,t)   \phi(x,t)^2 -  \tilde \phi(x,t) ^2 \phi(x,t) \right] \punkt
\eea
Note the relative sign change w.r.t.\ \Eq{92aBIS}.
It has four renormalizations, one  for each of the  three  quadratic terms, plus one for the coupling constant $g$. 
This leads to three independent exponents given in section \ref{s:A short summary on directed percolation}. Results at 2-loop order can be found in Refs.~\cite{Janssen1981,BronzanDash1974,CardySugar1980,JanssenTaeuber2005}. Partial 3-loop results are given in \cite{AdzhemyanHnaticKompanietsLucivjanskyMizisin2019}. The action \eq{Regge-action},  known as  Regge field theory  \cite{AbarbanelBronzanSugarWhite1975},   is also used as an effective field theory for deep inelastic scattering. There $\phi$ and $\tilde \phi$ are interpreted as   particle annihilation and creation operators${^{\footnotesize\ref{footnote-Regge}}}$.

\subsection{State variables of the Manna model}
\label{s:Manna}

\begin{figure}[t]
\centerline{\fig{8.3cm}{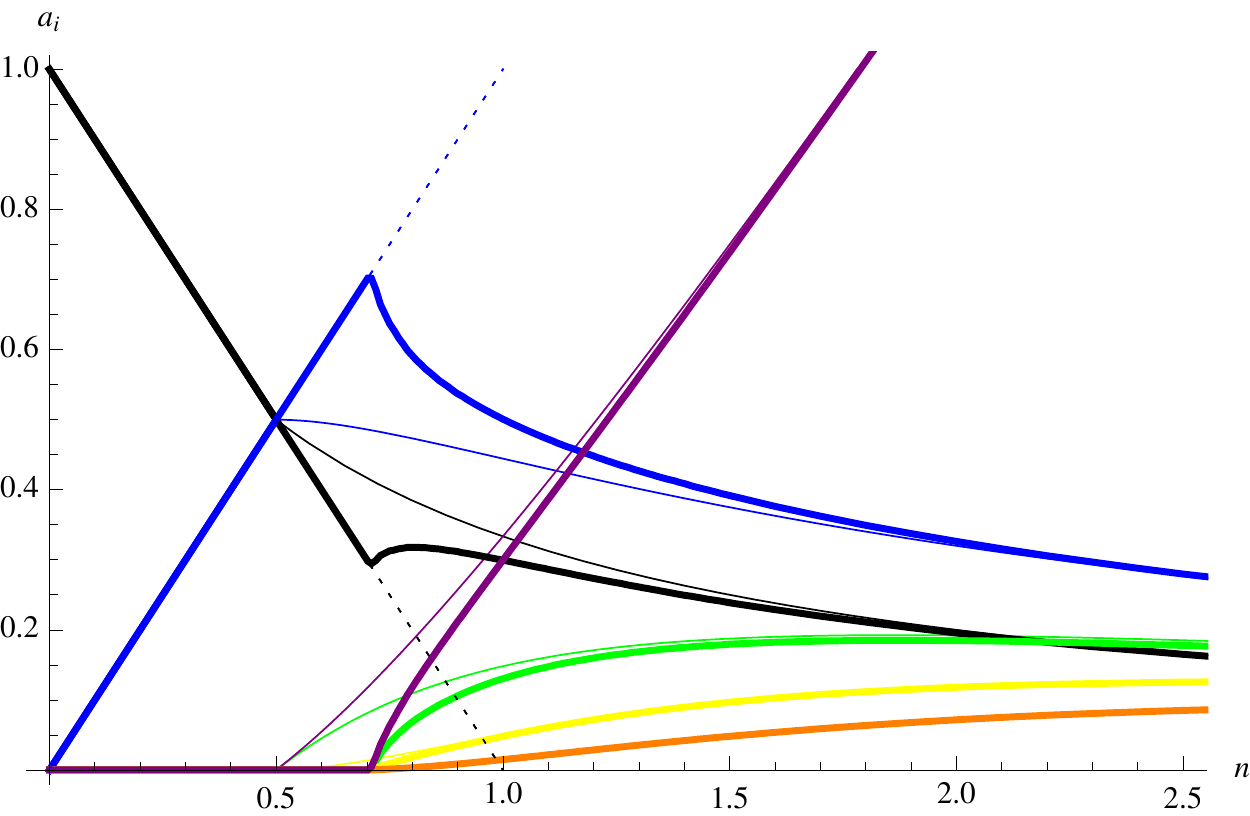}}
\caption{Thick lines: The order parameters of the Manna model, as a function of $n$, the average number of grains per site, obtained from a numerical simulation of the stochastic Manna model on a grid of size $150 \times 150$ with periodic boundary conditions. We randomly update a site for $10^{7}$ iterations, and then update the histogram $500$ times every $10^{5}$ iterations. Plotted are the fraction of sites that are: unoccupied  (black), singly occupied (blue), double occupied (green), triple occupied (yellow), quadruple occupied (orange). The activity $\rho=\sum_{i>1}a_{i}(i-1)$ is plotted in purple. No data were calculated for $n<0.5$, where $a_{0}=e=1-n$, $a_{1}=n$, and $a_{i>2}=0$ (inactive phase). Note that before the transition, $a_{0}=1-n$ and $a_{1}=n$. The transition is at $n=n_{\rm c}=0.702$. \newline Thin lines: The MF phase diagram, as given by Eqs.~(\ref{triv-sol})~ff.\ for $n\le \frac12$, and by Eqs.~(\ref{210})~ff.\ for $n\ge \frac12$. We checked the latter with a direct numerical simulation.  Fig.~from \cite{Wiese2015}.}
\label{fig:Manna:sim+MF}
\end{figure}

In this section, we apply our  considerations to a non-trivial example, the stochastic Manna model, following \cite{Wiese2015}. We will see that our formalism permits a systematic derivation of its effective stochastic equations of motion. While the result is known in the literature \cite{Pastor-SatorrasVespignani2000,VespignaniDickmanMunozZapperi1998,BonachelaAlavaMunoz2008,Alava2003}, it is  there derived  by symmetry principles, which are convincing   ``up to a certain degree''. Furthermore, they  leave undetermined all coefficients. While many of them can be eliminated by rescaling, our derivation   ``lands'' on a particular line of parameter space, characterized by the absence of additional memory terms, see section \ref{s:mapping}.

The Manna model, introduced in 1991 by S.S.~Manna \cite{Manna1991}, is 
  a stochastic version of the Bak-Tang-Wiesenfeld (BTW) sandpile \cite{BakTangWiesenfeld1987}. 
Let us recall its definition given   in section \ref{ss:Manna}:

\paragraph{Manna Model (MM).}
{Randomly throw grains on a lattice. If the height at one point is greater or equal to two, then with rate 1 move two grains from this site to randomly chosen neighboring sites.} Both grains may end up on the same  site. 

We  start by analyzing the phase diagram. We denote by $a_i$ the fraction of sites with $i$ grains. It satisfies the sum rule
\beq\label{normalization}
\sum_i a_i = 1\punkt
\eeq
In these variables, the number of grains $n$ per site can be written as
\beq\label{197a}
n := \sum_i a_i \,i
\punkt
\eeq
The empty sites are 
\beq
e:= a_0
\punkt
\eeq
The fraction of active sites is 
\beq
a := \sum_{i\ge 2} a_i \punkt
\eeq
We also define the (weighted) activity  as
\beq
\rho := \sum_{i\ge 2} a_i (i-1)\punkt
\eeq
Note that $\rho$  satisfies  the sum rule
\beq\label{sum-rule}
n-\rho+e=1
\punkt
\eeq
In order to take full advantage of this sum rule, we change the toppling rules of the Manna model to those of the 

\paragraph{Weighted Manna Model (wMM).} If a site contains $i\ge 2$ grains, randomly move these grains to neighboring sites with rate $(i-1)$.

\begin{widefigure}[t]
\Fig{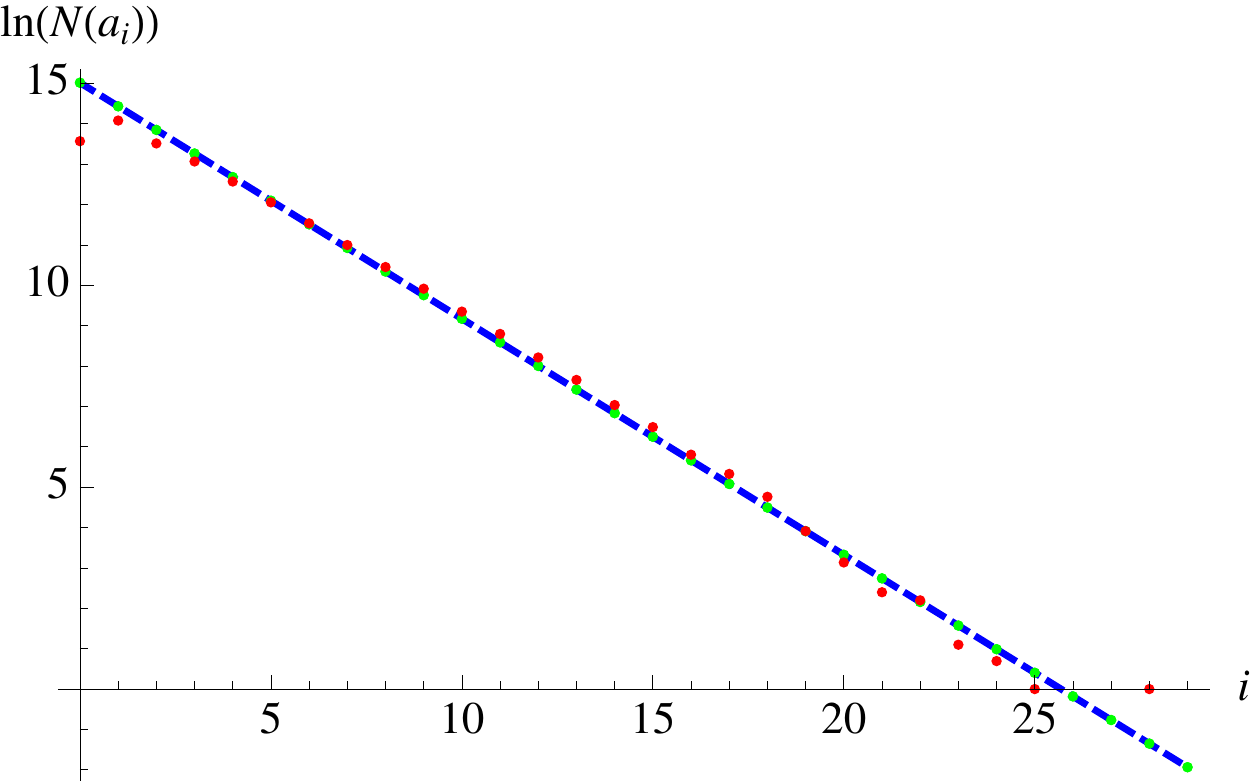}
\parbox{0mm}{\raisebox{25mm}[0mm][0mm]{\hspace*{-4cm}\includegraphics[width=4cm]{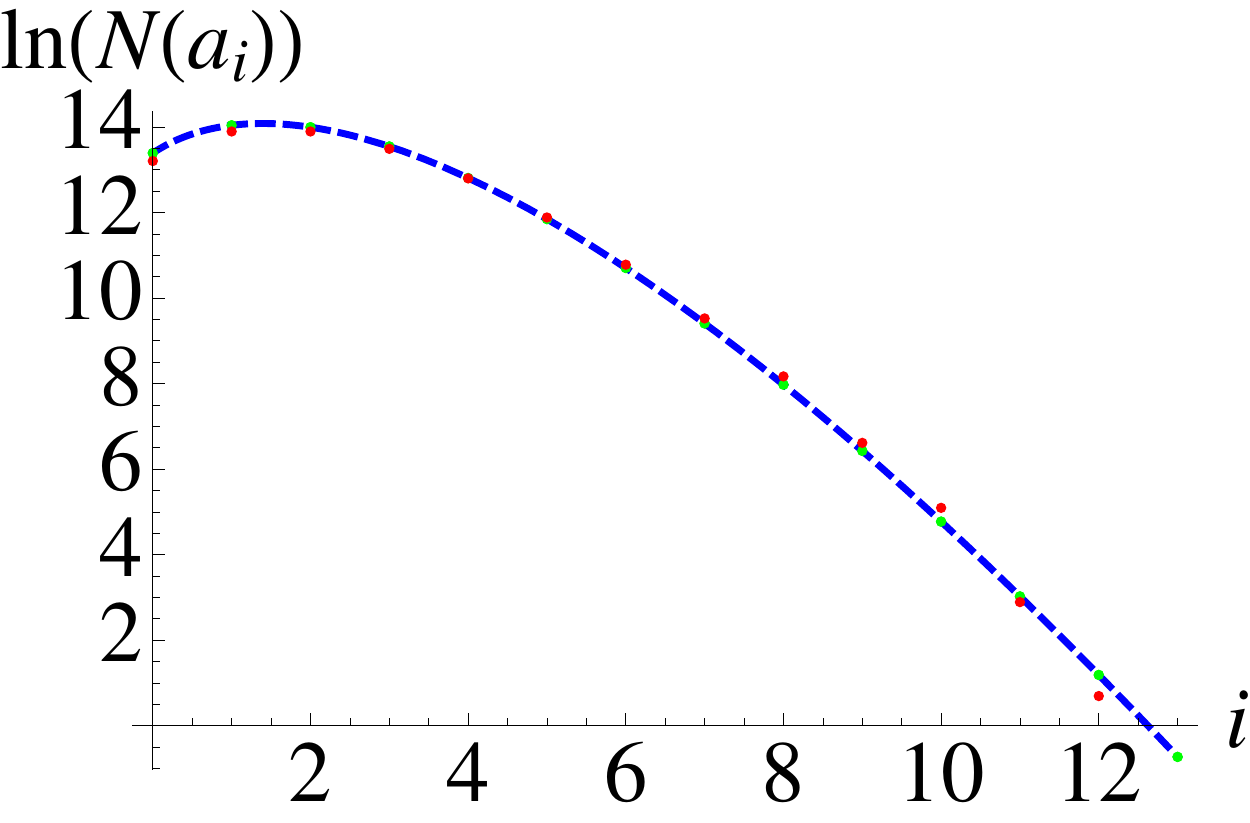}}}~~~
\Fig{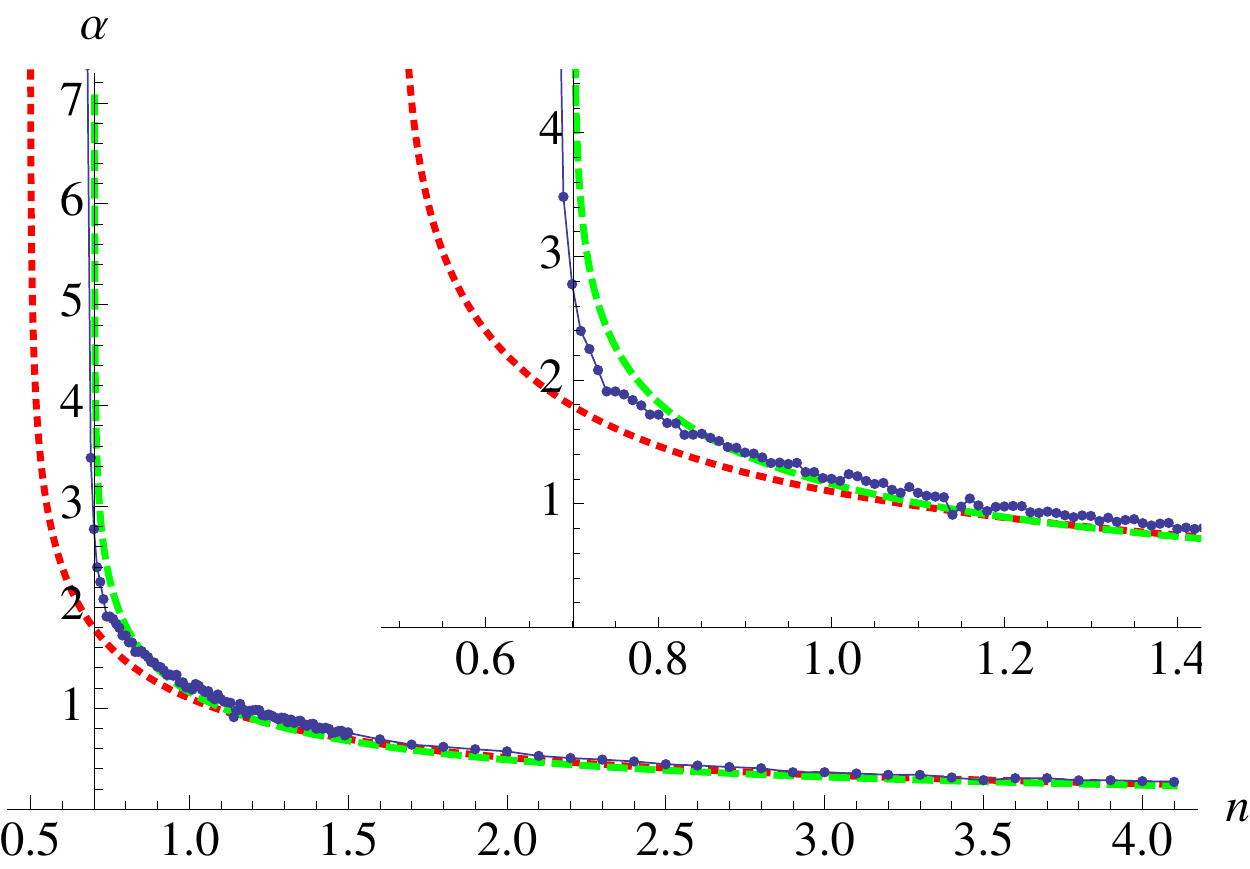}
\caption{
Left:  (Unnormalized) histogram  after many topplings for $n=2$; the probability that a site has $i$ grains decays as $\rme^{-0.585 i}$, for all $i\ge 1$. Inset: The initial distribution, a Poissonian. 
Right: The exponential decay coefficient $\alpha$ as a function of $n$. The dots are from a numerical simulation. The dashed red line is the MF result (\ref{212}). The green dashed line is a fit corresponding to $\alpha \approx \frac23 \ln\Big((n+n_{c})/(n-n_{c}) \Big)$. Inset: blow-up of main plot.  Figs.~from \cite{Wiese2015}.}
\label{f:gps=2}
\end{widefigure}
On figure \ref{fig:Manna:sim+MF} (thick lines), we show a numerical simulation of the Manna model in a  2-dimensional system of size $L \times L$, with $L=150$. There is a phase transition at $n=n_{\rm c}=0.702$. 
Close to $n_{c}$, the fraction of doubly occupied sites $a_{2}$ grows linearly with $n-n_{c}$, and higher occupancy is small. 
Indeed, we checked numerically that for  $n>n_{c}$ the probability $p_{i}$ to find $i$ grains on a site decays exponentially with $i$, i.e.\ $p_{i}\sim \exp(-\alpha_{n} i)$, where $\alpha_{n}$ depends on $n$, see figure \ref{f:gps=2}. This is to be contrasted with the initial condition,  where we randomly distribute $n\times L^{2}$ grains on the lattice of size $L\times L$. It yields a Poisson distribution, the coherent state $\ket n$, for the number of grains on each site, see inset of  figure \ref{f:gps=2} (left). 
This result suggests that coherent states may not be the best representation for this system. 
It further implies that 
close to the transition, $\rho\approx a$, and we expect that the  wMM and the original MM have the same critical behavior. We come back to  this question below.

\subsection{Mean-field solution of the Manna model}\label{s:Manna:MF}
\Mathematica{/tex/Vorlesungen/stochastic-field-theory/math/MF-solution.nb}
In order to make analytical progress, we  now study the {\em topple-away} or {\em mean field} (MF) solution of the stochastic Manna sandpile, which we can solve analytically: 

\paragraph{Mean-Field Manna Model (MF-MM).} If a site contains two or more grains,  move these grains to any randomly chosen   sites of the system.

\noindent
The rate equations
are, setting for convenience $a_{-1}:=0$:
\beq\label{rate:Manna}
\partial_t a_i = -a_i \Theta(i{\ge} 2) {+}a_{i+2}  {+} 2 \Big[ \sum_{j\ge 2} a_j \Big] ( a_{i-1} {-}a_i)
\punkt
\eeq
Using the sum rule \eq{normalization}, they can be rewritten as 
\beq\label{196}
\partial_t a_i {=} -a_i \Theta(i{\ge} 2) +a_{i+2}  +  2(1-a_0-a_1)(a_{i-1}-a_i)
\punkt~~
\eeq
We are interested in the steady state $\partial_{t} a_{i}=0$. 
One can solve these equations by introducing a generating function. A simpler approach consists in realizing that for $i\ge 2$, Eq.~(\ref{196}) admits a steady-state solution of the form\beq\label{213}
a_i = a_2 \kappa^{i-2}\komma  \qquad i>2
\punkt
\eeq
This reduces the number of independent equations $\partial_t a_i=0$ in Eq.~(\ref{196}) from infinity to three. Furthermore, there are the equations $\sum_{i=0}^\infty a_i = 1$, and $\sum_{i=0}^\infty i\,a_i = n$. Thus there are 5 equations for the 4 variables $a_0$, $a_1$, $a_2$, and $\kappa$. 
The reason we apparently have one redundant equation is  due to the fact that we  already used the normalization condition (\ref{normalization}) to go from Eq.~(\ref{rate:Manna}) to Eq.~(\ref{196}).

These equations  have two solutions:
For $0<n<1$, there is always the solution for the {\em inactive} or {\em absorbing state},  
\be
a_0=1-n \komma  \label{triv-sol}\quad a_1=n \komma  \quad
a_{i\ge 2}=0  \punkt
\ee
For $n>1/2$, there is a second non-trivial solution, 
\be\label{210}
a_{0} = \frac1{1+2n}\komma  \quad
a_{i>0} = \frac{4 n \left(\frac{2 n-1}{2 n+1}\right)^i}{4 n^2-1} \punkt
\ee
(Note that $a_{2}/a_{1}$ has the same geometric progression as $a_{i+1}/a_{i}$ for $i>2$, which we did note suppose in our ansatz.)
Thus the probability to find $i>0$ grains on a site is given by the exponential distribution
\beq\label{212}
p(i) =   \frac{4 n }{4 n^2{-}1} \exp\left( -i \alpha_{n}\right) \,, \quad \alpha_{n} = \ln\left(\frac{2 n{+}1}{2 n{-}1} \right) .
\eeq 
Using these two solutions, we get the  MF phase diagram plotted on figure \ref{fig:Manna:sim+MF} (thin lines). This has to be compared with the simulation of the Manna model on the same figure (thick lines).
One sees that for $n\ge2$, MF solution and simulation are   almost indistinguishable. We    also checked with simulations  that the Manna model  has a similar exponentially decaying distribution of grains per site, with a decay-constant $\alpha$ plotted on the right of figure \ref{f:gps=2}.

\subsection{Effective equations of motion for the Manna model: CDP theory}\label{CDP}
In this section, we   give the  effective equations of motion for the Manna model. Let us start from the mean-field equations for $\rho(t)$ and $n(t)$. For simplicity we use the weighted Manna model. The physics close to the transition should not depend on it.  
Let us start from the  hierarchy of MF equations for the weighted Manna model. These are similar to \Eq{196}, and   can  be rewritten as 
\beq\label{MF-w-Manna}
\partial_t a_i {=} (1-i)a_i \Theta(i{\ge} 2) +(i+1)a_{i+2} +  2\rho (a_{i-1}{-}a_i)
\punkt~~
\eeq
Let us  write explicitly the rate equation for the fraction of empty sites $e\equiv a_{0}$, 
\beq
\partial_t e = a_{2} -  2\rho e
\eeq
The first term, the gain $r_{+}=a_{2}$ comes from the sites with two grains, toppling away, and leaving an empty site. The second term, the loss term,   is the  rate at which one of the toppling grains lands on an empty site, $r_{-}=2\rho e$. 

The formalism developed in section \ref{s:discreteness}, Eqs.~(\ref{151})--(\ref{215b}),  demands to add an additional noise:
\beq
\partial_{t} e = a_{2} -  2\rho e + \sqrt{a_{2}+2\rho e}\, \bar\xi_{t}\komma 
\eeq
where $\left< \bar\xi_{t} \bar\xi_{t'}\right> = \delta(t-t')/\ell^{d}$, and $\ell$ is the size of the box which we consider.  
Close to the transition, $a_{2}\approx \rho$. Inserting this into the above equation, we arrive at
\beq\label{197}
\partial_{t} e \approx \rho(1-2 e) + \sqrt{\rho} \sqrt{1+2 e}\, \bar\xi_{t}\komma 
\eeq
Next we approximate $ \sqrt{1+2 e}$ by the value of $e$ at  the transition, i.e.\ $e\to e_{\rm c}^{\rm MF}=\frac12$, see the mean-field phase diagram in Fig.~\ref{fig:Manna:sim+MF}, leading to
\beq
\partial_{t} e   
\approx  \rho(1-2 e)  + \sqrt{2\rho} \, \bar\xi_{t}
\punkt
\eeq
This equation consistently gives back $e_{\rm c}^{\rm MF}=\frac12$, used above in the simplification of the noise term. 

As the number $n$ of grains  is conserved,    with the help of the sum rule $n+e=\rho+1$ we can write two more equations, 
\be
\partial_t n = 0 \komma  \quad \partial_t \rho = \partial_t e\punkt
\ee
Finally,    we do not have   a single box of size $\ell$, but  a lattice of boxes, indexed by a $d$-dimensional label $x$. Each toppling   moves two grains from a site to    neighboring sites, equivalent to a current 
\beq
J(x,t)  = - D \nabla \rho(x,t) + \sqrt{2 D \rho(x,t)} \xi(x,t) . 
\eeq
The diffusion constant is $D= 2 \times \frac 1{2d}= \frac1d$. The first factor of 2 is due to the fact that 
two grains topple. The factor of $\frac1{2d}$ is due to the fact that each grain can topple in any of the $2d$ directions, thus the rate $D$ per direction is $\frac1{2d}$, resulting into $D=1/d$.
As discussed above, we   drop the noise term as subdominant. 

This current corrects both the   activity $\rho(x,t)$, as   the number of grains $n(x,t)$, resulting into the same contribution for both  $\partial_{t}\rho(x,t)$, and  $\partial_{t}n(x,t) $. 
It does not couple to the density of empty sites. This is consistent with  the sum-rule (\ref{sum-rule}) $n-\rho+e=1$, which implies that
$
\partial_t \rho(x,t) \equiv \partial_t n(x,t) + \partial_t e(x,t)
$.

In conclusion, we have  the   set of equations
\begin{eqnarray}\label{233}
\partial_t e(x,t) =  [1{-} 2e(x,t)  ] \rho(x,t) + \sqrt{2\rho (x,t)}\,\xi(x,t) ,  \\
\partial_t \rho (x,t) =    \frac 1{d}  \nabla^2  \rho(x,t)  +\partial_t e(x,t), \label{234}\\
{\left< \xi(x,t)\xi(x',t') \right> =   \delta^d(x-x')\delta(t-t')}\punkt  
\end{eqnarray}
Instead of writing coupled equations for $e(x,t)$ and $\rho(x,t)$, with the help of the sum rule \eq{sum-rule} we can also write coupled equations for $\rho(x,t)$ and $n(x,t)$: 
\begin{eqnarray}\label{eff:1}
\partial_t \rho(x,t) &=&  \frac 1{d}  \nabla^2  \rho(x,t)+ \big[2n(x,t) {-}1\big] \rho(x,t)- 2 \rho(x,t)^2\nn\\
&&   + \sqrt{2\rho (x,t)}\,\xi(x,t) \komma \\
\partial_t n(x,t) &=&    \frac 1{d}  \nabla^2  \rho(x,t) \komma   \rule{0mm}{3.5ex} 
\label{eff:2}
\end{eqnarray}
\Eqs{eff:1}--\eq{eff:2} are known as the equations of motion for the conserved directed percolation (C-DP) class. 
They were obtained in the  literature \cite{Pastor-SatorrasVespignani2000,VespignaniDickmanMunozZapperi1998,BonachelaAlavaMunoz2008,Alava2003} by means of  symmetry principles. This  leaves all coefficients undefined, and does not ensure that      Eq.~(\ref{233}) is  {\em local}. This locality will prove  essential in the next section.
The  derivation above is due to Ref.~\cite{Wiese2015}.

\begin{figure}\setlength{\unitlength}{1mm}
\centerline{\mbox{\begin{picture}(84,56)
\put(3,2){\fig{8.1cm}{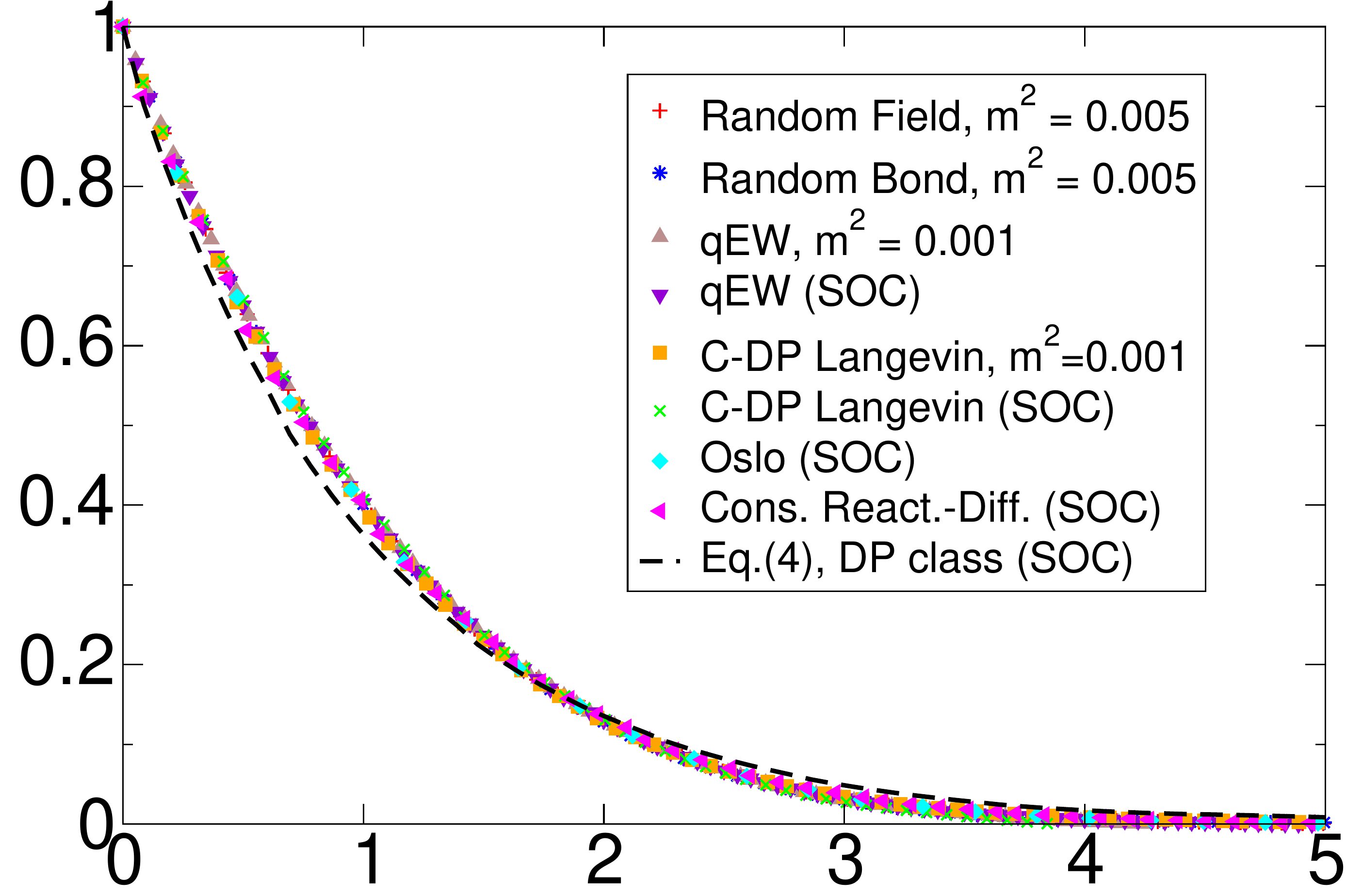}}
\put(82,0){$w$}
\put(0,53){$\Delta(w)$}
\end{picture}}}
\caption{The renormalized disorder correlator $\Delta(u)$, rescaled to $\Delta(0)=1$ and $\int_u \Delta(u)=1$, for several situations: RF and RB disorder for a disordered elastic manifolds, the Oslo and Manna models, as well as conserved directed percolation, all in $d=1$.  Reprinted from \protect\cite{BonachelaAlavaMunoz2008}, with kind permission.}  
\label{f:BonachelaAlavaMunoz2008}
\end{figure}

\subsection{Mapping of the Manna model to disordered elastic manifolds}
\label{s:mapping}
It had been conjectured for a long time that the Manna model and depinning of disordered elastic manifolds are equivalent, and much work was devoted to clarify this connection \cite{BonachelaChateDornicMunoz2007,BonachelaMunoz2008,BonachelaAlavaMunoz2008}.
The identification of fields which finally led to a simple proof of this equivalence is given in   \cite{LeDoussalWiese2014a}, followed by \cite{JanssenStenull2016}, 
\bea\label{235}
\rho(x,t) &=\partial_t u(x,t) &\mbox{\quad (the velocity of the interface),} \\
e(x,t)   &= {\cal F}(x,t) ~~~    &\mbox{\quad (the force acting on it).} 
\label{236}
\eea
The second equation (\ref{234}) is  the time derivative of the equation of motion of an interface, subject to a random force ${\cal F}(x,t)$, 
\beq\label{230}
\partial_t u(x,t) = \frac{1}{d} \nabla^2 u(x,t) + {\cal F}(x,t)\punkt
\eeq
Since $\rho(x,t)$ is positive for each $x$, $u(x,t)$ is  monotonously increasing. Instead of parameterizing ${\cal F}(x,t)$ by space $x$ and time $t$, it can be written as a function of space $x$ and {\em interface position} $u(x,t)$. Setting ${\cal F}(x,t) \to F\big(x,u(x,t)\big)$, the first equation (\ref{233}) becomes
\bea
\partial_t {\cal F}(x,t) &\to& \partial_t F\big(x,u(x,t)\big) \nn\\
&=& \partial_u F\big(x,u(u,t)\big) \partial_t u(x,t)  \nn \\
&=& \Big[1-2 F\big(x,u(x,t)\big)\Big] \partial_t u(x,t)  \nn\\
&&+ \sqrt{2 \partial_t u(x,t)} \xi(x,t)
\punkt
\eea
For each $x$, this equation is equivalent to the Ornstein-Uhlenbeck \cite{UhlenbeckOrnstein1930} process   $F(x,u)$, defined by  
\bea
\partial_u F(x,u) =  1-2 F(x,u) + \sqrt2\; \xi(x,u)\komma \\
\left< \xi(x,u) \xi(x',u') \right> = \delta^d(x-x') \delta(u-u')\punkt
\eea
It  is a Gaussian Markovian process with mean $\left< {F(x,u)}\right> =1/2$, and variance in the steady state of (see \Eq{FFOU})
\bea\label{cor:FF}
\left<{ \left[ F(x,u)-{\textstyle \half} \right] \left[ F(x',u')- {\textstyle \half} \right]}\right>\nn\\
 \quad = \frac12 \delta^d(x-x') \rme^{-2 |u-u'|}
\punkt
\eea 
Writing the equation of motion (\ref{230}) as  
\beq 
\partial_t u(x,t) = \frac{1}{d} \nabla^2 u(x,t) + F\big(x,u(x,t)\big)\komma 
\eeq
it is interpreted as the motion of an interface with {\em position} $u(x,t)$, subject to a disorder force $F\big(x,u(x,t)\big)$. The latter is $\delta$-correlated in the $x$-direction, and short-ranged correlated in the $u$-direction. In other words, this is a disordered elastic manifold subject to Random-Field disorder. 
As a consequence, the field-theoretic results of sections \ref{s:Field theory of the depinning transition}--\ref{s:dyn-2loop} are also valid for the Manna model. 

\Eq{eff:1} has a quite peculiar property, namely the factor of 2 in front of both $n(x,t)\rho(x,t)$ and $-\rho(x,t)^{2}$. As a consequence, \Eq{233} does not contain a term $\sim \rho^{2}(x,t)$, which would spoil the simple mapping presented above. The absence of this term {\em can not} be induced on symmetry arguments only. How this additional term, if present, can be treated is discussed in Ref.~\cite{LeDoussalWiese2014a}.

The mapping of the Manna model on disordered elastic manifolds implies that   properties of the latter should be measurable in the former.
As we discussed in sections \ref{s:FRG-fixed-points} to \ref{measurecusp}, and \ref{s:Field theory of the depinning transition} to \ref{s:dyn-2loop}, a key feature of the theory of disordered elastic manifolds is the existence of a  renormalized disorder correlator with a cusp. Its existence in the Manna model, and equivalence to the one measured at depinning was   established in the beautiful work
\cite{BonachelaAlavaMunoz2008}. The resulting (rescaled) disorder correlators $\Delta(w)$ are shown in Fig.~\ref{f:BonachelaAlavaMunoz2008}: it confirms the equivalence of depinning with both RB and RF disorder, C-DP, Oslo, and several sandpile automata. 
  
\subsubsection*{Remarks on the short-time dynamics of the Manna model}
The short-time dynamics of the Manna model has been measured in several publications \cite{DickmanAlavaMunozPeltolaVespignaniZapperi2001,KwonKim2016,TapaderPradhanDhar2020}, and was   interpreted as the dynamical exponent $z$ depending on the initial condition. We cannot follow this logic: the critical exponent $z$ is a bulk property of the system, and as such is defined only after memory of the initial state is erased. What is possible is that the initial-time critical exponent discussed in section \ref{s:Quench} depends on the initial condition. Simulations for much larger systems are needed to settle this question.

\section{KPZ, Burgers, and the directed polymer}
\label{s:KPZ, Burgers, and the directed polymer}
In this section we review basic properties for the non-linear surface growth known as the Kardar-Parisi-Zhang (KPZ) equation \cite{KPZ}. 
KPZ matters for disordered systems and the subject of this review for its multiple connections:
\begin{itemize}
\item  mapping of the $N$-dimensional KPZ equation  to the $N$-component directed polymer (random manifold with $d=1$),
\item  mapping of the $N$-dimensional decaying Burgers or KPZ equation to a particle  (formally a random manifold with $d=0$) in $N$ dimensions,
\item non-linear surface growth terms \`a la KPZ appear for disordered systems, producing the distinct quenched KPZ class discussed in section \ref{s:qKPZ}.
\end{itemize}
For further reading on non-linear surface growth we refer to the 1997 review by Krug \cite{Krug1997}.
A short summary of modern developments can be found in Ref.~\cite{Halpin-HealyTakeuchi2015}.

\paragraph{Notation.}
We use $N$ for the dimension of the KPZ equation instead of $d$, keeping $d$ for the random-manifold dimension with  $N$ components, i.e.\ living in $N$ dimensions. The $N$-dimensional KPZ equation will be shown to be equivalent to the directed polymer ($d=1$) in $N$ dimensions. 

\begin{widefigure}
\centerline{\Fig{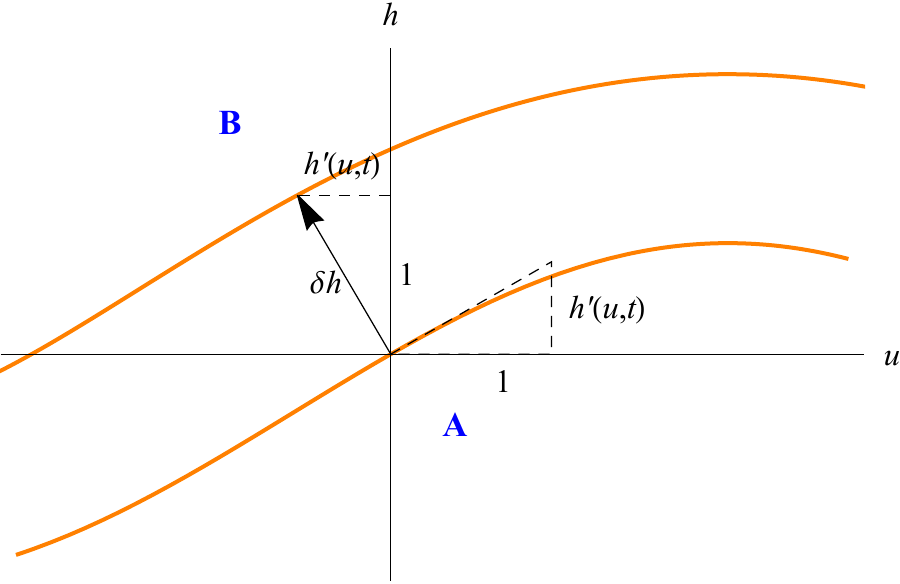}~~\hfill~~\Fig{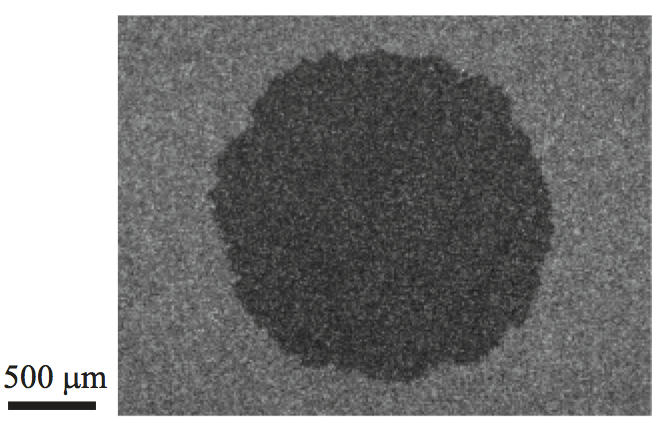}}
\caption{Left: An interface growing in its normal direction, with   phase A invading phase B. Right: An experimental realization using two phases of a nematic liquid crystal \cite{TakeuchiSano2012}.}
\label{f:KPZ-growth}
\end{widefigure}

\subsection{Non-linear surface growth: KPZ equation}
\label{s:KPZ}
Consider figure \ref{f:KPZ-growth}. What is seen is an interface between two phases, A and B. Phase A is stable, while phase B is unstable. The interface  grows with a velocity $\lambda(u,t)$ in its normal direction, increasing phase A, while diminishing phase B. 
Using the Monge representation $\{ u,h(u,t) \}$, $u \in \mathbb R^{N}$ the growth in $h$ direction is given by (see figure)
\be\label{Monge}
\delta h = \sqrt{1+ [\nabla h(u,t)]^{2}}\, \lambda(u,t) \delta t \punkt
\ee
Assume that the growth is due to a discrete process. Following the prescription in  section \ref{s:discreteness}, the growth velocity $\lambda (u,t)$ has a mean $\lambda$ plus fluctuations $\eta (u,t)$, 
\bea
\label{KPZ-noise}
\lambda (u,t)= \lambda + \eta (u,t)\komma  \nn\\
 \hl{ \left < \eta (u,t) \eta (u',t') \right> = 2 D \delta (t-t') \delta ^N(u-u')}\punkt
\eea
This leads to 
\be
\partial_{t} h(u,t) = \lambda + \frac{\lambda}2 \left[ \nabla h(u,t)\right]^{2}+ \eta(u,t) + ...\komma 
\ee
where the dots indicate higher-order terms in $(\nabla h)^{2}$. 
This is (almost) the famous KPZ \cite{KPZ} equation. To derive the latter, we first subtract the growth for a flat interface,  setting $h\to h-\lambda t$, and finally add one more term to the equation 
\be\label{KPZ}
\hl{  \partial_{t} h(u,t) =\nu \nabla^{2} h(u,t) + \frac{\lambda}2 \big[ \nabla h(u,t)\big]^{2} + \eta(u,t)}\punkt
\ee
The additional  term proportional to $\nu$ describes diffusion along the interface, rendering it smoother. 

One typically  measures the 2-point function 
\be\label{zeta-z-KPZ}
\left< [h(x,t)-h(x',t')]^2 \right> \simeq |x-x'|^{2\zeta_{\rm KPZ}} f(x^{z_{\rm KPZ}}/t),
\ee
where $f$ goes to a constant for $t\to 0$ ($x\to \infty$), and together with its prefactor becomes independent of $x$ for $x\to 0$. This defines   two exponents, the roughness $\zeta_{\rm KPZ}$ and the dynamic exponent $z_{\rm KPZ}$. The added index   allows us to distinguish it from the  exponents  of the directed polymer, especially since we will see later that $z_{\rm KPZ}=1/\zeta_{\rm directed~polymer}$.

\subsection{Burgers equation}
\label{s:Burgers}
Taking one spatial derivative of \Eq{KPZ} yields {\em Burgers'} equation \cite{Burgers74}. Define 
\be
  v(u,t):=\nabla h(u,t). 
\ee 
Burgers' equation reads
\bea\label{Burgers}
\partial_{t}   v(u,t) = { \nu}\nabla^{2}   v(u,t) +\frac\lambda 2 \nabla [   v(u,t)^{2}] + \nabla \eta (u,t)\komma    \\ 
\left < \eta (u,t) \eta (u',t') \right> = 2 D \delta (t-t') \delta ^N(u-u')\punkt
\eea
The non-linear term satisfies the identity
\bea
\half \sum_i \,\partial_{j}   \big[ v_{i}(u,t)^{2}\big] \equiv \half  \sum_i\,\partial_{j}   \big[ \partial_{i}h(u,t)^{2}\big] \\
 = \sum_i\big[  \partial_{i}h(u,t)\big] \big [ \partial_{j}  \partial_{i}h(u,t) \big]  \equiv  \sum_i\big[ v_{i}(u,t)  \partial_{i}  \big] v_{j}(u,t)\punkt
 \nn
\eea
\Eq{Burgers} can  be written as 
\bea
\partial_{t}   v(u,t) = { \nu}\nabla^{2}   v(u,t) +  \lambda\big[   v(u,t){\cdot} \nabla]         v(u,t)  + \nabla \eta (u,t)\punkt\nn\\
\eea
This is identical to Navier-Stokes' equation for incompressible fluids, with the crucial difference that     Burgers'  velocity is a total derivative, $v(u,t) = \nabla h(u,t)$, whereas for Navier-Stokes it is divergence free,  $\nabla v(u,t) =0$.  
For this reason, 
Burgers equation does not describe turbulence encountered e.g.\ in a fast-flowing river. It has, however, applications to the large-scale structure of galaxies \cite{GurbatovSaichevShandarin1989,Bertschinger1998,BernardeauColombiGaztanagaScoccimarro2002}.

\subsection{Cole-Hopf transformation}
\label{s:Cole-Hopf}
Consider the $N$-dimensional  KPZ  equation \eq{KPZ} in It\^o discretization, with noise as given in \Eq{KPZ-noise}.
We can eliminate the non-linear term by the so-called {\em Cole-Hopf transformation}  \cite{Hopf1950,Cole1951}
\bea\label{ColeHopf}
Z(u,t):= \rme^{\frac \lambda{ 2 \nu} h(u,t)-D \frac{\lambda^{2}}{4\nu^2} t}   \nn \\
\Longleftrightarrow \quad h(u,t) = \frac{2 \nu}\lambda \ln  Z(u,t) + \frac{D \lambda}{2\nu} t\punkt
\eea
(The reader might see this transformation without the term $-D\lambda^2 t/(4\nu^2)$; this is then done in mid-point, i.e.\ Stratonovich discretization \cite{Zinn,Janssen1985,Tauber2012}, see section  \ref{MSR-formalism}).
Using It\^o calculus (section \ref{s:Ito calculus}), we obtain
\bea
\rmd Z(u,t) = \frac\lambda{ 2\nu} Z(u,t) \rmd h(u,t)  + \frac{\lambda^{2}}{8 \nu^2} Z(u,t) \,\rmd h(u,t)^{2} \nn\\
\hp{\rmd Z(u,t) = }-\frac{ D \lambda ^{2}}{4 \nu^2} Z(u,t) \rmd t \nn\\
= \frac{ \lambda}{ 2\nu} Z(u,t) \Big\{ \! \Big( { \nu}\nabla^{2} h(u,t) {+} \frac\lambda 2[\nabla h(u,t)]^{2}\Big)\rmd t  + \rmd \eta (u,t)   \Big\} \nn\\
= { \nu} \nabla^{2}Z(u,t)\rmd t + Z(u,t) \frac{\lambda}{ 2\nu}\rmd \eta (u,t)\punkt
\eea
Noting $  \lambda  \eta(u,t) \equiv V(u,t)$ this can be  written as 
\begin{eqnarray}
\label{KPZ-CH}
&&\hl{\partial_{t}Z(u,t) ={ \nu} \nabla^{2}Z(u,t) + { \frac1{2\nu}}V(u,t) Z(u,t)}\komma    \\
&& \left < V (u,t) V (u',t') \right> =   \delta (t-t') R (u-u')\komma  \\ 
&& R(u) =  { 2 \lambda^{2} D}  \,  \delta^N (u)\punkt
\label{495}
\end{eqnarray}

\subsection{KPZ as a directed polymer}
\label{s:KPZ as a directed polymer}
The equation of motion \eq{KPZ-CH} can be solved by 
\bea\label{directed-polymer}
Z(u,t|V) =   \int^{u = u(t)}_{{u(\ti)}=u_{\rm i}} {\cal D}[u] \, \rme^{-\frac1T\int_{\ti}^{t} \rmd \tau\, \frac12 {u'(\tau)^{2}}- V(u(\tau),\tau) }  \komma  \nn\\
T=2 \nu \punkt \label{T-nu}
\eea
This is the path integral of a directed polymer in the quenched random potential $V(u)$, also referred to as the Feynman-Kac formula \cite{Feynman1948,Kac1949}. To average over disorder, we use the   formalism with $n$ replicas introduced in section \ref{s:replicas},
\bea
  Z(u_1,...,u_n,t) \nn\\
=  \prod_{\alpha=1}^n \int^{ u_\alpha(t)=u_\alpha }_{{u_\alpha(\ti)}=u_{\alpha,\rm i}} {\cal D}[u_\alpha]\int {\cal D}[V] \, \nn\\
~~~~~~\rme^{-\frac 1T \int_{\ti}^{t} \rmd \tau \sum_{\alpha=1}^n  \left[ \frac12{u_\alpha'(\tau)^{2}} + V(u_\alpha(\tau)) \right]} \rme^{-\frac1{ 4 \lambda^{2}D}\int_{u,\tau} V(u,\tau)^{2}}\nn\\
 =  \prod_{\alpha=1}^n   \int^{ u_\alpha(t)=u_\alpha }_{{u_\alpha(\ti)}=u_{\alpha,\rm i}} {\cal D}[u_\alpha]  \, \nn\\
~~~~~~ \rme^{- \int_{\ti}^{t} \rmd \tau\,\sum_{\alpha} \frac1{2T}  {u_\alpha'(\tau)^{2}} -\frac1{2T^2} \sum_{\alpha,\beta} R(u_\alpha(\tau) - u_\beta(\tau))  } \punkt
\eea
We had discussed its solution in the $T\to 0 $-limit in section \ref{s:replicas}, see \Eq{H}.

Using \Eqs{T-nu} and \eq{ColeHopf}, 
the free energy of a directed polymer is related to the KPZ height field $h(u,t)$ via
\be\label{F-h}
\ca F(u,t) := - T \ln  Z(u,t) \equiv - \lambda h(u,t) + \frac{D \lambda^2}{2\nu} t.
\ee
Apart from the (last) drift term which is due to the discretization scheme and which can always be subtracted, this relation is valid in the inviscid limit $\nu \to 0$, equivalent to $T\to 0$, i.e.\ for the ground state of the directed polymer. 
For further reading we refer to  \cite{BrunetDerrida2000a}.

\subsection{Galilean invariance, and scaling relations}
\label{s:KPZ-Galilean-invariance}
A scaling analysis of the KPZ equation \eq{KPZ} starts at \cite{Krug1997}
\be\label{KPZ-coarse-grain}
\tilde h(u,t) = b^{-\zeta_{\KPZ}} h(b u , b^{z_{\KPZ}} t)
\ee
with a roughness exponent $\zeta_{\KPZ}$ and a dynamical exponent $z_{\KPZ}$ defined  in \Eq{zeta-z-KPZ}.
The rescaled field $\tilde h$ satisfies a KPZ equation \eq{KPZ} with rescaled coefficients 
\bea\label{KPZ-scaling}
&&\tilde \nu = b^{z_{\KPZ}-2} \nu, \quad \tilde D = b^{z_{\KPZ}-d-2\zeta_{\KPZ}} D, \nn \\
&&  \tilde \lambda = b^{z_{\KPZ}+\zeta_{\KPZ}-2} \lambda.
\eea
If $\lambda=0$, the scaling of the diffusion equation $\zeta_{\KPZ} = (2-N)/2$, and $z_{\KPZ}=2$ yields a fixed point of the coarse graining transformation \eq{KPZ-coarse-grain}. For $\lambda\neq 0$, the non-linearity $\lambda$   grows if the combination 
$\zeta_{\KPZ}+z_{\KPZ}-2 \to \frac{2-N}2$ is positive. This is always the case in dimension $N<2$. 
As we will see below in section \ref{KPZ-all-orders} for dimension $N>2$ there is a transition between a weak-coupling and a strong-coupling regime.

The KPZ equation has an important invariance   in any dimension $N$ \cite{ForsterNelsonStephen1977,KPZ,Krug1997}.
Consider the tilt transformation
\noindent parameterized by a $N$-dimensional vector $c$, 
\be\label{Galileo}
h'(u, t) = h(u +\lambda c t, t) +  c u + \frac\lambda{2} c^2 t. 
\ee
For $\lambda\to 0 $, this reduces to  the statistical tilt symmetry  \eq{STS}.
It is reminiscent of the full rotational invariance of the growing surface before passing to the Monge representation \eq{Monge}. As a consequence, we expect the KPZ equation to remain invariant under this transformation.  Indeed, $h'$ satisfies the same KPZ equation \eq{KPZ} as $h$, with a shifted noise 
\be
\eta'(u, t) = \eta(u +\lambda c t, t). 
\ee
As long as the temporal correlations of $\eta$ are sufficiently short ranged, the shift does not affect the statistical properties of the noise \cite{MedinaHwaKardarZhang1989}, and the statistics of   $h$ is invariant under the transformation \eq{Galileo}. In the literature this property is   referred to as Galilean invariance, as in the context of  the  stirred Burgers equation \eq{Burgers}, where it was first discussed in Ref.~\cite{ForsterNelsonStephen1977},   it appears as a shift in the velocity, $v\to v'=v+\lambda c$.

As the tilt transformation explicitly contains the non-linearity $\lambda$, the latter should not change under rescaling. 
From \Eq{KPZ-scaling} we   conclude that at a fixed point with $ \lambda \neq 0$ \cite{MeakinRamanlalSanderBall1986,Krug1987,MedinaHwaKardarZhang1989}
\be\label{Galileo-exp-relation}
\zeta_{\KPZ} + z_{\KPZ} = 2 .
\ee
To make contact with the scaling properties of the directed polymer, we remind the exponent relation  \eq{free-energy-scaling} for the  free energy of an elastic manifold, 
$\ca F \sim L^\theta$, with $\theta = d-2+2 \zeta$. The directed polymer has internal dimension $d=1$. Using  that according to \Eq{ColeHopf} the free energy identifies with $h$, and that the length $L$ of the directed polymer is the time in the KPZ equation, we arrive at 
$h \,\widehat=\, \ca F \sim L^\theta  \,\widehat=\,  t^\theta $. 
On the other hand
$
h \sim u^{\zeta_{\KPZ}}$ and $u\sim t^{1/z_{\KPZ}}$, such that 
\be
\theta = 2\zeta-1 = \frac{\zeta_{\rm \scriptscriptstyle KPZ}}{z_{\KPZ}} = \frac{2-z_{\KPZ}}{z_{\KPZ}} \komma 
\ee
where for the last identity \Eq{Galileo-exp-relation} was used. 
This implies that 
\be\label{zKPZ=1zeta}
z_{\KPZ} = \frac1\zeta.
\ee
While the notations in the literature are somehow divergent, the prevailing  ones seem to be
\be
\alpha \equiv \chi\equiv \zeta_{\rm KPZ}, \quad  \beta =  \frac{\zeta_{\rm \scriptscriptstyle KPZ}}{z_{\KPZ}} \equiv \theta, \quad z = z_{\rm KPZ}\equiv \frac 1\zeta.
\ee

\begin{widefigure}
\centerline{\Fig{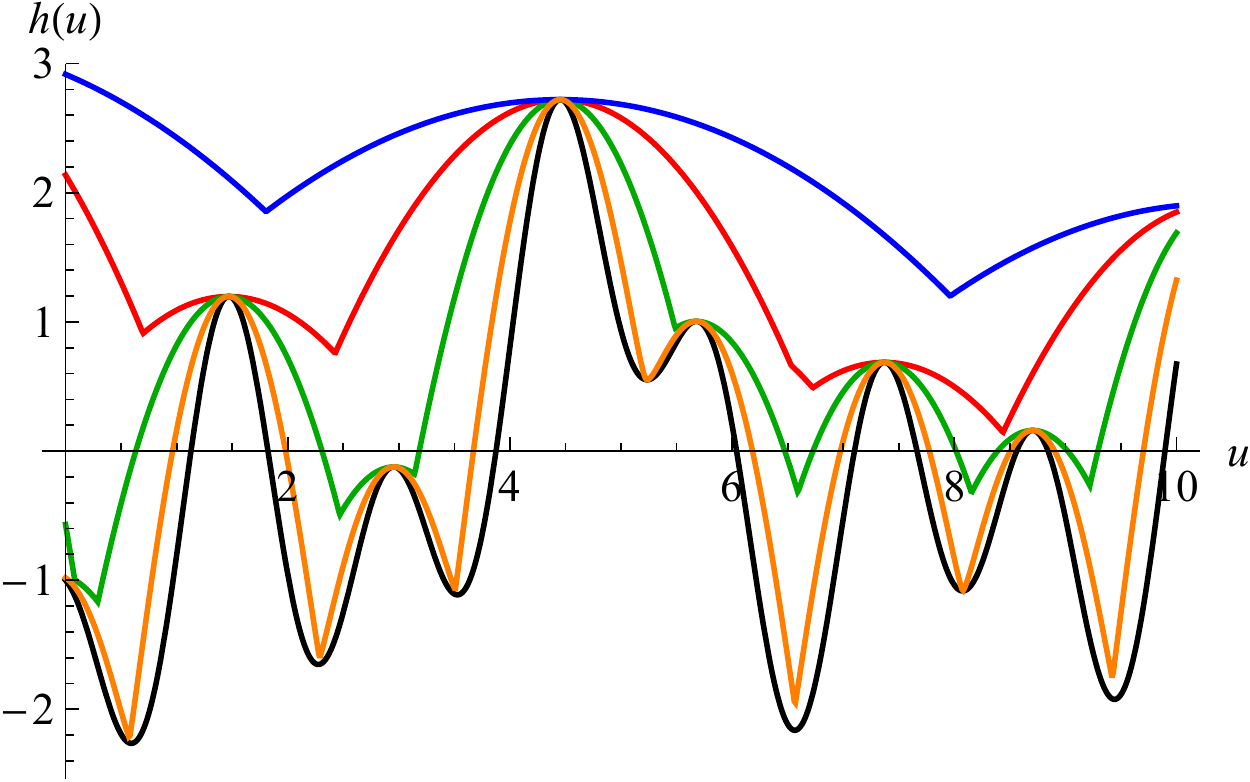}~~\Fig{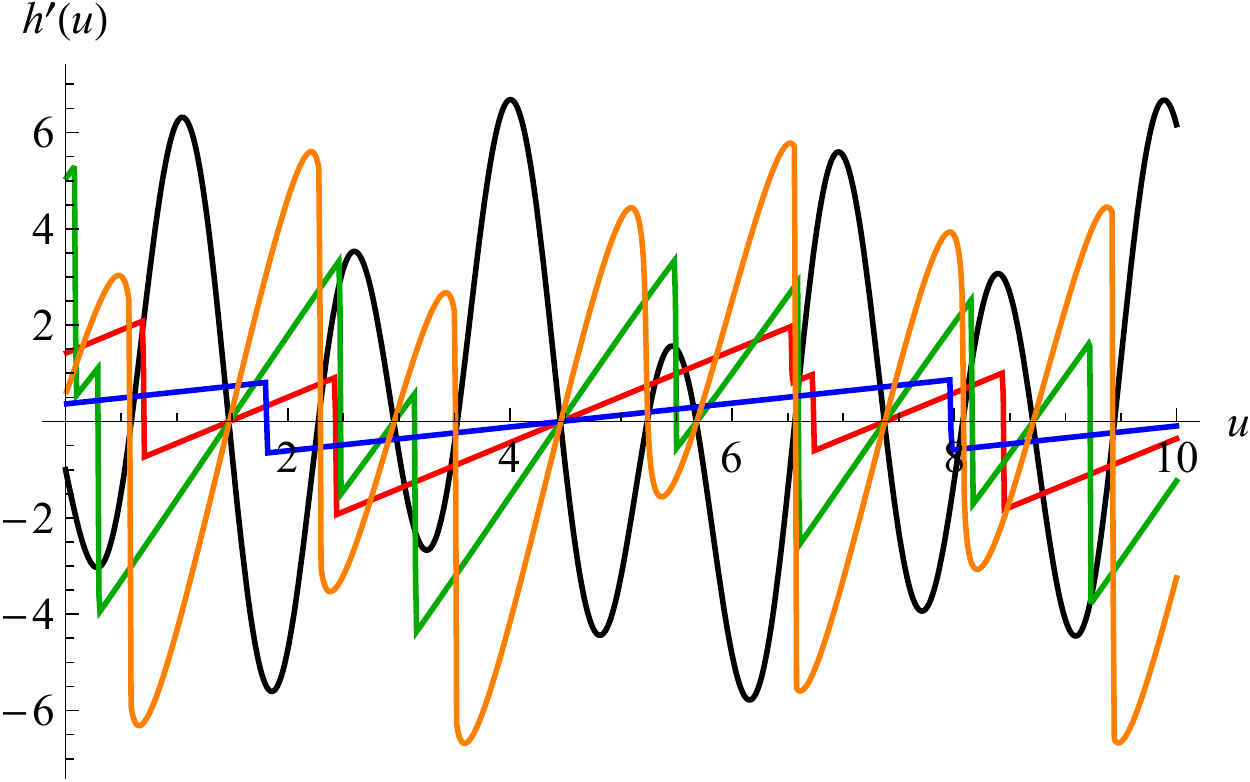}}
\caption{Left: Evolution of a random initial condition (black) at times $\lambda t = 0$ (black, bottom), $1/16$ (orange), $1/4$ (green), $1$ (red), and $4$ (blue). Right: ibid.\ for the Burgers velocity $v(u):=h'(u)$.}
\label{f:shocks}
\end{widefigure}
\begin{figure}
\Fig{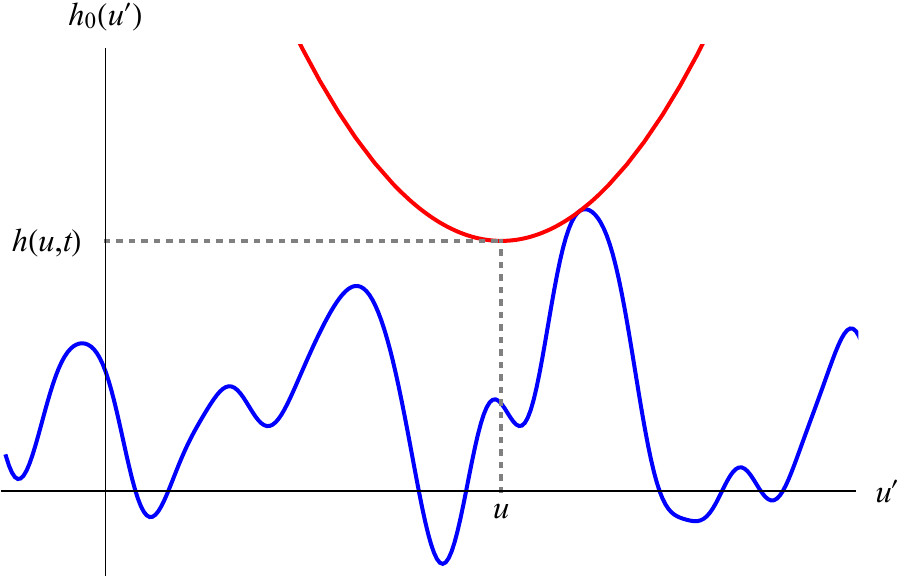}
\caption{A geometrical solution to \Eq{11.26} is obtained by moving a parabola of curvature $1/(2\lambda t)$  and centered at $u$ (in red) down until it hits $h_0(u')$ in blue. Its minimum is then at $h(u,t)$. (This construction is already discussed in the 1979 paper by Kida \cite{Kida1979}.)}
\label{f:Kida-construction}
\end{figure}

\subsection{A field theory for the Cole-Hopf transform of KPZ}
Another possibility  to obtain a field theory for  \Eqs{KPZ-CH}--\eq{495} is to write the partition function for the $n$-times replicated field $Z_\alpha$, $\alpha=1,...,n$, i.e.\ $Z\in \mathbb R^{n}$, as 
\be
{\cal Z} = \int {\cal D}[Z] \, {\cal D}[\tilde Z] \, {\cal D}[V] \,\rme^{{-\cal S}_{\rm CH} [Z,\tilde Z,V]}\komma 
\ee
\bea
{\cal S_{{\rm CH}}}[Z,\tilde Z,V] \nn\\
= 
\int\limits_{u,t} \tilde Z(u,t)\bigg[  \partial_{t }Z(u,t) 
     {-}{  \nu}\nabla^{2}Z(u,t) {-}{  \frac1{2 \nu}} V(u,t)Z(u,t)\bigg] \nn\\
~~~+\frac{  1}{  4\lambda^{2}D}\int_{u,t} V(u,t)^{2}\punkt
\eea
Performing the integral over $V$ we obtain
\bea
{\ca Z} = \int {\cal D}[Z]\, {\cal D}[\tilde Z] \,\rme^{{-\cal S}_{\rm CH} [Z,\tilde Z]}\komma   \\
{\cal S_{{\rm CH}}}[Z,\tilde Z] = \int_{u,t} \tilde Z(u,t)\left[\partial_{t }Z(u,t) -{  \nu}\nabla^{2}Z(u,t)\right] 
\nn\\
\hp{{\cal S_{{\rm CH}}}[Z,\tilde Z] =}~-{  \frac{  \lambda ^{2 }D}{4\nu^2}}\left[ \tilde Z(u,t) Z(u,t) \right] ^{2} \punkt \label{427}
\eea
Replacing $t\to t/\nu$, we arrive at \bea\label{428}
{\cal S_{{\rm CH}}}[Z,\tilde Z] = \int_{u,t} \tilde Z(u,t)\left[\partial_{t }Z(u,t) - \nabla^{2}Z(u,t)\right] \nn\\
\hp{{\cal S_{{\rm CH}}}[Z,\tilde Z] =} -\frac g2\left[ \tilde Z(u,t) Z(u,t) \right] ^{2} \komma \\ \label{429}
g = \frac{  \lambda^2 D}{2 \nu^3}\punkt
\eea
Note that we do not need to take the limit of $n\to 0$ at the end. This is allowed as in It\^o calculus  the partition function ${Z}=1$. 
To study the flow of the effective coupling constant 
$g$, 
we need at least two distinct ``replicas'', which can be thought of as   ``worldlines'' of two particles, starting at different initial positions.

\subsection{Decaying KPZ, and shocks}
\label{s:Decaying KPZ, and shocks}
To better understand the behavior of the KPZ equation \eq{KPZ}, let us consider \Eq{KPZ} for given initial condition $h_0(u):=h(u,t=0)$,   in   absence of the noise $\eta(u,t)$, i.e.\ $D=0$.
The Cole-Hopf transformed KPZ equation \eq{KPZ-CH}   reduces to a diffusion equation, solved as  
\be
Z(u,t) = \int_{u'}\frac{ \rme^{-\frac{(u-u')^2}{4 \nu t}}}{(4\pi \nu t) ^{d/2}} \,Z(u',0).
\ee
Putting back the definition \eq{ColeHopf} of $Z$ in terms of $h_0(u):=h(u,t=0)$, we obtain, since $D=0$,
\be\label{KPZ:746}
\rme^{\frac \lambda{ 2 \nu} h(u,t) } = \int_{u'}\frac{ \rme^{\frac\lambda{ 2 \nu} \left[ h_0(u') - \frac{(u-u')^2}{2 \lambda  t} \right] }}{(4\pi \nu t) ^{d/2}} \punkt
\ee
It is interesting to consider the limit of $\nu\to 0$, equivalent to $T\to 0$ for the directed polymer \eq{directed-polymer}. 
Then the solution to \Eq{KPZ:746} is
\be\label{11.26}
 h(u,t) = \max_{u'} \left[ h_0(u') - \frac{(u-u')^2}{2 \lambda  t} \right]\punkt
\ee
This solution is formally equivalent to the solution \eq{5.37} of the toy model introduced in section \ref{s:shocks}, replacing 
\bea
-h_0(u)  \to   V(u)  \mbox{~~~~~(microscopic disorder)}\\
\frac1{ \lambda t} \to  m^2 \\
-h(u,t)  \to   \hat V(u)   \mbox{~~~~~(effective disorder at scale $m^2= \frac1{ \lambda t}$)}\punkt\nn\\
\eea
As observed there and shown in Fig.~\ref{f:shocks} for a random initial condition, the function  $h(u,t)$ is composed of almost parabolic pieces, continuous everywhere but not differentiable at the junctures. Geometrically, this can be obtained by approaching a parabola of curvature $m^2=1/( \lambda t)$ from the top, and reporting as a function of its center $u$ the position $h(u,t)$ at which it first touches the initial condition $h_0(u)$. This construction is shown in Fig.~\ref{f:Kida-construction}. More can be learned via this approach, see Ref.~\cite{LeDoussal2008}. 

\subsection{All-order $\beta$-function for KPZ}
\label{KPZ-all-orders}
{
\newlength{\ppcm}
\setlength{\ppcm}{0.5cm}
\newlength{\ppmm}
\setlength{\ppmm}{0.1\ppcm}
\newcommand{\vertex}{\,\parbox{1.5\ppcm}{\includegraphics[width=1.5\ppcm]{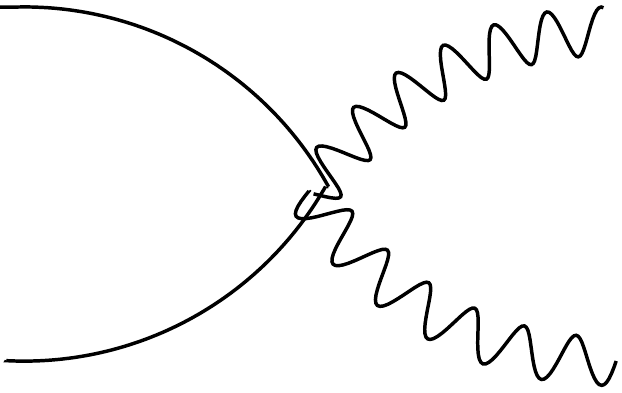}}\,}
\newcommand{\vertexzwei} {\,\parbox{0.75\ppcm}{\includegraphics[width=0.75\ppcm]{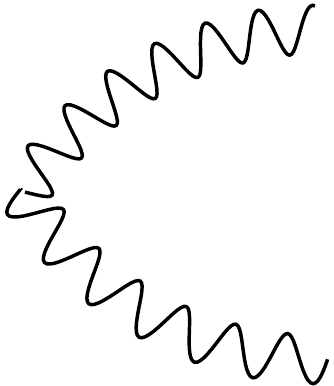}}\,}
\newcommand{\rightvertex} {\,\parbox{0.75\ppcm}{\includegraphics[width=0.75\ppcm]{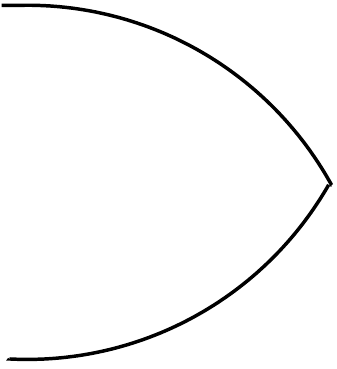}}\,}
\newcommand{\kleb}{\hspace{-3.0\ppmm}}
\newcommand{\smallvertex} {\,\parbox{1.0\ppcm}{\includegraphics[width=1.\ppcm]{KPZ}}\,}
\newcommand{\largevertexzwei} {\,\parbox{1.4\ppcm}{\includegraphics[width=1.4\ppcm]{KPZ2}}\,}
\newcommand{\rightlargevertex} {\,\parbox{1.4\ppcm}{\includegraphics[width=1.4\ppcm]{KPZ4}}\,}
As we had seen after \Eq{KPZ-scaling}, the Gaussian fixed point ($\lambda=0$) is stable for weak disorder as long as the number $N$ of dimensions is larger than 2. We now show that for $N>2$ there exists a phase transition between the weak-coupling phase (Gaussian fixed point), and a strong-coupling phase.
This transition is accessible to a standard perturbative RG treatment, contrary to the strong-coupling phase which is not.

Perturbative RG treatments of the KPZ equation are numerous, starting with the original work \cite{KPZ}. They were extended to 2-loop order in Refs.~\cite{FreyTaeuber1994,FreyTaeuber1994b,SunPlischke1994,Sun1995,Teodorovich1996}, leading to some controversy finally resolved in Ref.~\cite{Wiese1997c}. The treatment is much easier for the Cole-Hopf transformed version \cite{Laessig1995,Wiese1998a}, allowing us to resum perturbation theory to all orders. 
We now calculate the $\beta$-function associated to the model \eq{428}-\eq{429}, following Ref.~\cite{Wiese1998a}. 
To this aim we introduce the graphical notation 
\be
\vertex := \int_{u,t}\left[ \tilde Z(u,t) Z(u,t) \right] ^{2}\punkt
\ee
Then the only  diverging diagrams  are chains of $\smallvertex$, of the form $\smallvertex\kleb\smallvertex$, $\smallvertex\kleb\smallvertex\kleb\smallvertex$ and so on. Higher-order
vertices  are irrelevant in perturbation theory. As a result, the effective coupling constant is
\bea  
g_{\rm eff} \vertex &=&  g \vertex + g^2 \vertex\kleb\vertex + g^3 \vertex\kleb\vertex\kleb\vertex  
\label{pert-exp} \nn\\
 & &+ g^4 \vertex\kleb\vertex\kleb\vertex\kleb\vertex + \ldots \nn\\
 &&+ \mbox{higher order vertices} \punkt 
\eea
In Fourier-representation with incoming momentum $p$ and frequency $\omega$,  each chain in \Eq{pert-exp} factorizes, i.e.\ can 
be written as a product of the vertex $\smallvertex$ 
times a power
of the elementary loop diagram (which is a function of $p$ and
$\omega$):
\bea \label{fac-prop}
\stackrel{p,\omega}\longrightarrow
\underbrace{\vertex\kleb\vertex\kleb\vertex\kleb\vertex\kleb\vertex}_{\mbox{\scriptsize $n$ loops}} \nn\\
 = \left( \vertexzwei\kleb\rightvertex_{p,\omega} \right)^n   \vertex \punkt
\eea
\Eq{pert-exp} is a geometric sum which yields the effective 4-point function 
\be\label{eff Gamma 4}
g_{\rm eff}= \Gamma_{ZZ\tilde Z \tilde Z}\ts_{p,\omega} =   \frac g{1-g\left. \vertexzwei\kleb\rightvertex_{p,\omega} \right.} 
 \punkt 
\ee
The elementary diagram is 
\bea
&& \stackrel{p,\omega}\longrightarrow \ 
\raisebox{1.1\ppcm}[1.5\ppcm][0mm]{\parbox{0mm}{$\hspace{-.1\ppcm}{\frac \omega 2 +\nu,\ \frac p2+k\atop}$}}
{\raisebox{-1.4\ppcm}[-1.9\ppcm][0mm]{\parbox{0mm}{$\hspace{-.1\ppcm}{\frac \omega 2-\nu,\ \frac p2-k\atop}$}}}
\largevertexzwei\kleb\hspace{-0.18mm}\rightlargevertex
\ 
\stackrel{p,\omega}\longrightarrow  \ \ \ \  \rule[-7mm]{0mm}{15mm} \nn\\
&&=\int\!\!\frac{\rmd^N k}{(2\pi)^N} \int\!\frac{\rmd \nu}{2\pi} \frac{1}{\left(\frac p2{+}k\right)^2 +i \left( \frac \omega 2
{+}\nu\right)}
\frac{1}{\left(\frac p2{-}k\right)^2 {+}i \left( \frac \omega 2
{-}\nu\right)}    \nn\\
&&= a \left(\half p^2 {+}i\omega\right)^{\!N/2-1}\!, \quad a= \frac{1}{(8\pi)^{N/2}}
\,\Gamma\!\left(1{-}\frac N2\right) . \label{div}
\eea
This  integral   is divergent for any 
$p$ and $\omega$ when $N \to 2$. 
Renormalization means to absorb this divergence into
a reparametrization of the coupling constant $g$: 
We claim
 that   the  4-point function (the effective coupling $g_{\rm eff}$)
 is finite (renormalized) 
as a function of $g_{\rm r}$ instead of $g$, upon setting
\begin{equation} \label{ren-bare}
g = Z_g g_{\rm r}\mu^{-\varepsilon}\ , \quad 
Z_g = \frac {1} {1+a g_{\rm r}} \komma  \quad \E=N-2 \punkt
\end{equation}
$\mu$ is an arbitrary scale, the  {\em renormalization scale}. 
As a function of $g_{\rm r}$, the 4-point function reads 
\be \label{Gamma4R}
\Gamma_{ZZ\tilde Z \tilde Z}\ts_{p,\omega} =
	\frac{g_{\rm r}\mu^{-\E}}{1+\left(  a  -
\mu^{-\E}\,\vertexzwei\kleb\rightvertex_{p,\omega} 
 \right) g_{\rm r}} \punkt
\ee
Since 
$
\frac1\E  \left(\half p^2 +i\omega\right)^{\E/2} \mu^{-\E} 
$
is finite for $\E> 0$ as long as the  
combination   $\half p^2 +i\omega$ is finite, 
it can be read off from \Eq{Gamma4R} that    
$\Gamma_{ZZ\tilde Z \tilde Z}\ts_{p,\omega}$ is finite even in the limit of $\E\to 0$. 
(If useful, either $p=0$ or $\omega=0$ may safely be  taken.)
This completes the proof. Note that this ensures that the 
model is renormalizable to all orders in perturbation-theory,
what is normally a formidable task to show \cite{BogoliubovParasiuk1957,Hepp1966,Zimmermann1969,BergereLam1975,DDG2,DDG4,WieseHabil}.

The $\beta$-function that  we   calculate now is
exact to all orders in perturbation theory. It is defined
as the variation of the renormalized coupling constant, keeping the 
bare one fixed
\be
	\beta(g_{\rm r}) =- \mu \frac{\partial}{\partial \mu}\lts_{g} g_{\rm r} \punkt
\ee
From \Eq{Gamma4R} we see that it gives the 
dependence of the 4-point function on $p$ and $\omega$ for fixed bare
coupling.
The relation between $g$ and $g_{\rm r}$ is 
\be
g= \frac{g_{\rm r} \mu^{-\E}}{1+ a g_{\rm r}}
\quad \Leftrightarrow \quad 
g_{\rm r}=\frac{g}{\mu^{-\E}- a g} \komma 
\ee 
and hence
\be
\beta(g_{\rm r}) =-\E g_{\rm r} (1+ a g_{\rm r}) \punkt
\ee
Using $a$ from \Eq{div}, our final result is
\be \label{final result}
\beta(g_{\rm r})= (2-N) g_{\rm r} + \frac{2}{(8\pi)^{N/2}} \Gamma\left(2-\frac N2\right) g_{\rm r}^2 \punkt
\ee
This equation has a perturbative, IR repulsive, fixed point at 
\be
g^*_{\rm r} =\frac{2(8\pi)^{N/2}}{(N-2)\Gamma\!\left(2-\frac N2\right)}.
\ee
For $N>2$ it describes the phase transition between the weak-coupling phase  where the KPZ term is irrelevant ($g=0$), and a strong coupling phase, for which $g\to \infty$. This is the only fixed point available for $N\le 2$; especially, the perturbative treatment above does not describe KPZ in dimension $N=1$.

As a consequence, 
standard
perturbation theory fails to produce a strong-coupling fixed
point, a result which cannot be overemphasized. 
This means that any treatment of the strong coupling regime
has to rely on {\em non-perturbative methods}. The FRG approach discussed above qualifies as non-perturbative in this sense, since FRG follows more than the flow of a single coupling constant. It does of course
not rule out the possibility  to find an exactly solvable 
model, distinct from  KPZ, for which it is possible 
to expand towards the strong-coupling regime of KPZ. 

Let us also note that the
 $\beta$-function is divergent at $N=4$, and therefore 
our perturbation expansion breaks down at $N=4$.
To cure the problem, a lattice regularized version of 
\Eq{KPZ-CH} may be used. However, then the lattice 
cut-off $a$ will enter into the equations and the result 
is no longer model-independent. 
This may be interpreted as  $N=4$ being the upper
critical dimension of KPZ, or as a sign for
a simple technical problem.  See  \cite{BundschuhLaessig1996,DobrinevskiPhD}, and the discussion in section \ref{s:An upper critical dimension for KPZ?}.

\subsection{Anisotropic KPZ}
A special case is the anisotropic KPZ equation in two dimensions, for which  the KPZ-nonlinearity is positive in one direction, and negative in the other.  This competition produces  a perturbative fixed point  \cite{FreyTauberJanssen1999}.

\subsection{KPZ with spatially  correlated noise} 
When the noise $\eta(u,t)$ in \Eq{KPZ} is long-range correlated,
\be
\left< \eta(u,t) \eta(u',t')\right> = \delta(t-t') |u-u'|^{2\rho-N}, \quad \rho>0,
\ee  
a different exponent is expected. Using \Eq{a4} with $\gamma=N-2\rho$ and $d=1$,  we find $\zeta^{\rm LR}_{\rm Flory} = \frac3{4+N-2\rho}$, and as a consequence of \Eqs{zKPZ=1zeta} and \eq{Galileo-exp-relation}
\be
z^{\rm LR}_{\rm KPZ}= \frac{4+N-2\rho}3, \quad \zeta_{\rm KPZ}^{\rm LR}
=\frac{2-N+2\rho}3. 
\ee
This LR fixed point, which is exact as long as the disorder does not get renormalized,
 is in competition with the SR (random bond) fixed point for which disorder renormalizes. As a rule of thumb, the fixed point with the larger $\zeta$ or $\zeta_{\rm KPZ}$, and smaller $z_{\rm KPZ}$ dominates. In dimension $N=1$ where $\zeta_{\rm KPZ} =1/2$, the LR fixed point dominates for $\rho>\rho_c=\frac14$. 
These results can already be found in \cite{Kardar1987b,Nattermann1987}, and were reanalyzed via  RG in 
\cite{JanssenTauberFrey1999,TauberFrey2002}.

\subsection{An upper critical dimension for KPZ?}
\label{s:An upper critical dimension for KPZ?}
A lot of work has been devoted to either proving  or disproving 
 the existence of an upper critical dimension $N_{\rm c}\approx 4$. The arguments in favor are via {\em proof by consistency or contradiction} \cite{LassigKinzelbach1997,Fogedby2005,NewmanKallabis1996}, $N_{\rm c}=4$, or  mode-coupling: $N_{\rm c}=4$  \cite{Bhattacharjee1998,ColaioriMoore2001,CanetMoore2007},  $N_{\rm c}=3.7$ or $N_{\rm c}=4.3$, depending on the UV regularization \cite{KatzavSchwartz2002}. If these are wrong, presumably  one of the underlying assumptions fails. 
 
 The arguments against are  mostly from numerical simulations, either directly on the KPZ in its RSOS representation \cite{MarinariPagnaniParisi2000,MarinariPagnaniParisiRacz2002,AlvesOliveiraFerreira2014,GomesPennaOliveira2019,Ala-Nissila1998a,Ala-Nissila1998} or on the directed polymer \cite{SchwartzPerlsman2012}. The criticism voiced is that  they  are not  in the asymptotic regime, or  break the rotational symmetry of the KPZ equation. 
 Mode-coupling solutions without an upper critical dimension have been proposed \cite{Tu1994}, as well as approximate RG schemes \cite{CastellanoMarsiliPietronero1998}, or NPRG \cite{CanetChateDelamotteWschebor2010}.
 
 The issue is far from settled, and only few distinct arguments can be found: FRG for the directed polymer favors a critical dimension   $N_{\rm c}\approx 2.5$ \cite{LeDoussalWiese2005a}, approximately also found in \cite{BouchaudCates1993}. 
 While the work by \cite{Laessig1995,Wiese1998a} indicates the necessity for an   UV-cutoff in dimension $N=4$ (see above), an additional scale   may   appear at all even  $N$, i.e.\ $N=6$, $8$, etc.\ \cite{DobrinevskiPhD}. Closure relations in CFT lead to simple fractions for the critical exponents \cite{Lassig1998,Laessig1998}, not favored by numerical simulations. 
From the newer developments, let us mention the mapping to $d$-mer diffusion  \cite{OdorLiedkeHeinig2010}, which seems to give rather precise numerical estimates, see table \ref{KPZ-tab}.
Let us conclude with a quote from the recent review \cite{Halpin-HealyTakeuchi2015}: The ``equation proposed nearly three decades ago by Kardar, Parisi and Zhang continues to inspire, intrigue and confound its many admirers.''

\begin{table}[h]
\setlength{\tabcolsep}{3.1pt}
\begin{tabular}{|c|c|c|c|c|}
\hline
$d$   & $\zeta_{\KPZ}$($d$-mer) & $\frac{\zeta_{\KPZ}}{z_{\KPZ}}$($d$-mer)\rule[-1.5ex]{0mm}{4ex} & $z_{\KPZ}$($d$-mer) & $\zeta_{\KPZ}$(RSOS)\\
\hline
1     & $1/2$		&  $1/3$ & $3/2$& $1/2$ \\
2     & $0.395(5)$ & $0.245(5)$ & $1.58(10) $& $0.393(3)$\\
3     & $0.29(1)$  & $0.184(5)$ & $1.60(10)$ &   $\!0.3135(15)\!$ \\
4     & $0.245(5)$ & $0.15(1) $ & $1.91(10)$ &  $0.2537(8)$\\
5     & $0.22(1)$  &$ 0.115(5) $& $1.95(15)$ & $0.205(15)$ \\
\hline
\end{tabular}
\caption{Growth exponent estimates of the $d$-mer model  (d-mer) \cite{OdorLiedkeHeinig2010}; the results in $d=1$ are exact. The last column represents the results for the RSOS model, always taking the most recent and precise values \cite{Ala-Nissila1998a,Ala-Nissila1998,MarinariPagnaniParisi2000,MarinariPagnaniParisiRacz2002,PagnaniParisi2013,AlvesOliveiraFerreira2014}. At least some of the error bars seem overly optimistic.}
\label{KPZ-tab}
\end{table}

}

\subsection*{Quenched KPZ}
The quenched KPZ equation was discussed  in section \ref{s:qKPZ}.

\begin{figure*}[t]
\parbox{0.485\textwidth}{\vspace{6mm} \Fig{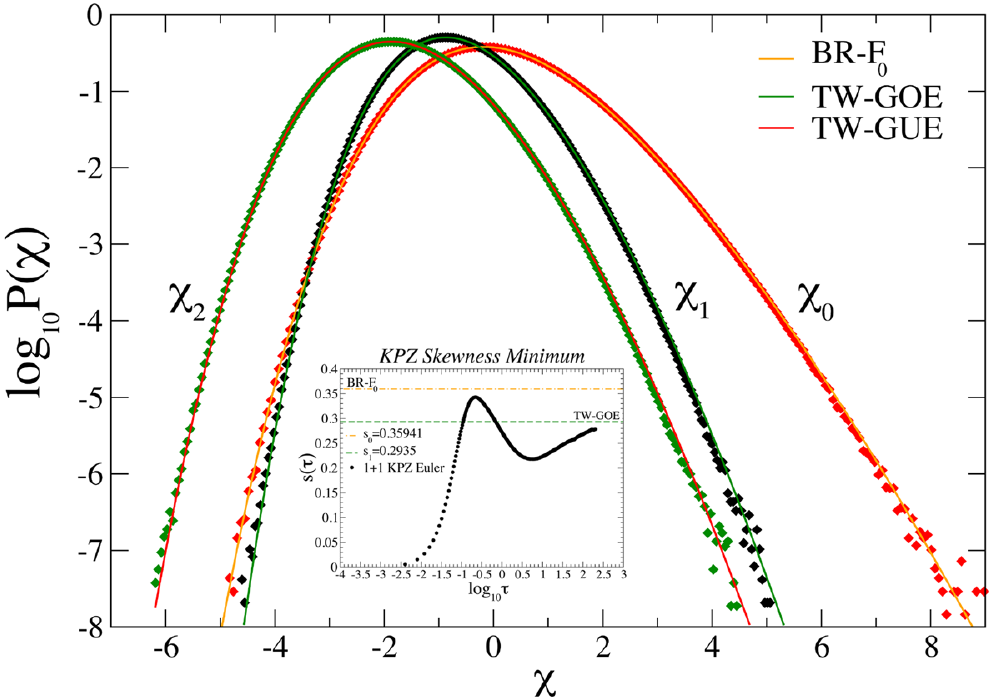}}\hfill\parbox{0.485\textwidth}{\Fig{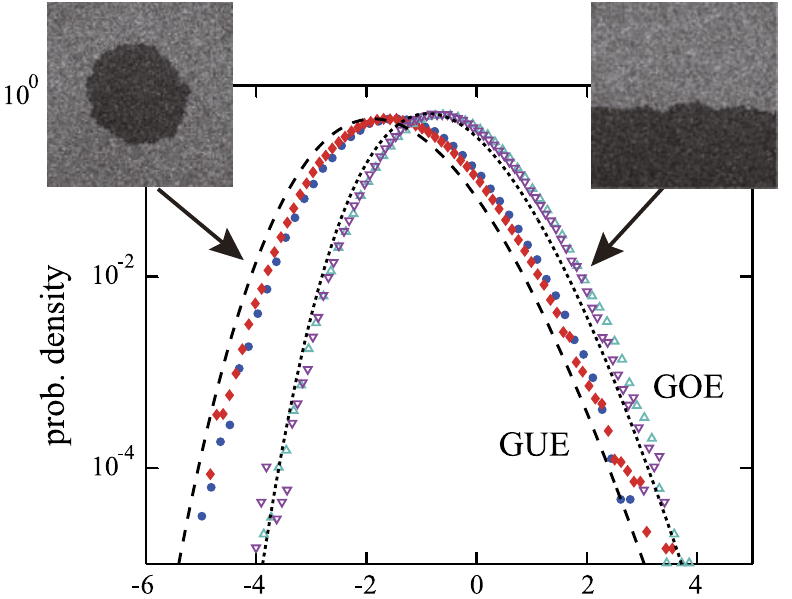}\vspace*{-2mm} \centerline{$~~~~~~~\chi$}}
\caption{Left: Numerical verification of the universal distributions for KPZ in one dimension as explained in the text. Figure from \cite{Halpin-HealyTakeuchi2015}. Right: experimental verification in Ref.~\cite{TakeuchiSano2012}. The blue circles and red diamonds display the histograms for the circular interfaces at $t = 10$s and $30$s, respectively, while the turquoise up-triangles and purple down-triangles are for the flat interfaces at $t = 20 $s and $60 $s, respectively.}
\label{f:KPZ-TW}
\end{figure*}
\subsection{The KPZ equation in  dimension  $d=1$}
The KPZ equation \eq{KPZ} is formally a Langevin equation. 
The corresponding Fokker-Planck equation for the evolution of its measure $P_t[h]$, derived in \Eq{forwardFP3},   reads
\bea\label{FP-KPZ}
&&\partial_t P_t[h] = D  \int_u \,  \frac{\delta^2}{\delta h(u)^2} P_t[h] \\
&& - \int_u \,  \frac{\delta }{\delta h(u)} \Big( \nu \nabla^{2} h(u) + \frac{\lambda}2 \big[ \nabla h(u)\big]^{2}  P_t[h] \Big) \punkt \nn
\eea
At least for $\lambda=0$, a steady-state solution can be found by asking that 
\be
D\frac{\delta}{\delta h(u)} P_t[h]  =  \nu \nabla^{2} h(u) P_t[h]. 
\ee
This is solved by\footnote{Note that some authors \cite{Krug1997} use a different normalization for \Eq{KPZ-noise} $2D_{\rm here} = D_{\rm there}$, reflected in the  invariant measure.}
\be\label{KPZ-eq-measure}
P_t[h] = {\ca N}  \exp \left(-\frac{\nu}{2D} \int_u \big[ \nabla h(u)\big]^2\right).
\ee
Unsurprisingly, this is the measure for  a diffusing elastic string. 
What are the additional terms for $\lambda \neq 0$?
Inserting the measure \eq{KPZ-eq-measure} into \Eq{FP-KPZ} yields 
\bea\label{FP-KPZ2}
&&\partial_t P_t[h] = 
 - \int_u \,  \frac{\delta }{\delta h(u)} \Big(  \frac{\lambda}2 \big[ \nabla h(u,t)\big]^{2}  P_t[h] \Big)  \nn\\
&& = \frac{\lambda \nu}{2D}\int_u \, \big[ \nabla h(u)\big]^{2}  \nabla^2 h(u) P_t[h].
\eea
While written in   continuous notation, the calculation should be made on the discretized version, with proper symmetrization of the $[\nabla h]^2$ term. To go to the second line,  we dropped the direct derivative of the latter, as it integrates to 0. 
In dimension $N=1$, the integrand is a total derivative, thus integrates to zero. In higher dimensions, this is not the case. The simplest explicit counterexample for periodic boundary conditions ($L=2\pi$) in dimension $N=2$ we found is $h(u_1,u_2)=[a + \cos (u_1)][b+\cos (u_2)]$, for which the last integral in \Eq{FP-KPZ2} evaluates to $- a b L^2$.

The measure \eq{KPZ-eq-measure} implies that equal-time correlation functions of the nonlinear ($\lambda>0$) theory are given by those of the linear theory ($\lambda=0$), first in Fourier and then in real space, 
\bea
&&\big< \tilde h(q) \tilde h(q') \big> = 2\pi \delta (q+q') \frac{D q^2}{\nu } \\
&&  \Leftrightarrow \quad   \big<  [h(x) - h(x')]^2 \big> = \frac D \nu |x-x'|^{2 \zeta_{\rm KPZ}}, \\
&&\zeta_{\rm KPZ}^{d=1}=\frac12.
\label{zeta-KPZ-d=1}
\eea
Moreover, the measure \eq{KPZ-eq-measure} is Gaussian. This can   be viewed as due to a fluctuation-dissipation theorem (FDT) \cite{LvovLebedevPatonProcaccia1993}.
\Eq{Galileo-exp-relation} further implies 
\be z_{\rm KPZ}^{d=1}=\frac 3 2.
\ee
As a consequence, 
\be
\zeta_{\rm RB}^{d=1} = \frac 1{z_{\rm KPZ}^{d=1}} =\frac 2 3
\ee
is the roughness exponent of a directed polymer in $1+1$ dimensions, and the roughness of  domain walls in   dirty 2d magnets.
Their energy-fluctuation exponent is 
\be
\theta = d-2 +2 \zeta  =\frac 1 3. 
\ee
This can  also be obtained via Bethe ansatz \cite{Kardar1987,MedinaHwaKardarZhang1989}.

\subsection{KPZ, polynuclear growth,  Tracey-Widom and Baik-Rains distributions}

Since the introduction of the KPZ equation in  1986 \cite{KPZ}, much progress has been made in one dimension. This  started in   2000 with the groundbreaking work by Pr\"ahofer and Spohn \cite{PraehoferSpohn2000,PraehoferSpohn2000a,PrahoferSpohn2002}  introducing the polynuclear growth model (PNG): One starts from a flat configuration. Steps of vanishing size are deposited as a Poisson process upon the already constructed surface. Steps then  grow with unit velocity at both ends. When   steps meet, they merge. Heuristically it seems clear that this process belongs to the KPZ universality class, similar to its discrete cousin, the RSOS model. 
Independently, Johansson introduced the single-step (SS) model \cite{Johansson2001}, relating surface growth to the combinatorial problem of finding the longest increasing subsequence in a random permutation \cite{BaikDeiftJohansson1999} and random matrix theory \cite{Johansson2000},   relating  to older work in this domain \cite{BaerBrock1968}.
The key observables   are constructed from the spatially averaged mean height $h(t)$, which for large times is assumed to grow with velocity $v_\infty$,  
$
\lim_{t\to \infty } \left < h(t) \right> - v_\infty t = \mbox{const}. 
$
As  the mean height is proportional to the free energy of a directed polymer, see \Eq{F-h},   results for the height fluctuation have an immediate interpretation in terms of free-energy fluctuations of a directed polymer. 

Key observables  are \bea
&&\chi_0 =\frac{ h(t_0+t)-h(t_0) - v_\infty t}{\left(\frac{D^2 \lambda}{2\nu^2} t\right)^{1/3}},   \\
&& \chi_1 = \frac{ h(t)-v_\infty t}{\left(\frac{D^2 \lambda}{2\nu^2} t\right)^{1/3}}, \quad \mbox{(flat initial conditions)},\\
&& \chi_2 = \frac{ h(t)-v_\infty t}{\left(\frac{D^2 \lambda}{2\nu^2} t\right)^{1/3}} , \quad \mbox{(circular initial conditions)}.
\eea
Each of these observables has a unique universal distribution:
\begin{itemize}
\item[$\chi_0$] is distributed according to the Baik-Rains $F_0$ distribution,

\item[$\chi_1$] is distributed according to the Tracy-Widom (TW) Gaussian Orthogonal Ensemble (GOE) distribution,

\item[$\chi_2$] is distributed according to the Tracy-Widom Gaussian Unitary Ensemble (GUE) distribution.
\end{itemize}
The reader wishing to test these laws himself can find a Mathematica implementation online \cite{Dobrinevski2014}. 

The
 Bethe-ansatz for the directed polymer was  introduced in 1987  by Kardar  to obtain the roughness exponent of a directed polymer, $\zeta=2/3$ and  $\theta=1/3$. 
  A revival started in 2010, when physicists succeeded \cite{CalabreseLeDoussalRosso2010,Dotsenko2010,Dotsenko2010b,CalabreseLeDoussal2011,LeDoussalCalabrese2012} to first reproduce the work of \cite{PraehoferSpohn2000,PraehoferSpohn2000a,PrahoferSpohn2002} via the Bethe ansatz, and then extend it   to other situations \cite{GueudreLeDoussal2012}.
At the same time, a new generation of mathematicians   developed complementary tools \cite{Corwin2012,AmirCorwinQuastel2011,BorodinCorwin2014,ImamuraSasamoto2012}, joining the work of Spohn and collaborators \cite{SasamotoSpohn2010,SasamotoSpohn2010b}.

Extraordinarily,  Takeuchi and Sano \cite{TakeuchiSano2010,TakeuchiSano2012} succeeded to extract the universal distributions from a turbulent liquid-crystal experiment. A snapshot is shown in Fig.~\ref{f:KPZ-growth}, and the two Tracey-Widom distributions for circular and flat initial conditions in Fig.~\ref{f:KPZ-TW}.

Let us conclude by mentioning   
pedagogical presentations  \cite{KriecherbauerKrug2010,Quastel2011,QuastelSpohn2015}, as well as  
attempts to port this at least numerically to higher dimensions 
\cite{Halpin-Healy2012,Halpin-Healy2013}.

\subsection{Models in the KPZ universality class, and experimental realizations}
In all dimensions:
\begin{itemize}
\item KPZ \cite{KPZ},
\item polynuclear growth (PNG) \cite{PraehoferSpohn2000,PraehoferSpohn2000a,PrahoferSpohn2002},
\item Rigid-solid-on-solid models (RSOS) \cite{MarinariPagnaniParisi2000,MarinariPagnaniParisiRacz2002,AlvesOliveiraFerreira2014,GomesPennaOliveira2019,Ala-Nissila1998a,Ala-Nissila1998}.
\item directed polymer in quenched disorder (section \ref{s:KPZ as a directed polymer}).
\end{itemize}
In one dimension: 
\begin{itemize}
\item longest growing subsequence in a random permutation \cite{BaikDeiftJohansson1999},
\item the asymmetric simple exclusion process (ASEP)  \cite{Spitzer1970,Krug1991,Derrida1998}.
\end{itemize}
Experimental realizations (one dimension only): 
\begin{itemize}
\item slow combustion of paper \cite{MyllysMaunukselaAlavaAla-NissilaMerikoskiTimonen2001,MiettinenMyllysMerikoskiTimonen2005},
\item   turbulent liquid crystals  \cite{TakeuchiSano2010,TakeuchiSano2012}, 
\item particle deposition (with crossover to qKPZ) \cite{DiasYunkerYodhAraujoTelo-da-Gama2018},
\item \new bacterial growth \cite{HallatschekHersenRamanathanNelson2007},
\item chemical reaction fronts \cite{AtisDubeySalinTalonLeDoussalWiese2014}.
\end{itemize}

\subsection{From Burgers' turbulence to Navier-Stokes turbulence?}

In Burgers' turbulence velocity profiles are locally linear (see e.g.~the right of Fig.~\ref{f:shocks}), interrupted by jumps.  Phenomenologically it is similar to the force field in disordered systems. This implies that, if non-vanishing, at  small distances  
\be
\left< [ v(u,t) - v(u',t)]^{n}\right> \simeq \ca A_n |u-u'|^{\zeta_n},
\ee 
and for Burger $  \zeta_n = 1$, 
independent of $n$. 
In Navier-Stokes turbulence, the exponent $\zeta_3 = 1$, as predicted by Kolmogorov in 1941 \cite{Kolmogorov1941}.
Other moments obey 
\be\label{569}
\zeta_n = \frac n 3 + \delta \zeta_n,
\ee
where $\delta \zeta_n$ is small. Calculating $\delta \zeta_n$ analytically is the outstanding   problem of turbulence research. 
One can try to use FRG for this problem \cite{FedorenkoLeDoussalWiese2012}, but something crucial is missing: While FRG correctly deals with shocks, the weaker singularities responsible for \Eq{569} are not captured by FRG. (It may work in dimension $d=2$, though \cite{FedorenkoLeDoussalWiese2012}.)
It is possible that the FRG fixed point which typically has a cusp, and which usually is implemented for the second cumulant, instead  applies to the third cumulant, as $\zeta_3=1$. How to implement this idea remains an open problem.

\section{Links between loop-erased random walks, CDWs, sandpiles, and scalar field theories}
\label{s:links}
  
\subsection{Supermathematics}\label{a:susy}
{\em Supermathematics}, introduced in \cite{Berezin1965} and nicely reviewed in \cite{Wegner2016} is an alternative  way to average over disorder. In this technique,  additional fermionic or {\em Grassmannian} degrees of freedom are introduced to normalize  the partition function to  $ Z = 1$, even before averaging over   disorder. 
We start by reviewing basic properties of Grassmann variables, before using them for disorder averages. 
Most of the material is standard, and the reader familiar with it, or wishing to advance may safely do so. 

Two points should  be retained:  While {\em supermathematics} is usually referred to as {\em supersymmetry technique}, this is a misnomer  as supersymmetry is broken at the Larkin scale, i.e.\ when a cusp appears. The name  is due to  historical reasons, stemming from a time when  people   believed that supersymmetry is not broken. To avoid this confusion, we prefer the term {\em supermathematics}.

Braking of supersymmetry, and the cusp,  can be found in this framework, as long as one considers at least two physically distinct copies. The technical reason   is that to assess the $n$-th cumulant of a distribution, one needs  at least $n$ distinct copies. Even when supposing the disorder to be Gaussian distributed, the variance, i.e.\ the second cumulant needs to be assessed, thus two  distinct replicas. 

Apart from these more formal considerations, the technique has proven powerful in the mapping of charge-density waves onto a  
$\phi^4$-type theory (section \ref{s:CDWs and their mapping onto phi4-theory}).

\subsection{Basic rules for   Grassmann variables}
\label{s:Grassmann}
Grassmann variables are anticommuting variables which allow one to write a  path-integral for fermions, in the same way as one does for bosons. 
There are only few rules to remember. If $\chi$ and $\psi$ are Grassmann variables, then 
\be
\hl {\chi \,\psi = - \psi\, \chi }\punkt
\ee
This immediately implies that 
\be
\hl {\chi^2 =0}\punkt
\ee
One introduces derivatives, and integrals through the same formula, known as Berezin integral \cite{Berezin1965},
\be
\hl { \int \rmd \chi\,  \chi \equiv \frac{\rmd}{\rmd \chi} \chi =1}\punkt
\ee
One checks that they  satisfy the usual properties associated to ``normal'' derivatives, and integrals. 
An important property is   
\be\label{829}
\int\rmd \bar \chi \,\rmd \chi\, \rme^{-a \bar \chi \chi} = a\punkt 
\ee
This is easily proven upon Taylor expansion. The minus sign in the exponential cancels   the minus sign obtained when exchanging $\bar \chi$ with $\chi$, which is necessary since an integral or derivative is defined to act directly on the variable following it. 
\Eq{829} can   be generalized to integrals over an $n$-component pair of vectors $\vec {\bar\chi}$, and $\vec \chi$:
\bea\label{B5}
\hl{ \int \rmd \vec{ \bar\chi}\,\rmd  \vec \chi \, \rme^{-\vec {\bar \chi} {\mathbb A}\vec \chi}:= \prod_{a=1}^n \int\rmd\bar  \chi^a \, \rmd\chi^a\, \rme^{-\vec {\bar \chi} {\mathbb A}\vec \chi} = \det ({\mathbb A})}\punkt\nn\\
\eea
It is proven by changing coordinates s.t.\ $\mathbb A$ becomes diagonal. 
For comparison we give the corresponding formula for normal (bosonic) fields, noting $\phi^a := \phi^a_x+i \phi^a_y$, $\tilde  \phi^a := \phi^a_x-i \phi^a_y$,
\bea\label{B6}
\int \rmd \vec{ \tilde\phi}\,\rmd  \vec \phi \, \rme^{-\vec {\tilde \phi} {\mathbb A}\vec \phi}:= \prod_{a=1}^n \int \frac{\rmd \phi^a_x   \, \rmd\phi^a_y}{\pi}\, \rme^{-\vec {\tilde \phi} {\mathbb A}\vec \phi} = \frac1 {\det ({\mathbb A})}\punkt\nn\\
\eea
When combining normal and Grassmannian integrals over the same number of variables into a product, the contributions from  \Eqs{B5} and \eq{B6}   cancel. This will be used below.

\subsection{Disorder averages with bosons and fermions}\label{a:susy-RM} 

The above formulas permit a different approach to  
average over disorder. For concreteness, define
\bea
{\cal H}[u ,V] =  \int_{x} \Big\{ &\half  \left[ \nabla u  (x)
\right]^{2} + { \frac{m^2}2} u(x)^2 + {\cal U}\big(u(x)\big)  \nn\\
&+V \big(x,u (x)\big)\Big\}
\punkt
\eea
The  disorder-average of an observable $\ca O$ is {\em defined} as
\begin{equation}\label{su2}
\overline {{\cal O}[u]} := \overline { \frac{ \ds \int \prod_{a=1}^r {\cal D}[{u_{a}}
] {{\cal O}[u_1]}  \rme^{-\frac{1}{T}{\cal H} [{u_{a}}, {V}]}}{\ds \int \prod_{a=1}^r {\cal
D}[{u_{a} } ]\rme^{-\frac{1}{T}{\cal H} [{u_{a} } ,{V} ] }} }
\punkt
\end{equation}
The function ${\cal U}(u)$ is an arbitrary potential, e.g.\ the non-linearity in $\phi^4$-theory, $\ca U(u)=g \,u^4$. 
The random potential $V(x,u)$  is the same one used in section \ref{model},   with correlations  given by \Eq{DOcorrelR}. Its average is indicated by the overline. 
We remind that the difficulty in evaluating \eq{su2} comes from the denominator. 
The replica trick used in section \ref{s:replicas} allowed us to set $r=0$, effectively  discarding the denominator. 
Here we follow a different strategy. 

In the limit of $T\to 0$ only configurations which minimize the energy
survive.  These configurations satisfy $\frac{\delta{\cal H}[{u_{a} },{V} ] }{\delta {u_{a} } (x) }=0$, and  we want to insert a
$\delta$-distribution enforcing this condition into the path-integral. This has to be accompanied
by a factor of $\det\! \left[\frac{\delta^{2}{\cal H}[{u_{a} },V ] }{\delta {u_{a} } (x)\delta {u_{a} } (y) }\right]$, such
that the path integral   is normalized to 1. 
The latter can be achieved by an additional integral over Grassmann variables, i.e.\ fermionic degrees of freedom, using that
\bea
\det\! \left(\frac{\delta^{2}{\cal H}[{u },V ] }{\delta {u} (x)\delta {u } (y) }\right)  \\
= \int{\cal D}[{\bar \psi_{a} }] {\cal D}[{\psi}_{a}] \exp\left( -\int_x \bar \psi(x)  \frac{\delta^{2}{\cal H}[{u },V ] }{\delta {u} (x)\delta {u } (y) } \psi(x) \right) . \nn
\eea
This
allows us to write the disorder average of any observable ${\cal O}[u]$ as 
\bea\label{su5}
\overline {{\cal O}[u]}  = \int\prod_{a=1}^r {\cal D}[{\tilde u_{a} }] {\cal D}[{ 
u_{a} }] {\cal D}[{\bar \psi_{a} }] {\cal D}[{\psi}_{a}]\, {{\cal O}[u]} \nn\\
\times\overline{\,\exp\!\left[-{ \int_{x}\tilde u_{a} (x)\frac{\delta{\cal H}[{u_{a} }]} {\delta {u_{a} } (x) }+\bar \psi_{a}(x)\frac{\delta^{2}
{\cal H}[{u_{a} } ] }{\delta {u_{a} } (x)\delta {u_{a} }
(y) } \psi_{a} (y) } \right]} \rule{0mm}{4ex} \punkt\nn\\
\eea
This method was first introduced in \cite{ParisiSourlas1979,ParisiSourlas1982}. 
An alternative derivation and   insight are offered by Cardy \cite{Cardy1983,Cardy1985,CardyMcKane1985}, see also \cite{KavirajRychkovTrevisani2019,KavirajRychkovTrevisani2020}.

Averaging over disorder  with the force-force correlator $\Delta
(u):=-R'' (u)$ yields
\begin{eqnarray}\label{su6a}
\overline{{\cal O}[u]}=\int\prod_{a} {\cal D}[{u_{a} }] {\cal
D}[{\tilde u_{a} }] {\cal D}[{\bar \psi_{a} }] {\cal D}[{\psi}_{a}]
\nn\\
~~~~~~~~~~~~~~~~~~~~ \exp \left(-{\cal S}[u_a,\tilde u_{a},\bar \psi_{a}, \psi_{a} ]
\right),
  \nonumber \\
\label{su6b} {\cal S}[\tilde u_a, u_{a},\bar \psi_{a}, \psi_{a}]  \nn\\
= \int_{x}  \sum_{a=1}^r \Big\{\tilde u_{a} (x) \Big[ (-\nabla^{2}{+}m^2) u_{a} (x) +{\cal U}'\big(u_a(x)\big)\Big] 
\nn\\ \qquad\quad~~ + \bar \psi_{a} (x)
\Big[{-}\nabla^{2}{+}m^2{+}{\cal U}''\big(u_a(x) \big)\Big]\psi_{a} (x) \Big\}\nonumber \\
- \sum_{a,b=1}^r \Big[ \half \tilde u_{a} (x)\Delta \big(u_{a} (x)-u_{b}
(x)\big)\tilde u_{b} (x)\nn\\
 \quad- \tilde u_{a} (x)
\Delta' \big(u_{a} (x)-u_{b} (x)\big) \bar \psi_{b} (x)\psi_{b} (x)\nonumber \\
 \quad   -\half \bar \psi_{a} (x)\psi _{a} (x)\Delta''
\big(u_{a} (x)-u_{b} (x)\big)\bar \psi_{b} (x)\psi_{b} (x)  \Big]
\punkt
\end{eqnarray}
We first analyze the special case of $n=1$. Suppose that $\Delta (u)$ is even and analytic
to start with, then   few terms survive from Eq.~(\ref{su6b}),
\bea\label{reallySusy}
{\cal S}_{\mathrm{Susy}}[u,\tilde u,\bar\psi,\psi ] =\nn\\
 \int_{x}\Big\{ \tilde u
(x) \Big[ ({-}\nabla^{2}{+}m^2) u  (x) +{\cal U}'\big(u(x) \big)\Big] \nn\\
 + \bar \psi (x)
\Big[{-}\nabla^{2}{+}m^2{+}{\cal U}''\big(u (x) \big)\Big]\psi (x) 
 - \half \tilde u (x)\Delta (0)\tilde
u (x)\Big\} \punkt\nn\\
\eea
(We have used that $\bar \psi_{a}^{2}=\psi_{a}^{2}=0$ to eliminate the
4-fermion-term.) 
A particularly simple case are random manifolds, for which  ${\cal U}(u) =0$. 
Then {\em bosons} $\tilde u$ and $ u$,  and {\em fermions} $\bar \psi$
and $\psi$ do not interact, all expectation values are
  Gaussian, perturbation theory gives \Eq{332}, and dimensional reduction holds. 
When ${\cal U}(u)\neq 0$, things are more complicated, but as we will see in the next section,   dimensional reduction   still holds, at least formally. 

The reason    is that the  action \eq{reallySusy} possesses an apparent supersymmetry,  made manifest by grouping all fields   into a (bosonic) superfield, 
\begin{equation}\label{superfielddef}
\Phi (x,\bar \Theta ,\Theta) := u (x)+ \bar \Theta \psi (x)+\bar \psi (x)
\Theta -\bar \Theta \Theta \tilde u (x)
\punkt
\end{equation}
Both $\bar \Theta$ and  $\Theta$ are Grassmann numbers. 
The action (\ref{reallySusy}) can then be written with the
super Laplacian $ \Delta_{\rm s}$ as 
\bea\label{S-Susy}
{\cal S}_{\mathrm{Susy}}= \int \rmd \bar\Theta \rmd  \Theta\int_{x} {  \frac12} \Phi
(x,\bar \Theta ,\Theta) (- \Delta_{\rm s}+m^2) \Phi (x,\bar \Theta ,\Theta)\nn\\
\hphantom{{\cal S}_{\mathrm{Susy}}= \int \rmd \bar\Theta \rmd  \Theta\int_{x}}+ {\cal U}(\Phi(x,\bar \Theta,\Theta))\komma 
\\
\label{superLaplacian}
  \Delta_{\rm s} := \nabla^{2}+ \Delta (0) \frac{\partial}{\partial
 \Theta}\frac{\partial}{\partial\bar \Theta}\punkt
\eea
As we will see   in   section \ref{a:dim-red}, the action is invariant when applying the two supergenerators 
\bea
Q := x \frac{\partial}{\partial \Theta}-\frac{2}{\Delta (0)} \bar
\Theta \nabla \komma  \qquad \bar Q:=x \frac{\partial}{\partial \bar
\Theta}+\frac{2}{\Delta (0)} \Theta \nabla\komma \nn\\
 \left\{Q,\bar Q \right\} = 0\punkt
\eea
This is sufficient to ``prove'' dimensional reduction.

\subsection{Renormalization of the disorder}
\label{s:susy:renormalization-of-disorder}
For more than $r=1$ replicas, the theory is richer, 
 and we can   
recover the renormalization of $\Delta (u)$ itself. To simplify matters, set ${\cal U}(u)=0$, and  
write 
\begin{eqnarray}\label{s1}
{\cal S}[\tilde u_{a},  u_{a},\bar\psi_{a},\psi_{a}] =\sum_{a} \int_{x} \Big[ \tilde u_{a} (x)  ({-}\nabla^{2}+m^2) u_{a} (x)   \nn\\
~~~~~~~~~~ + \bar \psi_{a} (x)
(-\nabla^{2}+m^2)\psi_{a} (x)- \half \tilde u_a (x)\Delta (0)\tilde
u_a (x) \Big] \nonumber \\
-\sum_{a\neq b}\int_{x}\Big[\half\tilde u_{a} (x)
\Delta \big(u_{a} (x)-u_{b}
(x)\big)\tilde u_{b} (x) 
\nn\\
 \qquad \quad- \tilde u_{a} (x)
\Delta' \big(u_{a} (x)-u_{b} (x)\big) \bar \psi_{b} (x)\psi_{b} (x) \nonumber \\
 \qquad \quad  - \half \bar \psi_{a} (x)\psi _{a} (x)\Delta''
\big(u_{a} (x)-u_{b} (x)\big)\bar \psi_{b} (x)\psi_{b} (x) \Big]
\punkt \nn\\
\end{eqnarray}
Corrections to $\Delta (u)$ are  
constructed by remarking that the interaction term quadratic in
$\tilde u$ is almost identical to the treatment of the dynamics in the
static limit (i.e.\ after integration over times). The diagrams in question are
\bea\label{deltaDeltafromSusy}
{\parbox{2.1cm}{{\begin{tikzpicture}
\coordinate (x1t1) at  (0,0) ; 
\coordinate (x1t2) at  (0,.5) ; 
\coordinate (x2t3) at  (1.5,0) ; 
\coordinate (x2t4) at  (1.5,0.5) ; 
\fill (x1t1) circle (2pt);
\fill (x1t2) circle (2pt);
\fill (x2t3) circle (2pt);
\fill (x2t4) circle (2pt);
\draw [directed] (x1t1) -- (x2t3);
\draw [directed] (x1t2) -- (x2t4);
\draw [dashed,thick] (x1t1) -- (x1t2);
\draw [dashed,thick] (x2t3) -- (x2t4);
\draw [enddirected]  (x2t3)--(2,0);
\draw [enddirected]  (x2t4)--(2,0.5);
\end{tikzpicture}}}} ~+~  {\parbox{2.5cm}{{\begin{tikzpicture}
\coordinate (x1t1) at  (0,0) ; 
\coordinate (x1t2) at  (0,.5) ; 
\coordinate (x2t3) at  (1.5,0) ; 
\coordinate (x2t4) at  (1.5,0.5) ; 
\fill (x1t1) circle (2pt);
\fill (x1t2) circle (2pt);
\fill (x2t3) circle (2pt);
\fill (x2t4) circle (2pt);
\draw [directed]  (x2t3)--(x1t1) ;
\draw [directed] (x1t2) -- (x2t4);
\draw [dashed,thick] (x1t1) -- (x1t2);
\draw [dashed,thick] (x2t3) -- (x2t4);
\draw [enddirected]  (x1t1)--(-.5,0);
\draw [enddirected]  (x2t4)--(2,0.5);
\end{tikzpicture}}}}
~+ ~ {\parbox{2.1cm}{{\begin{tikzpicture}
\coordinate (x1t1) at  (0,0) ; 
\coordinate (x1t2) at  (0,.5) ; 
\coordinate (x2t3) at  (1.5,0) ; 
\coordinate (x2t4) at  (1.5,0.5) ; 
\fill (x1t1) circle (2pt);
\fill (x1t2) circle (2pt);
\fill (x2t3) circle (2pt);
\fill (x2t4) circle (2pt);
\draw [directed] (x1t1) -- (x2t4);
\draw [directed] (x1t2) -- (x2t4);
\draw [dashed,thick] (x1t1) -- (x1t2);
\draw [dashed,thick] (x2t3) -- (x2t4);
\draw [enddirected]  (x2t3)--(2,0);
\draw [enddirected]  (x2t4)--(2,0.5);
\end{tikzpicture}}}} ~\nn\\
+ ~  {\parbox{2.5cm}{{\begin{tikzpicture}
\coordinate (x1t1) at  (0,0) ; 
\coordinate (x1t2) at  (0,.5) ; 
\coordinate (x2t3) at  (1.5,0) ; 
\coordinate (x2t4) at  (1.5,0.5) ; 
\fill (x1t1) circle (2pt);
\fill (x1t2) circle (2pt);
\fill (x2t3) circle (2pt);
\fill (x2t4) circle (2pt);
\draw [directed]  (x2t4)--(x1t1) ;
\draw [directed] (x1t2) -- (x2t4);
\draw [dashed,thick] (x1t1) -- (x1t2);
\draw [dashed,thick] (x2t3) -- (x2t4);
\draw [enddirected]  (x1t1)--(-.5,0);
\draw [enddirected]  (x2t3)--(2,0.);
\end{tikzpicture}}}}\komma 
\eea
where an arrow indicates the correlation-function,
$\mbox{${_{x}}-\!\!\!\!\!\rightarrow\!\!\!\!-\!\!\!\!-\!\!\!\!-{_{y}}\,$}$ $ =
\left< \tilde u (x)u (y) \right>=C (x-y)$. This leads to (in the order given
above)
\bea\label{s2}
\delta \Delta (u) &=& \left[ - \Delta (u) \Delta'' (u)  - \Delta' (u)^{2} +
\Delta'' (u)\Delta (0) \right] \nn\\
&& \times \int_{x-y}C (x-y)^{2}\punkt
\eea
The last term of \Eq{deltaDeltafromSusy} is odd, and  vanishes. \Eq{s2} is equal to  the  results of \Eqs{7.16}--\eq{7.24}.

A non-trivial ingredient is the cancellation of the acausal loop \eq{7.25} in
the dynamics, equivalent to the  3-replica term in the statics:
\begin{equation}\label{sloop}
{\parbox{2.5cm}{{\begin{tikzpicture}
\coordinate (x1t1) at  (0,0) ; 
\coordinate (x1t2) at  (0,.5) ; 
\coordinate (x2t3) at  (1.5,0) ; 
\coordinate (x2t4) at  (1.5,0.5) ; 
\fill (x1t1) circle (2pt);
\fill (x1t2) circle (2pt);
\fill (x2t3) circle (2pt);
\fill (x2t4) circle (2pt);
\draw [directed] (x2t4) arc(60:120:1.5);
\draw [directed] (x1t2) arc(-120:-60:1.5);
\draw [dashed,thick] (x1t1) -- (x1t2);
\draw [dashed,thick] (x2t3) -- (x2t4);
\draw [enddirected]  (x1t1)--(-.5,0);
\draw [enddirected]  (x2t3)--(2,0.);
\end{tikzpicture}}}}
~+~{\parbox{2.5cm}{{\begin{tikzpicture}
\coordinate (x1t1) at  (0,0) ; 
\coordinate (x1t2) at  (0,.5) ; 
\coordinate (x2t3) at  (1.5,0) ; 
\coordinate (x2t4) at  (1.5,0.5) ; 
\fill (x1t1) circle (2pt);
\fill (x1t2) circle (2pt);
\fill (x2t3) circle (2pt);
\fill (x2t4) circle (2pt);
\draw [snake it] (x2t4) arc(60:120:1.5);
\draw [directed] (0.7,0.83)--(0.69,0.83);
\draw [snake it] (x1t2) arc(-120:-60:1.5);
\draw [directed] (0.8,0.173)--(0.81,0.173);
\draw [dashed,thick] (x1t1) -- (x1t2);
\draw [dashed,thick] (x2t3) -- (x2t4);
\draw [enddirected]  (x1t1)--(-.5,0);
\draw [enddirected]  (x2t3)--(2,0.);
\end{tikzpicture}}}}
=0.
\end{equation}
The first diagram comes from the contraction of two terms proportional to $\tilde u_a \Delta (u_a-u_b)\tilde
u_b$. 
The second is obtained from contracting all fermions in two terms proportional to $\tilde u_{a}\Delta'
(u_{a}-u_{b})\bar \psi_{b}\psi_{b}$. 
Since the fermionic loop (oriented wiggly line in the second diagram)
contributes a factor of $-1$, both cancel.

One can treat the interacting theory completely in a superspace
formulation. The action is
\bea
{\cal S}[\Phi] = \sum_{a}\int\limits_{\bar\Theta,  \Theta }\int\limits_{x} \half  
\Phi_{a} (x,\bar \Theta ,\Theta) (- \Delta_{\rm s}{+}m^2) \Phi_{a} (x,\bar \Theta
,\Theta) \nn\\
~~~~~ -\half  \sum_{a\neq b} \int\limits_{x}\int\limits_{\bar \Theta ,\Theta} \int\limits_{
 \bar \Theta', \Theta'} R \big(\Phi_{a} (x,\bar \Theta ,\Theta)-\Phi_{b}
(x,\bar \Theta ',\Theta' )\big)
\punkt\nn\\
\eea
``Non-locality'' in replica-space or in time is replaced by
``non-locality'' in superspace. 
Corrections to $R (u)$ all stem from {\em superdiagrams},
which result into bilocal interactions in superspace, not trilocal, or
higher. The latter find their equivalent in 3-local terms in
replica-space in the replica-formulation, and 3-local terms in time,
in the dynamic formulation. 

Supersymmetry is broken, once $\Delta (0)$ changes, i.e.\ at the
Larkin length. A seemingly ``effective
supersymmetry'', or ``scale-dependent supersymmetry'' appears, in
which the parameter $\Delta (0)$, which appears in the
Susy-transformation, changes with scale, according to the FRG flow
equations \eq{RG1loop} for $\Delta (u)$, continued to $u=0$.

\subsection{Supersymmetry and dimensional reduction}
\label{a:dim-red}
Let us study invariants of the action. Since total derivatives both in $x$ and $\theta$ or $\bar \theta$ vanish,  the crucial term
to focus on is the super-Laplacian. To simplify notations, we set
\be
\rho := \Delta(0)\punkt
\ee
By explicit inspection, we find that the two generators of
super-translations
\be \label{su13}
\hl{  Q := x\frac{\partial }{\partial \Theta } + \frac{2}{\rho }\bar
\Theta\nabla \komma  \qquad 
\bar Q := x\frac{\partial }{\partial \bar \Theta } - \frac{2}{\rho }
\Theta \nabla}
\ee
both commute with the Super-Laplacian, and anti-commute with each other, 
\begin{equation}\label{su14}
\left[ \Delta_{\rm s},Q \right] = \left[ \Delta_{\rm s},\bar Q \right] = 0 \komma  \qquad \left\{Q,\bar Q \right\} = 0\punkt
\end{equation}
The following combination is  invariant under the action of $Q$ and $\bar Q$, 
\begin{equation}\label{su15}
\bar Q \left(x^{2}+\frac{4}{\rho }\bar \Theta \Theta \right) = Q
\left(x^{2}+\frac{4}{\rho }\bar \Theta \Theta \right) = 0\punkt
\end{equation} Applying the Super-Laplacian \eq{superLaplacian} gives\footnote{This relation comes out incorrectly in \cite{ParisiSourlas1979}.}
\begin{equation}
\label{su16}
 \Delta_{\rm s}  \left(x^{2}+\frac{4}{\rho }\bar \Theta \Theta \right) = 2 (d-2)\punkt
\end{equation}
To obtain the super-propagator,   inverse of the
super-Laplacian plus mass term in \Eq{S-Susy}, we remark that
\bea\label{su16b}
\left(m^2- \nabla^2 - \rho \frac{\partial }{\partial \bar \Theta }\frac{\partial
}{\partial \Theta }\right)  \left(m^2- \nabla^2 + \rho \frac{\partial }{\partial \bar \Theta }\frac{\partial
}{\partial \Theta }\right)\nn\\
= (m^2-\nabla^2)^{2}\punkt
\eea
This implies that 
\begin{equation}\label{su17}
\left(m^2- \Delta_{\rm s} \right)^{-1} = \frac{\ds m^2-\nabla^2+{\rho
}\frac{\partial }{\partial \bar \Theta }\frac{\partial }{\partial
\Theta }}{ \displaystyle (m^2-\nabla^2)^{2}}
\punkt
\end{equation}
Therefore
\bea\label{su18}
C (x-x',\Theta -\Theta ',\bar \Theta -\bar \Theta ') \\
= \frac{\ds m^2-\nabla^2+{\rho
}\frac{\partial }{\partial \bar \Theta }\frac{\partial }{\partial
\Theta }}{ \displaystyle (m^2-\nabla^2)^{2}} \delta (x-x')\delta (\Theta
-\Theta ')\delta (\bar \Theta -\bar \Theta ')  \punkt \nn
\eea
The Grassmanian $\delta $-functions are defined as
\begin{equation}\label{su19}
\int \rmd \Theta \delta (\Theta-\Theta ' ) f (\Theta ) = f (\Theta ')\punkt
\end{equation}
By direct calculation one finds
\begin{equation}\label{su20}
\delta (\Theta -\Theta ') = \Theta '-\Theta = \int \rmd \bar \chi\, \rme^{\bar \chi (\Theta
'-\Theta )}
\punkt
\end{equation}
One can   transform (\ref{su18})   into a
representation in dual spaces of momentum ($k$-space) and
super-coordinates ($\chi $-space) as
\bea\label{su22}
C (k,\bar \chi ,\chi ) = \frac{m^2+k^{2}+{\rho }\bar \chi \chi
}{(m^2+k^{2})^{2}} \equiv \frac{1}{m^2+k^{2}+{\rho }\chi \bar \chi } \nn\\
\equiv \int_{0}^{\infty} \rmd s \, \rme^{-s (m^2+k^{2}+{\rho }\chi \bar \chi  )}
\punkt
\eea
The final proof of dimensional reduction is   performed with
this representation of the super-correlator\footnote{Note that one can also
work in position-space \cite{ParisiSourlas1979}. Then the super-correlator is
explicitly  $d$-dependent, and one should check that this 
$d$-dependence comes out correctly.}. Any diagram can be written as
\bea\label{su23}
\int_{k_{1}}\!\int_{\bar \chi_{1}\chi_{1} } ...  \int_{k_{n}}\!\int_{\bar
\chi_{n}\chi_{n} }\  \prod _{i=1}^{n} \left[\int_{0}^{\infty } \rmd
s_{i}\, \rme^{-s_{i}
(m^2+k_{i}^{2}+{\rho } \chi _{{i}}\bar \chi_{i} )} \right]
\komma \nn\\
\eea
where some $\delta $-distributions have already been used to eliminate
integrations over $k$'s, i.e.\ some of the $k_{i} $'s appearing in the
exponential are not independent variables, but linear combinations of
other $k_{j}$'s, and the same holds for the corresponding $\chi _{i}$ and
$\bar \chi _{i}$.   The product of   exponential factors can be
written as
\bea\label{su24}
 \prod _{i=1}^{n} \left[ \rme^{-s_{i}
(m^2+k_{i}^{2}+{\rho } \chi _{{i}}\bar \chi_{i} )} \right]  = \exp
\!\left({- }\left(\begin{array}{c}k_{1}\\
k_{2}\\
\dots \\
k_{n}
\end{array} \right) {\mathbb W}\left(\begin{array}{c}k_{1}\\
k_{2}\\
\dots \\
k_{n}
\end{array} \right) \right) \nn\\
\times \exp\!
\left({-} \left(\begin{array}{c}\chi _{1}\\
\chi _{2}\\
\dots \\
\chi _{n}
\end{array} \right) {\mathbb W}\left(\begin{array}{c}\bar\chi _{1}\\
\bar \chi _{2}\\
\dots \\
\bar \chi _{n}
\end{array} \right) \right)
\punkt
\eea
Integration over the $k$'s gives
\begin{equation}\label{su25}
 \int_{k_{1}} \dots \int_{k_{n}} \rme^{- \vec k \cdot{\mathbb W}\cdot \vec
 {k}}   = \left(\frac{1}{4\pi}\right)^{l d/2}\det ({\mathbb W})^{-d/2} \komma 
\end{equation}
where $l$ is the number of loops.
Integration over $\bar  \chi $ and $\chi $ gives
\begin{equation}\label{su26}
\int_{\bar \chi_{1} \chi_{1} } \dots \int_{\bar \chi_{n} \chi_{n} }
\rme^{- \rho \vec \chi \cdot {\mathbb W} \cdot \vec {\bar\chi} } = (\rho
)^{l} \det ({\mathbb W})\punkt
\end{equation}
The product of the two  factors \eq{su25} and \eq{su26} is the same as for a  standard bosonic diagram  in dimension
$d-2$. Remarking that  the expansion is  in  powers of 
$T$, and
combining these relations, we obtain after integration  over
the $s_{i}$: 
\bea\label{su27}
\hl{ \mbox{$l$-loop super-diagram in dimension $d$} \\
= \left( \frac{\rho
}{4\pi T }\right) ^{\!\!l} \times \mbox{$l$-loop standard-diagram in
dimension $d-2$}}\punkt \nn
\eea
This implies that for any observable $\ca O (T)$
\begin{equation}\label{sufinal}
\hl{\ca  O_{\mathrm{disordered}}^{d} (\rho ) = \ca O_{\mathrm{thermal}}^{d-2} \left(T=
\frac{\rho }{4\pi}\right)}
\punkt
\end{equation}
The above proof can be extended to theories with derivative
couplings. The rules are as follows:
Consider ${\cal H}_{s}[\Phi]$ given in Eq.~(\ref{S-Susy}). To this we can
add an interaction in derivatives, for a total of
\bea\label{p1}
{\cal H}_{s} [\Phi ] &= \int \limits_{\bar \Theta,\Theta} \int\limits_{x}\bigg[ \half
\Phi (x,\bar \Theta,\Theta ) (- \Delta_{\rm s}+m^2) \Phi (x,\bar \Theta,\Theta
) \nn\\
& + {\cal U}\big( \Phi (x,\bar \Theta,\Theta )\big) \nn\\
& + {\cal A}_{1} \big(\Phi (x,\bar
\Theta ,\Theta) \big) (- \Delta_{\rm s}+m^2) {\cal A}_{2} \big(\Phi (x,\bar \Theta ,\Theta) \big)\bigg]\punkt
\nn\\
\eea
This theory is supersymmetric: Calculating diagrams in perturbation
theory, we get additional vertices. The corresponding diagrams can   be
calculated from the same type of generating function
(\ref{su24}). The trick is to use instead of an integral w.r.t.\ $s$
a derivative w.r.t.\ $s$, taken at $s=0$,
\begin{equation}
\left(m^2+k^{2} +\rho \chi \bar \chi \right) = - \frac{\rmd}{\rmd
s}\bigg|_{s=0} \rme^{-s \left(m^2+k^{2} +\rho \chi \bar \chi \right)}\punkt
\end{equation}

\subsection{CDWs and their mapping onto $\phi^4$-theory with two  fermions and one  boson}
\label{s:CDWs and their mapping onto phi4-theory}
Consider the fixed point \eq{RP-fixed-point} for CDWs. It has the form 
\bea
\Delta (u) = \frac g {12}{-} \frac g 2 u(1{-}u) = \frac g2 u^2 + \mbox{lower-order terms in }u\punkt\nn \\
\eea
The renormalization, encoded in $g$, can be gotten by retaining only terms of order $u^2$, and dropping lower-order terms which do not feed back into terms of order $u^2$.  
To this aim,  consider the action \eq{s1} with {\em two replicas}, replacing $\Delta(u)\to \frac{g}2 u^2$. We further go to center-of-mass coordinates for the bosons by introducing 
\begin{eqnarray}
u_1(x) &=& u(x) + \frac12 \phi(x),~~   u_2(x)= u(x) - \frac12  \phi(x)\komma \\
\tilde u_1(x) &=& \frac12 \tilde u(x) + \tilde \phi(x),~~   \tilde u_2(x)= \frac12  \tilde u(x) -  \tilde \phi (x)\punkt
\end{eqnarray}
The action \eq{su6b} can then be rewritten as  
\begin{eqnarray}
\label{CDW=phi4} {\cal S} &=&   \int_{x} \tilde \phi(x) (-\nabla^2 +m^2)\phi(x)+ \tilde  u(x) (-\nabla^2 +m^2) u(x) \nn\\
&& ~~~ + \sum_{a=1}^2\bar \psi_{a} (x)
(-\nabla^{2}+m^2)\psi_{a} (x) \nonumber \\
&& ~~~
+ \frac g 2\tilde u(x) \phi(x)\left[\bar \psi_2(x)\psi_2(x) -\bar \psi_1(x)\psi_1(x)\right]\nn\\
&& ~~~ - \frac g 8 \tilde u(x) ^2 \phi(x)^2
\nonumber \\
&&~~~+ \frac g 2\left[ \tilde \phi(x)\phi(x)+\bar \psi_1(x)\psi_1(x) +\bar \psi_2(x)\psi_2(x) \right]^2 \punkt\nn\\
 \end{eqnarray}
 Note that only $\tilde u(x)$, but not the center-of-mass $u(x)$ appears in the interaction. While $u(x)$  may have non-trivial expectations, it does not contribute to the renormalization of $g$, and the latter can be obtained by considering solely the fist and  last line of \Eq{CDW=phi4}: This is a $\phi^4$-type theory, with one complex bosonic, and two complex fermionic fields. It can equivalently be viewed as complex $\phi^4$-theory at $N=-1$, or real $\phi^4$-theory with $n=-2$ \cite{WieseFedorenko2018}.

\subsection{Supermathematics: A critical discussion}
When supersymmetry was first proposed \cite{ParisiSourlas1979,ParisiSourlas1982}, it was believed to produce an exact result, namely dimensional reduction. While the latter was found earlier \cite{ImryMa1975,AharonyImryMa1976,Young1977} by inspection of diagrams and a combinatorial analysis, supermathematics proved to be a clever tool to show it efficiently.  Supermathematics  got {\em discredited} when it was realized  that dimensional reduction breaks down at  the Larkin scale. As a remedy, 
 breaking of replica symmetry was invoked, or FRG. As we have seen in section \ref{s:RSB}, RSB and FRG fit together, and the applied field inherent to  FRG  explicitly breaks replica symmetry, and as a consequence supersymmetry. 
 As shown in section \ref{s:susy:renormalization-of-disorder}, supermathematics can be used to obtain the FRG flow equation for the disorder. Thus dimensional reduction beyond the Larkin length is not a problem of supermathematics, but of its improper application.
 
It has been argued that \Eq{su5} is inappropriate as it sums over all saddle points, not only the minima, and for this 
reason it fails. The objection {\em per se} can not be discarded. But does it invalidate the formalism put in place above? 
We believe not: the formalism makes numerous predictions, especially for  such non-trivial observables as the FRG fixed-point
function $\Delta(w)$ (sections \ref{s:deriveRG}--\ref{beyond1loop}). Our intuition is  that each RG step merges pairs of  
close minima, without   invoking any of the higher-lying states present in the above argument.  And that by this the objection becomes irrelevant. 

\subsection{Mapping loop-erased random walks onto $\phi^4$-theory with two   fermions and one   boson}  
\begin{figure}[bt]
\Fig{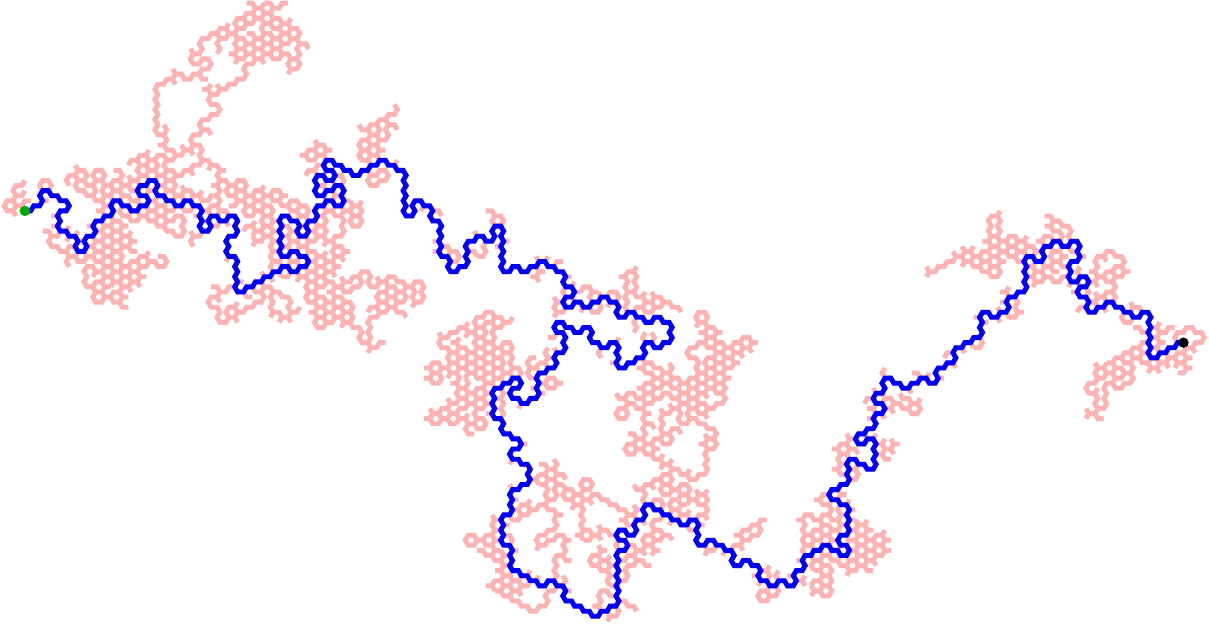}
\caption{Example of a loop-erased random walk on the hexagonal lattice with 3000 steps, starting at the black point to the right and arriving at the green point to the left.}
\label{f:LERW}
\end{figure}  
\subsubsection*{Introduction.}
A loop-erased random walk (LERW) is defined as the trajectory
of a random walk (RW) in which any loop is erased as soon as it is formed \cite{Lawler1980}. An example is shown on figure \ref{f:LERW}, where the underlying RW is drawn in red, and the  LERW remaining after erasure in blue.
Similar to a self-avoiding walk it has a scaling limit
in all dimensions, e.g.~the end-to-end distance $R$ scales with the intrinsic length $\ell$
as $R \sim  \ell^{1/z} $, where $z$ is the fractal dimension~\cite{Kozma2007}.
It is   crucial to note that while both LERWs and SAWs are non-selfintersecting, their  fractal dimensions do not agree since they have a different statistics on the same set of allowed trajectories.
LERWs appear in many combinatorial problems, e.g.\ the shortest path on a uniform spanning tree is an LERW. We have collected the many connections  in Fig.~\ref{CDW-relations}, together with other identities,   which we discuss in the next section.

In contrast to SAWs,
 LERWs have no obvious field-theoretic description.
In three dimensions  LERWs have been
studied   numerically~\cite{GuttmannBursill1990,AgrawalDhar2001,Grassberger2009,Wilson2010}, while
in two dimensions  they are described by  SLE with $\kappa=2$   \cite{Schramm2000,LawlerSchrammWerner2004},   predicting a fractal dimension $z_{\rm LERW}(d=2)=\frac54$. 
Coulomb-gas techniques link this to the  2d $O(n)$-model at $n=-2$ \cite{Nienhuis1982,Duplantier1992}. It was recently shown that 
 LERWs can  be mapped {\em in all dimensions} to the  theory of two complex fermions and one complex boson, equivalent to the $O(n)$ model at $n=-2$ \cite{WieseFedorenko2018,WieseFedorenko2019,HelmuthShapira2020,ShapiraWiese2020}. 

\subsubsection*{Perturbative argument.}

This mapping was first established perturbatively \cite{WieseFedorenko2018,WieseFedorenko2019}. Consider perturbation theory for the complex $N$-component $\phi^4$ theory
{\setlength{\unitlength}{1cm}
\bea
&&
{\parbox{2.25\unitlength}{\begin{picture}(2.15,1)\put(0.15,0){\fig{2\unitlength}{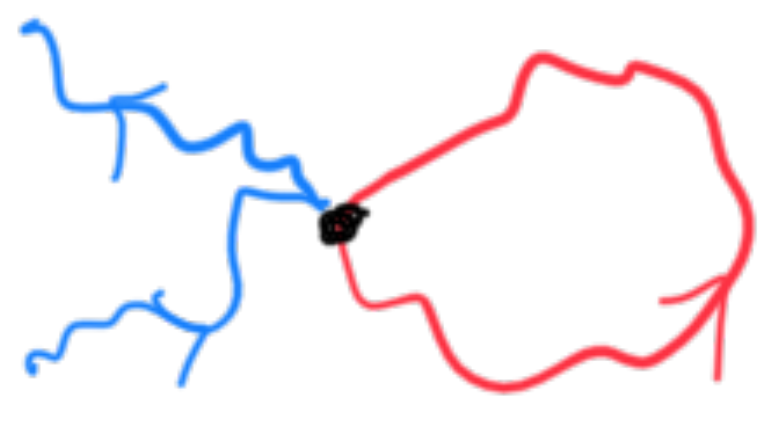}}
\put(0.,0){$\scriptstyle x$}
\put(1.0,0.7){$\scriptstyle y$}
\put(0.45,0){\scriptsize 1}
\put(1.9,0.5){\scriptsize 2}
\put(0.45,0.925){\scriptsize 3}
\put(0.03,0.95){$\scriptstyle z$}
\end{picture}}}\ \ \ \longrightarrow \ \ \
{\parbox{1.1cm}{{\begin{tikzpicture}
\coordinate (v1) at  (0,1.25) ; \coordinate (v2) at  (0,-.25) ;  \node (x) at  (0,0)    {$\!\!\!\parbox{0mm}{$\raisebox{-3mm}[0mm][0mm]{$\scriptstyle x$}$}$};
\coordinate (x1) at  (0.5,0);\coordinate (y) at  (1.5,0.5); \coordinate (y1) at  (0.5,1) ;\node (z) at  (0,1)    {$\!\!\!\parbox{0mm}{$\raisebox{1mm}[0mm][0mm]{$\scriptstyle z$}$}$};
\fill (x) circle (1.5pt);
\fill (z) circle (1.5pt);
\draw [blue] (x) -- (x1);
\draw [blue] (y1) -- (z);
\draw [blue,directed](0.5,0) arc (-90:90:0.5);
\end{tikzpicture}}}}~
-g
{\parbox{1.6cm}{{\begin{tikzpicture}
\node (v1) at  (0,1.25){} ;
\node (v2) at  (0,-.25){} ;
\node (x) at  (0,0)    {$\!\!\!\parbox{0mm}{$\raisebox{-3mm}[0mm][0mm]{$\scriptstyle x$}$}$};
\coordinate (x1) at  (1,0) ;\coordinate (y) at  (1.5,0.5); \coordinate (y1) at  (1,1);\node (z) at  (0,1)    {$\!\!\!\parbox{0mm}{$\raisebox{1mm}[0mm][0mm]{$\scriptstyle z$}$}$};
\fill (x) circle (1.5pt);
\fill (z) circle (1.5pt);
\fill (x1) circle (1.5pt);
\fill (y1) circle (1.5pt);
\draw [blue,directed] (x) -- (x1);
\draw [blue,directed] (y1) -- (z);
\draw [blue,directed](1,0) arc (-90:90:0.5);
\draw [dashed] (x1) -- (y1);
\end{tikzpicture}}}}\nn\\
&&
\hspace{3.3\unitlength}-g N
{\parbox{2.6cm}{{\begin{tikzpicture}
\node (v1) at  (0,1.25){} ;
\node (v2) at  (0,-.25){} ;
\node (x) at  (0,0)    {$\!\!\!\parbox{0mm}{$\raisebox{-3mm}[0mm][0mm]{$\scriptstyle x$}$}$};
\coordinate (x1) at  (0.5,0) ;\coordinate (y) at  (2.5,0.5);\coordinate (y1) at  (0.5,1);\coordinate (y2) at  (1.5,1) ; \node (z) at  (0,1)    {$\!\!\!\parbox{0mm}{$\raisebox{1mm}[0mm][0mm]{$\scriptstyle z$}$}$};
\coordinate (h1) at  (1,0.5) ;
\coordinate (h2) at  (1.5,0.5) ;
\fill (x) circle (1.5pt);
\fill (z) circle (1.5pt);
\fill (h1) circle (1.5pt);
\fill (h2) circle (1.5pt);
\draw [blue,directed] (x) -- (x1);
\draw [blue,directed] (y1) -- (z);
\draw [blue](0.5,0) arc (-90:90:0.5);
\draw [red,directed](1.5,0.5) arc (-180:180:0.5);
\draw [dashed] (h1) -- (h2);
\end{tikzpicture}}}}\ \ .  \label{eq:LERW-diag}
\eea}The drawing on the l.h.s.\ of \Eq{eq:LERW-diag}
is an LERW  starting at $x$, ending in $z$, and passing through the  segments numbered 1 to 3. Due to the crossing at $y$, the loop labeled 2 is erased; we draw it in red. The r.h.s of \Eq{eq:LERW-diag}  gives all   diagrams   of $\phi^4$ theory  up to order $g^{s}$.
The first term is the free-theory result, proportional to $g^{0}$. The second term $\sim g$ cancels the first term, if one puts $g\to 1$. Here it is crucial to have the same   regularization for the interaction as for the  conditioning. The third term is proportional to $N$, due to the loop, indicated in red. Setting  $N\to -1$ compensates for the
subtracted second term. Thus setting $g\to 1$ and $N\to -1$,   the probability to go from $x$ to $z$ remains  unchanged as compared to the free theory. This is a necessary condition to be    satisfied. Since the first two terms cancel, what remains is the last diagram,  corresponding   to the drawing for the trajectory of the LERW we started with.

Continuing to higher orders, one  establishes a one-to-one correspondence between traces of LERWs and diagrams in perturbation theory.
We still need an observable which is $1$ when inserted into a blue part of the trace, and $0$ within a red part.
This can be achieved by  the operator  
\be\label{calO}
{\cal O}(y) :=  \Phi^{*} _{1}(y) \Phi _{1}(y) - \Phi^{*} _{2}(y) \Phi  _{2}(y) .
\ee
When inserted into a loop, it   cancels, whereas inserted into the backbone (LERW, blue), it yields one for each point. 
The fractal dimension $z$ of an LERW is extracted from the length of the walk after erasure (blue part) as
\bea\label{LERW-10}
\frac{\left< \int_{{y,z}}  \Phi_1^{*}(x) {\cal O}(y) \Phi_1(z)  \right>}{\left< \int_{{z}} \Phi_1^{*}(x)   \Phi_1(z) \right>} \equiv
m^2  {\left< \int_{{y,z}} \Phi_1^{*}(x) {\cal O}(y) \Phi_1(z)  \right>} \nn\\
\sim m^{-z}\ .
\eea

\subsubsection*{Proof via Viennot's theorem.}
This section is a shortened version of \cite{ShapiraWiese2020}, itself   inspired by \cite{HelmuthShapira2020}. 
The main tool  is a combinatorial theorem due to Viennot \cite{Viennot1986}. It is part of the general theory of {\em heaps of pieces} (online lectures    \cite{ViennotLectures}).  Here it reduces to a relation between loop-erased random walks, and collections of loops.
To state the theorem, we need some definitions.

Consider a random walk on a directed graph $\ca G$.  The walk moves from vertex $x$ to $y$ with rate $\beta_{xy}$, and dies out with rate $\lambda_x = m_x^2$. The coefficients $\{\beta_{xy}\}_{x,y\in \ca G}$ are weights on the graph. In particular, when  $\beta_{xy}$ is positive, $\ca G$ contains an edge from $x$ to $y$. Denote by $r_x := \lambda_x+\sum_y \beta_{xy} $ the total rate at which the walk exits from vertex $x$.

A {\em path} is  a sequence of vertices, denoted $\omega=(\omega_1,\dots,\omega_n)$.  
The probability $\mathbb P(\omega)$ that the random walk selects the path $\omega$ and then stops is
\begin{eqnarray}\label{LERW-1}
\mathbb P(\omega) &=& \frac{\lambda_{\omega_n}}{r_{\omega_n}} q(\omega ) , \\
 q(\omega ) &=& \frac{\beta_{\omega_1 \omega_2}}{r_{\omega_1}} \frac{\beta_{\omega_2 \omega_3}}{r_{\omega_2}}  \dots \frac{\beta_{\omega_{n-1} \omega_n}}{r_{\omega_{n-1}}}.
\end{eqnarray}

A {\em loop} is a path  $\omega=(\omega_1,\dots,\omega_{n-1},\omega_n=\omega_1)$ where the first and last points are identical, and all other vertices distinct, so it cannot be decomposed into smaller loops. Loops obtained from each other via cyclic permutations  are considered identical.

A {\em collection of disjoint loops} is  a set $L=\{C_1,C_2,\dots\}$ of mutually non-intersecting loops. We denote the {\em set of all such  collections} by $\ca L$.

To formulate the theorem,     fix a self-avoiding\footnote{\textbf{Warning:} a {\em self-avoiding path} is a cominatorial object. The loop-erased random walk is one possible distribution on the set of self-avoiding paths. It should not be confused with the {\em self-avoiding walk} or {\em self-avoiding polymer}, a distinct distribution on the same set of self-avoiding paths.} path $\gamma$.
Define the set $\ca  L_\gamma$ to consist of all collections of disjoint loops in which no loop intersects $\gamma$.
Then Viennot's theorem can be written as ($|L|$ being the  number of loops)
\bea\label{eq:Viennot}
\label{A(gamma)}
  A(\gamma):=q(\gamma) \sum_{L\in \ca  L_\gamma} (-1)^{|L|}\prod_{C\in L}q(C) =  P(\gamma) \times  Z, \\
  P(\gamma) = \sum_{\omega:{\ca L}(\omega)=\gamma} q(\omega),\\
 Z =  \sum_{L\in \ca  L_\gamma}(-1)^{|L|}\prod_{C\in L}q(C).\label{ZLERW}
\eea
For $  A(\gamma)$ one sums over the ensemble of collections of loops which do not intersect $\gamma$, giving each collection a weight $(-1)^{|L|}\prod_{C\in L}q(C)$.
The r.h.s.\ contains two factors. The first, $  P(\gamma)$, is the weight to find the LERW path $\gamma$, our object of interest. The second is the partition function $ Z$.
Conditioning  the walk to stop at $x$, this relation can   be read as $  P(\gamma) =   A(\gamma)/ Z$.

To prove \Eq{eq:Viennot}  consider a pair  $ \{\omega, L\}$ constructed as follows: Take a path $\omega$ such that ${\ca L}(\omega)=\gamma$ and an {\em arbitrary} collection $L$ of disjoint loops. Our goal is to construct another pair $\{ \omega', L'\}$ by transferring a loop from $L$ to $\omega$ or vice versa, depending on where the loop originally was. For example,
\begin{equation}
\label{LERW-8}
\mbox{\hspace*{-0.7cm}\parbox{6.5cm}{{\begin{tikzpicture}[scale=0.38,every node/.style={scale=1}]
		\node  (1) at (-4.25, 7) {};
		\node  (2) at (-6.75, 7) {};
		\node  (3) at (-1.75, 7) {};
		\node  (5) at (-3, 9.25) {};
		\node  (7) at (-3, 8.5) {};
		\node  (8) at (4.5, 7) {};
		\node  (9) at (2, 7) {};
		\node  (10) at (7, 7) {};
		\node  (11) at (5.75, 9.25) {};
		\node  (12) at (5.75, 8.5) {};
		\node  (13) at (3.25, 7.75) {};
		\node  (14) at (3.25, 9.25) {};
		\node  (15) at (3.25, 7.85) {};
\node  (13-15) at (3.25, 7.8) {};
\fill [green] (13-15) circle (4pt);
\node  (new) at (-5.5, 7.85) {};
\fill [green] (new) circle (4pt);
		\node  at (0,8.25) {$+$};
		\node at (-7.2,7) {\rot  $\omega$};
		\node at (-2.3,9.45) {\blue  $L$};

		\node at (1.5,7) {\rot  $\omega'$};
		\node at (4.6,9) {\blue  $L'$};

		\draw [bend left, looseness=6.25, red] (2.center) to (1.center);
		\draw [bend right, red] (1.center) to (3.center);
		\draw [bend left=90, looseness=1.75, blue] (5.center) to (7.center);
		\draw [bend right=90, looseness=1.50, blue] (5.center) to (7.center);
		\draw [bend right, red] (8.center) to (10.center);
		\draw [bend left=90, looseness=1.75,blue] (11.center) to (12.center);
		\draw [bend right=90, looseness=1.50,blue] (11.center) to (12.center);
		\draw [red] (9.center) to (13.center);
		\draw [red] (13.center) to (8.center);
		\draw [in=-5, out=35, looseness=1.75,blue] (15.center) to (14.center);
		\draw [in=145, out=-175, looseness=1.75,blue] (14.center) to (15.center);
\end{tikzpicture}}}}=0.
\end{equation}
In the first drawing, the left loop is part of $\omega$, whereas in the second one it is part of $L'$. 
These terms cancel, as  $(-1)^{|L|} = - (-1)^{|L'|}$, and all other factors are identical.
After each such pair is canceled, we are left with the terms in which it is impossible to transfer a loop from $\omega$ to $L$ or vice versa. These are exactly the terms on the l.h.s of \Eq{eq:Viennot}.

For this procedure to work we need to ensure that we cannot obtain the same pair $\{\omega',L'\}$ starting form two different pairs $\{\omega,L\}$.
In order to  achieve this,  we use the following prescription. 
Start walking along $\omega$, until  
\begin{enumerate}
\item we reach a vertex $\omega_i$ that belongs to some $C = (\omega_i=c_1,c_2\,\dots,c_m=\omega_i)\in L$,  or
\item we reach a vertex $\omega_i$ that does not belong to any $C$, but that we have already seen before, i.e., $\omega_j=\omega_i$ for $j<i$.
\end{enumerate}
In the first case, we   transfer $C$ to $\omega$, i.e.,
\bea
\omega' = (\omega_1,\dots,\omega_i,c_2,\dots,c_{m-1},\omega_i,\dots,\omega_n), \nn\\
L' = L \setminus \{C\}.
\eea
In the second case, we apply the one-loop erasure to $\omega$, and transfer the erased loop to $L$, 
\bea
\omega' = (\omega_1,\dots,\omega_j,\omega_{i+1},\dots,\omega_n), \nn\\
L' = L \cup \{(\omega_j,\omega_{j+1},\dots,\omega_i=\omega_j)\}.
\eea
Note that disjointness of the loop collections is  preserved under the transfer, and that the loop erasure of $\omega'$ remains $\gamma$.
This completes the proof. 
For examples see \cite{ShapiraWiese2020}.

\subsubsection*{A lattice action with two complex fermions and one complex boson.}
Our   goal is to write a lattice action which generates $\ca A(\gamma)$, the l.h.s.\ of \Eq{eq:Viennot}.
This can be achieved with an action based on one pair   of complex conjugate fermionic fields.
While this theory   sums over all paths $\gamma$, yielding back  the random-walk propagator, it contains no information on the erasure. In order to answer   whether the resulting loop-erased path passes through a given point $y$ it is necessary to use more fields. The simplest such setting consists of two pairs of complex conjugate fermionic fields $(\phi_1,\phi_1^*)$, 
and $(\phi_2,\phi_2^*)$, as well as a pair of complex conjugate bosonic fields $(\phi_3,\phi_3^*)$. When appearing in a loop, the latter   cancels one of the fermions. 

We   define the action as
\bea
\label{LERW-10-bis}
\phi^*(y) \phi(x)   := \sum_{i=1}^3  \phi^*_i(y) \phi_i(x) , \\
\label{eq:expS}
e^{-\ca S} = \prod_x \rme^{ -r_x \phi^*(x) \phi(x) }  \Big[1 + \sum_y \beta_{xy} \phi^*(y) \phi(x) \Big]  .
\eea
The path integral is defined by integrating over the $n_{\rm f}=2 $ families of fermionic fields, $(\phi^*_i,\phi_i)$, $i=1,2$, and  $n_{\rm b}=1 $ family of bosonic fields, $i=3$. 
For $\beta_{xy}=0$, we obtain 
\be
 Z_0 = \prod_x r_x^{n_{\rm f} - n_{\rm b}} =  \prod_x r_x.
\ee
Define (normalized) expectation values $\left< \ca O(\phi^*, \phi) \right>$ w.r.t.\ the action \eq{eq:expS} and the (normalized) partition function $ Z$ as 
\bea\label{cal-O}
\left< \ca O(\phi^*, \phi) \right> := \frac1{ Z_0} \int \cal D[\phi] \cal D[\phi^*]  \, \rme^{-\ca S} \,  \ca O(\phi^*, \phi) , 
\\
\label{cal-Z}
{ Z}:=\left< 1\right>.
\eea
Calculating $ Z$ by expansion in $\beta_{xy}$ is best done graphically: 
Due to the square bracket in  \Eq{eq:expS}, at each $x$ one can place exactly one outgoing arrow   to one of the neighbors $y$, with {\em color} $i$, or no arrow. 
Summing   over all possible colorings and all graphs, we obtain
$ Z$ as given in \Eq{ZLERW}.

In order to assess whether a point $b$ belongs to a  loop-erased random walk  from $a$ to $c$ after erasure, we fix the three vertices $a,b$ and $c$, and consider the observable
\begin{equation}\label{eq:U}
U(a,b,c) = \lambda_c r_b^2 r_c \left\langle \phi_2(c) \phi_2^*(b) \phi_1(b) \phi_1^*(a) \right\rangle ,
\end{equation}
 defined by \Eq{cal-O}.

The graphs that contribute   consist of a self-avoiding path $\gamma$ and a collection $L$ of disjoint self-avoiding colored loops such that (see Fig.~\ref{fig:Udiagrams}):
\begin{enumerate}
\item $\gamma$ is a path from $a$ to $c$ passing through $b$. The edges of $\gamma$ between $a$ and $b$ have color $1$, and the edges between $b$ and $c$ have color $2$.
\item Fix $C \in L$. If the color of $C$ is $2$ then it cannot intersect $\gamma$. If its color is $1$ or $3$, it can only intersect $\gamma$ at the (final) point $c$.
\end{enumerate}
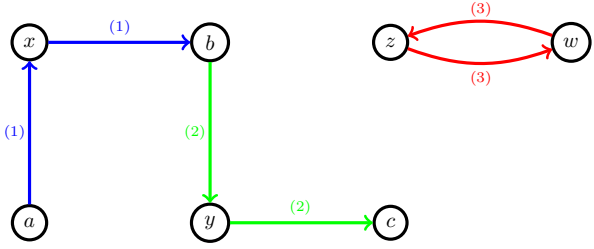
\begin{figure}
\begin{tikzpicture}[scale=0.8,every node/.style={scale=0.8},every path/.style={line width=1.2pt}]
	\node[circle, draw] (a) at (0,0) {$a$};
	\node[circle, draw]  (b) at (3,3) {$b$};
	\node[circle, draw]  (c) at (6,0) {$c$};

	\node[circle, draw] (x) at (0,3) {$x$};
	\node[circle, draw] (y) at (3,0) {$y$};
	\node[circle, draw] (z) at (6,3) {$z$};
	\node[circle, draw] (w) at (9,3) {$w$};

	\draw[->,bend left=0,blue] (a) to (x);
	\node[blue] at (-0.25,1.5) {\scriptsize $(1)$};

	\draw[->,bend left=0,blue] (x) to (b);
	\node[blue] at (1.5,3.25) {\scriptsize $(1)$};

	\draw[->,bend left=0,green] (b) to (y);
	\node[green] at (2.75,1.5) {\scriptsize $(2)$};

	\draw[->,bend left=0,green] (y) to (c);
	\node[green] at (4.5,0.25) {\scriptsize $(2)$};

	\draw[->,bend right=20, red] (w) to (z);
	\node[red] at (7.5,3.55) {\scriptsize $(3)$};

	\draw[->,bend right=20, red] (z) to (w);
	\node[red] at (7.5,2.4) {\scriptsize $(3)$};

\end{tikzpicture}
\caption{An example of a diagram that contributes to $U$. 
Our coloring conventions are blue for   $(\phi_1,\phi_1^*)$, green for $(\phi_2,\phi_2^*)$, and red for $(\phi_3,\phi_3^*)$.
 Fig.~from \cite{ShapiraWiese2020}.} 
\label{fig:Udiagrams}
\end{figure}
In the latter case, the contribution to $U(a,b,c) $ is
\begin{equation}
 (-1)^{\# \text{fermionic loops}}   q(\gamma) \prod_{C \in L} q(C).
\end{equation}
We   now sum over all possible colorings of the loops. Since loops that intersect $c$ may have either color $1$ or $3$, one fermionic and one bosonic, they cancel, leaving only   graphs in which loops do not intersect $\gamma$. The other loops, as before, give a factor of $-1$.\begin{widefigure}[h]
\begin{center}
{\parbox{16.5cm}{{\begin{tikzpicture}
\node (t0) at (-1,4) {Eulerian Circuits} ;
\node (t1) at (4,2.5) {LERW} ;
\node (t2) at (-1,2.5) {Laplacian Walks} ;
\node (t3) at (4,4) {UST} ;
\node (t4) at (9,4) {ASM} ;
\node (t5) at (13,4) {Potts$|_{q\to 0}$} ;
\node (t6) at (9,2) {CDW} ;
\node (t7) at (9,0) {FRG at depinning} ;
\node (t8) at (4,0) {{\parbox{3.cm}{\begin{center}2 fermions + 1 boson\\$\phi^{4}\big|_{n=-2}$\end{center}}}};
\node (t9) at (-1,0) {{\parbox{3.cm}{\begin{center}spin system with \\ 2 fermions + 1 boson\end{center}}}};
\draw [thick,blue,<->] (t0) -- (t3);
\draw [thick,blue,<->] (t1) -- (t2);
\draw [thick,blue,<->] (t3) -- (t4);
\draw [thick,blue,<->] (t5) -- (t4);
\draw [thick,blue,<->] (t6) -- (t4);
\draw [thick,blue,<->] (t3) -- (t1);
\draw [thick,blue,->] (t6) -- (t7) ;
\draw [thick,blue,->]  (t7) -- (t8);

\draw [thick,blue,->]  (t1) -- (t8);
\draw [thick,blue,->]  (t1) -- (t9);
\draw [thick,blue,->]  (t9) -- (t8);

\path (t7) -- node [text width=1cm,midway,above ] {~\,SUSY} (t8);
\path (t7) -- node [text width=1cm,midway,below ] {~~\cite{WieseFedorenko2018}} (t8);
\path (t8) -- node [text width=0.8cm,midway,above ] {\cite{ShapiraWiese2020}} (t9);
\path (t1) -- node [text width=0.5cm,near end,above ] {\rotatebox{26}{~~\cite{HelmuthShapira2020}}} (t9);
\path (t1) -- node [text width=2cm,midway,right ] {KW \cite{KenyonWilson2015}} (t3);
\path (t1) -- node [text width=1cm,midway,above ] {L\;\cite{Lawler2006}} (t2);
\path (t0) -- node [text width=1cm,midway,above ] {L\;\cite{Kasteleyn1967}} (t3);
\path (t1) -- node [text width=2cm,midway,right ] {pert.\,\cite{WieseFedorenko2018}
\\exact~\cite{ShapiraWiese2020}} (t8);
\path (t4) -- node [text width=1cm,midway,above ] {MD\,\cite{MajumdarDhar1992}} (t5);
\path (t4) -- node [text width=3cm,midway,right ] {NM conj.\,\cite{NarayanMiddleton1994}} (t6);
\path (t6) -- node [text width=3cm,midway,right ] {FLW test\,\cite{FedorenkoLeDoussalWiese2008a}} (t7);
\path (t3) -- node [text width=.5cm,midway,above ] {M\;\cite{Majumdar1992}} (t4);
\end{tikzpicture}}}}
\end{center}
\vspace*{-1ex}
\caption{Relations between Laplacian walks, loop-erased random walks (LERW), uniform spanning trees (UST), Eulerian Circuits, the Abelian Sandpile Model (ASM), the Potts-model in the limit of $q\to 0$, charge-density waves at depinning (CDW), mapping onto the FRG field theory at depinning, reducing to $\phi^4$-theory at $n=-2$, and equivalent to an interacting theory of 2 complex fermions and one complex boson.}
\label{CDW-relations}
\end{widefigure}We   therefore established that
\bea\label{eq:4pntfuntion}
U(a,b,c) =      \sum_{{\gamma \in {\rm SA}(a,b,c)}} \!
q(\gamma) \sum_{L \in \ca L_\gamma} (-1)^{|L|} \prod_{C \in L} q(C),
\eea
where the sum is over all self-avoiding paths from $a$ to $c$ passing through $b$, and denoted ${\rm SA}(a,b,c)$.
In view of \Eq{A(gamma)} this can be written as\be
U(a,b,c) =     \sum_{{\gamma \in {\rm SA}(a,b,c)}}   A(\gamma).
\ee
It   implies that our object of interest, the probability that an LERW starting at $a$ and ending in $c$ passes through $b$ is
\be
\sum_{{\gamma \in {\rm SA}(a,b,c)}}     P(\gamma) =   \frac{U(a,b,c)}{ Z}.\\
\ee
\subsubsection*{Continuous limit of the lattice action.}
Let us rewrite the action $\ca S$ explicitly, 
\be\label{831}
\ca S = \!\sum_x \!\Big[ r_x \phi^*(x) \phi(x) -\ln\! \Big(1 + \sum_y \beta_{xy} \phi^*(y) \phi(x) \Big) \Big].~
\ee
The leading  term in $\ca S$ reads
\bea
\sum_x \Big[ r_x   \phi^*(x) \phi(x) -\sum_y \beta_{xy} \phi^*(y) \phi(x)  \Big] \nn \\
= \sum_x\phi^*(x) [ m_x^2-\nabla_\beta^2] \phi(x),\\
m_x^2 = r_x {-}\sum_y \beta_{yx} , \quad \nabla_{\beta}^2 \phi(x) = \sum _y \beta_{yx} [\phi(y) {-}\phi(x)].\nn
\eea
The subleading term in $\ca S$ is 
\bea
  \frac{1}{2}\sum_x  \Big[\sum_y \beta_{xy} \phi^*(y) \phi(x)\Big]^2 = \frac g 2  \sum_x \Big[\phi^*(x) \phi(x)\Big]^2 + ... \nn \\
g:=  \Big[ \sum_y \beta_{xy} \Big]^2, 
\eea
where the dropped terms contain at least one lattice Laplacian $\nabla_\beta^2$. 
Standard arguments \cite{Zinn} show that the latter are irrelevant in an RG analysis, as are higher-order terms in the expansion of the $\ln$ in \Eq{831}. Taking the continuum limit, we arrive at the theory defined in \Eq{CDW=phi4}, setting there $u,\tilde u\to 0$, and identifying 
$(\bar \psi_i,\psi_i) $, $i=1,2$ there with $(\phi^*_i,\phi_i)$ here, and $(\tilde \phi,\phi)$ there with $(\phi^*_3,\phi_3)$ here.

\subsubsection*{Perturbative results.}
Using $\phi^4$-theory at $n=-2$ allows us to obtain  an extremely precise estimate of the fractal dimension $z$, which can be compared to an even more precise Monte Carlo simulation, 
\bea
z=  1.6243(10) \quad \mbox{ (6 loops) }~~~~~~~~~~~\;\mbox{\cite{KompanietsWiese2019}},\\
z= 1.62400 (5)  \quad \mbox{ (Monte Carlo) }~~~ \mbox{\cite{Wilson2010}} .
\eea
The agreement is quite impressive.

\subsection{Other models equivalent to loop-erased random walks, and  CDWs}
\label{s:Loop-erased random walks, and other models equivalent to CDWs}
There is a plethora of further relations relating CDWs or LERWs to other critical systems, see Fig.~\ref{CDW-relations}. Let us discuss at least some of them:
 While LERWs are non-Markovian RWs, their traces are equivalent to those of the {\em Laplacian Random Walk} \cite{LyklemaEvertszPietronero1986,Lawler2006}, which is Markovian, if one considers the whole trace as the state variable. It is constructed on the lattice by solving the Laplace equation $\nabla^2 \Phi(x)=0$ with boundary conditions $\Phi(x)=0$ on the already constructed curve, and $\Phi(x)=1$ at the destination of the walk, either a chosen point, or infinity. The walk then advances from its tip $x$ to a neighbouring point $y$, with probability proportional to $\Phi(y) $. As $\Phi(x)=0$,    $ \Phi(y)\equiv      \Phi(y) -\Phi(x) $ can be interpreted as the {\em electric field} of the {\em potential $\Phi(y)$}.
 
 In a variant of this model    growth is allowed not only from the tip, but from any point on the already constructed object, with a probability $\sim \Phi(y)$. This is known as the {\em dielectric breakdown model}
\cite{NiemeyerPietroneroWiesmann1984}, the simplest model for lightning. The same construction  pertains to diffusion-limited aggregation \cite{WittenSander1981}.

The shortest path on a uniform spanning tree is an LERW \cite{KenyonWilson2015}.  The latter are equivalent to Eulerian circuits \cite{Lawler2006}, and Abelian sandpiles \cite{Majumdar1992}. Abelian sandpiles are equivalent to the Potts-model in the limit of $q\to 0$ \cite{MajumdarDhar1992}. Many of these exact mappings can be found in the lecture \cite{Dhar1999}.
It was conjectured long ago that Abelian sandpiles  map on charge-density waves
\cite{NarayanMiddleton1994}.  A test on the FRG field theory was performed in \REF{FedorenkoLeDoussalWiese2008a}, and validated in \REF{Grassberger2009}.
 
It would be interesting to generalize loops   to higher-dimensional surfaces, as was done for self-avoiding manifolds in Refs.~\cite{WieseKardar1998a,WieseKardar1998b}.

\subsection{Conformal field theory  for critical curves}

In $d=2$, {\em all critical exponents} should be accessible via conformal field theory (CFT). The latter  is 
based on ideas proposed in the 80s by Belavin, Polyakov and Zamolodchikov \cite{BelavinPolyakovZamolodchikov1984}. They constructed a series of minimal models, indexed by an integer $m\ge 3$, starting with the Ising model at $m=3$. These models are conformally invariant and  unitary, equivalent to reflection positive in Euklidean theories. For details, see one of the many excellent textbooks on CFT \cite{DotsenkoCFT,DiFrancescoMathieuSenechal,ItzyksonDrouffe2,HenkelCFT}.
Their conformal charge\footnote{The conformal charge is the coefficient in the leading term of the OPE of the stress-energy tensor. It also gives the amplitude of finite-size corrections \cite{CardyBook}.} is given by
\be
c=   1- \frac{6}{m(m+1)}\ .
\ee
The list of conformal dimensions for allowed operators at a given $m$ is given by the Kac formula with integers $r,s$
\bea\label{36}
h_{r,s}= \frac{[r (m+1)-sm]^{2}-1}{4 m (m+1)} , \  1\le r < m, 1 \le s \le m. \nn\\
\eea
It was later realized that other values of $m$ also correspond to physical systems, in particular $m=1$ (loop-erased random walks), and $m=2$ (self-avoiding walks). These values  can further be extended to non-integer $n$ and $m$, using the identification
\be
n = 2 \cos\left(\frac \pi m\right)\ .
\ee
More strikingly, the table of dimensions allowed by \Eq{36} has to be extended to half-integer values, including $0$. 
This yields: the fractal dimension of the propagator line \cite{RushkinBettelheimGruzbergWiegmann2007,BloteKnopsNienhuis1992,JankeSchakel2010} \be\label{df-CFT}
d_{\rm f} = 2 -2 h_{1,0} = 1+ \frac{\pi }{2 \left(\arccos\left(\frac{n}{2}\right)+\pi
   \right)} \ .
\ee
$\nu$, i.e.\ the inverse fractal dimension of all lines, be it propagator or loops (\cite{JankeSchakel2010}, inline after Eq.~(2))
\be\label{nu-CFT}
\nu =\frac1{2-2 h_{1,3}} = \frac14\left(1+\frac \pi{\arccos(\frac n2)}\right) \ .
\ee
For $\eta$, there are two suggestive candidates from the Ising model,  $\eta = 4 h_{1,2}= 4 h_{2,2}$, which do  not work for other values of $n$; instead   
\cite{RushkinBettelheimGruzbergWiegmann2007,BloteKnopsNienhuis1992,JankeSchakel2010}
\be\label{eta-CFT}
\eta = 4 h_{\frac 1 2,0} = \frac{5}{4} -\frac{3 \arccos \left(\frac{n}{2}\right)}{4 \pi
   }-\frac{\pi }{\arccos  \left(\frac{n}{2}\right)+\pi
   }\ .
\ee
It has a square-root singularity both for $n=-2$ and $n=2$. There is no clear candidate for the exponent $\omega$ \cite{KompanietsWiese2019}. 
The crossover exponent $\phi_{\rm c}$ \cite{Amit,Kirkham1981,KompanietsWiese2019} (explained in \cite{KompanietsWiese2019}, page 7) becomes
\be\label{phic-CFT}
\phi_{\rm c} = \nu d_{\rm f} = \frac{1-h_{1,0}}{1-h_{1,3}} = \frac14 + \frac {3\pi}{8 \arccos (\frac n2)}\ .
\ee
To conclude, we remark that
ideas identifying symplectic fermions with the ASM 
\cite{MoghimiAraghiRajabpourRouhani2005}
are overly simplistic, as they do not catch any of the above exponents.

\begin{reviewKay}

\section{Further developments and ideas}
\label{s:Further developments and ideas}

\subsection{Non-perturbative   RG (NPRG)}
\label{s:NPRG}
The   renormalization  transformation  originally proposed by K.~Wilson \cite{WilsonKogut1974} consists in integrating out a specific range of fast modes, and following the effective action of the remaining modes under this transformation. While this procedure is exact by construction, its implementation is  infeasible, and one has to rely on approximation schemes. 
Several such schemes have been proposed:
\begin{enumerate}
\item expansion in  $\epsilon=4-d$ \cite{WilsonFisher1972}. This scheme produces a divergent, albeit Borel-resummable series in $\epsilon$, with high predictive power \cite{Zinn,KompanietsPanzer2017,MeraPedersenNikolic2018,KompanietsWiese2019}.
\item expansion in the number of components $N$ \cite{Hooft1974,Zinn}. This works in general well for  $N\gtrapprox 5$, but gives mostly qualitative information at  $N=1$.
\item the non-perturbative RG approach (NPRG). This technique can be formulated for the free energy  $\ca F(J)=\ln Z(J)$  \cite{Polchinski1984}, or the  effective action $\Gamma[\phi]$ \cite{Wetterich1993}. For  the effective action it has the structure 
\be\label{exactRG}
\partial_\ell \Gamma[\phi] = \half \tr \left [\left(\frac{\delta^2 \Gamma [\phi]}{\delta \phi^2} +R_\ell \right)^{{\!-1}} \partial_\ell R_\ell \right].
\ee
The function $R_l$ is a momentum cutoff function, optimized for   convergence. 
The simplest truncation of \Eq{exactRG} is the {\em local potential approximation} (LPA), sometimes followed by a {\em gradient expansion} (GE). 
(For historical work and a  recent review see  \cite{HazenfratzHasenfratz1968,WegnerHoughton1973,Polchinski1984,Wetterich1993,Morris1994,BergesTetradisWetterich2002,DupuisCanetEichhornMetznerPawlowskiTissierWschebor2021}).
\end{enumerate}
The FRG technique     used in this review is perturbative in $\epsilon=4-d$; keeping the exact  field dependence is crucial. 
The NPRG  is an approximation (truncation) in the momentum dependence. Based on a numeric integration of the flow equations,   keeping only the leading orders in the field is often sufficient and accelerates the implementation. The non-perturbative FRG (NPFRG) (used  for the RF Ising model in section \ref{s:Random-field magnets}) is approximate in the momentum, but aims at keeping the full field dependence as does   perturbative FRG. One sometimes encounters the  term {\em exact RG} instead of NPRG, a notion better reserved for the   concept of RG than any of its approximate implementations. 
The Heidelberg school \cite{Wetterich1993} now uses FRG instead of NPRG, a notion we reserve to situations when the exact functional form is required.

\subsection{Random-field magnets}\label{s:Random-field magnets}
Another domain of application of the Functional RG are spin models in
a random field (for an introduction see \cite{NattermannBookYoung}). The model usually studied is
\begin{eqnarray}
{\cal H} = \int \rmd^d x\, \half (\nabla \vec S)^2 + \vec h(x)   \vec
S(x) \komma  \label{rf}
\end{eqnarray}
where $\vec S(x)$ is a unit vector  with
$N$-components, and $\vec S(x)^2=1$. This is the so-called $O(N)$ sigma model, to which has
been added a random field, which can be taken Gaussian
$\overline{h_i(x) h_j(x')} = \sigma \delta_{ij} \delta^d(x-x')$. In
the absence of disorder the model has a ferromagnetic phase for
$T<T_{\mathrm{f}}$ and a paramagnetic phase above $T_{\mathrm{f}}$.
The lower critical dimension is $d=2$ for any $N \geq 2$, meaning that
below $d=2$ no ordered phase exists.  In $d=2$ solely a paramagnetic
phase exists for $N>2$; for $N=2$ (XY model) quasi long-range
order exists at low temperature, with $\overline{\vec S(x) \vec
S(x')}$ decaying as a power law of $x-x'$. This is the RP fixed point of sections \ref{s:Charge-density wave (CDW) fixed point} and \ref{s:Sine-Gordon model, Kosterlitz-Thouless transition}. 
Here we wish to study the model directly at $T=0$. 
The first step is to rewrite the hard-spin constraint $\vec S(x)^2=1$ as a field theory. This yields an energy before disorder-averaging
\bea\label{613}
&&{\cal H} = \int \rmd^d x\, \half \big[\nabla \vec \phi(x) \big]^2 + \ca V\big( \vec\phi(x) \big) + \vec h(x)   \vec
\phi(x) .
\eea
The potential $\ca V(\vec \phi)$ is the typical double-well potential, as e.g.\ $\ca V(\vec \phi) \simeq (\vec\phi^2-1)^2$. 
The dimensional-reduction
theorem in section \ref{dimred}, written for this energy indicates that the effect of a quenched random field in
dimension $d$  equals the one 
for a pure model at a temperature $T \sim \sigma$ in dimension $d-2$. Hence one   expects a  
transition from a ferromagnetic phase to a disordered phase at $\sigma_{\rm c}$ as
the disorder increases in any dimension $d>4$, and no order   for
$d<4$ and $N \geq 2$. Not surprisingly, this  is again incorrect, as can be seen
using FRG.

It was noticed by Fisher \cite{Fisher1985b} that an infinity of
relevant operators are generated. These operators, which correspond to
an infinite set of random anisotropies, are irrelevant by naive power
counting near $d=6$ \cite{Feldman2000,Feldman2001}, the
naive upper critical dimension (corresponding to $d=4$ for the pure
$O(N)$ model via   dimensional reduction).  Indeed many early
studies concentrating on $d$ around $d=6$ missed the anisotropies mentioned above. 

A controlled $\epsilon$-expansion using FRG can be  constructed around dimension $d=4$, the naive lower critical dimension, using  the reformulation of the Hamiltonian \eq{rf} in terms of a non-linear $\sigma$-model, first at 1-loop order
\cite{Fisher1985b,Feldman2000,Feldman2001}, and then  extended to two loops
\cite{LeDoussalWiese2005b,TarjusTissier2006}. 
The   FRG  includes all   operators
which are marginal in $d=4$. Its action  
 in replicated form reads
\begin{eqnarray} \label{action}
 {\cal S} =\!\! \int \rmd^d x \, \frac{1}{2 T} \sum_a  \big[\nabla
\vec S_a(x)\big]^2  {-} \frac{1}{2 T^2} \sum_{a b} \hat R\big(\vec S_a(x) \vec S_b(x)\big).\nn\\ 
\end{eqnarray}
The function $\hat R(z)$ parameterizes the disorder. 
The term $\hat R(z) \sim z $ is a direct result of the disorder average of \Eq{613}; higher-order terms are generated within perturbation theory.  The FRG flow equation has been
obtained to order $R^2$ (one loop)
\cite{Fisher1985b,Feldman2000,Feldman2001} and $R^3$ (two loops)
\cite{LeDoussalWiese2005b,TarjusTissier2006}. It is best 
 parameterized in terms of the variable $\phi$, 
  the angle between the two replicas,  defining $R(\phi)=\hat
R(z=\cos \phi)$. Since the vectors
are of unit norm, $z=\cos \phi$ lies in the interval $[-1,1]$. 
\begin{eqnarray}
\partial_{\ell} R (\phi ) = \epsilon R (\phi )+ \half R''
(\phi)^2-R''(0)R''(\phi) \nn
\\
+ (N{-}2)\left[\frac 1 2
\frac{R'(\phi)^2}{\sin^2 \phi }-
 \cot \phi R'(\phi)R''(0)\right] \nonumber \\
 +  \half \big[ R''(\phi)-R''(0) \big] R'''(\phi)^2 \nn\\
+ (N{-}2) \bigg[ \frac{\cot \phi}{\sin^4 \phi} R'(\phi)^3
- \frac{5+ \cos 2 \phi}{4 \sin^4 \phi} R'(\phi)^2 R''(\phi)
\nonumber \\
 + \frac{1}{2 \sin^2 \phi} R''(\phi)^3   - \frac{1}{4
\sin^4 \phi} R''(0) \Big( 2 (2 {+} \cos 2 \phi) R'(\phi)^2 \nn\\
- 6 \sin 2
\phi
R'(\phi) R''(\phi)  +(5{+} \cos 2 \phi) \sin^2 \phi R''(\phi)^2 \Big) \bigg] \nonumber \\
 - \frac{N{+}2} 8  R'''(0^+)^2 R'' (\phi ) - \frac{N{-}2}{4} \cot
\phi R'''(0^+)^2 R' (\phi )
\nonumber \\
  - 2 (N{-}2) \Big[R'' (0) - R'' (0)^{2} + \gamma_{a} R'''
(0^+)^{2} \Big]
   R (\phi ) . \qquad  \label{beta}
\end{eqnarray}
The last factor proportional to $R (\phi )$ takes into account the
renormalization of temperature,  absent in the
manifold problem\footnote{The constant $\gamma_a$ is discussed in \REF{LeDoussalWiese2005b}.}.  The full analysis of this equation is quite
involved.
The key observation is that under FRG 
again a cusp   develops near $z=1$. Analysis of the FRG fixed points 
 shows interesting
features already at 1-loop order. For $N=2$, the fixed point corresponds to the Bragg-glass
phase of the XY model with quasi-long range order accessible via a
$d=4-\epsilon$ expansion below $d=4$ \cite{GiamarchiLeDoussal1995}. Hence for $N=2$ the lower
critical dimension is $d_{\mathrm{lc}} < 4$, and conjectured to be
$d_{\mathrm{lc}} < 3$  \cite{GiamarchiLeDoussal1995}. On the other
hand Feldman \cite{Feldman2000,Feldman2001} found that for $N=3,
4,\dots$ there is a fixed point in dimension $d=4+\epsilon >4$.  This
fixed point has exactly one unstable direction,    
  corresponding to the ferromagnetic-to-disorder transition. The
situation at one loop is thus rather strange: For $N=2$, only a stable
FP which describes a {\em unique} phase exists, while for $N=3$ only
an unstable FP exists, describing the transition between two
phases. The question is: Where does the disordered phase go as $N$
decreases? 
\begin{figure}\centerline{\Fig{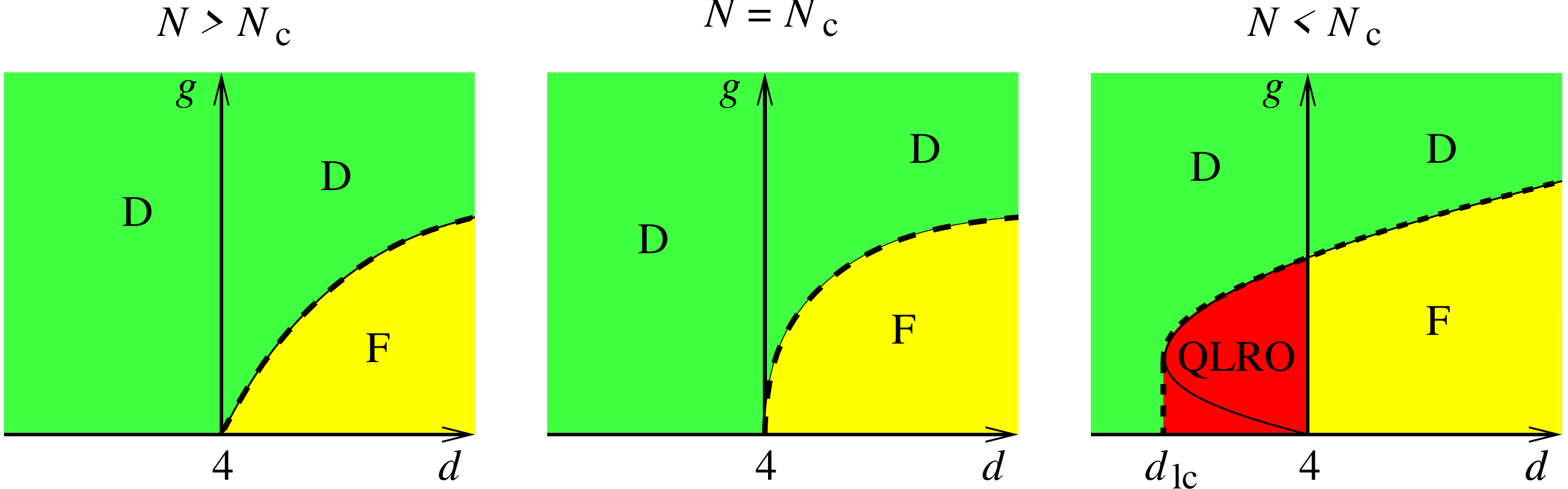}}\caption{Phase diagram of the RF non-linear sigma model. D $=$ disordered, F $=$ ferromagnetic,
QLRO $=$ quasi long-range order. Reprinted from \cite{LeDoussalWiese2005b}.}\label{f:phases}\end{figure}The complete analysis at 2-loop order \cite{LeDoussalWiese2005b} shows that there is a
critical value of $N$, $N_{\rm c}=2.8347408$, below which the lower critical
dimension $d_{\mathrm{lc}}$ of the quasi-ordered phase plunges below
$d=4$, resulting into   two new fixed points below $d=4$. 
This is schematically shown in Fig.~\ref{f:phases}.
For $N>N_{\rm c}$ a
ferromagnetic phase exists with lower critical dimension
$d_{\mathrm{lc}}=4$. For $N<N_{\rm c}$ one finds the expansion
\begin{equation}\label{expansiondc}
d_{\mathrm{lc}}^{\mathrm{RF}} = 4-\epsilon_{c}\approx 4 - 0.1268
(N-N_{c})^{2}+ \ca O   (N-N_{c})^{3} .~~ 
\end{equation}
One can then compute the critical exponents at this fixed point
\cite{Feldman2000,Feldman2001,LeDoussalWiese2005b,TarjusTissier2006}.

The expansions discussed above were either in $d=6-\epsilon$, neglecting by construction  FRG corrections of the disorder with the physics of the cusp, or in $d=4+ \epsilon$, neglecting amplitude fluctuations in $\vec \phi(x):= \int_{\rm box } \vec S(x)$, as they were formulated in terms of a non-linear $\sigma$-model.  
To find a consistent renormalization-group treatment in the full $(N, d)$-plane is much more complicated, and can to date only be achieved within the non-perturbative FRG approach (NP-FRG),  i.e.\ the RG must be both non-perturbative (NP) and functional (FRG). For this formalism to work, and to correctly encounter shocks, i.e.\ the physics of the cusp, one has to allow for a cusp in the effective disorder correlator. It is the merit of G.~Tarjus and M.~Tissier to have transformed this general idea into a predictive framework 
\cite{TarjusTissier2004,TarjusTissier2005,TarjusTissier2008,TissierTarjus2008b,TissierTarjus2011,TarjusTissier2020}.
We show in Fig.~\ref{f:Tarjus+Tissier} their phase diagram. The behavior in the region around $d=4$ and $N=N_{\rm c}$ was obtained above from  the non-linear $\sigma$ model.
\begin{figure}\centerline{\Fig{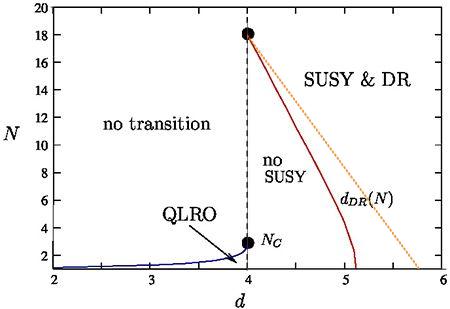}}
\caption{The phase diagram of  Tarjus and Tissier  \cite{TarjusTissier2020}, reproduced with kind permission.
The added   line given by \Eq{N*}   qualitatively agrees with the NP-FRG prediction.  }
\label{f:Tarjus+Tissier}
\end{figure}A novel prediction is that for $d=4+\epsilon$ and    $N>N^*=18$  there is a solution without cusp, and as a consequence dimensional reduction and super-symmetry are restored \cite{TissierTarjus2011,TissierTarjus2012,TissierTarjus2012b,BaczykTarjusTissierBalog2014}. The critical line starting at $d=4$ and  $N^*=18$ can be obtained in an $\epsilon$-expansion
\cite{LeDoussalWiese2006b,TarjusTissier2006}, 
\be\label{N*}
N^*(d)=18 - \frac{49}5 (d-4) + \ca O(d-4)^2.
\ee
The FRG in its perturbative and non-perturbative versions can be applied to a variety of disordered systems in and out of equilibrium, see e.g.~\cite{TarjusTissier2020}. In particular, $O(N)$ models with a random anisotropy can be treated.
For   this universality class,  details of the phase diagram, the critical exponents, and the many subtleties involved,   the  reader is referred  to Refs.~\cite{TarjusTissier2004,LeDoussalWiese2005b,TarjusTissier2005,TarjusTissier2006,LeDoussalWiese2006b,FedorenkoKuehnel2007,TarjusTissier2008,TissierTarjus2008b,TissierTarjus2011,TarjusBaczykTissier2012,MouhannaTarjus2016,TarjusTissier2020}.

For random field, 
the NP-FRG solution qualitatively agrees with the perturbative FRG solution,  but systematically predicts a smaller $N^*(d)$, terminating for $N=1$ at $d=5.1$, while the analytic solution favors $d=5.74$. 
We remind that   NP-FRG is based on a truncation of the functional form of the effective action. By construction it includes loop corrections at all orders, but in an approximate way beyond one loop.
Thus for the Ising model, dimensional reduction is valid near dimension  $d=6$, whereas a non-trivial   ordered phase exists down to $d=2$. 
This has been confirmed numerically in Refs.~\cite{MiddletonFisher2002,FytasMartin-MayorPiccoSourlas2017,FytasMartin-MayorPiccoSourlas2018,FytasMartin-MayorParisiPiccoSourlas2019}, the most remarkable  test being the comparison of diverse correlation functions in the 5-dimensional RF model, as compared to their pure 3-dimensional counterparts at $T=T_{\rm c}$ \cite{FytasMartin-MayorParisiPiccoSourlas2019}.

There is renewed interest into the RF Ising model \cite{KavirajRychkovTrevisani2019,KavirajRychkovTrevisani2020}. The authors follow the proposition of Cardy \cite{Cardy1985} to use
$n$ bosonic replicas $\phi_i$, $i=1,...,n$, to introduce fields 
\bea
u:= \half \left[\phi_1+(n-1)^{-1}(\phi_2+ ...+ \phi_n) \right], \\
\tilde u:= \half \left[\phi_1-\frac{T}{\Delta(0)}(n-1)^{-1}(\phi_2+ ...+ \phi_n) \right],
\eea
together with $(n-2)$ fields $\psi_j$ which are linear combinations of $\frac T{\Delta(0)}(\phi_2,..., \phi_n)$ chosen to be orthogonal to $\phi_2+...+\phi_n$. As Cardy showed, this choice of fields allows one to write an action which (apart from terms irrelevant in $6-\epsilon$ dimensions) is {\em formally} equivalent to \Eq{reallySusy}, but containing more fields, thus a  breaking of super-symmetry becomes possible. 
The authors then identify \cite{KavirajRychkovTrevisani2020} such perturbations which destabilize the supersymmetric dimensional-reduction  fixed point below $d_{\rm c}\approx 4.2$.
In section \ref{s:susy:renormalization-of-disorder} we showed that in order to see the renormalization of the disorder, one needs more than one physical copy. To be precise, the cusp appears in the renormalized disorder correlations, as a function of the difference between the two physical copies. In such a calculation, the critical dimension moves up to $d_{\rm c}\approx 4.6$ \cite{TarjusTissier2016}. We do not see how this difference between replicas is present in the above choice of coordinates, but we believe that by doubling the set of Cardy variables this can be achieved. 

We would like to conclude by some general   remarks on the form of the effective action necessary for a proper RG treatment of the RF Ising model. 
As in all disordered systems, it should {\em at least} contain   a 1-replica {\em and} a 2-replica contribution. Its  general form should be as given in \Eq{H} in a replica formulation, in \Eq{dyn-action} in a dynamical formulation, or in \Eq{su6a} in the Susy formulation. While the 1-replica part may contain an arbitrary function of $u$ and $\nabla u$, let us concentrate on the 2-replica part parameterizing the disorder correlations. For disordered elastic manifolds, this is achieved by the  function $\Delta(u_1-u_2)$, where we remind that $\Delta(u_1-u_2)$ has only one argument due to the statistical tilt symmetry \eq{STS}. As the latter is absent for the RF Ising model, one needs to make a more general ansatz, see e.g.~\cite{TarjusTissier2005,TarjusBaczykTissier2012}, 
\bea
\Delta(u_1,u_2) =\hat \Delta(\bar u, \delta u),\\
\bar u:= \frac12 (u_1+u_2)\ , \quad \delta u := |u_2 - u_1| .
\eea
The absolute value appears since $\Delta(u_1,u_2)= \Delta(u_2,u_1)$. 
Let us apply as in sections \ref{s:shocks} and \ref{measurecusp} a field $h=m^2 w$, and denote 
$u_i \equiv  u(w_i)$ the expectation of $u$ given $w_i$. Both in the statics and at depinning $u(w_i)$ is unique.  
The  connected correlation function $\left<u_1 u_2 \right>^{\rm c} $ is \be\label{<u(h1)u(h2)>}
\overline{ u(w_1) u(w_2) }^{\rm c} = \Gamma_1''(u_1)^{-1} \Delta(u_1,u_2)  \Gamma_1''(u_2)^{-1} .
\ee
Here $\Gamma_1''(u)$ is the second (functional) derivative of the 1-replica contribution to the effective action. Formally, the l.h.s.\ which depends on $w_1$ and $w_2$  is the Legendre transform of  the second cumulant $\Delta(u_1,u_2)$ in the effective action depending on $u_1$ and $u_2$. Graphically, the prescription amounts to amputating the 1-particle irreducible contributions to   \eq{<u(h1)u(h2)>}. 
The key point is that the   observable on the l.h.s.\ can be measured. For small $w_1-w_2>0$ it behaves with $\bar w:=(w_1+w_2)/2$ as 
\bea
&&\half \overline{[u(w_1) -u(w_2) ]^2}^{\rm c}\simeq \ca A(\bar w) |w_1-w_2|+ \ca O(w_1{-}w_2)^2,\nn\\
\\
\label{cusp-amp-RF}
&&\ca A (\bar w) := \frac{\left<S^2\right>}{2\left< S \right>}\Bigg|_{\bar w} \times   \overline{ u'(\bar w) } . 
\eea
As indicated, the ratio ${\left<S^2\right>}/(2\left< S \right>)$ depends on $\bar w$.
These relations are derived similar to \Eq{Delta'(0+)}, except that when writing \Eq{w-w'} as
\be
\overline{u(w_1) {-} u(w_2)} = \left< S\right> \rho_{\rm shock} |w_1{-}w_2| + \ca O(w_1{-}w_2)^2, 
\ee
the l.h.s.\ becomes 
\be
\overline{u(w_1) {-} u(w_2)}\simeq \overline{u'(\bar w)} (w_1{-}w_2)+ \ca O(w_1{-}w_2)^2.
\ee
Solving \Eq{<u(h1)u(h2)>} for $\Delta(u_1,u_2)$ proves that   it has a cusp as a function of $u_1-u_2$, with amplitude given in \Eq{cusp-amp-RF}.

\end{reviewKay}

\subsection{Dynamical selection of critical exponents}
\begin{figure}[bt]
\centerline{\fig{0.5\textwidth}{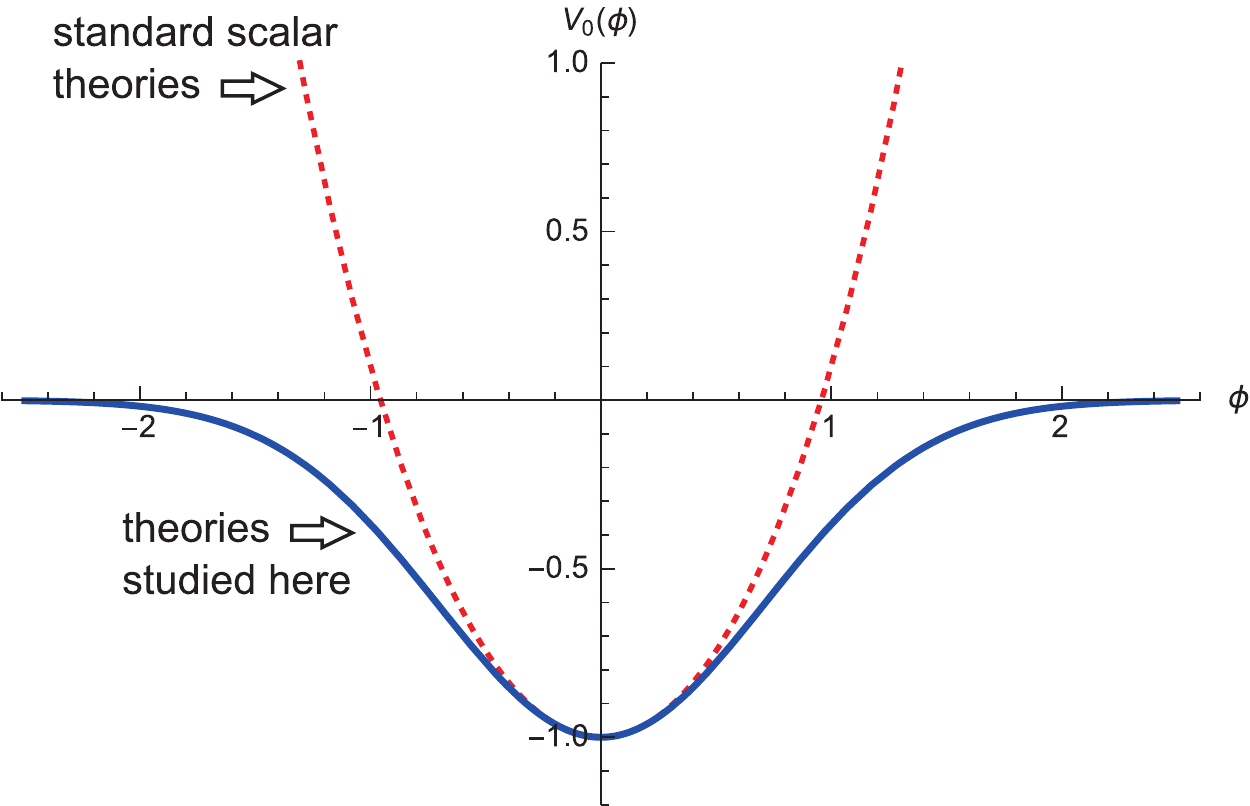}}
\caption{The function ${\cal V}_0(\phi)$, for $\phi^4$ theory (top, red, dashed), and a bounded potential (bottom, blue, solid). Fig.~from \cite{Wiese2016}.}
\label{f:VofPhi0}
\end{figure}
Evaluating  the partition function of a field theory in presence of a potential ${\cal V}_0(u)$ at constant background field $u$ to 1-loop order, and normalizing with its counterpart at ${\cal V}_0= 0$, one typically gets a partition function of the form
\be\label{3}
\ln\! \left( \frac{{  Z}[u]}{{  Z}_0[u]}\right) = - \int^{\Lambda} \frac{\rmd^d k}{(2\pi)^d} \ln\!\left(1+ \frac{{\cal V}_0''(u)}{k^2+m^2} \right) \punkt
\ee
We have explicitly written an  UV cutoff $\Lambda$. 
This equation is at the origin of {\em non-perturbative} renormalization group (NPRG) schemes  (section \ref{s:NPRG}). 
To leading order, the effective action is $\Gamma(u) = -\ln  ( {{  Z}[u]}/{{  Z}_0[u]}) $, and denoting its local part  by ${\cal V}(u)$, we arrive at the following {\em functional} flow equation for the {\em renormalized} potential ${\cal V}(u)$
\be\label{4}
-m \partial_m {\cal V}(u)  = - m \partial_m \int\limits^{\Lambda} \frac{\rmd^d k}{(2\pi)^d} \ln\!\left(1{+} \frac{{\cal V}_0''(u)}{k^2+m^2} \right) \punkt
\ee
Keeping only the leading non-linear term \cite{Wiese2016}
leads to the  simple flow equation
\be
-m \partial_m {\cal V}(u)  = -  m^{d-4}  \frac12  {\cal V}''(u)^2  + ...   \punkt
\ee
Note that this equation is very similar to the FRG flow equation \eq{RG1loop-bare} for disordered elastic manifolds. 
It reproduces the standard RG-equation for   $\phi^4$ theory; indeed, setting 
\be\label{phi4-potential}
{\cal V}(u) = m^{4-d}  \frac{u^4}{72}\, g\komma 
\ee
we arrive with  $\epsilon:=4-d$ at 
\be \label{projection}
-m \partial_m \, g = \epsilon g -  g^2 + ...
\punkt
\ee
This is the standard flow equation of $\phi^4$ theory, with fixed point $g_*=\epsilon$. One knows that the potential (\ref{phi4-potential}) at $g=g_*$ is attractive, i.e.\ perturbing it with a perturbation $\phi^{2n}$, $n>2$, the flow   brings it back to its fixed-point form.

This fixed point, and its treatment with the projected simplified flow equation (\ref{projection}) is relevant in many situations, the most famous being the Ising model. The form of its microscopic potential, which is plotted in figure \ref{f:VofPhi0} (red dashed curve), grows unboundedly for large $\phi$. This is indeed expected for the Ising model, for which the spin, of which $\phi$ is the coarse-grained version, is bounded.

There are, however, situations, where this is not the case. An example is the attraction of a domain wall by a defect. In this situation, one expects that the potential at large $\phi$ vanishes, as  plotted on  figure \ref{f:VofPhi0} (solid blue line). The question to be asked is then: Where does the RG flow lead? 

As one   sees from figure  \ref{f:VofPhi0},  the {\em bounded} potential ${\cal V}_0$ is negative. In order to deal   with positive quantities,     set ${\cal V}(u) \equiv - {\cal R}(u)$. The flow equation to be studied is
\be\label{flow-bare}
-m \partial_m {\cal R} (u)  = m^{-\epsilon}  \frac12  {\cal R}''(u)^2  + ...   
\punkt
\ee
As shown in  \cite{Wiese2016},   for generic smooth initial conditions as plotted on figure \ref{f:VofPhi0}: 
\begin{enumerate}
\item [(i)] The flow equation  (\ref{flow-bare})   develops a cusp at $u=0$, and a cubic singularity at $u=u_{\rm c}>0$.
\item [(ii)] The rescaled flow equation for the dimensionless function $\tilde R(u)$ (to be compared to \Eq{RG1loop}) 
\bea\label{flow-rescaled}
-m \partial_m \tilde R (u)  &=& (\epsilon-4 \zeta ) \tilde R(u)+ \zeta u \tilde R'(u)\nn\\
&&+ \frac12 \tilde R''(u)^2  + ...   \punkt
\eea
 has an infinity of fixed points $-m \partial_m \tilde R(u)=0$, indexed by $\zeta\in [\frac\epsilon4,\infty]$.
\item [(iii)] The solution   chosen dynamically when starting from smooth initial conditions is $\zeta=\frac\epsilon3$. Its analytic expression for $0\le u\le 1$ reads
\beq
\label{R-zeta=1/3}
\tilde R_{{  \zeta}=\frac\epsilon 3}(u) = \epsilon \left[ 
\frac{1}{18} (1-u)^3-\frac{1}{72}
   (1-u)^4 \right] \punkt
\eeq
It vanishes for $u>1$, and is continued symmetrically to $u<0$.  
\end{enumerate}
This scenario is quite unusual: Normally, the perturbatively accessible fixed points of the RG flow  have   only  one fixed point. In the few cases where there is more than one fixed point, the spectrum of    fixed points is at least {\em discrete}. In contrast, here is a spectrum of  fixed points. On the other hand, only one of them seems to be chosen. Thus experiments would only see this one fixed point. 

It is yet not clear to which physical system it applies. As   discussed in the literature \cite{BrezinHalperinLeibler1983a,BrezinHalperinLeibler1983,DSFisherHuse1985,BrezinHalpin-Healy1983,LipowskyFisher1987,ForgasLipowskyNieuwenhuizenInDombGreen}, 
 experiments describing wetting are usually attributed to a flow equation {\em linear} in ${  R}(u)$.  Is a nonlinear fixed point possible?
Let me cite   Thierry Giamarchi, one of the pioneers of FRG: ``Whenever there is a simple equation, it is somewhere realized  in nature."

\begin{reviewKay}

\begin{reviewKay}
\subsection{Conclusion and perspectives}\label{perspectives}
The aim of this review was to give a thorough overview over the physics of disordered elastic manifolds, with its numerous connections to  systems as diverse as sandpiles and  loop-erased random walks. We covered all theoretical tools developed to date, including FRG, replicas, replica-symmetry breaking, MSR dynamics, and super-symmetry. We put emphasis on applications, giving experimentalists   the necessary tools to verify the theoretical concepts, and going beyond critical exponents. 

Our aim at completeness was seriously challenged by the shear amount of publications on the subject, and we apologize for any omissions. Please let us know, and we will try to remedy. 

While the presented  methods are powerful, fundamental questions remain:
Can FRG be applied to other systems such as spin glasses, sheared colloids, real glasses, or Navier-Stokes turbulence? 
Can  FRG be applied beyond the elastic limit, i.e.\ to systems with overhangs or topological defects, or to     fractal curves  that can not be represented by directed interfaces? 

The author is looking forward to exciting new   discoveries, and the contributions of today's PhD students and postdocs. As for this review, it needs to stop here.

\section*{Acknowledgments}
It is a pleasure to thank all my collaborators over the past years: Pierre Le Doussal for introducing me to the subject and all his enthusiasm in the many projects we undertook together. A.~Rosso, A.~Kolton,  L. Laurson,	
	 and A.~Middleton for putting the theoretical concepts to test. M.~M\"uller for all his insights  in the connection to replica-symmetry breaking. Many thanks go   to my students M.~Delorme, A.~Dobrinevski, C.~ter Burg, G.~Mukerjee, T.~Thierry, Z.~Zhu, and  postdocs C.~Husemann,  A.\ Petkovic, Z.\ Ristivojevic and A.\ Shapira, for their help and pertinent questions.  I am grateful to A.A.~Fedorenko for the many enjoyable common projects. 
 S.~Atis, F.~Bohn, M.~Correa, A.~Douin, A.K.~Dubey,  G.\ Durin, F.~Lechenault, S.\ Moulinet, M.R.~Past\'o, F.~Ritort, P. Rissone,  E.\ Rolley, D.~Salin, R.~Sommer, and L.~Talon were essential in testing the concepts in experiments. I have benefitted from   collaborations with L.~Aragon, C.~Bachas, P.~Chauve,  R.~Golestanian,  E.~Jagla,  M.~Kardar, M.~Kompaniets, W.~Krauth, A.~Ludwig, C.~Marchetti,  A.~Perret,  E.~Raphael,   G.~Schehr and J.~Vannimenus.  
Many thanks for  discussions  go to  M. Alava, C.~Aron, E.~Br\'ezin, L.~Balents, E.~Bouchaud, J.P.~Bouchaud, J.~Cardy,  F.~David, H.W.~Diehl, T.~Giamarchi, P.~Goldbart,  D.~Gross, F.~Haake, A.~Hartmann, J.~Jacobsen, J.~Krug,   S.~Majumdar, M.~M\'ezard,  T.~Nattermann, G.~Parisi, H.~Rieger,  L.~Ponson,   S.~Rychkov, S.~Santucci, L.~Sch\"afer, S.~Stepanov, L.~S\"utterlin, G.~Tarjus, M.~Tissier, F.~Wegner,   J.~Zinn-Justin and A.~Zippelius.  
This review is based on lecture notes for the ICTP master program at ENS, and I thank all students for their    feedback. 
\end{reviewKay}

\section{Appendix: Basic Tools}
\label{s:Appendix: Basic Tools}
\subsection{Markov chains,  Langevin equation, inertia}
\label{s:Markov chains,  Langevin equation, inertia}
In Markov chains  the state at time $t_N:= N \tau$ is given by the product of {\em transition probabilities}
\be
{\cal P}(x_N,x_{N-1},... ,x_1,x_0) =\prod_{i=1}^N {\cal P}_{\tau} (x_i|x_{i-1})\punkt
\ee
Transition probabilities are drawn from  a Gaussian distribution. The probability to be at $x$ (the variable) given $x'$ (prime as previous), reads
\be\label{Ptransition}
{\cal P}_{\tau} (x|x') = \frac1{\sqrt{4\pi \tau D(x')}}{\rme^{-\frac{[\eta(x-x')-\tau F(x')]^2}{4\eta \tau D(x')}}}\punkt
\ee
Both $F$ and $D$ depend on the previous position (It\^o discretization).  
As a stochastic process, this reads
\bea\label{3.3}
&&\eta (x_{i+1}- x_{i} )= \tau F(x_i) + \sqrt{\tau} \xi_i\komma  \\
&& \left< \xi_i   \right>= 0\komma  \quad  \left< \xi_i \xi_j  \right> = 2 \delta_{ij} \eta D(x_i)
\eea
The formal limit of $\tau\to 0$ is the {\em It\^o-Langevin equation}, 
\bea\label{185bis}
&&\eta \dot x(t) = F \big(x(t)\big) + \xi(t)\komma  \\
&& \left< \xi(t)   \right>= 0\komma  \quad  \left< \xi(t) \xi(t')  \right> = 2\eta \delta(t-t') D\big(x(t)\big)\punkt
\label{185cis}
\eea
The factor of $\eta$ is the friction coefficient in Newton's equation of motion. Indeed, for the problem at hand the latter reads
\be
M \partial_t \dot x(t) = F\big(x(t)\big) + \xi(t) - \eta  \dot x(t)\punkt
\ee
On the l.h.s.\ is the mass $M$ (or inertia) of the particle (not to be confounded with the mass $m$ in field theory), times its acceleration. 
This defines a characteristic time scale 
\be
\tau_{M} =  \frac  M \eta \punkt 
\ee
For times $t\gg \tau_{M}$, inertia    plays no role, $M$ can be set to 0, and we arrive at  \Eq{185bis}.

The situation is different, when the noise is correlated on a time scale $\tau\gg \tau_{M} $.  Then in the equation of motion $F(x(t))$ changes, since $x(t)$ changes, and it is better to discretize this limit as 
\be
\eta (x_{i+1}- x_{i} )= \tau F\!\left(\frac{x_i+x_{i+1}}2\right) + \sqrt{\tau} \xi_i\punkt
\ee
This prescription is known as mid-point or Stratonovich discretization. 

Let us finally rescale time, $t\to \eta t$, which effectively sets  $\eta\to 1$. The {\em friction coefficient}   $\eta$  can always be restored by multiplying each time derivative with $\eta$.

\subsection{It\^o calculus}
\label{s:Ito calculus}
Consider (with $\eta=1$)
\bea
&&{g (x_{i+1} )- g(x_i)} = g\big(x_i+ \tau F(x_i) + \sqrt{\tau} \xi_i\big) -g(x_i)  \nn\\
&&= g'(x_i) \left[ \tau F(x_i) + \sqrt{\tau} \xi_i \right] + \half g''(x_i) \tau \xi_i^2 + {\cal O}(\tau^{3/2}) \nn \\
&&= g'(x_i) \left[ \tau F(x_i) + \sqrt{\tau} \xi_i \right] +   g''(x_i) \tau D(x_i) + {\cal O}(\tau^{3/2}).\nn\\
\eea
The last relation is justified since in any time slice maximally two powers of $\xi_i$ can appear. (If there could be 4 then one would have to use Wick's theorem to decouple them pairwise.)
It is  implicitly understood that the noise is independent of $x$, thus $\left<  g(x_i)   \xi_i\right> =0  $, and $\left<g(x_i)   \xi_i^2 \right> = 2 g(x_i) D(x_i)\rmd t$.

Mathematicians prefer setting $x_i\to x$, $\tau \to \rmd t$, $\xi_i \sqrt\tau \to \rmd \xi$, and write the Langevin equation  as 
\bea\label{3.11}
&&\rmd x = F(x) \rmd t + \rmd \xi\komma \\
&& \left< \rmd \xi \right> =0 , \quad \rmd \xi^2 = \left< \rmd \xi^2 \right> =2 D(x)\rmd t\punkt
\eea
The stochastic evolution of a  function $g(x)$ can  then be written with these ``differentials'' as
\bea\label{3.12}
&&\rmd g(x) = g'(x)\rmd x + \half g''(x)\rmd x^2+ ...\nn\\
&&=   g'(x) [ F(x) \rmd t +\rmd \xi]  + \frac12 {g''(x)} [ F(x) \rmd t +\rmd \xi]^{2} + ... \nn\\
&&= \left[ g'(x)F(x) + g''(x) D(x)\right]  \rmd t + g'(x)\rmd \xi 
\eea
This is known as {\em It\^o calculus}. The rule of thumb to remember is that when expanding to first order in the time differential $\rmd t$, as $\rmd \xi\sim \sqrt {\rmd t}$, one has to keep all terms up to second order in $\rmd \xi$.

\subsection{Fokker-Planck equation}
\label{s:FP}

\paragraph{Derivation of (forward) Fokker-Planck equation using It\^o's formalism:}
The forward Fokker-Planck equation can   be derived from It\^o's formalism. 
Consider the expectation of a test function $g(x)$ at time $t$:
\be \label{191}
\left< g(x_t) \right> \equiv \int_x g(x) P_t(x) \punkt
\ee 
Taking the expectation of the first line of \Eq{3.12} yields
\be
\left < \rmd g(x_t) \right>  = \left< g'(x_t) \rmd x\right> + \half \left< g''(x_t) \rmd x^2\right> + ...
\ee
Averaging over the noise gives
\be
\frac{\rmd }{\rmd t}\left <   g(x_t) \right> =\left< g'(x_t) F(x_t)\right>   + \left< g''(x_t) D(x_t) \right>  \punkt
\ee
Expressing the expectation values   with the help of \Eq{191}, we obtain
\bea
&&\int _x g(x) \partial_t P_t(x)   \nn\\
&& =  \int _x  g'(x) F(x)  P_t(x)+g''(x) D(x)  P_t(x)  \punkt
\eea
Integrating by part, and using that $g(x)$ is an arbitrary test function, we obtain 
the {\em forward Fokker-Planck equation}
\be\label{forwardFP}
\hl {\partial_t P_t(x) =  \frac{\partial ^2}{\partial x^2} \Big( D(x) P_t(x) \Big)- \frac{\partial }{\partial x} \Big( F(x) P_t(x) \Big)} \punkt
\ee
Our derivation is valid for any initial condition, thus the propagator $P(\xfi,\tf|\xin,\ti)$ also satisfies the forward 
Fokker Planck-equation as a function of $x=\xfi$, $t=\tf$. 

If there are several degrees of freedom $x_u$, $u=1,...,L$, then \Eq{forwardFP} generalizes to an equation for the joint probability $P_t[x]\equiv P_t(x_1,x_2,...,x_L)$
\bea\label{forwardFP2}
&&\partial_t P_t[x] =  \sum_{u=1}^L \frac{\partial ^2}{\partial x_u^2} \Big( D_u[x] P_t[x] \Big)- \frac{\partial }{\partial x_u} \Big( F_u[x] P_t[x] \Big) \punkt\nn\\
\eea
Passing to the continuum limit, this yields the functional Fokker-Planck equation
\bea\label{forwardFP3}
&&\partial_t P_t[x] \\
&&=  \int \rmd u  \frac{\delta^2}{\delta x(u)^2} \Big( D_u[x] P_t[x] \Big)- \frac{\delta }{\delta x(u)} \Big( F_u[x] P_t[x] \Big) \punkt \nn
\eea

\paragraph{The backward Fokker-Planck equation:}

Let us   study  $P(\xfi,\tf|\xin,\ti)$ as a function of its initial time and position. To this purpose, 
 write down the exact equation, using the notations of \Eq{3.11},  
\be\label{3.18}
P(\xfi,\tf|x,t) = \left< P(\xfi,\tf|x +\rmd x,t+\rmd t)\right> .
\ee
The average is over all realizations of the noise $\eta$ during a time step $\rmd t$. 
Expanding inside the expectation value to first order in $\rmd t$ and second order in $\rmd x$, and   taking the expectation, we find 
\bea
\lefteqn{\left< P(\xfi,\tf|x +\rmd x,t+\rmd t) \right>}\nn\\
 &&= \Big< P(\xfi,\tf|x  ,t ) + 
\rmd t \,\partial_{t }P(\xfi,\tf|x  ,t )  \nn\\
&& \qquad + \rmd x\,  \partial_{x} P(\xfi,\tf|x  ,t )  + \frac{\rmd x ^{2}}2 \partial_{x}^{2} P(\xfi,\tf|x  ,t ) \Big> \nn\\
 &&=  P(\xfi,\tf|x  ,t ) + 
 \rmd t\Big[  \partial_{t }P(\xfi,\tf|x  ,t )  \nn\\
 && \qquad + F(x)  \, \partial_{x} P(\xfi,\tf|x  ,t ) +   D(x) \partial_{x}^{2} P(\xfi,\tf|x  ,t ) \Big] . \nn\\
 \eea
Comparing to \Eq{3.18} implies that 
the term of order $\rmd t$ vanishes, thus
\bea\label{backwardFP}
\lefteqn{ - \partial _t P(\xfi,\tf|x,t)  }\nn\\
\lefteqn{= F(x)\frac{\partial}{\partial x} P(\xfi,\tf|x,t) + D(x) \frac{\partial^2}{\partial x^2}   P(\xfi,\tf|x,t).}
\eea
This is the {\em backward Fokker-Planck equation}. Note that contrary to the forward equation, all derivatives act   on $P(\xfi,\tf|y,t)$,   not on $F$ or $D$. 

\paragraph{Remark on Consistency:}
The   form of the backward and forward equations is constraint by an important consistency relation: Using that the process is Markovian, we can write the {\em Chapman-Kolmogorov equation}
\be
P(\xfi,\tf|\xin,\ti) = \int_x P(\xfi,\tf|x,t)  P(x,t|\xin,\ti) .
\ee
This relation must hold for all $t$ between $\ti$ and $\tf$. Taking a $t$ derivative and using the backward Fokker-Planck equation for the first propagator $ P(\xfi,\tf|x,t) $, and the forward equation for the second $ P(x,t|\xin,\ti) $, we find cancelation of all tems upon partial integration in $x$, due to the specific arrangement of the derivatives in \Eqs{forwardFP} and \eq{backwardFP}.

\paragraph{Remark on Steady State:} Let us find a steady-state solution of \Eq{forwardFP}, i.e.\ a solution which does not depend on time. Integrating once and dropping the time argument yields
\be
\frac{\partial}{\partial x} \big[ D(x) P(x) \big]=    F(x) P(x) + \mbox{const.}
\ee  
Let us   suppose that the probability $P(x)$ vanishes when $x\to \infty$. This implies that  the constant vanishes. The solution is  obtained as ($x_{0}$ is arbitrary)
\bea\label{FP-ss1}
&&P(x) = \frac{\ca N}{D(x)}\exp \left(\int_{x_0}^{x} \frac{F(y)}{D(y)}\rmd y \right),\\
&& {\cal N}^{{-1}}= \int_{-\infty}^{{\infty}}\rmd x\,  \frac{1}{D(x)}\exp \left(\int_{x_0}^{x} \frac{F(y)}{D(y)}\rmd y\right).
\eea
The simplest case is obtained for thermal  noise, i.e.\ $D(x)=T$, and when the force $F(x)$ is the derivative of a potential, $F(x) = -V'(x)$. \Eq{FP-ss1} can then be written as
\be\label{FP-ss2}
P(x) = {\cal N} \rme^{-V(x)/T}, \quad {\cal N}^{{-1}}= \int_{-\infty}^{{\infty}}\rmd x\, \rme^{-V(x)/T}.
\ee
This is Boltzmann's law \cite{Boltzmann1868}.

\subsection{Martin-Siggia-Rose (MSR) formalism}
\label{MSR-formalism}
\paragraph{The path integral:}

The transition probability \eq{Ptransition} from $x'$ to $x$  was 
\be\label{3.25}
{\cal P}_{\tau} (x|x') \rmd x = \frac{\rmd x}{\sqrt{4\pi \tau D(x')}}{\rme^{-\frac{[x-x'-\tau F(x')]^2}{4 \tau D(x')}}}.
\ee
This is ugly: our standard field-theory calculations work with polynomials in the exponential.
We therefore rewrite this measure as
\bea\label{P-slice}
&& {\cal P}_{\tau} (x|x') \rmd x = \rmd x \int_{-i \infty}^{i \infty}\frac{\rmd \tilde x}{2\pi i} \rme^{{-{\cal S}_\tau}[x,\tilde x]},\\
&& {\cal S}_\tau[x,\tilde x] = \tilde x \big(  x-x' - \tau F(x') \big) - \tau \tilde x^2 D(x').
\eea
The term ${\cal S}_\tau[x,\tilde x] $ is termed {\em action}. Reassembling all time slices, 
it is normally written in the limit of $\tau\to 0 $ as 
\bea
&& {\cal P} (x |x_0)  = \!\int_{x(0)=x_0 }^{x(N)=x} {\cal D} [x] {\cal D} [\tilde x] \rme^{-{\cal S}[x,\tilde x]},\\
&& {\cal S}[x,\tilde x] =\! \int_t \tilde x(t)\left[ \dot x(t) {-} F\big(x(t)\big) \right] - \tilde x(t)^2 D\big(x(t)\big),\\
&& {\cal D}[x]{\cal D}[\tilde x] = \prod_{i=1}^N \int_{-\infty}^{\infty}\rmd x_i \int_{-i \infty}^{i \infty}\frac{\rmd \tilde x_i}{2\pi i}.
\eea
This is known as the MSR formalism (Martin-Siggia-Rose) \cite{MSR}, the action also as Martin-Siggia-Rose-Janssen-DeDominicis action, in honor of their respective work \cite{Janssen1976,Janssen1985,Janssen1992,DeDominicis1976} .

\paragraph{Changing the discretization:}
Let us   turn back to a single time slice, as given in \Eq{P-slice}. 
The variables $\tilde x$ and $x$ are conjugate, i.e.
\bea
&&\int \frac{\rmd x\,\rmd \tilde x}{2\pi i} \rme^{-\tilde x (x-x')}\tilde x^n f(x,\tilde x) \nn\\
&&=\int \frac{\rmd x\,\rmd \tilde x}{2\pi i} \rme^{-\tilde x (x-x')}  \partial_x^n f(x,\tilde x )\ , \\
&&\int \frac{\rmd x\,\rmd \tilde x}{2\pi i} \rme^{-\tilde x (x-x')}  (x-x')^n f(x,\tilde x) \nn\\
&&=\int \frac{\rmd x\,\rmd \tilde x}{2\pi i} \rme^{-\tilde x (x-x')} \partial_{\tilde x}^n f(x,\tilde x) .
\eea
We can thus change our discretization scheme, i.e.\ replace
$F(x')\to F(\bar x)$, $D(x')\to D(\bar x)$, where
\be\label{xbar}
\bar x = \alpha x +(1-\alpha) x'\ , \qquad \ 0\le \alpha \le 1.
\ee 
There are  cases where this change is advantageous. On the other hand, the microscopic dynamics may be such that $F$ and $D$ depend on $\bar x$ instead of $x'$ (see end of section \ref{s:Markov chains,  Langevin equation, inertia}).

The consequences of the reparametrization \eq{xbar} is   understood  from the following example: Expand $\rme^{\tau \tilde x F(\bar x)}-1$ to linear order in $\tau$, 
\bea
&&\tau \int \frac{\rmd x\,\rmd \tilde x}{2\pi i} \rme^{-\tilde x (x-x')} \tilde x F(\bar x) f(x,\tilde x) \nn\\
&&= \tau \int \frac{\rmd x\,\rmd \tilde x}{2\pi i} \rme^{-\tilde x (x-x')} \partial_x \left[  F(\bar x) f(x,\tilde x) \right]. 
\eea
As the derivative   acts  on $F(\bar x)$, this depends on $\alpha$, as $\partial_x \bar x = \alpha$. Luckily, we can compensate this by an explicit $x$ derivative: Wherever we   change $x\to \bar x$, we also  replace  $\tilde x$ by $\tilde x- \partial_x $. This yields for the action of a single time slice
\bea\label{e:MSR}
&&{\cal S}_\tau[x,\tilde x] \nn\\
&&=\displaystyle \tilde x \big(  x-x')-\tau (\tilde x-\partial_x)    F(\bar x)  - \tau (\tilde x-\partial_x)^2 D(\bar x) \nn \\
&&= \displaystyle \alpha \tau F'(\bar x) - \alpha^2 \tau D''(\bar x) \nn\\
&& + \tilde x \big[  x-x'-\tau    F(\bar x) +2 \alpha \tau D'(\bar x) \big]  - \tau \tilde x^2 D(\bar x)  .
\eea
The noise-correlator $D(x)$ did not change, but there is an additional contribution to the force 
\be
F(x') \to F(\bar x) - 2\alpha D'(\bar x) .
\ee
The first two terms, $\alpha   F'(\bar x) - \alpha^2   D''(\bar x)$ can be interpreted as a change in the integration measure.  
Let us stress that the change in the action {\em leaves the physics of the problem invariant}. One may arrive at $\alpha = \frac12$ also when the bath is evolving more slowly than the time scale set by viscosity (see end of section \ref{s:Markov chains,  Langevin equation, inertia}). Then  the choice $\alpha=1/2$, known as {\em Stratanovich discretization}, is natural;  one can use the above procedure to get back to It\^o's discretization. 

\paragraph{Interpretation of the field $\tilde x(t)$:} 
Let us now turn to an interpretation of the two fields $x(t)$ and $\tilde x(t)$, and modify equation \eq{185bis} to 
\be\label{658}
\dot x(t) = F\big(x(t)\big) + \xi(t) + f \delta (t-t_0).
\ee 
Thus at time $t=t_0$, we kick the system with an infinitely small force $f$. 
Then, the probability changes by 
\bea
&&\partial_f\Big|_{f=0} {\cal P} (x |x_0)  = \partial_f\Big|_{f=0} \int_{ x(0) = x_0 }^{x(t)=x} {\cal D} [x] {\cal D} [\tilde x] \,\rme^{-{\cal S}[x,\tilde x]} \nn \\
&&=   \int_{x_0 = x(0)}^{x(t)=x} {\cal D} [x] {\cal D} [\tilde x] \, \tilde x(t_0) \rme^{-{\cal S}[x,\tilde x]}
\eea
Multiplying with $x(t)$ and integrating over all final configurations, we obtain
\be\label{response}
\highlight{
R(t,t_0) = \partial_f\Big|_{f=0} \left< x(t) \right>  = \left <x(t)  \tilde x(t_0) \right> .}
\ee
The expectation is w.r.t\ the measure  ${\cal D} [x] {\cal D} [\tilde x] \,  \rme^{-{\cal S}[x,\tilde x]}$. 
As \Eq{response} is the response of the system to a change in force, $\tilde x$ is called {\em response field}, and $R(t,t_0)$ {\em  response function}. 
{\em Correlation functions} are similarly obtained as 
\be
C(t,t') = \left< x(t)x(t')\right> .
\ee
Since the probability is normalized,
\be
\int   {\cal P} (x |x_0)\, \rmd x = 1,
\ee
for all forces, one shows by taking derivatives w.r.t.\ forces at different times that expectations of the sole response field vanish, 
\be
 \left< \tilde x(t) \right> = \left< \tilde x(t)\tilde x(t') \right>  =  \left< \tilde x(t)\tilde x(t') \tilde x(t'')\right> = ... = 0.
\ee

\subsection{Gaussian theory with spatial degrees of freedom}
Consider theories with spatial dependence, and let us suppose that the energy is given by 
\be
{\cal H}[u] = \int_x \half \left[ \nabla u(x)\right]^2 + \frac {m^2}2 u(x)^2.
\ee
This corresponds to an elastic manifold inside a confining potential of curvature $m^2$. 
The {\em elastic forces} acting on a piece of the manifold at position $x$ are given by 
\be
F(x) = -\frac{\delta {\cal H}[u]}{\delta u(x)} = \left(  \nabla^2 - m^2 \right) u(x).
\ee
Its Langevin dynamics   reads 
\bea\label{Edwards-Wilkinson}
&&  \partial_{t} u (x,t) =  \left(   \nabla^2 - m^2 \right) u(x,t) +  \xi(x,t), \\ 
&&\left< \xi(t) \xi(t') \right> = 2  T \delta (t-t')\delta (x-x').
\eea
The action in It\^o discretization is
\bea\label{Ito-S0}
 {\cal S}[u,\tilde u] = \int_{x,t} \tilde u(x,t)\Big[ \partial_t  {-} \nabla^2{+}m^2 \Big]u(x,t) - 
T \tilde u(x,t)^2 . \nn\\
\eea
It can be diagonalized in momentum and frequency space, 
\bea
 {\cal S}[u,\tilde u] &=& \int_{k,\omega } \tilde u(-k,-\omega)\left[i \omega +k^2+m^2 \right]u(k,\omega) \nn\\
 && \quad ~~~ - 
T \tilde u(-k,-\omega) \tilde u(k,\omega)  \nn\\
&=&\half \int_{k,\omega} {\left( u(-k,-\omega) \atop \tilde u(-k,-\omega) \right)  }  {\cal M} {\left( u(k,\omega) \atop \tilde u(k,\omega)  \right) },
\eea
\bea
&& {\cal M} = \left( 
\begin{array}{cc}
0 & -i \omega + k^2 + m^2\\
i \omega + k^2 + m^2 &- 2T 
\end{array}
 \right) .
 \eea
 This implies  
\be{\cal M}^{-1} = \left( 
\begin{array}{cc}
\frac{2T}{(i \omega + k^2 + m^2)(-i \omega + k^2 + m^2)} & \frac1{i \omega + k^2 + m^2} \\
\frac{1\rule{0mm}{3mm}}{-i \omega + k^2 + m^2} & 0 
\end{array}
 \right) .
\ee
As a consequence, 
\bea
&&\highlight{
R(k,\omega) := \left<  u(-k,-\omega) \tilde u(k,\omega)\right> = \frac1{i \omega {+} k^2 {+} m^2}},\\
&&\highlight{C(k,\omega) := \left<  u(-k,-\omega)   u(k,\omega)\right> = \frac{2T}{|i \omega + k^2 + m^2|^2}} .\nn\\
\eea
Inverse Fourier transforming $R$ leads to 
\bea\label{R(k,t)}
&&\highlight{R(k,t) =} \left<  u(-k,t) \tilde u(k,0)\right> \nn\\
&&= \int_{-\infty}^{\infty}\frac{\rmd \omega}{2\pi }\frac{\rme^{i \omega t}}{i \omega + k^2 + m^2} = \highlight{\rme^{-( k^2+m^2 )t}\Theta(t).}
\eea
We used the residue theorem to evaluate the integral: There is a pole at $\omega = i (k^2+m^2)$, i.e.\ in the upper complex half-plane. If $t>0$, then the integral converges in the upper half plane, and closing the contour there yields the residuum as written. For $t<0$, one has to close the path in the lower half-plane, and there is no contribution,  thus the $\Theta(t)$. 

The response function \eq{R(k,t)} satisfies the {\em massive  diffusion equation}, 
\be
(\partial_t + k^2 + m^2 )R(k,t) = \delta(t).
\ee
Performing one more inverse Fourier transform yields the response function in real space, a.k.a.\ the diffusion kernel (we complete the square)
\bea
R(x,t) &=& \int_{-\infty}^\infty\frac{\rmd ^d k}{(2\pi)^d}\, \rme^{i k x -(k^2+m^2) t}\Theta(t) \nn\\
& =& \frac{\rme^{-m^2 t - \frac{x^2}{4t}}}{(4 \pi t)^{d/2}}\Theta(t).
\eea
Setting $d=1$ this is identical to \Eq{3.25} (setting there $\eta=D=1$,   $F=x'=0$, and $\tau =t$). 
Correlation functions can   be obtained as
\bea\label{Ckt}
\highlight {C(k,t-t')} &=& \left< u(k,t) u(-k,t')\right>  \nn\\
&=& 2 T\int_{-\infty}^\infty \rmd {\tau}\, R(k,t-\tau)R(k,t'-\tau) \nn\\
&=&2 T\int_{-\infty}^{\min (t,t')} \rmd {\tau}\,  \rme^{-(k^2+m^2)(t+t'-2 \tau)} \nn\\
&=& { \frac{T}{k^2+m^2}\, \rme^{-(k^2+m^2)|t-t'|}}.
\eea\begin{figure}
\Fig{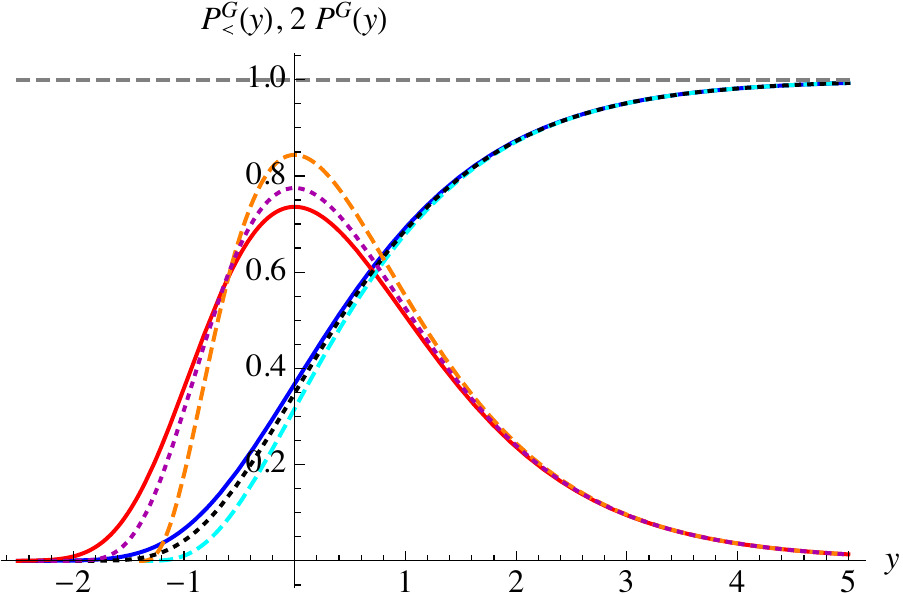}
\caption{The cumulative Gumbel distribution $P_<^{\rm G}(y)$ (blue, solid) and its derivative $P^{\rm G}(y)$ (red, solid, rescaled by a factor of 2 for better readability). This is compared to the law exact law \eq{PmaxN} for $N=4$ (dashed, note the bounded support) and $N=10$  (dotted). For $N=100$ no difference would be visible on this plot.}
\label{f:Gumbel}
\end{figure}For equal times one recovers the equilibrium correlator, 
\be
C(k,0) = \left< u(k,t) u(-k,t)\right> = \frac{T}{k^2+m^2}.
\ee 
The correlation function \eq{Ckt} satisfies the differential equation 
\be\label{pert-FDT}
\hl{\partial_t C(k,t-t') = T \left[ R(k,t'-t)-R(k,t-t') \right]. }
\ee
This relation is known as fluctuation-dissipation theorem. It is   more generally valid, see e.g.\ \cite{Janssen1985,Janssen1992}.

\subsection{The inverse of the Laplace operator.}\label{inv lap}
\begin{equation}\label{il1}
\nabla^2 \left[\frac{1}{ (2-d)S_{d}}|\vec z|^{2-d} \right]=\delta^{d} (\vec z),
\end{equation}
where the volume of the unitsphere is defined as
\be
S_d=  \frac{2 \pi^{d/2}}{\Gamma{(d/2)}}.
\ee
Proof:   $\nabla^2 \, |\vec z|^{2-d+\eta } = (2-d+\eta )\eta |\vec z|^{{-d+\eta
}} $. Integrating the last term against a test function $f(\vec z)$ yields $(2-d+\eta)S_df(0)$. Taking the limit of $\eta\to 0$   completes the proof.

\paragraph{The inverse Laplacian in $d=2$.} In $d=2$,  we set   $\vec z:=(x,y)$, and  $z=x+i y$, $\bar z = x- iy$. \Eq{il1} reduces to
\begin{equation}\label{inv lap d=2}
\nabla^2 \frac{\ln (\vec z^{\,2})}{4\pi } = \delta ^2 (\vec z) \ .
\end{equation}
Our notations imply $\ln (\vec z^{\,2})=   \ln (z \bar z)=    \ln z +
\ln \bar z  $. 
Since $\bar \partial \partial =\frac{1}{4}\nabla^2$ (check of norm: $\partial \bar\partial (z \bar z)=1$, $\nabla^2 (x^2{+}y^2)=4$), 
\begin{equation}\label{partial inv}
\frac{1}{\partial } =4\frac{\bar \partial }{\nabla^2} = \frac{ \bar \partial
\ln (z \bar z)}{\pi  } = \frac{1}{\pi \bar z}\ .
\end{equation}
As a consequence
\be
\partial \frac{1}{\pi \bar z} =  \bar \partial \frac{1}{\pi   z} = \delta^2(\vec z)   \equiv \delta(x)\delta(y).
\ee

\subsection{Extreme-value statistics:   Gumbel, Weibull and Fr\'echet distributions}
\subsubsection*{Generalities:}
Consider a random variable $x$ with probability distribution $P(x)$, and cumulative distributions 
\bea
 P_>(x) := \int_{x}^\infty P(y)\, \rmd y ,\\
P_<(x) := \int_{-\infty }^x P(y)\, \rmd y = 1-P_>(x).
\eea
Suppose $x_i$, $i=1,...,N$ are drawn from the measure $P(x)$. We are interested in the law of their maximum $m$, 
\be
m:= \max(x_1,...,x_N).
\ee
The probability that the maximum is smaller than $m$ is equivalent to the probability that $x_i<m$ for all $i$, 
\be\label{PmaxN}
P_<^{\rm max}(m) = P_<(m)^N = \left[1-  P_>(m)\right]^N.
\ee
For large $N$, this can be approximated by 
\be
P_<^{\rm max}(m)  \simeq \rme^{-N P_>(m)}, 
\ee
with density 
\be
P^{\rm max}(m) =  \partial_m P_<^{\rm max}(m) \simeq  N  P(m) \rme^{-N P_>(m)}.
\ee

\begin{figure}
\Fig{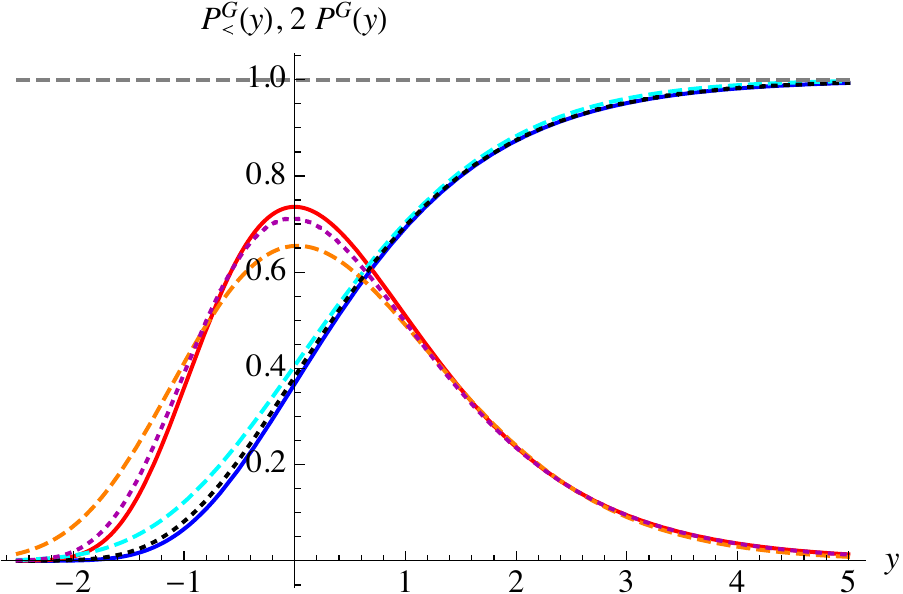}
\caption{The cumulative Gumbel distribution $P_<^{\rm G}(y)$ (blue, solid) and its derivative $P^{\rm G}(y)$ (red, solid).   This is compared to the   exact law \eq{PmaxN} for a Gauss-distribution, using the equality in \Eq{764},    for $N=100$ (dashed) and $N=10^{13}$  (dotted). }
\label{f:Gumbel-Gauss}
\end{figure}

\subsubsection*{Gumbel distribution:} Suppose that 
\be
P(x) = \rme^{-x}\Theta(x ) ~~ \Leftrightarrow ~~ P_>(x) = \rme^{-x}\Theta(x).
\ee
This implies that for large $N$
\be
P_<^{\rm max}(m) \simeq \rme^{-N \rme^{-m}} \Theta(m) = \rme^{-\rme^{-m+\ln( N)}}  \Theta(m).
\ee
The variable 
\be
y= m- \ln (N)  \ee
is distributed according to a Gumbel distribution \cite{Gumbel1935}
\bea
P_<^{\rm G}(y) = \rme^{- \rme^{-y}}, \quad 
 P^{\rm G}(y) = \partial_y P_<^{\rm G}(y) =  \rme^{-y- \rme^{-y}} . 
\eea
A plot elucidating the convergence is shown in Fig.~\ref{f:Gumbel}.
The Gumbel class has a large basin of attraction, encompassing all distributions which decay as $P_>(m) \sim \rme^{-x^\alpha}$, $\alpha>0$, including in particular the Gauss distribution. The idea is that a particular point $x_{\rm c}$ in the distribution of $P(x)$ will dominate $P^{\rm max}(m)$; it then suffices to approximate $\ln P_>(x)$ by a linear fit at $m=m_{\rm c}$. 
For the standard Gauss-distribution 
\bea\label{764}
P(x)= \frac{\rme^{-x^2/2}}{\sqrt{2\pi}}, \\
 P_>(x)= \frac{1}{2}
   \text{erfc}\Big(\frac{x}{\sqrt{2}}\Big)\simeq \frac{\rme^{-x^2/2}}{\sqrt{2\pi}x} .
\eea
A strategy is to replace $x^{-1}\rme^{-x^2/2} \to x_{\rm c}^{-1} \rme^{-x_{\rm c}^2/2-x x_{\rm c}}$, and then to find the best $x_{\rm c}$ to eliminate
the $N$-dependence. This yields after some algebra
\be\label{756}
y \simeq x \sqrt{  \ln (N^2)} -  \ln (N^2) + \frac12 \ln \big(2\pi \ln (N^2) \big).
\ee 
A numerical check reveals a very slow convergence to the asymptotic form: while the right tail and the center of the density are correct even for small $N$,  the left tail converges   very slowly (from above), while the peak amplitude converges   slowly from below. Note that this could not be repaired by changing the parameters in \Eq{756}, which work for the peak-position and the right tail. 

\begin{figure}
\Fig{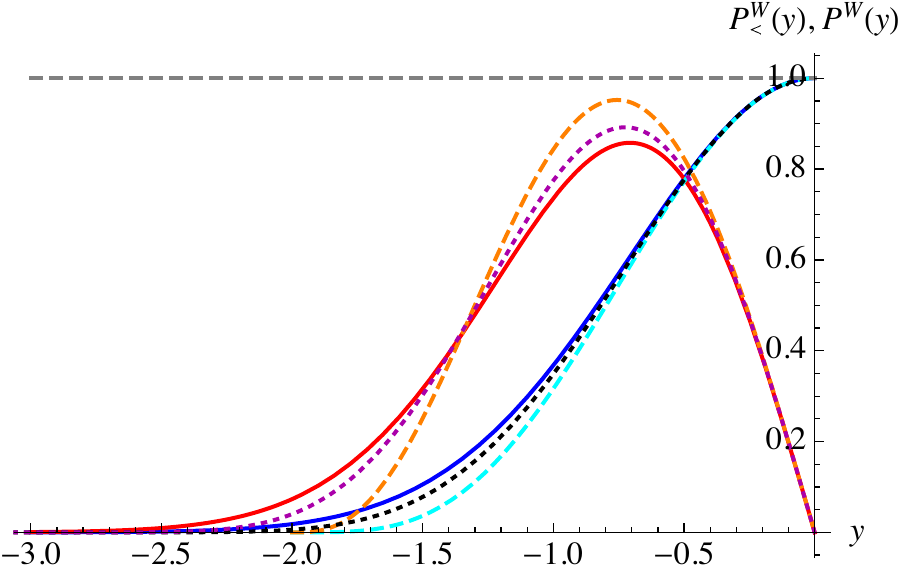}
\caption{The cumulative Weibull distribution $P_<^{\rm W}(y)$ (blue, solid) and its derivative $P^{\rm W}(y)$ (red, solid) for $\alpha=2$. This is compared to the   exact law \eq{PmaxN} for $N=4$ (dashed) and $N=10$  (dotted). For $N=100$ virtually no difference would be visible on this plot.}
\label{f:Weibull}
\end{figure}
\subsubsection*{Weibull distribution:}
Suppose that $P(x)$ is distributed according to a power-law,  {\em  bounded from above} by  $x=0$, 
\bea
P(x) = \alpha (-x)^{\alpha-1} \, \Theta(-1\le x\le 0), \\
 P_>(x) = (-x)^\alpha \,\Theta(-1 \le x\le 0) .
\eea
This implies that for large $N$ (we suppress the lower bound at $x=-1$ for compactness of notation, and since it does not matter in the final result)
\bea
P_<^{\rm max}(m) \simeq \rme^{-N (-m)^\alpha} \Theta(m) ,\nn\\
= \exp\left(- \left[{- m N^{ \frac 1 \alpha}}\right]^\alpha \right) \Theta(-m).
\eea
The variable 
\be
y= m {N^{\frac 1 \alpha}}  \ee
is distributed according to a Weibull distribution \cite{Weibull1951} with index $\alpha$
\bea
P_{\alpha,<}^{\rm W}(y) = \rme^{- (-y)^\alpha} \Theta(-y), 
\\ 
 P_{\alpha}^{\rm W}(y) = \partial_y P_<^{\rm W}(y) = \alpha (-y)^{\alpha-1} \rme^{-(-y)^\alpha} \Theta(-y). 
\eea

\subsubsection*{Fr\'echet distribution:}
Suppose that $P(x)$ is distributed according to an unbounded power law, $\alpha>0$, 
\bea
P(x) = \alpha x^{\alpha-1} \, \Theta(x>1), \\
 P_>(x) =  x^{-\alpha} \,\Theta(x>1) .
\eea
This implies that for large $N$ 
\bea
P_<^{\rm max}(m) \simeq \rme^{-N m^{-\alpha}} \Theta(m>1) \nn\\
= \exp\left(- \left[{m N^{ -\frac 1 \alpha}}\right]^{-\alpha} \right) \Theta(m>1).
\eea
The variable 
\be
y= m {N^{-\frac 1 \alpha}}  \ee
is distributed according to a Fr\'echet distribution \cite{Frechet1927} with index $\alpha$, 
\bea
P_{\alpha,<}^{\rm F}(y) = \rme^{- y^{-\alpha}} \Theta(y), 
\\ 
 P_{\alpha}^{\rm F}(y) = \partial_y P_<^{\rm W}(y) = \alpha y^{-\alpha-1} \rme^{-y^{-\alpha}} \Theta(y). 
\eea
Note that $P_{\alpha,<}^{\rm F}(y)$ has an algebraic tail $\sim y^{-\alpha}$, thus  decays much more slowly than the Gumbel
distribution for large $y$.

\end{reviewKay}

\begin{figure}[t]
\Fig{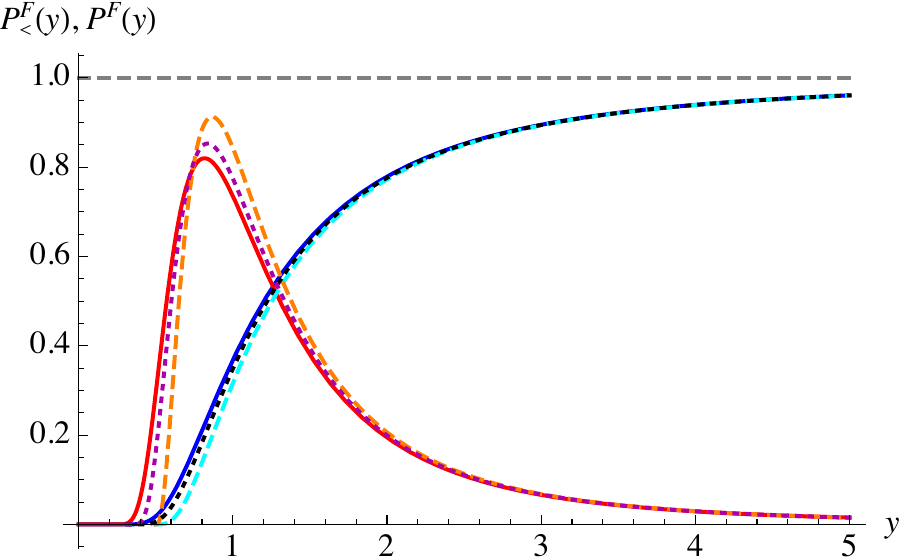}
\caption{The cumulative Fr\'echet distribution $P_<^{\rm W}(y)$ (blue, solid) and its derivative $P^{\rm W}(y)$ (red, solid) for $\alpha=2$. This is compared to the   exact law \eq{PmaxN} for $N=4$ (dashed) and $N=10$  (dotted). For $N=100$ virtually no difference would be visible on this plot.}
\label{f:Frechet}
\end{figure}

\subsection{Gel'fand Yaglom method}
\label{s:Gelfand-Yaglom}
We want to  compute   functional determinants of the  form
\begin{eqnarray}
f(\alpha,m^2):= \frac{\det [ - \nabla^2 + \alpha V(x) + m^2 ] }{
  \det [ - \nabla^2 + m^2 ] } ,
\end{eqnarray}
with  Dirichlet boundary conditions at $x=0$ and $x=L$, at $\alpha=1$. In order for the problem to be well-defined,  $-\nabla^2 +\alpha V(x)+m^2$ must have a discrete spectrum.

In dimension $d=1$, 
this can  efficiently be  calculated   using the Gel'fand Yaglom method \cite{GelfandYaglom1960}.
Consider   solutions of the   ODE
\begin{eqnarray}
&&[ - \nabla^2 + \alpha V(x) + m^2 ]\psi_\alpha(x)=0, \label{362} 
\end{eqnarray}
with   boundary conditions
\begin{equation}\label{bc}
\psi_\alpha(0)=0,\qquad
\psi_\alpha'(0)=1\punkt
\end{equation}
Define 
\be\label{364}
g(\alpha,m^2) :=  \frac{\psi_\alpha(L)}{{\psi}_0(L)} \punkt
\ee
Then the ratio of   determinants is given  by
\begin{equation}
f(\alpha,m^2)   =   g(\alpha,m^2)\punkt
\end{equation}
\underline{\bf  Proof:} 
Set $\Lambda_\alpha:= - \nabla^2 + \alpha V(x) + m^2 $. Call its eigenvalues $\lambda_i(\alpha)$, ordered, and non-degenerate. 
Consider the analytic structure of $f(\alpha, m^2-\lambda)$ and $g(\alpha,m^2-\lambda)$, as a function of $\lambda$. 
By definition $f$ is a product over   eigenvalues, 
\be\label{366}
f(\alpha, m^2-\lambda) = \prod_{i}\frac{\lambda_i(\alpha)-\lambda}{\lambda_i(0)-\lambda}\punkt
\ee
Note that for large $i$ the ratio $\lambda_i(\alpha)/\lambda_i(0)$ goes to 1, thus the product should converge; that was the reason why the ratio of determinants was introduced in the first place. If we want to make the proof   rigorous, we can put the system on a lattice, replacing the Laplacian by its lattice version. Then the spectrum is finite, and the product converges. 
As a consequence of \Eq{366}, $f(\alpha, m^2-\lambda)$ is an analytic function of $\lambda$, which vanishes at  $\lambda=\lambda_i(\alpha)$. 
Now consider $g(\alpha, m^2-\lambda)$. If $\lambda$ is an eigenvalue, $\lambda=\lambda_i(\alpha)$, then the solution of \Eq{362}   vanishes at $x=L$. Playing around with solutions of differential equations, we can convince ourselves that for  $\lambda-\lambda_i(\alpha)\to 0$,
\be
\psi_\alpha(L) \sim \lambda-\lambda_i(\alpha)\punkt
\ee 
We further expect $g$ to be analytic in $\lambda$.
Thus, as a function of $\lambda$, $f$ and $g$ have the same analytic structure, i.e.\ the same zeros and poles. The latter cancel in  the ratio
\be
r(\lambda):=
\frac{f(\alpha,m^2-\lambda)}{g(\alpha,m^2-\lambda)}\punkt
\ee
The only possibility for a zero or a pole we have to check  is for $|\lambda|\to \infty$. 

Consider   $f$    in the limit of $\lambda\to -\infty$: Each factor in  \eq{366} will go to 1, s.t.\ also $f$   goes to 1. 
(For the discretized version, this is evident, and does not depend on the phase of $\lambda$; for the continuous version one has to work a little bit, and take the limit away from the positive real axes, where the spectrum lies.)

Now consider the differential equation \eq{362} with $m^2\to m^2-\lambda$, in the same limit $\lambda\to -\infty$. In this case, one can convince oneself that both solutions grow exponentially, and that $V(x)$ is a small perturbation, s.t.\ again $g(\alpha,m^2-\lambda) \to 1$. Thus    $r(\lambda)$ is a    function in the complex plane which has no poles. As a consequence,   $r(\lambda)$ is bounded. According to Liouville's theorem it is a constant. 
This constant can be extracted from both limits   $\lambda \to \infty$ and $\lambda=0$, which shows that 
$
r(\lambda) =1.
$
This concludes our proof.

A more rigorous proof can be found in \cite{LevitSmilansky1977}: The idea there is to show that $\partial_\alpha f(\alpha,m^2) =\partial_\alpha g(\alpha,m^2) $ for all $\alpha$, as both can be written as Green functions at the given value of $\alpha$. 
A proof similar to ours, using Fredholm-determinant theory, can be found in section 7, appendix 1 of \cite{ColemannBook}.

\section*{References}
\newcommand{\doi}[2]{\href{http://dx.doi.org/#1}{#2}}
\newcommand{\arxiv}[1]{\href{http://arxiv.org/abs/#1}{#1}}
\newcommand{\link}[2]{\href{http://#1}{#2}}


\begin{thebibliography}{100}

\bibitem{Amit}
D.J. Amit and V.~Martin-Mayor,
\newblock \doi{10.1142/5715}{\rm {\em Field Theory, the Renormalization Group,
  and Critical Phenomena}}\null,
\newblock World Scientific, Singapore, 3rd edition, 2005.

\bibitem{Zinn}
J.~Zinn-Justin,
\newblock \doi{10.1093/acprof:oso/9780199227198.001.0001}{\rm {\em Quantum
  Field Theory and Critical Phenomena}}\null,
\newblock Oxford University Press, Oxford, 1989.

\bibitem{CardyBook}
J.~Cardy,
\newblock \doi{10.1017/CBO9781316036440}{\rm {\em Scaling and Renormalization
  in Statistical Physics}}\null,
\newblock Cambridge University Press, 1996.

\bibitem{KardarBook}
M.~Kardar,
\newblock \doi{10.1017/CBO9780511815881}{\rm {\em Statistical Physics of
  Fields}}\null,
\newblock Cambridge University Press, 2007.

\bibitem{BrezinBook}
E.~Br\'ezin,
\newblock \doi{10.1017/CBO9780511761546}{\rm {\em Introduction to Statistical
  Field Theory}}\null,
\newblock Cambridge University Press, 2010.

\bibitem{Vasilev2004}
A.N. Vasil'ev,
\newblock \doi{10.1201/9780203483565}{\rm {\em The Field Theoretic
  Renormalization Group in Critical Behavior Theory and Stochastic
  Dynamics}}\null,
\newblock Chapman \& Hall/CRC, New York, 2004.

\bibitem{ParisiBook}
G.~Parisi,
\newblock {\em Statistical Field Theory},
\newblock Frontiers in Physics,
\newblock Addison-Wesely, 1988.

\bibitem{PelissettoVicari2002}
A.~Pelissetto and E.~Vicari,
\newblock {\em Critical phenomena and renormalization-group theory},
\newblock \doi{https://doi.org/10.1016/S0370-1573(02)00219-3}{\rm Phys. Rep.
  {\bf 368} (2002)   549--727}\null,
\newblock \arxiv{cond-mat/0012164}.

\bibitem{El-ShowkPaulosPolandRychkovSimmons-DuffinVichi2014}
S.~{El-Showk}, M.~F. {Paulos}, D.~{Poland}, S.~{Rychkov}, D.~Simmons-Duffin
  and A.~{Vichi},
\newblock {\em Solving the 3d {Ising} model with the conformal bootstrap ii.
  {$c$}-minimization and precise critical exponents},
\newblock \doi{10.1007/s10955-014-1042-7}{\rm J. Stat. Phys. {\bf 157} (2014)
  869--914}\null,
\newblock \arxiv{arXiv:1403.4545}.

\bibitem{El-ShowkPaulosPolandRychkovSimmons-DuffinVichi2012}
S.~El-Showk, M.F. Paulos, D.~Poland, S.~Rychkov, D.~Simmons-Duffin  and
  A.~Vichi,
\newblock {\em Solving the 3{D Ising} model with the conformal bootstrap},
\newblock \doi{10.1103/PhysRevD.86.025022}{\rm Phys. Rev. D {\bf 86} (2012)
  025022}\null.

\bibitem{ChesterLandryLiuPolandSimmons-DuffinNing-Su2019}
S.M. Chester, W.~Landry, J.~Liu, D.~Poland, D.~Simmons-Duffin  and A.~Vichi
  N.~Su,
\newblock {\em Carving out ope space and precise {$O(2)$} model critical
  exponents},
\newblock (2019),
\newblock \arxiv{arXiv:1912.03324}.

\bibitem{FerrenbergXuLandau2018}
A.~M. Ferrenberg, J.~Xu  and D.P. Landau,
\newblock {\em Pushing the limits of monte carlo simulations for the
  three-dimensional ising model},
\newblock \doi{10.1103/PhysRevE.97.043301}{\rm Phys. Rev. E {\bf 97} (2018)
  043301}\null.

\bibitem{ClisbyDunweg2016}
N.~Clisby and B.~D\"unweg,
\newblock {\em High-precision estimate of the hydrodynamic radius for
  self-avoiding walks},
\newblock \doi{10.1103/PhysRevE.94.052102}{\rm Phys. Rev. E {\bf 94} (2016)
  052102}\null.

\bibitem{Clisby2017}
N.~Clisby,
\newblock {\em Scale-free {Monte Carlo} method for calculating the critical
  exponent $\gamma$ of self-avoiding walks},
\newblock \doi{10.1088/1751-8121/aa7231}{\rm J. Phys. A {\bf 50} (2017)
  264003}\null.

\bibitem{LipaSwansonNissenGengWilliamsonStrickerChuiIsraelssonLarson2000}
J.A. Lipa, D.R. Swanson, J.A. Nissen, Z.K. Geng, P.R. Williamson, D.A.
  Stricker, T.C.P. Chui, U.E. Israelsson  and M.~Larson,
\newblock {\em Specific heat of helium confined to a 57-${\mu}m$ planar
  geometry near the lambda point},
\newblock \doi{10.1103/PhysRevLett.84.4894}{\rm Phys. Rev. Lett. {\bf 84}
  (2000)   4894--4897}\null.

\bibitem{Hasenbusch2019}
M.~Hasenbusch,
\newblock {\em {Monte Carlo} study of an improved clock model in three
  dimensions},
\newblock \doi{10.1103/PhysRevB.100.224517}{\rm Phys. Rev. B {\bf 100} (2019)
  224517}\null.

\bibitem{FranzJacquinParisiUrbaniZamponi2012}
S.~Franz, H.~Jacquin, G.~Parisi, P.~Urbani  and F.~Zamponi,
\newblock {\em Quantitative field theory of the glass transition},
\newblock \doi{10.1073/pnas.1216578109}{\rm PNAS {\bf 109} (2012)
  18725--18730}\null.

\bibitem{MuellerWyart2014}
M.~M\"uller and M.~Wyart,
\newblock {\em Marginal stability in structural, spin, and electron glasses},
\newblock \doi{10.1146/annurev-conmatphys-031214-014614}{\rm Annu. Rev.
  Condens. Matter Phys. {\bf 6} (2014)   177--200}\null.

\bibitem{NattermannScheidl2000}
T.~Nattermann and S.~Scheidl,
\newblock {\em Vortex-glass phases in type-{II} superconductors},
\newblock \doi{10.1080/000187300412257}{\rm Adv. Phys. {\bf 49} (2000)
  607--704}\null,
\newblock \arxiv{cond-mat/0003052}.

\bibitem{KierfeldNattermannHwa1997}
J.~Kierfeld, T.~Nattermann  and T.~Hwa,
\newblock {\em Topological order in the vortex-glass phase of high-temperature
  superconductors},
\newblock \doi{10.1103/PhysRevB.55.626}{\rm Phys. Rev. B {\bf 55} (1997)
  626--629}\null.

\bibitem{CarpentierLedoussalGiamarchi1996}
D.~Carpentier, P.~Ledoussal  and T.~Giamarchi,
\newblock {\em Stability of the {Bragg} glass phase in a layered geometry},
\newblock \doi{10.1209/epl/i1996-00123-2}{\rm Europhys. Lett. {\bf 35} (1996)
  379--384}\null.

\bibitem{CuleShapir1995}
D.~Cule and Y.~Shapir,
\newblock {\em Nonergodic dynamics of the two-dimensional random-phase
  sine-{Gordon} model: Applications to vortex-glass arrays and
  disordered-substrate surfaces},
\newblock \doi{10.1103/PhysRevB.51.3305}{\rm Phys. Rev {\bf B 51} (1995)
  3305}\null.

\bibitem{HwaFisher1994a}
T.~Hwa and D.S. Fisher,
\newblock {\em Vortex glass phase and universal susceptibility variations in
  planar arrays of flux lines},
\newblock \doi{10.1103/PhysRevLett.72.2466}{\rm Phys. Rev. Lett. {\bf 72}
  (1994)   2466--9}\null.

\bibitem{HwaFisher1994b}
T.~Hwa and D.S. Fisher,
\newblock {\em Anomalous fluctuations of directed polymers in random media},
\newblock \doi{10.1103/PhysRevB.49.3136}{\rm Phys. Rev. B {\bf 49} (1994)
  3136--54}\null,
\newblock \arxiv{cond-mat/9309016}.

\bibitem{Balents1993}
L.~Balents,
\newblock {\em Localization of elastic layers by correlated disorder},
\newblock \doi{10.1209/0295-5075/24/6/011}{\rm Europhys. Lett. {\bf 24} (1993)
   489--94}\null.

\bibitem{Feldman2002}
D.E. Feldman,
\newblock {\em Critical exponents of the random-field {O$(N)$} model},
\newblock \doi{10.1103/PhysRevLett.88.177202}{\rm Phys. Rev. Lett. {\bf 88}
  (2002)   177202}\null.

\bibitem{MiddletonFisher2002}
A.~Middleton and D.S. Fisher,
\newblock {\em Three-dimensional random-field {Ising} magnet: Interfaces,
  scaling, and the nature of states},
\newblock \doi{10.1103/PhysRevB.65.134411}{\rm Phys. Rev. B {\bf 65} (2002)
  134411}\null.

\bibitem{DahmenSethnaKuntzPerkovic2001}
K.A. Dahmen, J.P. Sethna, MC. Kuntz  and O.~Perkovic,
\newblock {\em Hysteresis and avalanches: phase transitions and critical
  phenomena in driven disordered systems},
\newblock \doi{10.1016/S0304-8853(00)00749-6}{\rm J. Magn. Magn. Mater. {\bf
  226} (2001)   1287--1292}\null.

\bibitem{DahmenSethnaPerkovic2000}
K.A. Dahmen, J.P. Sethna  and O.~Perkovic,
\newblock {\em Hysteresis, {Barkhausen} noise, and disorder induced critical
  behavior},
\newblock \doi{0.1109/20.908717}{\rm IEEE Transactions On Magnetics {\bf 36}
  (2000)   3150--3154}\null.

\bibitem{BricmontKupiainen1987}
J.~Bricmont and A.~Kupiainen,
\newblock {\em Lower critical dimension for the random-field {Ising} model},
\newblock \doi{10.1103/PhysRevLett.59.1829}{\rm Phys. Rev. Lett. {\bf 59}
  (1987)   1829--1832}\null.

\bibitem{Imbrie1984}
J.Z. Imbrie,
\newblock {\em Lower critical dimension of the random-field {Ising} model},
\newblock \doi{10.1103/PhysRevLett.53.1747}{\rm Phys. Rev. Lett. {\bf 53}
  (1984)   1747--50}\null.

\bibitem{ParisiSourlas1979}
G.~Parisi and N.~Sourlas,
\newblock {\em Random magnetic fields, supersymmetry, and negative dimensions},
\newblock \doi{10.1103/PhysRevLett.43.744}{\rm Phys. Rev. Lett. {\bf 43} (1979)
    744--5}\null.

\bibitem{LeDoussalWiese2005b}
P.~Le Doussal and K.J.\ Wiese,
\newblock {\em Random field spin models beyond one loop: a mechanism for
  decreasing the lower critical dimension},
\newblock \doi{10.1103/PhysRevLett.96.197202}{\rm Phys. Rev. Lett. {\bf 96}
  (2006)   197202}\null,
\newblock \arxiv{cond-mat/0510344}.

\bibitem{TissierTarjus2011}
M.~Tissier and G.~Tarjus,
\newblock {\em Supersymmetry and its spontaneous breaking in the random field
  ising model},
\newblock \doi{10.1103/PhysRevLett.107.041601}{\rm Phys. Rev. Lett. {\bf 107}
  (2011)   041601}\null.

\bibitem{TarjusTissier2008}
G.~Tarjus and M.~Tissier,
\newblock {\em Nonperturbative functional renormalization group for random
  field models and related disordered systems. {I. Effective} average action
  formalism},
\newblock \doi{10.1103/PhysRevB.78.024203}{\rm Phys. Rev. B {\bf 78} (2008)
  024203}\null.

\bibitem{TissierTarjus2008b}
M.~Tissier and G.~Tarjus,
\newblock {\em Nonperturbative functional renormalization group for random
  field models and related disordered systems. {II. Results} for the random
  field {$O(N)$} model},
\newblock \doi{10.1103/PhysRevB.78.024204}{\rm Phys. Rev. B {\bf 78} (2008)
  024204}\null.

\bibitem{TarjusTissier2005}
G.~Tarjus and M.~Tissier,
\newblock {\em A unified picture of ferromagnetism, quasi-long range order and
  criticality in random field models},
\newblock \doi{10.1103/PhysRevLett.96.087202}{\rm Phys. Rev. Lett. {\bf 96}
  (2006)   087202}\null,
\newblock \arxiv{cond-mat/0511096}.

\bibitem{TarjusTissier2006}
G.~Tarjus and M.~Tissier,
\newblock {\em Two-loop functional renormalization group of the random field
  and random anisotropy {$O(N)$} models},
\newblock \doi{10.1103/PhysRevB.74.214419}{\rm Phys. Rev. B {\bf 74} (2006)
  214419}\null,
\newblock \arxiv{cond-mat\slash{\bf 0606698}}.

\bibitem{TarjusTissier2004}
G.~Tarjus and M.~Tissier,
\newblock {\em Nonperturbative functional renormalization group for
  random-field models: The way out of dimensional reduction},
\newblock \doi{10.1103/PhysRevLett.93.267008}{\rm Phys. Rev. Lett. {\bf 93}
  (2004)   267008}\null,
\newblock \arxiv{cond-mat/0410118}.

\bibitem{HusemannWiese2017}
C.~Husemann and K.J. Wiese,
\newblock {\em Field theory of disordered elastic interfaces to 3-loop order:
  Results},
\newblock \doi{10.1016/j.nuclphysb.2018.04.015}{\rm Nucl. Phys. B {\bf 932}
  (2018)   589--618}\null,
\newblock \arxiv{arXiv:1707.09802}.

\bibitem{WieseHusemannLeDoussal2018}
K.J. Wiese, C.~Husemann  and P.~{Le Doussal},
\newblock {\em Field theory of disordered elastic interfaces at 3-loop order:
  The $\beta$-function},
\newblock \doi{10.1016/j.nuclphysb.2018.04.013}{\rm Nucl. Phys. B {\bf 932}
  (2018)   540--588}\null,
\newblock \arxiv{arXiv:1801.08483}.

\bibitem{WieseLeDoussal2006}
K.J. Wiese and P.~Le Doussal,
\newblock {\em Functional renormalization for disordered systems: Basic recipes
  and gourmet dishes},
\newblock \link{math-mprf.org/journal/articles/id1143/}{Markov Processes Relat.
  Fields} {\bf 13} (2007)   777--818,
\newblock \arxiv{cond-mat/0611346}.

\bibitem{HuiTang2006}
S.~Hui and L.-H. Tang,
\newblock {\em Ground state and glass transition of the {RNA} secondary
  structure},
\newblock \doi{10.1140/epjb/e2006-00347-x}{\rm Eur. Phys. J. B {\bf 53} (2006)
   77--84}\null.

\bibitem{FedorenkoLeDoussalWiese2006b}
A.~Fedorenko, P.~{Le~Doussal}  and K.J. Wiese,
\newblock {\em Statics and dynamics of elastic manifolds in media with
  long-range correlated disorder},
\newblock \doi{10.1103/PhysRevE.74.061109}{\rm Phys. Rev. E {\bf 74} (2006)
  061109}\null,
\newblock \arxiv{cond-mat/0609234}.

\bibitem{Wiese2004}
K.J.\ Wiese,
\newblock {\em Supersymmetry breaking in disordered systems and relation to
  functional renormalization and replica-symmetry breaking},
\newblock \doi{10.1088/0953-8984/17/20/016}{\rm J. Phys.: Condens. Matter. {\bf
  17} (2005)   S1889--S1898}\null,
\newblock \arxiv{cond-mat/0411656}.

\bibitem{Wiese2005}
K.J. Wiese,
\newblock {\em Why one needs a functional renormalization group to survive in a
  disordered world},
\newblock \link{www.ias.ac.in/article/fulltext/pram/064/05/0817-0827}{Pramana}
  {\bf 64} (2005)   817--827,
\newblock \arxiv{cond-mat/0511529}.

\bibitem{RepainBauerJametFerreMouginChappertBernas2004}
V.~Repain, M.~Bauer, J.-P. Jamet, J.~Ferr\'{e}, A.~Mougin, C.~Chappert  and
  H.~Bernas,
\newblock {\em Creep motion of a magnetic wall: Avalanche size divergence},
\newblock \doi{10.1209/epl/i2004-10213-7}{\rm EPL {\bf 68} (2004)
  460--466}\null.

\bibitem{BolechRosso2004}
C.J. Bolech and A.~Rosso,
\newblock {\em Universal statistics of the critical depinning force of elastic
  systems in random media},
\newblock \doi{10.1103/PhysRevLett.93.125701}{\rm Phys. Rev. Lett. {\bf 93}
  (2004)   125701}\null,
\newblock \arxiv{cond-mat/0403023}.

\bibitem{LeDoussalWiese2003b}
P.~Le Doussal and K.J. Wiese,
\newblock {\em Functional renormalization group at large ${N}$ for disordered
  elastic systems, and relation to replica symmetry breaking},
\newblock \doi{10.1103/PhysRevB.68.174202}{\rm Phys. Rev. B {\bf 68} (2003)
  174202}\null,
\newblock \arxiv{cond-mat/0305634}.

\bibitem{LeDoussalWiese2003a}
P.~Le Doussal and K.J. Wiese,
\newblock {\em Higher correlations, universal distributions and finite size
  scaling in the field theory of depinning},
\newblock \doi{10.1103/PhysRevE.68.046118}{\rm Phys. Rev. E {\bf 68} (2003)
  046118}\null,
\newblock \arxiv{cond-mat/0301465}.

\bibitem{Wiese2003a}
K.J. Wiese,
\newblock {\em The functional renormalization group treatment of disordered
  systems: a review},
\newblock \doi{10.1007/s00023-003-0940-z}{\rm Ann. Henri Poincar\'e {\bf 4}
  (2003)   473--496}\null,
\newblock \arxiv{cond-mat/0302322}.

\bibitem{Wiese2002}
K.J. Wiese,
\newblock {\em Disordered systems and the functional renormalization group: {A}
  pedagogical introduction},
\newblock Acta Physica Slovaca {\bf 52} (2002)   341,
\newblock \arxiv{cond-mat/0205116}.

\bibitem{RossoKrauth2001b}
A.~Rosso and W.~Krauth,
\newblock {\em Origin of the roughness exponent in elastic strings at the
  depinning threshold},
\newblock \doi{10.1103/PhysRevLett.87.187002}{\rm Phys. Rev. Lett. {\bf 87}
  (2001)   187002}\null,
\newblock \arxiv{cond-mat/0104198}.

\bibitem{CuleHwa1998}
D.~Cule and T.~Hwa,
\newblock {\em Static and dynamic properties of inhomogeneous elastic media on
  disordered substrate},
\newblock \doi{10.1103/PhysRevB.57.8235}{\rm Phys. Rev. B {\bf 57} (1998)
  8235--53}\null.

\bibitem{Derrida1980}
B.~Derrida,
\newblock {\em Random-energy model: Limit of a family of disordered models},
\newblock \doi{10.1103/PhysRevLett.45.79}{\rm Phys. Rev. Lett. {\bf 45} (1980)
   79--82}\null.

\bibitem{FisherHuse1986}
D.S. Fisher and D.A. Huse,
\newblock {\em Ordered phase of short-range {Ising} spin-glasses},
\newblock \doi{10.1103/PhysRevLett.56.1601}{\rm Phys. Rev. Lett. {\bf 56}
  (1986)   1601--4}\null.

\bibitem{MezardParisiVirasoroBook}
M.~M\'ezard, G.~Parisi  and M.A. Virasoro,
\newblock \doi{10.1142/0271}{\rm {\em Spin Glas Theory and Beyond}}\null,
\newblock World Scientific, Singapore, 1987.

\bibitem{KirkpatrickSherrington1978}
S.~Kirkpatrick and D.~Sherrington,
\newblock {\em Infinite-ranged models of spin-glasses},
\newblock \doi{10.1103/PhysRevB.17.4384}{\rm Phys. Rev. B {\bf 17} (1978)
  4384--4403}\null.

\bibitem{SherringtonKirkpatrick1975}
D.~Sherrington and S.~Kirkpatrick,
\newblock {\em Solvable model of a spin-glass},
\newblock \doi{10.1103/PhysRevLett.35.1792}{\rm Phys. Rev. Lett. {\bf 35}
  (1975)   1792--1796}\null.

\bibitem{Parisi1979}
G.~Parisi,
\newblock {\em Infinite number of order parameters for spin-glasses},
\newblock \doi{10.1103/PhysRevLett.43.1754}{\rm Phys. Rev. Lett. {\bf 43}
  (1979)   1754--1756}\null.

\bibitem{MezardParisiSourlasToulouseVirasoro1984}
M.~M\'ezard, G.~Parisi, N.~Sourlas, G.~Toulouse  and M.~Virasoro,
\newblock {\em Nature of the spin-glass phase},
\newblock \doi{10.1103/PhysRevLett.52.1156}{\rm Phys. Rev. Lett. {\bf 52}
  (1984)   1156--1159}\null.

\bibitem{MezardParisiVirasoro1985}
M.~M\'ezard, G.~Parisi  and M.A. Virasoro,
\newblock {\em Random free energies in spin glasses},
\newblock \doi{10.1051/jphyslet:01985004606021700}{\rm J. Physique Lett. {\bf
  46} (1985)   217--222}\null.

\bibitem{CugliandoloKurchan1993}
L.F. Cugliandolo and J.~Kurchan,
\newblock {\em Analytical solution of the off-equilibrium dynamics of a
  long-range spin-glass model},
\newblock \doi{10.1103/PhysRevLett.71.173}{\rm Phys. Rev. Lett. {\bf 71} (1993)
    173--6}\null.

\bibitem{Guerra2003}
F.~Guerra,
\newblock {\em Broken replica symmetry bounds in the mean field spin glass
  model},
\newblock \doi{10.1007/s00220-002-0773-5}{\rm Commun, Math. Phys. {\bf 233}
  (2003)   1--12}\null.

\bibitem{Talagrand2011a}
M.~Talagrand,
\newblock \doi{10.1007/978-3-642-15202-3}{\rm {\em Mean Field Models for Spin
  Glasses}}\null, {\em {\em Volume} I: Basic Examples},
\newblock Springer Verlag, Berlin, Heidelberg, 2011.

\bibitem{Talagrand2011b}
M.~Talagrand,
\newblock \doi{10.1007/978-3-642-15202-3}{\rm {\em Mean Field Models for Spin
  Glasses}}\null, {\em {\em Volume} II: Advanced Replica-Symmetry and Low
  Temperature},
\newblock Springer Verlag, Berlin, Heidelberg, 2011.

\bibitem{Panchenko2013}
D.~Panchenko,
\newblock \doi{10.1007/978-1-4614-6289-7}{\rm {\em {The
  Sherrington-Kirkpatrick} Model}}\null,
\newblock Springer, Berlin, Heidelberg, 2013.

\bibitem{Barkhausen1919}
H.~Barkhausen,
\newblock {\em {Zwei mit Hilfe der neuen Verst\"arker entdeckte
  Erscheinungen}},
\newblock Phys. Ztschr. {\bf 20} (1919)   401--403.

\bibitem{CizeauZapperiDurinStanley1997}
P.~Cizeau, S.~Zapperi, G.~Durin  and H.~Stanley,
\newblock {\em {Dynamics of a Ferromagnetic Domain Wall and the Barkhausen
  Effect}},
\newblock \doi{10.1103/PhysRevLett.79.4669}{\rm Phys. Rev. Lett. {\bf 79}
  (1997)   4669--4672}\null.

\bibitem{DurinBohnCorreaSommerDoussalWiese2016}
G.~Durin, F.~Bohn, M.A. Correa, R.L. Sommer, P.~Le Doussal  and K.J. Wiese,
\newblock {\em Quantitative scaling of magnetic avalanches},
\newblock \doi{10.1103/PhysRevLett.117.087201}{\rm Phys. Rev. Lett. {\bf 117}
  (2016)   087201}\null,
\newblock \arxiv{arXiv:1601.01331}.

\bibitem{LeDoussalWieseMoulinetRolley2009}
P.~Le Doussal, K.J. Wiese, S.~Moulinet  and E.~Rolley,
\newblock {\em Height fluctuations of a contact line: {A} direct measurement of
  the renormalized disorder correlator},
\newblock \doi{10.1209/0295-5075/87/56001}{\rm EPL {\bf 87} (2009)
  56001}\null,
\newblock \arxiv{arXiv:0904.4156}.

\bibitem{PonsonBonamyBouchaud2007}
L.~Ponson, D.~Bonamy  and E.~Bouchaud,
\newblock {\em Method and system for determining the propagation path of at
  least one crack from one or more fracture surfaces created by said crack(s)},
\newblock French patent {\bf FR:2892811} (2007).

\bibitem{BonamyPonsonPradesBouchaudGuillot2006}
D.~Bonamy, L.~Ponson, S.~Prades, E.~Bouchaud  and C.~Guillot,
\newblock {\em Scaling exponents for fracture surfaces in homogenous glass and
  glassy ceramics},
\newblock \doi{10.1103/PhysRevLett.97.135504}{\rm Phys. Rev. Lett. {\bf 97}
  (2006)   135504}\null.

\bibitem{PonsonBonamyBouchaud2006}
L.~Ponson, D.~Bonamy  and E.~Bouchaud,
\newblock {\em Two-dimensional scaling properties of experimental fracture
  surfaces},
\newblock \doi{10.1103/PhysRevLett.96.035506}{\rm Phys. Rev. Lett. {\bf 96}
  (2006)   035506}\null.

\bibitem{GutenbergRichter1956}
B.~Gutenberg and C.F. Richter,
\newblock {\em {Earthquake magnitude, intensity, energy, and acceleration}},
\newblock Bulletin of the Seismological Society of America {\bf 46} (1956)
  105--145.

\bibitem{BinderYoung1986}
K.~Binder and A.P. Young,
\newblock {\em Spin glasses: Experimental facts, theoretical concepts, and open
  questions},
\newblock \doi{10.1103/RevModPhys.58.801}{\rm Rev. Mod. Phys. {\bf 58} (1986)
  801}\null.

\bibitem{NATOASISeries1995}
A.~McKane, M.~Droz, J.~Vannimenus  and D.~Wolf, editors,
\newblock {\em Scale Invariance, Interfaces, and Non-Equilibrium Dynamics},
\newblock NATO ASI Series 344,
\newblock Springer US, 1995.

\bibitem{Kardar1997}
M.~Kardar,
\newblock {\em Nonequilibrium dynamics of interfaces and lines},
\newblock \doi{10.1016/S0370-1573(98)00007-6}{\rm Phys. Rep. {\bf 301} (1998)
  85--112}\null.

\bibitem{GiamarchiLeDoussalBookYoung}
T.~Giamarchi and P.~{Le~Doussal},
\newblock {\em Statics and dynamics of disordered elastic systems},
\newblock in A.P. Young, editor, {\em Spin glasses and random fields}, World
  Scientific, Singapore, 1997,
\newblock \arxiv{cond-mat/9705096}.

\bibitem{DSFisher1998}
D.S. Fisher,
\newblock {\em Collective transport in random media: {From} superconductors to
  earthquakes},
\newblock \doi{10.1016/S0370-1573(98)00008-8}{\rm Phys. Rep. {\bf 301} (1998)
  113--150}\null.

\bibitem{BrazovskiiNattermann2004}
S.~Brazovskii and T.~Nattermann,
\newblock {\em Pinning and sliding of driven elastic systems: from domain walls
  to charge density waves},
\newblock \doi{10.1080/00018730410001684197}{\rm Adv. Phys. {\bf 53} (2004)
  177--252}\null,
\newblock \arxiv{cond-mat/0312375}.

\bibitem{LeDoussal2008}
P.~{Le Doussal},
\newblock {\em Exact results and open questions in first principle {functional
  RG}},
\newblock \doi{10.1016/j.aop.2009.10.010}{\rm Ann. Phys. (NY) {\bf 325} (2009)
   49--150}\null,
\newblock \arxiv{arXiv:0809.1192}.

\bibitem{PruessnerBook}
G.~Pruessner,
\newblock \doi{10.1017/CBO9780511977671}{\rm {\em Self-Organised Criticality:
  Theory, Models and Characterisation}}\null,
\newblock Cambridge University Press, 2012.

\bibitem{LemerleFerreChappertMathetGiamarchiLeDoussal1998}
S.~Lemerle, J.~{Ferr\'e}, C.~Chappert, V.~Mathet, T.~Giamarchi  and P.~{Le
  Doussal},
\newblock {\em Domain wall creep in an {Ising} ultrathin magnetic film},
\newblock \doi{10.1103/PhysRevLett.80.849}{\rm Phys. Rev. Lett. {\bf 80} (1998)
    849}\null.

\bibitem{MoulinetGuthmannRolley2002}
S.~Moulinet, C.~Guthmann  and E.~Rolley,
\newblock {\em Roughness and dynamics of a contact line of a viscous fluid on a
  disordered substrate},
\newblock \doi{10.1140/epje/i2002-10032-2}{\rm Eur. Phys. J. E {\bf 8} (2002)
  437--443}\null.

\bibitem{Peierls1955}
R.E. Peierls,
\newblock {\em
  \link{https://global.oup.com/academic/product/quantum-theory-of-solids-9780198507819?cc=fr&lang=en&}{Quantum
  Theory of Solids}},
\newblock Oxford University Press, London, 1955.

\bibitem{FukuyamaLee1978}
H.~Fukuyama and P.A. Lee,
\newblock {\em Dynamics of the charge-density wave. {I. Impurity} pinning in a
  single chain},
\newblock \doi{10.1103/PhysRevB.17.535}{\rm Phys. Rev. {\bf B 17} (1978)
  535}\null.

\bibitem{LeeRice1979}
P.A. Lee and T.M. Rice,
\newblock {\em Electric-field depinning of charge-density waves},
\newblock \doi{10.1103/PhysRevB.19.3970}{\rm Phys. Rev. B {\bf 19} (1979)
  3970--3980}\null.

\bibitem{Gruner1988}
G.~Gr\"uner,
\newblock {\em The dynamics of charge-density waves},
\newblock \doi{10.1103/RevModPhys.60.1129}{\rm Rev. Mod. Phys. {\bf 60} (1988)
   1129--81}\null.

\bibitem{Monceau2012}
P.~Monceau,
\newblock {\em Electronic crystals: an experimental overview},
\newblock \doi{10.1080/00018732.2012.719674}{\rm Adv. Phys. {\bf 61} (2012)
  325}\null.

\bibitem{KardarLH1994}
M.~Kardar,
\newblock {\em Lectures on directed paths in random media},
\newblock in F.~David, P.~Ginsparg  and J.~Zinn-Justin, editors, {\em
  Fluctuating Geometries in Statistical Mechanics and Field Theory}, {\em {\em
  Volume} LXII} of {\em Les Houches, \'ecole d'\'et\'e de physique th\'eorique
  1994}, Elsevier Science, Amsterdam, 1996,
\newblock \arxiv{cond-mat/9411022}.

\bibitem{BrochardGennes1991}
F.~Brochard and P.G.~De Gennes,
\newblock {\em Collective modes of a contact line},
\newblock \doi{10.1021/la00060a049}{\rm Langmuir {\bf 7} (1991)
  3216--3218}\null.

\bibitem{Rice1985}
J.R. Rice,
\newblock {\em First-order variation in elastic fields due to variation in
  location of a planar crack front},
\newblock \doi{10.1115/1.3169103}{\rm J. Appl. Mech. {\bf 52} (1985)
  571--579}\null.

\bibitem{BachasLeDoussalWiese2006}
C.~Bachas, P.~{Le Doussal}  and K.J. Wiese,
\newblock {\em Wetting and minimal surfaces},
\newblock \doi{10.1103/PhysRevE.75.031601}{\rm Phys. Rev. E {\bf 75} (2007)
  031601}\null,
\newblock \arxiv{hep-th/0606247}.

\bibitem{LeDoussalWieseRaphaelGolestanian2004}
P.~Le Doussal, K.J.\ Wiese, E.~Raphael  and Ramin Golestanian,
\newblock {\em Can non-linear elasticity explain contact-line roughness at
  depinning?},
\newblock \doi{10.1103/PhysRevLett.96.015702}{\rm Phys. Rev. Lett. {\bf 96}
  (2006)   015702}\null,
\newblock \arxiv{cond-mat/0411652}.

\bibitem{LeDoussalWiese2009a}
P.~{Le Doussal} and K.J. Wiese,
\newblock {\em Elasticity of a contact-line and avalanche-size distribution at
  depinning},
\newblock \doi{10.1103/PhysRevE.82.011108}{\rm Phys. Rev. E {\bf 82} (2010)
  011108}\null,
\newblock \arxiv{arXiv:0908.4001}.

\bibitem{GutenbergRichter1944}
B.~Gutenberg and C.F. Richter,
\newblock {\em {Frequency of earthquakes in California}},
\newblock Bulletin of the Seismological Society of America {\bf 34} (1944)
  185.

\bibitem{ZapperiCizeauDurinStanley1998}
S.~Zapperi, P.~Cizeau, G.~Durin  and H.E. Stanley,
\newblock {\em Dynamics of a ferromagnetic domain wall: Avalanches, depinning
  transition, and the {Barkhausen} effect},
\newblock \doi{10.1103/PhysRevB.58.6353}{\rm Phys. Rev. B {\bf 58} (1998)
  6353--6366}\null.

\bibitem{Flory1953}
P.J. Flory,
\newblock {\em
  \link{https://archive.org/details/FloryF.1953PrinciplesOfPolymerChemistry/mode/2up}{Principles
  of Polymer Chemistry}},
\newblock Cornell University Press, 1953.

\bibitem{Harris1974}
A.B. Harris,
\newblock {\em Effect of random defects on the critical behaviour of {Ising}
  models},
\newblock \doi{10.1088/0022-3719/7/9/009}{\rm J. Phys. C {\bf 7} (1974)
  1671--1692}\null.

\bibitem{ImryMa1975}
Y.~Imry and S.K. Ma,
\newblock {\em Random-field instability of the ordered state of continuous
  symmetry},
\newblock \doi{10.1103/PhysRevLett.35.1399}{\rm Phys. Rev. Lett. {\bf 355}
  (1975)   1399--1401}\null.

\bibitem{Brout1959}
R.~Brout,
\newblock {\em Statistical mechanical theory of a random ferromagnetic system},
\newblock \doi{10.1103/PhysRev.115.824}{\rm Phys. Rev. {\bf 115} (1959)
  824--835}\null.

\bibitem{EdwardsAnderson1975}
S.F. Edwards and P.W. Anderson,
\newblock {\em Theory of spin glasses},
\newblock \doi{10.1088/0305-4608/5/5/017}{\rm J. Phys. F Met. Phys. {\bf 5}
  (1975)   965--974}\null.

\bibitem{AharonyImryMa1976}
A.~Aharony, J.~Imry  and S.-K. Ma,
\newblock {\em Lowering of dimensionality in phase transitions with random
  fields},
\newblock \doi{10.1103/PhysRevLett.37.1364}{\rm Phys. Rev. Lett. {\bf 37}
  (1976)   1364--1367}\null.

\bibitem{EfetovLarkin1977}
K.B. Efetov and A.I. Larkin,
\newblock {\em
  \link{http://www.jetp.ac.ru/cgi-bin/e/index/e/45/6/p1236?a=list}{Charge-density
  wave in a random potential}},
\newblock Sov. Phys. JETP {\bf 45} (1977)   1236.

\bibitem{Young1977}
A.P. Young,
\newblock {\em On the lowering of dimensionality in phase transitions with
  random fields},
\newblock \doi{10.1088/0022-3719/10/9/007}{\rm J. Phys. C {\bf 10} (1977)
  L257--L256}\null.

\bibitem{Ising1925}
E.~Ising,
\newblock {\em Beitrag zur {Theorie} des {Ferromagnetismus}},
\newblock \doi{10.1007/BF02980577}{\rm Z. Phys. {\bf 31} (1925)
  253--258}\null.

\bibitem{Kardar1987}
M.~Kardar,
\newblock {\em Replica {Bethe} ansatz studies of two-dimensional interfaces
  with quenched random impurities},
\newblock \doi{10.1016/0550-3213(87)90203-3}{\rm Nucl. Phys. B {\bf 290} (1987)
    582--602}\null.

\bibitem{Larkin1970}
A.I. Larkin,
\newblock Sov. Phys. JETP {\bf 31} (1970)   784.

\bibitem{Nattermann1985}
T.~Nattermann,
\newblock {\em Ising domain wall in a random pinning potential},
\newblock \doi{10.1088/0022-3719/18/36/021}{\rm J. Phys. C {\bf 18} (1985)
  6661--79}\null.

\bibitem{WilsonFisher1972}
K.G. Wilson and M.E. Fisher,
\newblock {\em Critical exponents in 3.99 dimensions},
\newblock \doi{10.1103/PhysRevLett.28.240}{\rm Phys. Rev. Lett. {\bf 28} (1972)
    240--243}\null.

\bibitem{BogoliubovParasiuk1957}
N.N. Bogoliubov and O.S. Parasiuk,
\newblock {\em {\"Uber die Multiplikation der Kausalfunktionen in der
  Quantentheorie der Felder}},
\newblock \doi{10.1007/BF02392399}{\rm Acta Math. {\bf 97} (1957)   227}\null.

\bibitem{Hepp1966}
K.~Hepp,
\newblock {\em Proof of the {Bogoliubov-Parasiuk} theorem on renormalization},
\newblock \doi{10.1007/BF01773358}{\rm Comm. Math. Phys. {\bf 2} (1966)
  301--326}\null.

\bibitem{Zimmermann1969}
W.~Zimmermann,
\newblock {\em Convergence of {Bogoliubov's} method of renormalization in
  monmentum space},
\newblock \doi{10.1007/BF01645676}{\rm Commun. Math. Phys. {\bf 15} (1969)
  208--234}\null.

\bibitem{BergereLam1975}
M.C. Bergere and Y.-M.P. Lam,
\newblock {\em {Bogoliubov-Parasiuk} theorem in the $\alpha$-parametric
  representation},
\newblock \doi{10.1063/1.523078}{\rm J. Math. Phys. {\bf 17} (1976)
  1546--1557}\null.

\bibitem{RivasseauBook}
V.~Rivasseau,
\newblock {\em From perturbative to constructive renormalization},
\newblock Princeton university press, Princeton, New Jersey, 1991.

\bibitem{WilsonKogut1974}
K.~Wilson and J.~Kogut,
\newblock {\em The renormalization group and the $\varepsilon$-expansion},
\newblock \doi{10.1016/0370-1573(74)90023-4}{\rm Phys.~Rep. {\bf 12} (1974)
  75--200}\null.

\bibitem{WieseHabil}
K.J. Wiese,
\newblock \doi{10.1016/S1062-7901(01)80016-1}{\rm {\em Polymerized membranes, a
  review}}\null, {\em {\em Volume}~19} of {\em Phase Transitions and Critical
  Phenomena},
\newblock Acadamic Press, London, 1999.

\bibitem{DSFisher1986}
D.S. Fisher,
\newblock {\em Interface fluctuations in disordered systems: {$5-\epsilon$}
  expansion},
\newblock \doi{10.1103/PhysRevLett.56.1964}{\rm Phys. Rev. Lett. {\bf 56}
  (1986)   1964--97}\null.

\bibitem{NarayanMiddleton1994}
O.~Narayan and A.A. Middleton,
\newblock {\em Avalanches and the renormalization-group for pinned
  charge-density waves},
\newblock \doi{10.1103/PhysRevB.49.244}{\rm Phys. Rev. B {\bf 49} (1994)
  244--256}\null.

\bibitem{NarayanDSFisher1992b}
O.~Narayan and D.S. Fisher,
\newblock {\em Critical behavior of sliding charge-density waves in 4-epsilon
  dimensions},
\newblock \doi{10.1103/PhysRevB.46.11520}{\rm Phys. Rev. B {\bf 46} (1992)
  11520--49}\null.

\bibitem{NarayanDSFisher1992a}
O.~Narayan and D.S. Fisher,
\newblock {\em Dynamics of sliding charge-density waves in 4-epsilon
  dimensions},
\newblock \doi{10.1103/PhysRevLett.68.3615}{\rm Phys. Rev. Lett. {\bf 68}
  (1992)   3615--18}\null.

\bibitem{LeDoussalWieseChauve2003}
P.~Le Doussal, K.J. Wiese  and P.~Chauve,
\newblock {\em Functional renormalization group and the field theory of
  disordered elastic systems},
\newblock \doi{10.1103/PhysRevE.69.026112}{\rm Phys. Rev. E {\bf 69} (2004)
  026112}\null,
\newblock \arxiv{cond-mat/0304614}.

\bibitem{LeDoussalWieseChauve2002}
P.~Le Doussal, K.J. Wiese  and P.~Chauve,
\newblock {\em 2-loop functional renormalization group analysis of the
  depinning transition},
\newblock \doi{10.1103/PhysRevB.66.174201}{\rm Phys. Rev. B {\bf 66} (2002)
  174201}\null,
\newblock \arxiv{cond-mat/0205108}.

\bibitem{ChauveLeDoussalWiese2000a}
P.~Chauve, P.~Le Doussal  and K.J. Wiese,
\newblock {\em Renormalization of pinned elastic systems: How does it work
  beyond one loop?},
\newblock \doi{10.1103/PhysRevLett.86.1785}{\rm Phys. Rev. Lett. {\bf 86}
  (2001)   1785--1788}\null,
\newblock \arxiv{cond-mat/0006056}.

\bibitem{BalentsBouchaudMezard1996}
L.~Balents, J.P. Bouchaud  and M.~M\'ezard,
\newblock {\em The large scale energy landscape of randomly pinned objects},
\newblock \doi{10.1051/jp1:1996112}{\rm J. Phys. I (France) {\bf 6} (1996)
  1007--20}\null,
\newblock \arxiv{cond-mat/9601137}.

\bibitem{LeDoussal2006b}
P.~{Le Doussal},
\newblock {\em Finite temperature {Functional RG}, droplets and decaying
  {Burgers} turbulence},
\newblock \doi{10.1209/epl/i2006-10295-1}{\rm Europhys. Lett. {\bf 76} (2006)
  457--463}\null,
\newblock \arxiv{cond-mat/0605490}.

\bibitem{MiddletonLeDoussalWiese2006}
A.A. Middleton, P.~{Le~Doussal}  and K.J. Wiese,
\newblock {\em Measuring functional renormalization group fixed-point functions
  for pinned manifolds},
\newblock \doi{10.1103/PhysRevLett.98.155701}{\rm Phys. Rev. Lett. {\bf 98}
  (2007)   155701}\null,
\newblock \arxiv{cond-mat/0606160}.

\bibitem{LeDoussalWiese2006a}
P.~{Le Doussal} and K.J. Wiese,
\newblock {\em How to measure {Functional RG} fixed-point functions for
  dynamics and at depinning},
\newblock \doi{10.1209/0295-5075/77/66001}{\rm EPL {\bf 77} (2007)
  66001}\null,
\newblock \arxiv{cond-mat/0610525}.

\bibitem{WieseLeDoussal2007}
K.J. Wiese and P.~{Le Doussal},
\newblock {\em How to measure the effective action for disordered systems},
\newblock in Wolfhard Janke and Axel Pelster, editors, {\em Path Integrals -
  New Trends and Perspectives}, World Scientific, 2008,
\newblock \arxiv{arXiv:0712.4286}.

\bibitem{terBurgWiese2020}
C.~ter Burg and K.J. Wiese,
\newblock {\em Mean-field theories for depinning and their experimental
  signatures},
\newblock \doi{10.1103/PhysRevE.103.052114}{\rm Phys. Rev. E {\bf 103} (2021)
  052114}\null,
\newblock \arxiv{arXiv:2010.16372}.

\bibitem{BalentsDSFisher1993}
L.~Balents and D.S. Fisher,
\newblock {\em Large-{$N$} expansion of $4-\varepsilon$-dimensional oriented
  manifolds in random media},
\newblock \doi{10.1103/PhysRevB.48.5949}{\rm Phys. Rev. {\bf B 48} (1993)
  5949--5963}\null.

\bibitem{WagnerGeshkenbeinLarkinBlatter1999}
O.S. Wagner, V.B. Geshkenbein, A.I. Larkin  and G.~Blatter,
\newblock {\em Renormalization-group analysis of weak collective pinning in
  type-{II} superconductors},
\newblock \doi{10.1103/PhysRevB.59.11551}{\rm Phys. Rev. B {\bf 59} (1999)
  11551--62}\null.

\bibitem{Scheidl2loopPrivate}
S.~Scheidl,
\newblock Private communication about 2-loop calculations for the random
  manifold problem. 2000-2004.

\bibitem{DincerDiplom}
Y.~Dincer,
\newblock {\em Zur Universalit\"at der Struktur elastischer Mannigfaltigkeiten
  in Unordnung},
\newblock Master's thesis, Universit\"at K\"oln, 1999.

\bibitem{ChauveLeDoussal2001}
P.~Chauve and P.~Le Doussal,
\newblock {\em Exact multilocal renormalization group and applications to
  disordered problems},
\newblock \doi{10.1103/PhysRevE.64.051102}{\rm Phys. Rev. E {\bf 64} (2001)
  051102/1--27}\null,
\newblock \arxiv{cond-mat/0006057}.

\bibitem{Middleton1995}
A.A. Middleton,
\newblock {\em Numerical results for the ground-state interface in a random
  medium},
\newblock \doi{10.1103/PhysRevE.52.R3337}{\rm Phys. Rev. E {\bf 52} (1995)
  R3337--40}\null.

\bibitem{AlavaDuxbury1996}
M.J. Alava and P.M. Duxbury,
\newblock {\em Disorder-induced roughening in the three-dimensional {Ising}
  model},
\newblock \doi{10.1103/PhysRevB.54.14990}{\rm Phys. Rev. B {\bf 54} (1996)
  14990--14993}\null.

\bibitem{KardarHuseHenleyFisher1985}
M.~Kardar, D.A. Huse, C.L. Henley  and D.S. Fisher,
\newblock {\em Roughening by impurities at finite temperatures (comment and
  reply)},
\newblock \doi{10.1103/PhysRevLett.55.2923}{\rm Phys. Rev. Lett. {\bf 55}
  (1985)   2923--4}\null.

\bibitem{KompanietsWiese2019}
M.~Kompaniets and K.J. Wiese,
\newblock {\em Fractal dimension of critical curves in the {$O(n)$}-symmetric
  $\phi^4$-model and crossover exponent at 6-loop order: Loop-erased random
  walks, self-avoiding walks, {Ising, XY and Heisenberg} models},
\newblock \doi{10.1103/PhysRevE.101.012104}{\rm Phys. Rev. E {\bf 101} (2019)
  012104}\null,
\newblock \arxiv{arXiv:1908.07502}.

\bibitem{Wegner1974}
F.J. Wegner,
\newblock {\em Some invariance properties of the renormalization group},
\newblock \doi{10.1088/0022-3719/7/12/004}{\rm J. Phys. C {\bf 7} (1974)
  2098--2108}\null.

\bibitem{PolandRychkovVichi2019}
D.~Poland, S.~Rychkov  and A.~Vichi,
\newblock {\em The conformal bootstrap: Theory, numerical techniques, and
  applications},
\newblock \doi{10.1103/RevModPhys.91.015002}{\rm Rev. Mod. Phys. {\bf 91}
  (2019)   015002}\null.

\bibitem{ChauveGiamarchiLeDoussal2000}
P.~Chauve, T.~Giamarchi  and P.~Le Doussal,
\newblock {\em Creep and depinning in disordered media},
\newblock \doi{10.1103/PhysRevB.62.6241}{\rm Phys. Rev. B {\bf 62} (2000)
  6241--67}\null,
\newblock \arxiv{cond-mat/0002299}.

\bibitem{BalentsLeDoussal2004}
L.~Balents and P.~Le Doussal,
\newblock {\em Thermal fluctuations in pinned elastic systems: field theory of
  rare events and droplets},
\newblock \doi{10.1016/j.aop.2004.10.001}{\rm Ann. Phys. (NY) {\bf 315} (2005)
   213--303}\null,
\newblock \arxiv{cond-mat/0408048}.

\bibitem{BalentsLeDoussal2003}
L.~Balents and P.~Le Doussal,
\newblock {\em Broad relaxation spectrum and the field theory of glassy
  dynamics for pinned elastic systems},
\newblock \doi{10.1103/PhysRevE.69.061107}{\rm Phys. Rev. E {\bf 69} (2004)
  061107}\null,
\newblock \arxiv{cond-mat/0312338}.

\bibitem{Wasow1965}
W.~Wasow,
\newblock {\em {Asymptotic expansions for ordinary differential equations}},
\newblock Pure and Applied Mathematics, Vol. XIV,
\newblock Interscience Publishers John Wiley \& Sons, Inc., New
  York-London-Sydney, 1965.

\bibitem{Bogolyubov2011}
{N.N.~Bogolyubov, jr. (originator)},
\newblock {\em Encyclopedia of Mathematics:
  \link{http://www.encyclopediaofmath.org/index.php?title=Perturbation_theory&oldid=11676}{Perturbation
  Theory}}.
\newblock 2011.

\bibitem{Smith1985}
D.R. Smith,
\newblock {\em
  \link{http://services.cambridge.org/us/academic/subjects/mathematics/differential-and-integral-equations-dynamical-systems-and-co/singular-perturbation-theory-introduction-applications?format=HB&isbn=9780521300421}{Singular-Perturbation
  Theory}},
\newblock Cambridge University Press, 1985.

\bibitem{HairerWanner1996}
E.~Hairer and G.~Wanner,
\newblock \doi{10.1007/978-3-662-09947-6}{\rm {\em Solving Ordinary
  Differential Equations II: Stiff and Differential-Algebra Problems}}\null,
\newblock Springer, Berlin, Heidelberg, 1996.

\bibitem{LeDoussalWiese2005a}
P.~Le Doussal and K.J.\ Wiese,
\newblock {\em 2-loop functional renormalization for elastic manifolds pinned
  by disorder in {$N$} dimensions},
\newblock \doi{10.1103/PhysRevE.72.035101}{\rm Phys. Rev. E {\bf 72} (2005)
  035101 (R)}\null,
\newblock \arxiv{cond-mat/0501315}.

\bibitem{BustingorryLeDoussalRosso2010}
A.~Rosso S.~Bustingorry, P. Le~Doussal,
\newblock {\em The universal high temperature regime of pinned elastic
  objects},
\newblock \doi{10.1103/PhysRevB.82.140201}{\rm Phys. Rev. B {\bf 82} (2010)
  140201}\null,
\newblock \arxiv{arXiv:1006.0603}.

\bibitem{LeDoussal2006a}
P.~{Le~Doussal},
\newblock {\em Chaos and residual correlations in pinned disordered systems},
\newblock \doi{10.1103/PhysRevLett.96.235702}{\rm Phys. Rev. Lett. {\bf 96}
  (2006)}\null,
\newblock \arxiv{cond-mat/0505679}.

\bibitem{DuemmerLeDoussal2007}
O.~Duemmer and P.~Le Doussal,
\newblock {\em Chaos in the thermal regime for pinned manifolds via functional
  {RG}},
\newblock (2007),
\newblock \arxiv{arXiv:0709.1378}.

\bibitem{LeDoussalWiese2001}
P.~Le Doussal and K.J. Wiese,
\newblock {\em Functional renormalization group at large {$N$} for random
  manifolds},
\newblock \doi{10.1103/PhysRevLett.89.125702}{\rm Phys. Rev. Lett. {\bf 89}
  (2002)   125702}\null,
\newblock \arxiv{cond-mat/0109204}.

\bibitem{MezardParisi1991}
M.~M\'ezard and G.~Parisi,
\newblock {\em Replica field theory for random manifolds},
\newblock \doi{10.1051/jp1:1991171}{\rm J. Phys. I (France) {\bf 1} (1991)
  809--837}\null.

\bibitem{LeDoussalWiese2004a}
P.~{Le Doussal} and K.J. Wiese,
\newblock {\em Derivation of the functional renormalization group
  $\beta$-function at order $1/{N}$ for manifolds pinned by disorder},
\newblock \doi{10.1016/j.nuclphysb.2004.08.022}{\rm Nucl. Phys. B {\bf 701}
  (2004)   409--480}\null,
\newblock \arxiv{cond-mat/0406297}.

\bibitem{Parisi1980c}
G~Parisi,
\newblock {\em A sequence of approximated solutions to the {SK} model for spin
  glasses},
\newblock \doi{10.1088/0305-4470/13/4/009}{\rm J. Phys. A {\bf 13} (1980)
  L115--L121}\null.

\bibitem{Parisi1980b}
G~Parisi,
\newblock {\em The order parameter for spin glasses: a function on the interval
  0-1},
\newblock \doi{10.1088/0305-4470/13/3/042}{\rm J. Phys. A {\bf 13} (1980)
  1101--1112}\null.

\bibitem{Parisi1980d}
G~Parisi,
\newblock {\em Magnetic properties of spin glasses in a new mean field theory},
\newblock \doi{10.1088/0305-4470/13/5/047}{\rm J. Phys. A {\bf 13} (1980)
  1887--1895}\null.

\bibitem{ParisiToulouse1980}
G.~Parisi and G.~Toulouse,
\newblock {\em A simple hypothesis for the spin glass phase of the
  infinite-ranged {SK} model},
\newblock \doi{10.1051/jphyslet:019800041015036100}{\rm J. Phys. (France) {\bf
  41} (1980)   L361--4}\null.

\bibitem{ParisiToulouse1980b}
G.~Parisi and G.~Toulouse,
\newblock {\em {Erratum: A} simple hypothesis for the spin glass phase of the
  infinite-ranged {SK} model},
\newblock \doi{10.1051/jphyslet:0198100420307100}{\rm J. Physique Lett. {\bf
  42} (1981)  ~71}\null.

\bibitem{MezardParisiSourlasToulouseVirasoro1984b}
M.~M\'ezard, G.~Parisi, N.~Sourlas, G.~Toulouse  and M.~Virasoro,
\newblock {\em Replica symmetry breaking and the nature of the spin glass
  phase},
\newblock \doi{10.1051/jphys:01984004505084300}{\rm J. Phys. France {\bf 45}
  (1984)   843--854}\null.

\bibitem{Duplantier1992}
B.~Duplantier,
\newblock {\em Loop-erased self-avoiding walks in two dimensions: exact
  critical exponents and winding numbers},
\newblock \doi{10.1016/0378-4371(92)90575-B}{\rm Physica A {\bf 191} (1992)
  516--522}\null.

\bibitem{LeDoussalMuellerWiese2007}
P.~{Le Doussal}, M.~M\"uller  and K.J. Wiese,
\newblock {\em Cusps and shocks in the renormalized potential of glassy random
  manifolds: How functional renormalization group and replica symmetry breaking
  fit together},
\newblock \doi{10.1103/PhysRevB.77.064203}{\rm Phys. Rev. B {\bf 77} (2007)
  064203}\null,
\newblock \arxiv{arXiv:0711.3929}.

\bibitem{AgoritsasLecomte2017}
E.~Agoritsas and V.~Lecomte,
\newblock {\em Power countings versus physical scalings in disordered elastic
  systems{\textemdash}case study of the one-dimensional interface},
\newblock \doi{10.1088/1751-8121/aa5753}{\rm J. Phys. A {\bf 50} (2017)
  104001}\null.

\bibitem{AgoritsasLecomteUnpublished}
E.~Agoritsas and V.~Lecomte,
\newblock unpublished.

\bibitem{BovierFrohlich1986}
A.~Bovier and J.~Fr\"ohlich,
\newblock {\em A heuristic theory of the spin glass phase},
\newblock \doi{10.1007/BF01011303}{\rm J. Stat. Phys. {\bf 44} (1986)
  347--391}\null.

\bibitem{FisherHuse1987}
D.S. Fisher and D.A. Huse,
\newblock {\em Absence of many states in realistic spin glasses},
\newblock \doi{10.1088/0305-4470/20/15/013}{\rm J. Phys. A {\bf 20} (1987)
  L1005--10}\null.

\bibitem{HuseFisher1987a}
D.A. Huse and D.S. Fisher,
\newblock {\em Dynamics of droplet fluctuations in pure and random {Ising}
  systems},
\newblock \doi{10.1103/PhysRevB.35.6841}{\rm Phys. Rev. B {\bf 35} (1987)
  6841--6}\null.

\bibitem{FisherHuse1988}
D.S. Fisher and D.A. Huse,
\newblock {\em Equilibrium behavior of the spin-glass ordered phase},
\newblock \doi{10.1103/PhysRevB.38.386}{\rm Phys. Rev. B {\bf 38} (1988)
  386--11}\null.

\bibitem{FisherHuse1988a}
D.S. Fisher and D.A. Huse,
\newblock {\em Nonequilibrium dynamics of spin glasses},
\newblock \doi{10.1103/PhysRevB.38.373}{\rm Phys. Rev. B {\bf 38} (1988)
  373--85}\null.

\bibitem{BrayMoore1985}
A.J. Bray and M~A. Moore,
\newblock {\em Critical behavior of the three-dimensional {Ising} spin glass},
\newblock \doi{10.1103/PhysRevB.31.631}{\rm Phys. Rev. B {\bf 31} (1985)
  631--633}\null.

\bibitem{NewmanStein1997}
C.M. Newman and D.L. Stein,
\newblock {\em Metastate approach to thermodynamic chaos},
\newblock \doi{10.1103/PhysRevE.55.5194}{\rm Phys. Rev. E {\bf 55} (1997)
  5194--211}\null.

\bibitem{MooreBokilDrossel1998}
M.A. Moore, H.~Bokil  and B.~Drossel,
\newblock {\em Evidence for the droplet picture of spin glasses},
\newblock \doi{10.1103/PhysRevLett.81.4252}{\rm Phys. Rev. Lett. {\bf 81}
  (1998)   4252--4255}\null.

\bibitem{MarinariParisiRuizLorenzoRitort1996}
E.~Marinari, G.~Parisi, J.~Ruiz-Lorenzo  and F.~Ritort,
\newblock {\em Numerical evidence for spontaneously broken replica symmetry in
  3d spin glasses},
\newblock \doi{10.1103/PhysRevLett.76.843}{\rm Phys. Rev. Lett. {\bf 76} (1996)
    843--846}\null.

\bibitem{MarinariParisiRicci-TersenghiRuiz-LorenzoZuliani2000}
E.~Marinari, G.~Parisi, F.~Ricci-Tersenghi, J.~J. Ruiz-Lorenzo  and F.~Zuliani,
\newblock {\em Replica symmetry breaking in short-range spin glasses:
  Theoretical foundations and numerical evidences},
\newblock \doi{10.1023/A:1018607809852}{\rm J. Stat. Phys. {\bf 98} (2000)
  973--1074}\null.

\bibitem{AspelmeierWangMooreKatzgraber2016}
T.~Aspelmeier, Wenlong Wang, M.~A. Moore  and Helmut~G. Katzgraber,
\newblock {\em Interface free-energy exponent in the one-dimensional ising spin
  glass with long-range interactions in both the droplet and broken replica
  symmetry regions},
\newblock \doi{10.1103/PhysRevE.94.022116}{\rm Phys. Rev. E {\bf 94} (2016)
  022116}\null.

\bibitem{CharbonneauYaida2017}
P.~Charbonneau and S.~Yaida,
\newblock {\em Nontrivial critical fixed point for replica-symmetry-breaking
  transitions},
\newblock \doi{10.1103/PhysRevLett.118.215701}{\rm Phys. Rev. Lett. {\bf 118}
  (2017)   215701}\null.

\bibitem{Moore2019}
M.A. Moore,
\newblock {\em Why replica symmetry breaking does not occur below six
  dimensions in ising spin glasses},
\newblock (2019),
\newblock \arxiv{arXiv:1902.07099}.

\bibitem{YeoMoore2020}
J.~Yeo and M.~A. Moore,
\newblock {\em Possible instability of one-step replica symmetry breaking in
  $p$-spin ising models outside mean-field theory},
\newblock \doi{10.1103/PhysRevE.101.032127}{\rm Phys. Rev. E {\bf 101} (2020)
  032127}\null.

\bibitem{HollerRead2020}
J.~H\"oller and N.~Read,
\newblock {\em One-step replica-symmetry-breaking phase below the de
  {Almeida--Thouless} line in low-dimensional spin glasses},
\newblock \doi{10.1103/PhysRevE.101.042114}{\rm Phys. Rev. E {\bf 101} (2020)
  042114}\null.

\bibitem{BalentsLeDoussal2002}
L.~Balents and P.~Le Doussal,
\newblock {\em Field theory of statics and dynamics of glasses: Rare events and
  barrier distributions},
\newblock \doi{10.1209/epl/i2003-10170-7}{\rm Europhys. Lett. {\bf 65} (2004)
  685--691}\null,
\newblock \arxiv{cond-mat/0205358}.

\bibitem{DrosselKardar1995}
B.~Drossel and M.~Kardar,
\newblock {\em Scaling of energy barriers for flux lines and other random
  systems},
\newblock \doi{10.1103/PhysRevE.52.4841}{\rm Phys. Rev. E {\bf 52} (1995)
  4841--4852}\null.

\bibitem{MikheevDrosselKardar1995}
L.V. Mikheev, B.~Drossel  and M.~Kardar,
\newblock {\em Energy barriers to motion of flux lines in random media},
\newblock \doi{10.1103/PhysRevLett.75.1170}{\rm Phys. Rev. Lett. {\bf 75}
  (1995)   1170--1173}\null.

\bibitem{terBurgPhD}
C.~ter Burg,
\newblock {\em to appear},
\newblock PhD thesis, PSL Research University, 2022.

\bibitem{terBurgWieseUnpublished}
C.~ter Burg and K.J. Wiese,
\newblock unpublished.

\bibitem{Kida1979}
S.~Kida,
\newblock {\em Asymptotic properties of {Burgers} turbulence},
\newblock \doi{10.1017/S0022112079001932}{\rm J. Fluid Mech. {\bf 93} (1979)
  337--377}\null.

\bibitem{BouchaudMezard1997}
J.-P. Bouchaud and M.~M\'ezard,
\newblock {\em Universality classes for extreme-value statistics},
\newblock \doi{10.1088/0305-4470/30/23/004}{\rm J. Phys. A {\bf 30} (1997)
  7997--8015}\null,
\newblock \arxiv{arXiv:cond-mat/9707047}.

\bibitem{Sinai1983}
Y.G. Sinai,
\newblock {\em The limiting behaviour of a one-dimensional random walk in a
  random environments},
\newblock \doi{10.1137/1127028}{\rm Theory Probab. Appl. {\bf 27} (1983)
  256--268}\null.

\bibitem{LeDoussalMonthus2003}
P.~Le Doussal and C.~Monthus,
\newblock {\em Exact solutions for the statistics of extrema of some random 1d
  landscapes, application to the equilibrium and the dynamics of the toy
  model},
\newblock \doi{10.1016/S0378-4371(02)01317-1}{\rm Physica A {\bf 317} (2003)
  140--98}\null,
\newblock \arxiv{cond-mat/0204168}.

\bibitem{Derrida1981}
B.~Derrida,
\newblock {\em Random-energy model: An exactly solvable model of disordered
  systems},
\newblock \doi{10.1103/PhysRevB.24.2613}{\rm Phys. Rev. B {\bf 24} (1981)
  2613--2626}\null.

\bibitem{DerridaToulouse1985}
B.~Derrida and G.~Toulouse,
\newblock {\em Sample to sample fluctuations in the random energy model},
\newblock \doi{10.1051/jphyslet:01985004606022300}{\rm J. Physique Lett. {\bf
  46} (1985)   223--228}\null.

\bibitem{Ruelle1987}
D.~Ruelle,
\newblock {\em A mathematical reformulation of derrida's rem and grem},
\newblock \doi{10.1007/BF01210613}{\rm Commun. Math. Phys. {\bf 108} (1987)
  225--239}\null.

\bibitem{Mukaida2015}
H.~Mukaida,
\newblock {\em Non-differentiability of the effective potential and the replica
  symmetry breaking in the random energy model},
\newblock \doi{10.1088/1751-8113/49/4/045002}{\rm J. Phys. A {\bf 49} (2015)
  045002}\null.

\bibitem{GrossMezard1984}
D.J. Gross and M.~Mezard,
\newblock {\em The simplest spin glass},
\newblock \doi{https://doi.org/10.1016/0550-3213(84)90237-2}{\rm Nucl. Phys. B
  {\bf 240} (1984)   431 -- 452}\null.

\bibitem{Derrida1991}
B.~Derrida,
\newblock {\em The zeroes of the partition function of the random energy
  model},
\newblock \doi{https://doi.org/10.1016/0378-4371(91)90130-5}{\rm Physica A {\bf
  177} (1991)   31 -- 37}\null.

\bibitem{DobrinevskiLeDoussalWiese2011}
A.~Dobrinevski, P.~{Le Doussal}  and K.J. Wiese,
\newblock {\em Interference in disordered systems: A particle in a complex
  random landscape},
\newblock \doi{10.1103/PhysRevE.83.061116}{\rm Phys. Rev. E {\bf 83} (2011)
  061116}\null,
\newblock \arxiv{arXiv:1101.2411}.

\bibitem{GorokhovFisherBlatter2002}
D.A. Gorokhov, D.S. Fisher  and G.~Blatter,
\newblock {\em Quantum collective creep: a quasiclassical {Langevin} equation
  approach},
\newblock \doi{10.1103/PhysRevB.66.214203}{\rm Phys. Rev. B {\bf 66} (2002)
  214203}\null.

\bibitem{NguenSpivakShklovskii1985}
V.L. Nguen, B.Z. Spivak  and B.I. Shklovskii,
\newblock {\em \link{www.jetp.ac.ru/cgi-bin/dn/e_062_05_1021.pdf}{Tunnel
  hopping in disordered systems}},
\newblock JETP {\bf 62} (1985)   1021.

\bibitem{MedinaKardar1991}
E.~Medina and M.~Kardar,
\newblock {\em Spin-orbit scattering and magnetoconductance of strongly
  localized electrons},
\newblock \doi{10.1103/PhysRevLett.66.3187}{\rm Phys. Rev. Lett. {\bf 66}
  (1991)   3187--3190}\null.

\bibitem{MedinaKardar1992}
E.~Medina and M.~Kardar,
\newblock {\em Quantum interference effects for strongly localized electrons},
\newblock \doi{10.1103/PhysRevB.46.9984}{\rm Phys. Rev. B {\bf 46} (1992)
  9984--10006}\null.

\bibitem{MedinaKardarShapirWang1989}
E.~Medina, M.~Kardar, Y.~Shapir  and X.R. Wang,
\newblock {\em Interference of directed paths in disordered systems},
\newblock \doi{10.1103/PhysRevLett.62.941}{\rm Phys. Rev. Lett. {\bf 62} (1989)
    941--944}\null.

\bibitem{MedinaKardarShapirWang1990}
E.~Medina, M.~Kardar, Y.~Shapir  and X.R. Wang,
\newblock {\em Magnetic-field effects on strongly localized electrons},
\newblock \doi{10.1103/PhysRevLett.64.1816}{\rm Phys. Rev. Lett. {\bf 64}
  (1990)   1816--1819}\null.

\bibitem{SomozaOrtunoPrior2007}
A.~M. Somoza, M.~Ortu\~no  and J.~Prior,
\newblock {\em Universal distribution functions in two-dimensional localized
  systems},
\newblock \doi{10.1103/PhysRevLett.99.116602}{\rm Phys. Rev. Lett. {\bf 99}
  (2007)   116602}\null.

\bibitem{PriorSomozaOrtuno2009}
J.~Prior, A.~M. Somoza  and M.~Ortu{\~n}o,
\newblock {\em Conductance distribution in two-dimensional localized systems
  with and without magnetic fields},
\newblock \doi{10.1140/epjb/e2009-00244-x}{\rm EPJB {\bf 70} (2009)
  513--521}\null.

\bibitem{ChalkerCoddington1988}
J.T. Chalker and P.D. Coddington,
\newblock {\em Percolation, quantum tunnelling and the integer hall effect},
\newblock \doi{10.1088/0022-3719/21/14/008}{\rm {\bf 21} (1988)
  2665--2679}\null.

\bibitem{Cardy2010}
J.~Cardy,
\newblock {\em Quantum network models and classical localization problems},
\newblock \doi{10.1142/S0217979210064678}{\rm Int. J. Mod. Phys. B {\bf 24}
  (2010)   1989--2014}\null,
\newblock \arxiv{https://doi.org/10.1142/S0217979210064678}.

\bibitem{BeamondCardyChalker2002}
E.J. Beamond, J.~Cardy  and J.T. Chalker,
\newblock {\em Quantum and classical localization, the spin quantum hall
  effect, and generalizations},
\newblock \doi{10.1103/PhysRevB.65.214301}{\rm Phys. Rev. B {\bf 65} (2002)
  214301}\null.

\bibitem{CookDerrida1990}
J.~Cook and B.~Derrida,
\newblock {\em Lyapunov exponents of large, sparse random matrices and the
  problem of directed polymers with complex random weights},
\newblock \doi{10.1007/BF01014363}{\rm Journal of Statistical Physics {\bf 61}
  (1990)   961--986}\null.

\bibitem{Derrida1991b}
B.~Derrida,
\newblock {\em Mean field theory of directed polymers in a random medium and
  beyond},
\newblock \doi{10.1088/0031-8949/1991/t38/002}{\rm Physica Scripta {\bf T38}
  (1991)   6--12}\null.

\bibitem{DerridaEvansSpeer1993}
B.~Derrida, M.R. Evans  and E.R. Speer,
\newblock {\em Mean field theory of directed polymers with random complex
  weights},
\newblock \doi{10.1007/BF02098482}{\rm Comm. Math. Phys. {\bf 156} (1993)
  221--244}\null.

\bibitem{Toft-PetersenAbrahamsenBalogPorcarLaver2018}
R.~Toft-Petersen, A.B. Abrahamsen, S.~Balog, L.~Porcar  and M.~Laver,
\newblock {\em Decomposing the bragg glass and the peak effect in a type-ii
  superconductor},
\newblock \doi{10.1038/s41467-018-03267-z}{\rm Nat. Commun. {\bf 9} (2018)
  901}\null.

\bibitem{MeissnerOchsenfeld1933}
W.~Meissner and R.~Ochsenfeld,
\newblock {\em Ein neuer {Effekt bei Eintritt der Supraleitf\"ahigkeit}},
\newblock \doi{10.1007/BF01504252}{\rm Naturwissenschaften {\bf 21} (1933)
  787--788}\null.

\bibitem{GiamarchiLeDoussal1994}
T.~Giamarchi and P.~Le Doussal,
\newblock {\em Elastic theory of pinned flux lattices},
\newblock \doi{10.1103/PhysRevLett.72.1530}{\rm Phys. Rev. Lett. {\bf 72}
  (1994)   1530--3}\null.

\bibitem{GiamarchiLeDoussal1996b}
T.~Giamarchi and P.~Le Doussal,
\newblock {\em Variational theory of elastic manifolds with correlated disorder
  and localization of interacting quantum particles},
\newblock \doi{10.1103/PhysRevB.53.15206}{\rm Phys. Rev. B {\bf 53} (1996)
  15206}\null,
\newblock \arxiv{cond-mat/9509008}.

\bibitem{KleinJoumardBlanchardMarcusCubittGiamarchiLeDoussal2001}
T.~Klein, I.~Joumard, S.~Blanchard, J.~Marcus, R.~Cubitt, T.~Giamarchi  and
  P.~Le Doussal,
\newblock {\em A {Bragg} glass phase in the vortex lattice of a type {II}
  superconductor},
\newblock \doi{10.1038/35096534}{\rm Nature {\bf 413} (2001)   404--406}\null.

\bibitem{Fisher1989}
M.P.A. Fisher,
\newblock {\em Vortex-glass superconductivity: A possible new phase in bulk
  high-${\mathrm{t}}_{\mathrm{c}}$ oxides},
\newblock \doi{10.1103/PhysRevLett.62.1415}{\rm Phys. Rev. Lett. {\bf 62}
  (1989)   1415--1418}\null.

\bibitem{KochFogliettiGallagherKorenGuptaFisher1989}
R.~H. Koch, V.~Foglietti, W.~J. Gallagher, G.~Koren, A.~Gupta  and M.~P.~A.
  Fisher,
\newblock {\em Experimental evidence for vortex-glass superconductivity in
  {Y-Ba-Cu-O}},
\newblock \doi{10.1103/PhysRevLett.63.1511}{\rm Phys. Rev. Lett. {\bf 63}
  (1989)   1511--1514}\null.

\bibitem{RegerTokuyasuYoungFisher1991}
J.~D. Reger, T.~A Tokuyasu, A.~P. Young  and Matthew P.~A. Fisher,
\newblock {\em Vortex-glass transition in three dimensions},
\newblock \doi{10.1103/PhysRevB.44.7147}{\rm Phys. Rev. B {\bf 44} (1991)
  7147--7150}\null.

\bibitem{MoserHugParashikovStiefelFritzThomasBaratoffGuntherodtChaudhari1995}
A.~Moser, H.~J. Hug, I.~Parashikov, B.~Stiefel, O.~Fritz, H.~Thomas,
  A.~Baratoff, H.-J. G\"untherodt  and P.~Chaudhari,
\newblock {\em Observation of single vortices condensed into a vortex-glass
  phase by magnetic force microscopy},
\newblock \doi{10.1103/PhysRevLett.74.1847}{\rm Phys. Rev. Lett. {\bf 74}
  (1995)   1847--1850}\null.

\bibitem{BalentsMarchettiRadzihovsky1998}
L.~Balents, MC. Marchetti  and L.~Radzihovsky,
\newblock {\em Nonequilibrium steady states of driven periodic media},
\newblock \doi{10.1103/PhysRevB.57.7705}{\rm Phys. Rev. B {\bf 57} (1998)
  7705--7739}\null,
\newblock \arxiv{cond-mat/9707302}.

\bibitem{AransonScheidlVinokur1998}
I.S. Aranson, S.~Scheidl  and V.M. Vinokur,
\newblock {\em Nonequilibrium dislocation dynamics and instability of driven
  vortex lattices in two dimensions},
\newblock \doi{10.1103/PhysRevB.58.14541}{\rm Phys. Rev. B {\bf 58} (1998)
  14541--14547}\null,
\newblock \arxiv{cond-mat/9805400}.

\bibitem{ScheidlVinokur1998}
S.~Scheidl and V.M. Vinokur,
\newblock {\em Driven dynamics of periodic elastic media in disorder},
\newblock \doi{10.1103/PhysRevE.57.2574}{\rm Phys. Rev. E {\bf 57} (1998)
  2574--2593}\null,
\newblock \arxiv{cond-mat/9708222}.

\bibitem{PfeifferReiger1999}
F.O. Pfeiffer and H.~Reiger,
\newblock {\em Numerical study of the strongly screened vortex-glass model in
  an external field},
\newblock \doi{10.1103/PhysRevB.60.6304}{\rm Phys. Rev. B {\bf 60} (1999)
  6304--6307}\null.

\bibitem{Fedorenko2008}
A.A. Fedorenko,
\newblock {\em Elastic systems with correlated disorder: {Response} to tilt and
  application to surface growth},
\newblock \doi{10.1103/PhysRevB.77.094203}{\rm Phys. Rev. B {\bf 77} (2008)
  094203}\null,
\newblock \arxiv{arXiv:0712.0801}.

\bibitem{Dupuis2019}
N.~Dupuis,
\newblock {\em Glassy properties of the {Bose}-glass phase of a one-dimensional
  disordered {Bose} fluid},
\newblock \doi{10.1103/PhysRevE.100.030102}{\rm Phys. Rev. E {\bf 100} (2019)
  030102}\null.

\bibitem{DupuisDaviet2020}
N.~Dupuis and R.~Daviet,
\newblock {\em Bose-glass phase of a one-dimensional disordered {Bose} fluid:
  Metastable states, quantum tunneling, and droplets},
\newblock \doi{10.1103/PhysRevE.101.042139}{\rm Phys. Rev. E {\bf 101} (2020)
  042139}\null.

\bibitem{Dupuis2020}
N.~Dupuis,
\newblock {\em Is there a {Mott}-glass phase in a one-dimensional disordered
  quantum fluid with linearly confining interactions?},
\newblock \doi{10.1209/0295-5075/130/56002}{\rm {EPL} {\bf 130} (2020)
  56002}\null.

\bibitem{DavietDupuis2020}
R.~Daviet and N.~Dupuis,
\newblock {\em Mott-glass phase of a one-dimensional quantum fluid with
  long-range interactions},
\newblock \doi{10.1103/PhysRevLett.125.235301}{\rm Phys. Rev. Lett. {\bf 125}
  (2020)   235301}\null.

\bibitem{DavietDupuis2021}
R.~Daviet and N.~Dupuis,
\newblock {\em Chaos in the {Bose}-glass phase of a one-dimensional disordered
  {Bose} fluid},
\newblock (2021),
\newblock \arxiv{arXiv:2101.12487}.

\bibitem{EmigNattermann2006}
T.~Emig and T.~Nattermann,
\newblock {\em Effect of planar defects on the stability of the {Bragg} glass
  phase of type-{II} superconductors},
\newblock \doi{10.1103/PhysRevLett.97.177002}{\rm Phys. Rev. Lett. {\bf 97}
  (2006)   177002}\null.

\bibitem{EmigNattermann1997}
T.~Emig and T.~Nattermann,
\newblock {\em A new disorder-driven roughening transition of charge-density
  waves and flux-line lattices},
\newblock \doi{10.1103/PhysRevLett.79.5090}{\rm Phys. Rev. Lett. {\bf 79}
  (1997)   5090--5093}\null,
\newblock \arxiv{cond-mat/9708116}.

\bibitem{LeDoussal2010Book}
P.~Le Doussal,
\newblock {\em Novel phases of vortices in superconductors},
\newblock \doi{10.1142/9789814304665_0013}{\rm Int. J. Mod. Phys. B {\bf 24}
  (2010)   3855--3914}\null.

\bibitem{DiFrancescoMathieuSenechal}
P.~Di Francesco, P.~Mathieu  and D.~S\'en\'echal,
\newblock \doi{10.1007/978-1-4612-2256-9}{\rm {\em Conformal Field
  Theory}}\null,
\newblock Springer, New York, 1997.

\bibitem{vonDelftSchoeller1998}
J.~von Delft and H.~Schoeller,
\newblock {\em Bosonization for beginners -- refermionization for experts},
\newblock
  \doi{10.1002/(SICI)1521-3889(199811)7:4<225::AID-ANDP225>3.0.CO;2-L}{\rm Ann.
  Phys. {\bf 7} (1998)   225--305}\null,
\newblock \arxiv{cond-mat/9805275}.

\bibitem{GiamarchiBook}
T.~Giamarchi,
\newblock \doi{10.1093/acprof:oso/9780198525004.001.0001}{\rm {\em Quantum
  Physics in One Dimension}}\null,
\newblock Oxford University Press, 2004.

\bibitem{DotsenkoCFT}
Vl.~S. Dotsenko,
\newblock \link{cel.archives-ouvertes.fr/cel-00092929}{S\'erie de cours sur la
  th\'eorie conforme},
\newblock Lecture notes, Universit\'es Paris VI and VII.

\bibitem{HenkelCFT}
M.~Henkel,
\newblock \doi{10.1007/978-3-662-03937-3}{\rm {\em Conformal Invariance and
  Critical Phenomena}}\null,
\newblock Springer, Berlin, Heidelberg, 1999.

\bibitem{Thirring1958}
W.E. Thirring,
\newblock {\em A soluble relativistic field theory},
\newblock \doi{https://doi.org/10.1016/0003-4916(58)90015-0}{\rm Annals of
  Physics {\bf 3} (1958)   91 -- 112}\null.

\bibitem{Coleman1975}
S.~Coleman,
\newblock {\em Quantum {sine-Gordon} equation as the massive {Thirring} model},
\newblock \doi{10.1103/PhysRevD.11.2088}{\rm Phys. Rev. D {\bf 11} (1975)
  2088--2097}\null.

\bibitem{KosterlitzThouless1973}
J.M. Kosterlitz and D.J. Thouless,
\newblock {\em Ordering, metastability and phase transitions in two-dimensional
  systems},
\newblock \doi{10.1088/0022-3719/6/7/010}{\rm {\bf 6} (1973)
  1181--1203}\null.

\bibitem{BalogHegedus2000}
J.~Balog and A.~Hegedus,
\newblock {\em Two-loop beta-functions of the sine-{Gordon} model},
\newblock \doi{10.1088/0305-4470/33/37/306}{\rm J. Phys. A {\bf 33} (2000)
  6543}\null.

\bibitem{AmitGoldschmidtGrinstein1980}
D.J. Amit, Y.Y. Goldschmidt  and S.~Grinstein,
\newblock {\em Renormalisation group analysis of the phase transition in the 2d
  {Coulomb} gas, sine-{Gordon} theory and {XY}-model},
\newblock \doi{10.1088/0305-4470/13/2/024}{\rm J. Phys. A {\bf 13} (1980)
  585}\null.

\bibitem{Lovelace1986}
C.~Lovelace,
\newblock {\em Stability of string vacua: (i). a new picture of the
  renormalization group},
\newblock \doi{10.1016/0550-3213(86)90253-1}{\rm Nucl. Phys. B {\bf 273} (1986)
    413--467}\null.

\bibitem{Boyanovsky1989}
D.~Boyanovsky,
\newblock {\em Field-theoretical renormalisation and fixed-point structure of a
  generalised coulomb gas},
\newblock \doi{10.1088/0305-4470/22/13/051}{\rm J. Phys. A {\bf 22} (1989)
  2601--2614}\null.

\bibitem{Naik1993}
S.~Naik,
\newblock {\em The exact mass gap of the chiral {SU(n)$\times$SU(n)} model},
\newblock \doi{10.1016/0920-5632(93)90197-E}{\rm Nucl. Phys. B Proc. Suppl.
  {\bf 30} (1993)   232--235}\null.

\bibitem{KonikLeClair1996}
R.M. Konik and A.~LeClair,
\newblock {\em Short-distance expansions of correlation functions in the
  {sine-Gordon} theory},
\newblock \doi{https://doi.org/10.1016/0550-3213(96)00279-9}{\rm Nucl. Phys. B
  {\bf 479} (1996)   619--653}\null.

\bibitem{Kehrein2001}
S.~Kehrein,
\newblock {\em Flow equation approach to the {sine-Gordon} model},
\newblock \doi{https://doi.org/10.1016/S0550-3213(00)00507-1}{\rm Nucl. Phys. B
  {\bf 592} (2001)   512 -- 562}\null.

\bibitem{DavietDupuis2019}
R.~Daviet and N.~Dupuis,
\newblock {\em Nonperturbative functional renormalization-group approach to the
  sine-{Gordon} model and the {Lukyanov-Zamolodchikov} conjecture},
\newblock \doi{10.1103/PhysRevLett.122.155301}{\rm Phys. Rev. Lett. {\bf 122}
  (2019)   155301}\null.

\bibitem{RiegerBlasum1997}
H.~Rieger and U.~Blasum,
\newblock {\em Ground-state properties of solid-on-solid models with disordered
  substrates},
\newblock \doi{10.1103/PhysRevB.55.R7394}{\rm Phys. Rev. B {\bf 55} (1997)
  R7394--R7397}\null.

\bibitem{ZengMiddletonShapir1996}
C.~Zeng, A.A. Middleton  and Y.~Shapir,
\newblock {\em Ground-state roughness of the disordered substrate and flux
  lines in $\mathit{d}\phantom{\rule{0ex}{0ex}}=\phantom{\rule{0ex}{0ex}}2$},
\newblock \doi{10.1103/PhysRevLett.77.3204}{\rm Phys. Rev. Lett. {\bf 77}
  (1996)   3204--3207}\null.

\bibitem{Kenyon2001}
R.~Kenyon,
\newblock {\em Dominos and the gaussian free field},
\newblock \doi{doi:10.1214/aop/1015345599}{\rm Ann. Probab. {\bf 29} (2001)
  1128--1137}\null.

\bibitem{LeDoussalSchehr2007}
P.~{Le~Doussal} and G.~Schehr,
\newblock {\em Disordered free fermions and the {Cardy-Ostlund} fixed line at
  low temperature},
\newblock \doi{10.1103/PhysRevB.75.184401}{\rm Phys. Rev. B {\bf 75} (2007)
  184401}\null,
\newblock \arxiv{cond-mat/0607657}.

\bibitem{PerretRistivojevicLeDoussalSchehrWiese2012}
A.~Perret, A.~Ristivojevic, P.~Le~Doussal, Gr\'egory Schehr  and K.~J. Wiese,
\newblock {\em Super-rough glassy phase of the random field {$XY$} model in two
  dimensions},
\newblock \doi{10.1103/PhysRevLett.109.157205}{\rm Phys. Rev. Lett. {\bf 109}
  (2012)   157205}\null,
\newblock \arxiv{arXiv:1204.5685}.

\bibitem{Propp2003}
J.~Propp,
\newblock {\em Generalized domino-shuffling},
\newblock \doi{https://doi.org/10.1016/S0304-3975(02)00815-0}{\rm Theoretical
  Computer Science {\bf 303} (2003)   267 -- 301}\null.

\bibitem{GuruswamyLeClairLudwig2000}
S.~Guruswamy, A.~LeClair  and A.W.W. Ludwig,
\newblock {\em {$\mbox{gl}(N|N)$} super-current algebras for disordered {Dirac}
  fermions in two dimensions},
\newblock \doi{10.1016/S0550-3213(00)00245-5}{\rm Nucl. Phys. {\bf B 583}
  (2000)   475--512}\null,
\newblock \arxiv{cond-mat/9909143}.

\bibitem{RistivojevicLeDoussalWiese2012}
Z.~Ristivojevic, P.~Le~Doussal  and K.J. Wiese,
\newblock {\em Super-rough phase of the random-phase sine-{Gordon} model:
  Two-loop results},
\newblock \doi{10.1103/PhysRevB.86.054201}{\rm Phys. Rev. B {\bf 86} (2012)
  054201}\null,
\newblock \arxiv{arXiv:1204.6221}.

\bibitem{CarpentierLeDoussal1998}
D.~Carpentier and P.~Le Doussal,
\newblock {\em Disordered {XY} models and {Coulomb} gases: renormalization via
  traveling waves},
\newblock \doi{10.1103/PhysRevLett.81.2558}{\rm Phys. Rev. Lett. {\bf 81}
  (1998)   2558--61}\null.

\bibitem{CarpentierLeDoussal2000}
D.~Carpentier and P.~Le Doussal,
\newblock {\em Topological transitions and freezing in {XY models and Coulomb}
  gases with quenched disorder: renormalization via traveling waves},
\newblock \doi{10.1016/S0550-3213(00)00468-5}{\rm Nucl. Phys. B {\bf 588}
  (2000)   565--629}\null,
\newblock \arxiv{cond-mat/9908335}.

\bibitem{CarpentierLeDoussal2001}
D.~Carpentier and P.~Le Doussal,
\newblock {\em Glass transition of a particle in a random potential, front
  selection in nonlinear renormalization group, and entropic phenomena in
  {Liouville and sinh-Gordon models}},
\newblock \doi{10.1103/PhysRevE.63.026110}{\rm Phys. Rev. E {\bf 63} (2001)
  026110}\null,
\newblock \arxiv{cond-mat/0003281}.

\bibitem{CarpentierLeDoussal2007}
D.~Carpentier and P.~{Le~Doussal},
\newblock {\em Electromagnetic {Coulomb} gas with vector charges and
  ``elastic'' potentials: Renormalization group equations},
\newblock \doi{10.1016/j.nuclphysb.2007.10.019}{\rm Nucl. Phys. B {\bf 795}
  (2008)   491--518}\null,
\newblock \arxiv{arXiv:0707.2667}.

\bibitem{FrischBook}
U.~Frisch,
\newblock {\em Turbulence},
\newblock Cambridge University Press, 1995.

\bibitem{LesieurBook}
M.~Lesieur,
\newblock {\em Turbulence in Fluids},
\newblock Kluwer Academic Publishers, Dordrecht/Boston/London, third edition,
  1995.

\bibitem{Gawedzki2008}
K.~Gawedzki,
\newblock {\em Stochastic processes in turbulent transport},
\newblock (2008),
\newblock \arxiv{arXiv:0806.1949}.

\bibitem{KraichnanYakhotChen1995}
R.H. Kraichnan, V.~Yakhot  and S.~Chen,
\newblock {\em Scaling relations for a randomly advected passive scalar field},
\newblock \doi{10.1103/PhysRevLett.75.240}{\rm Phys. Rev. Lett. {\bf 75} (1995)
    240--3}\null.

\bibitem{GawedzkiKupiainen1995}
K.~Gaw\c edzki and A.~Kupiainen,
\newblock {\em Anomalous scaling of the passive scalar},
\newblock \doi{10.1103/PhysRevLett.75.3834}{\rm Phys. Rev. Lett. {\bf 75}
  (1995)   3834}\null.

\bibitem{BernardGawedzkiKupiainen1998}
D.~Bernard, K.~Gawedzki  and A.~Kupiainen,
\newblock {\em Slow modes in passive advection},
\newblock \doi{10.1023/A:1023212600779}{\rm J. Stat. Phys. {\bf 90} (1998)
  519--69}\null.

\bibitem{Antonov1999}
N.V. Antonov,
\newblock {\em Anomalous scaling regimes of a passive scalar advected by the
  synthetic velocity field},
\newblock \doi{10.1103/PhysRevE.60.6691}{\rm Phys. Rev. E {\bf 60} (1999)
  6691--707}\null.

\bibitem{AdzhemyanAntonovVasilev1998}
L.T. Adzhemyan, N.V. Antonov  and A.N. Vasil'ev,
\newblock {\em Renormalization group, operator product expansion, and anomalous
  scaling in a model of advected passive scalar},
\newblock \doi{10.1103/PhysRevE.58.1823}{\rm Phys. Rev. {\bf E 58} (1998)
  1823--1835}\null.

\bibitem{AdzhemyanAntonovBarinovKabritsVasilev2001}
L.Ts. Adzhemyan, N.V. Antonov, V.A. Barinov, Yu.~S. Kabrits  and A.N. Vasil'ev,
\newblock {\em Anomalous exponents to order ${\ensuremath{\varepsilon}}^{3}$ in
  the rapid-change model of passive scalar advection},
\newblock \doi{10.1103/PhysRevE.63.025303}{\rm Phys. Rev. E {\bf 63} (2001)
  025303}\null.

\bibitem{AdzhemyanAntonovBarinovKabritsVasilev2001b}
L.~Ts. Adzhemyan, N.~V. Antonov, V.~A. Barinov, Yu.~S. Kabrits  and A.~N.
  Vasil'ev,
\newblock {\em Erratum: Anomalous exponents to order
  ${\ensuremath{\varepsilon}}^{3}$ in the rapid-change model of passive scalar
  advection {[Phys. Rev. E 63, 025303 (2001)]}},
\newblock \doi{10.1103/PhysRevE.64.019901}{\rm Phys. Rev. E {\bf 64} (2001)
  019901}\null.

\bibitem{Wiese1999}
K.J. Wiese,
\newblock {\em The passive polymer problem},
\newblock \doi{10.1023/A:1026473504422}{\rm J. Stat. Phys. {\bf 101} (2000)
  843--891}\null,
\newblock \arxiv{chao-dyn/9911005}.

\bibitem{FosterRyuLudwig2009}
M.S. Foster, S.~Ryu  and A.W.W. Ludwig,
\newblock {\em Termination of typical wave-function multifractal spectra at the
  {Anderson} metal-insulator transition: Field theory description using the
  functional renormalization group},
\newblock \doi{10.1103/PhysRevB.80.075101}{\rm Phys. Rev. B {\bf 80} (2009)
  075101}\null.

\bibitem{HalseyJensenKadanoffProcacciaShraiman1986}
T.C. Halsey, M.H. Jensen, L.P. Kadanoff, I.~Procaccia  and B.I. Shraiman,
\newblock {\em Fractal measures and their singularities: The characterization
  of strange sets},
\newblock \doi{10.1103/PhysRevA.33.1141}{\rm Phys. Rev. A {\bf 33} (1986)
  1141--1151}\null.

\bibitem{FedorenkoLeDoussalWiese2014}
A.A. Fedorenko, P.~Le Doussal  and K.J. Wiese,
\newblock {\em {Non-Gaussian} effects and multifractality in the {Bragg}
  glass},
\newblock \doi{10.1209/0295-5075/105/16002}{\rm EPL {\bf 105} (2014)
  16002}\null,
\newblock \arxiv{arXiv:1309.6529}.

\bibitem{LeDoussalRistivojevicWiese2013}
P.~Le~Doussal, Z.~Ristivojevic  and K.J. Wiese,
\newblock {\em Exact form of the exponential correlation function in the glassy
  super-rough phase},
\newblock \doi{10.1103/PhysRevB.87.214201}{\rm Phys. Rev. B {\bf 87} (2013)
  214201}\null,
\newblock \arxiv{arXiv:1304.4612}.

\bibitem{FedorenkoLeDoussalWiese2006}
A.~Fedorenko, P.~{Le~Doussal}  and K.J. Wiese,
\newblock {\em Universal distribution of threshold forces at the depinning
  transition},
\newblock \doi{10.1103/PhysRevE.74.041110}{\rm Phys. Rev. E {\bf 74} (2006)
  041110}\null,
\newblock \arxiv{cond-mat/0607229}.

\bibitem{GelfandYaglom1960}
I.~M. Gel'fand and A.~M. Yaglom,
\newblock {\em Integration in functional spaces and its applications in quantum
  physics},
\newblock \doi{10.1063/1.1703636}{\rm J. Phys. A {\bf 1} (1960)  ~48}\null.

\bibitem{HartmannRiegerBook}
A.K. Hartmann and H.~Rieger,
\newblock \doi{10.1002/3527600876}{\rm {\em Optimization Algorithms in
  Physics}}\null,
\newblock WILEY-VCH Verlag, Berlin, 2002.

\bibitem{KrauthBook}
W.~Krauth,
\newblock {\em Statistical Mechanics: Algorithms and Computations},
\newblock Oxford University Press, 2006.

\bibitem{HartmannRiegerBook2}
A.K. Hartmann and H.~Rieger, editors,
\newblock \doi{10.1002/3527603794}{\rm {\em New Optimization Algorithms in
  Physics}}\null,
\newblock Wiley-VCH Verlag, Berlin, 2002.

\bibitem{TinocoBustamante1999}
I.~Tinoco and C.~Bustamante,
\newblock {\em How {RNA} folds},
\newblock \doi{DOI: 10.1006/jmbi.1999.3001}{\rm Journal of Molecular Biology
  {\bf 293} (1999)   271--281}\null.

\bibitem{McCaskill1990}
J.S. McCaskill,
\newblock {\em The equilibrium partition function and base pair binding
  probabilities for rna secondary structure},
\newblock \doi{10.1002/bip.360290621}{\rm Biopolymers {\bf 29} (1990)
  1105--1119}\null.

\bibitem{BundschuhHwa1999}
R.~Bundschuh and T.~Hwa,
\newblock {\em {RNA} secondary structure formation: a solvable model of
  heteropolymer folding},
\newblock \doi{10.1103/PhysRevLett.83.1479}{\rm Phys. Rev. Lett. {\bf 83}
  (1999)   1479--82}\null,
\newblock \arxiv{cond-mat/9903089}.

\bibitem{Higgs2000}
P.G. Higgs,
\newblock {\em {RNA} secondary structure: physical and computational aspects},
\newblock \doi{10.1017/s0033583500003620}{\rm Q. Rev. Biophys. {\bf 33} (2000)
   199}\null.

\bibitem{BaezWieseBundschuh2019}
William~D. Baez, Kay~J\"org Wiese  and Ralf Bundschuh,
\newblock {\em Behavior of random rna secondary structures near the glass
  transition},
\newblock \doi{10.1103/PhysRevE.99.022415}{\rm Phys. Rev. E {\bf 99} (2019)
  022415}\null,
\newblock \arxiv{arXiv:1808.02351}.

\bibitem{Sedgewick1990}
R.~Sedgewick,
\newblock {\em Algorithms in {C}},
\newblock Addison-Wesely, 1990.

\bibitem{LeDoussalMiddletonWiese2008}
P.~{Le~Doussal}, A.A. Middleton  and K.J.\ Wiese,
\newblock {\em Statistics of static avalanches in a random pinning landscape},
\newblock \doi{10.1103/PhysRevE.79.050101}{\rm Phys. Rev. E {\bf 79} (2009)
  050101 (R)}\null,
\newblock \arxiv{arXiv:0803.1142}.

\bibitem{Rieger1998}
H.~Rieger,
\newblock {\em Ground state properties of fluxlines in a disordered
  environment},
\newblock \doi{10.1103/PhysRevLett.81.4488}{\rm Phys. Rev. Lett. {\bf 81}
  (1998)   4488--4491}\null.

\bibitem{NohRieger2001}
N.D. Jae and H.~Rieger,
\newblock {\em Disorder-driven critical behavior of periodic elastic media in a
  crystal potential},
\newblock \doi{10.1103/PhysRevLett.87.176102}{\rm Phys. Rev. Lett. {\bf 87}
  (2001)   176102}\null.

\bibitem{Diaz-PardoMoisanAlbornozLemaitreCurialeJeudy2019}
R.~D\'{\i}az~Pardo, N.~Moisan, L.~J. Albornoz, A.~Lema\^{\i}tre, J.~Curiale
  and V.~Jeudy,
\newblock {\em Common universal behavior of magnetic domain walls driven by
  spin-polarized electrical current and magnetic field},
\newblock \doi{10.1103/PhysRevB.100.184420}{\rm Phys. Rev. B {\bf 100} (2019)
  184420}\null.

\bibitem{Diaz-PardoPhD}
R.~Diaz~Pardo,
\newblock {\em Universal behaviors of magnetic domain walls in thin
  ferromagnets},
\newblock \link{tel.archives-ouvertes.fr/tel-01935833}{\rm PhD thesis,
  Universit\'e Paris Saclay, 2018},
\newblock \null.

\bibitem{HuguetFornsRitort2009}
J.M. Huguet, N.~Forns  and F.~Ritort,
\newblock {\em Statistical properties of metastable intermediates in {DNA}
  unzipping},
\newblock \doi{10.1103/PhysRevLett.103.248106}{\rm Phys. Rev. Lett. {\bf 103}
  (2009)   248106}\null.

\bibitem{NattermannStepanowTangLeschhorn1992}
T.~Nattermann, S.~Stepanow, L.-H. Tang  and H.~Leschhorn,
\newblock {\em Dynamics of interface depinning in a disordered medium},
\newblock \doi{10.1051/jp2:1992214}{\rm J. Phys. II (France) {\bf 2} (1992)
  1483--8}\null.

\bibitem{MSR}
P.C. Martin, E.D. Siggia  and H.A. Rose,
\newblock {\em Statistical dynamics of classical systems},
\newblock \doi{10.1103/PhysRevA.8.423}{\rm Phys. Rev. {\bf A 8} (1973)
  423--437}\null.

\bibitem{Janssen1976}
H.-K. Janssen,
\newblock {\em On a {Lagrangean} for classical field dynamics and
  renormalization group calculations of dynamical critical properties},
\newblock \doi{10.1007/BF01316547}{\rm Z. Phys. B {\bf 23} (1976)
  377--380}\null.

\bibitem{DeDominicis1976}
C.~De Dominicis,
\newblock {\em Techniques de renormalisation de la th\'eorie des champs et
  dynamique des ph\'enom\`enes critiques},
\newblock \doi{10.1051/jphyscol:1976138}{\rm J. Phys. Colloques {\bf 37} (1976)
    C1--247--253}\null.

\bibitem{Janssen1985}
H.K. Janssen,
\newblock {\em {Feldtheoretische Methoden in der Statistischen Mechanik}},
\newblock Vorlesungsmanuskript Uni D\"usseldorf (1985).

\bibitem{Tauber2012}
U.~T{\"a}uber,
\newblock \doi{10.1017/CBO9781139046213}{\rm {\em Critical dynamics: A field
  theory approach to equilibrium and non-equilibrium scaling behavior}}\null,
\newblock Cambridge University Press, 2012.

\bibitem{Middleton1992}
A.A. Middleton,
\newblock {\em Asymptotic uniqueness of the sliding state for charge-density
  waves},
\newblock \doi{10.1103/PhysRevLett.68.670}{\rm Phys. Rev. Lett. {\bf 68} (1992)
    670--673}\null.

\bibitem{HohenbergHalperin1977}
P.C. Hohenberg and B.I. Halperin,
\newblock {\em Theory of dynamical critical phenomena},
\newblock \doi{10.1103/RevModPhys.49.435}{\rm Rev. Mod. Phys. {\bf 49} (1977)
  435}\null.

\bibitem{LeschhornNattermannStepanowTang1997}
H.~Leschhorn, T.~Nattermann, S.~Stepanow  and L.-H. Tang,
\newblock {\em Driven interface depinning in a disordered medium},
\newblock \doi{10.1002/andp.19975090102}{\rm Annalen der Physik {\bf 509}
  (1997)   1--34}\null,
\newblock \arxiv{arXiv:cond-mat/9603114}.

\bibitem{NarayanDSFisher1993a}
O.~Narayan and D.S. Fisher,
\newblock {\em Threshold critical dynamics of driven interfaces in random
  media},
\newblock \doi{10.1103/PhysRevB.48.7030}{\rm Phys. Rev. B {\bf 48} (1993)
  7030--42}\null.

\bibitem{FerreroBustingorryKolton2012}
E.E. Ferrero, S.~Bustingorry  and A.B. Kolton,
\newblock {\em Non-steady relaxation and critical exponents at the depinning
  transition},
\newblock \doi{10.1103/PhysRevE.87.032122}{\rm Phys. Rev. E {\bf 87} (2013)
  032122}\null,
\newblock \arxiv{arXiv:1211.7275}.

\bibitem{GrassbergerDharMohanty2016}
P.~Grassberger, D.~Dhar  and P.~K. Mohanty,
\newblock {\em Oslo model, hyperuniformity, and the quenched
  {Edwards-Wilkinson} model},
\newblock \doi{10.1103/PhysRevE.94.042314}{\rm Phys. Rev. E {\bf 94} (2016)
  042314}\null.

\bibitem{RossoLeDoussalWiese2006a}
A.~Rosso, P.~{Le~Doussal}  and K.J. Wiese,
\newblock {\em Numerical calculation of the functional renormalization group
  fixed-point functions at the depinning transition},
\newblock \doi{10.1103/PhysRevB.75.220201}{\rm Phys. Rev. B {\bf 75} (2007)
  220201}\null,
\newblock \arxiv{cond-mat/0610821}.

\bibitem{RossoHartmannKrauth2002}
A.~Rosso, A.K. Hartmann  and W.~Krauth,
\newblock {\em Depinning of elastic manifolds},
\newblock \doi{10.1103/PhysRevE.67.021602}{\rm Phys. Rev. E {\bf 67} (2003)
  021602}\null,
\newblock \arxiv{cond-mat/0207288}.

\bibitem{RotersHuchtLubeckNowakUsadel1999}
L.~Roters, A.~Hucht, S.~L\"ubeck, U.~Nowak  and K.D. Usadel,
\newblock {\em Depinning transition and thermal fluctuations in the
  random-field {Ising} model},
\newblock \doi{10.1103/PhysRevE.60.5202}{\rm Phys. Rev. E {\bf 60} (1999)
  5202}\null.

\bibitem{KasparMungan2013}
D.C. Kaspar and M.~Mungan,
\newblock {\em Subthreshold behavior and avalanches in an exactly solvable
  charge density wave system},
\newblock \doi{10.1209/0295-5075/103/46002}{\rm EPL {\bf 103} (2013)
  46002}\null.

\bibitem{WieseFedorenko2018}
{K.J.} Wiese and A.A. Fedorenko,
\newblock {\em Field theories for loop-erased random walks},
\newblock \doi{10.1016/j.nuclphysb.2019.114696}{\rm Nucl. Phys. B {\bf 946}
  (2019)   114696}\null,
\newblock \arxiv{arXiv:1802.08830}.

\bibitem{WieseFedorenko2019}
K.J. Wiese and A.A. Fedorenko,
\newblock {\em Depinning transition of charge-density waves: Mapping onto
  ${O}(n)$ symmetric $\phi^4$ theory with $n\to -2$ and loop-erased random
  walks},
\newblock \doi{10.1103/PhysRevLett.123.197601}{\rm Phys. Rev. Lett. {\bf 123}
  (2019)   197601}\null,
\newblock \arxiv{arXiv:1908.11721}.

\bibitem{ShapiraWiese2020}
A.~Shapira and K.J. Wiese,
\newblock {\em An exact mapping between loop-erased random walks and an
  interacting field theory with two fermions and one boson},
\newblock \doi{10.21468/SciPostPhys.9.5.063}{\rm SciPost Phys. {\bf 9} (2020)
  063}\null,
\newblock \arxiv{arXiv:2006.07899}.

\bibitem{BalogTarjusTissier2019}
I.~Balog, G.~Tarjus  and M.~Tissier,
\newblock {\em Benchmarking the nonperturbative functional renormalization
  group approach on the random elastic manifold model in and out of
  equilibrium},
\newblock \doi{10.1088/1742-5468/ab3da5}{\rm J. Stat. Mech. {\bf 2019} (2019)
  103301}\null.

\bibitem{FedorenkoStepanow2002}
A.~Fedorenko and S.~Stepanow,
\newblock {\em Depinning transition at the upper critical dimension},
\newblock \doi{10.1103/PhysRevE.67.057104}{\rm Phys. Rev. E {\bf 67} (2003)
  057104}\null,
\newblock \arxiv{cond-mat/0209171}.

\bibitem{LeDoussalWiese2008a}
P.~Le Doussal and K.J. Wiese,
\newblock {\em Driven particle in a random landscape: disorder correlator,
  avalanche distribution and extreme value statistics of records},
\newblock \doi{10.1103/PhysRevE.79.051105}{\rm Phys. Rev. E {\bf 79} (2009)
  051105}\null,
\newblock \arxiv{arXiv:0808.3217}.

\bibitem{AlessandroBeatriceBertottiMontorsi1990}
B.~Alessandro, C.~Beatrice, G.~Bertotti  and A.~Montorsi,
\newblock {\em Domain-wall dynamics and {Barkhausen} effect in metallic
  ferromagnetic materials. {I. Theory}},
\newblock \doi{10.1063/1.346423}{\rm J. Appl. Phys. {\bf 68} (1990)
  2901}\null.

\bibitem{AlessandroBeatriceBertottiMontorsi1990b}
B.~Alessandro, C.~Beatrice, G.~Bertotti  and A.~Montorsi,
\newblock {\em {Domain-wall dynamics and Barkhausen effect in metallic
  ferromagnetic materials. II. Experiments}},
\newblock \doi{10.1063/1.346424}{\rm J. Appl. Phys. {\bf 68} (1990)
  2908}\null.

\bibitem{VergneCotillardPorteseil1981}
R.~Vergne, J.C. Cotillard  and J.L. Porteseil,
\newblock {\em Quelques aspects statistiques des processus d'aimantation dans
  les corps ferromagn{\'e}tiques. {Cas} du d\'eplacement d'une seule paroi de
  {Bloch} \`a $180^\circ$ dans un milieu monocristallin al\'eatoirement
  perturb\'e},
\newblock \doi{10.1051/rphysap:01981001609044900}{\rm Rev. Phys. Appl. (Paris)
  {\bf 16} (1981)   449--476}\null.

\bibitem{DurinZapperi2006b}
G.~Durin and S.~Zapperi,
\newblock {\em {The Barkhausen effect}},
\newblock in G.~Bertotti and I.~Mayergoyz, editors, {\em The Science of
  Hysteresis}, page~51, Amsterdam, 2006,
\newblock \arxiv{cond-mat/0404512}.

\bibitem{CsikorMotzWeygandZaiserZapperi2007}
F.~F. Csikor, C.~Motz, D.~Weygand, M.~Zaiser  and S.~Zapperi,
\newblock {\em Dislocation avalanches, strain bursts, and the problem of
  plastic forming at the micrometer scale},
\newblock \doi{10.1126/science.1143719}{\rm Science {\bf 318} (2007)
  251--254}\null.

\bibitem{LeDoussalWiese2012a}
P.~Le~Doussal and K.J. Wiese,
\newblock {\em Avalanche dynamics of elastic interfaces},
\newblock \doi{10.1103/PhysRevE.88.022106}{\rm Phys. Rev. E {\bf 88} (2013)
  022106}\null,
\newblock \arxiv{arXiv:1302.4316}.

\bibitem{ZhuWiese2017}
Z.~Zhu and {K.J.} Wiese,
\newblock {\em The spatial shape of avalanches},
\newblock \doi{10.1103/PhysRevE.96.062116}{\rm Phys. Rev. E {\bf 96} (2017)
  062116}\null,
\newblock \arxiv{arXiv:1708.01078}.

\bibitem{terBurgBohnDurinSommerWiese2021}
C.~ter Burg, F.~Bohn, F.~Durin, R.L. Sommer  and {K.J.} Wiese,
\newblock {\em Force-force correlations in disordered magnets},
\newblock (2021),
\newblock \arxiv{arXiv:2109.01197}.



\bibitem{DouinterBurgLechenaultWieseUnpublished}
A.~Douin, C.~ter Burg, F.~Lechenault  and K.J. Wiese,
\newblock unpublished.

\bibitem{BustingorryKoltonGiamarchi2010}
S.~Bustingorry, A.~B. Kolton  and T.~Giamarchi,
\newblock {\em Random-manifold to random-periodic depinning of an elastic
  interface},
\newblock \doi{10.1103/PhysRevB.82.094202}{\rm Phys. Rev. B {\bf 82} (2010)
  094202}\null.

\bibitem{BustingorryKolton2010}
S.~Bustingorry and A.~Kolton,
\newblock {\em Anisotropic finite-size scaling of an elastic string at the
  depinning threshold in a random-periodic medium},
\newblock \doi{10.4279/pip.020008}{\rm Papers in Physics {\bf 020008}
  (2010)}\null.

\bibitem{KoltonBustingorryFerreroRosso2013}
A.B. Kolton, S.~Bustingorry, E.E. Ferrero  and A.~Rosso,
\newblock {\em Uniqueness of the thermodynamic limit for driven disordered
  elastic interfaces},
\newblock \doi{10.1088/1742-5468/2013/12/p12004}{\rm J. Stat. Mech. {\bf 2013}
  (2013)   P12004}\null,
\newblock \arxiv{arXiv:1308.4329}.
%

\bibitem{RossoKrauth2001a}
A.~Rosso and W.~Krauth,
\newblock {\em {Monte Carlo} dynamics of driven strings in disordered media},
\newblock \doi{10.1103/PhysRevB.65.012202}{\rm Phys. Rev. B {\bf 65} (2001)
  012202}\null,
\newblock \arxiv{cond-mat/0102017}.

\bibitem{RossoKrauth2002}
A.~Rosso and W.~Krauth,
\newblock {\em Roughness at the depinning threshold for a long-range elastic
  string},
\newblock \doi{10.1103/PhysRevE.65.025101}{\rm Phys. Rev. E {\bf 65} (2002)
  025101}\null.

\bibitem{Rosso2002}
A.~Rosso,
\newblock {\em D\'epi\'egeage de vari\'etes \'elastiques en milieu
  al\'eatoire},
\newblock \link{http://tel.archives-ouvertes.fr/tel-00002097/}{\rm PhD thesis,
  Universit{\'e} Pierre et Marie Curie, tel.archives-ouvertes.fr/tel-00002097,
  2002},
\newblock \null.

\bibitem{SparfelWiese2021}
J.~Sparfel and K.J. Wiese,
\newblock {\em Skewness at depinning, and conformal invariance},
\newblock unpublished (2021).

\bibitem{GinspargLH1988}
P.~Ginsparg,
\newblock {\em Applied conformal field theory},
\newblock in E.~Br\'ezin and J.~Zinn-Justin, editors, {\em Fields, strings and
  critical phenomena}, {\em {\em Volume} XLIX} of {\em Les Houches, \'ecole
  d'\'et\'e de physique th\'eorique 1988}, North Holland, Amsterdam, 1988.

\bibitem{CardyInDombGreen}
J.~Cardy,
\newblock {\em Conformal invariance}.
\newblock {\em {\em Volume}~11} of {\em Phase Transitions and Critical
  Phenomena}, pages 55--126, Academic Press London, 1987.

\bibitem{LeschhornTang1993}
H.~Leschhorn and L.-H. Tang,
\newblock {\em Comment on ``{Elastic} string in a random potential''},
\newblock \doi{10.1103/PhysRevLett.70.2973}{\rm Phys. Rev. Lett. {\bf 70}
  (1993)   2973--2973}\null.

\bibitem{RossoKrauthPrivate}
A.~Rosso and W.~Krauth,
\newblock private communication.

\bibitem{KoltonPrivate}
A.B. Kolton,
\newblock private communication.

\bibitem{DummerKrauth2007}
O.~D\"ummer and W.~Krauth,
\newblock {\em Depinning exponents of the driven long-range elastic string},
\newblock \doi{10.1088/1742-5468/2007/01/P01019}{\rm J. Stat. Mech. (2007)
  P01019}\null,
\newblock \arxiv{cond-mat/0612323}.

\bibitem{ErtasKardar1994b}
D.~Ertas and M.~Kardar,
\newblock {\em Critical dynamics of contact line depinning},
\newblock \doi{10.1103/PhysRevE.49.R2532}{\rm Phys. Rev. E {\bf 49} (1994)
  2532}\null.

\bibitem{WieseOriginalReview}
K.J. Wiese,
\newblock {\em Original reserch done for this review, yet unpublished.}

\bibitem{RolleyGuthmannGombrowiczRepain1998}
E.~Rolley, C.~Guthmann, R.~Gombrowicz  and V.~Repain,
\newblock {\em {Roughness of the Contact Line on a Disordered Substrate}},
\newblock \doi{10.1103/PhysRevLett.80.2865}{\rm Phys. Rev. Lett. {\bf 80}
  (1998)   2865--2868}\null.

\bibitem{IlievPeshevaIliev2018}
P.~Iliev, N.~Pesheva  and S.~Iliev,
\newblock {\em Roughness of the contact line on random self-affine rough
  surfaces},
\newblock \doi{10.1103/PhysRevE.98.060801}{\rm Phys. Rev. E {\bf 98} (2018)
  060801}\null.

\bibitem{SantucciGrobToussaintSchmittbuhlHansenMaloy2010}
S.~Santucci, M.~Grob, R.~Toussaint, J.~Schmittbuhl, A.~Hansen  and K.~J.
  Mal{\o}y,
\newblock {\em Fracture roughness scaling: A case study on planar cracks},
\newblock \doi{10.1209/0295-5075/92/44001}{\rm {EPL} {\bf 92} (2010)
  44001}\null,
\newblock \arxiv{arXiv:1007.1188}.

\bibitem{RamanathanFisher1997}
S.~Ramanathan and DS. Fisher,
\newblock {\em Dynamics and instabilities of planar tensile cracks in
  heterogeneous media},
\newblock \doi{10.1103/PhysRevLett.79.877}{\rm Phys. Rev. Lett. {\bf 79} (1997)
    877--880}\null.

\bibitem{KatzavAdda-BediaBenAmarBoudaoud2007}
E.~Katzav, M.~Adda-Bedia, M.~Ben Amar  and A.~Boudaoud,
\newblock {\em Roughness of moving elastic lines: Crack and wetting fronts},
\newblock \doi{10.1103/PhysRevE.76.051601}{\rm Phys. Rev. E {\bf 76} (2007)
  051601}\null.

\bibitem{SantucciMaaloyDelaplaceMathiesenHansenHaavigBakkeSchmittbuhlVanelRay2007}
S.~Santucci, K.J. M\aa{}l\o{}y, A.~Delaplace, J.~Mathiesen, A.~Hansen, J.{\O}.
  {Haavig Bakke}, J~Schmittbuhl, L.~Vanel  and P.~Ray,
\newblock {\em Statistics of fracture surfaces},
\newblock \doi{10.1103/PhysRevE.75.016104}{\rm Phys. Rev. E {\bf 75} (2007)
  016104}\null.

\bibitem{BouchaudLapassetPlanes1990}
E~Bouchaud, G~Lapasset  and J~Plan{\`{e}}s,
\newblock {\em Fractal dimension of fractured surfaces: A universal value?},
\newblock \doi{10.1209/0295-5075/13/1/013}{\rm EPL {\bf 13} (1990)
  73--79}\null.

\bibitem{LawnBook1993}
B.~Lawn,
\newblock {\em Fracture of Brittle Solids},
\newblock Cambridge University Press, Cambridge, UK, 2nd edition, 1993.

\bibitem{ParisiCaldarelliPiotronero2000}
A.~Parisi, G.~Caldarelli  and L.~Piotronero,
\newblock {\em Roughness of fracture surfaces},
\newblock \doi{10.1209/epl/i2000-00439-9}{\rm Europhys. Lett. {\bf 52} (2000)
  304--10}\null.

\bibitem{ArndtNattermann2001}
P.F. Arndt and T.~Nattermann,
\newblock {\em Criterion for crack formation in disordered materials},
\newblock \doi{10.1103/PhysRevB.63.134204}{\rm Phys. Rev. B {\bf 63} (2001)
  134204}\null.

\bibitem{Ponson2007}
L.~Ponson,
\newblock {\em Crack propagation in disordered materials: How to decipher
  fracture surfaces},
\newblock \doi{10.1051/anphys:2008044}{\rm Ann. Phys. {\bf 32} (2007)
  1--128}\null.

\bibitem{Ponson2008}
L.~Ponson,
\newblock {\em Depinning transition in failure of inhomogeneous brittle
  materials},
\newblock \doi{10.1103/PhysRevLett.103.055501}{\rm Phys. Rev. Lett. {\bf 103}
  (2009)   055501}\null,
\newblock \arxiv{arXiv:0805.1802}.

\bibitem{TallakstadToussaintSantucciMaaloy2013}
R.~Tallakstad, K.T.and~Toussaint, S.~Santucci  and K.J. M\aa{}l\o{}y,
\newblock {\em Non-gaussian nature of fracture and the survival of fat-tail
  exponents},
\newblock \doi{10.1103/PhysRevLett.110.145501}{\rm Phys. Rev. Lett. {\bf 110}
  (2013)   145501}\null.

\bibitem{Bouchaud1997}
E.~Bouchaud,
\newblock {\em Scaling properties of cracks},
\newblock \doi{10.1088/0953-8984/9/21/002}{\rm J. Phys. Cond. Mat. {\bf 9}
  (1997)   4319--4344}\null.

\bibitem{PonsonPrivate}
L.~Ponson,
\newblock private communication.

\bibitem{RamanathanErtasFisher1997}
S.~Ramanathan, D.~Ertas  and D.S. Fisher,
\newblock {\em Quasistatic crack propagation in heterogeneous media},
\newblock \doi{10.1103/PhysRevLett.79.873}{\rm Phys. Rev. Lett. {\bf 79} (1997)
    873--876}\null.

\bibitem{BonamySantucciPonson2008}
D.~Bonamy, S.~Santucci  and L.~Ponson,
\newblock {\em Crackling dynamics in material failure as the signature of a
  self-organized dynamic phase transition},
\newblock \doi{10.1103/PhysRevLett.101.045501}{\rm Phys. Rev. Lett. {\bf 101}
  (2008)   045501}\null.

\bibitem{ErtasKardar1994}
D.~Ertas and M.~Kardar,
\newblock {\em Anisotropic scaling in depinning of a flux line},
\newblock \doi{10.1103/PhysRevLett.73.1703}{\rm Phys. Rev. Lett. {\bf 73}
  (1994)   1703--6}\null.

\bibitem{ErtasKardar1996}
D.~Ertas and M.~Kardar,
\newblock {\em Anisotropic scaling in threshold critical dynamics of driven
  directed lines},
\newblock \doi{10.1103/PhysRevB.53.3520}{\rm Phys. Rev. {\bf B 53} (1996)
3520--42}\null.



\bibitem{EliasWieseKolton2022}
F.~Elias, K.J. Wiese  and A.B. Kolton,
\newblock {\em Depinning and flow of a vortex line in an uniaxial random
  medium},
\newblock \doi{10.1103/PhysRevB.105.224209}{\rm Phys. Rev. B {\bf 105} (2022)
  224209}\null,
\newblock \arxiv{arXiv:2204.09003}.




\bibitem{DalmasLelargeVandembroucq2008}
D.~Dalmas, A.~Lelarge  and D.~Vandembroucq,
\newblock {\em Crack propagation through phase-separated glasses: Effect of the
  characteristic size of disorder},
\newblock \doi{10.1103/PhysRevLett.101.255501}{\rm Phys. Rev. Lett. {\bf 101}
  (2008)   255501}\null.

\bibitem{VernedePonsonBouchaud2015}
S.~Vern\`ede, L.~Ponson  and J.-P. Bouchaud,
\newblock {\em Turbulent fracture surfaces: A footprint of damage
  percolation?},
\newblock \doi{10.1103/PhysRevLett.114.215501}{\rm Phys. Rev. Lett. {\bf 114}
  (2015)   215501}\null.

\bibitem{Griffith1921}
A.A. Griffith,
\newblock {\em The phenomena of rupture and flow in solids},
\newblock \doi{10.1098/rsta.1921.0006}{\rm Phil. Trans. R. Soc. A (1921)}\null.

\bibitem{Irwin1957}
G~Irwin,
\newblock {\em Analysis of stresses and strains near the end of a crack
  traversing a plate},
\newblock Journal of Applied Mechanics {\bf 24} (1957)   361--364.

\bibitem{WieseBercyMelkonyanBizebard2019}
K.J. Wiese, M.~Bercy, L.~Melkonyan  and T.~Bizebard,
\newblock {\em Universal force correlations in an {RNA-DNA} unzipping
  experiment},
\newblock \doi{10.1103/PhysRevResearch.2.043385}{\rm Phys. Rev. Research {\bf
  2} (2020)   043385}\null,
\newblock \arxiv{arXiv:1909.01319}.

\bibitem{Ponson2016}
L.~Ponson,
\newblock {\em Statistical aspects in crack growth phenomena: how the
  fluctuations reveal the failure mechanisms},
\newblock \doi{10.1007/s10704-016-0117-7}{\rm Int. J. Fract. {\bf 201} (2016)
  11--27}\null.

\bibitem{deArcangelisRednerHerrmann1985}
L.~de~Arcangelis, S.~Redner  and H.J. Herrmann,
\newblock {\em A random fuse model for breaking processes},
\newblock \doi{10.1051/jphyslet:019850046013058500}{\rm J. Physique Lett. {\bf
  46} (1985)   585--590}\null.

\bibitem{BatrouniHansen1998}
G.G. Batrouni and A.~Hansen,
\newblock {\em Fracture in three-dimensional fuse networks},
\newblock \doi{10.1103/PhysRevLett.80.325}{\rm Phys. Rev. Lett. {\bf 80} (1998)
    325--328}\null.

\bibitem{NukalaimunoviZapperi2004}
P.K.V.V. Nukala, S.~Simunovi  and S.~Zapperi,
\newblock {\em Percolation and localization in the random fuse model},
\newblock \doi{10.1088/1742-5468/2004/08/p08001}{\rm J. Stat. Mech. {\bf 2004}
  (2004)   P08001}\null.

\bibitem{ZapperiNukalaSimunovic2005}
S.~Zapperi, P.K.V.V. Nukala  and S.~Simunovic,
\newblock {\em Crack avalanches in the three-dimensional random fuse model},
\newblock \doi{https://doi.org/10.1016/j.physa.2005.05.071}{\rm Physica A {\bf
  357} (2005)   129 -- 133}\null.

\bibitem{ZapperiNukala2006}
S.~Zapperi and P.K.V.V Nukala,
\newblock {\em Fracture statistics in the three-dimensional random fuse model},
\newblock \doi{10.1007/s10704-005-4659-3}{\rm Int. J. Fract. {\bf 140} (2006)
  99--111}\null.

\bibitem{GjerdenStormoHansen2014}
K.S. Gjerden, A.~Stormo  and A.~Hansen,
\newblock {\em Local dynamics of a randomly pinned crack front: a numerical
  study},
\newblock \doi{10.3389/fphy.2014.00066}{\rm Frontiers in Physics {\bf 2} (2014)
   ~66}\null.

\bibitem{StormoLenglineSchmittbuhlHansen2016}
A.~Stormo, O.~Lenglin\'e, J.~Schmittbuhl  and A.~Hansen,
\newblock {\em Soft-clamp fiber bundle model and interfacial crack propagation:
  Comparison using a non-linear imposed displacement},
\newblock \doi{10.3389/fphy.2016.00018}{\rm Frontiers in Physics {\bf 4} (2016)
   ~18}\null.

\bibitem{IoffeVinokur1987}
L.B. Ioffe and V.M. Vinokur,
\newblock {\em Dynamics of interfaces and dislocations in disordered media},
\newblock \doi{10.1088/0022-3719/20/36/016}{\rm J. Phys. C {\bf 20} (1987)
  6149--6158}\null.

\bibitem{Nattermann1990}
T.~Nattermann,
\newblock {\em Scaling approach to pinning - charge-density waves and giant
  flux creep in superconductors},
\newblock \doi{10.1103/PhysRevLett.64.2454}{\rm Phys. Rev. Lett. {\bf 64}
  (1990)   2454--2457}\null.

\bibitem{ChauveGiamarchiLeDoussal1998}
P.~Chauve, T.~Giamarchi  and P.~Le Doussal,
\newblock {\em Creep via dynamical functional renormalization group},
\newblock \doi{10.1209/epl/i1998-00443-7}{\rm Europhys. Lett. {\bf 44} (1998)
  110--15}\null.

\bibitem{FerreroFoiniGiamarchiKoltonRosso2020}
E.E. Ferrero, L.~Foini, T.~Giamarchi, A.B. Kolton  and A.~Rosso,
\newblock {\em Creep motion of elastic interfaces driven in a disordered
  landscape},
\newblock Ann. Rev. Cond. Mat. Phys. {\bf 12} (2020)   111--134,
\newblock \arxiv{arXiv:2001.11464}.

\bibitem{KoltonRossoGiamarchi2005}
A.B. Kolton, A.~Rosso  and T.~Giamarchi,
\newblock {\em Creep motion of an elastic string in a random potential},
\newblock \doi{10.1103/PhysRevLett.94.047002}{\rm Phys. Rev. Lett. {\bf 94}
  (2005)   047002}\null,
\newblock \arxiv{cond-mat/0408284}.

\bibitem{KoltonRossoGiamarchiKrauth2006}
A.B. Kolton, A.~Rosso, T.~Giamarchi  and W.~Krauth,
\newblock {\em Dynamics below the depinning threshold in disordered elastic
  systems},
\newblock \doi{10.1103/PhysRevLett.97.057001}{\rm Phys. Rev. Lett. {\bf 97}
  (2006)   057001}\null.

\bibitem{KoltonRossoGiamarchiKrauth2009}
A.B. Kolton, A.~Rosso, T.~Giamarchi  and W.~Krauth,
\newblock {\em Creep dynamics of elastic manifolds via exact transition
  pathways},
\newblock \doi{10.1103/PhysRevB.79.184207}{\rm Phys. Rev. B {\bf 79} (2009)
  184207}\null.

\bibitem{FerreroBustingorryKoltonRosso2013}
E.E. Ferrero, S.~Bustingorry, A.B. Kolton  and A.~Rosso,
\newblock {\em Numerical approaches on driven elastic interfaces in random
  media},
\newblock \doi{https://doi.org/10.1016/j.crhy.2013.08.002}{\rm Comptes Rendus
  Physique {\bf 14} (2013)   641 -- 650}\null.

\bibitem{FerreroFoiniGiamarchiKoltonRosso2017}
E.E. Ferrero, L.~Foini, T.~Giamarchi, A.B. Kolton  and A.~Rosso,
\newblock {\em Spatiotemporal patterns in ultraslow domain wall creep
  dynamics},
\newblock \doi{10.1103/PhysRevLett.118.147208}{\rm Phys. Rev. Lett. {\bf 118}
  (2017)   147208}\null.

\bibitem{MetaxasJametMouginCormierFerreBaltzRodmacqDienyStamps2007}
P.~J. Metaxas, J.~P. Jamet, A.~Mougin, M.~Cormier, J.~Ferre, V.~Baltz,
  B.~Rodmacq, B.~Dieny  and R.~L. Stamps,
\newblock {\em Creep and flow regimes of magnetic domain-wall motion in
  ultrathin {Pt/Co/Pt} films with perpendicular anisotropy},
\newblock \doi{10.1103/PhysRevLett.99.217208}{\rm Phys. Rev. Lett. {\bf 99}
  (2007)   217208}\null.

\bibitem{GorchonBustingorryFerreJeudyKoltonGiamarchi2014}
J.~Gorchon, S.~Bustingorry, J.~Ferr\'e, V.~Jeudy, A.B. Kolton  and
  T.~Giamarchi,
\newblock {\em Pinning-dependent field-driven domain wall dynamics and thermal
  scaling in an ultrathin $\mathrm{Pt}/\mathrm{Co}/\mathrm{Pt}$ magnetic film},
\newblock \doi{10.1103/PhysRevLett.113.027205}{\rm Phys. Rev. Lett. {\bf 113}
  (2014)   027205}\null.

\bibitem{JeudyMouginBustingorrySavero-TorresGorchonKoltonLemaitreJamet2016}
V.~Jeudy, A.~Mougin, S.~Bustingorry, W.~Savero~Torres, J.~Gorchon, A.~B.
  Kolton, A.~Lema\^{\i}tre  and J.-P. Jamet,
\newblock {\em Universal pinning energy barrier for driven domain walls in thin
  ferromagnetic films},
\newblock \doi{10.1103/PhysRevLett.117.057201}{\rm Phys. Rev. Lett. {\bf 117}
  (2016)   057201}\null.

\bibitem{Diaz-PardoSavero-TorresKoltonBustingorryJeudy2017}
R.~Diaz~Pardo, W.~Savero~Torres, A.B. Kolton, S.~Bustingorry  and V.~Jeudy,
\newblock {\em Universal depinning transition of domain walls in ultrathin
  ferromagnets},
\newblock \doi{10.1103/PhysRevB.95.184434}{\rm Phys. Rev. B {\bf 95} (2017)
  184434}\null.

\bibitem{TroyanovskiAartsKes1999}
AM. Troyanovski, J.~Aarts  and P.H. Kes,
\newblock {\em Collective and plastic vortex motion in superconductors at high
  flux densities},
\newblock \doi{10.1038/21385}{\rm Nature {\bf 399} (1999)   665--668}\null.

\bibitem{TallakstadToussaintSantucciSchmittbuhlMaaloy2011}
K.T. Tallakstad, R.~Toussaint, S.~Santucci, J.~Schmittbuhl  and K.J.
  M\aa{}l\o{}y,
\newblock {\em Local dynamics of a randomly pinned crack front during creep and
  forced propagation: An experimental study},
\newblock \doi{10.1103/PhysRevE.83.046108}{\rm Phys. Rev. E {\bf 83} (2011)
  046108}\null.

\bibitem{Vincent-DospitalCochardSantucciMaloyToussaint2020}
T.~Vincent-Dospital, A.~Cochard, S.~Santucci, K.J. Maloy  and R.~Toussaint,
\newblock Thermally activated intermittent dynamics of creeping crack fronts
  along disordered interfaces 2020,
\newblock \arxiv{arXiv:2010.06865}.

\bibitem{NattermannGiamarchiLeDoussal2003}
T.~Nattermann, T.~Giamarchi  and P.~Le Doussal,
\newblock {\em Variable-range hopping and quantum creep in one dimension},
\newblock \doi{10.1103/PhysRevLett.91.056603}{\rm Phys. Rev. Lett. {\bf 91}
  (2003)   056603}\null,
\newblock \arxiv{cond-mat/0303233}.

\bibitem{AndreanovFedorenko2014}
A.~Andreanov and A.A. Fedorenko,
\newblock {\em Localization of spin waves in disordered quantum rotors},
\newblock \doi{10.1103/PhysRevB.90.014205}{\rm Phys. Rev. B {\bf 90} (2014)
  014205}\null.

\bibitem{KoltonJagla2020}
A.~B. Kolton and E.~A. Jagla,
\newblock {\em Thermally rounded depinning of an elastic interface on a
  washboard potential},
\newblock (2020),
\newblock \arxiv{arXiv:2008.00534}.

\bibitem{JanssenSchaubSchmittmann1989}
H.~K. Janssen, B.~Schaub  and B.~Schmittmann,
\newblock {\em New universal short-time scaling behaviour of critical
  relaxation processes},
\newblock \doi{10.1007/BF01319383}{\rm Z. Phys. B {\bf 73} (1989)
  539--549}\null.

\bibitem{ChenGuoLiMarculescuSchulke2000}
Y.~Chen, S.H. Guo, Z.B. Li, S.~Marculescu  and L.~Sch\"ulke,
\newblock {\em The short-time critical behaviour of the {Ginzburg-Landau} model
  with long-range interaction},
\newblock \doi{10.1007/s100510070060}{\rm EPJB {\bf 18} (2000)
  289--296}\null.

\bibitem{SchehrLeDoussal2005}
G.~Schehr and P.~Le Doussal,
\newblock {\em Functional renormalization for pinned elastic systems away from
  their steady states},
\newblock \doi{10.1209/epl/i2005-10074-6}{\rm Europhys. Lett. {\bf 71} (2005)
  290--296}\null,
\newblock \arxiv{cond-mat/0501199}.

\bibitem{KoltonSchehrLeDoussal2009}
A.B. Kolton, G.~Schehr  and P.~Le Doussal,
\newblock {\em Universal non-stationary dynamics at the depinning transition},
\newblock \doi{10.1103/PhysRevLett.103.160602}{\rm Phys. Rev. Lett. {\bf 103}
  (2009)   160602}\null,
\newblock \arxiv{arXiv:0906.2494}.

\bibitem{DickmanAlavaMunozPeltolaVespignaniZapperi2001}
R.~Dickman, M.~Alava, M.A. Mu\~noz, J.~Peltola, A.~Vespignani  and S.~Zapperi,
\newblock {\em Critical behavior of a one-dimensional fixed-energy stochastic
  sandpile},
\newblock \doi{10.1103/PhysRevE.64.056104}{\rm Phys. Rev. E {\bf 64} (2001)
  056104}\null.

\bibitem{KwonKim2016}
S.~Kwon and Ji.M. Kim,
\newblock {\em Critical behavior for random initial conditions in the
  one-dimensional fixed-energy manna sandpile model},
\newblock \doi{10.1103/PhysRevE.94.012113}{\rm Phys. Rev. E {\bf 94} (2016)
  012113}\null.

\bibitem{TapaderPradhanDhar2020}
D.~Tapader, P.~Pradhan  and D.~Dhar,
\newblock {\em Density relaxation in conserved {Manna} sandpiles},
\newblock (2020),
\newblock \arxiv{arXiv:2011.01173}.

\bibitem{TheScienceOfHysteresis}
G.~Bertotti and I.~Mayergoyz, editors,
\newblock {\em The science of hysteresis}, {\em {\em Volume} 1-3},
\newblock Elsevier, 2005.

\bibitem{GrassiKoltonJeudyMouginBustingorryCuriale2018}
M.P. Grassi, A~B. Kolton, V.~Jeudy, A.~Mougin, S.~Bustingorry  and J.~Curiale,
\newblock {\em Intermittent collective dynamics of domain walls in the creep
  regime},
\newblock \doi{10.1103/PhysRevB.98.224201}{\rm Phys. Rev. B {\bf 98} (2018)
  224201}\null.

\bibitem{AlbornozFerreroKoltonJeudyBustingorryCuriale2021}
L.J. Albornoz, E.E. Ferrero, A.B. Kolton, V.~Jeudy, S.~Bustingorry  and
  J.~Curiale,
\newblock {\em Universal critical exponents of the magnetic domain wall
  depinning transition},
\newblock (2021),
\newblock \arxiv{arXiv:2101.06555}.

\bibitem{JeudyDiaz-PardoSavero-TorresBustingorryKolton2018}
V.~Jeudy, R.~D\'{\i}az~Pardo, W.~Savero~Torres, S.~Bustingorry  and A.~B.
  Kolton,
\newblock {\em Pinning of domain walls in thin ferromagnetic films},
\newblock \doi{10.1103/PhysRevB.98.054406}{\rm Phys. Rev. B {\bf 98} (2018)
  054406}\null.

\bibitem{ShibauchiKrusin-ElbaumVinokurArgyleWellerTerris2001}
T.~Shibauchi, L.~Krusin-Elbaum, V.~M. Vinokur, B.~Argyle, D.~Weller  and B.~D.
  Terris,
\newblock {\em Deroughening of a 1d domain wall in an ultrathin magnetic film
  by a correlated defect},
\newblock \doi{10.1103/PhysRevLett.87.267201}{\rm Phys. Rev. Lett. {\bf 87}
  (2001)   267201}\null.

\bibitem{BauerMouginJametRepainFerreStampsBernasChappert2005}
M.~Bauer, A.~Mougin, J.~P. Jamet, V.~Repain, J.~Ferr\'e, R.~L. Stamps,
  H.~Bernas  and C.~Chappert,
\newblock {\em Deroughening of domain wall pairs by dipolar repulsion},
\newblock \doi{10.1103/PhysRevLett.94.207211}{\rm Phys. Rev. Lett. {\bf 94}
  (2005)   207211}\null.

\bibitem{MoonKimYooChoHwangKahngMinShinChoe2013}
K.-W. Moon, D.-H. Kim, S.-C.~Cheol Yoo, C.-G. Cho, S.~Hwang, B.~Kahng, B.-C.
  Min, K.-H. Shin  and S.-B. Choe,
\newblock {\em Distinct universality classes of domain wall roughness in
  two-dimensional $\mathrm{Pt}/\mathrm{Co}/\mathrm{Pt}$ films},
\newblock \doi{10.1103/PhysRevLett.110.107203}{\rm Phys. Rev. Lett. {\bf 110}
  (2013)   107203}\null.

\bibitem{DomenichiniQuinterosGranadaCollinGeorgeCurialeBustingorryCapelutoPasquini2019}
P.~Domenichini, C.P. Quinteros, M.~Granada, S.~Collin, J.-M. George,
  J.~Curiale, S.~Bustingorry, M.G. Capeluto  and G.~Pasquini,
\newblock {\em Transient magnetic-domain-wall ac dynamics by means of
  magneto-optical {Kerr} effect microscopy},
\newblock \doi{10.1103/PhysRevB.99.214401}{\rm Phys. Rev. B {\bf 99} (2019)
  214401}\null.

\bibitem{FerreMetaxasMouginJametGorchonJeudy2013}
J.~Ferr\'{e}, P.J. Metaxas, A.~Mougin, J.-P. Jamet, J.~Gorchon  and V.~Jeudy,
\newblock {\em Universal magnetic domain wall dynamics in the presence of weak
  disorder},
\newblock \doi{https://doi.org/10.1016/j.crhy.2013.08.001}{\rm Comptes Rendus
  Physique {\bf 14} (2013)   651 -- 666}\null.

\bibitem{Albornoz2021}
L.J. Albornoz,
\newblock {\em Dynamics and morphology of driven domain walls in magnetic thin
  films from the standpoint of statistical physics},
\newblock PhD thesis, Universit\'e Paris-Saclay and Universidad Nacional de
  Cuyo, 2021.

\bibitem{LyuksyutovNattermannPokrovsky1999}
I.F. Lyuksyutov, T.~Nattermann  and V.~Pokrovsky,
\newblock {\em Theory of the hysteresis loop in ferromagnets},
\newblock \doi{10.1103/PhysRevB.59.4260}{\rm Phys. Rev. B {\bf 59} (1999)
  4260--4272}\null.

\bibitem{NattermannPokrovskyVinokur2001}
T.~Nattermann, V.~Pokrovsky  and V.M. Vinokur,
\newblock {\em Hysteretic dynamics of domain walls at finite temperatures},
\newblock \doi{10.1103/PhysRevLett.87.197005}{\rm Phys. Rev. Lett. {\bf 87}
  (2001)   197005}\null.

\bibitem{GlatzNattermannPokrovsky2002}
A.~Glatz, T.~Nattermann  and V.~Pokrovsky,
\newblock {\em Domain wall depinning in random media by ac fields},
\newblock \doi{10.1103/PhysRevLett.90.047201}{\rm Phys. Rev. Lett. {\bf 90}
  (2000)   047201}\null.

\bibitem{KleemannRhensiusPetracicFerreJametBernas2007}
W.~Kleemann, J.~Rhensius, O.~Petracic, J.~Ferre, J.~P. Jamet  and H.~Bernas,
\newblock {\em Modes of periodic domain wall motion in ultrathin ferromagnetic
  layers},
\newblock \doi{10.1103/PhysRevLett.99.097203}{\rm Phys. Rev. Lett. {\bf 99}
  (2007)   097203}\null.

\bibitem{DobrinevskiPhD}
A.~Dobrinevski,
\newblock {\em Field theory of disordered systems -- avalanches of an elastic
  interface in a random medium},
\newblock PhD Thesis, ENS Paris (2013),
\newblock \arxiv{arXiv:1312.7156}.

\bibitem{SchwarzFisher2001}
J.M. Schwarz and D.S. Fisher,
\newblock {\em Depinning with dynamic stress overshoots: Mean field theory},
\newblock \doi{10.1103/PhysRevLett.87.096107}{\rm Phys. Rev. Lett. {\bf 87}
  (2001)   096107}\null.

\bibitem{LeDoussalPetkovicWiese2012}
P.~Le~Doussal, A.~Petkovi\ifmmode~\acute{c}\else \'{c}\fi{}  and K.J. Wiese,
\newblock {\em Distribution of velocities and acceleration for a particle in
  {Brownian} correlated disorder: Inertial case},
\newblock \doi{10.1103/PhysRevE.85.061116}{\rm Phys. Rev. E {\bf 85} (2012)
  061116}\null,
\newblock \arxiv{arXiv:1203.5620}.

\bibitem{LebowitzSpohn1999}
J.L. Lebowitz and H.~Spohn,
\newblock {\em A {Gallavotti--Cohen}-type symmetry in the large deviation
  functional for stochastic dynamics},
\newblock \doi{10.1023/A:1004589714161}{\rm J. Stat. Phys. {\bf 95} (1999)
  333--365}\null.

\bibitem{MajumdarSchehr2014}
S.N. Majumdar and G.~Schehr,
\newblock {\em Top eigenvalue of a random matrix: large deviations and third
  order phase transition},
\newblock \doi{10.1088/1742-5468/2014/01/P01012}{\rm J. Stat. Mech. (2014)
  P01012}\null.

\bibitem{KrapivskyMallickSadhu2014}
P.~L. Krapivsky, K.~Mallick  and T.~Sadhu,
\newblock {\em Large deviations in single-file diffusion},
\newblock \doi{10.1103/PhysRevLett.113.078101}{\rm Phys. Rev. Lett. {\bf 113}
  (2014)   078101}\null.

\bibitem{SadhuDerrida2015}
T.~Sadhu and B.~Derrida,
\newblock {\em Large deviation function of a tracer position in single file
  diffusion},
\newblock \doi{10.1088/1742-5468/2015/09/P09008}{\rm J. Stat. Mech. {\bf 2015}
  (2015)   P09008}\null.

\bibitem{VinokurNattermann1997}
V.M. Vinokur and T.~Nattermann,
\newblock {\em Hysteretic depinning of anisotropic charge density waves},
\newblock \doi{10.1103/PhysRevLett.79.3471}{\rm Phys. Rev. Lett. {\bf 79}
  (1997)   3471--3474}\null.

\bibitem{MarchettiMiddletonPrellberg2000}
MC. Marchetti, AA. Middleton  and T.~Prellberg,
\newblock {\em Viscoelastic depinning of driven systems: Mean-field plastic
  scallops},
\newblock \doi{10.1103/PhysRevLett.85.1104}{\rm Phys. Rev. Lett. {\bf 85}
  (2000)   1104--1107}\null.

\bibitem{MarchettiSaunders2002}
M.C. Marchetti and K.~Saunders,
\newblock {\em Viscoelasticity from a microscopic model of dislocation
  dynamics},
\newblock \doi{10.1103/PhysRevB.66.224113}{\rm Phys. Rev. B {\bf 66} (2002)
  224113}\null.

\bibitem{MarchettiDahmen2002}
M.C. Marchetti and K.A. Dahmen,
\newblock {\em Hysteresis in driven disordered systems: From plastic depinning
  to magnets},
\newblock \doi{10.1103/PhysRevB.66.214201}{\rm Phys. Rev. B {\bf 66} (2002)
  214201}\null.

\bibitem{SaundersSchwarzMarchettiMiddleton2004}
K.~Saunders, J.M. Schwarz, M.C. Marchetti  and A.A. Middleton,
\newblock {\em Mean-field theory of collective transport with phase slips},
\newblock \doi{10.1103/PhysRevB.70.024205}{\rm Phys. Rev. B {\bf 70} (2004)
  024205}\null.

\bibitem{Marchetti2005}
M.C. Marchetti,
\newblock {\em Models of plastic depinning of driven disordered systems},
\newblock \doi{10.1007/BF02704171}{\rm Pramana {\bf 64} (2005)
  1097--1107}\null.

\bibitem{Marchetti2006}
M.C. Marchetti,
\newblock \doi{10.1007/3-540-33204-9_9}{\rm {\em Depinning and plasticity of
  driven disordered lattices}}\null,
\newblock in {\em Jamming, Yielding, and Irreversible Deformations in Condensed
  Matter}, Springer-Verlag, Berlin, 2006.

\bibitem{LeDoussalMarchettiWiese2008}
P.~{Le Doussal}, M.C. Marchetti  and K.J. Wiese,
\newblock {\em Depinning in a two-layer model of plastic flow},
\newblock \doi{10.1103/PhysRevB.78.224201}{\rm Phys. Rev. B {\bf 78} (2008)
  224201}\null,
\newblock \arxiv{arXiv:0801.0137}.

\bibitem{FerreroJagla2019}
E.~E. Ferrero and E.~A. Jagla,
\newblock {\em Elastic interfaces on disordered substrates: From mean-field
  depinning to yielding},
\newblock \doi{10.1103/PhysRevLett.123.218002}{\rm Phys. Rev. Lett. {\bf 123}
  (2019)   218002}\null.

\bibitem{NicolasMartensBocquetBarrat2014}
A.~Nicolas, K.~Martens, L.~Bocquet  and J.-L. Barrat,
\newblock {\em Universal and non-universal features in coarse-grained models of
  flow in disordered solids},
\newblock \doi{10.1039/C4SM00395K}{\rm Soft Matter {\bf 10} (2014)
  4648--4661}\null.

\bibitem{AgoritsasBertinMartensBarrat2015}
E.~Agoritsas, E.~Bertin, K.~Martens  and J.-L. Barrat,
\newblock {\em On the relevance of disorder in athermal amorphous materials
  under shear},
\newblock \doi{10.1140/epje/i2015-15071-x}{\rm EPJB {\bf 38} (2015)  ~71}\null.

\bibitem{VasishtGoffMartensBarrat2018}
V.V. Vasisht, M.~Le Goff, K.~Martens  and J.-L. Barrat,
\newblock {\em Permanent shear localization in dense disordered materials due
  to microscopic inertia},
\newblock (2018),
\newblock \arxiv{arXiv:1812.03948}.

\bibitem{TyukodiPatinetRouxVandembroucq2016}
B.~Tyukodi, S.~Patinet, S.~Roux  and D.~Vandembroucq,
\newblock {\em From depinning transition to plastic yielding of amorphous
  media: A soft-modes perspective},
\newblock \doi{10.1103/PhysRevE.93.063005}{\rm Phys. Rev. E {\bf 93} (2016)
  063005}\null.

\bibitem{NicolasFerreroMartensBarrat2018}
A.~Nicolas, E.E. Ferrero, K.~Martens  and J.-L. Barrat,
\newblock {\em Deformation and flow of amorphous solids: Insights from
  elastoplastic models},
\newblock \doi{10.1103/RevModPhys.90.045006}{\rm Rev. Mod. Phys. {\bf 90}
  (2018)   045006}\null.

\bibitem{BalentsKardar1993}
L.~Balents and M.~Kardar,
\newblock {\em Delocalization of flux lines from extended defects by bulk
  randomness},
\newblock \doi{10.1209/0295-5075/23/7/007}{\rm Europhys. Lett. {\bf 23} (1993)
   503--509}\null,
\newblock \arxiv{cond-mat/9303015}.

\bibitem{ChauveLeDoussalGiamarchi2000}
P.~Chauve, P.~Le Doussal  and T.~Giamarchi,
\newblock {\em Dynamical transverse {Meissner} effect and transition in moving
  {Bose} glass},
\newblock \doi{10.1103/PhysRevB.61.R11906}{\rm Phys. Rev. B {\bf 61} (2000)
  11906--11909}\null.

\bibitem{OliveSoretDoussalGiamarchi2003}
E.~Olive, J.C. Soret, P.L. Doussal  and T.~Giamarchi,
\newblock {\em Numerical simulation evidence of dynamical transverse {Meissner}
  effect and moving {Bose} glass phase},
\newblock \doi{10.1103/PhysRevLett.91.037005}{\rm Phys. Rev. Lett. {\bf 91}
  (2003)   037005}\null,
\newblock \arxiv{cond-mat/0301471}.

\bibitem{ChenBalentsFisherMarchetti1996}
L.W. Chen, L.~Balents, M.P.A. Fisher  and M.C. Marchetti,
\newblock {\em Dynamical transition in sliding charge-density waves with
  quenched disorder},
\newblock \doi{10.1103/PhysRevB.54.12798}{\rm Phys. Rev. B {\bf 54} (1996)
  12798--12806}\null,
\newblock \arxiv{cond-mat/9605007}.

\bibitem{LeDoussalCugliandoloPeliti1997}
P.~{Le Doussal}, L.F. Cugliandolo  and L.~Peliti,
\newblock {\em Dynamics of particles and manifolds in a quenched random force
  field},
\newblock \doi{10.1209/epl/i1997-00323-8}{\rm Europhys. Lett. {\bf 39} (1997)
  111}\null,
\newblock \arxiv{cond-mat/9612079}.

\bibitem{LeDoussalWiese1997}
P.~Le Doussal and K.J. Wiese,
\newblock {\em Glassy trapping of elastic manifolds in nonpotential static
  random flows},
\newblock \doi{10.1103/PhysRevLett.80.2362}{\rm Phys. Rev. Lett. {\bf 80}
  (1998)   2362}\null,
\newblock \arxiv{cond-mat/9708112}.

\bibitem{WieseLedoussal1998}
K.J. Wiese and P.~Le Doussal,
\newblock {\em Polymers and manifolds in static random flows: a {RG} study},
\newblock \doi{10.1016/S0550-3213(99)00179-0}{\rm Nucl. Phys. {\bf B 552}
  (1999)   529--598}\null,
\newblock \arxiv{cond-mat/9808330}.

\bibitem{GiamarchiLeDoussal1996}
T.~Giamarchi and P.~{Le Doussal},
\newblock {\em Moving glass phases of driven lattices},
\newblock \doi{10.1103/PhysRevLett.76.3408}{\rm Phys. Rev. Lett. {\bf 76}
  (1996)   3408}\null.

\bibitem{LeDoussalGiamarchi1997}
P.~Le Doussal and T.~Giamarchi,
\newblock {\em Moving glass theory of driven lattices with disorder},
\newblock \doi{10.1103/PhysRevB.57.11356}{\rm Phys. Rev. {\bf B 57} (1998)
  11356--11403}\null,
\newblock \arxiv{cond-mat/9708085}.

\bibitem{BalentsFisher1995}
L.~Balents and M.P.A. Fisher,
\newblock {\em Temporal order in dirty driven periodic media},
\newblock \doi{10.1103/PhysRevLett.75.4270}{\rm Phys. Rev. Lett. {\bf 75}
  (1995)   4270}\null,
\newblock \arxiv{cond-mat/9504082}.

\bibitem{BalentsMarchettiRadzihovsky1997}
L.~Balents, M.C. Marchetti  and L.~Radzihovsky,
\newblock {\em Moving glass phase of driven lattices - comment},
\newblock \doi{10.1103/PhysRevLett.78.751}{\rm Phys. Rev. Lett. {\bf 78} (1997)
    751--751}\null.

\bibitem{RossoKrauthLeDoussalVannimenusWiese2003}
A.~Rosso, W.~Krauth, P.~Le Doussal, J.~Vannimenus  and K.J. Wiese,
\newblock {\em Universal interface width distributions at the depinning
  threshold},
\newblock \doi{10.1103/PhysRevE.68.036128}{\rm Phys. Rev. E {\bf 68} (2003)
  036128}\null,
\newblock \arxiv{cond-mat/0301464}.

\bibitem{MoulinetRossoKrauthRolley2004}
S.~Moulinet, A.~Rosso, W.~Krauth  and E.~Rolley,
\newblock {\em Width distribution of contact lines on a disordered substrate},
\newblock \doi{10.1103/PhysRevE.69.035103}{\rm Phys. Rev. E {\bf 69} (2004)
  035103}\null,
\newblock \arxiv{cond-mat/0310173}.

\bibitem{FedorenkoStepanow2003}
A.~Fedorenko and S.~Stepanow,
\newblock {\em Universal energy distribution for interfaces in a random-field
  environment},
\newblock \doi{10.1103/PhysRevE.68.056115}{\rm Phys. Rev. E {\bf 68} (2003)
  056115}\null.

\bibitem{KadanoffNagelWuZhou1989}
L.P. Kadanoff, S.R. Nagel, L.~Wu  and S.~Zhou,
\newblock {\em Scaling and universality in avalanches},
\newblock \doi{10.1103/PhysRevA.39.6524}{\rm Phys. Rev. A {\bf 39} (1989)
  6524--6537}\null.

\bibitem{AragonKoltonDoussalWieseJagla2016}
L.E. Aragon, A.B. Kolton, P.~Le Doussal, K.J. Wiese  and E.~Jagla,
\newblock {\em Avalanches in tip-driven interfaces in random media},
\newblock \doi{10.1209/0295-5075/113/10002}{\rm EPL {\bf 113} (2016)
  10002}\null,
\newblock \arxiv{arXiv:1510.06795}.

\bibitem{PaczuskiBoettcher1996}
M.~Paczuski and S.~Boettcher,
\newblock {\em Universality in sandpiles, interface depinning, and earthquake
  models},
\newblock \doi{10.1103/PhysRevLett.77.111}{\rm Phys. Rev. Lett. {\bf 77} (1996)
    111}\null.

\bibitem{NakanishiSneppen1997}
H.~Nakanishi and K.~Sneppen,
\newblock {\em Universal versus drive-dependent exponents for sandpile models},
\newblock \doi{10.1103/PhysRevE.55.4012}{\rm Phys. Rev. E {\bf 55} (1997)
  4012--4016}\null.

\bibitem{DelormeLeDoussalWiese2016}
M.~Delorme, P.~Le Doussal  and K.J. Wiese,
\newblock {\em Distribution of joint local and total size and of extension for
  avalanches in the {Brownian} force model},
\newblock \doi{10.1103/PhysRevE.93.052142}{\rm Phys. Rev. E {\bf 93} (2016)
  052142}\null,
\newblock \arxiv{arXiv:1601.04940}.

\bibitem{Colaiori2008}
F.~Colaiori,
\newblock {\em Exactly solvable model of avalanches dynamics for {Barkhausen}
  crackling noise},
\newblock \doi{10.1080/00018730802420614}{\rm Adv. Phys. {\bf 57} (2008)
  287}\null,
\newblock \arxiv{arXiv:0902.3173}.

\bibitem{DobrinevskiLeDoussalWiese2011b}
A.~Dobrinevski, P.~{Le Doussal}  and K.J. Wiese,
\newblock {\em Non-stationary dynamics of the
  {Alessandro-Beatrice-Bertotti-Montorsi} model},
\newblock \doi{10.1103/PhysRevE.85.031105}{\rm Phys. Rev. E {\bf 85} (2012)
  031105}\null,
\newblock \arxiv{arXiv:1112.6307}.

\bibitem{Munoz2004}
M.A. Mu{\~n}oz,
\newblock {\em Multiplicative noise in non-equilibrium phase transitions: A
  tutorial},
\newblock in {\em Advances in Condensed Matter and Statistical Physics}, pages
  37--68, Nova Science Publishers, Inc., E. Korutcheva and R. Cuerno eds.,
  2004.

\bibitem{DornicChateMunoz2005}
I.~Dornic, H.~Chat\'e  and M.A. Mu\~noz,
\newblock {\em Integration of {Langevin} equations with multiplicative noise
  and the viability of field theories for absorbing phase transitions},
\newblock \doi{10.1103/PhysRevLett.94.100601}{\rm Phys. Rev. Lett. {\bf 94}
  (2005)   100601}\null.

\bibitem{WatsonGalton1875}
H.W. Watson and F.~Galton,
\newblock {\em On the probability of the extinction of families},
\newblock Journal of the Anthropological Institute of Great Britain {\bf 4}
  (1875)   138--144.

\bibitem{DobrinevskiLeDoussalWiese2014a}
A.~Dobrinevski, P.~{Le Doussal}  and K.J. Wiese,
\newblock {\em Avalanche shape and exponents beyond mean-field theory},
\newblock \doi{10.1209/0295-5075/108/66002}{\rm EPL {\bf 108} (2014)
  66002}\null,
\newblock \arxiv{arXiv:1407.7353}.

\bibitem{ThieryLeDoussalWiese2015}
T.~Thiery, P.~Le Doussal  and K.J. Wiese,
\newblock {\em Spatial shape of avalanches in the {Brownian} force model},
\newblock \doi{10.1088/1742-5468/2015/08/P08019}{\rm J. Stat. Mech. {\bf 2015}
  (2015)   P08019}\null,
\newblock \arxiv{arXiv:1504.05342}.

\bibitem{LeDoussalWiese2008c}
P.~{Le~Doussal} and K.J. Wiese,
\newblock {\em Size distributions of shocks and static avalanches from the
  functional renormalization group},
\newblock \doi{10.1103/PhysRevE.79.051106}{\rm Phys. Rev. E {\bf 79} (2009)
  051106}\null,
\newblock \arxiv{arXiv:0812.1893}.

\bibitem{LeDoussalWiese2011b}
P.~{Le Doussal} and K.J. Wiese,
\newblock {\em First-principle derivation of static avalanche-size
  distribution},
\newblock \doi{10.1103/PhysRevE.85.061102}{\rm Phys. Rev. E {\bf 85} (2011)
  061102}\null,
\newblock \arxiv{arXiv:1111.3172}.

\bibitem{RossoLeDoussalWiese2009a}
A.~Rosso, P.~{Le~Doussal}  and K.J.\ Wiese,
\newblock {\em Avalanche-size distribution at the depinning transition: A
  numerical test of the theory},
\newblock \doi{10.1103/PhysRevB.80.144204}{\rm Phys. Rev. B {\bf 80} (2009)
  144204}\null,
\newblock \arxiv{arXiv:0904.1123}.

\bibitem{LaursonPrivate}
L.~Laurson,
\newblock private communication.

\bibitem{LaursonIllaSantucciTallakstadyAlava2013}
L.~Laurson, X.~Illa, S.~Santucci, K.T. Tallakstad, K.J.~M\aa l\o y  and M.J.
  Alava,
\newblock {\em Evolution of the average avalanche shape with the universality
  class},
\newblock \doi{10.1038/ncomms3927}{\rm Nat. Commun. {\bf 4} (2013)
  2927}\null.

\bibitem{DobrinevskiLeDoussalWiese2013}
A.~Dobrinevski, P.~Le~Doussal  and K.J. Wiese,
\newblock {\em Statistics of avalanches with relaxation and {Barkhausen} noise:
  A solvable model},
\newblock \doi{10.1103/PhysRevE.88.032106}{\rm Phys. Rev. E {\bf 88} (2013)
  032106}\null,
\newblock \arxiv{arXiv:1304.7219}.

\bibitem{ThieryLeDoussal2016}
T.~Thiery and P.~Le Doussal,
\newblock {\em Universality in the mean spatial shape of avalanches},
\newblock \doi{10.1209/0295-5075/114/36003}{\rm EPL {\bf 114} (2016)
  36003}\null,
\newblock \arxiv{arXiv:1601.00174}.

\bibitem{KoltonLeDoussalWiese2019}
A.~Kolton, P.~{Le Doussal}  and K.J. Wiese,
\newblock {\em Distribution of velocities in an avalanche, and related
  quantities: Theory and numerical verification},
\newblock \doi{10.1209/0295-5075/127/46001}{\rm EPL {\bf 127} (2019)
  46001}\null,
\newblock \arxiv{arXiv:1904.08657}.

\bibitem{LeDoussalWiese2011a}
P.~{Le Doussal} and K.J. Wiese,
\newblock {\em Distribution of velocities in an avalanche},
\newblock \doi{10.1209/0295-5075/97/46004}{\rm EPL {\bf 97} (2012)
  46004}\null,
\newblock \arxiv{arXiv:1104.2629}.

\bibitem{ThieryLeDoussalWiese2016}
T.~Thiery, P.~Le Doussal  and K.J. Wiese,
\newblock {\em Universal correlations between shocks in the ground state of
  elastic interfaces in disordered media},
\newblock \doi{10.1103/PhysRevE.94.012110}{\rm Phys. Rev. E {\bf 94} (2016)
  012110}\null,
\newblock \arxiv{arXiv:1604.05556}.

\bibitem{LeDoussalThiery2020}
P.~Le~Doussal and T.~Thiery,
\newblock {\em Correlations between avalanches in the depinning dynamics of
  elastic interfaces},
\newblock \doi{10.1103/PhysRevE.101.032108}{\rm Phys. Rev. E {\bf 101} (2020)
  032108}\null.

\bibitem{ZapperiCastellanoColaioriDurin2005}
S.~Zapperi, C.~Castellano, F.~Colaiori  and G.~Durin,
\newblock {\em Signature of effective mass in crackling-noise asymmetry},
\newblock \doi{10.1038/nphys101}{\rm Nat. Phys. {\bf 1} (2005)   46--49}\null.

\bibitem{PiterbargBook1995}
V.I. Piterbarg,
\newblock {\em Asymptotic Methods in the Theory of Gaussian Processes and
  Fields},
\newblock Translations of Mathematical Monographs, v.148,
\newblock American Mathematical Society, Providence, Rhode Island, 1995.

\bibitem{PiterbargBook2015}
V.I. Piterbarg,
\newblock {\em Twenty Lectures About {Gaussian} Processes},
\newblock Atlantic Financial Press, London, New York, 2015.

\bibitem{Michna2009}
Z.~Michna,
\newblock {\em Remarks on {Pickands} theorem},
\newblock \doi{10.19195/0208-4147.37.2.10}{\rm Probability and Mathematical
  Statististics {\bf 37} (2017)   373--393}\null,
\newblock \arxiv{arXiv:0904.3832}.

\bibitem{DelormeWiese2015}
M.~Delorme and {K.J.} Wiese,
\newblock {\em Maximum of a fractional {Brownian} motion: Analytic results from
  perturbation theory},
\newblock \doi{10.1103/PhysRevLett.115.210601}{\rm Phys. Rev. Lett. {\bf 115}
  (2015)   210601}\null,
\newblock \arxiv{arXiv:1507.06238}.

\bibitem{WieseMajumdarRosso2010}
K.J. Wiese, S.N. Majumdar  and A.~Rosso,
\newblock {\em Perturbation theory for fractional {Brownian} motion in presence
  of absorbing boundaries},
\newblock \doi{10.1103/PhysRevE.83.061141}{\rm Phys. Rev. E {\bf 83} (2011)
  061141}\null,
\newblock \arxiv{arXiv:1011.4807}.

\bibitem{DelormeWiese2016b}
M.~Delorme and K.J. Wiese,
\newblock {\em Extreme-value statistics of fractional {Brownian} motion
  bridges},
\newblock \doi{10.1103/PhysRevE.94.052105}{\rm Phys. Rev. E {\bf 94} (2016)
  052105}\null,
\newblock \arxiv{arXiv:1605.04132}.

\bibitem{DelormeWiese2016}
M.~Delorme and K.J. Wiese,
\newblock {\em Perturbative expansion for the maximum of fractional {Brownian}
  motion},
\newblock \doi{10.1103/PhysRevE.94.012134}{\rm Phys. Rev. E {\bf 94} (2016)
  012134}\null,
\newblock \arxiv{arXiv:1603.00651}.

\bibitem{DelormeRossoWiese2017}
M.~Delorme, A.~Rosso  and K.J. Wiese,
\newblock {\em Pickands' constant at first order in an expansion around
  {Brownian} motion},
\newblock \doi{10.1088/1751-8121/aa5c98}{\rm J. Phys. A {\bf 50} (2017)
  16LT04}\null,
\newblock \arxiv{arXiv:1609.07909}.

\bibitem{Wiese2018}
{K.J.} Wiese,
\newblock {\em First passage in an interval for fractional {Brownian} motion},
\newblock \doi{10.1103/PhysRevE.99.032106}{\rm Phys. Rev. E {\bf 99} (2018)
  032106}\null,
\newblock \arxiv{arXiv:1807.08807}.

\bibitem{SadhuDelormeWiese2017}
T.~Sadhu, M.~Delorme  and {K.J.} Wiese,
\newblock {\em Generalized arcsine laws for fractional {Brownian} motion},
\newblock \doi{10.1103/PhysRevLett.120.040603}{\rm Phys. Rev. Lett. {\bf 120}
  (2018)   040603}\null,
\newblock \arxiv{arXiv:1706.01675}.

\bibitem{BenigniCoscoShapiraWiese2017}
L.~Benigni, C.~Cosco, A.~Shapira  and K.J. Wiese,
\newblock {\em Hausdorff dimension of the record set of a fractional {Brownian}
  motion},
\newblock \doi{10.1214/18-ECP121}{\rm Electron. Commun. Probab. {\bf 23} (2018)
    1--8}\null,
\newblock \arxiv{arXiv:1706.09726}.

\bibitem{WalterWiese2019a}
B.~Walter and K.J. Wiese,
\newblock {\em Monte {Carlo} sampler of first passage times for fractional
  {Brownian} motion using adaptive bisections: Source code},
\newblock \link{https://hal.archives-ouvertes.fr/hal-02270046}{hal-02270046}
  (2019).

\bibitem{WalterWiese2019b}
B.~Walter and K.J. Wiese,
\newblock {\em Sampling first-passage times of fractional {Brownian} motion
  using adaptive bisections},
\newblock \doi{10.1103/PhysRevE.101.043312}{\rm Phys. Rev. E {\bf 101} (2020)
  043312}\null,
\newblock \arxiv{arXiv:1908.11634}.

\bibitem{ArutkinWalterWiese2020}
M.~Arutkin, B.~Walter  and K.J. Wiese,
\newblock {\em Extreme events for fractional {Brownian} motion with drift:
  Theory and numerical validation},
\newblock \doi{10.1103/PhysRevE.102.022102}{\rm Phys. Rev. E {\bf 102} (2020)
  022102}\null,
\newblock \arxiv{arXiv:1908.10801}.

\bibitem{RambeauBustingorryKoltonSchehr2011}
J.~Rambeau, S.~Bustingorry, A.B. Kolton  and G.~Schehr,
\newblock {\em Maximum relative height of elastic interfaces in random media},
\newblock \doi{10.1103/PhysRevE.84.041131}{\rm Phys. Rev. E {\bf 84} (2011)
  041131}\null.

\bibitem{LeDoussalWieseUnpublished}
P.~{Le Doussal} and K.J. Wiese,
\newblock unpublished.

\bibitem{LeDoussalRossoWiese2011}
P.~Le Doussal, A.~Rosso  and K.J. Wiese,
\newblock {\em Shock statistics in higher-dimensional {Burgers} turbulence},
\newblock \doi{10.1209/0295-5075/96/14005}{\rm EPL {\bf 96} (2011)
  14005}\null,
\newblock \arxiv{arXiv:1104.5048}.

\bibitem{MaaloySantucciSchmittbuhlToussaint2006}
K.J. M\aa{}l\o{}y, S.~Santucci, J.~Schmittbuhl  and R.~Toussaint,
\newblock {\em Local waiting time fluctuations along a randomly pinned crack
  front},
\newblock \doi{10.1103/PhysRevLett.96.045501}{\rm Phys. Rev. Lett. {\bf 96}
  (2006)   045501}\null.

\bibitem{LaursonSantucciZapperi2010}
L.~Laurson, S.~Santucci  and S.~Zapperi,
\newblock {\em Avalanches and clusters in planar crack front propagation},
\newblock \doi{10.1103/PhysRevE.81.046116}{\rm Phys. Rev. E {\bf 81} (2010)
  046116}\null.

\bibitem{BudrikisZapperi2013}
Z.~Budrikis and S.~Zapperi,
\newblock {\em Avalanche localization and crossover scaling in amorphous
  plasticity},
\newblock \doi{10.1103/PhysRevE.88.062403}{\rm Phys. Rev. E {\bf 88} (2013)
  062403}\null.

\bibitem{LePriolChopinLeDoussalPonsonRosso2020}
C.~Le~Priol, J.~Chopin, P.~Le~Doussal, L.~Ponson  and A.~Rosso,
\newblock {\em Universal scaling of the velocity field in crack front
  propagation},
\newblock \doi{10.1103/PhysRevLett.124.065501}{\rm Phys. Rev. Lett. {\bf 124}
  (2020)   065501}\null.

\bibitem{LePriolLeDoussalRosso2020}
C.~Le~Priol, P.~Le~Doussal  and A.~Rosso,
\newblock {\em Spatial clustering of depinning avalanches in presence of
  long-range interactions},
\newblock (2020),
\newblock \arxiv{arXiv:2008.10025}.

\bibitem{LePriolThesis}
C.~Le Priol,
\newblock {\em Long-range interactions in the avalanches of elastic
  interfaces},
\newblock PhD thesis, PSL Research University, 2020,
\newblock \arxiv{arXiv:2103.07701}.

\bibitem{Terrot2021}
W.~Terrot,
\newblock {\em Avalanches en pr\'esence d'interactions \`a longue port\'ee},
\newblock internship report, universit\'e Paris-Saclay (2021).

\bibitem{Nature1945}
{\em Frequency of earthquakes in {California}},
\newblock \doi{10.1038/156371a0}{\rm Nature {\bf 156} (1945)   371--371}\null.

\bibitem{Dieterich1992}
J.H. Dieterich,
\newblock {\em {Earthquake nucleation on faults with rate-and state-dependent
  strength}},
\newblock \doi{10.1016/0040-1951(92)90055-B}{\rm Tectonophysics {\bf 211}
  (1992)   115--134}\null.

\bibitem{Omori1894}
F.~Omori,
\newblock {\em On the aftershocks of earthquakes},
\newblock Journal of the College of Science, Imperial University of Tokyo. {\bf
  7} (1894)   111--200.

\bibitem{BurridgeKnopoff1967}
R.~Burridge and L.~Knopoff,
\newblock {\em Model and theoretical seismicity},
\newblock Bulletin of the Seismological Society of America {\bf 57} (1967)
  341--371.

\bibitem{BenZionRice1993}
Y.~Ben-Zion and J.R. Rice,
\newblock {\em Earthquake failure sequences along a cellular fault zone in a
  three-dimensional elastic solid containing asperity and nonasperity regions},
\newblock \doi{10.1029/93JB01096}{\rm J. Geophys. Res. {\bf 98} (1993)
  14109--14131}\null.

\bibitem{Ruina1983}
A.~Ruina,
\newblock {\em {Slip instability and state variable friction laws}},
\newblock \doi{10.1029/JB088iB12p10359}{\rm J. Geophys. Res. {\bf 88} (1983)
  359--10}\null.

\bibitem{CarlsonLangerShaw1994}
J.M. Carlson, J.S. Langer  and B.E. Shaw,
\newblock {\em Dynamics of earthquake faults},
\newblock \doi{10.1103/RevModPhys.66.657}{\rm Rev. Mod. Phys. {\bf 66} (1994)
  657--670}\null.

\bibitem{BenZionRice1997}
Y.~Ben-Zion and J.R. Rice,
\newblock {\em Dynamic simulations of slip on a smooth fault in an elastic
  solid},
\newblock \doi{10.1029/97JB01341}{\rm J. Geophys. Res. {\bf 102} (1997)
  17--17}\null.

\bibitem{FisherDahmenRamanathanBenZion1997}
D.~Fisher, K.~Dahmen, S.~Ramanathan  and Y.~Ben-Zion,
\newblock {\em {Statistics of Earthquakes in Simple Models of Heterogeneous
  Faults}},
\newblock \doi{10.1103/PhysRevLett.78.4885}{\rm Phys. Rev. Lett. {\bf 78}
  (1997)   4885--4888}\null.

\bibitem{Scholz1998}
C.H. Scholz,
\newblock {\em {Earthquakes and friction laws}},
\newblock \doi{10.1038/34097}{\rm Nature {\bf 391} (1998)   37--42}\null.

\bibitem{ShomeCornellBazzurroCarballo1998}
N.~Shome, C.A. Cornell, P.~Bazzurro  and J.E. Carballo,
\newblock {\em Earthquakes, records, and nonlinear responses},
\newblock \doi{10.1193/1.1586011}{\rm Earthquake Spectra {\bf 14} (1998)
  469--500}\null.

\bibitem{Monte-MorenoHernandez-Pajares2014}
E.~Monte-Moreno and M.~Hern\'andez-Pajares,
\newblock {\em Occurrence of solar flares viewed with {GPS}: Statistics and
  fractal nature},
\newblock \doi{10.1002/2014JA020206}{\rm J. Geophys. Res. {\bf 119} (2014)
  9216--9227}\null.

\bibitem{Kagan2002}
Y.Y. Kagan,
\newblock {\em {Seismic moment distribution revisited: I. Statistical
  results}},
\newblock \doi{10.1046/j.1365-246x.2002.01594.x}{\rm Geophys. J. Int. {\bf 148}
  (2002)   520--541}\null.

\bibitem{SchwarzFisher2003}
J.M. Schwarz and D.S. Fisher,
\newblock {\em Depinning with dynamic stress overshoots: A hybrid of critical
  and pseudohysteretic behavior},
\newblock \doi{10.1103/PhysRevE.67.021603}{\rm Phys. Rev. E {\bf 67} (2003)
  021603}\null,
\newblock \arxiv{cond-mat/0204623}.

\bibitem{JaglaKolton2009}
E.A. Jagla and A.B. Kolton,
\newblock {\em The mechanisms of spatial and temporal earthquake clustering},
\newblock \doi{10.1029/2009JB006974}{\rm J. Geophys. Res. {\bf 115} (2009)
  B05312}\null,
\newblock \arxiv{arXiv:0901.1907}.

\bibitem{LeDoussalMuellerWiese2010}
P.~Le Doussal, M.~M\"uller  and K.J. Wiese,
\newblock {\em Avalanches in mean-field models and the {Barkhausen} noise in
  spin-glasses},
\newblock \doi{10.1209/0295-5075/91/57004}{\rm EPL {\bf 91} (2010)
  57004}\null,
\newblock \arxiv{arXiv:1007.2069}.

\bibitem{LeDoussalMuellerWiese2011}
P.~Le Doussal, M.~M\"uller  and K.J. Wiese,
\newblock {\em Equilibrium avalanches in spin glasses},
\newblock \doi{10.1103/PhysRevB.85.214402}{\rm Phys. Rev. B {\bf 85} (2012)
  214402}\null,
\newblock \arxiv{arXiv:1110.2011}.

\bibitem{PazmandiZarandZimanyi1999}
F.~P\'azm\'andi, G.~Zar\'and  and G.T. Zim\'anyi,
\newblock {\em Self-organized criticality in the hysteresis of the
  {Sherrington-Kirkpatrick} model},
\newblock \doi{10.1103/PhysRevLett.83.1034}{\rm Phys. Rev. Lett. {\bf 83}
  (1999)   1034--1037}\null.

\bibitem{BakTangWiesenfeld1987}
P.~Bak, C.~Tang  and K.~Wiesenfeld,
\newblock {\em Self-organized criticality - an explanation of 1/f noise},
\newblock \doi{10.1103/PhysRevLett.59.381}{\rm Phys. Rev. Lett. {\bf 59} (1987)
    381--384}\null.

\bibitem{MajumdarDhar1992}
S.N. Majumdar and D.~Dhar,
\newblock {\em Equivalence between the {Abelian} sandpile model and the {$q\to
  0$} limit of the {Potts}-model},
\newblock \doi{10.1016/0378-4371(92)90447-X}{\rm Physica A {\bf 185} (1992)
  129--145}\null.

\bibitem{Dhar1999}
D.~Dhar,
\newblock {\em Studying self-organized criticality with exactly solved models},
\newblock (1999),
\newblock \arxiv{cond-mat/9909009}.

\bibitem{Dhar1999b}
D.~Dhar,
\newblock {\em The {Abelian} sandpile and related models},
\newblock \doi{10.1016/S0378-4371(98)00493-2}{\rm Physica A {\bf 263} (1999)
  ~4}\null,
\newblock \arxiv{ond-mat/9808047}.

\bibitem{Dhar2006}
D.~Dhar,
\newblock {\em Theoretical studies of self-organized criticality},
\newblock \doi{10.1016/j.physa.2006.04.004}{\rm Physica A {\bf 369} (2006)
  29--70}\null.

\bibitem{BonachelaAlavaMunoz2008}
J.A. Bonachela, M.~Alava  and M.A. Mu{\~n}oz,
\newblock {\em Cusps in systems with (many) absorbing states},
\newblock \doi{10.1103/PhysRevE.79.050106}{\rm Phys. Rev. E {\bf 79} (2009)
  050106(R)}\null,
\newblock \arxiv{arXiv:0810.4395}.

\bibitem{BonachelaPhD}
J.A. Bonachela,
\newblock {\em Universality in self-organized criticality},
\newblock PhD thesis, University of Granada, Spain, 2008.

\bibitem{BonachelaChateDornicMunoz2007}
J.A. Bonachela, H.~Chate, I.~Dornic  and M.A. Mu{\~n}oz,
\newblock {\em Absorbing states and elastic interfaces in random media: Two
  equivalent descriptions of self-organized criticality},
\newblock \doi{10.1103/PhysRevLett.98.155702}{\rm Phys. Rev. Lett. {\bf 98}
  (2007)   155702}\null.

\bibitem{UritskyPaczuskiDavilaJones2007}
V.M. Uritsky, M.~Paczuski, J.M. Davila  and S.I. Jones,
\newblock {\em Coexistence of self-organized criticality and intermittent
  turbulence in the solar corona},
\newblock \doi{10.1103/PhysRevLett.99.025001}{\rm Phys. Rev. Lett. {\bf 99}
  (2007)   025001}\null.

\bibitem{Jeng2005}
M.~Jeng,
\newblock {\em Conformal field theory correlations in the {Abelian} sandpile
  model},
\newblock \doi{10.1103/PhysRevE.71.016140}{\rm Phys. Rev. E {\bf 71} (2005)
  016140}\null,
\newblock \arxiv{cond-mat/0407115}.

\bibitem{StapletonChristensen2005}
M.~Stapleton and K.~Christensen,
\newblock {\em Mean-field theory and sandpile models},
\newblock (2005),
\newblock \arxiv{cond-mat/0510626}.

\bibitem{Dhar2004}
D.~Dhar,
\newblock {\em Steady state and relaxation spectrum of the {Oslo} rice-pile
  model},
\newblock \doi{https://doi.org/10.1016/j.physa.2004.05.003}{\rm Physica A {\bf
  340} (2004)   535--543}\null.

\bibitem{Alava2003}
M.~Alava,
\newblock {\em Self-organized criticality as a phase transition}, pages
  69--102,
\newblock Nova Science Publishers, New York, NY, USA, 2003,
\newblock \arxiv{cond-mat/0307688}.

\bibitem{Alava2002}
M.~Alava,
\newblock {\em Scaling in self-organized criticality from interface
  depinning?},
\newblock \doi{10.1088/0953-8984/14/9/324}{\rm J. Phys. Cond. Mat. {\bf 14}
  (2002)   2353}\null,
\newblock \arxiv{cond-mat/0204226}.

\bibitem{BasslerPaczuskiAltshuler2001}
K.E. Bassler, M.~Paczuski  and E.~Altshuler,
\newblock {\em Simple model for plastic dynamics of a disordered flux-line
  lattice},
\newblock \doi{10.1103/PhysRevB.64.224517}{\rm Phys. Rev. B {\bf 64} (2001)
  224517}\null,
\newblock \arxiv{cond-mat/0009278}.

\bibitem{DickmanMunozVespignaniZapperi2000}
R.~Dickman, M.A. Mu{\~n}oz, A.~Vespignani  and S.~Zapperi,
\newblock {\em Paths to self-organized criticality},
\newblock \doi{10.1590/S0103-97332000000100004}{\rm Braz. J. Phys. {\bf 30}
  (2000)   27--41}\null,
\newblock \arxiv{cond-mat/9910454}.

\bibitem{DickmanVespignaniZapperi1998}
R.~Dickman, A.~Vespignani  and S.~Zapperi,
\newblock {\em Self-organized criticality as an absorbing-state phase
  transition},
\newblock \doi{10.1103/PhysRevE.57.5095}{\rm Phys. Rev. E {\bf 57} (1998)
  5095--5105}\null.

\bibitem{BasslerPaczuski1998}
K.E. Bassler and M.~Paczuski,
\newblock {\em Simple model of superconducting vortex avalanches},
\newblock \doi{10.1103/PhysRevLett.81.3761}{\rm Phys. Rev. Lett. {\bf 81}
  (1998)   3761--3764}\null,
\newblock \arxiv{cond-mat/9804249}.

\bibitem{Jensen1998}
H.J. Jensen,
\newblock {\em Self-Organized Criticality},
\newblock Cambridge University Press, Cambridge, UK, 1998.

\bibitem{TanguyGounelleRoux1998}
A.~Tanguy, M.~Gounelle  and S.~Roux,
\newblock {\em From individual to collective pinning: Effect of long-range
  elastic interactions},
\newblock \doi{10.1103/PhysRevE.58.1577}{\rm Phys. Rev. E {\bf 58} (1998)
  1577--90}\null,
\newblock \arxiv{cond-mat/9804105}.

\bibitem{ChristensenCorralFretteFederJossang1996}
K.~Christensen, \'A. Corral, V.~Frette, J.~Feder  and T.~J\o{}ssang,
\newblock {\em Tracer dispersion in a self-organized critical system},
\newblock \doi{10.1103/PhysRevLett.77.107}{\rm Phys. Rev. Lett. {\bf 77} (1996)
    107--110}\null.

\bibitem{FretteChristensenMalthe-SorenssenFederJossangMeakin1996}
V.~Frette, K.~Christensen, A.~Malthe-S{\o}renssen, J.~Feder, T.~J{\o}ssang  and
  P.~Meakin,
\newblock {\em Avalanche dynamics in a pile of rice},
\newblock \doi{10.1038/379049a0}{\rm Nature {\bf 379} (1996)   49--52}\null.

\bibitem{UrbachMadisonMarkert1995}
J.S. Urbach, R.C. Madison  and J.T. Markert,
\newblock {\em Interface depinning, self-organized criticality, and the
  {Barkhausen} effect},
\newblock \doi{10.1103/PhysRevLett.75.276}{\rm Phys. Rev. Lett. {\bf 75} (1995)
    276--279}\null.

\bibitem{Sneppen1992}
K.~Sneppen,
\newblock {\em Self-organized pinning and interface growth in a random medium},
\newblock \doi{10.1103/PhysRevLett.69.3539}{\rm Phys. Rev. Lett. {\bf 69}
  (1992)   3539--3542}\null.

\bibitem{Manna1991}
S.S. Manna,
\newblock {\em Two-state model of self-organized critical phenomena},
\newblock \doi{10.1088/0305-4470/24/7/009}{\rm J.~Phys.~A {\bf 24} (1991)
  L363--L369}\null.

\bibitem{DharMajumdar1990}
D.~Dhar and S.N. Majumdar,
\newblock {\em Abelian sandpile model on the {Bethe} lattice},
\newblock \doi{10.1088/0305-4470/23/19/018}{\rm J. Phys. A {\bf 23} (1990)
  4333--4350}\null.

\bibitem{DharRamaswamy1989}
D.~Dhar and R.~Ramaswamy,
\newblock {\em Exactly solved model of self-organized critical phenomena},
\newblock \doi{10.1103/PhysRevLett.63.1659}{\rm Phys. Rev. Lett. {\bf 63}
  (1989)   1659--1662}\null.

\bibitem{TangBak1988}
C.~Tang and P.~Bak,
\newblock {\em Mean field-theory of self-organized critical phenomena},
\newblock \doi{10.1007/BF01014884}{\rm J. Stat. Phys. {\bf 51} (1988)
  797--802}\null.

\bibitem{Frette1993}
V.~Frette,
\newblock {\em Sandpile models with dynamically varying critical slopes},
\newblock \doi{10.1103/PhysRevLett.70.2762}{\rm Phys. Rev. Lett. {\bf 70}
  (1993)   2762--2765}\null.

\bibitem{HuynhPruessnerChew2011}
H.N. Huynh, G.~Pruessner  and L.Y. Chew,
\newblock {\em The {Abelian Manna} model on various lattices in one and two
  dimensions},
\newblock \doi{10.1088/1742-5468/2011/09/P09024}{\rm J. Stat. Mech. {\bf 2011}
  (2011)   P09024}\null,
\newblock \arxiv{arXiv:1106.0406}.

\bibitem{Hinrichsen2000}
H.~Hinrichsen,
\newblock {\em Non-equilibrium critical phenomena and phase transitions into
  absorbing states},
\newblock \doi{10.1080/00018730050198152}{\rm Adv. Phys. {\bf 49} (2000)
  815--958}\null,
\newblock \arxiv{cond-mat/0001070}.

\bibitem{HenkelHinrichsenLubeck2008}
M.~Henkel, H.~Hinrichsen  and S.~L\"ubeck,
\newblock \doi{10.1007/978-1-4020-8765-3}{\rm {\em Non-Equilibrium Phase
  Transitions}}\null,
\newblock Springer, Dordrecht, 2008.

\bibitem{WeiBechingerLeiderer2000}
Q.-H. Wei, C.~Bechinger  and P.~Leiderer,
\newblock {\em Single-file diffusion of colloids in one-dimensional channels},
\newblock \doi{10.1126/science.287.5453.625}{\rm Science {\bf 287} (2000)
  625--627}\null.

\bibitem{KrapivskyMallickSadhu2015}
P.L. Krapivsky, K.~Mallick  and T.~Sadhu,
\newblock {\em Dynamical properties of single-file diffusion},
\newblock \doi{10.1088/1742-5468/2015/09/P09007}{\rm J. Stat. Mech. (2015)
  P09007}\null,
\newblock \arxiv{arXiv:1505.01287}.

\bibitem{KrapivskyMallickSadhu2015a}
P.~L. Krapivsky, K.~Mallick  and T.~Sadhu,
\newblock {\em Tagged particle in single-file diffusion},
\newblock \doi{10.1007/s10955-015-1291-0}{\rm J. Stat. Phys. {\bf 160} (2015)
  885--925}\null.

\bibitem{ShapiraWieseUnpublished}
A.~Shapira and K.J. Wiese,
\newblock unpublished.

\bibitem{BasuBasuBondyopadhyayMohantyHinrichsen2012}
M.~Basu, U.~Basu, S.~Bondyopadhyay, P.~Mohanty  and H.~Hinrichsen,
\newblock {\em Fixed-energy sandpiles belong generically to directed
  percolation},
\newblock \doi{10.1103/PhysRevLett.109.015702}{\rm Phys. Rev. Lett. {\bf 109}
  (2012)   015702}\null.

\bibitem{HexnerLevine2015}
D.~Hexner and D.~Levine,
\newblock {\em Hyperuniformity of critical absorbing states},
\newblock \doi{10.1103/PhysRevLett.114.110602}{\rm Phys. Rev. Lett. {\bf 114}
  (2015)   110602}\null.

\bibitem{Lee2014}
S.B. Lee,
\newblock {\em Universality class of the conserved {Manna} model in one
  dimension},
\newblock \doi{10.1103/PhysRevE.89.060101}{\rm Phys. Rev. E {\bf 89} (2014)
  060101}\null.

\bibitem{DickmanCunha2015}
R.~Dickman and S.D. da~Cunha,
\newblock {\em Particle-density fluctuations and universality in the conserved
  stochastic sandpile},
\newblock \doi{10.1103/PhysRevE.92.020104}{\rm Phys. Rev. E {\bf 92} (2015)
  020104}\null.

\bibitem{Garcia-MillanPruessnerPickeringChristensen2018}
R.~Garcia-Millan, G.~Pruessner, L.~Pickering  and K.~Christensen,
\newblock {\em Correlations and hyperuniformity in the avalanche size of the
  {Oslo} model},
\newblock \doi{10.1209/0295-5075/122/50003}{\rm EPL {\bf 122} (2018)
  50003}\null.

\bibitem{BerthierChaudhuriCoulaisDauchotSollich2011}
L.~Berthier, P.~Chaudhuri, C.~Coulais, O.~Dauchot  and P.~Sollich,
\newblock {\em Suppressed compressibility at large scale in jammed packings of
  size-disperse spheres},
\newblock \doi{10.1103/PhysRevLett.106.120601}{\rm Phys. Rev. Lett. {\bf 106}
  (2011)   120601}\null.

\bibitem{TangLeschhorn1992}
L.-H. Tang and H.~Leschhorn,
\newblock {\em Pinning by directed percolation},
\newblock \doi{10.1103/PhysRevA.45.R8309}{\rm Phys. Rev. A {\bf 45} (1992)
  R8309--12}\null.

\bibitem{BuldyrevBarabasiCasertaHavlinStanleyVicsek1992}
S.V. Buldyrev, A.-L. Barabasi, F.~Caserta, S.~Havlin, H.E. Stanley  and
  T.~Vicsek,
\newblock {\em Anomalous interface roughening in porous media: experiment and
  model},
\newblock \doi{10.1103/PhysRevA.45.R8313}{\rm Phys. Rev. A {\bf 45} (1992)
  R8313--16}\null.

\bibitem{GlotzerGyureSciortinoConiglioStanley1994}
S.C. Glotzer, M.F. Gyure, F.~Sciortino, A.~Coniglio  and H.E. Stanley,
\newblock {\em Pinning in phase-separating systems},
\newblock \doi{10.1103/PhysRevE.49.247}{\rm Phys. Rev. E {\bf 49} (1994)
  247--58}\null.

\bibitem{BarabasiGrinsteinMunoz1996}
A.-L. Barabasi, G.~Grinstein  and M.A. Mu{\~n}oz,
\newblock {\em Directed surfaces in disordered media},
\newblock \doi{10.1103/PhysRevLett.76.1481}{\rm Phys. Rev. Lett. {\bf 76}
  (1996)   1481--4}\null.

\bibitem{AmaralBarabasiStanley1994}
L.A.N. Amaral, A.-L. Barabasi  and H.E. Stanley,
\newblock {\em Critical dynamics of contact line depinning},
\newblock \doi{10.1103/PhysRevLett.73.62}{\rm Phys. Rev. Lett. {\bf 73} (1994)
  ~62}\null.

\bibitem{KPZ}
M.~Kardar, G.~Parisi  and Y.-C. Zhang,
\newblock {\em Dynamic scaling of growing interfaces},
\newblock \doi{10.1103/PhysRevLett.56.889}{\rm Phys. Rev. Lett. {\bf 56} (1986)
    889--892}\null.

\bibitem{LeeKim2005}
C.~Lee and J.M. Kim,
\newblock {\em Depinning transition of the quenched {Kardar-Parisi-Zhang}
  equation},
\newblock
  \link{www.jkps.or.kr/journal/download_pdf.php?spage=13&volume=47&number=1}{J.
  Korean Phys. Soc.} {\bf 47} (2005)   13--17.

\bibitem{TangKardarDhar1995}
L.-H. Tang, M.~Kardar  and D.~Dhar,
\newblock {\em Driven depinning in anisotropic media},
\newblock \doi{10.1103/PhysRevLett.74.920}{\rm Phys. Rev. Lett. {\bf 74} (1995)
    920--3}\null.

\bibitem{AraujoGrassbergerKahngSchrenkZiff2014}
N.~Ara{\'u}jo, P.~Grassberger, B.~Kahng, K.~J. Schrenk  and R.~M. Ziff,
\newblock {\em Recent advances and open challenges in percolation},
\newblock \doi{10.1140/epjst/e2014-02266-y}{\rm Eur. Phys. J. Spec. Top. {\bf
  223} (2014)   2307--2321}\null,
\newblock \arxiv{arXiv:1404.5325}.

\bibitem{Dhar2017}
D.~Dhar,
\newblock {\em Directed percolation and directed animals},
\newblock (2017),
\newblock \arxiv{arXiv:1703.07541}.

\bibitem{Janssen1981}
H.K. Janssen,
\newblock {\em On the nonequilibrium phase transition in reaction-diffusion
  systems with an absorbing stationary state},
\newblock \doi{10.1007/BF01319549}{\rm Z. Phys. B {\bf 42} (1981)
  151--154}\null.

\bibitem{BronzanDash1974}
J.B. Bronzan and J.W. Dash,
\newblock {\em Higher order $\epsilon$-terms in the renormalization group
  approach to {Reggeon} field theory},
\newblock \doi{10.1016/0370-2693(74)90319-0}{\rm Phys. Lett. B {\bf 51} (1974)
   496 -- 498}\null.

\bibitem{CardySugar1980}
J.L. Cardy and R.L. Sugar,
\newblock {\em Directed percolation and {Reggeon} field theory},
\newblock \doi{10.1088/0305-4470/13/12/002}{\rm J. Phys. A {\bf 13} (1980)
  L423}\null.

\bibitem{JanssenTaeuber2005}
H.K. Janssen and U.C. T\"auber,
\newblock {\em The field theory approach to percolation processes},
\newblock \doi{10.1016/j.aop.2004.09.011}{\rm Ann. Phys. (NY) {\bf 315} (2005)
   147 -- 192}\null.

\bibitem{AbarbanelBronzanSugarWhite1975}
H.D.I. Abarbanel, J.D. Bronzan, R.L. Sugar  and A.R. White,
\newblock {\em Reggeon field theory: Formulation and use},
\newblock \doi{10.1016/0370-1573(75)90034-4}{\rm Phys. Rep. {\bf 21} (1975)
  119 -- 182}\null.

\bibitem{AdzhemyanHnaticKompanietsLucivjanskyMizisin2019}
L.T. Adzhemyan, M.~Hnati{\v c}, M.V. Kompaniets, T.~Lu{\v c}ivjansk{\'y}  and
  L.~Mi{\v z}i{\v s}in,
\newblock {\em \link{doi.org/10.1007/978-3-030-15297-0_18}{Renormalization
  Group Approach of Directed Percolation: Three-Loop Approximation}},
\newblock in Christos~H. Skiadas and Ihor Lubashevsky, editors, {\em 11th
  Chaotic Modeling and Simulation International Conference}, pages 195--204,
  Springer International Publishing, Cham, 2019.

\bibitem{HavlinAmaralBuldyrevHarringtonStanley1995}
S.~Havlin, L.A.N. Amaral, S.V. Buldyrev, S.T. Harrington  and H.E. Stanley,
\newblock {\em Dynamics of surface roughening with quenched disorder},
\newblock \doi{10.1103/PhysRevLett.74.4205}{\rm Phys. Rev. Lett. {\bf 74}
  (1995)   4205--8}\null.

\bibitem{LeDoussalWiese2002a}
P.~Le Doussal and K.J. Wiese,
\newblock {\em Functional renormalization group for anisotropic depinning and
  relation to branching processes},
\newblock \doi{10.1103/PhysRevE.67.016121}{\rm Phys. Rev. E {\bf 67} (2003)
  016121}\null,
\newblock \arxiv{cond-mat/0208204}.

\bibitem{AtisDubeySalinTalonLeDoussalWiese2014}
S.~Atis, A.K. Dubey, D.~Salin, L.~Talon, P.~Le Doussal  and K.J. Wiese,
\newblock {\em Experimental evidence for three universality classes for
  reaction fronts in disordered flows},
\newblock \doi{10.1103/PhysRevLett.114.234502}{\rm Phys. Rev. Lett. {\bf 114}
  (2015)   234502}\null,
\newblock \arxiv{arXiv:1410.1097}.

\bibitem{Grassberger1982}
P.~Grassberger,
\newblock {\em On phase-transitions in {Schl\"ogl's} second model},
\newblock Z. Phys. B {\bf 47} (1982)   365.

\bibitem{Jensen1993}
I.~Jensen,
\newblock {\em Critical behavior of the pair contact process},
\newblock \doi{10.1103/PhysRevLett.70.1465}{\rm Phys. Rev. Lett. {\bf 70}
  (1993)   1465--1468}\null.

\bibitem{MunozGrinsteinDickmanLivi1996}
M.A. Mu\~noz, G.~Grinstein, R.~Dickman  and R.~Livi,
\newblock {\em Critical behavior of systems with many absorbing states},
\newblock \doi{10.1103/PhysRevLett.76.451}{\rm Phys. Rev. Lett. {\bf 76} (1996)
    451--454}\null.

\bibitem{MunozGrinsteinDickmanLivi1997}
M.A. Mu{\~n}oz, G.~Grinstein, R.~Dickman  and R.~Livi,
\newblock {\em Infinite numbers of absorbing states: Critical behavior},
\newblock \doi{https://doi.org/10.1016/S0167-2789(96)00280-1}{\rm Physica D
  {\bf 103} (1997)   485--490}\null.

\bibitem{MunozGrinsteinDickman1998}
M.A. Mu{\~n}oz, G.~Grinstein  and R.~Dickman,
\newblock {\em Phase structure of systems with infinite numbers of absorbing
  states},
\newblock \doi{10.1023/A:1023021409588}{\rm J. Stat. Phys. {\bf 91} (1998)
  541--569}\null.

\bibitem{Munoz1998}
M.A. Mu\~noz,
\newblock {\em Nature of different types of absorbing states},
\newblock \doi{10.1103/PhysRevE.57.1377}{\rm Phys. Rev. E {\bf 57} (1998)
  1377--1383}\null.

\bibitem{VespignaniDickmanMunozZapperi1998}
A.~Vespignani, R.~Dickman, M.A. Mu\~noz  and S.~Zapperi,
\newblock {\em Driving, conservation, and absorbing states in sandpiles},
\newblock \doi{10.1103/PhysRevLett.81.5676}{\rm Phys. Rev. Lett. {\bf 81}
  (1998)   5676--5679}\null.

\bibitem{VespignaniDickmanMunozZapperi2000}
A.~Vespignani, R.~Dickman, M.A. Mu\~noz  and S.~Zapperi,
\newblock {\em Absorbing-state phase transitions in fixed-energy sandpiles},
\newblock \doi{10.1103/PhysRevE.62.4564}{\rm Phys. Rev. E {\bf 62} (2000)
  4564--4582}\null.

\bibitem{AlavaMunoz2002}
M.~Alava and M.A. Mu\~noz,
\newblock {\em Interface depinning versus absorbing-state phase transitions},
\newblock \doi{10.1103/PhysRevE.65.026145}{\rm Phys. Rev. E {\bf 65} (2002)
  026145}\null.

\bibitem{JeongKahngKim1996}
H.~Jeong, B.~Kahng  and D.~Kim,
\newblock {\em Anisotropic surface growth model in disordered media},
\newblock \doi{10.1103/PhysRevLett.77.5094}{\rm Phys. Rev. Lett. {\bf 77}
  (1996)   5094--5097}\null.

\bibitem{JeongKahngKim1999}
H.~Jeong, B.~Kahng  and D.~Kim,
\newblock {\em Facet formation in the negative quenched {Kardar-Parisi-Zhang}
  equation},
\newblock \doi{10.1103/PhysRevE.59.1570}{\rm Phys. Rev. E {\bf 59} (1999)
  1570--1573}\null.

\bibitem{TakeuchiKurodaChateSano2007}
K.A. Takeuchi, M.~Kuroda, H.~Chat\'e  and M.~Sano,
\newblock {\em Directed percolation criticality in turbulent liquid crystals},
\newblock \doi{10.1103/PhysRevLett.99.234503}{\rm Phys. Rev. Lett. {\bf 99}
  (2007)   234503}\null.

\bibitem{TakeuchiKurodaChateSano2009}
K.A. Takeuchi, F.~Kuroda, H.~Chat\'e  and M.~Sano,
\newblock {\em Experimental realization of directed percolation criticality in
  turbulent liquid crystals},
\newblock \doi{10.1103/PhysRevE.80.051116}{\rm Phys. Rev. E {\bf 80} (2009)
  051116}\null.

\bibitem{Wiese2015}
K.J. Wiese,
\newblock {\em Coherent-state path integral versus coarse-grained effective
  stochastic equation of motion: From reaction diffusion to stochastic
  sandpiles},
\newblock \doi{10.1103/PhysRevE.93.042117}{\rm Phys. Rev. E {\bf 93} (2016)
  042117}\null,
\newblock \arxiv{arXiv:1501.06514}.

\bibitem{Doi1976a}
M.~Doi,
\newblock {\em Second quantization representation for classical many-particle
  system},
\newblock \doi{10.1088/0305-4470/9/9/008}{\rm J. Phys. A {\bf 9} (1976)
  1465--1477}\null.

\bibitem{Doi1976b}
M.~Doi,
\newblock {\em Stochastic theory of diffusion-controlled reaction},
\newblock \doi{10.1088/0305-4470/9/9/009}{\rm J. Phys. A {\bf 9} (1976)
  1479}\null.

\bibitem{Peliti1985}
L.~Peliti,
\newblock {\em Path integral approach to birth-death processes on a lattice},
\newblock \doi{10.1051/jphys:019850046090146900}{\rm J. Phys. (France) {\bf 46}
  (1985)   1469--1483}\null.

\bibitem{Cardy2006}
J.~Cardy,
\newblock {\em Reaction-diffusion processes},
\newblock Lectures held at Warwick University, unpublished (2006).

\bibitem{AndreanovBiroliBouchaudLefevre2006}
A.~Andreanov, G.~Biroli, J.P. Bouchaud  and A.~Lef\`evre,
\newblock {\em Field theories and exact stochastic equations for interacting
  particle systems},
\newblock \doi{10.1103/PhysRevE.74.030101}{\rm Phys. Rev. E {\bf 74} (2006)
  030101}\null.

\bibitem{GredatDornicLuck2011}
D.~Gredat, I.~Dornic  and J.M. Luck,
\newblock {\em On an imaginary exponential functional of {Brownian} motion},
\newblock \doi{10.1088/1751-8113/44/17/175003}{\rm J. Phys. A {\bf 44} (2011)
  175003}\null,
\newblock \arxiv{arXiv:1101.1173}.

\bibitem{TaeuberBook}
U.C. T\"auber,
\newblock {\em Critical Dynamics: A Field Theory Approach to Equilibrium and
  Non-Equilibrium Scaling Behavior},
\newblock Cambridge University Press, 2014.

\bibitem{DeloubriereFrachebourgHilhorstKitahara2002}
O.~Deloubri\`ere, L.~Frachebourg, H.J. Hilhorst  and K.~Kitahara,
\newblock {\em Imaginary noise and parity conservation in the reaction {$A+A
  \leftrightarrow 0$}},
\newblock \doi{10.1016/S0378-4371(02)00548-4}{\rm Physica A {\bf 308} (2002)
  135--147}\null.

\bibitem{GardinerMcNeilWallsMatheson1976}
C.W. Gardiner, K.J. McNeil, D.F. Walls  and I.S. Matheson,
\newblock {\em Correlations in stochastic theories of chemical reactions},
\newblock \doi{10.1007/BF01030197}{\rm J. Stat. Phys. {\bf 14} (1976)
  307--331}\null.

\bibitem{WilliamsTuckey1992}
R.M. Williams and P.A. Tuckey,
\newblock {\em Regge calculus: a brief review and bibliography},
\newblock \doi{10.1088/0264-9381/9/5/021}{\rm {\bf 9} (1992)
  1409--1422}\null.

\bibitem{RasettiRegge1975}
M.~Rasetti and T.~Regge,
\newblock {\em Vortices in {He II}, current algebras and quantum knots},
\newblock \doi{https://doi.org/10.1016/0378-4371(75)90105-3}{\rm Physica A {\bf
  80} (1975)   217--233}\null.

\bibitem{Pastor-SatorrasVespignani2000}
R.~Pastor-Satorras and A.~Vespignani,
\newblock {\em Field theory of absorbing phase transitions with a nondiffusive
  conserved field},
\newblock \doi{10.1103/PhysRevE.62.R5875}{\rm Phys. Rev. E {\bf 62} (2000)
  R5875--R5878}\null.

\bibitem{BonachelaMunoz2008}
J.A. Bonachela and M.A. Mu\~noz,
\newblock {\em Confirming and extending the hypothesis of universality in
  sandpiles},
\newblock \doi{10.1103/PhysRevE.78.041102}{\rm Phys. Rev. E {\bf 78} (2008)
  041102}\null.

\bibitem{LeDoussalWiese2014a}
P.~Le Doussal and K.J. Wiese,
\newblock {\em An exact mapping of the stochastic field theory for {Manna}
  sandpiles to interfaces in random media},
\newblock \doi{10.1103/PhysRevLett.114.110601}{\rm Phys. Rev. Lett. {\bf 114}
  (2014)   110601}\null,
\newblock \arxiv{arXiv:1410.1930}.

\bibitem{JanssenStenull2016}
H.-K. Janssen and O.~Stenull,
\newblock {\em Directed percolation with a conserved field and the depinning
  transition},
\newblock \doi{10.1103/PhysRevE.94.042138}{\rm Phys. Rev. E (2016)
  042138}\null,
\newblock \arxiv{arXiv:607.01635}.

\bibitem{UhlenbeckOrnstein1930}
G.~E. Uhlenbeck and L.~S. Ornstein,
\newblock {\em On the theory of the {Brownian} motion},
\newblock \doi{10.1103/PhysRev.36.823}{\rm Phys. Rev. {\bf 36} (1930)
  823--841}\null.

\bibitem{Krug1997}
J.~Krug,
\newblock {\em Origins of scale invariance in growth processes},
\newblock \doi{10.1080/00018739700101498}{\rm Adv. Phys. {\bf 46} (1997)
  139--282}\null.

\bibitem{Halpin-HealyTakeuchi2015}
T.~Halpin-Healy and K.A. Takeuchi,
\newblock {\em A {KPZ} cocktail -- shaken, not stirred...},
\newblock \doi{10.1007/s10955-015-1282-1}{\rm J. Stat. Phys. {\bf 160} (2015)
  794--814}\null.

\bibitem{TakeuchiSano2012}
K.A. Takeuchi and M.~Sano,
\newblock {\em Evidence for geometry-dependent universal fluctuations of the
  {Kardar-Parisi-Zhang} interfaces in liquid-crystal turbulence},
\newblock \doi{10.1007/s10955-012-0503-0}{\rm J. Stat. Phys. {\bf 147} (2012)
  853--890}\null.

\bibitem{Burgers74}
J.M. Burgers,
\newblock {\em The non-linear diffusion equation},
\newblock Dordrecht, 1974.

\bibitem{GurbatovSaichevShandarin1989}
S.N. Gurbatov, A.I. Saichev  and S.F. Shandarin,
\newblock {\em The large-scale structure of the universe in the frame of the
  model equation of non-linear diffusion},
\newblock \doi{10.1093/mnras/236.2.385}{\rm Monthly Notices of the Royal
  Astronomical Society {\bf 236} (1989)   385--402}\null.

\bibitem{Bertschinger1998}
E.~Bertschinger,
\newblock {\em Simulations of structure formation in the universe},
\newblock \doi{10.1146/annurev.astro.36.1.599}{\rm Annu. Rev. Astron.
  Astrophys. {\bf 36} (1998)   599--654}\null.

\bibitem{BernardeauColombiGaztanagaScoccimarro2002}
F.~Bernardeau, S.~Colombi, E.~Gazta{\~n}aga  and R.~Scoccimarro,
\newblock {\em Large-scale structure of the universe and cosmological
  perturbation theory},
\newblock \doi{https://doi.org/10.1016/S0370-1573(02)00135-7}{\rm Phys. Rep.
  {\bf 367} (2002)   1--248}\null.

\bibitem{Hopf1950}
E.~Hopf,
\newblock {\em The partial differential equation $u_t + uu_x = \mu xx$},
\newblock \doi{10.1002/cpa.3160030302}{\rm Comm. Pure Appl. Math. {\bf 3}
  (1950)   201--230}\null.

\bibitem{Cole1951}
J.D. Cole,
\newblock {\em On a quasi-linear parabolic equation occurring in aerodynamics},
\newblock \doi{10.1090/qam/42889}{\rm Q. Appl. Math. {\bf 9} (1951)
  225--236}\null.

\bibitem{Feynman1948}
R.P. Feynman,
\newblock {\em Space-time approach to non-relativistic quantum mechanics},
\newblock \doi{10.1103/RevModPhys.20.367}{\rm Rev. Mod. Phys. {\bf 20} (1948)
  367--387}\null.

\bibitem{Kac1949}
M.~Kac,
\newblock {\em On distributions of certain {Wiener} functionals},
\newblock \doi{10.1090/S0002-9947-1949-0027960-X}{\rm Trans. Amer. Math. Soc.
  {\bf 65} (1949)   1--13}\null.

\bibitem{BrunetDerrida2000a}
E.~Brunet and B.~Derrida,
\newblock {\em Probability distribution of the free energy of a directed
  polymer in a random medium},
\newblock \doi{10.1103/PhysRevE.61.6789}{\rm Phys. Rev. E {\bf 61} (2000)
  6789--801}\null.

\bibitem{ForsterNelsonStephen1977}
D.~Forster, D.R. Nelson  and M.J. Stephen,
\newblock {\em Large-distance and long-time properties of a randomly stirred
  fluid},
\newblock \doi{10.1103/PhysRevA.16.732}{\rm Phys. Rev. A {\bf 16} (1977)
  732--749}\null.

\bibitem{MedinaHwaKardarZhang1989}
E.~Medina, T.~Hwa, M.~Kardar  and Y.C. Zhang,
\newblock {\em Burgers equation with correlated noise: Renormalization-group
  analysis and applications to directed polymers and interface growth},
\newblock \doi{10.1103/PhysRevA.39.3053}{\rm Phys. Rev. {\bf A 39} (1989)
  3053}\null.

\bibitem{MeakinRamanlalSanderBall1986}
P.~Meakin, P.~Ramanlal, L.M. Sander  and R.C. Ball,
\newblock {\em Ballistic deposition on surfaces},
\newblock \doi{10.1103/PhysRevA.34.5091}{\rm Phys. Rev. A {\bf 34} (1986)
  5091--5103}\null.

\bibitem{Krug1987}
J.~Krug,
\newblock {\em Scaling relation for a growing interface},
\newblock \doi{10.1103/PhysRevA.36.5465}{\rm Phys. Rev. A {\bf 36} (1987)
  5465--5466}\null.

\bibitem{FreyTaeuber1994}
E.~Frey and U.C. T{\"a}uber,
\newblock {\em Two-loop renormalization group analysis of the
  {Burgers-Kardar-Parisi-Zhang} equation},
\newblock \doi{10.1103/PhysRevE.50.1024}{\rm Phys. Rev. E {\bf 50} (1994)
  1024--1045}\null.

\bibitem{FreyTaeuber1994b}
E.~Frey and U.C. T{\"a}uber,
\newblock {\em Reply to ``{Comment} on `{Two-loop} renormalization group
  analysis of the {Burgers-Kardar-Parisi-Zhang} equation' ''},
\newblock \doi{10.1103/PhysRevE.51.6319}{\rm Phys. Rev. E {\bf 51} (1995)
  6319--6322}\null.

\bibitem{SunPlischke1994}
T.~Sun and M.~Plischke,
\newblock {\em Field-theory renormalization approach to the
  {Kardar-Parisi-Zhang} equation},
\newblock \doi{10.1103/PhysRevE.49.5046}{\rm Phys. Rev. E {\bf 49} (1994)
  5046--5057}\null.

\bibitem{Sun1995}
T.~Sun,
\newblock {\em {Comment} on `{Two-loop} renormalization group analysis of the
  {Burgers-Kardar-Parisi-Zhang} equation''},
\newblock \doi{10.1103/PhysRevE.51.6316}{\rm Phys. Rev. E {\bf 51} (1995)
  6316--6318}\null.

\bibitem{Teodorovich1996}
E.V. Teodorovich,
\newblock {\em \link{www.jetp.ac.ru/cgi-bin/dn/e_082_02_0268.pdf}{Anomalous
  dimensions in the {Burgers-Kardar-Parisi-Zhang} equation}},
\newblock JETP {\bf 82} (1996)   268--277.

\bibitem{Wiese1997c}
K.J. Wiese,
\newblock {\em Critical discussion of the 2-loop calculations for the
  {KPZ}-equation},
\newblock \doi{10.1103/PhysRevE.56.5013}{\rm Phys. Rev. {\bf E 56} (1997)
  5013--5017}\null,
\newblock \arxiv{cond-mat/9706009}.

\bibitem{Laessig1995}
M.~L{\"a}ssig,
\newblock {\em On the renormalization of the {Kardar-Parisi-Zhang} equation},
\newblock \doi{10.1016/0550-3213(95)00268-W}{\rm Nucl. Phys. {\bf B 448} (1995)
    559--574}\null,
\newblock \arxiv{cond-mat/9501094}.

\bibitem{Wiese1998a}
K.J. Wiese,
\newblock {\em On the perturbation expansion of the {KPZ}-equation},
\newblock \doi{10.1023/B:JOSS.0000026730.76868.c4}{\rm J. Stat. Phys. {\bf 93}
  (1998)   143--154}\null,
\newblock \arxiv{cond-mat/9802068}.

\bibitem{DDG2}
F.~David, B.~Duplantier  and E.~Guitter,
\newblock {\em Renormalization theory for interacting crumpled manifolds},
\newblock \doi{10.1016/0550-3213(93)90226-F}{\rm Nucl. Phys. {\bf B 394} (1993)
    555--664}\null,
\newblock \arxiv{hep-th/9211038}.

\bibitem{DDG4}
F.~David, B.~Duplantier  and E.~Guitter,
\newblock {\em Renormalization theory for the self-avoiding polymerized
  membranes},
\newblock (1997),
\newblock \arxiv{cond-mat/9702136}.

\bibitem{BundschuhLaessig1996}
R.~Bundschuh and M.~L{\"a}ssig,
\newblock {\em Directed polymers in high dimensions},
\newblock \doi{10.1103/PhysRevE.54.304}{\rm Phys. Rev. {\bf E 54} (1996)
  304--320}\null,
\newblock \arxiv{cond-mat/9602045}.

\bibitem{FreyTauberJanssen1999}
E.~Frey, U.C. Tauber  and H.K. Janssen,
\newblock {\em Scaling regimes and critical dimensions in the
  {Kardar-Parisi-Zhang} problem},
\newblock \doi{10.1209/epl/i1999-00343-4}{\rm Europhys. Lett. {\bf 47} (1999)
  14--20}\null.

\bibitem{Kardar1987b}
M.~Kardar,
\newblock {\em Domain walls subject to quenched impurities (invited)},
\newblock \doi{10.1063/1.338687}{\rm J. Appl. Phys. {\bf 61} (1987)
  3601--3604}\null,
\newblock \arxiv{https://doi.org/10.1063/1.338687}.

\bibitem{Nattermann1987}
T.~Nattermann,
\newblock {\em Interface roughening in systems with quenched random
  impurities},
\newblock \doi{10.1209/0295-5075/4/11/005}{\rm Europhys. Lett. {\bf 4} (1987)
  1241--6}\null.

\bibitem{JanssenTauberFrey1999}
H.K. Janssen, U.C. Tauber  and E.~Frey,
\newblock {\em Exact results for the {Kardar-Parisi-Zhang} equation with
  spatially correlated noise},
\newblock \doi{10.1007/s100510050790}{\rm Eur. Phys. J. B {\bf 9} (1999)
  491--511}\null.

\bibitem{TauberFrey2002}
U.C. T\"auber and E.~Frey,
\newblock {\em Universality classes in the anisotropic {Kardar-Parisi-Zhang}
  model},
\newblock \doi{10.1209/epl/i2002-00175-8}{\rm Europhys. Lett. {\bf 59} (2002)
  655--61}\null.

\bibitem{LassigKinzelbach1997}
M.~L\"assig and H.~Kinzelbach,
\newblock {\em Upper critical dimension of the {Kardar-Parisi-Zhang} equation},
\newblock \doi{10.1103/PhysRevLett.78.903}{\rm Phys. Rev. Lett. {\bf 78} (1997)
    903--6}\null.

\bibitem{Fogedby2005}
H.C. Fogedby,
\newblock {\em Localized growth modes, dynamic textures, and upper critical
  dimension for the {Kardar-Parisi-Zhang} equation in the weak-noise limit},
\newblock \doi{10.1103/PhysRevLett.94.195702}{\rm Phys. Rev. Lett. {\bf 94}
  (2005)   195702}\null.

\bibitem{NewmanKallabis1996}
T.J. Newman and H.~Kallabis,
\newblock {\em Strong coupling probe for the {Kardar-Parisi-Zhang} equation},
\newblock \doi{10.1051/jp1:1996162}{\rm J. Phys. I France {\bf 6} (1996)
  373--383}\null.

\bibitem{Bhattacharjee1998}
J.K. Bhattacharjee,
\newblock {\em Upper critical dimension of the {Kardar-Parisi-Zhang} equation},
\newblock \doi{10.1088/0305-4470/31/5/001}{\rm J. Phys. A {\bf 31} (1998)
  L93--L96}\null.

\bibitem{ColaioriMoore2001}
F.~Colaiori and M.A. Moore,
\newblock {\em Upper critical dimension, dynamic exponent, and scaling
  functions in the mode-coupling theory for the {Kardar-Parisi-Zhang}
  equation},
\newblock \doi{10.1103/PhysRevLett.86.3946}{\rm Phys. Rev. Lett. {\bf 86}
  (2001)   3946--3949}\null.

\bibitem{CanetMoore2007}
L.~Canet and M.A. Moore,
\newblock {\em Universality classes of the {Kardar-Parisi-Zhang} equation},
\newblock \doi{10.1103/PhysRevLett.98.200602}{\rm Phys. Rev. Lett. {\bf 98}
  (2007)   200602}\null.

\bibitem{KatzavSchwartz2002}
E.~Katzav and M.~Schwartz,
\newblock {\em Existence of the upper critical dimension of the
  {Kardar--Parisi--Zhang} equation},
\newblock \doi{https://doi.org/10.1016/S0378-4371(02)00553-8}{\rm Physica A
  {\bf 309} (2002)   69--78}\null,
\newblock \arxiv{cond-mat/0012103}.

\bibitem{MarinariPagnaniParisi2000}
E.~Marinari, A.~Pagnani  and G.~Parisi,
\newblock {\em Critical exponents of the {KPZ} equation via multi-surface
  coding numerical simulations},
\newblock \doi{10.1088/0305-4470/33/46/303}{\rm J. Phys. A {\bf 33} (2000)
  8181--92}\null.

\bibitem{MarinariPagnaniParisiRacz2002}
E.~Marinari, A.~Pagnani, G.~Parisi  and Z.~Racz,
\newblock {\em Width distributions and the upper critical dimension of
  {Kardar-Parisi-Zhang} interfaces},
\newblock \doi{10.1103/PhysRevE.65.026136}{\rm Phys. Rev. E {\bf 65} (2002)
  026136}\null.

\bibitem{AlvesOliveiraFerreira2014}
S.G. Alves, T.J. Oliveira  and S.C. Ferreira,
\newblock {\em Universality of fluctuations in the {Kardar-Parisi-Zhang} class
  in high dimensions and its upper critical dimension},
\newblock \doi{10.1103/PhysRevE.90.020103}{\rm Phys. Rev. E {\bf 90} (2014)
  020103}\null.

\bibitem{GomesPennaOliveira2019}
W.P. Gomes, A.L.A. Penna  and F.A. Oliveira,
\newblock {\em From cellular automata to growth dynamics: The
  {Kardar-Parisi-Zhang} universality class},
\newblock \doi{10.1103/PhysRevE.100.020101}{\rm Phys. Rev. E {\bf 100} (2019)
  020101}\null.

\bibitem{Ala-Nissila1998a}
T.~Ala-Nissila,
\newblock {\em Comment on ``{Upper} critical dimension of the
  {Kardar-Parisi-Zhang} equation''},
\newblock \doi{10.1103/PhysRevLett.80.887}{\rm Phys. Rev. Lett. {\bf 80} (1998)
    887--887}\null.

\bibitem{Ala-Nissila1998}
T.~Ala-Nissila,
\newblock {\em Erratum: Comment on {``Upper Critical Dimension of the
  {Kardar-Parisi-Zhang} Equation''} {[Phys. Rev. Lett. 80, 887 (1998)]}},
\newblock \doi{10.1103/PhysRevLett.80.5459}{\rm Phys. Rev. Lett. {\bf 80}
  (1998)   5459--5459}\null.

\bibitem{SchwartzPerlsman2012}
M.~Schwartz and E.~Perlsman,
\newblock {\em Upper critical dimension of the {Kardar-Parisi-Zhang} equation},
\newblock \doi{10.1103/PhysRevE.85.050103}{\rm Phys. Rev. E {\bf 85} (2012)
  050103}\null.

\bibitem{Tu1994}
Y.~Tu,
\newblock {\em Absence of finite upper critical dimension in the spherical
  {Kardar-Parisi-Zhang} model},
\newblock \doi{10.1103/PhysRevLett.73.3109}{\rm Phys. Rev. Lett. {\bf 73}
  (1994)   3109--3112}\null.

\bibitem{CastellanoMarsiliPietronero1998}
C.~Castellano, M.~Marsili  and L.~Pietronero,
\newblock {\em Nonperturbative renormalization of the {Kardar-Parisi-Zhang}
  growth dynamics},
\newblock \doi{10.1103/PhysRevLett.80.3527}{\rm Phys. Rev. Lett. {\bf 80}
  (1998)   3527--3530}\null.

\bibitem{CanetChateDelamotteWschebor2010}
L.~Canet, H.~Chat\'e, B.~Delamotte  and N.~Wschebor,
\newblock {\em Nonperturbative renormalization group for the
  {Kardar-Parisi-Zhang} equation},
\newblock \doi{10.1103/PhysRevLett.104.150601}{\rm Phys. Rev. Lett. {\bf 104}
  (2010)   150601}\null.

\bibitem{BouchaudCates1993}
J.P. Bouchaud and M.E. Cates,
\newblock {\em Self-consistent approach to the {Kardar-Parisi-Zhang} equation},
\newblock \doi{10.1103/PhysRevE.47.R1455}{\rm Phys. Rev. E {\bf 47} (1993)
  R1455--R1458}\null.

\bibitem{Lassig1998}
M.~L\"assig,
\newblock {\em Quantized scaling of growing surfaces},
\newblock \doi{10.1103/PhysRevLett.80.2366}{\rm Phys. Rev. Lett. {\bf 80}
  (1998)   2366--2369}\null.

\bibitem{Laessig1998}
M.~{L{\"a}ssig},
\newblock {\em On growth, disorder, and field theory},
\newblock \doi{10.1088/0953-8984/10/44/003}{\rm J. Phys. C {\bf 10} (1998)
  9905}\null.

\bibitem{OdorLiedkeHeinig2010}
G.~\'Odor, B.~Liedke  and K.-H. Heinig,
\newblock {\em Directed $d$-mer diffusion describing the
  {Kardar-Parisi-Zhang}-type surface growth},
\newblock \doi{10.1103/PhysRevE.81.031112}{\rm Phys. Rev. E {\bf 81} (2010)
  031112}\null.

\bibitem{PagnaniParisi2013}
A.~Pagnani and G.~Parisi,
\newblock {\em Multisurface coding simulations of the restricted solid-on-solid
  model in four dimensions},
\newblock \doi{10.1103/PhysRevE.87.010102}{\rm Phys. Rev. E {\bf 87} (2013)
  010102}\null.

\bibitem{LvovLebedevPatonProcaccia1993}
V.Ss L'vov, V.V. Lebedev, M.~Paton  and I.~Procaccia,
\newblock {\em Proof of scale invariant solutions in the {Kardar-Parisi-Zhang
  and Kuramoto-Sivashinsky} equations in 1+1 dimensions: analytical and
  numerical results},
\newblock \doi{10.1088/0951-7715/6/1/002}{\rm Nonlinearity {\bf 6} (1993)
  25--47}\null.

\bibitem{PraehoferSpohn2000}
M.~Pr{\"a}hofer and H.~Spohn,
\newblock {\em Statistical self-similarity of one-dimensional growth
  processes},
\newblock \doi{10.1016/S0378-4371(99)00517-8}{\rm Physica A {\bf 279} (2000)
  342--52}\null,
\newblock \arxiv{cond-mat/9910273}.

\bibitem{PraehoferSpohn2000a}
M.~Pr{\"a}hofer and H.~Spohn,
\newblock {\em Universal distributions for growth processes in 1 + 1 dimensions
  and random matrices},
\newblock \doi{10.1103/PhysRevLett.84.4882}{\rm Phys. Rev. Lett. {\bf 84}
  (2000)   4882--4885}\null,
\newblock \arxiv{cond-mat/9912264}.

\bibitem{PrahoferSpohn2002}
M.~Pr{\"a}hofer and H.~Spohn,
\newblock {\em Scale invariance of the {PNG} droplet and the {Airy} process},
\newblock \doi{10.1023/A:1019791415147}{\rm J. Stat. Phys. {\bf 108} (2002)
  1071--1106}\null.

\bibitem{Johansson2001}
K.~Johansson,
\newblock {\em Universality of the local spacing distribution in certain
  ensembles of hermitian wigner matrices},
\newblock \doi{10.1007/s002200000328}{\rm Commun, Math. Phys. {\bf 215} (2001)
   683--705}\null.

\bibitem{BaikDeiftJohansson1999}
J.~Baik, P.~Deift  and K.~Johansson,
\newblock {\em On the distribution of the length of the longest increasing
  subsequence of random permutations},
\newblock \doi{10.1090/S0894-0347-99-00307-0}{\rm J. Amer. Math. Soc. {\bf 12}
  (1999)   1119--1178}\null,
\newblock \arxiv{arXiv:math/9810105}.

\bibitem{Johansson2000}
K.~Johansson,
\newblock {\em Shape fluctuations and random matrices},
\newblock \doi{10.1007/s002200050027}{\rm Commun, Math. Phys. {\bf 209} (2000)
   437--76}\null,
\newblock \arxiv{math/9903134}.

\bibitem{BaerBrock1968}
R.M. Baer and P.~Brock,
\newblock {\em Natural sorting over permutation spaces},
\newblock \doi{doi:10.2307/2004668}{\rm Math. Comput. {\bf 22} (1968)
  385}\null.

\bibitem{Dobrinevski2014}
A.~Dobrinevski,
\newblock Simulating directed polymers and the {Tracy-Widom} distribution,
\newblock
  \link{inordinatum.wordpress.com/2014/01/22/simulating-directed-polymers-and-the-tracy-widom-distribution}{inordinatum.wordpress.com}.

\bibitem{CalabreseLeDoussalRosso2010}
P.~Calabrese, P.~Le~Doussal  and A.~Rosso,
\newblock {\em Free-energy distribution of the directed polymer at high
  temperature},
\newblock \doi{10.1209/0295-5075/90/20002}{\rm EPL {\bf 90} (2010)
  20002}\null,
\newblock \arxiv{arXiv:1002.4560}.

\bibitem{Dotsenko2010}
V.~Dotsenko,
\newblock {\em Bethe ansatz derivation of the {Tracy-Widom} distribution for
  one-dimensional directed polymers},
\newblock \doi{10.1209/0295-5075/90/20003}{\rm EPL {\bf 90} (2010)
  20003}\null.

\bibitem{Dotsenko2010b}
V.~Dotsenko,
\newblock {\em Replica {Bethe} ansatz derivation of the {Tracy-Widom}
  distribution of the free energy fluctuations in one-dimensional directed
  polymers},
\newblock \doi{10.1088/1742-5468/2010/07/p07010}{\rm J. Stat. Mech. {\bf 2010}
  (2010)   P07010}\null.

\bibitem{CalabreseLeDoussal2011}
P.~Calabrese and P.~Le~Doussal,
\newblock {\em Exact solution for the {Kardar-Parisi-Zhang} equation with flat
  initial conditions},
\newblock \doi{10.1103/PhysRevLett.106.250603}{\rm Phys. Rev. Lett. {\bf 106}
  (2011)   250603}\null.

\bibitem{LeDoussalCalabrese2012}
P.~Le~Doussal and P.~Calabrese,
\newblock {\em The {KPZ} equation with flat initial condition and the directed
  polymer with one free end},
\newblock \doi{10.1088/1742-5468/2012/06/p06001}{\rm {\bf 2012} (2012)
  P06001}\null.

\bibitem{GueudreLeDoussal2012}
T.~Gueudr{\'{e}} and P.~Le Doussal,
\newblock {\em Directed polymer near a hard wall and {KPZ} equation in the
  half-space},
\newblock \doi{10.1209/0295-5075/100/26006}{\rm EPL {\bf 100} (2012)
  26006}\null.

\bibitem{Corwin2012}
I.~Corwin,
\newblock {\em The {Kardar-Parisi-Zhang} equation and universality class},
\newblock \doi{10.1142/S2010326311300014}{\rm Random Matrices: Theory and
  Applications {\bf 01} (2012)   1130001}\null.

\bibitem{AmirCorwinQuastel2011}
G.~Amir, I.~Corwin  and J.~Quastel,
\newblock {\em Probability distribution of the free energy of the continuum
  directed random polymer in 1 + 1 dimensions},
\newblock \doi{10.1002/cpa.20347}{\rm Commun. Pur. Appl. Math. {\bf 64} (2011)
   466--537}\null.

\bibitem{BorodinCorwin2014}
A.~Borodin and I.~Corwin,
\newblock {\em Macdonald processes},
\newblock \doi{10.1007/s00440-013-0482-3}{\rm Probab. Th. Rel. Fields {\bf 158}
  (2014)   225--400}\null.

\bibitem{ImamuraSasamoto2012}
T.~Imamura and T.~Sasamoto,
\newblock {\em Exact solution for the stationary {Kardar-Parisi-Zhang}
  equation},
\newblock \doi{10.1103/PhysRevLett.108.190603}{\rm Phys. Rev. Lett. {\bf 108}
  (2012)   190603}\null.

\bibitem{SasamotoSpohn2010}
T.~Sasamoto and H.~Spohn,
\newblock {\em One-dimensional {Kardar-Parisi-Zhang} equation: An exact
  solution and its universality},
\newblock \doi{10.1103/PhysRevLett.104.230602}{\rm Phys. Rev. Lett. {\bf 104}
  (2010)   230602}\null.

\bibitem{SasamotoSpohn2010b}
T.~Sasamoto and H.~Spohn,
\newblock {\em Exact height distributions for the {KPZ} equation with narrow
  wedge initial condition},
\newblock \doi{https://doi.org/10.1016/j.nuclphysb.2010.03.026}{\rm Nucl. Phys.
  B {\bf 834} (2010)   523--542}\null.

\bibitem{TakeuchiSano2010}
K.A. Takeuchi and M.~Sano,
\newblock {\em Universal fluctuations of growing interfaces: Evidence in
  turbulent liquid crystals},
\newblock \doi{10.1103/PhysRevLett.104.230601}{\rm Phys. Rev. Lett. {\bf 104}
  (2010)   230601}\null.

\bibitem{KriecherbauerKrug2010}
T.~Kriecherbauer and J.~Krug,
\newblock {\em A pedestrian's view on interacting particle systems, {KPZ}
  universality and random matrices},
\newblock \doi{10.1088/1751-8113/43/40/403001}{\rm J. Phys. A {\bf 43} (2010)
  403001}\null.

\bibitem{Quastel2011}
J.~Quastel,
\newblock {\em Introduction to {KPZ}},
\newblock \doi{10.4310/CDM.2011.v2011.n1.a3}{\rm Current Developments in
  Mathematics {\bf 2011} (2013)}\null.

\bibitem{QuastelSpohn2015}
J.~Quastel and H.~Spohn,
\newblock {\em The one-dimensional {KPZ} equation and its universality class},
\newblock \doi{10.1007/s10955-015-1250-9}{\rm J. Stat. Phys. {\bf 160} (2015)
  965--984}\null.

\bibitem{Halpin-Healy2012}
T.~Halpin-Healy,
\newblock {\em ($2\mathbf{+}1$)-dimensional directed polymer in a random
  medium: Scaling phenomena and universal distributions},
\newblock \doi{10.1103/PhysRevLett.109.170602}{\rm Phys. Rev. Lett. {\bf 109}
  (2012)   170602}\null.

\bibitem{Halpin-Healy2013}
T.~Halpin-Healy,
\newblock {\em Extremal paths, the stochastic heat equation, and the
  three-dimensional {Kardar-Parisi-Zhang} universality class},
\newblock \doi{10.1103/PhysRevE.88.042118}{\rm Phys. Rev. E {\bf 88} (2013)
  042118}\null.

\bibitem{Spitzer1970}
F.~Spitzer,
\newblock {\em Interaction of {Markov} processes},
\newblock \doi{https://doi.org/10.1016/0001-8708(70)90034-4}{\rm Adv. Math.
  {\bf 5} (1970)   246 -- 290}\null.

\bibitem{Krug1991}
J.~Krug,
\newblock {\em Boundary-induced phase transitions in driven diffusive systems},
\newblock \doi{10.1103/PhysRevLett.67.1882}{\rm Phys. Rev. Lett. {\bf 67}
  (1991)   1882--1885}\null.

\bibitem{Derrida1998}
B.~Derrida,
\newblock {\em An exactly soluble non-equilibrium system: The asymmetric simple
  exclusion process},
\newblock \doi{https://doi.org/10.1016/S0370-1573(98)00006-4}{\rm Phys. Rep.
  {\bf 301} (1998)   65 -- 83}\null.

\bibitem{MyllysMaunukselaAlavaAla-NissilaMerikoskiTimonen2001}
M.~Myllys, J.~Maunuksela, M.~Alava, T.~Ala-Nissila, J.~Merikoski  and
  J.~Timonen,
\newblock {\em Kinetic roughening in slow combustion of paper},
\newblock \doi{10.1103/PhysRevE.64.036101}{\rm Phys. Rev. E {\bf 64} (2001)
  036101}\null.

\bibitem{MiettinenMyllysMerikoskiTimonen2005}
L.~Miettinen, M.~Myllys, J.~Merikoski  and J.~Timonen,
\newblock {\em Experimental determination of {KPZ} height-fluctuation
  distributions},
\newblock \doi{10.1140/epjb/e2005-00235-y}{\rm EPJB {\bf 46} (2005)
  55--60}\null.

\bibitem{DiasYunkerYodhAraujoTelo-da-Gama2018}
C.S. Dias, P.J. Yunker, A.G. Yodh, N.A.M. Ara{\'u}jo  and M.M. Telo~da Gama,
\newblock {\em Interaction anisotropy and the {KPZ to KPZQ} transition in
  particle deposition at the edges of drying drops},
\newblock \doi{10.1039/C7SM02136D}{\rm Soft Matter {\bf 14} (2018)
  1903--1907}\null.

\bibitem{HallatschekHersenRamanathanNelson2007}
O.~Hallatschek, P.~Hersen, S.~Ramanathan  and D.R. Nelson,
\newblock {\em Genetic drift at expanding frontiers promotes gene segregation},
\newblock \doi{10.1073/pnas.0710150104}{\rm PNAS {\bf 104} (2007)
  19926--19930}\null.

\bibitem{Kolmogorov1941}
A.N. Kolmogorov,
\newblock {\em On the energy distribution in the spectrum of a turbulent flow},
\newblock C.R. Acad. Sci. URSS {\bf 30} (1941)   301--305.

\bibitem{FedorenkoLeDoussalWiese2012}
A.~Fedorenko, P.~{Le~Doussal}  and K.J. Wiese,
\newblock {\em Functional renormalization-group approach to decaying
  turbulence},
\newblock \doi{10.1088/1742-5468/2013/04/P04014}{\rm J. Stat. Mech. (2013)
  P04014}\null,
\newblock \arxiv{arXiv:1212.2117}.

\bibitem{Berezin1965}
F.~Berezin,
\newblock \doi{10.1007/978-3-642-58150-2_2}{\rm {\em The Method of Second
  Quantization}}\null,
\newblock Academic Press, 1965.

\bibitem{Wegner2016}
F.~Wegner,
\newblock \doi{10.1007/978-3-662-49170-6}{\rm {\em Supermathematics and its
  Applications in Statistical Physics}}\null,
\newblock Springer-Verlag, Berlin, Heidelberg, 2016.

\bibitem{ParisiSourlas1982}
G.~Parisi and N.~Sourlas,
\newblock {\em Supersymmetric field theories and stochastic differential
  equations},
\newblock \doi{10.1016/0550-3213(82)90538-7}{\rm Nucl. Phys. B {\bf B206}
  (1982)   321--32}\null.

\bibitem{Cardy1983}
J.L. Cardy,
\newblock {\em Nonperturbative effects in a scalar supersymmetric theory},
\newblock \doi{10.1016/0370-2693(83)91328-X}{\rm Phys. Lett. {\bf 125 B} (1983)
    470--2}\null.

\bibitem{Cardy1985}
J.L. Cardy,
\newblock {\em Nonperturbative aspects of supersymmetry in statistical
  mechanics},
\newblock \doi{https://doi.org/10.1016/0167-2789(85)90154-X}{\rm Physica D {\bf
  15} (1985)   123--128}\null.

\bibitem{CardyMcKane1985}
J.L. Cardy and A.J. McKane,
\newblock {\em Field theoretic approach to the study of {Yang-Lee and
  Griffiths} singularities in the randomly diluted {Ising} model},
\newblock \doi{https://doi.org/10.1016/0550-3213(85)90352-9}{\rm Nucl. Phys. B
  {\bf 257} (1985)   383--396}\null.

\bibitem{KavirajRychkovTrevisani2019}
A.~Kaviraj, S.~Rychkov  and E.~Trevisani,
\newblock {\em Random field {Ising} model and {Parisi-Sourlas} supersymmetry.
  {Part I. Supersymmetric CFT}},
\newblock \doi{10.1007/JHEP04(2020)090}{\rm JHEP {\bf 2020} (2019)
  1--49}\null,
\newblock \arxiv{arXiv:1912.01617}.

\bibitem{KavirajRychkovTrevisani2020}
A.~Kaviraj, S.~Rychkov  and E.~Trevisani,
\newblock {\em Random field {Ising} model and {Parisi-Sourlas} supersymmetry
  {II. Renormalization} group},
\newblock (2020),
\newblock \arxiv{arXiv:2009.10087}.

\bibitem{Lawler1980}
G.F. Lawler,
\newblock {\em A self-avoiding random walk},
\newblock \doi{10.1215/S0012-7094-80-04741-9}{\rm Duke Math. J. {\bf 47} (1980)
    655--693}\null.

\bibitem{Kozma2007}
G.~Kozma,
\newblock {\em The scaling limit of loop-erased random walk in three
  dimensions},
\newblock \doi{10.1007/s11511-007-0018-8}{\rm Acta Mathematica {\bf 199} (2007)
    29--152}\null.

\bibitem{GuttmannBursill1990}
A.J. Guttmann and R.J. Bursill,
\newblock {\em Critical exponent for the loop erased self-avoiding walk by
  monte carlo methods},
\newblock \doi{10.1007/BF01015560}{\rm J. Stat. Phys. {\bf 59} (1990)
  1--9}\null.

\bibitem{AgrawalDhar2001}
H.~Agrawal and D.~Dhar,
\newblock {\em Distribution of sizes of erased loops of loop-erased random
  walks in two and three dimensions},
\newblock \doi{10.1103/PhysRevE.63.056115}{\rm Phys. Rev. E {\bf 63} (2001)
  056115}\null.

\bibitem{Grassberger2009}
P.~Grassberger,
\newblock {\em Scaling of loop-erased walks in 2 to 4 dimensions},
\newblock \doi{10.1007/s10955-009-9787-0}{\rm J. Stat. Phys. {\bf 136} (2009)
  399--404}\null,
\newblock \arxiv{arXiv:0905.3440}.

\bibitem{Wilson2010}
D.B. Wilson,
\newblock {\em Dimension of the loop-erased random walk in three dimensions},
\newblock \doi{10.1103/PhysRevE.82.062102}{\rm Phys. Rev. E {\bf 82} (2010)
  062102}\null,
\newblock \arxiv{arXiv:1008.1147}.

\bibitem{Schramm2000}
O.~Schramm,
\newblock {\em Scaling limits of loop-erased random walks and uniform spanning
  trees},
\newblock \doi{10.1007/BF02803524}{\rm Israel J. Math. {\bf 118} (2000)
  221--288}\null,
\newblock \arxiv{arXiv:math/9904022}.

\bibitem{LawlerSchrammWerner2004}
G.F. Lawler, O.~Schramm  and W.~Werner,
\newblock {\em Conformal invariance of planar loop-erased random walks and
  uniform spanning trees},
\newblock \doi{10.1007/978-1-4419-9675-6_30}{\rm Ann. Probab. {\bf 32} (2004)
  939--995}\null,
\newblock \arxiv{arXiv:math/0112234}.

\bibitem{Nienhuis1982}
B.~Nienhuis,
\newblock {\em Exact critical point and critical exponents of $\mathrm{O}(n)$
  models in two dimensions},
\newblock \doi{10.1103/PhysRevLett.49.1062}{\rm Phys. Rev. Lett. {\bf 49}
  (1982)   1062--1065}\null.

\bibitem{HelmuthShapira2020}
T.~Helmuth and A.~Shapira,
\newblock {\em Loop-erased random walk as a spin system observable},
\newblock \doi{10.1007/s10955-020-02628-7}{\rm J. Stat. Phys. {\bf 181} (2020)
   1306--1322}\null,
\newblock \arxiv{arXiv:2003.10928}.

\bibitem{Viennot1986}
G.X. Viennot,
\newblock {\em \doi{10.1007/BFb0072524}{Heaps of pieces, {I : Basic}
  definitions and combinatorial lemmas}},
\newblock in Gilbert Labelle and Pierre Leroux, editors, {\em Combinatoire
  {\'e}num{\'e}rative}, pages 321--350, Springer, Berlin, Heidelberg, 1986.

\bibitem{ViennotLectures}
X.~Viennot,
\newblock \link{viennot.org/abjc.html}{The Art of Bijective Combinatorics},
\newblock Online Lectures, 2017.

\bibitem{KenyonWilson2015}
R.W. Kenyon and D.B. Wilson,
\newblock {\em Spanning trees of graphs on surfaces and the intensity of
  loop-erased random walk on planar graphs},
\newblock \doi{10.1090/S0894-0347-2014-00819-5}{\rm J. Amer. Math. Soc. {\bf
  28} (2015)   985--1030}\null,
\newblock \arxiv{arXiv:1107.3377}.

\bibitem{Lawler2006}
G.F. Lawler,
\newblock {\em The {Laplacian}-$b$ random walk and the {Schramm-Loewner}
  evolution},
\newblock \doi{10.1215/ijm/1258059489}{\rm Illinois J. Math. {\bf 50} (2006)
  701--746}\null.

\bibitem{Kasteleyn1967}
P.W. Kasteleyn,
\newblock in F.~Harary, editor, {\em Graph Theory and Theoretical Physics},
  Academic Press, London and New York, 1967.

\bibitem{FedorenkoLeDoussalWiese2008a}
A.A. Fedorenko, P.~Le Doussal  and K.J. Wiese,
\newblock {\em Field theory conjecture for loop-erased random walks},
\newblock \doi{10.1007/s10955-008-9642-8}{\rm J. Stat. Phys. {\bf 133} (2008)
  805--812}\null,
\newblock \arxiv{arXiv:0803.2357}.

\bibitem{Majumdar1992}
S.N. Majumdar,
\newblock {\em Exact fractal dimension of the loop-erased self-avoiding walk in
  two dimensions},
\newblock \doi{10.1103/PhysRevLett.68.2329}{\rm Phys. Rev. Lett. {\bf 68}
  (1992)   2329--2331}\null.

\bibitem{LyklemaEvertszPietronero1986}
J.~W. Lyklema, C.~Evertsz  and L.~Pietronero,
\newblock {\em The {Laplacian} random walk},
\newblock \doi{0295-5075/2/i=2/a=001}{\rm EPL {\bf 2} (1986)  ~77}\null.

\bibitem{NiemeyerPietroneroWiesmann1984}
L.~Niemeyer, L.~Pietronero  and H.~J. Wiesmann,
\newblock {\em Fractal dimension of dielectric breakdown},
\newblock \doi{10.1103/PhysRevLett.52.1033}{\rm Phys. Rev. Lett. {\bf 52}
  (1984)   1033--1036}\null.

\bibitem{WittenSander1981}
T.A. Witten and L.M. Sander,
\newblock {\em Diffusion-limited aggregation, a kinetic critical phenomenon},
\newblock \doi{10.1103/PhysRevLett.47.1400}{\rm Phys. Rev. Lett. {\bf 47}
  (1981)   1400--1403}\null.

\bibitem{WieseKardar1998a}
K.J. Wiese and M.~Kardar,
\newblock {\em Generalizing the {$O(N)$}-field theory to {$N$}-colored
  manifolds of arbitrary internal dimension {$D$}},
\newblock \doi{10.1016/S0550-3213(98)00381-2}{\rm Nucl. Phys. {\bf B 528}
  (1998)   469--522}\null,
\newblock \arxiv{cond-mat/9803389}.

\bibitem{WieseKardar1998b}
K.J. Wiese and M.~Kardar,
\newblock {\em A geometric generalization of field theory to manifolds of
  arbitrary dimension},
\newblock \doi{10.1007/s100510050604}{\rm Eur. Phys. J. {\bf B 7} (1998)
  187--190}\null,
\newblock \arxiv{cond-mat/9803279}.

\bibitem{BelavinPolyakovZamolodchikov1984}
A.A. Belavin, A.M. Polyakov  and A.B. Zamolodchikov,
\newblock {\em Infinite conformal symmetry in two-dimensional quantum field
  theory},
\newblock \doi{10.1016/0550-3213(84)90052-X}{\rm Nucl. Phys. B {\bf 241} (1984)
    333--380}\null.

\bibitem{ItzyksonDrouffe2}
C.\ Itzykson and J.-M.\ Drouffe,
\newblock {\em Statistical Field Theory}, {\em {\em Volume}~2},
\newblock Cambridge University Press, 1989.

\bibitem{RushkinBettelheimGruzbergWiegmann2007}
I.~Rushkin, E.~Bettelheim, I.A. Gruzberg  and P.~Wiegmann,
\newblock {\em Critical curves in conformally invariant statistical systems},
\newblock \doi{10.1088/1751-8113/40/9/020}{\rm J. Phys. A {\bf 40} (2007)
  2165--2195}\null,
\newblock \arxiv{cond-mat/0610550}.

\bibitem{BloteKnopsNienhuis1992}
H.W.J. Bl\"ote, Y.M.M. Knops  and B.~Nienhuis,
\newblock {\em Geometrical aspects of critical {Ising} configurations in 2
  dimensions},
\newblock \doi{10.1103/PhysRevLett.68.3440}{\rm Phys. Rev. Lett. {\bf 68}
  (1992)   3440--3443}\null.

\bibitem{JankeSchakel2010}
W.~Janke and A.M.J. Schakel,
\newblock {\em Holographic interpretation of two-dimensional {$O(N)$} models
  coupled to quantum gravity},
\newblock (2010),
\newblock \arxiv{arXiv:1003.2878}.

\bibitem{Kirkham1981}
J.E. Kirkham,
\newblock {\em Calculation of crossover exponent from {Heisenberg} to {Ising}
  behaviour using the fourth-order $\epsilon$ expansion},
\newblock \doi{10.1088/0305-4470/14/11/004}{\rm J. Phys. A {\bf 14} (1981)
  L437--L442}\null.

\bibitem{MoghimiAraghiRajabpourRouhani2005}
S.~Moghimi-Araghi, M.A. Rajabpour  and S.~Rouhani,
\newblock {\em Abelian sandpile model: A conformal field theory point of view},
\newblock \doi{10.1016/j.nuclphysb.2005.04.002}{\rm Nucl. Phys. B {\bf 718}
  (2005)   362--370}\null,
\newblock \arxiv{cond-mat/0410434}.

\bibitem{KompanietsPanzer2017}
M.V. Kompaniets and E.~Panzer,
\newblock {\em Minimally subtracted six-loop renormalization of
  {$O(n)$}-symmetric ${\ensuremath{\phi}}^{4}$ theory and critical exponents},
\newblock \doi{10.1103/PhysRevD.96.036016}{\rm Phys. Rev. D {\bf 96} (2017)
  036016}\null,
\newblock \arxiv{arXiv:1705.06483}.

\bibitem{MeraPedersenNikolic2018}
H.~Mera, T.~G. Pedersen  and B.K. Nikoli\ifmmode~\acute{c}\else \'{c}\fi{},
\newblock {\em Fast summation of divergent series and resurgent transseries
  from {Meijer-$G$} approximants},
\newblock \doi{10.1103/PhysRevD.97.105027}{\rm Phys. Rev. D {\bf 97} (2018)
  105027}\null.

\bibitem{Hooft1974}
G.'t Hooft,
\newblock {\em A planar diagram theory for strong interactions},
\newblock \doi{https://doi.org/10.1016/0550-3213(74)90154-0}{\rm Nucl. Phys. B
  {\bf 72} (1974)   461--473}\null.

\bibitem{Polchinski1984}
J.~Polchinski,
\newblock {\em Renormalization and effective {Lagrangians}},
\newblock \doi{10.1016/0550-3213(84)90287-6}{\rm Nucl. Phys. B {\bf 231} (1984)
    269--95}\null.

\bibitem{Wetterich1993}
C.~Wetterich,
\newblock {\em Exact evolution equation for the effective potential},
\newblock \doi{https://doi.org/10.1016/0370-2693(93)90726-X}{\rm Phys. Lett. B
  {\bf 301} (1993)   90--94}\null.

\bibitem{HazenfratzHasenfratz1968}
A.~Hasenfratz and P.~Hasenfratz,
\newblock {\em Renormalization group study of scalar field theory},
\newblock \doi{10.1016/0550-3213(86)90573-0}{\rm Nucl. Phys. {\bf B 270} (1986)
    687--701}\null.

\bibitem{WegnerHoughton1973}
F.J. Wegner and A.~Houghton,
\newblock {\em Renormalization group equation for critical phenomena},
\newblock \doi{10.1103/PhysRevA.8.401}{\rm Phys. Rev. A {\bf 8} (1973)
  401--12}\null.

\bibitem{Morris1994}
T.R. Morris,
\newblock {\em The exact renormalization group and approximate solutions},
\newblock \doi{10.1142/S0217751X94000972}{\rm Int. J. Mod. Phys. A {\bf 09}
  (1994)   2411}\null.

\bibitem{BergesTetradisWetterich2002}
J.~Berges, N.~Tetradis  and C.~Wetterich,
\newblock {\em Non-perturbative renormalization flow in quantum field theory
  and statistical physics},
\newblock \doi{10.1016/S0370-1573(01)00098-9}{\rm Phys. Rep. {\bf 363} (2002)
  223--386}\null.

\bibitem{DupuisCanetEichhornMetznerPawlowskiTissierWschebor2021}
N.~Dupuis, L.~Canet, A.~Eichhorn, W.~Metzner, J.M. Pawlowski, M.~Tissier  and
  N.~Wschebor,
\newblock {\em The nonperturbative functional renormalization group and its
  applications},
\newblock \doi{https://doi.org/10.1016/j.physrep.2021.01.001}{\rm Phys. Rep.
  (2021) 1-114}\null,
\newblock \arxiv{arXiv:2006.04853}.

\bibitem{NattermannBookYoung}
T.~Nattermann,
\newblock {\em Theory of the random field {Ising} model},
\newblock in A.P. Young, editor, {\em Spin glasses and random fields}, World
  Scientific, Singapore, 1997,
\newblock \arxiv{cond-mat/9705295}.

\bibitem{Fisher1985b}
D.S. Fisher,
\newblock {\em Random fields, random anisotropies, nonlinear sigma models and
  dimensional reduction},
\newblock \doi{10.1103/PhysRevB.31.7233}{\rm Phys. Rev. B {\bf 31} (1985)
  7233--51}\null.

\bibitem{Feldman2000}
D.E. Feldman,
\newblock {\em Quasi-long-range order in the random anisotropy {Heisenberg}
  model: Functional renormalization group in 4- epsilon dimensions},
\newblock \doi{10.1103/PhysRevB.61.382}{\rm Phys. Rev. B {\bf 61} (2000)
  382--90}\null.

\bibitem{Feldman2001}
D.E. Feldman,
\newblock {\em Quasi-long range order in glass states of impure liquid
  crystals, magnets, and superconductors},
\newblock \doi{10.1142/S0217979201006641}{\rm Int. J. Mod. Phys. B {\bf 15}
  (2001)   2945}\null,
\newblock \arxiv{cond-mat/0201243}.

\bibitem{GiamarchiLeDoussal1995}
T.~Giamarchi and P.~Le Doussal,
\newblock {\em Elastic theory of flux lattices in the presence of weak
  disorder},
\newblock \doi{10.1103/PhysRevB.52.1242}{\rm Phys. Rev. B {\bf 52} (1995)
  1242--70}\null,
\newblock \arxiv{cond-mat/9501087}.

\bibitem{TarjusTissier2020}
G.~Tarjus and M.~Tissier,
\newblock {\em Random-field {Ising} and {$O(N)$} models: theoretical
  description through the functional renormalization group},
\newblock \doi{10.1140/epjb/e2020-100489-1}{\rm Eur. Phys. J. B {\bf 93} (2020)
   ~50}\null.

\bibitem{TissierTarjus2012}
M.~Tissier and G.~Tarjus,
\newblock {\em Nonperturbative functional renormalization group for random
  field models and related disordered systems. {III. Superfield} formalism and
  ground-state dominance},
\newblock \doi{10.1103/PhysRevB.85.104202}{\rm Phys. Rev. B {\bf 85} (2012)
  104202}\null.

\bibitem{TissierTarjus2012b}
M.~Tissier and G.~Tarjus,
\newblock {\em Nonperturbative functional renormalization group for random
  field models and related disordered systems. {IV. Supersymmetry} and its
  spontaneous breaking},
\newblock \doi{10.1103/PhysRevB.85.104203}{\rm Phys. Rev. B {\bf 85} (2012)
  104203}\null.

\bibitem{BaczykTarjusTissierBalog2014}
M.~Baczyk, G.~Tarjus, M.~Tissier  and I.~Balog,
\newblock {\em Fixed points and their stability in the functional
  renormalization group of random field models},
\newblock \doi{10.1088/1742-5468/2014/06/p06010}{\rm J. Stat. Mech. {\bf 2014}
  (2014)   P06010}\null.

\bibitem{LeDoussalWiese2006b}
P.~{Le Doussal} and K.J. Wiese,
\newblock {\em Stability of random-field and random-anisotropy fixed points at
  large {$N$}},
\newblock \doi{10.1103/PhysRevLett.98.269704}{\rm Phys. Rev. Lett. {\bf 98}
  (2007)   269704}\null,
\newblock \arxiv{cond-mat/0612310}.

\bibitem{FedorenkoKuehnel2007}
A.A. Fedorenko and F.~K\"uhnel,
\newblock {\em Long-range correlated random field and random anisotropy {O(N)}
  models: A functional renormalization group study},
\newblock \doi{10.1103/PhysRevB.75.174206}{\rm Phys. Rev. B {\bf 75} (2007)
  174206}\null,
\newblock \arxiv{cond-mat/0701256}.

\bibitem{TarjusBaczykTissier2012}
G.~{Tarjus}, M.~{Baczyk}  and M.~{Tissier},
\newblock {\em Avalanches and dimensional reduction breakdown in the critical
  behavior of disordered systems},
\newblock \doi{10.1103/PhysRevLett.110.135703}{\rm Phys. Rev. Lett. {\bf 110}
  (2013)   135703}\null,
\newblock \arxiv{arXiv:1209.3161}.

\bibitem{MouhannaTarjus2016}
D.~Mouhanna and G.~Tarjus,
\newblock {\em Phase diagram and criticality of the random anisotropy model in
  the large-$n$ limit},
\newblock \doi{10.1103/PhysRevB.94.214205}{\rm Phys. Rev. B {\bf 94} (2016)
  214205}\null.

\bibitem{FytasMartin-MayorPiccoSourlas2017}
N.G. Fytas, V.~Mart\'{\i}n-Mayor, M.~Picco  and N.~Sourlas,
\newblock {\em Restoration of dimensional reduction in the random-field {Ising}
  model at five dimensions},
\newblock \doi{10.1103/PhysRevE.95.042117}{\rm Phys. Rev. E {\bf 95} (2017)
  042117}\null.

\bibitem{FytasMartin-MayorPiccoSourlas2018}
N.G. Fytas, V.~Mart\'{\i}n-Mayor, M.~Picco  and N.~Sourlas,
\newblock {\em Review of recent developments in the random-field {Ising}
  model},
\newblock \doi{10.1007/s10955-018-1955-7}{\rm J. Stat. Phys. {\bf 172} (2018)
  665--672}\null.

\bibitem{FytasMartin-MayorParisiPiccoSourlas2019}
N.G. Fytas, V.~Mart\'{\i}n-Mayor, G.~Parisi, M.~Picco  and N.~Sourlas,
\newblock {\em Evidence for supersymmetry in the random-field {Ising} model at
  $d=5$},
\newblock \doi{10.1103/PhysRevLett.122.240603}{\rm Phys. Rev. Lett. {\bf 122}
  (2019)   240603}\null.

\bibitem{TarjusTissier2016}
G.~Tarjus and M.~Tissier,
\newblock {\em Avalanches and perturbation theory in the random-field {Ising}
  model},
\newblock \doi{10.1088/1742-5468/2016/02/023207}{\rm J. Stat. Mech. {\bf 2016}
  (2016)   023207}\null.

\bibitem{Wiese2016}
K.J. Wiese,
\newblock {\em Dynamical selection of critical exponents},
\newblock \doi{10.1103/PhysRevE.93.042105}{\rm Phys. Rev. E {\bf 93} (2016)
  042105}\null,
\newblock \arxiv{arXiv:1602.00601}.

\bibitem{BrezinHalperinLeibler1983a}
E.~Br{\'e}zin, B.I. Halperin  and S.~Leibler,
\newblock {\em Critical wetting: the domain of validity of mean field theory},
\newblock \doi{10.1051/jphys:01983004407077500}{\rm J. Phys. (France) {\bf 44}
  (1983)   775--783}\null.

\bibitem{BrezinHalperinLeibler1983}
E.~Br\'ezin, B.I. Halperin  and S.~Leibler,
\newblock {\em Critical wetting in three dimensions},
\newblock \doi{10.1103/PhysRevLett.50.1387}{\rm Phys.~Rev. Lett. {\bf 50}
  (1983)   1387}\null.

\bibitem{DSFisherHuse1985}
D.S. Fisher and D.A. Huse,
\newblock {\em Wetting transitions: A functional renormalization-group
  approach},
\newblock \doi{10.1103/PhysRevB.32.247}{\rm Phys. Rev. B {\bf 32} (1985)
  247--256}\null.

\bibitem{BrezinHalpin-Healy1983}
E.~Br{\'e}zin and T.~Halpin-Healy,
\newblock {\em Scaling functions for 3d critical wetting},
\newblock \doi{10.1051/jphys:01987004805075700}{\rm J. Phys. (France) {\bf 48}
  (1987)   757--761}\null.

\bibitem{LipowskyFisher1987}
R.~Lipowsky and M.E. Fisher,
\newblock {\em Scaling regimes and functional renormalization for wetting
  transitions},
\newblock \doi{10.1103/PhysRevB.36.2126}{\rm Phys. Rev. B {\bf 36} (1987)
  2126--2241}\null.

\bibitem{ForgasLipowskyNieuwenhuizenInDombGreen}
G.~Forgas, R.~Lipowsky  and T.M. Nieuwenhuizen,
\newblock {\em The behaviour of interfaces in ordered and disordered systems}.
\newblock {\em {\em Volume}~14} of {\em Phase Transitions and Critical
  Phenomena}, pages 136--376, Academic Press London, 1991.

\bibitem{Boltzmann1868}
L.~Boltzmann,
\newblock {\em {Studien \"uber das Gleichgewicht der lebendigen Kraft zwischen
  bewegten materiellen Punkten}},
\newblock Wiener Berichte {\bf 58} (1868)   517--560.

\bibitem{Janssen1992}
H.K. Janssen,
\newblock {\em On the renormalized field theory of nonlinear critical
  relaxation},
\newblock in {\em From Phase Transitions to Chaos}, Topics in Modern
  Statistical Physics, pages 68--117, World Scientific, Singapore, 1992.

\bibitem{Gumbel1935}
E.J. Gumbel,
\newblock {\em \link{archive.numdam.org/article/AIHP_1935__5_2_115_0.pdf}{Les
  valeurs extr\^emes des distributions statistiques}},
\newblock Ann. Inst. Henri Poincar\'e {\bf 5} (1935)   115--118.

\bibitem{Weibull1951}
W.~Weibull,
\newblock {\em
  \link{web.cecs.pdx.edu/~cgshirl/Documents/Weibull-ASME-Paper-1951.pdf}{A
  statistical distribution function of wide applicability}},
\newblock J. Appl. Mech. {\bf 18} (1951)   293--297.

\bibitem{Frechet1927}
M.~Fr\'echet,
\newblock {\em
  \link{cybra.lodz.pl/Content/6198/AnnSocPolMathe_t.VI_1927.pdf}{Sur la loi de
  probabilit\'e de l'\'ecart maximum}},
\newblock Ann. Soc. Math. Polon. {\bf 6} (1927)  ~93.

\bibitem{LevitSmilansky1977}
S.~Levit and U.~Smilansky,
\newblock {\em A theorem on infinite products of eigenvalues of
  {Sturm-Liouville} type operators},
\newblock \doi{10.2307/2041911}{\rm Proceedings of the American Mathematical
  Society {\bf 65} (1977)   299--302}\null.

\bibitem{ColemannBook}
S.~Colemann,
\newblock \doi{10.1017/CBO9780511565045.006}{\rm {\em Aspects of
  Symmetry}}\null,
\newblock Cambridge University Press, 1985.

\end{thebibliography}

\end{document}